\documentclass[twocolumn]{aastex701}

\usepackage{blindtext}
\usepackage[colorinlistoftodos]{todonotes}
\usepackage{tabularx}
\usepackage{caption}
\usepackage{lineno}
\usepackage{amsmath}
\usepackage{MnSymbol}
\usepackage{graphicx}
\usepackage{appendix}
\usepackage{multirow}
\usepackage{url}
\usepackage{gensymb}
\usepackage{wasysym}
\usepackage{natbib}
\usepackage{float}
\usepackage{rotating}

\begin{document}

\title{Powerful Radio Sources in the Southern Sky. IV. Observations of the G4Jy-3CRE Catalog with the Australian Square Kilometre Array Pathfinder}

\author[0009-0002-8146-6533]{Siegfried A.\ Gawenda}
\affiliation{Department of Physics and Astronomy, The University of Texas, R\'io Grande Valley, Brownsville, TX 78520, USA}
\email{siegfried.gawenda01@utrgv.edu}

\author{Juan P.\ Madrid}
\affiliation{Department of Physics and Astronomy, The University of Texas, R\'io Grande Valley, Brownsville, TX 78520, USA}
\affiliation{Astronomy Division, National Science Foundation, 2415 Eisenhower Avenue, Alexandria, VA 22314, USA}
\email{juan.madrid@utrgv.edu}

\author[0000-0002-1704-9850]{Francesco Massaro}  
\affiliation{Dipartimento di Fisica, Universit\`a degli Studi di Torino, via Pietro Giuria 1, I-10125 Torino, Italy}
\affiliation{INFN-Istituto Nazionale di Fisica Nucleare, Sezione di Torino, I-10125 Torino, Italy.}
\affiliation{INAF-Osservatorio Astrofisico di Torino, via Osservatorio 20, I-10025 Pino Torinese, Italy}
\affiliation{Consorzio Interuniversitario per la Fisica Spaziale, via Pietro Giuria 1, I-10125 Torino, Italy}
\email{fmassaro79@gmail.com}

\author[0000-0002-2340-8303]{Sarah V. White}
\affiliation{South African Astronomical Observatory (SAAO), PO Box 9, Observatory, 7935, South Africa}
\email{sarahwhite.astro@gmail.com}

\author[0000-0002-4377-0174]{C.C. Cheung}
\affiliation{Space Science Division, Naval Research Laboratory, Washington, DC 20375, USA}
\email{teddysunshine@gmail.com}

\author[0000-0002-5941-5214]{Chiara Mazzucchelli}
\affiliation{Instituto de Estudios Astrofísicos, Facultad de Ingeniería y Ciencias, Universidad Diego Portales, Avenida Ejercito Libertador 441, Santiago, Chile}
\email{chiara.mazzucchelli@mail.udp.cl}

\author[0000-0002-9896-6430]{Abigail García-Pérez}
\affiliation{Instituto Nacional de Astrofísica, Óptica y Electrónica, Luis Enrique Erro 1, Tonantzintla, Puebla 72840, M\'exico}
\affiliation{Dipartimento di Fisica, Universit\`a degli Studi di Torino, via Pietro Giuria 1, I-10125 Torino, Italy}
\affiliation{Center for Astrophysics | Harvard \& Smithsonian, 60 Garden Street, Cambridge, MA 02138, USA}
\email{a96garciapz@gmail.com}

\author[0000-0003-1562-5188]{I. Andruchow}
\affiliation{Instituto Argentino de Radioastronomía, CONICET-CICPBA-UNLP, CC5 (1897) Villa Elisa, Provincia de Buenos Aires, Argentina}
\affiliation{Facultad de Ciencias Astronómicas y Geofísicas, Universidad Nacional de La Plata, Paseo del Bosque, B1900FWA La Plata, Argentina}
\email{ileandru@gmail.com}

\author[0000-0002-2558-0967]{Vahram Chavushyan}
\affiliation{Instituto Nacional de Astrofísica, Óptica y Electrónica, Luis Enrique Erro 1, Tonantzintla, Puebla 72840, Mexico}
\email{vahram@inaoep.mx}

\author[0000-0002-0765-0511]{Ralph Kraft}
\affiliation{Center for Astrophysics | Harvard \& Smithsonian, 60 Garden Street, Cambridge, MA 02138, USA}
\email{rkraft@cfa.harvard.edu}

\author[0000-0002-6472-6711]{Victoria Reynaldi}
\affiliation{Facultad de Ciencias Astronómicas y Geof\'isicas, Universidad Nacional de La Plata, Paseo del Bosque, B1900FWA La Plata, Argentina}
\affiliation{Instituto de Astrofísica de La Plata (CCT La Plata-CONICET-UNLP), La Plata, Argentina}
\email{victoria.reynaldi@gmail.com}

\author[0000-0003-4413-7722]{Ana Jimenez-Gallardo}
\affiliation{European Southern Observatory, Alonso de Córdova 3107, Vitacura, Región Metropolitana, Chile}
\email{anaj1610@gmail.com}

\author[0000-0003-3684-4275]{Alessandro Capetti}
\affiliation{Istituto Nazionale di Astrofisica (INAF) - Osservatorio Astrofisico di Torino, via Osservatorio 20, I-10025 Pino Torinese, Italy}
\email{alessandro.capetti@inaf.it}

\author[0000-0002-0690-0638]{Barbara Balmaverde}
\affiliation{Istituto Nazionale di Astrofisica (INAF) - Osservatorio Astrofisico di Torino, via Osservatorio 20, I-10025 Pino Torinese, Italy}
\email{barbara.balmaverde@inaf.it}

\author[0000-0002-9478-1682]{William R. Forman}
\affiliation{Center for Astrophysics | Harvard \& Smithsonian, 60 Garden Street, Cambridge, MA 02138, USA}
\email{wforman@cfa.harvard.edu}

\author[0000-0003-0032-9538]{H. A. Peña-Herazo}
\affiliation{East Asian Observatory (EAO), 660 North A’ohōkū Place, Hilo, HI 96720, USA}
\email{harold.penah@gmail.com}

\author[0000-0001-5783-6544]{Nicole Nesvadba}
\affiliation{Université de la Côte d’Azur, Observatoire de la Côte d’Azur, CNRS, Laboratoire Lagrange, Boulevard de l’Observatoire, CS 34229, 06304 Nice CEDEX 4, France}
\email{nicole.nesvadba@oca.eu}

\author[0000-0002-3866-2726]{Sergio A. Cellone}
\affiliation{Facultad de Ciencias Astronómicas y Geofísicas, Universidad Nacional de La Plata, Paseo del Bosque, B1900FWA La Plata, Argentina}
\affiliation{Complejo Astronómico El Leoncito (CASLEO), CONICET-UNLP-UNC-UNSJ, Avenida España 1512 (sur), J5402DSP San Juan, Argentina}
\email{scellone@casleo.gov.ar}

\author[0000-0003-3471-7459]{Romana Grossová}
\affiliation{Department of Theoretical Physics and Astrophysics, Faculty of Science, Masaryk University, Kotlářská 2, Brno, CZ-611 37, Czech Republic}
\affiliation{Astronomical Institute of the Czech Academy of Sciences, Bocní II 1401, 141 00, Prague, Czech Republic}
\email{romana.grossova@gmail.com}

\author[0000-0002-5646-2410]{Alessandro Paggi}
\affiliation{Dipartimento di Fisica, Università degli Studi di Torino, via Pietro Giuria 1, I-10125 Torino, Italy}
\affiliation{Istituto Nazionale di Astrofisica (INAF) - Osservatorio Astrofisico di Torino, via Osservatorio 20, I-10025 Pino Torinese, Italy}
\affiliation{Istituto Nazionale di Fisica Nucleare (INFN) - Sezione di Torino, via Pietro Giuria 1, I-10125 Torino, Italy}
\email{alessandro.paggi@gmail.com}

\author[0000-0002-3140-4070]{Eleonora Sani}
\affiliation{European Southern Observatory, Alonso de Córdova 3107, Vitacura, Región Metropolitana, Chile}
\email{esani@eso.org}

\author{C. Leto}
\affiliation{ASI—Agenzia Spaziale Italiana, Via del Politecnico snc, I-00133 Roma, Italy}
\email{cristina.leto@asi.it}

                         
\begin{abstract}
A recent 2023 paper by Massaro et al.\ introduced the G4Jy-3CRE, a new catalog of the brightest radio sources in the southern hemisphere that serve as a southern equivalent to the Third Cambridge Catalog Revised (3CR). The G4Jy-3CRE catalog selected 264 sources from the GLEAM-4Jy survey based on the same criteria used to select the sources in the 3CR. In this paper, we present new Australian Square Kilometre Array Pathfinder (ASKAP) continuuum imaging of the  G4Jy-3CRE catalog. We use the three most recent data releases from the Rapid ASKAP Continuum Survey (RACS), covering the sky south of +30\degree decl.: RACS-low1, RACS-mid, and RACS-high. Together, these data releases cover a range of frequencies from 600 to 1800 MHz. The RACS surveys have improved spatial resolution and sensitivity over archival surveys at the same frequency, enabling us to classify 173 sources (66\% of the sample) with morphologies indicative of the presence of jets, 37 of which did not show jet activity on archival radio maps. We were able to effectively classify FRI/FRII galaxies up to a redshift of $z = 1.35$. Moreover, we identified six optical counterparts for sources that were either previously unidentified or ambiguous. 
\end{abstract}

\keywords{Radio active galactic nuclei (2134); Radio jets (1347); AGN host
galaxies (2017); Fanaroff-Riley radio galaxies (526); Radio loud quasars (1349); Sky surveys (1464)Radio active galactic nuclei (2134); Radio jets (1347); AGN host galaxies (2017); Fanaroff-Riley radio galaxies (526); Radio loud quasars (1349); Sky surveys (1464)}


\section{Introduction}
Our understanding of AGNs and other extragalactic radio sources has been shaped by surveys conducted from the northern hemisphere. The most influential of which was the Third Cambridge Catalog (3C; \citealt{3C,Edge1959}) and its revised edition (3CR; \citealt{3CR, Bennet1962}). The 3C contains observations of radio sources above a decl. of $\delta = -22 \degree $ at 159 MHz and the 3CR contains observations of radio sources above 9 Jy north of a decl. of $\delta = -5 \degree $ at 178 MHz, produced using the Cambridge Interferometer in 1959 \citep{3C} and 1962 \citep{3CR} for each frequency. 

While 3C was instrumental in shaping our current understanding of extragalactic sources, the many observation campaigns of 3C across different frequencies were subject to fundamental limitations.

First, the 3CR is over 60 yr old and it took decades of work to achieve the level of multifrequency coverage of 3CR sources we now have. Many of these observations were made using relatively old instruments and have rarely been repeated or updated. 

Second, 3CR is limited to sources visible in the southern hemisphere, meaning many of the sources are outside the range of new and upcoming state-of-the-art radio facilities like VLT, ELT, or ALMA. Even what will be the world's most powerful radio telescope, the planned Square Kilometer Array (SKA; \cite{SKA2009}), will only be able to observe a limited number of 3CR sources. 

Lastly, the vast majority of 3CR sources ($\sim$80\%), lie at redshifts $z<1$ and less than 10\% at $z>1.5$ \citep{3CR}, limiting our ability to study the cosmological evolution of these sources. Thus, a larger sample will be necessary to effectively study these powerful radio sources lying at moderate redshifts, and may give us better insight into extragalactic sources at different stages of their development.

Given these limitations, there is a need for new observations across a wide variety of frequencies, taking advantage of new state-of-the-art telescopes in the southern hemisphere. Additionally, a  multi-frequency approach will always be necessary to effectively study radio galaxies, their environments, and the mechanisms that govern their evolution. The 3CR had several large follow-up surveys conducted across different frequencies: The HST Snapshot Survey of 3CR Radio Source Counterparts\footnote{https://archive.stsci.edu/prepds/3cr/} \citep{Madrid2006, Privon2008}, the 3CR \textit{Chandra} Snapshot Survey \citep{Massaro2010,Massaro2012,Massaro2015,Stuardi2018,Jimenez-Gallardo2020}, and the MUse RAdio Loud Emission line Snapshot survey \citep{Balmaverde2018,Balmaverde2019,Balmaverde2021}. A similarly comprehensive set of surveys will be essential for facilitating future investigations into southern hemisphere radio sources.

Historically, AGN survey science targeting southern hemisphere sources has drawn from one of a few catalogs of radio sources. One of the largest of these catalogs is the Molonglo Reference catalog (MRC; \citealp{Large1981}), containing nearly 12,000 discrete sources with flux densities greater than 0.7 Jy at 408 MHz in the decl. range between +18.5$^{\circ}$ and -85$^{\circ}$ and excluding regions within 3$^{\circ}$ of the Galactic equator. There have been multiple attempts at constructing a complete sample similar to 3CR (or one of its later editions) from the MRC, most notably the 5 Jy sample and the MS4 sample created by \cite{Best1999} and \cite{Burgess_2006} respectively. Though comprehensive, these samples use relatively low resolution MRC data. At the time, the Molonglo Cross Telescope was operating at an angular resolution of $2.62' \times 2.86'$ \citep{Large1981}. In addition, the 5 Jy sample was limited in its coverage, using only sources lying between $-30 \degree \leq \delta \leq 10 \degree$ \citep{Best1999}. With the establishment of new, more advanced radio telescopes in the southern hemisphere like the Australian Square Kilometre Array (ASKAP; \citealp{ASKAP}), there is now ample opportunity to expand the database of bright radio sources to those visible from the southern hemisphere in a wide range of frequencies and at much higher resolutions.

This paper presents new ASKAP data of the  G4Jy-3CRE catalog recently published by \cite{G4Jy-3CRE_Catalogue}. The G4Jy-3CRE is a new catalog of powerful radio sources visible from the southern hemisphere using observations from the Galactic and Extragalactic All-Sky Murchison Widefield Array (MWA; \cite{MWA}) survey (GLEAM; \cite{GLEAM}), obtained from Western Australia and conducted in frequencies between 72 and 231 MHz. This catalog builds on the work of \cite{White2020A,White2020B}, who presented the GLEAM 4Jy sample (G4Jy) in 2020. The  G4Jy-3CRE catalog is the GLEAM 4 Jy equivalent of the Third Cambridge catalog.   G4Jy-3CRE lists the 264 brightest radio sources in the GLEAM 4 Jy survey with the same selection criteria as the 3CR catalog ($S_{178}~ \geq $ 9 Jy), focused on sources from the southern hemisphere visible below a decl. of $\delta = -5 \degree $. \cite{G4Jy-3CRE_Catalogue} presents archival radio contours at various frequencies and compares them with images from different state-of-the-art optical surveys to identify optical counterparts corresponding to each radio source as well as redshift estimates. At the time of publishing, 226 of the 264 sources have an identified optical counterpart and 207 of the 264 have a firm redshift estimate, 168 of which have an existing optical spectroscopic campaign that allowed classification of the source (García-Pérez et al.\ 2026 submitted; private\ comm.). The archival radio maps used by  \citet{G4Jy-3CRE_Catalogue} come from higher angular resolution observations than what GLEAM provides, such as those retrieved from the Very Large Array (VLA) Sky Survey (VLASS; \citealp{Lacy2020}) conducted at 3 GHz, the National Radio Astronomy Observatory (NRAO) VLA Archive Sky Survey (NVSS; \citealp{Condon1998_NVSS}) conducted at 1.4 GHz, the Sydney University Molonglo Sky Survey (SUMSS; \citealt{SUMSS_Mauch2003}), and the TIFR GMRT Sky Survey (TGSS; \citealt{TGSS_Intema2017}). The majority of sources belonging to the G4Jy-3CRE catalog are either not visible or poorly suited for follow-ups with northern hemisphere
telescopes like the VLA. 

With the recent data releases of the Rapid ASKAP Continuum Survey (RACS; \citealp{RACS_1}) conducted in the southern sky, we have the opportunity to study the G4Jy-3CRE catalog with data at additional frequencies and better spatial resolution than archival surveys can provide. RACS covers three frequency ranges, all of which have been made available across three data releases: 887.5 MHz, 1367.5 MHz, and 1655.5 MHz. These data releases hereon will be referred to as RACS-low \citep{RACS-low}, RACS-mid \citep{RACS-mid}, and RACS-high \citep{RACS_High} respectively. The RACS survey was conducted using a single ASKAP antenna configuration, designed to maximize sensitivity for extragalactic HI surveys with additional elements added to improve surface brightness sensitivity and spatial resolution \citep{RACS_1,Gupta2008}. In the following sections, we present a catalog of the G4Jy-3CRE sources using data from RACS-low, RACS-mid, and RACS-high. 

The construction of this G4Jy-3CRE ASKAP catalog allows the identification of previously undiscovered AGN jets as well as facilitate deeper analyses into evolution and morphology of AGN jets across multiple frequencies as they interact with the intergalactic medium (IGM). Additionally, using new and archival radio observations across a range of frequencies, we can more accurately classify each radio source based on morphology as well as define morphological AGN properties such as jet length. The availability of all three RACS data sets will also allow us to classify these sources uniformly. We can then measure how the intrinsic properties of the AGN relate to these morphological properties and classifications of each source. We are especially interested in classifying FRI and FRII galaxies (defined in Table \ref{classification_def}). FRI/FRII galaxies are associated with many open questions on jet evolution and the conditions under which FRI/FRII jets develop. Furthermore, as stated by \cite{G4Jy-3CRE_Catalogue}, collecting observations for these brightest sources across multiple frequencies will facilitate future investigations into individual sources by larger telescopes like the planned Square Kilometre Array (SKA; \citealt{SKA2009}) for which ASKAP is a precursor. 

The paper is structured as follows. The Section 2 will cover each of the three RACS data releases, discussing telescope properties and other relevant details from each survey. In addition, we also discuss the selection of optical images. The second section will discuss the construction of the radio contours, improvements over archival radio maps, and the identification of optical counterparts. Section 3 will also discuss our process for characterizing jet properties and classifying sources based on the radio morphologies. We then finish off with a discussion of the catalog population and a comparison with past surveys that do similar work. The final section summarizes our findings and detail any conclusions that could be drawn from the results. The full suite of radio and optical images in the catalog is presented in Appendix A as a Figure set (available in the online version of the paper) in Figure 6. Appendix B contains a discussion of notable sources within our sample, and lastly, provides a sample of the master table (see Table \ref{MT}).


\section{Observations}

ASKAP, located at Inyarrimanha Ilgar Bundara, the CSIRO Murchison Radio-astronomy Observatory in Western Australia, is a survey radio telescope designed to conduct quick, all-sky surveys \citep{ASKAP}. 
It has an operational frequency range of 700 - 1800 MHz and an instantaneous bandwidth restricted to a contiguous 288 MHz band within this frequency range \citep{RACS_1}. It has the ability to cover a total of 31 deg$^2$ in a single pointing with a sensitivity of ~0.35 mJy beam$^{-1}$. 

RACS is split into three observation campaigns across three frequency bands ranging between 700 and 1800 MHz: RACS-low, RACS-mid, and RACS-high. Each of the three observation campaigns covers 90\% of the sky south of $\delta = + 41 \deg$ over approximately two weeks of total integration time. Thus far, data from RACS-low, RACS-mid, and RACS-high has been made available on the CSIRO ASKAP Science DATA Archive \footnote{https://data.csiro.au/domain/casdaCutoutService} (CASDA; \cite{Chapman2017, Huynh2020}). Two more observation campaigns were completed after RACS-high: an additional epoch at 887.5 MHz (RACS-low2) and an epoch at 943.5 Mhz (RACS-low3; \cite{RACS_High}). Data from these epochs will be released at a later date and may serve to supplement this catalog.

\begin{table*}[t]

    \centering
    
    \resizebox{\linewidth}{!}{%
    \begin{tabular}{ccccccc}
        \hline 
          & Frequency & Bandwidth & Median $\sigma_{rms}$ & Median PSF & Dec Limit & \\
         Epoch name & (MHz) & (MHz) & ($\mu$Jy $PSF^{-1}$) & ($\arcsec \times \arcsec$) & ( $\degree$ ) & References \\
         \hline
         RACS-low & 887.5 & 288 & 266 & $18.4 \times 11.6$ & [-80, +30] & \cite{RACS_1}\\
         RACS-mid & 1367.5 & 288 & 198 & $10.1 \times 8.1$ & +49 & \cite{RACS-mid}\\
         RACS-high &  1655.5 & 288 & 209 & $8.6 \times 6.3$ & +49 & \cite{RACS_High}\\
         RACS-low2&  887.5 & 288 & - & - & - & -\\
         RACS-low3 & 943.5 & 288 & 205 & $13.4 \times 11.0$ & - & - \\
         \hline
    \end{tabular}%
    }

    \caption{The completed and planned observation campaigns for the RACS survey. The data for each survey can be accessed using the CSIRO data access portal: https://data.csiro.au/domain/casdaObservation}
    \label{RACS_prop}

\end{table*}

\subsection{RACS-low}
The RACS-low survey covers approximately 2,123,638 radio sources between the decl.s $-80\degree \leq \delta \leq +30\degree$ \citep{RACS-low} and was centered at 887.5 MHz with a 288 MHz bandwidth and a median PSF of $18.4\arcsec \times 11.6\arcsec$ \citep{RACS_High}.  RACS-low has both variable resolution images up to approximately $15 \arcsec$ and common resolution images convolved to $25 \arcsec$. These observations were made over a 12 day period in April and May 2019. Subsequent reobservations of selected fields were conducted in August-November 2019 and March-June 2020 in order to reduce the PSF variation within each individual tile. These images have better resolution than those in the GLEAM survey (approximately 15$\arcsec$ compared to 120$\arcsec$) and ASKAP observations were conducted at an improved sensitivity (0.25-0.3 mJy beam$^{-1}$ compared to a 3.5 mJy beam$^{-1}$ for GLEAM, \citet{RACS_1}). It should be noted that the GLEAM survey was conducted at a lower frequency than RACS-low (72-231 MHz vs. 887.5 MHz; \citealp{GLEAM,RACS-low}).

\subsection{RACS-mid}
The RACS-mid survey, centered at 1367.5 MHz, was completed in 2021 with follow-up observations taking place over 2021-2022 \citep{RACS-mid}. This survey covered the sky south of decl. $\delta = +49 \deg$ and has a median angular resolution of $11.2\arcsec \times 9.3\arcsec$. The primary RACS-mid catalog contains 3,105,668 radio sources, 2,861,923 of which are beyond the galactic plane $|b|>5\deg$ \citep{RACS_High}.

\subsection{RACS-high}
The RACS-high survey, centered at 1655.5 MHz, was completed in 2022. This survey covered the sky south of decl. $\delta = +49 \deg$ and has a median angular resolution of $11.8\arcsec \times 8.1\arcsec$. RACS-high contains 2,677,509 sources \citep{RACS_High}. A summary of the RACS survey properties can be found in Table 1. \\

\subsection{Optical Images}
The images presented in this paper consist of radio contours from the RACS overlaid on $r$-band optical images from the best available optical survey: Dark Energy Survey (DES\footnote{\url{https://datalab.noirlab.edu/data-explorer}}; \citealp{DES2018}), the Digitized Sky Survey II (DSS2\footnote{\url{https://irsa.ipac.caltech.edu/data/DSS/index_cutouts.html}}; \citealp{DSS}), and the Panoramic Survey Telescope $\&$ Rapid Response System database (PanSTARRS\footnote{\url{https://outerspace.stsci.edu/display/PANSTARRS/PS1+Image+Cutout+Service\#PS1ImageCutoutService-Thewebinterface}}; \citealp{PanSTARRS2020}). These optical images were obtained using the image cutout services provided on each survey's respective website or through a provided Jupyter notebook.

\section{Analysis and Results}

\subsection{Cross-matching datasets}

We began by cross-matching G4Jy-3CRE with each of the three data releases of RACS. As of writing this paper, RACS-low data is unavailable for five sources that fall below the RACS-low decl. limit of $-80 \degree$: G4Jy 350, G4Jy 718, G4Jy 1284, G4Jy 1411, and G4Jy 1723. Consequently, RACS-low data for these sources will not be included in the catalog. Additionally, several sources needed their RACS-low flux values recalculated, as the flux values provided in the RACS-low catalogue did not encapsulate all the gaussian components of the entire source. The affected sources were G4Jy 133, G4Jy 680, G4Jy 1079, and G4Jy 1613. 

\subsection{Radio Contours}

The radio contours for each source were initially generated on a logarithmic scale ranging from approximately 10 $\times$ the rms of the background as defined using the photutils python package (these values are listed in the master table for each frequency), to 6 counts/pixel. We extracted the median background rms from each radio image using a sample size of [100,100] pixels and a sigma clip value of 5 to define the background rms for each image. In some cases the lowest one or two contours pick up a portion of the background signal. In those cases, we did not consider these contour levels when measuring the jet length or largest angular size for each source, instead using the next highest contour level without background signal.

\subsection{Improvement Over Archival Maps}
\begin{figure*}[ht]
    \centering
    \includegraphics[width=0.49\linewidth]{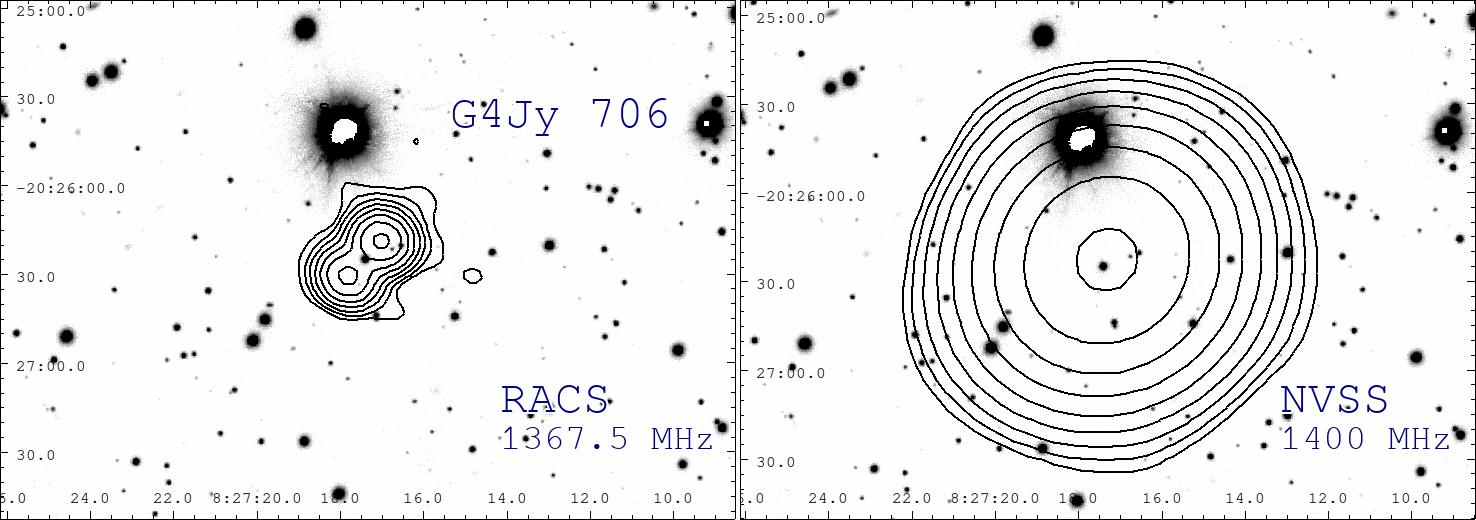}
    \includegraphics[width=0.49\linewidth]{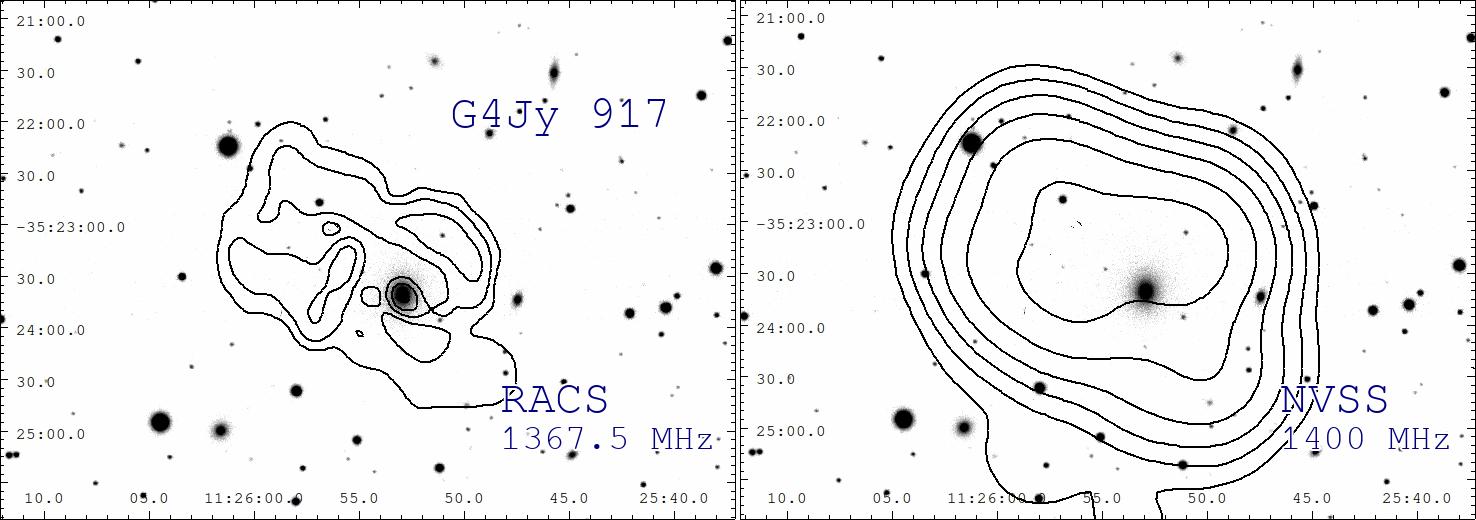}
    \includegraphics[width=0.49\linewidth]{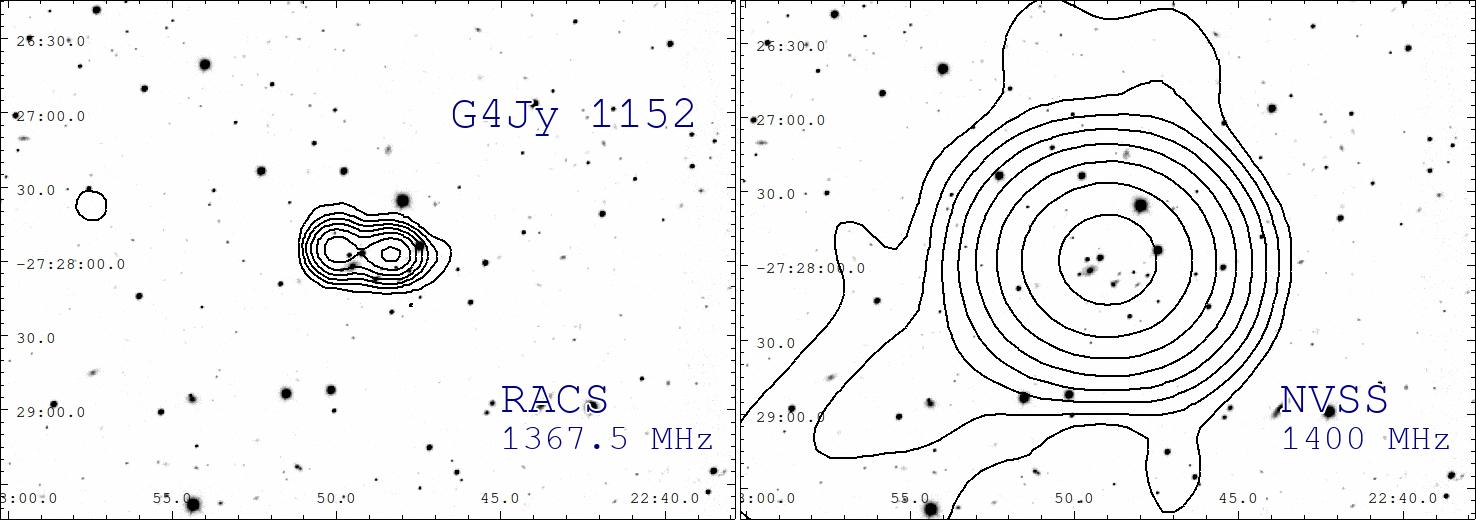}
    \includegraphics[width=0.49\linewidth]{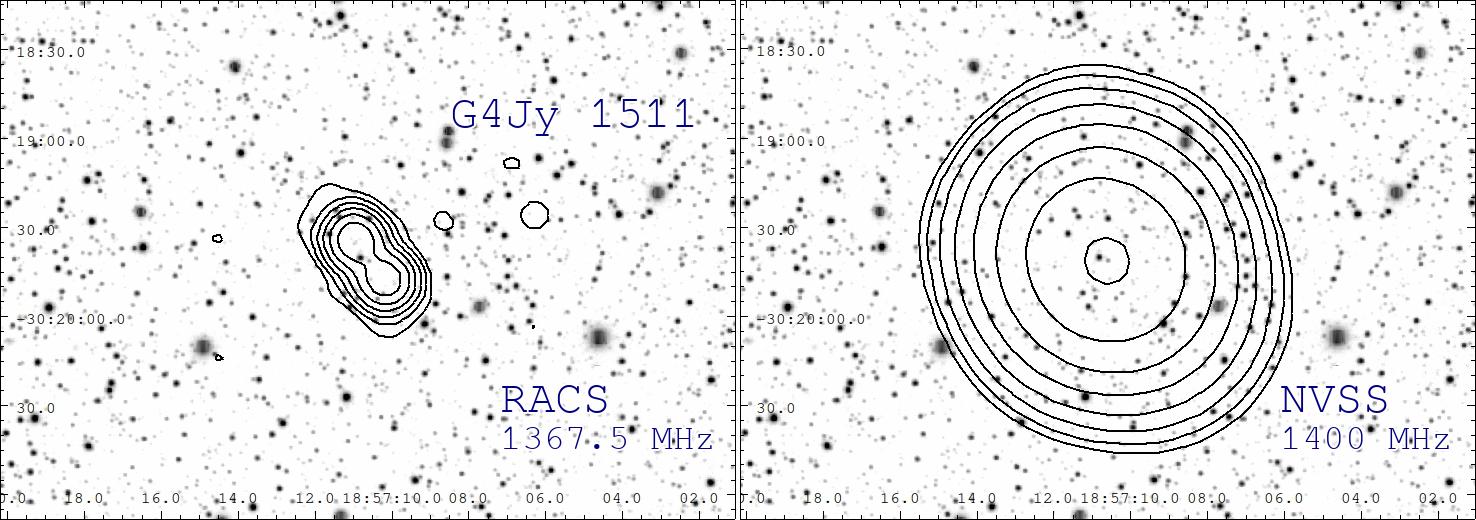}
    \includegraphics[width=0.49\linewidth]{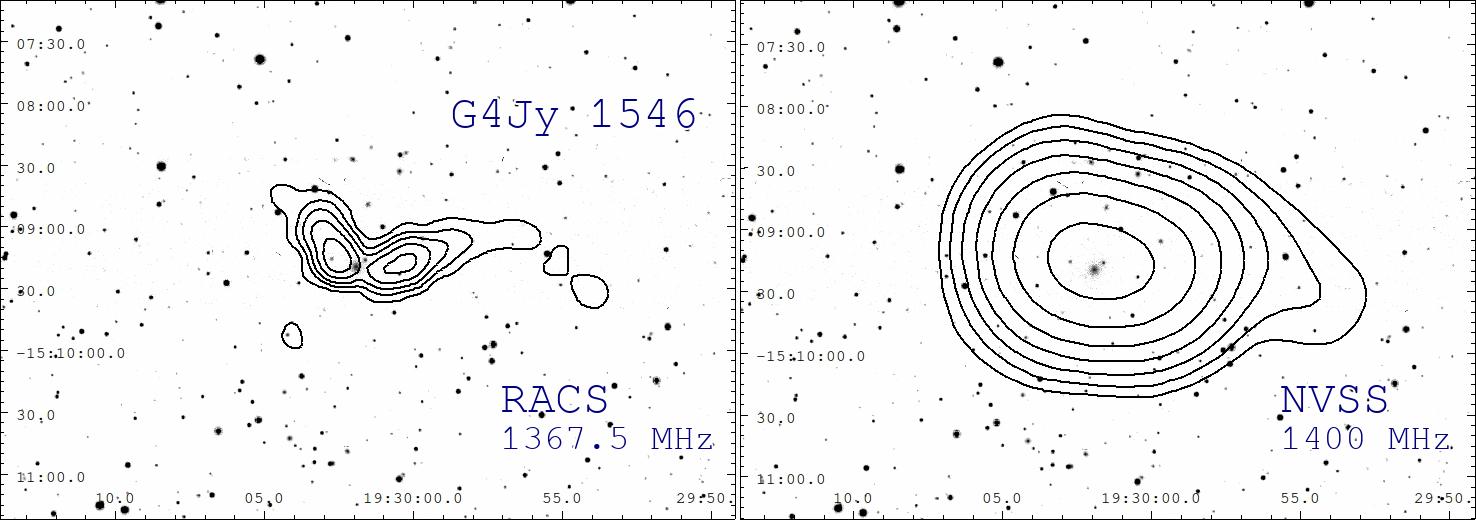}
    \includegraphics[width=0.49\linewidth]{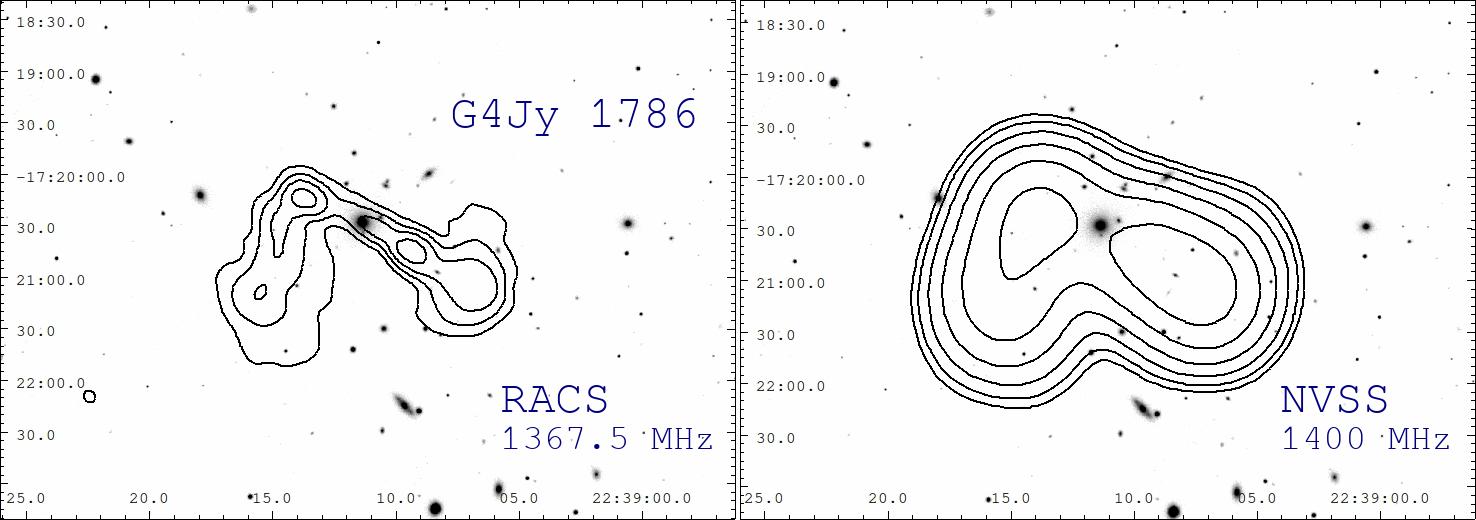}
    \caption{A comparison of radio contours from NVSS (1400 MHz) and RACS-mid (1367.5 MHz). NVSS has a spatial resolution of approximately $45 \arcsec$ and RACS-mid has a spatial resolution of $10.1 \arcsec \times 8.1 \arcsec$ \citep{Condon1998_NVSS, RACS-mid}. The NVSS Radio contours were defined similarly to the other images in the catalog, with the lowest contour level being approximately $5 \times$ the background rms and the highest contour level set to a value of 6 counts/beam. These examples show the marked improvement in resolution of the new ASKAP observations over the archival data.}
    \label{racs-sumss}
\end{figure*}

RACS operated at a spatial resolution ranging from $18.4\arcsec \times 11.6 \arcsec$ for RACS-low to $8.6 \arcsec \times 6.3 \arcsec$ for RACS-high (see Table \ref{RACS_prop}). Comparatively, archival radio surveys like the NRAO VLA Sky Survey (NVSS; \citealt{Condon1998_NVSS}) and SUMSS \citep{SUMSS_Mauch2003} both had spatial resolutions of approximately $45 \arcsec$. TGSS has an approximate spatial resolution of $25 \arcsec \times 25 \arcsec$ \citep{TGSS_Intema2017}. The RACS survey offers a significant improvement in spatial resolution over NVSS, SUMSS, and TGSS. This allows us to see detail in the radio structure more clearly. Figure \ref{racs-sumss} shows a comparison of contours from RACS-mid and NVSS (1367.5 MHz and 1400 MHz respectively). The six sources shown in this figure demonstrate the clear improvement in resolution of the ASKAP data with respect to archival radio maps. For 40 sources, among them those shown in Fig. 1, the ASKAP data allows the proper identification of their true morphology as these sources show clear evidence of double lobes and linear jets, unidentified in archival NVSS data.

\subsection{Optical Counterpart Identification}
Thanks to the improved resolution and detail of the RACS radio contours, we were able to identify/disambiguate potential optical counterparts for six radio sources. These sources had either no associated counterpart or an ambiguous counterpart \citep{White2026A,White2026B}. Our primary point of comparison was the work done by \cite{G4Jy-3CRE_Catalogue} and \cite{White2020B} and the optical counterparts identified therein. These six radio sources (pictured in Fig. \ref{counterparts}) are: G4Jy 168, G4Jy 182, G4Jy 453, G4Jy 854, G4Jy 1302, and G4Jy 1513. G4Jy 168, G4Jy 182, and G4Jy 1513 are all newly identified counterparts. G4Jy 1302's counterpart was previously identified, though it was not subject to follow-up observation. With the RACS data, we are able to confirm the identification of the counterpart. Information on G4Jy 453 and G4Jy 854 has been passed on to our collaboration and optical spectra has been collected for further analysis (Garcia-Perez et al. 2026 submitted, private\ comm.\ ). Details on the coordinates of the newly identified counterparts can be found in Table \ref{counterparts}.

\begin{table}[h]
    \begin{tabular*}{\linewidth}{l@{\extracolsep{\fill}}ccr}
        \hline
         Source Name & RA & Dec & Survey\\
         \hline
         G4Jy 168 & 01:31:36.358 & $-7:03:58.11$ & DES\\
         G4Jy 182 & 01:41:55.216 & $-69:41:36.32$ & DSS2\\
         G4Jy 453 & 04:23:57.388 & $-72:46:02.05$ & DES\\
         G4Jy 854 & 10:33:13.149 & $-34:18:45.44$ & DSS2\\
         G4Jy 1302 & 16:05:13.052 & $-28:59:14.88$ & PanSTARRS\\
         G4Jy 1513 & 19:02:49.268 & $-23:29:51.25$ & PanSTARRS\\
         \hline
    \end{tabular*}
    \caption{Sources with newly identified optical counterparts and their associated J2000 coordinates.  All optical images are in the $r$-band.}
    \label{counterparts}
\end{table}

\begin{figure*}
    \centering
    \includegraphics[width=0.32\linewidth]{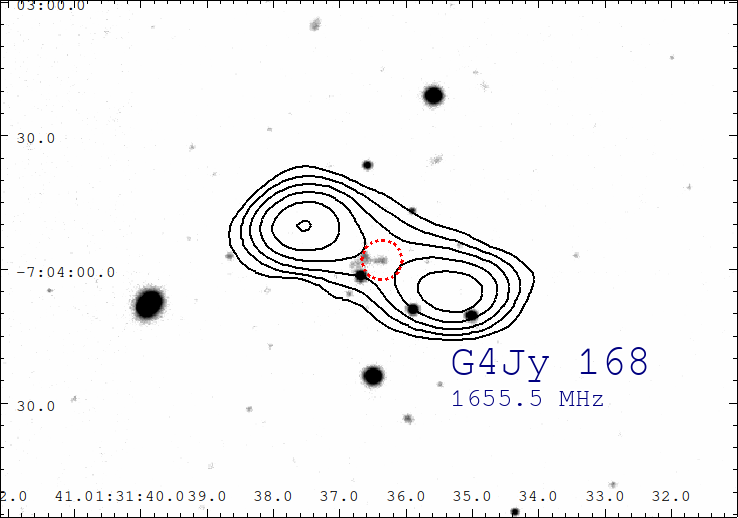}
    \includegraphics[width=0.32\linewidth]{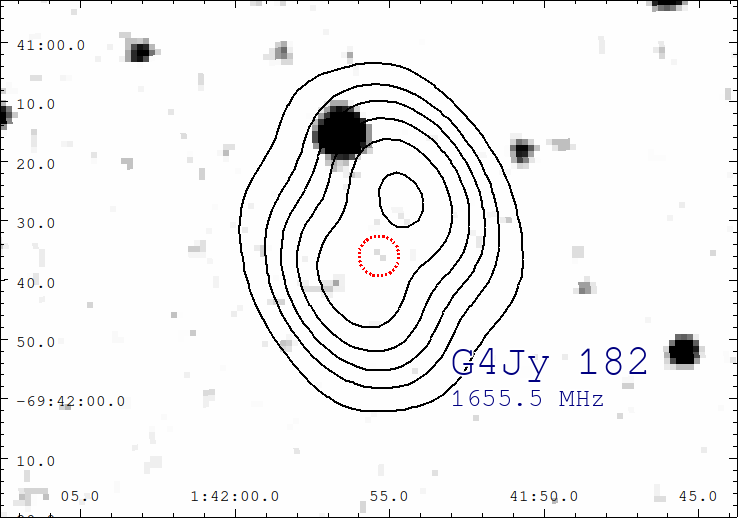}
    \includegraphics[width=0.32\linewidth]{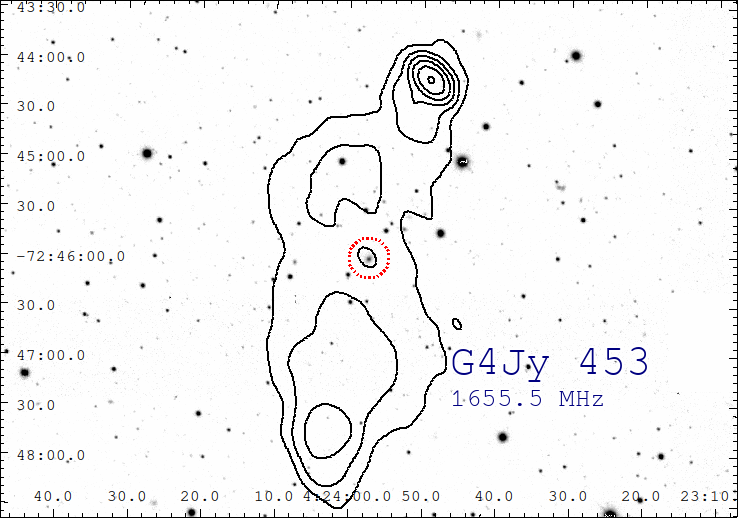}
    \includegraphics[width=0.32\linewidth]{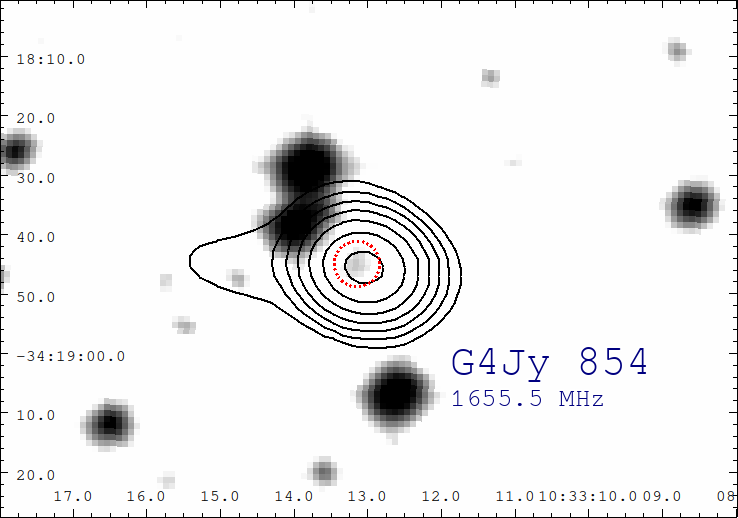}
    \includegraphics[width=0.32\linewidth]{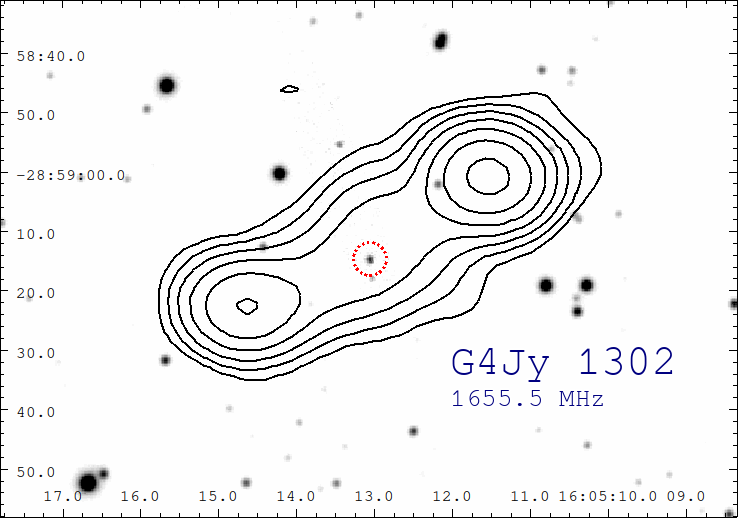}
    \includegraphics[width=0.32\linewidth]{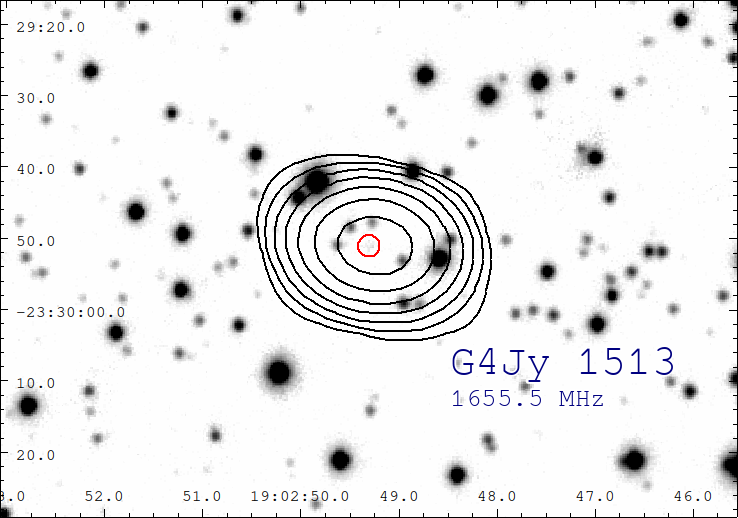}
    
    \caption{The six newly identified optical counterparts are marked with a dashed circle in each image.}
    \label{optical}
\end{figure*}

\subsection{Measuring LAS and Jet Length}
We defined the largest angular size (LAS) ($b$) of each source as the diameter of the circle around the AGN's lowest contour at 1655.5 MHz (RACS-high) in arcseconds.

We define the jet length ($a$) for sources with a FR morphology \citep{Fanaroff_Riley} as the distance in arcseconds between the brightest pixels at 1655.5 MHz (RACS-high) on opposite side of the central host galaxy. We applied this process only to those FR sources with minimal jet bending (jets with an angle of more than $170 \degree$ between them \citep{Norris2025}. 

We chose to use RACS-high for both the LAS and jet length measurements because it contains the highest resolution data of the three frequencies and to maintain consistency in our classifications and with other papers that only analyzed sources in a single frequency \citep{Norris2025, Mingo2019}. We wanted to avoid any possible biases introduced by mixing measurements from images of different frequencies. The only exceptions are G4Jy 77, G4Jy 517, G4Jy 1605, and G4Jy 1613 whose LAS and jet length were measured in RACS-low due to insufficient signal in higher frequencies. It is worth noting that G4Jy 77 is the well known radio phoenix associated with the \textbf{Abell 85 galaxy cluster \citep{G4Jy-3CRe_II, Bagchi1998,Kempner2004,Ichinohe2015, Rahman2022}}, and G4Jy 1605 is the radio relic associated with the Abell 3667 galaxy cluster \citep{G4Jy-3CRe_II, Owen1976, Johnston-Hollitt2008}.

We then used the redshifts provided in \cite{G4Jy-3CRE_Catalogue, G4Jy-3CRe_III, White2025} where available to derive physical sizes and jet lengths from angular measurements using Ned Wright's cosmology calculator. We adopt a flat general cosmology with $H_0 = 69.6 $, $\Omega_M = 0.286$, and $\Omega_{vac} = 0.714$ \citep{WrightCalc}.  It should be noted that the upper limit of the redshifts for which we were able to reliably classify a radio source is $z = 1.35$ ($1 \arcsec = 8.54$ kpc). Beyond this redshift, we are unable to distinguish between FRI and FRII radio sources. Of the 264 sources, 45 did not have any redshift data and thus were excluded from this analysis.

\begin{table*}
    \begin{tabularx}{\linewidth}{l|c|l}
        \hline
        Tag & \# sources & Description \\ \hline
        FRI & 29 & Fanaroff-Riley type I\\
        FRII & 72 & Fanaroff-Riley type II\\
        FRX & 68 & An FR I or FR II type galaxy, but data is inadequate to distinguish between them\\
        HyMoRS & 8 & A radio source which is FR I on one side of the host and FR II on the other\\
        BT & 12 & Bent-tail sources. This category includes any FR-type sources with one or both jets bent \\
        & & greater than $10 \degree$ with respect to one another \\
        DD & 17 & A double-double radio source which is sometimes interpreted as a ``restarted" radio galaxy\\
        WRG & 24 & Winged radio galaxies, those galaxies with peculiar lobe morphologies that give a non-linear \\
         & & shape. These can be further broken down into Z-shaped, X-shaped, and T-shaped radio galaxies\\
        CPLX & 13 & a radio source or group of radio sources too complex or unusual to classify unambiguously \\ 
        E & 28 & A source with an elongated morphology but no clear jet structure or otherwise defining features. \\
        U & 52 & An unresolved source with no visible structure. \\\hline
    \end{tabularx}
    \caption{This classification methodology follows the same criteria as \cite{Norris2025}. Due to our use of a tag classification system, these numbers will add up to a sum greater than the total number of sources. In addition, we elected to omit classifications found in  \cite{Norris2025} that were not present in this sample.}
    \label{classification_def}

\end{table*}

\subsection{Classification}

The morphology of each G4Jy-3CRE source was determined according to the criteria defined by \cite{Norris2025} in their 2025 catalog of Double Radio AGNs (DRAGNs). Their work used data from the EMU  survey \citep{EMU2021}, a different ASKAP survey conducted at 947 MHz. Due to the similarity of the work and goals of \cite{Norris2025} to our own, their paper serves as an ideal point of comparison. We used a combination of visual inspection using all three RACS data releases and jet angular size/largest angular size comparison to classify all 264 sources. The availability of radio images across three different frequencies allows us to pick up on detail that images from one or two frequencies may not contain. The following paragraphs will summarize the classification criteria and definitions. Like \citet{Norris2025}, we employed a tag system for classification. This means each source may have multiple classifications.

-- \textit{FRI/FRII/FRX:}
Fanaroff-Riley galaxies (FR) are characterized by a central source flanked on two sides by symmetrical jets. FR galaxy characterization is further split into two more categories: FRI and FRII. FRI galaxies are characterized by diffuse jets emanating from the central core (see G4Jy 1633 for an example). FRII galaxies are characterized by long, collimated jets extending from a central source ending in a jet lobe with a radio hotspot and jet-induced shocks where the hot plasma of the jet interacts with the surrounding IGM. 

To differentiate FR I and FR II morphologies, we used the method of Norris et al. of taking the ratio of the jet length ($a$) and the Largest Angular Size ($b$) of each source with FR morphology. We use the same values for the minimum length and error as used in their paper. $b_{min} = 50 \arcsec$ and $\epsilon = 10 \arcsec$. Norris et al. chose this criteria based on the EMU pilot survey, which operated at a similar spatial resolution to RACS ($11\arcsec - 18 \arcsec$; \cite{EMU2021}). The FR definitions are as follows:
\begin{itemize}
    \item $FRI: a/(b-\epsilon)<0.5 ~AND~ b > b_{min}$
    \item $FRII: a/(b-\epsilon)>0.5 ~AND~ b > b_{min}$
    \item $FRX: a/(b+\epsilon)<0.5 < a/(b-\epsilon) ~OR~ b < b_{min}$
\end{itemize}

Where FRX sources are those with FR morphology, but the spatial resolution of the image is not enough to differentiate between FRI or FRII.

-- \textit{HyMoRS:}
A Hybrid Morphology Radio Source (HyMoRS) is radio galaxy that has FRI morphology on one side and FRII morphology on the other \cite{Gopal2000}. 

-- \textit{Bent Tail:}
A Bent Tail source (BT) is a radio source where one jet is bent off-axis from the other jet. This includes wide-angle tail (WAT) galaxies and narrow angle tail (NAT) galaxies \citep{Norris2025}. In the most extreme case in which the tails are bent to the point of being parallel or even overlapping, these galaxies are generally classified as Head Tail (HT) galaxies \citep{Schoenmakers2000}.

-- \textit{Double Double:}
A `double-double' radio galaxy consists of a pair of double radio sources with a common center. Furthermore, the two lobes of the inner radio source must have a clearly extended, edge-brightened radio morphology \citep{Schoenmakers2000}. These are often referred to as ``restarted" radio sources, as the second pair of hotspots is thought to develop after a second epoch of jet activity.

-- \textit{Winged Radio Galaxy:}
Winged radio galaxies (WRGs) are sources with lobe morphologies that have a non-linear shape. They can be further broken down into different morphological categories, X-shaped, Z-shaped, T-shaped, depending on the location of the plume. Though there is no definitive explanation for the mechanisms by which WRGs develop, one proposed explanation is the plumes of WRGs are caused by backflow of jet material from the jet shock, wherein local conditions determine the exact location of the protrusion \citep{Cotton2020, Norris2025}.

-- \textit{Complex Sources:}
A radio source or group of radio sources too complex (CPLX) or unusual to categorize in any straightforward manner. One example is G4Jy 1677 and G4Jy 1678. Their proximity to one another makes it difficult to categorize either of the sources definitively. 

-- \textit{Elongated and Unresolved:}
The Elongated and Unresolved classifications were saved for those sources with little to no defining features. Elongated sources are those sources with no defining radio morphology other than an elongation along one axis, almost like an amoeba or pill shape. This elongation is at least indicative of jet activity, however ASKAP did not have the spatial resolution to make out any further morphological detail.

Unresolved sources are similar to elongated sources in their lack of morphological detail, however there is no elongation present.

\subsection{Catalog Population}
This catalog of the brightest radio sources in the southern hemisphere, contains 787 images, each containing contours from one of the three data releases of RACS. All of these images can be found in the online figure set, with several examples shown in Figures \ref{racs-sumss} and \ref{counterparts}. The ASKAP data shows that of these 264 sources, 173 (66\%) have morphologies that clearly indicate jet activity. Of these 173 sources, 72 (42\%) are classified as FRII, making them the majority of AGNs within the sample. This is expected, since FRIs are much more populous at lower luminosities and FRIIs at higher luminosities \citep{Mingo2019, Clews2025}. FRI sources make up 29 of the 173 sources (17\%), much lower than the percentages in other archival surveys \citep{Fanaroff_Riley, Mingo2019}. The 9 Jy flux threshold on our sample has limited the number of low luminosity sources, and thus FRI AGNs that appear in our sample. Lastly, FRX sources make up 68 of the 173 sources (39\%).

\begin{figure}
    \centering
    \includegraphics[width=\linewidth]{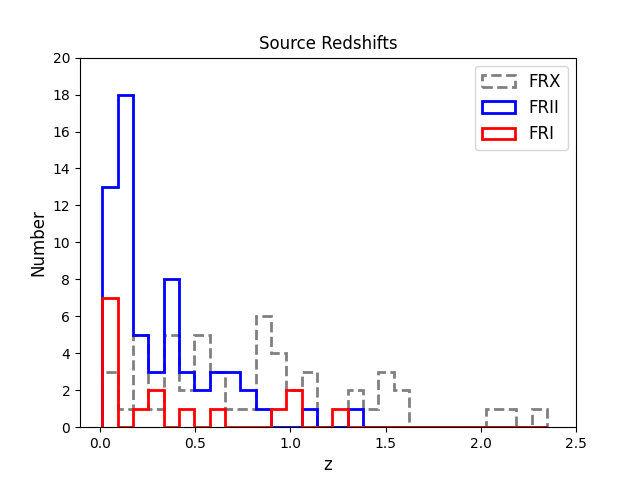}
    \includegraphics[width=\linewidth]{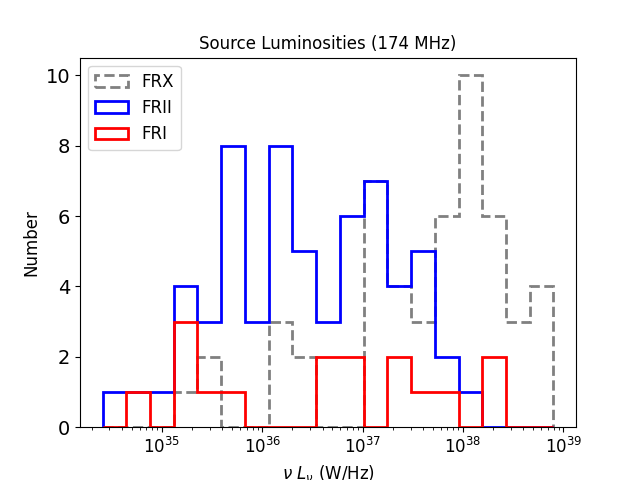}
    \includegraphics[width=\linewidth]{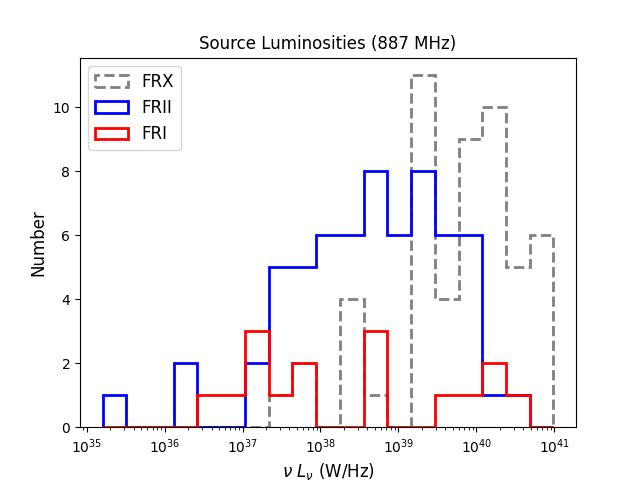}
    \caption{Distribution of AGNs by FR type across redshift and radio luminosity. The classification method we used was effective in classifying sources up to a redshift of $z = 1.35$.}
    \label{hist}
\end{figure}

In Fig \ref{hist}, we show the distribution of sources across redshift and luminosity in both GLEAM (174 MHz) and RACS-low (887 MHz). The primary obstacle in classifying FR sources was angular size of the source, which is influenced by the physical size of the source and its redshift. Low redshift, physically small sources and more extensive high redshift sources became too small in their angular size to resolve. Unsurprisingly, FRX sources make up most of the higher luminosity sources in the sample. Higher luminosity sources are generally the most distant, leading to the aforementioned issue with angular size. \textbf{This speaks to the limitations of the current configuration of ASKAP in effectively analyzing the morphologies of more distant AGNs, as the classification method we used was only effective in classifying sources up to a redshift of $z = 1.35$ \citep{Norris2025}.}

\begin{table}[h]

    \begin{tabular*}{\linewidth}{l@{\extracolsep{\fill}}cccc}
         \hline
         Source & FR Type & Size (kpc) & z & L887 (W/Hz)\\
         \hline
         G4Jy 133 & FRII & 1484.39 & 0.1478 & $7.15 \times 10^{35}$\\
         G4Jy 347 & FRII & 1925.65 & 0.0624 & $2.95 \times 10^{37}$\\
         G4Jy 517 & FRII & 1560.33 & 0.0381 & $1.60 \times 10^{35}$\\
         G4Jy 1079 & FRII & 1401.40 & 0.0836& $2.21 \times 10^{36}$\\
         G4Jy 1613 & FRI & 1425.70 & 0.06063 & $3.36 \times 10^{36}$\\
         \hline
    \end{tabular*}
    \caption{Sources in the sample larger than 1 Mpc. "Size" in this table refers to jet length for FRII sources and region size for FRI sources.}
    \label{FRSize}
\end{table}

\subsection{AGN Jet Properties}

\begin{figure}
    \centering
    \includegraphics[width=1\linewidth]{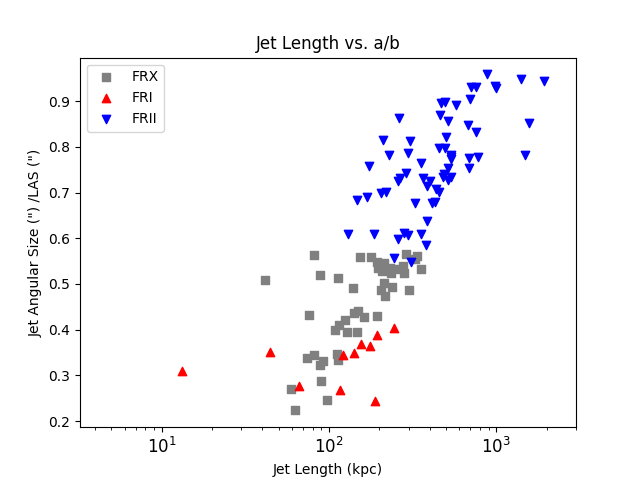}
    \caption{The jet lengths of Fanaroff \& Riley AGNs in the G4Jy-3CRE sample. Four FRI sources, nine FRII sources, and 15 FRX sources were excluded due to missing redshift data.}
    \label{FRIIFRI}
\end{figure}

Within our sample of FRI and FRII AGNs (not counting FRX sources), all jets are at least 127 kpc in length with the only exception being G4Jy 1749 (Figure \ref{FRIIFRI}). Jet lengths for the FRI population range from 42.708 kpc to 1425.70 kpc, with a median jet length of $\mu$=407.31 kpc and a standard deviation of $\sigma = 320.97$ kpc. For the FRII population, their jet lengths range from 129.21 kpc to 1925.65 kpc, with a median length of $\mu$=436.15 kpc, and a standard deviation of $\sigma = 301.62$ kpc. 

Due to the small sample size, it is difficult to identify a trend with the distribution of FRI sources. It is worth mentioning the "flattened" distribution is likely a result of many of the smaller potential FRI sources being automatically disqualified in the classification process. As discussed in section 3.6, the minimum angular size required for classification as either FRI or FRII ($b_{min} > 50 \arcsec$) meant many sources that would otherwise be obvious FRI candidates (with a low a/b ratio) were classified as FRX sources. With higher resolution images or a relaxation of our minimum size requirement, many FRX sources would be classified as FRI or FRII. 

We compared the region size of all FRI, FRII, and FRX sources in kiloparsecs with the luminosity of each source for both GLEAM and RACS-low (shown in Figure \ref{Luminosity}). The flux used to calculate the RACS-low luminosity is taken from the data release for RACS-low \citep{RACS_1}. The flux for RACS-low was estimated by summing over all the Gaussian components of each source. The GLEAM \citep{GLEAM} luminosities were estimated by Massaro et al. We found no correlation between jet length and luminosity within this sample with a Pearson coefficient of $r=0.13$. Note that the small sample size may obscure a correlation across a wider range of jet sizes. There is an upper limit to the jet length for the vast majority of FRII sources around 1 Mpc, which cannot be accounted for by the limited sample size. 

Several past relativistic magnetohydrodynamic (RMHD) simulations \citep{Hardcastle2018, Palau2025_IP} of AGN jet evolution show a significant drop-off in cocoon pressure in the jet at a distance of 1 Mpc, right around the limit found in our sample. In the simulations, only the largest, most luminous sources have enough jet power to maintain pressure to expand the jet beyond this limit. Only five AGNs in our sample, G4Jy 133, G4Jy 347, and G4Jy 517, G4Jy 1079, and G4Jy 1613 have jet lengths beyond this limit, detailed in Table \ref{FRSize}. All five of these sources are on the lower end of the luminosity range of our sample, ranging from $1.60 \times 10^{35}$ W/Hz to $2.95 \times 10^{37}$ W/Hz at 887 MHz. Four of the five sources have redshifts $z < 0.1$. Only G4Jy 133 is well beyond this redshift at $z = 0.1478$. Lastly, 46 sources within the whole sample are missing redshift data. These sources were excluded from plot A of Figure \ref{hist}, as well as Figures \ref{FRIIFRI} and \ref{Luminosity}. The FRII, FRI, and FRX samples have 4, 9, and 15 sources without redshift, respectively.

\subsection{Comparison with Past Catalogs}
\begin{figure}
    \centering
    \includegraphics[width=\linewidth]{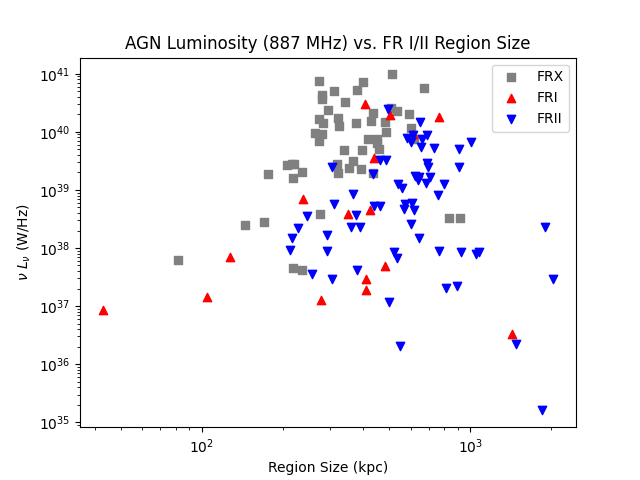}
    \includegraphics[width=\linewidth]{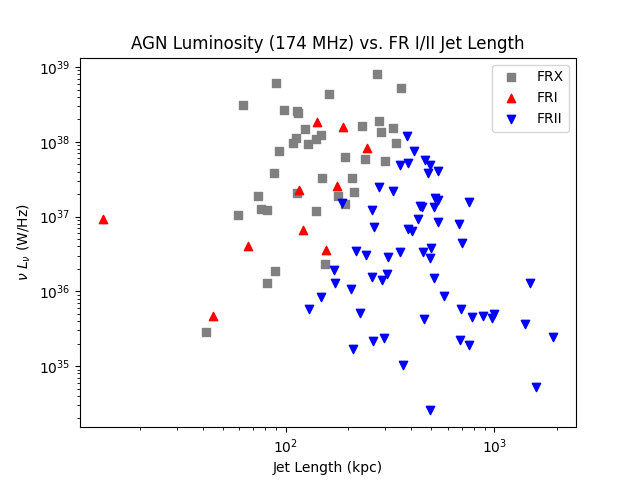}
    \caption{Radio luminosity vs. region size using both RACS-low (887.5 MHz) and GLEAM data (174 MHz). The jet length for $83.5 \%$ of these sources lie between 100 kpc and 1 Mpc. The number of sources in each plot will differ depending on whether the source has luminosity data at 887 MHz and 174 MHz.}
    \label{Luminosity}
\end{figure}

The work of \cite{Mingo2019} serves as a point of comparison to our own, as they constructed a catalog of AGNs using data from the LOFAR Two-Metre Sky Survey (LOTSS; \citealt{Shimewell2019}). Their catalog of 5805 AGNs was used to investigate the relationship between AGN luminosity and jet morphology, an objective we share in our own work. The most relevant finding of their work to this paper is that luminosity does not reliably predict whether a source has FRI or FRII morphology, as FRII sources were found at luminosities below the traditional ``luminosity break" separating FRI and FRII sources observed in past works \citep{Fanaroff_Riley}. We compared the plots in Figure \ref{Luminosity} to the ``Luminosity vs. Size" plot (Figure 5 in \cite{Mingo2019}). Though their sample size is much larger and their luminosity range is different, the same trends can be seen as in our plot. Our sample of FRI and FRII sources populate the same range of luminosities with neither population showing any particular luminosity bias, just as was seen in the LOTSS sample by Mingo et al. While our flux-limited sample is not large enough to fully corroborate the findings of Mingo et al., it is at least indicative that their findings hold across a range of frequencies. 
In addition, The vast majority of FR galaxies in their sample are smaller than 1 Mpc, with only a handful exceeding this limit.

\section{Summary and Future Prospects}
Overall, the data presented in the current work has painted a clear picture of the high resolution capabilities of ASKAP when it comes to investigating AGNs at relatively low redshift. ASKAP observations of our sample showed significant improvement in morphological detail over all the archival radio maps of the Sydney University Molonglo Sky Survey (SUMSS; \citealt{SUMSS_Mauch2003}) and a few for NVSS. Furthermore, we were able to use the more detailed radio maps to identify the optical counterparts for six sources that either had no previous counterpart or a previously misidentified or ambiguous counterpart.

For the purposes of classification, ASKAP was capable of identifying 173 sources ($66\%$ of the sample) with morphologies indicative of jet activity, $42\%$ of which are FRII sources, $17\%$ of which are FRI sources, and $39\%$ of which are FRX sources. Using the classification method outlined by \cite{Norris2025}, we were able to effectively classify FR-type radio sources up to a redshift of $z = 1.35$. 

Our sample showed a soft limit in the jet length of FR radio sources around 1 Mpc. Recent simulations of AGNs with similar luminosity profiles show that this is the distance at which internal pressure in the jet lobe sharply falls off, leading to their expansion slowing significantly. Our sample only contains two jetted AGN larger than this limit. Our catalog also shows an overlap in the FRI and FRII population in this luminosity range, contrary to the traditional FRI/FRII break just as was found in \cite{Mingo2019}.

ASKAP shows significant  promise in its capability of high resolution imaging for low redshift AGNs. Future work includes a comparison of ASKAP data for the entire G4Jy catalogue (S. V. White et al., in prep.). In addition, follow-up observations of the identified optical counterparts are already taking place and will be published in a subsequent work. Future papers in the G4Jy-3CRE series are currently in preparation, including both an infrared and X-ray analysis of the G4Jy-3CRE catalog and an on optical spectra classification for the previously identified optical counterparts.


\section{Acknowledgments}

This research has made use of the CIRADA cutout service at URL cutouts.cirada.ca, operated by the Canadian Initiative for Radio Astronomy Data Analysis (CIRADA). This research uses services or data provided by the Astro Data Lab, which is part of the Community Science and Data Center (CSDC) Program of NSF NOIRLab. NOIRLab is operated by the Association of Universities for Research in Astronomy (AURA), Inc. under a cooperative agreement with the U.S. National Science Foundation. 

This research has made use of the NASA/IPAC Infrared Science Archive, which is funded by the National Aeronautics and Space Administration and operated by the California Institute of Technology. This research has made use of SAOImageDS9, developed by Smithsonian Astrophysical Observatory. C.M. acknowledges support from Fondecyt Iniciacion grant 11240336 and the ANID BASAL project FB210003.

JPM would like to thank B\"arbel Koribalski for the original opportunity to work with ASKAP data. WF acknowledges support from from the Smithsonian Institution, the Chandra High Resolution Camera Project through NASA contract NAS8-03060, and NASA Grants 80NSSC19K0116 and GO1-22132X.

\facilities  ASKAP{}.


\bibliography{citations}
\bibliographystyle{aasjournal}

\appendix

\section{Notable Sources}
The following section contains details on several sources we felt warranted further discussion or for which ASKAP provides some new insight. We primarily discuss differences in classification when comparing to archival surveys, interesting morphological details, archival designations for the radio source, identified optical counterparts (or a lack thereof), or otherwise interesting features in the radio image. A similar discussion of sources can be found in \cite{G4Jy-3CRE_Catalogue}, which may contain information not covered in this section.\\

\underline{G4Jy 9:} also known as PKS 0003-56, this source has a double radio lobe morphology in in RACS-mid and RACS-high. This source was previously identified as a normal galaxy lying at z=0.2912 \citep{Jones2009}.

\underline{G4Jy 12:} also known as PKS 0003-83 \citep{Burgess_2006}, the radio contours of this source are indicative of the jets of an AGN,. The RACS data shows significant improvement over archival SUMSS contours, giving a clearer picture of the jet's morphology.

\underline{G4Jy 20:} also known as PKS 0008-44, is a double-lobed radio source with a typical FR II morphology. The RACS data shows significant improvement over the archival SUMSS data, showing a clear jet morphology where it was not previously visible in the SUMSS data. 

\underline{G4Jy 78:} also known as PKS 0039-44, has a much more constrained morphology in RACS than archival maps. The RACS-high contours also show an elongation of the source

\underline{G4Jy 90:} also known as PKS 0048-447. Much like G4Jy 78, this source has a more constrained morphology and an elongation in RACS-high.

\underline{G4Jy 120:} this source has an AGN jet structure typical of an FRII galaxy. As mentioned in \cite{G4Jy-3CRE_Catalogue}, the associated optical counterpart does not line up with the center of the radio source between the two lobes. The RACS data does little to make the counterpart clearer.

\underline{G4Jy 129:} also known as PKS 0110-69 (or AT20G J01114343-690016), this source has a hybrid FR morphology, with hot spots at each end of the respective jets and an additional hot spot (possibly a jet knot) in the southern part of the northern jet, presumably closer to the AGN. No optical counterpart is clearly identifiable. 

\underline{G4Jy 133:} also known as PKS 0114-47, this source shows evidence of multiepoch activity. This is clearly shown by the secondary hot spots closer to the identified radio core. While these hot spots were somewhat visible in archival maps, they are much more visible in RACS.

\underline{G4Jy 168:} also known as PKS 0129-073 (also MRC 0129-073 and PMN J0131-0703), this is a double lobed radio source with an FR II structure. Though RACS does not provide any new information on the morphology, the RACS contours indicate that the optical counterpart was previously misidentified. We have identified a different optical counterpart that aligns with both the RACS radio contours (along the jet axis) and lines up with the radio centroid shown.

\underline{G4Jy 241:} also known as ESO 198-1, this source was previously identified as an FSRQ. The RACS contours show this source to have an FR I morphology with a bent tail toward the end of the northern jet. 

\underline{G4Jy 247:} also known as AT20G J021902-362607, this source is an FR II radio galaxy with clear jet hotspots to the north and south of the radio core. This is a significant improvement over SUMSS, which showed only an elongated radio morphology and only a single elongated hot spot to the north of the radio core.

\underline{G4Jy 290:} a radio source with an FR II morphology. The jet lobes and hot spots shown in RACS were not previously identifiable in SUMSS.

\underline{G4Jy 347:} a 2.5 Mpc giant radio galaxy (also known as MRC 0319-454, PMN J0321-4510, and MSH 03-43 \citep{Burgess_2006, Malarecki2015}) located within the Horologium Reticulum Supercluster. The RACS data shows two hot spots of activity toward the end of the jet plume. This is possibly indicative of two epochs of jet activity, though there may also be a result of interaction with the surrounding IGM. 

\underline{G4Jy 350:} also known as PKS 0352-88 (also MRC 0352-884, SUMSS J032359-881618), this source has an FR II morphology. While the morphology visible in RACS is not too different from SUMSS, we have a much clearer picture of the overall morphology and the hot spots of the jets are now clearly identifiable.

\underline{G4Jy 381:} this source is an FR II radio galaxy \citep{Scarpa1996,Drinkwater2001}. Only two jet plumes were previously identified using 3GHz VLASS contours \citep{G4Jy-3CRE_Catalogue}, however, RACS-low radio contours outline a radio morphology that extends beyond the eastern jet plume.

\underline{G4Jy 453:} this source is an FR II galaxy and well known radio source whose optical counterpart we were able to identify. The extension of the plume beyond the central axis of the jet also classifies this object as a "winged" radio source, more specifically, an X-shaped radio galaxy (XRG) \cite{Leahy1984}.

\underline{G4Jy 462:} this source is an FR I radio galaxy \citep{Ekers1969,McAdam1988,Morganti1993}, also known as IC2082 and PKS -427-53 \citep{Carter1981, Lilly1987}.

\underline{G4Jy 507:} also known as NGC 1692 \citep{NGC_Sulentic_1973}, this is an unresolved radio source. One interesting feature in this image is the trace radio emission from ESO 552-22, a Seyfert galaxy southeast of G4Jy 507. This galaxy seems to be undergoing ram pressure stripping from the local galaxy group, making it a jellyfish galaxy with an increased rate of star formation \citep{DES2018, Jellyfish_2025Vendhan}. 

\underline{G4Jy 513:} also known as PKS 0456-30, this source has an amorphous morphology with no obvious jet structure. 

\underline{G4Jy 530:} also known as PKS 0511-48, this source has previously been classified as a Seyfert 2 galaxy \citep{Veron2010}. In the RACS data, this source has a hybrid morphology (HyMor), as it has a radio hotspot near the core in addition to the FR II jet structure.

\underline{G4Jy 531:} this source is listed in the MS4 sample as MRC 0511-305 \citep{Burgess_2006} and has since been classified as a Seyfert 2 galaxy \citep{Chen2022}. The RACS-low radio contours show two distinct radio jets with the southwestern jet diverging from the axis of jet activity. The RACS contours also clearly identify this AGN as having an FR I morphology, with the tails of the jet being bent off-axis (likely by some interaction with the IGM).

\underline{G4Jy 580:} this source is an FR II radio galaxy with a counterpart previously identified in the MS4 sample. In previous works there was some ambiguity about the optical counterpart due to multiple sources proximity to an IR counterpart, however the RACS data eliminates this ambiguity \citet{G4Jy-3CRE_Catalogue}.

\underline{G4Jy 607:} this source is a radio galaxy \citep{Tritton1972,Wills2004} with a flat radio spectrum \citep{Healey2007}. The RACS-low radio contours exhibit a WAT/FRI morphology undergoing significant ram pressure in its southwestern lobe. The hotspots of the two jets can be clearly identified in the RACS data. 

\underline{G4Jy 611:} this source is an LERG hosted in a dumbbell galaxy \citep{Frank2013,Almeida2011,Ineson2015} located at the center of the Abell 3391 galaxy cluster. The RACS-low radio contours outline two radio tails extending outwards from the optical center. This source has previously been identified as having WAT radio structure \citep{Morganti1999}, which is supported by the RACS radio contours. RACS also shows that this source has an FR I morphology, though this fact is somewhat obscured by the bending of the jets from interaction with the IGM.

\underline{G4Jy 613:} this source is the BCG of the galaxy cluster Abell3395 \citep{Brown1991} and has an associated detection of x-ray emission \citep{Sun2009}. The source has an FR I type radio jet morphology as well as a WAT structure.

\underline{G4Jy 619:} this source is a radio galaxy \citep{Storchi1996,Jones2009} that has also been detected in the x-ray \citep{Cusumano2010,Oh2018}. This source has an FR II radio morphology across all frequencies of RACS

\underline{G4Jy 718:} also known as MRC 0842-835, this source has an elongated radio morphology indicative of AGN jet activity. Higher resolution data is needed to make a conclusive determination.

\underline{G4Jy 854:} this source is a radio source whose optical counterpart we have managed to identify. Archival radio maps from NVSS show the emission as misaligned with the counterpart identified using RACS, however this is likely due to an astrometric problem.

\underline{G4Jy 957:} this source is a radio galaxy with multiple nearby optical and mid-IR sources \citep{Danziger1983, G4Jy-3CRE_Catalogue}. It has a typical FR II morphology.

\underline{G4Jy 987:} this source was identified by Massaro et al. as a candidate Hybrid Morphology Radio Source (HyMoR) \citep{G4Jy-3CRE_Catalogue}, meaning that this source has a FR I radio morphology on one side and FR II on the other side of the AGN \citep{Gopal2000,Cheung2009}. However, the RACS-low radio contours contradict this description, as both ends of the AGN jet exhibit FR II type morphology.

\underline{G4Jy 1080:} this source is the radio galaxy IC 4296 \citep{Younis1985,Killeen1986,Killeen1988,Smith2000,Wegner2003,Grossova2019,Condon2021,Grossova2022}. This source has an FR II morphology.  

\underline{G4Jy 1135:} this source is a radio galaxy (a.k.a. PKS 1413-36) previously identified to have an FRII morphology \citep{Burgess_2006,G4Jy-3CRE_Catalogue}. While archival radio images from NVSS are of a much higher resolution, RACS does a better job of showing the entire radio structure of the jets. 

\underline{G4Jy 1172:} this is a source with a hybrid FR morphology. Though the overall structure is that of an FRII, the lack of visible radio hot spots at the end of either lobe makes the exact classification unclear.

\underline{G4Jy 1262:} this source is a radio galaxy listed in the MS4 sample (a.k.a. PMN J1530-4231) \citep{Burgess_2006}. This source has an FR II morphology. The RACS data has also allowed us to pinpoint the optical counterpart.

\underline{G4Jy 1279:} this is a radio source with an FR II morphology. The RACS data has revealed more of the morphology than what was visible in archival NVSS radio maps. The jet plumes extend beyond the previously observed hotspots, revealing multiple epochs of jet activity from the galaxy.

\underline{G4Jy 1289:} this is an as of yet unclassified radio source. The morphology of the source seems to be that of an FR II galaxy, however the asymmetry of the two hot spots and lack of connecting emission between them makes us hesitate to classify this source conclusively.

\underline{G4Jy 1302:} this is a radio source with an FR II structure. Previous works by \cite{G4Jy-3CRE_Catalogue} and \cite{White2020B}  showed disagreement between the identified optical and infrared counterparts for this source, however the RACS data strongly suggests an agreement with the counterpart identified by Massaro et al. 

\underline{G4Jy 1350:} this source is a radio source with an FR II morphology.

\underline{G4Jy 1360:} this source is an FR II galaxy with an X shaped morphology. The hotspots of the different epochs of jet activity are visible across all three frequency ranges in RACS.

\underline{G4Jy 1423:} this source is a FR II radio galaxy \citep{Tadhunter1993} previously associated with the optical source reported in the literature \citep{G4Jy-3CRE_Catalogue,Morganti1993, Jauncey1989, Wall1985}. The RACS data outlines the radio core as well as multiple hot spots in the southwestern jet, likely caused by a knot in the jet.

\underline{G4Jy 1432:} this is a radio source previously identified as a blazar \citep{Maselli2013} and associated with an X-ray source \citep{Oh2018}. The RACS-low radio contours have an FR II morphology, which contradicts the source's previous identification as a blazar. In addition, the optical source previously identified as the counterpart to this galaxy is misaligned with the radio jets, making it an unlikely candidate.

\underline{G4Jy 1498:} this source is a radio galaxy with an FR II and WAT structure. With RACS improved resolution, we are able to discern the exact morphology and direction of the jets.

\underline{G4Jy 1504:} this source is an FR I galaxy with WAT structure. RACS shows significant improvement over SUMMS across all frequencies for the purposes of determining AGN morphology. 

\underline{G4Jy 1505:} this source is a radio galaxy listed in the SUMSS catalog, this radio source has an FR II morphology.

\underline{G4Jy 1513:} also known as PKS 1859-23, this source has a newly identified optical counterpart.

\underline{G4Jy 1569:} this is a radio source with an FR II morphology. Much more of the detail can be seen in the RACS radio contours, including hot spots along and at the far end of the jets.

\underline{G4Jy 1590:} this is a radio source with an Z-shaped morphology and 2 hot spots in each jet. This morphology is likely due to two separate epochs of activity with the jets firing at different angles.

\underline{G4Jy 1613:} this is a radio source with an X-shaped morphology. This source has an FR I structure and distinctive and symmetrical jet plumes emanating from the radio core.

\underline{G4Jy 1618:} this is a radio source with an FR II morphology. Unlike in the SUMSS radio map, the jet hot spots are discernible.

\underline{G4Jy 1677 \& 1678:} two or more radio galaxies in the A3744 cluster \citep{Abell1989} whose interactions seem to be heavily distorting the jets. Previous observations by \cite{Birkinshaw_Rawes_Worrall_2018} seem to indicate a pairing of an FRI galaxy (G4Jy 1678, the northern source) and an FRII galaxy (G4Jy 1677, the southern source). RACS data seems to show G4Jy 1678 undergoing significant jet bending due to either ram pressure or jet backflow as in WRGs \cite{Norris2025}.

\underline{G4Jy 1847:} also known as PKS 2338-58, this source is a radio galaxy with a double lobe morphology. The optical counterpart for this source was previously identified by Massaro et al.; however, the lobed morphology indicative of jet activity was not previously visible in the GLEAM radio images \citep{G4Jy-3CRE_Catalogue}.

\section{Master Table}

\onecolumngrid

\begin{sidewaystable*}
    \centering
    \caption{Catalog Table Sample}
    \begin{tabular}{|c|c|c|c|c|c|c|c|c|c|}
        \hline
            Object&RA (J2000)&Dec (J2000)&z&Total Flux (mJy)&L174 (W/Hz)&L887  (W/Hz)&Region Size (kpc)&Jet Length (kpc)&Classification \\
            (i) & (ii) & (iii) & (iv) & (v) & (vi) & (vii) & (xiii) & (xiv) & (xvi) \\
            \hline
            G4Jy 4&0.841671&-17.4532&1.467&3382&$2.37\times 10^{38}$&$4.17\times 10^{40}$&267.7& - &U \\
            G4Jy 9&1.49021&-56.4754&0.2912&2820&$4.04\times 10^{36}$&$6.86\times 10^{38}$&239.20&66.2&FRI \\
            G4Jy 12&1.54995&-83.0991&0.2347&3602&$3.11\times 10^{36}$&$1.09\times 10^{39}$&438.9&244.2&FRII \\
            G4Jy 20&2.62729&-44.3825&1&3075& - &$1.47\times 10^{40}$&479.6&255.2&FRX \\
            G4Jy 26&3.85118&-38.0765&0.899&2065&$6.24\times 10^{37}$&$2.47\times 10^{39}$&448.8&192.5&FRX \\
            G4Jy 27&4.01117&-63.1687& - &2973& - & - & - & - &FRI \\
            G4Jy 33&4.71408&-12.7093&1.599&3587&$3.18\times 10^{38}$&$5.4\times 10^{40}$&302.6& - &U \\
            G4Jy 43&5.78892&-25.0432&0.354&3303&$6.9\times 10^{36}$&$1.23\times 10^{39}$&539.2&384.6&FRII \\
            G4Jy 45&6.1255&-29.4802&0.4065&4447&$1.49\times 10^{37}$&$2.36\times 10^{39}$&353.1&193.2&FRX \\
            G4Jy 48&6.45487&-26.0369&0.3219&11626&$1.05\times 10^{37}$&$3.57\times 10^{39}$&165.5& - &U \\
        \hline
    \end{tabular}
    \label{MT}
    \caption{A sample of the master table containing all the relevant information for each source. Column (i) contains the object source name in the G4Jy catalogue. Columns (ii,iii) contain the right ascension and decl. in degrees (Equinox J2000) of the brightness-weighted radio centroid. Column (iv) contains the redshift (\citealt{G4Jy-3CRE_Catalogue,G4Jy-3CRe_III}, García-Pérez et al. private comm). Column (v) contains the total flux of the gaussian components associated with the source in mJy, taken from the data release for RACS-low, \citep{RACS_1}. Columns (vi, vii) contain the luminosity of each source in W/Hz L174 is taken from \cite{GLEAM} and L887 was derived from the RACS-low flux values and redshifts where available. Columns (viii, ix, x) contain the background rms values used to define the lowest contour levels of each image in counts/pixel. Columns (xi, xii) contain the angular measurements for region size and jet length in arcseconds. Columns (xiii,xiv) contain the region diameter and jet length of each source in kpc. Column (xv) contains the survey origin of the background optical image for each source. Column (xvi) contains the tags we used to classify each source. The full table with all the columns is available in a machine readable format online.}
\end{sidewaystable*}

\pagebreak
\clearpage

\section{Catalog Images}

\begin{figure*}
    \centering
    \includegraphics[scale=0.2]{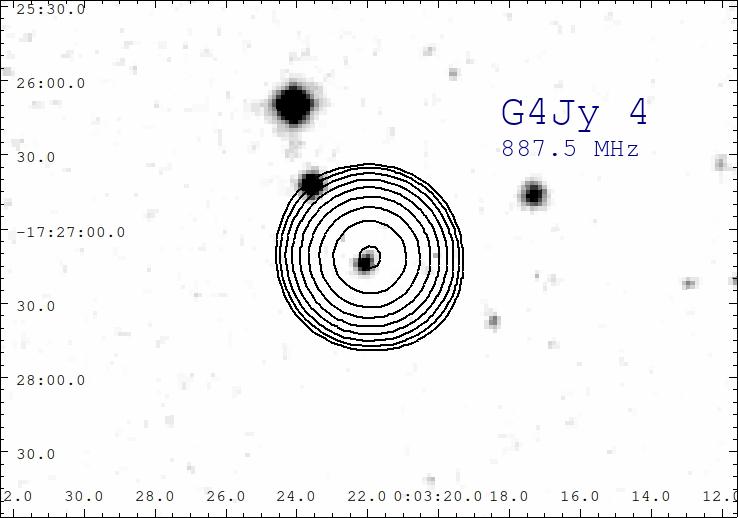}
    \includegraphics[scale=0.2]{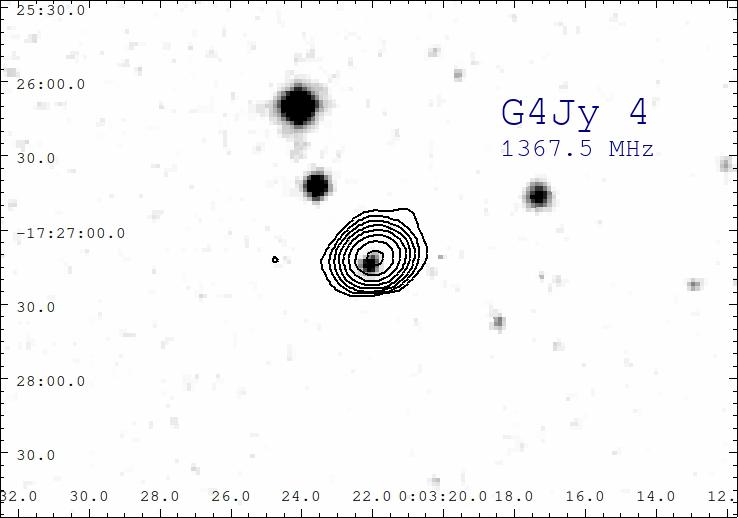}
    \includegraphics[scale=0.2]{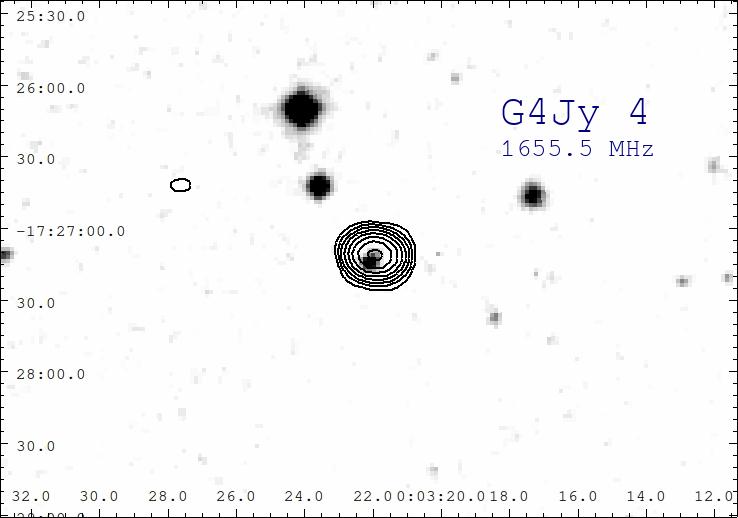}
    \includegraphics[scale=0.2]{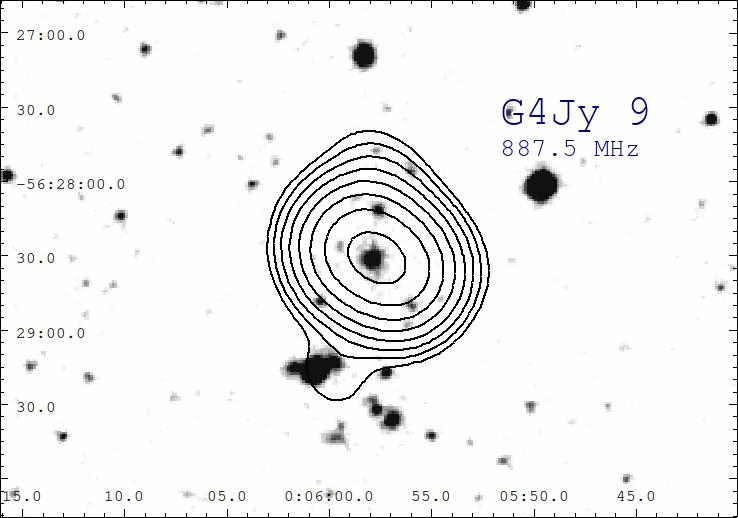}
    \includegraphics[scale=0.2]{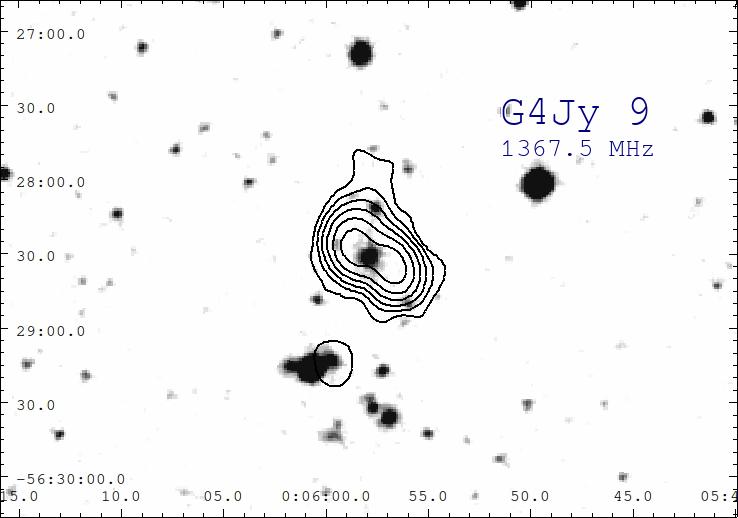}
    \includegraphics[scale=0.2]{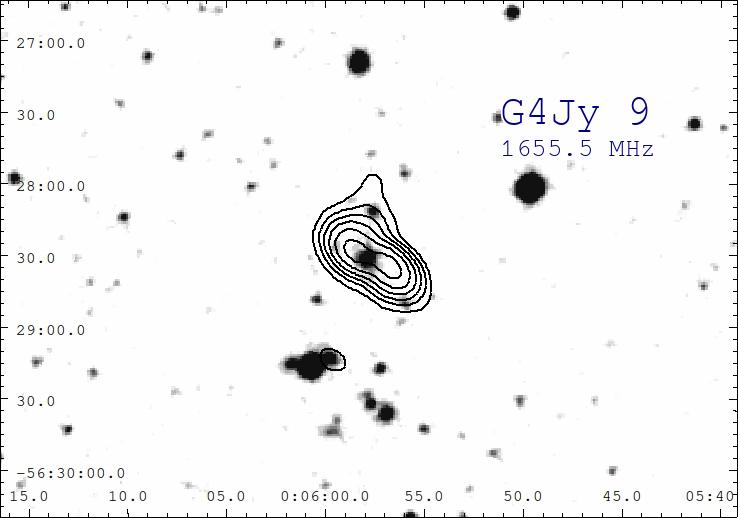}
    \includegraphics[scale=0.2]{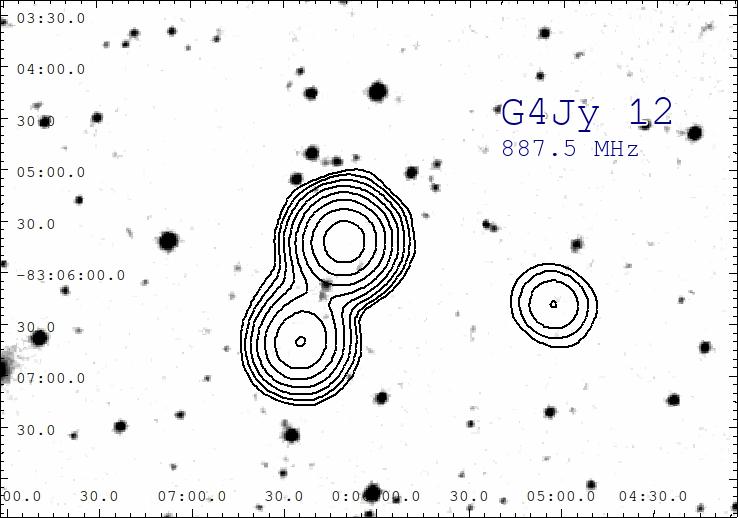}
    \includegraphics[scale=0.2]{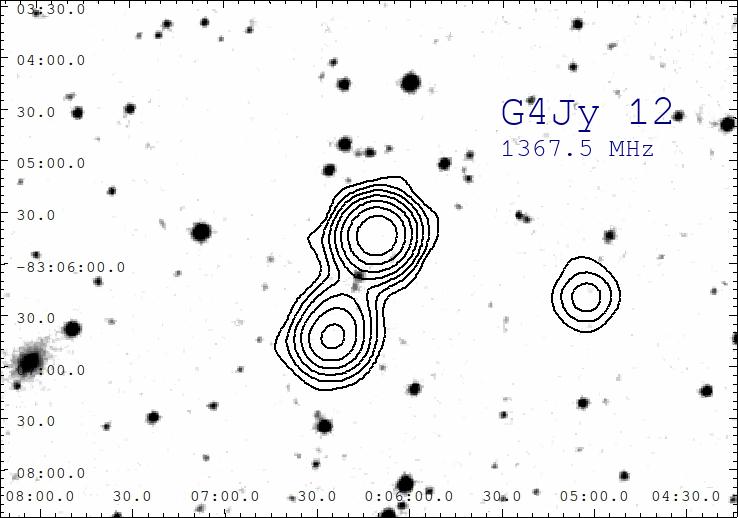}
    \includegraphics[scale=0.2]{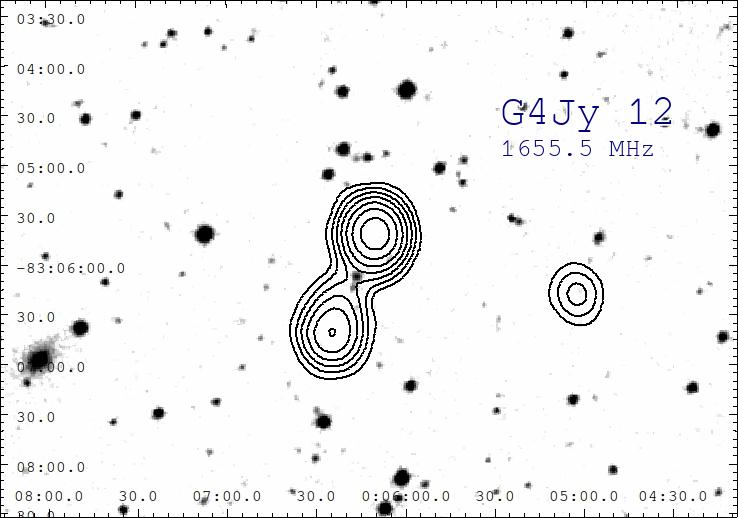}
    \includegraphics[scale=0.2]{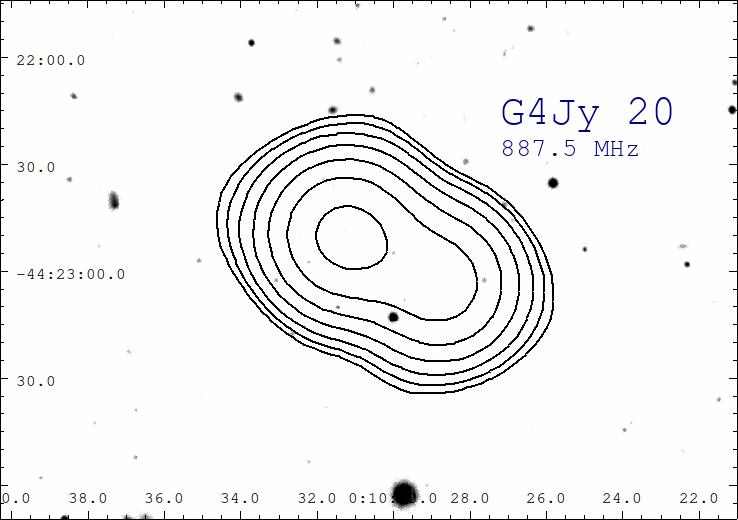}
    \includegraphics[scale=0.2]{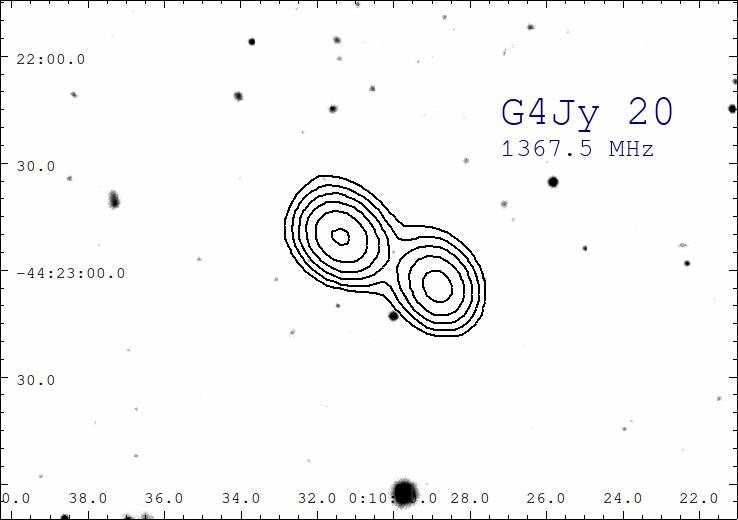}
    \includegraphics[scale=0.2]{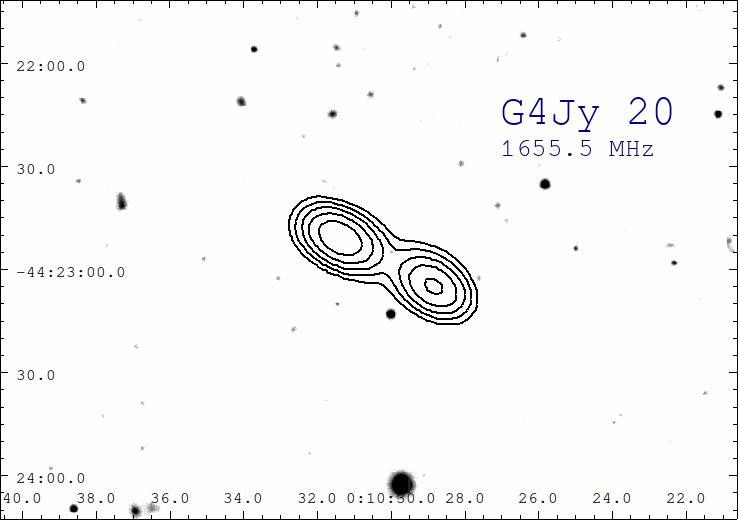}
    \includegraphics[scale=0.2]{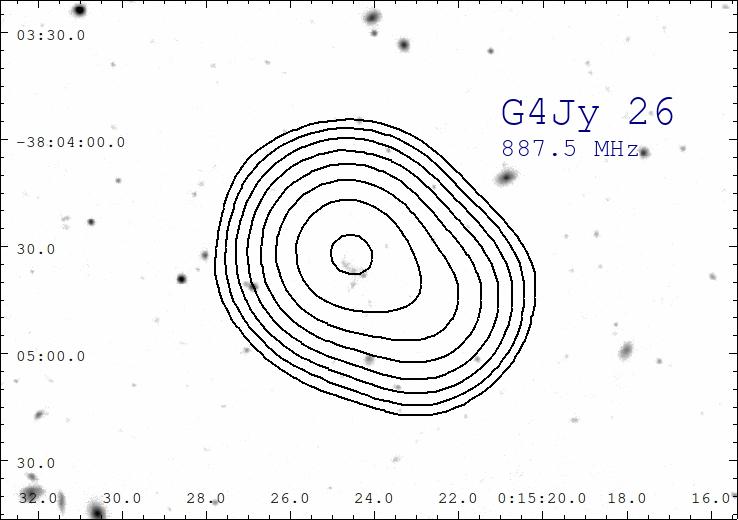}
    \includegraphics[scale=0.2]{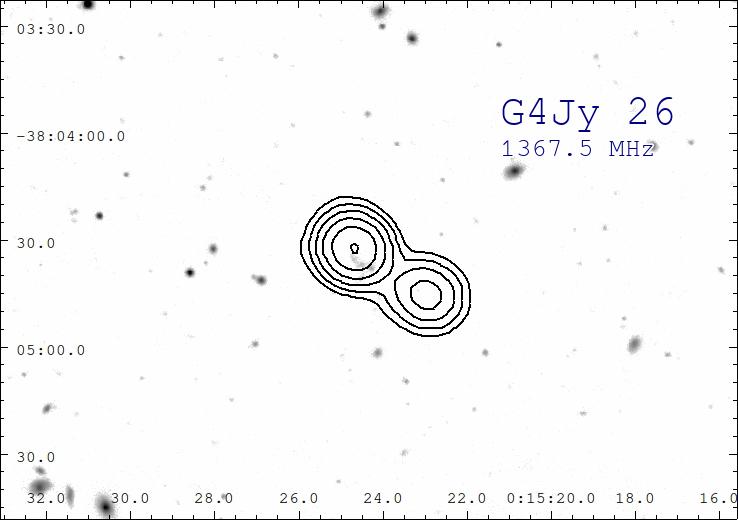}
    \includegraphics[scale=0.2]{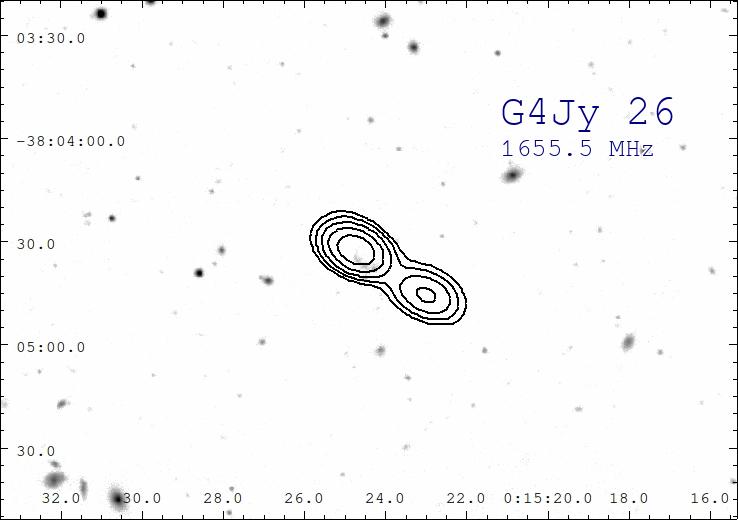}
    \caption{The images created for the catalogue. Each images corresponds to a single G4Jy-3CRE source at one of three frequencies, RACS-low, RACS-mid, and RACS-high. Each image consists of RACS radio contours overlaid an optical image corresponding corresponding to the radio source. The background images are cutouts corresponding to the coordinates of the radio source from DES, PanSTARRS, or DSS2 (based on availability and best image quality) \citep{DES2018, PanSTARRS2020, DSS}.}
    \label{A}
\end{figure*}
\clearpage
\begin{figure*}
    \centering
    \includegraphics[scale=0.225]{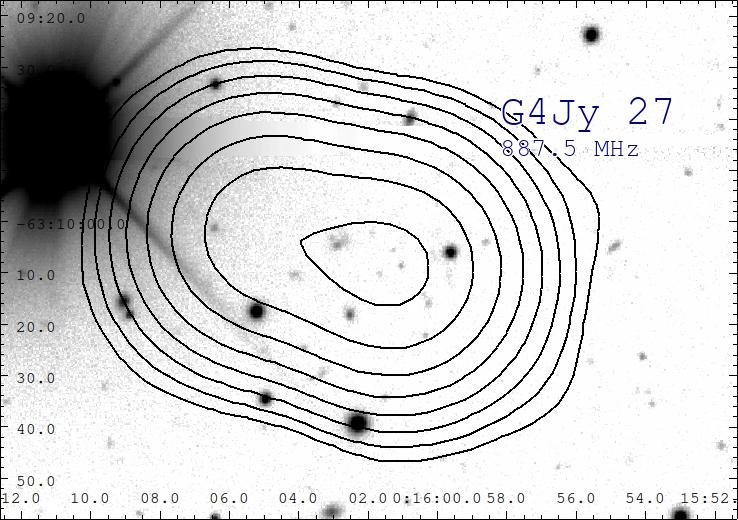}
    \includegraphics[scale=0.225]{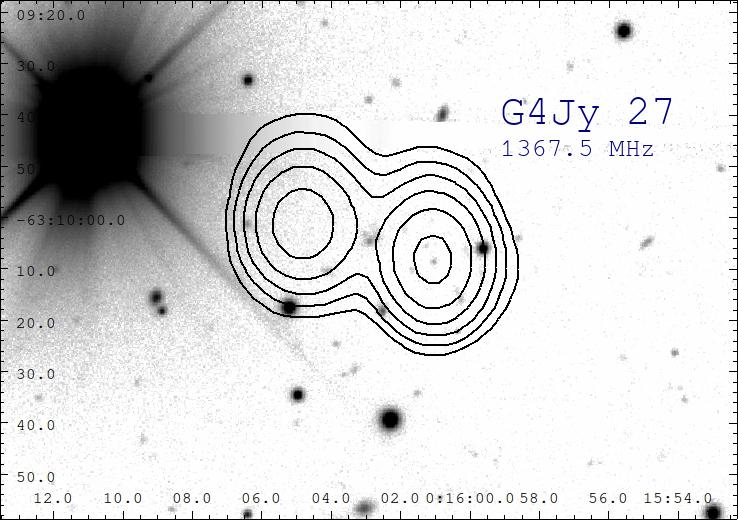}
    \includegraphics[scale=0.225]{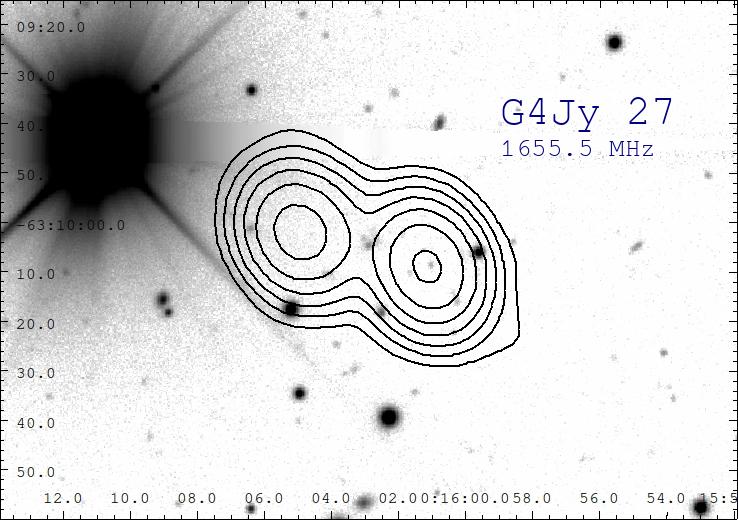}
    \includegraphics[scale=0.225]{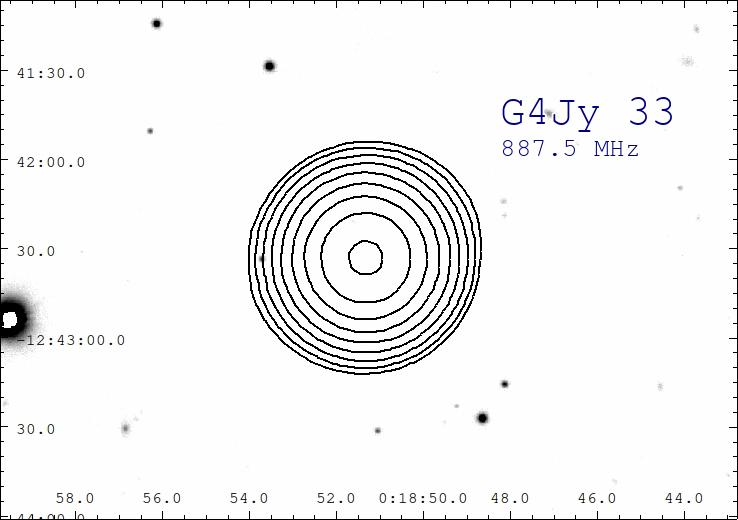}
    \includegraphics[scale=0.225]{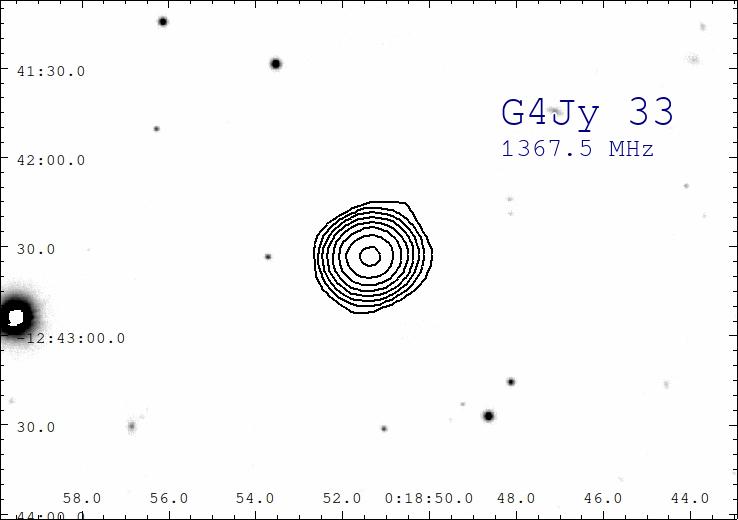}
    \includegraphics[scale=0.225]{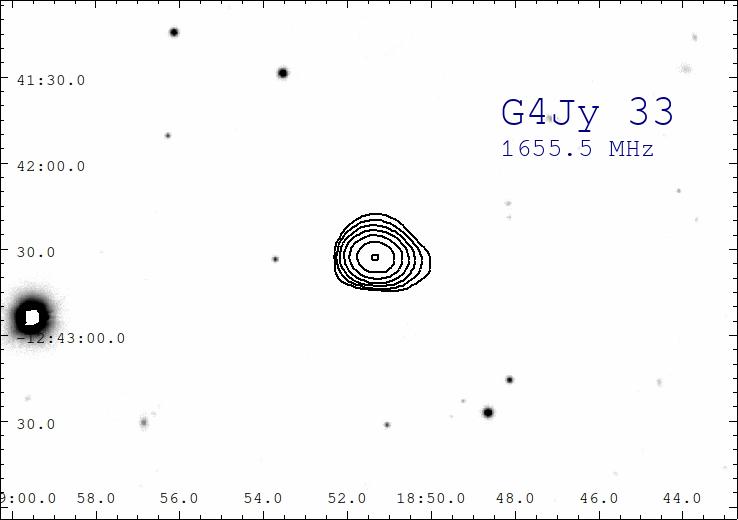}
    \includegraphics[scale=0.225]{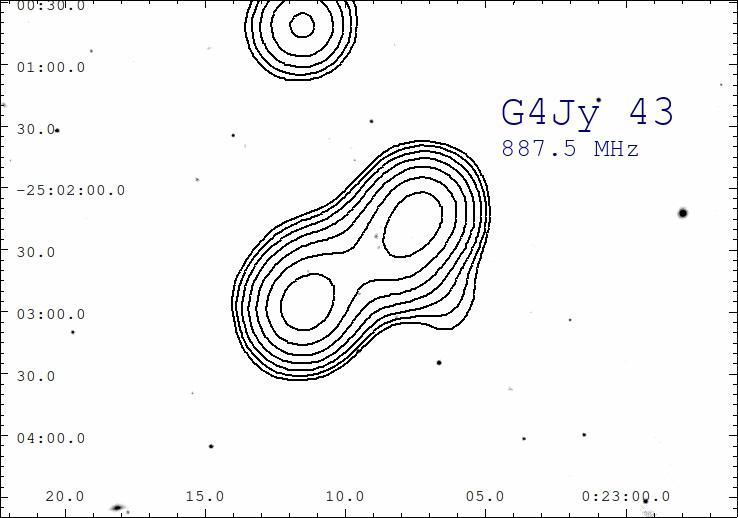}
    \includegraphics[scale=0.225]{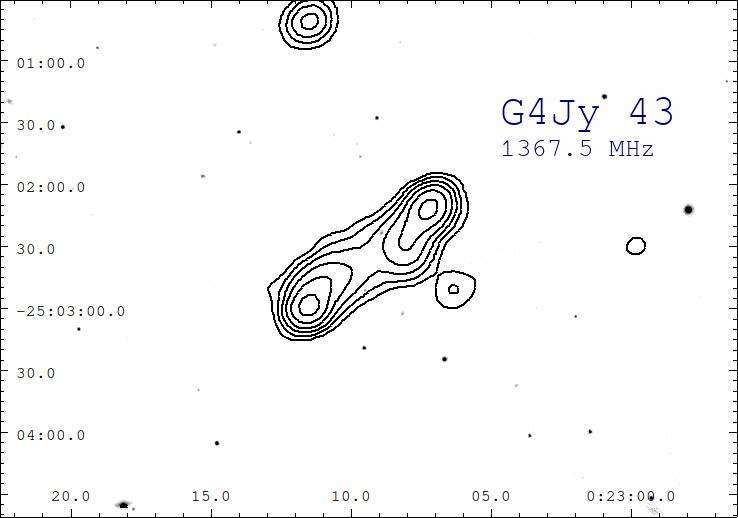}
    \includegraphics[scale=0.225]{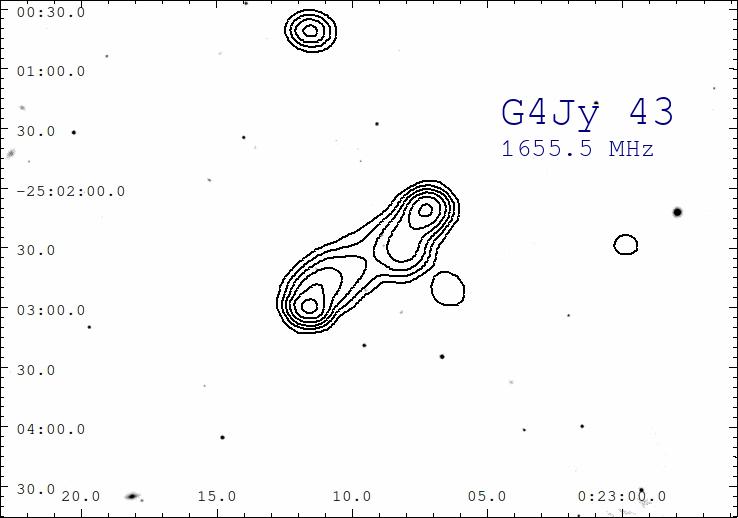}
    \includegraphics[scale=0.225]{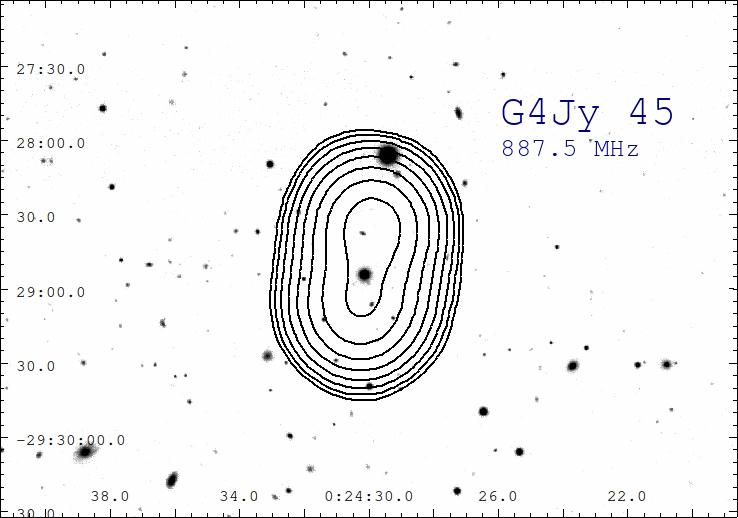}
    \includegraphics[scale=0.225]{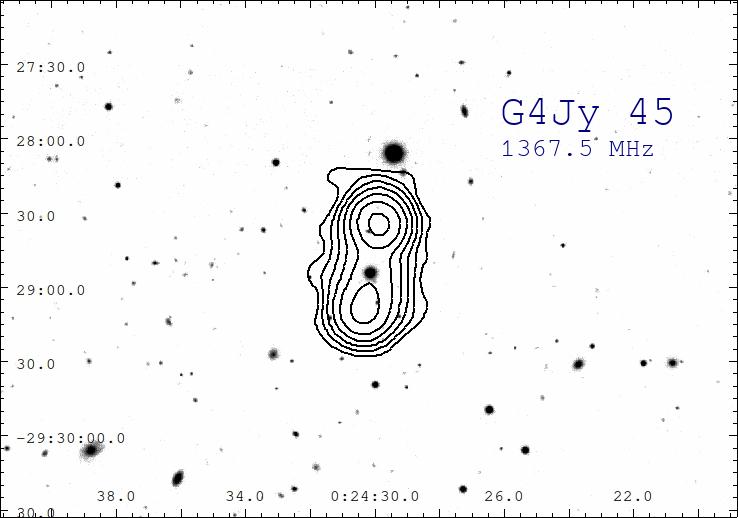}
    \includegraphics[scale=0.225]{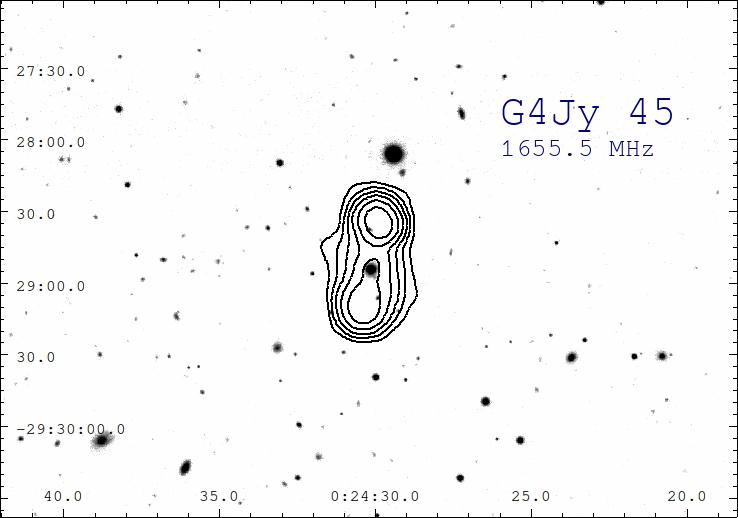}
    \includegraphics[scale=0.225]{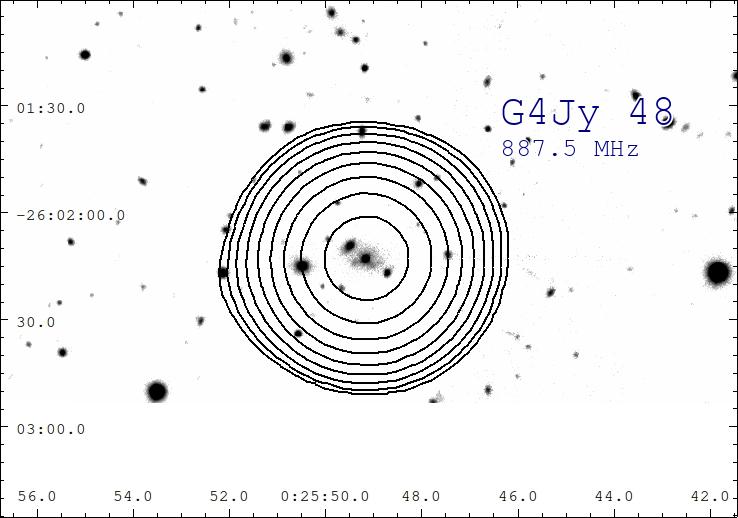}
    \includegraphics[scale=0.225]{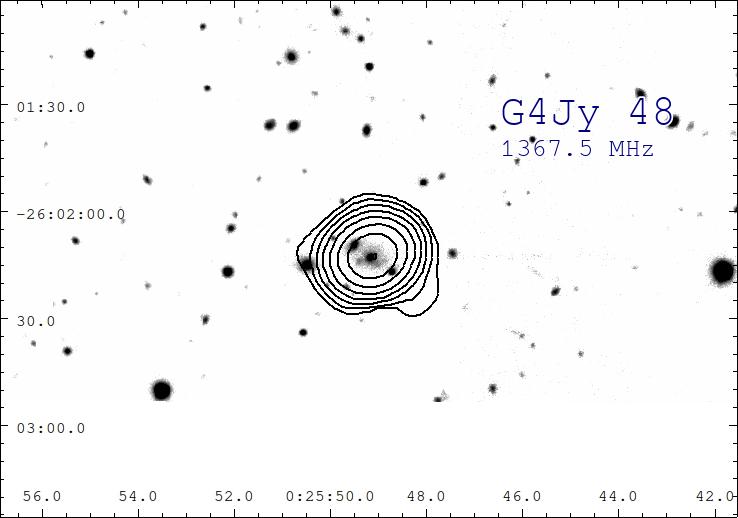}
    \includegraphics[scale=0.225]{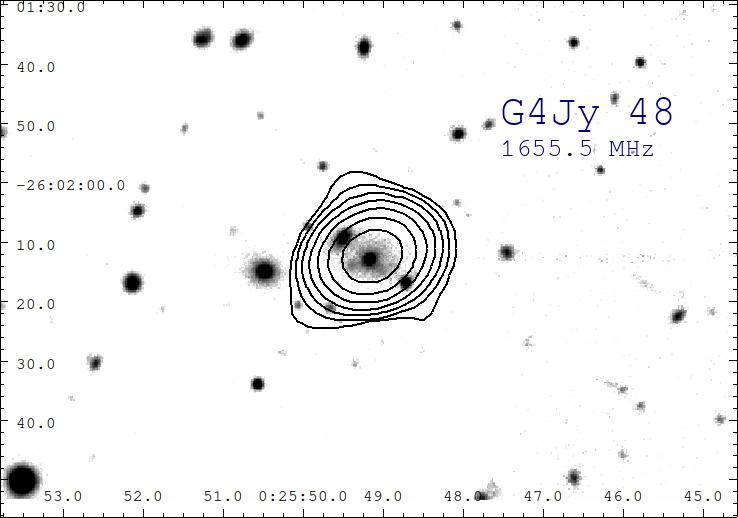}
    \caption{}
    \label{B}
\end{figure*}
\clearpage
\begin{figure*}
    \centering
    \includegraphics[scale=0.225]{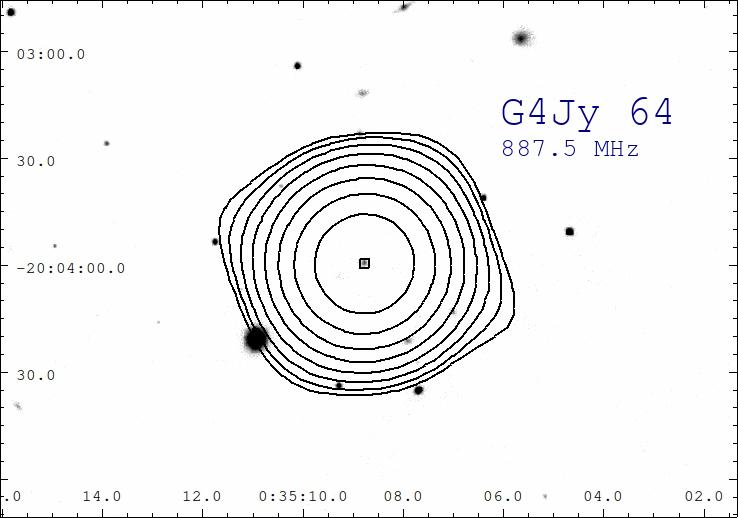}
    \includegraphics[scale=0.225]{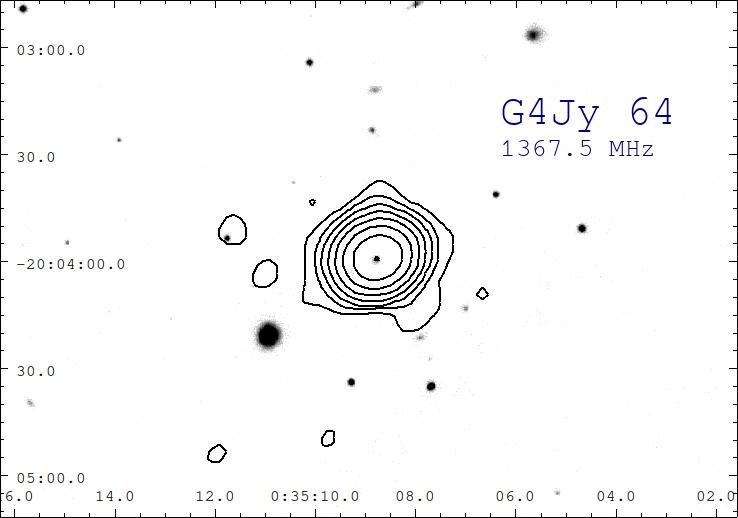}
    \includegraphics[scale=0.225]{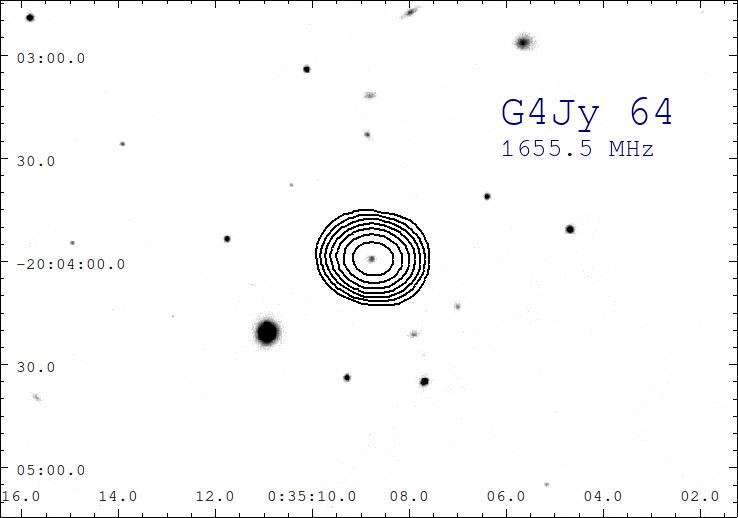}
    \includegraphics[scale=0.225]{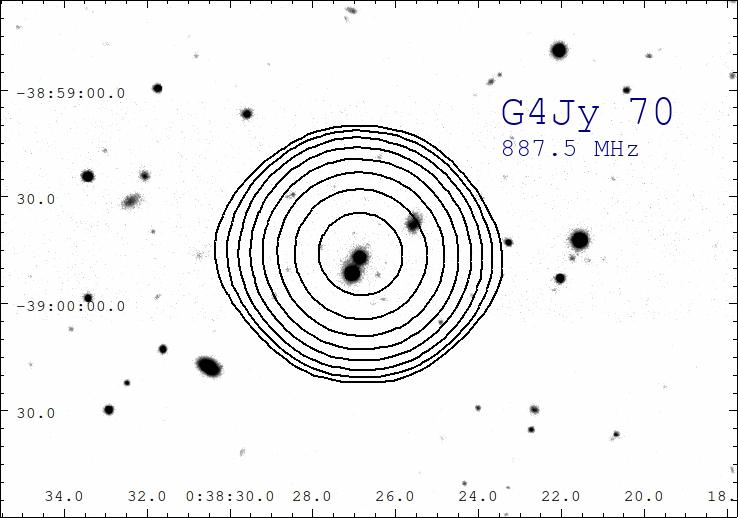}
    \includegraphics[scale=0.225]{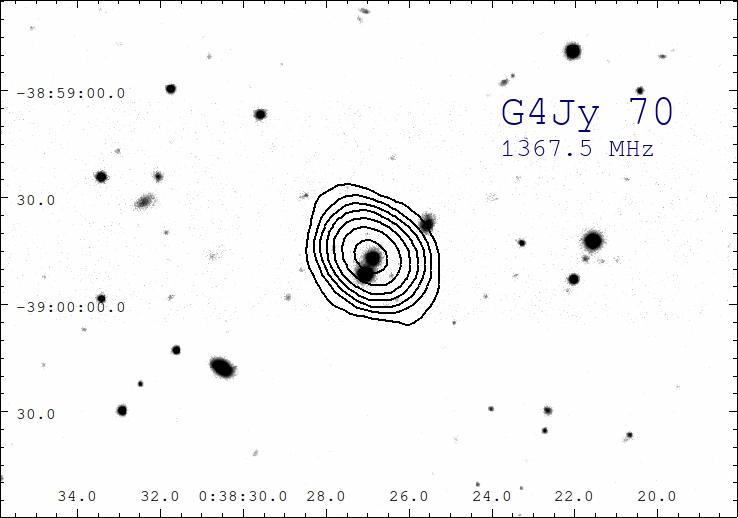}
    \includegraphics[scale=0.225]{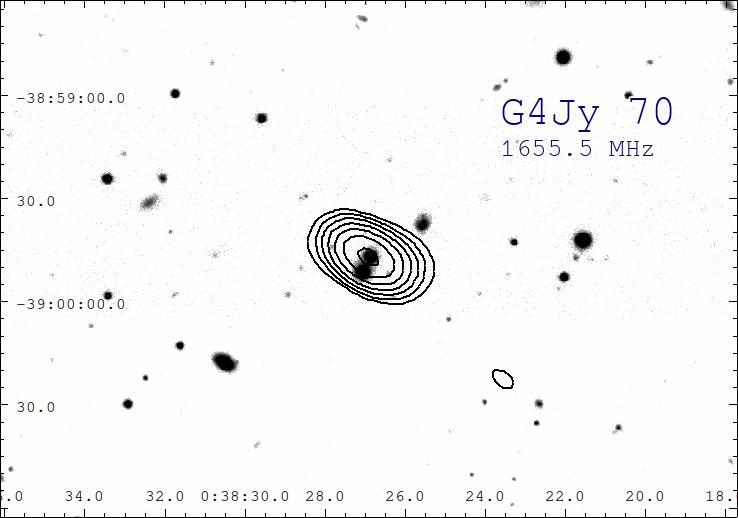}
    \includegraphics[scale=0.225]{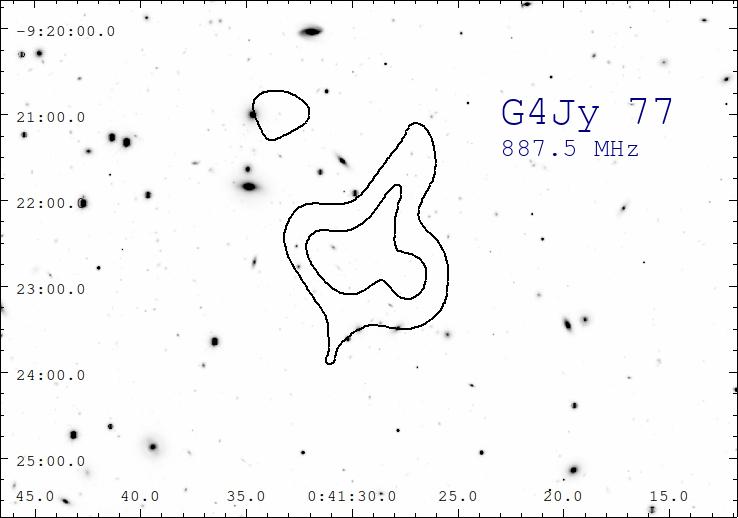}
    \includegraphics[scale=0.225]{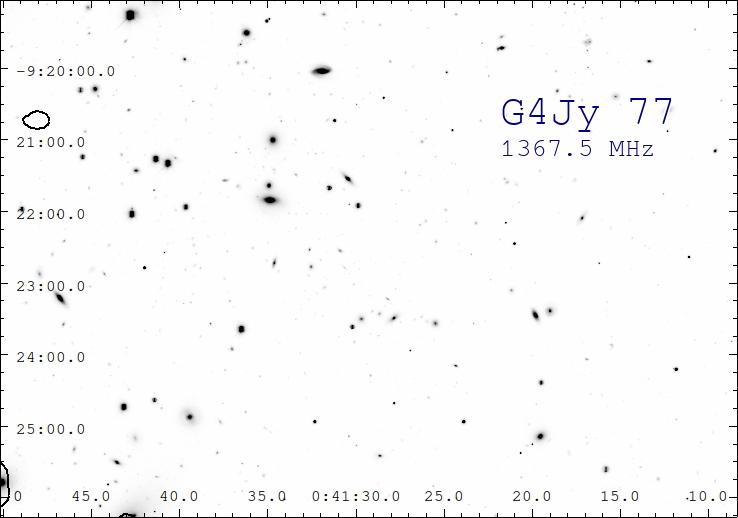}
    \includegraphics[scale=0.225]{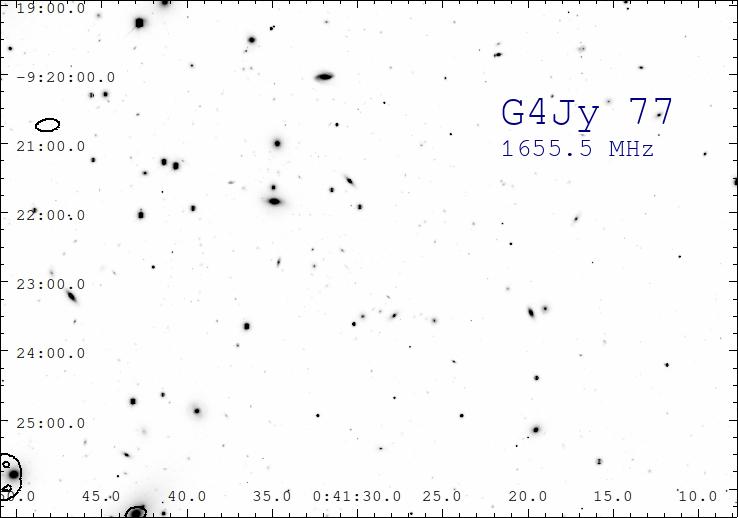}
    \includegraphics[scale=0.225]{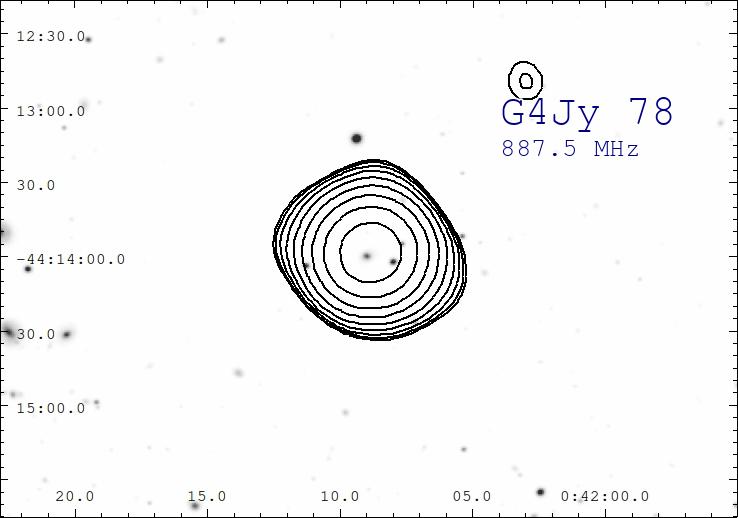}
    \includegraphics[scale=0.225]{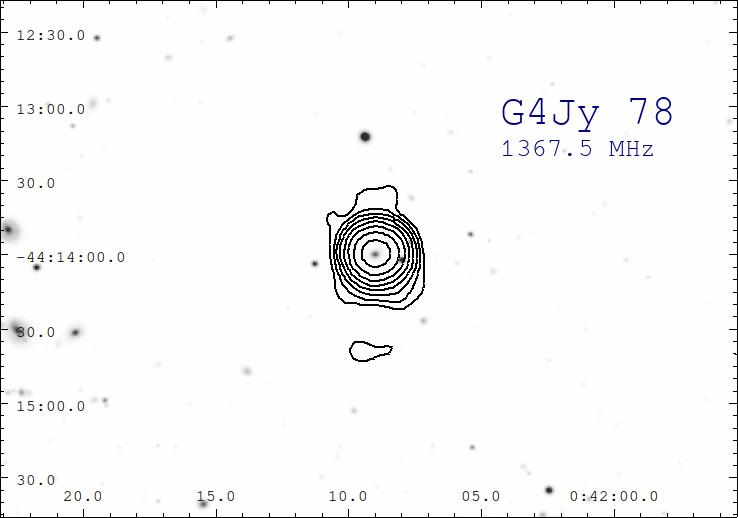}
    \includegraphics[scale=0.225]{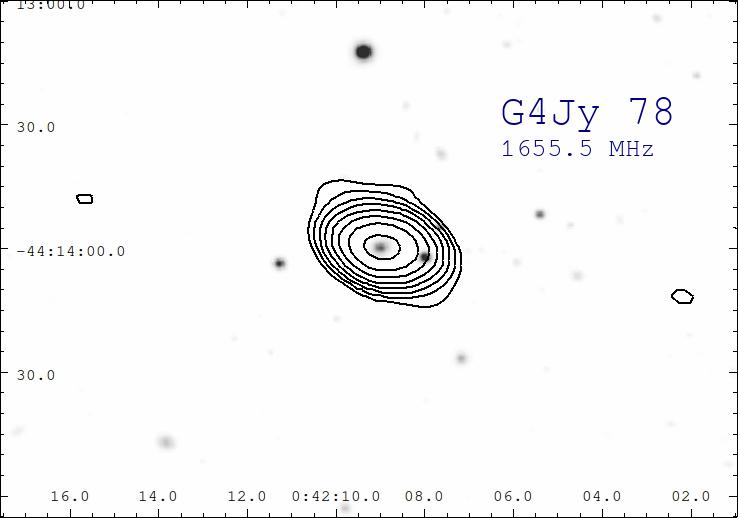}
    \includegraphics[scale=0.225]{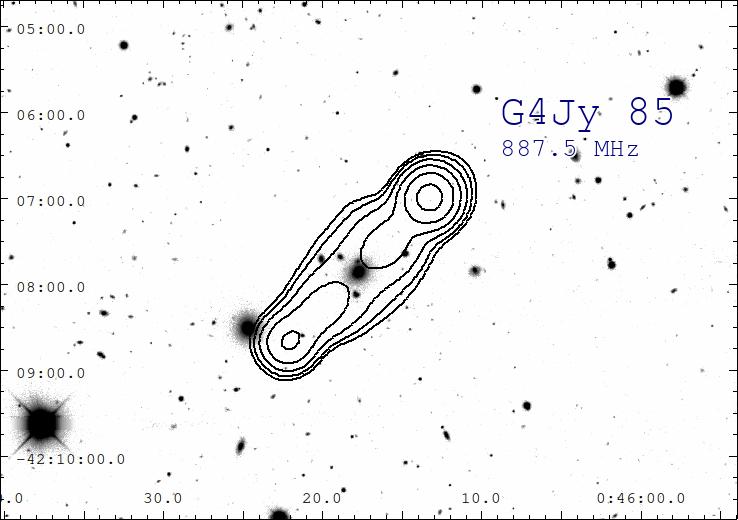}
    \includegraphics[scale=0.225]{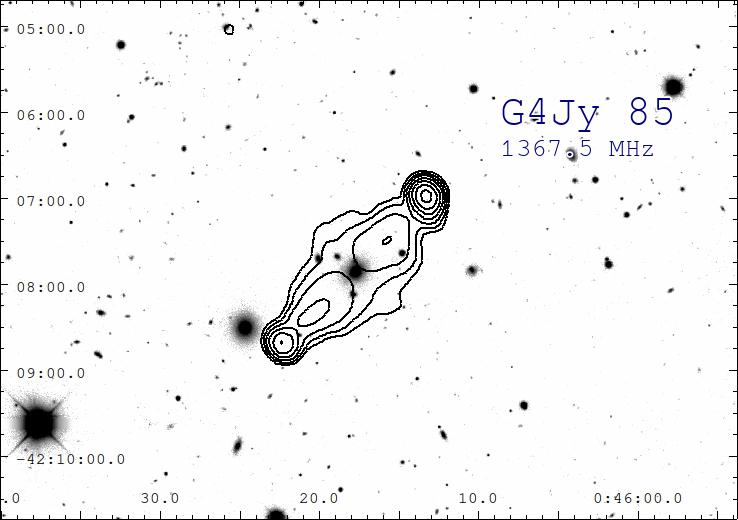}
    \includegraphics[scale=0.225]{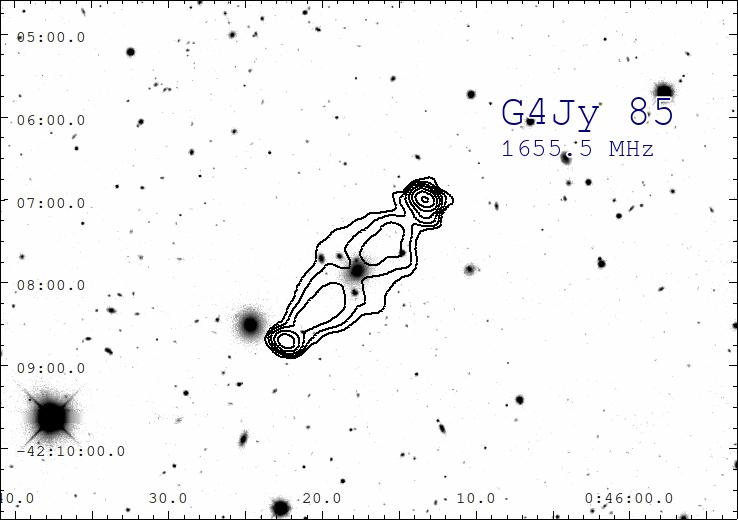}
    \caption{No signal was detected in RACS-mid and RACS-high for G4Jy 77.}
    \label{C}
\end{figure*}
\clearpage
\begin{figure*}
    \centering
    \includegraphics[scale=0.225]{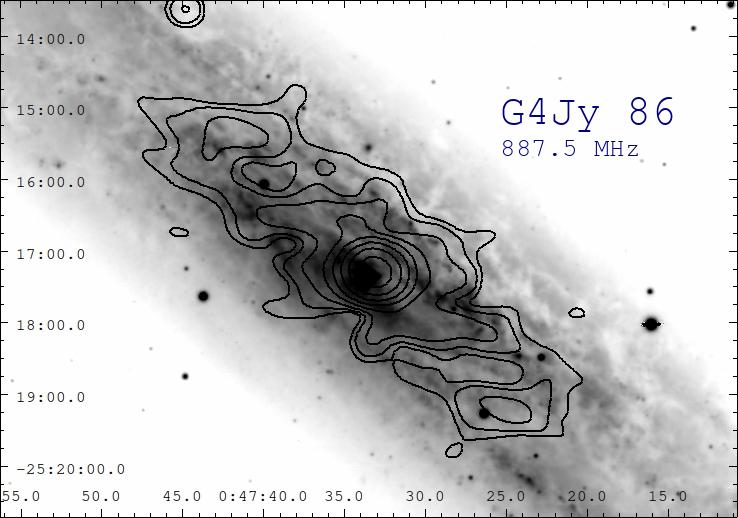}
    \includegraphics[scale=0.225]{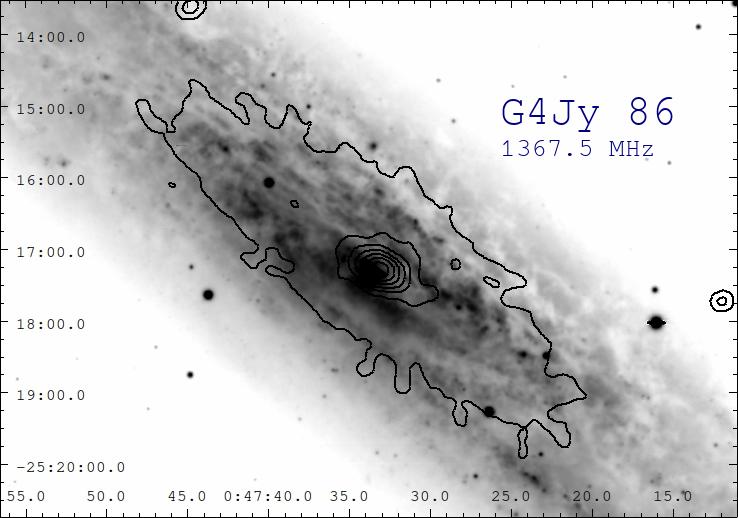}
    \includegraphics[scale=0.225]{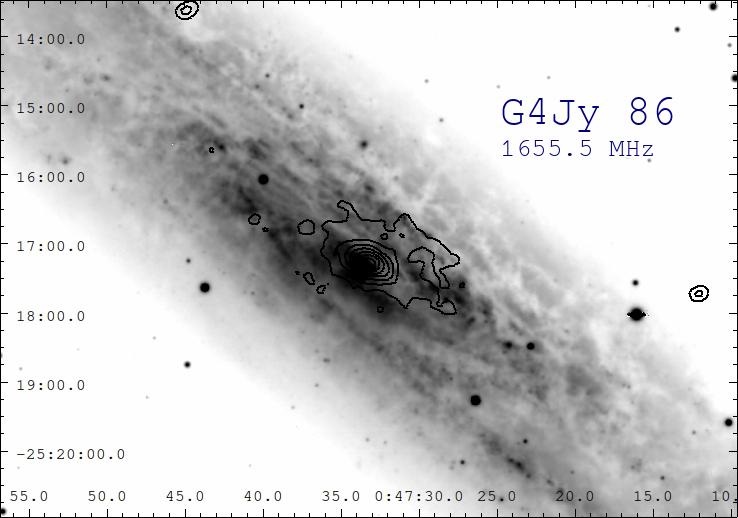}
    \includegraphics[scale=0.225]{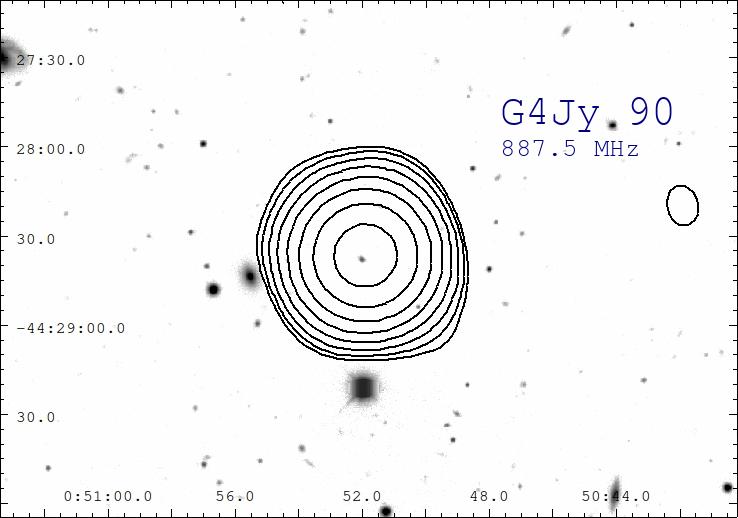}
    \includegraphics[scale=0.225]{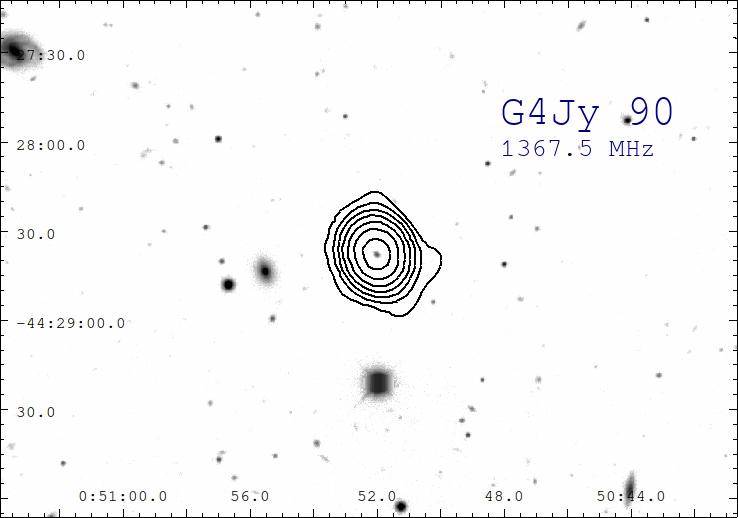}
    \includegraphics[scale=0.225]{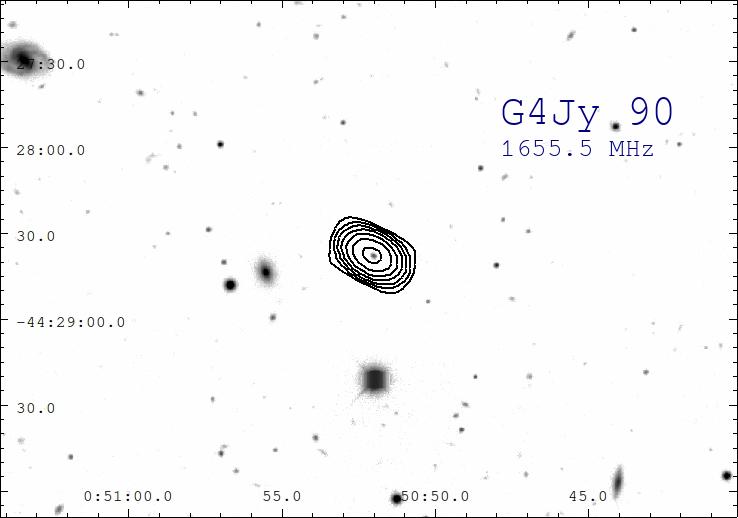}
    \includegraphics[scale=0.225]{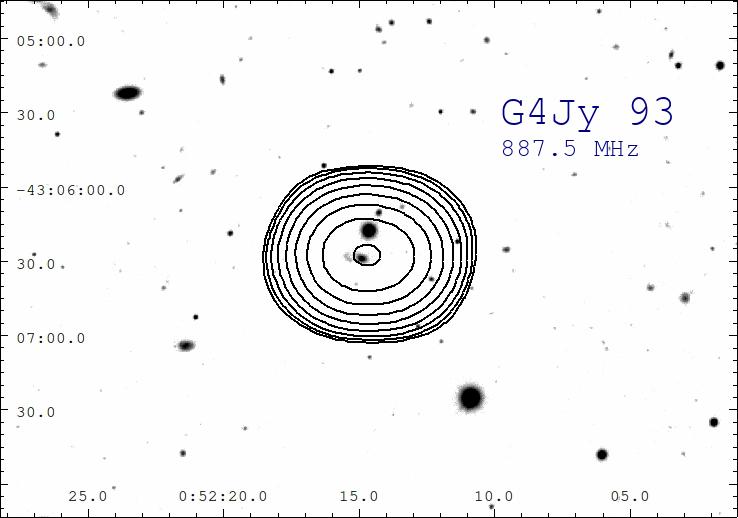}
    \includegraphics[scale=0.225]{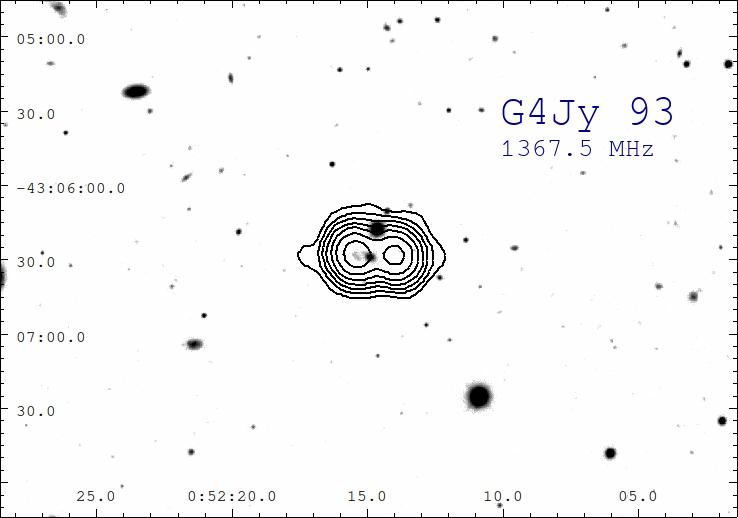}
    \includegraphics[scale=0.225]{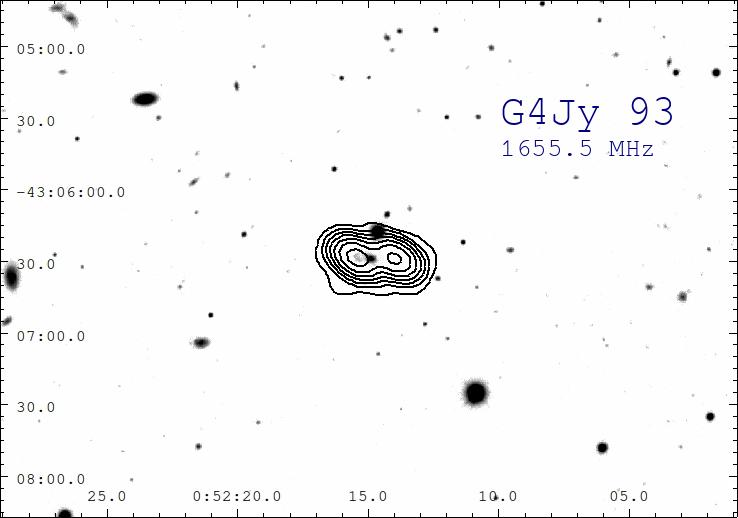}
    \includegraphics[scale=0.225]{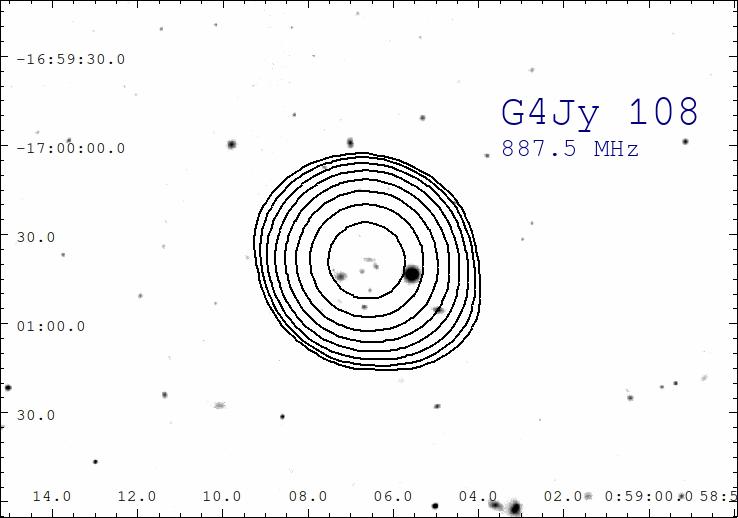}
    \includegraphics[scale=0.225]{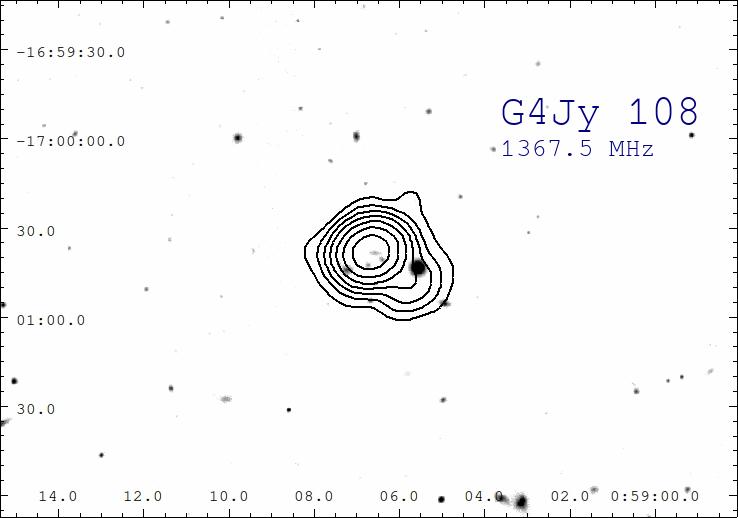}
    \includegraphics[scale=0.225]{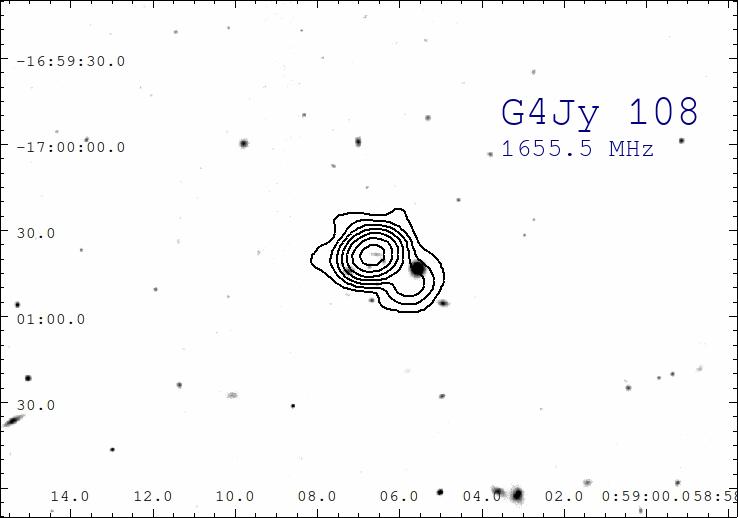}
    \includegraphics[scale=0.225]{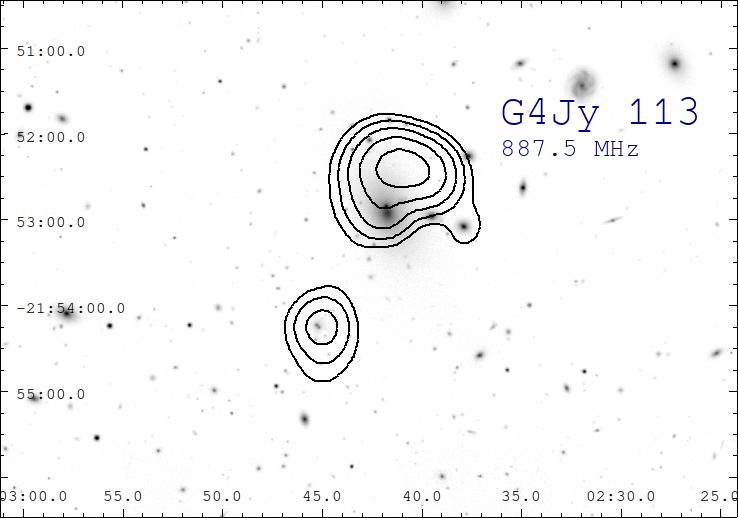}
    \includegraphics[scale=0.225]{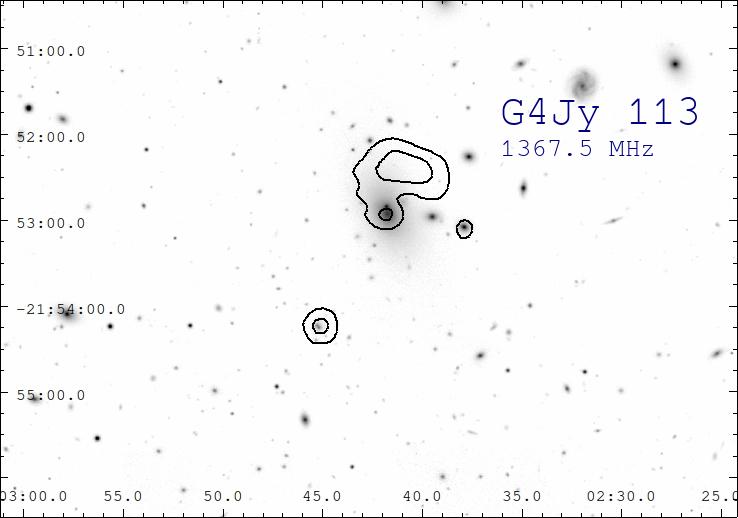}
    \includegraphics[scale=0.225]{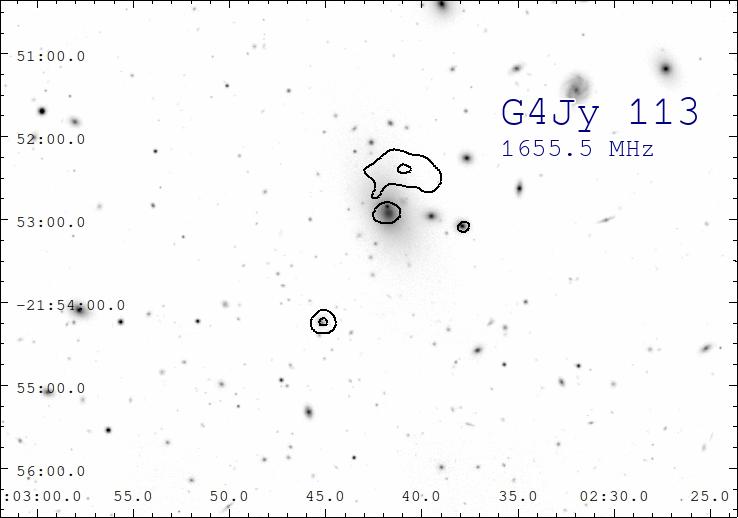}
    \caption{}
    \label{D}
\end{figure*}
\clearpage
\begin{figure*}
    \centering
    \includegraphics[scale=0.225]{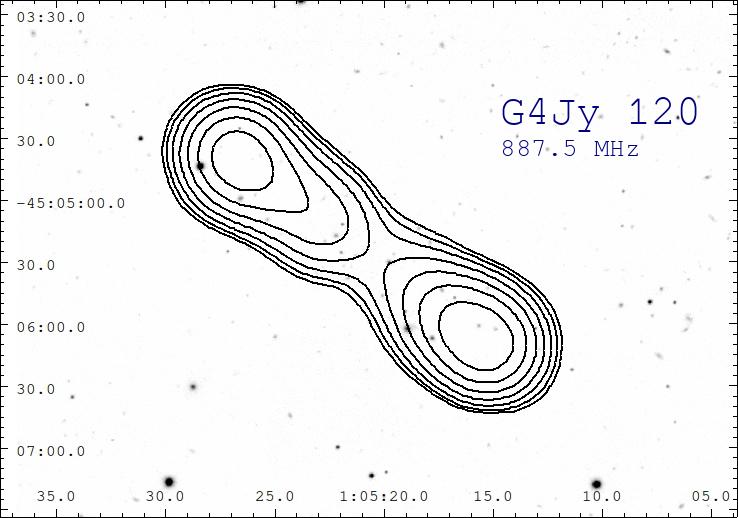}
    \includegraphics[scale=0.225]{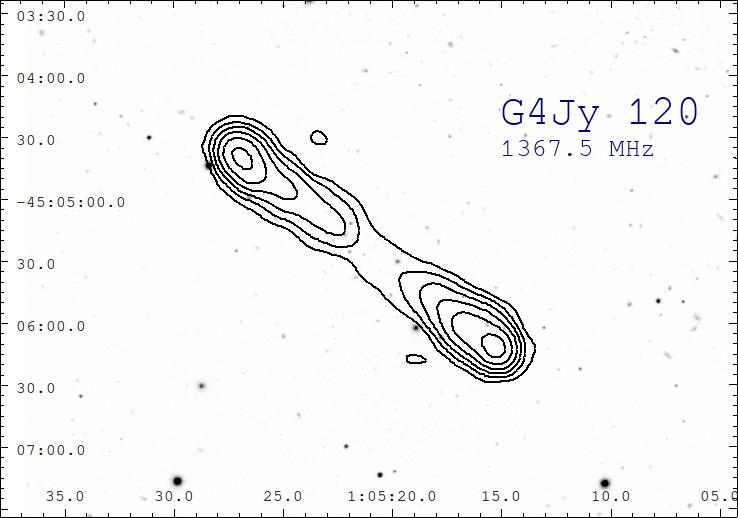}
    \includegraphics[scale=0.225]{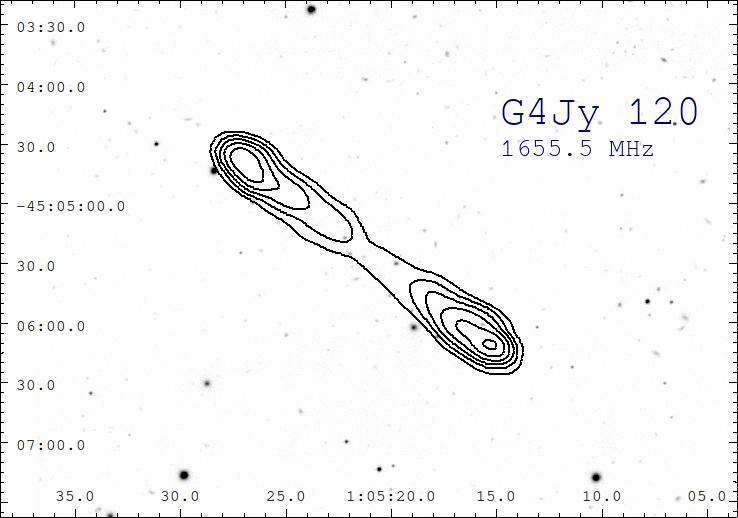}
    \includegraphics[scale=0.225]{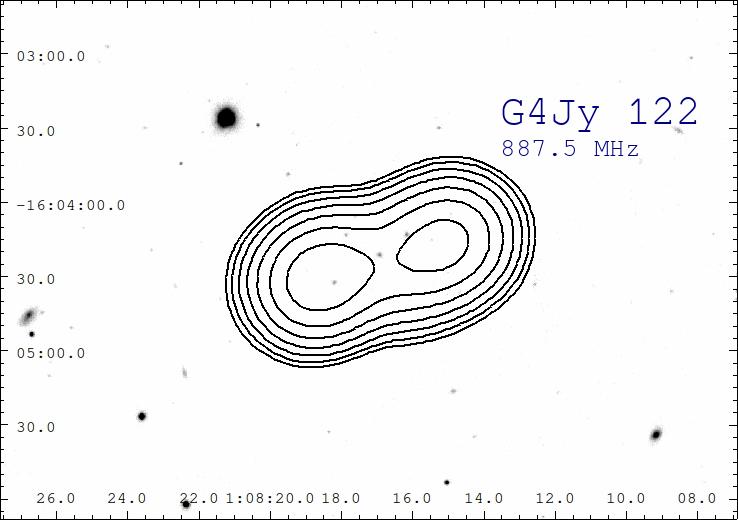}
    \includegraphics[scale=0.225]{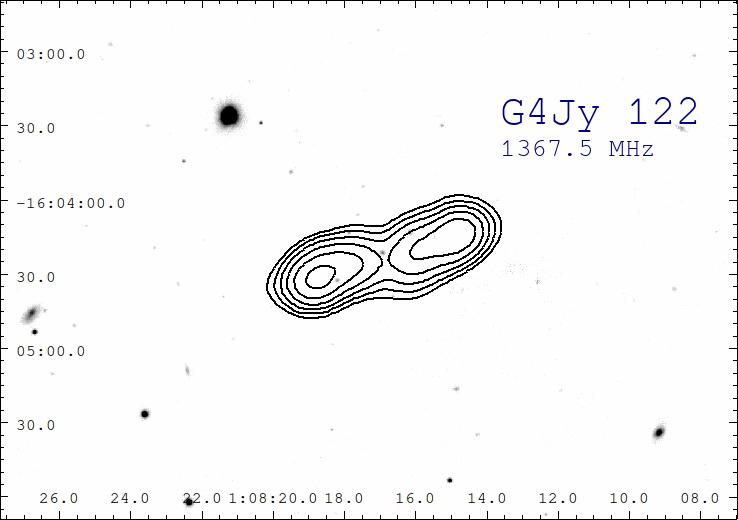}
    \includegraphics[scale=0.225]{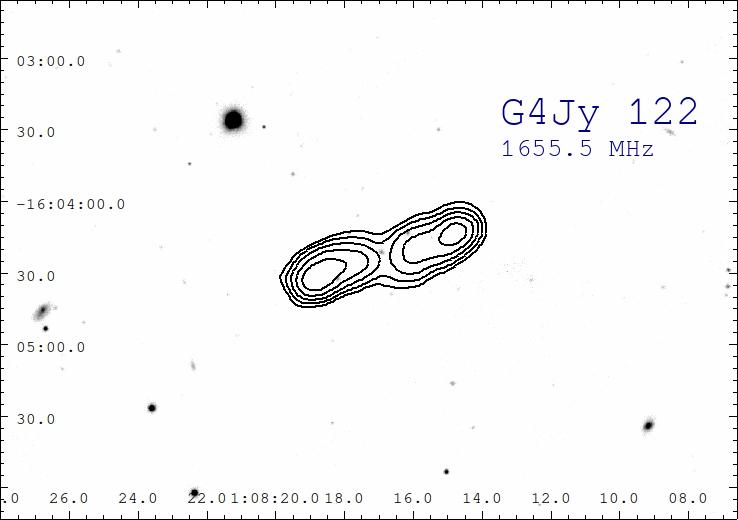}
    \includegraphics[scale=0.225]{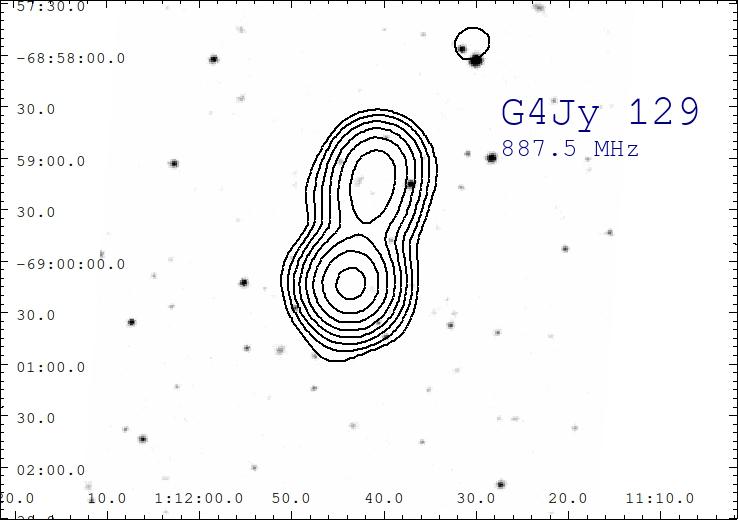}
    \includegraphics[scale=0.225]{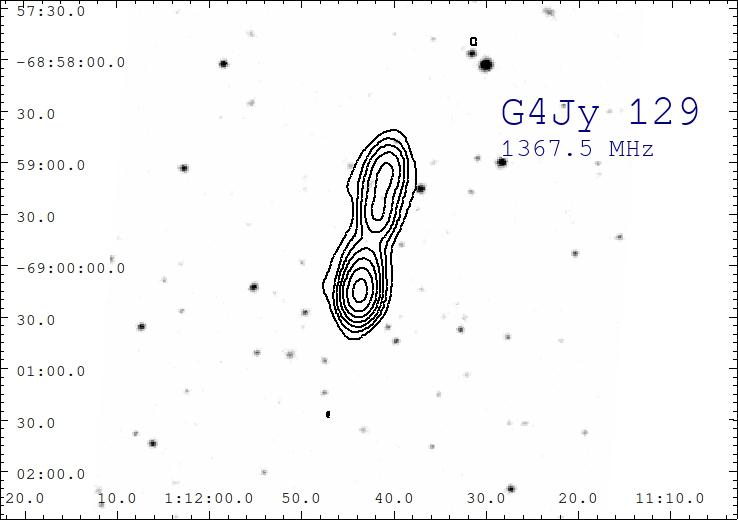}
    \includegraphics[scale=0.225]{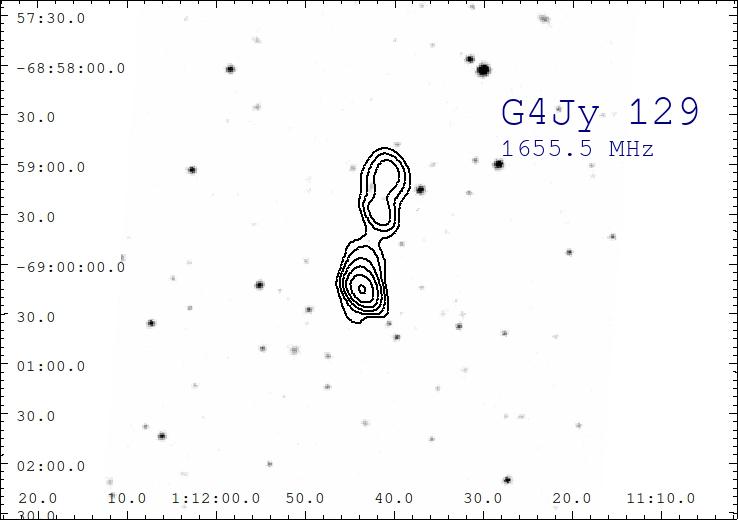}
    \includegraphics[scale=0.225]{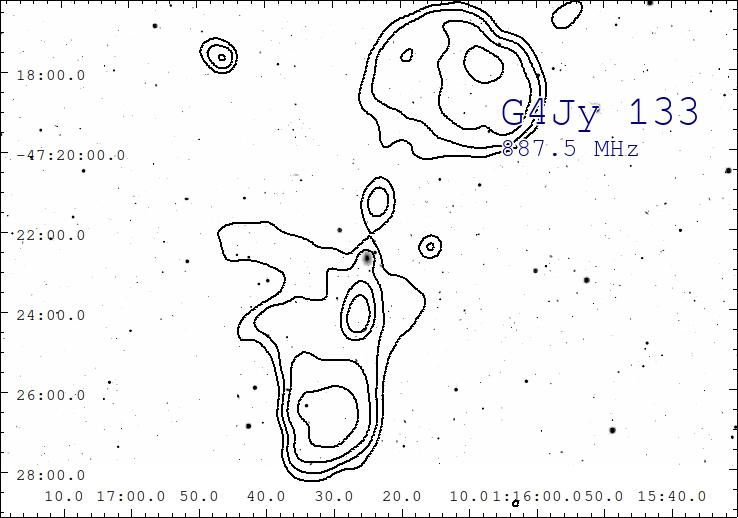}
    \includegraphics[scale=0.225]{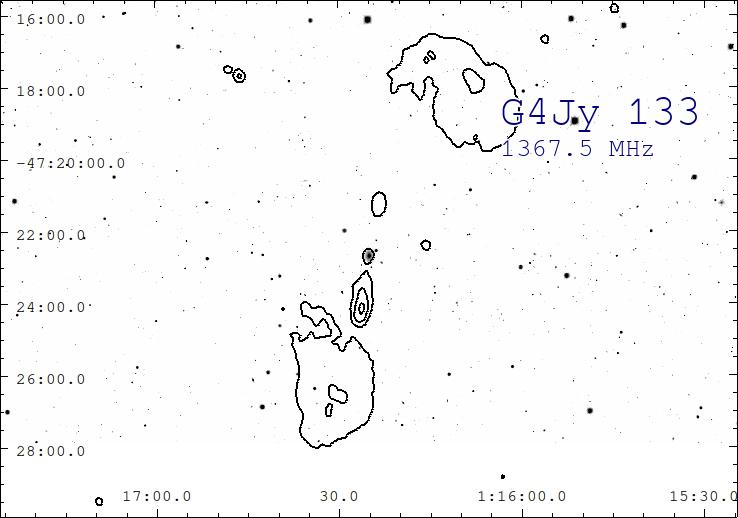}
    \includegraphics[scale=0.225]{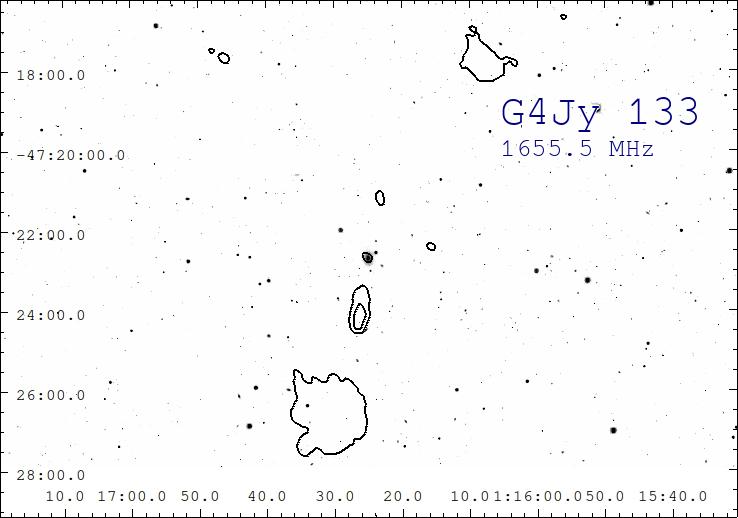}
    \includegraphics[scale=0.225]{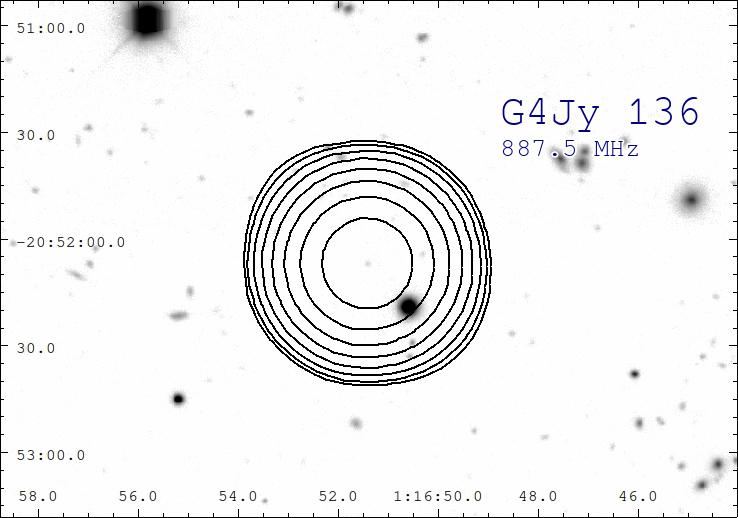}
    \includegraphics[scale=0.225]{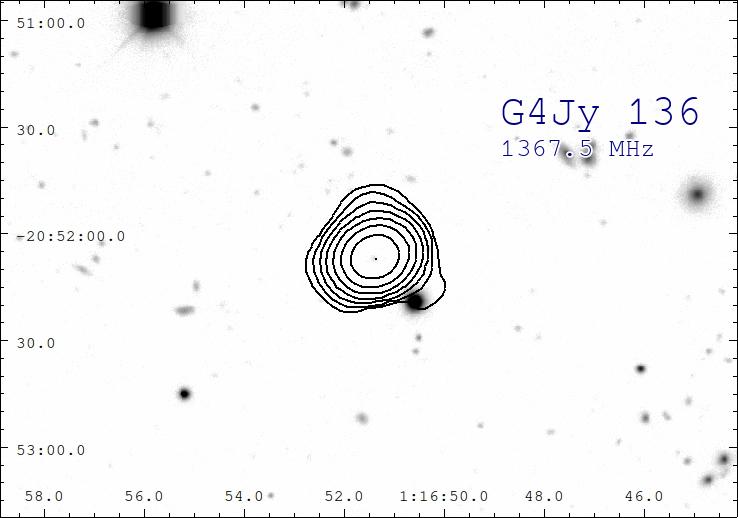}
    \includegraphics[scale=0.225]{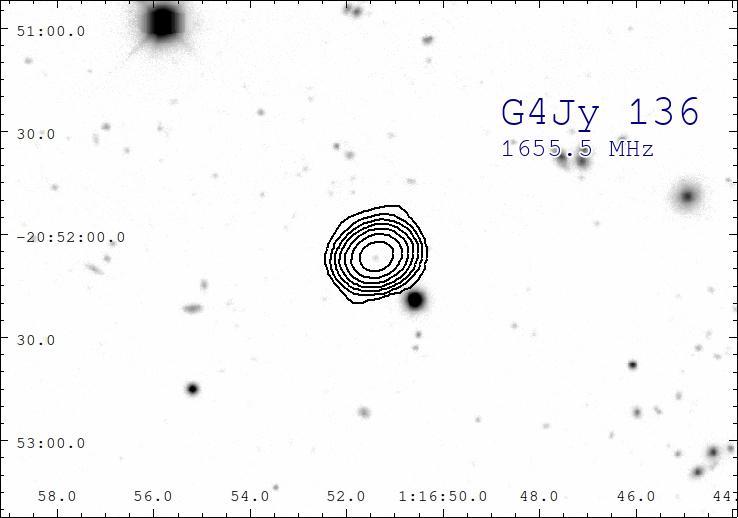}
    \caption{}
    \label{E}
\end{figure*}
\clearpage
\begin{figure*}
    \centering
    \includegraphics[scale=0.225]{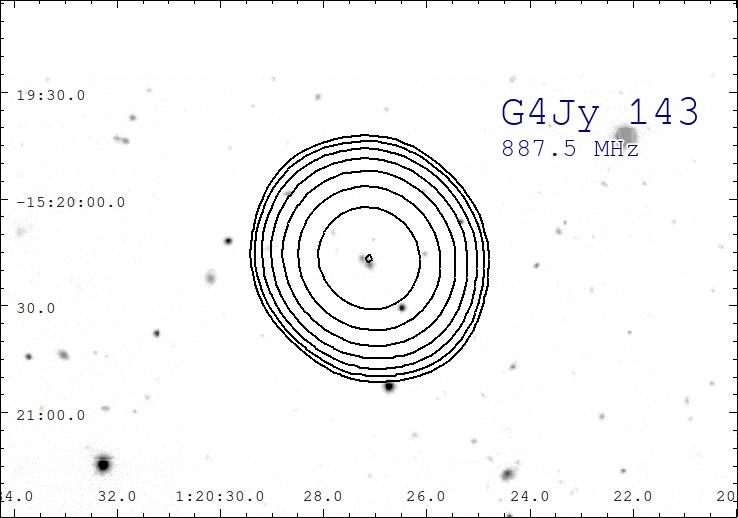}
    \includegraphics[scale=0.225]{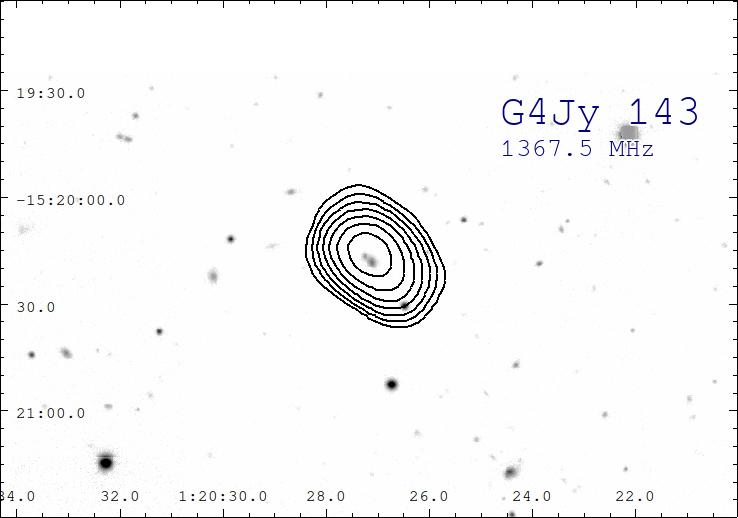}
    \includegraphics[scale=0.225]{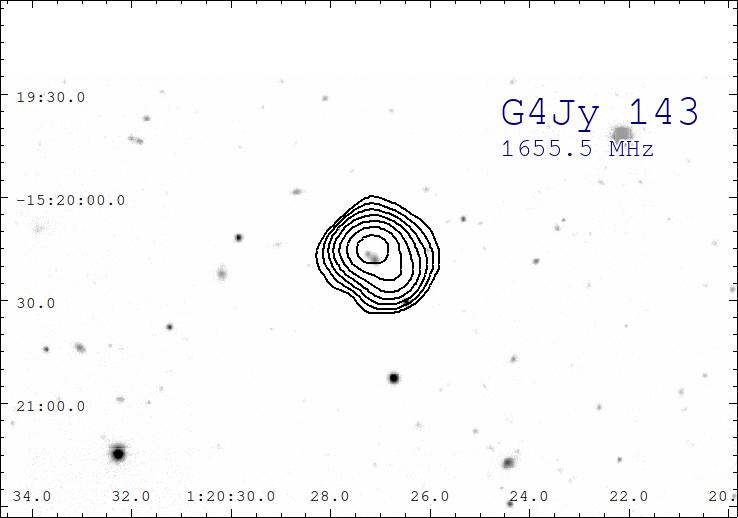}
    \includegraphics[scale=0.225]{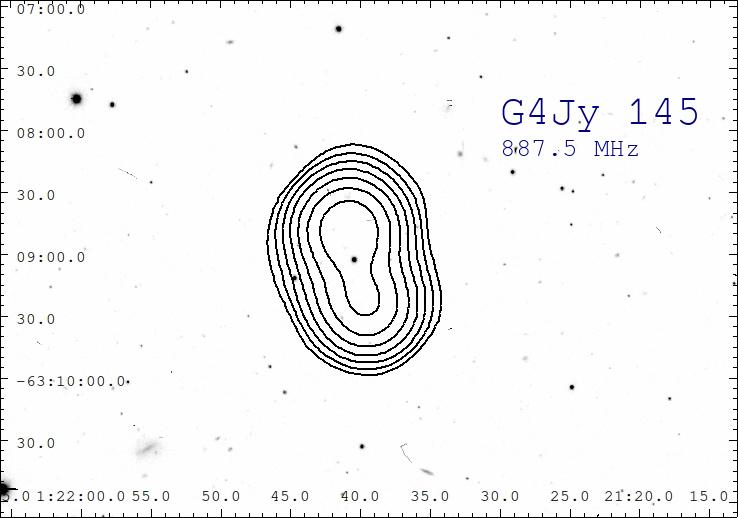}
    \includegraphics[scale=0.225]{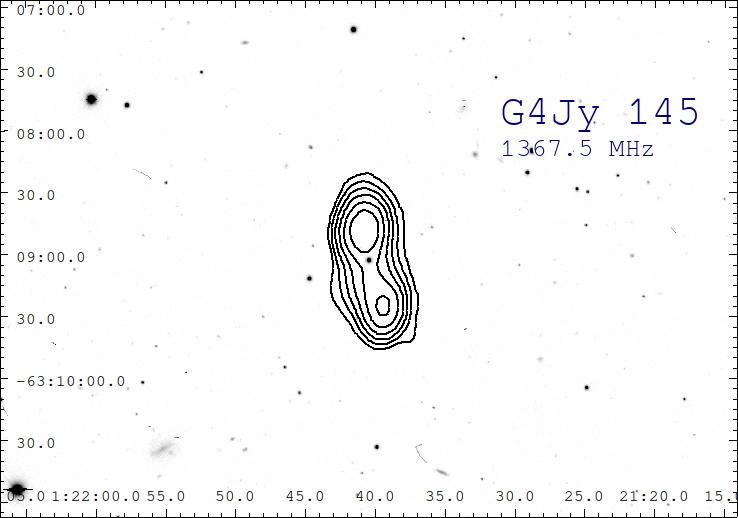}
    \includegraphics[scale=0.225]{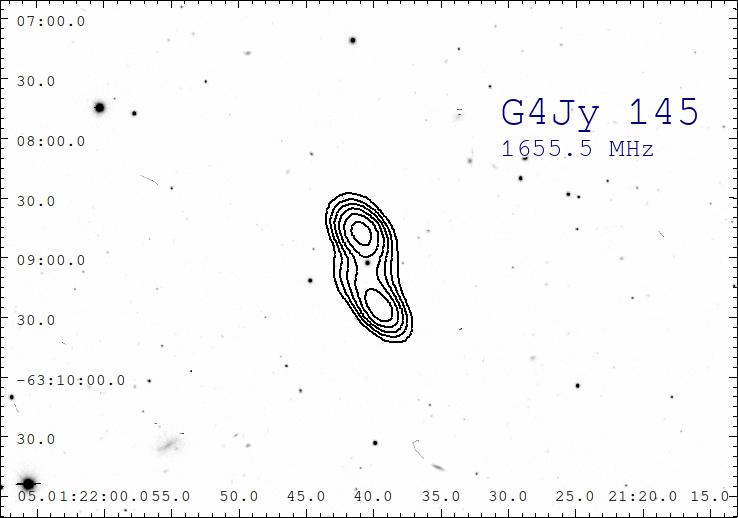}
    \includegraphics[scale=0.225]{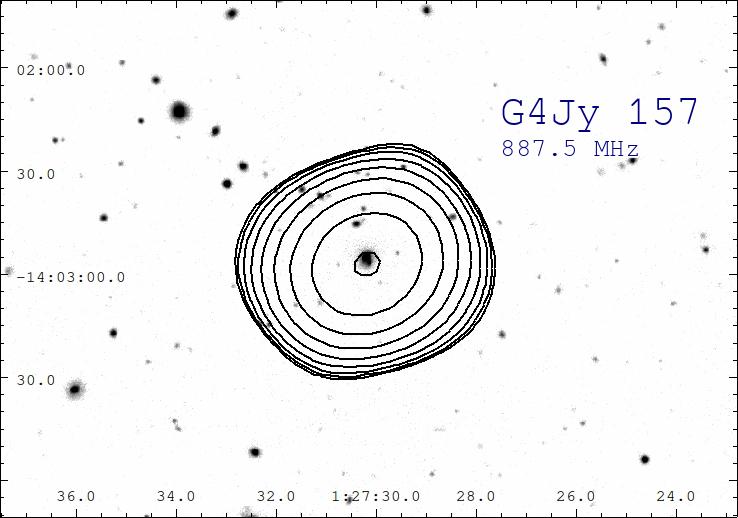}
    \includegraphics[scale=0.225]{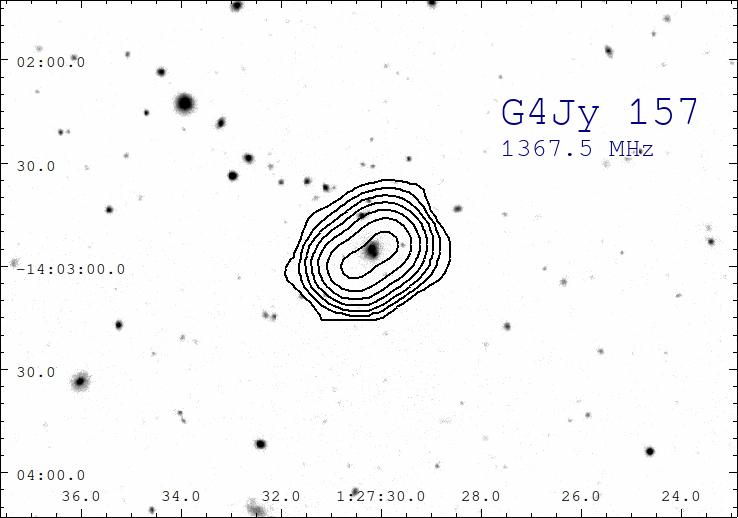}
    \includegraphics[scale=0.225]{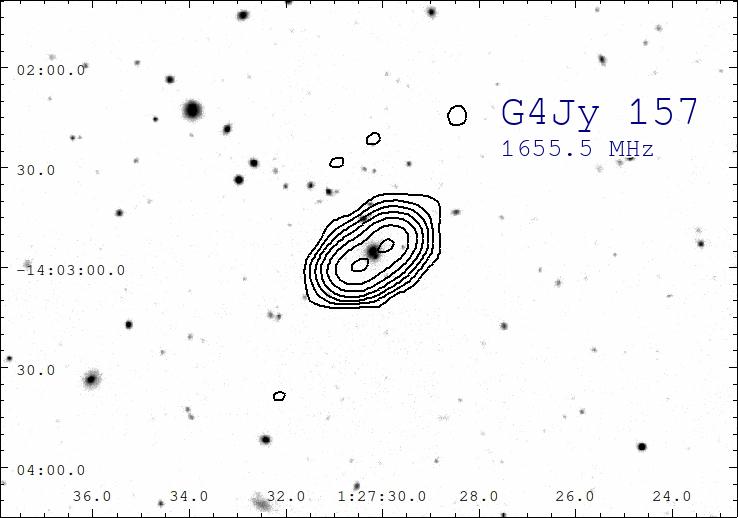}
    \includegraphics[scale=0.225]{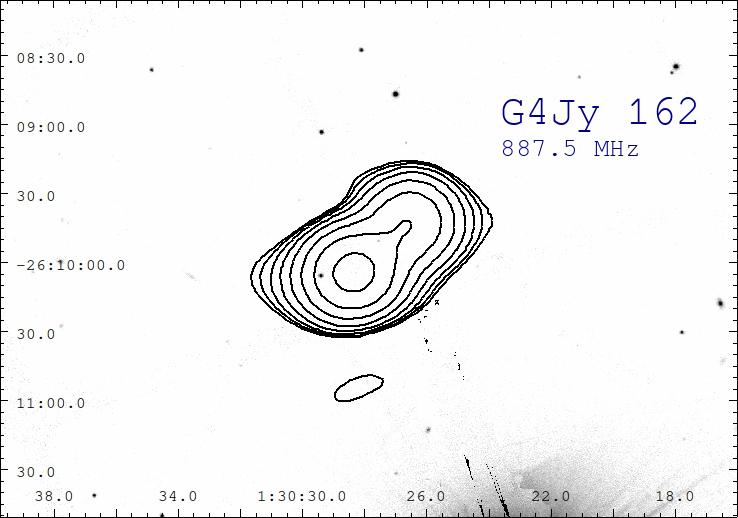}
    \includegraphics[scale=0.225]{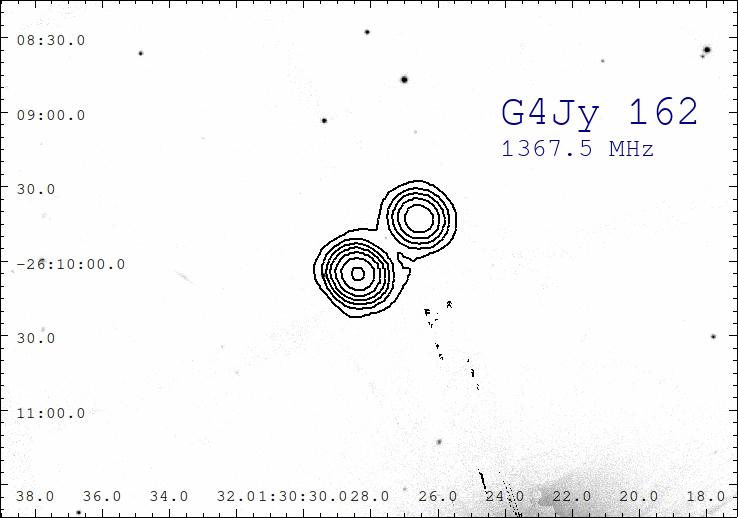}
    \includegraphics[scale=0.225]{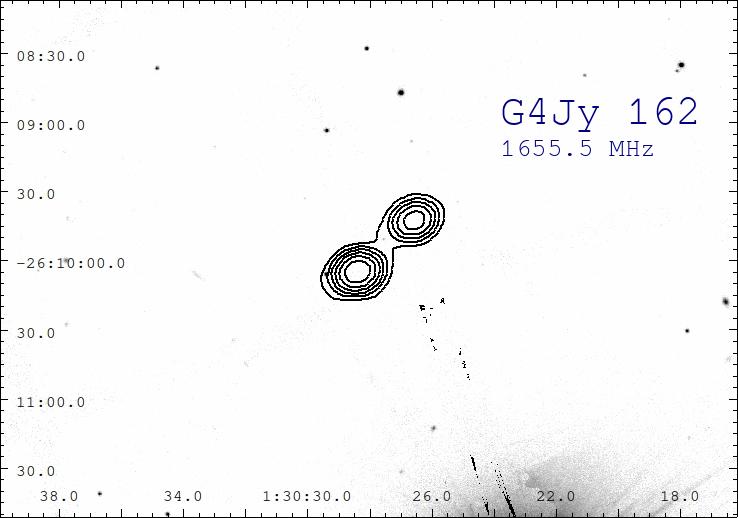}
    \includegraphics[scale=0.225]{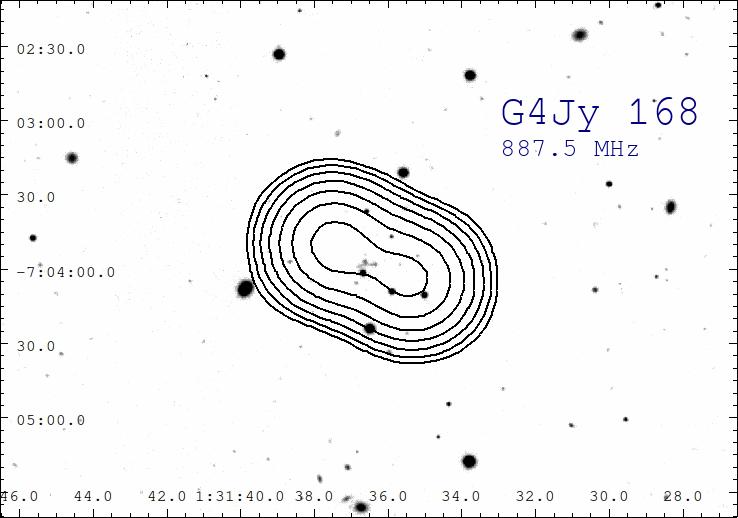}
    \includegraphics[scale=0.225]{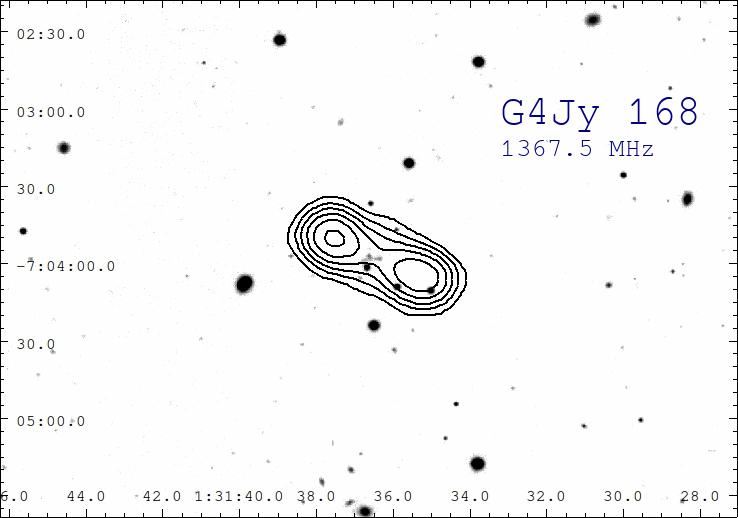}
    \includegraphics[scale=0.225]{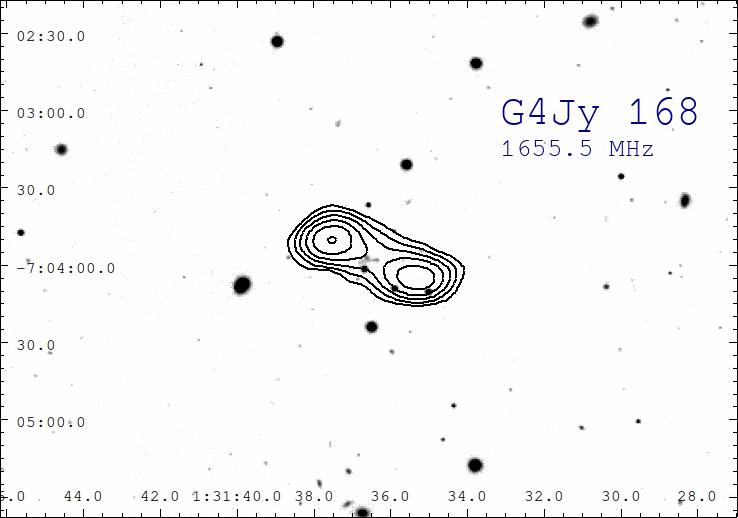}
    \caption{}
    \label{F}
\end{figure*}
\clearpage
\begin{figure*}
    \centering
    \includegraphics[scale=0.225]{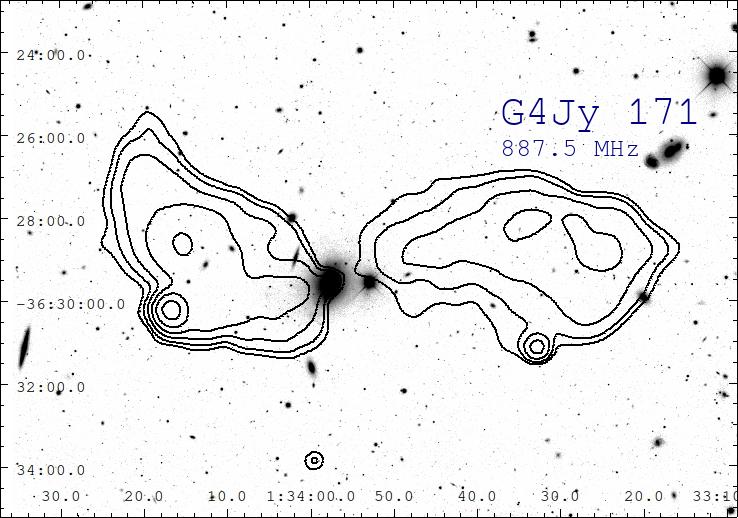}
    \includegraphics[scale=0.225]{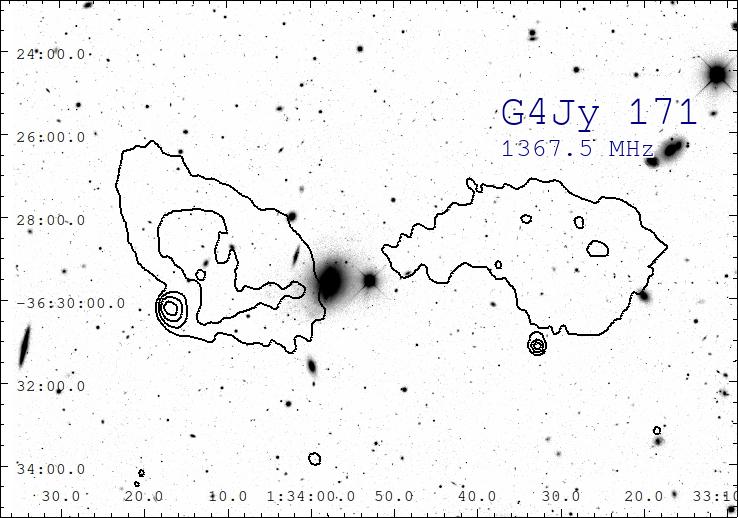}
    \includegraphics[scale=0.225]{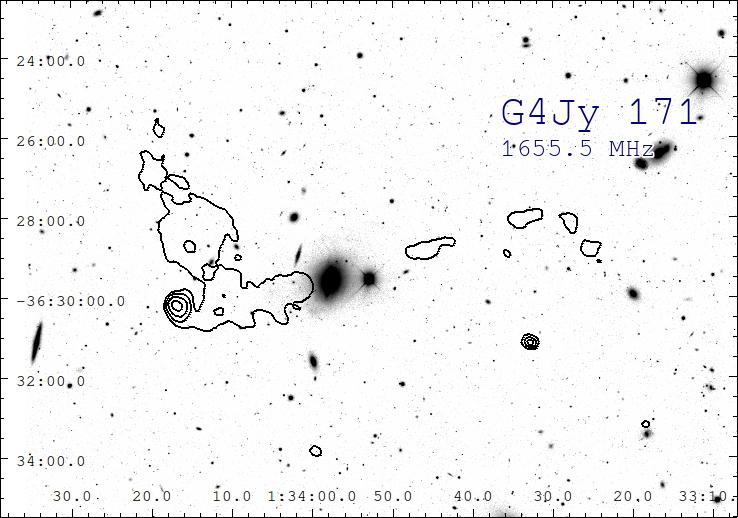}
    \includegraphics[scale=0.225]{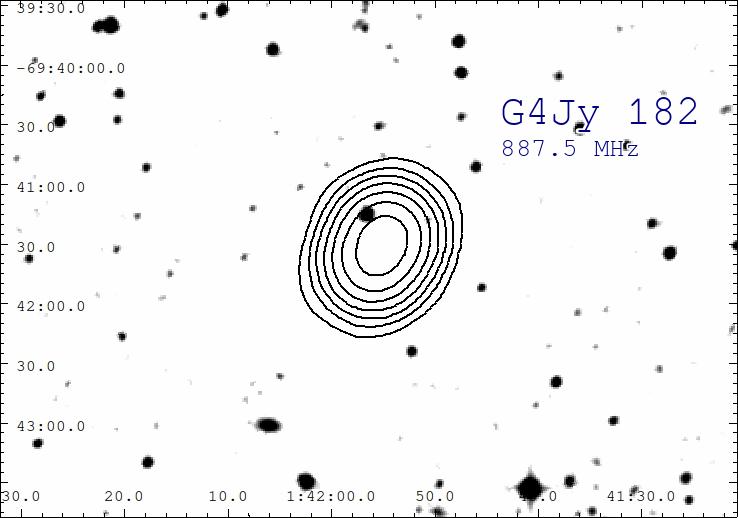}
    \includegraphics[scale=0.225]{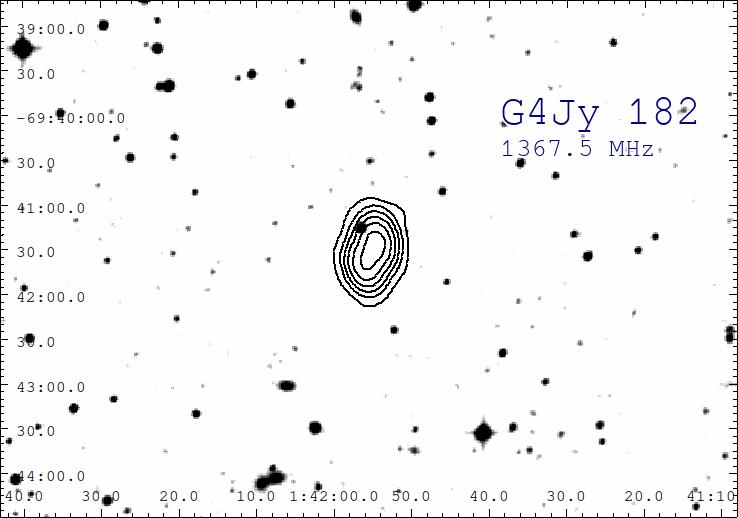}
    \includegraphics[scale=0.225]{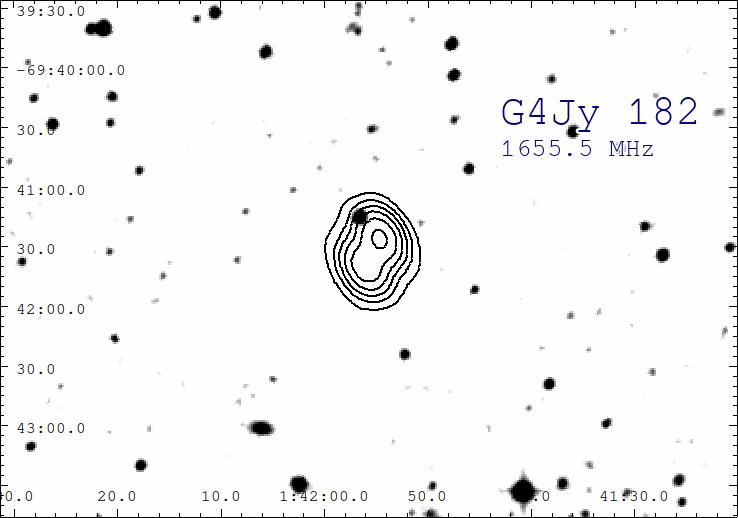}
    \includegraphics[scale=0.225]{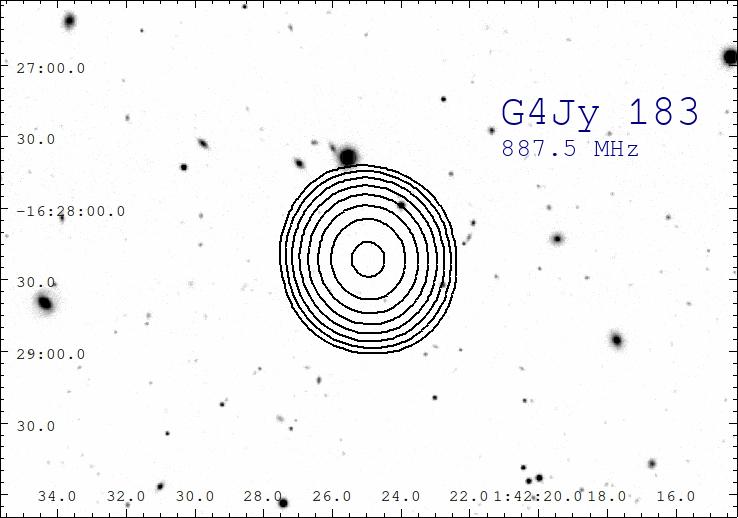}
    \includegraphics[scale=0.225]{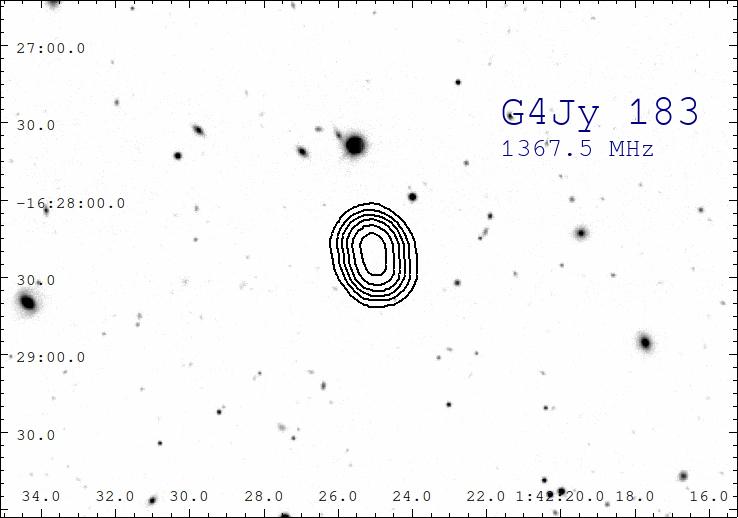}
    \includegraphics[scale=0.225]{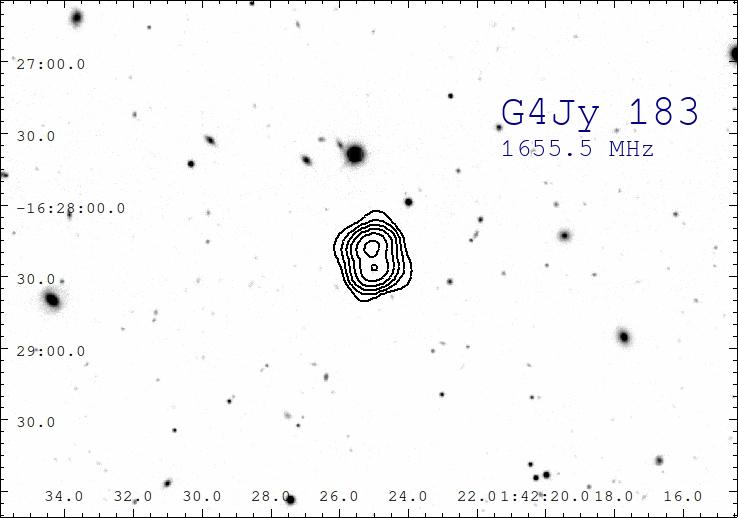}
    \includegraphics[scale=0.225]{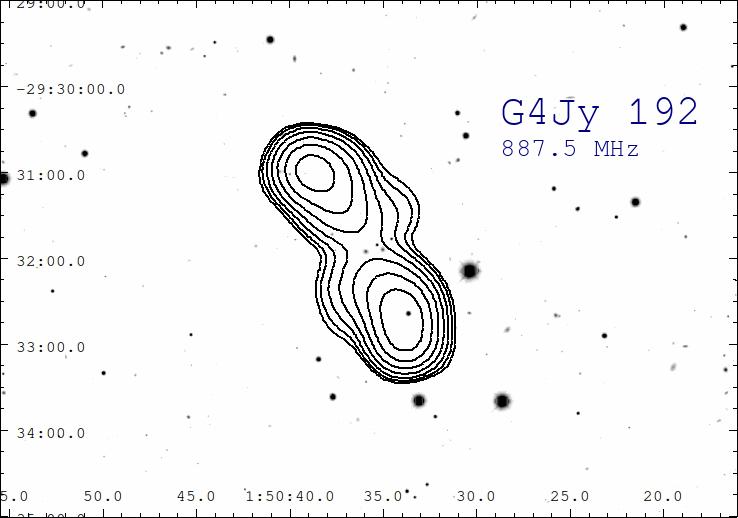}
    \includegraphics[scale=0.225]{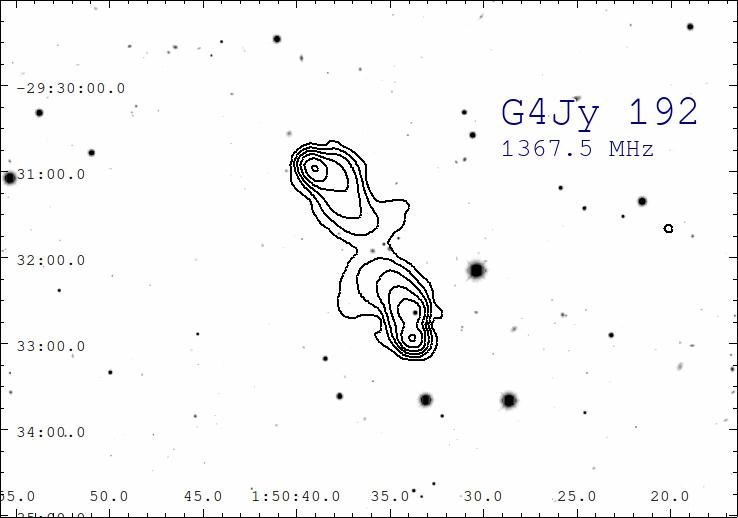}
    \includegraphics[scale=0.225]{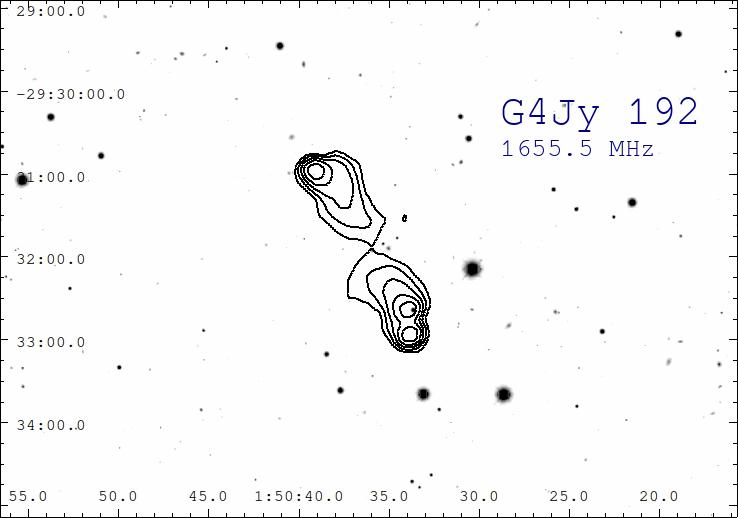}
    \includegraphics[scale=0.225]{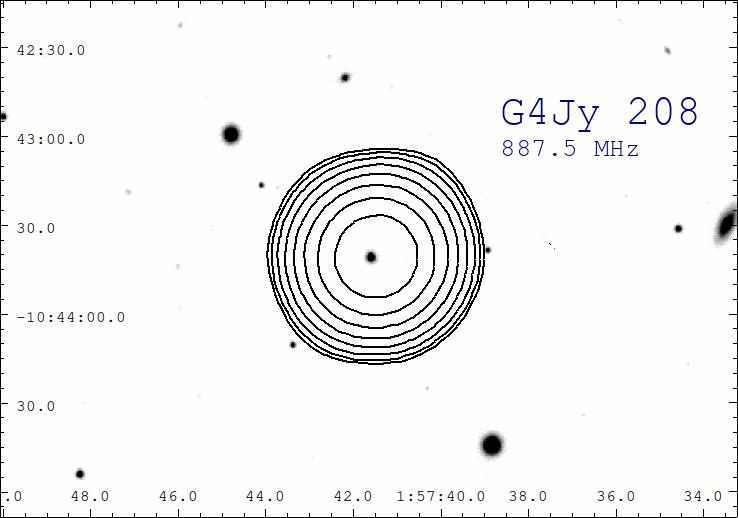}
    \includegraphics[scale=0.225]{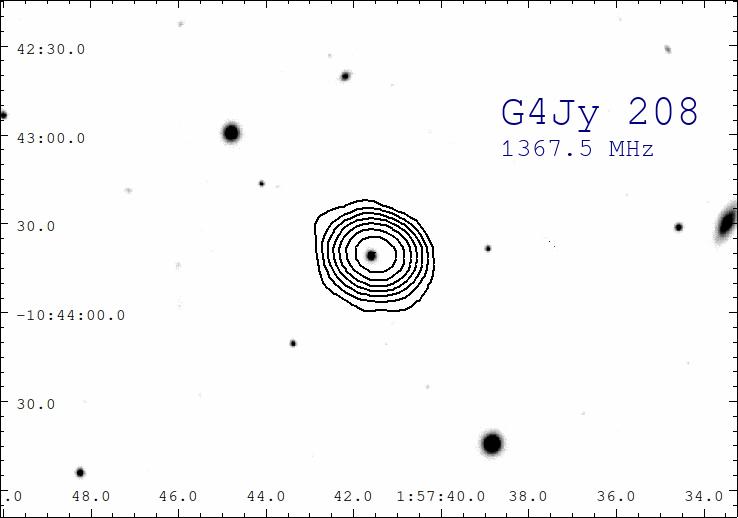}
    \includegraphics[scale=0.225]{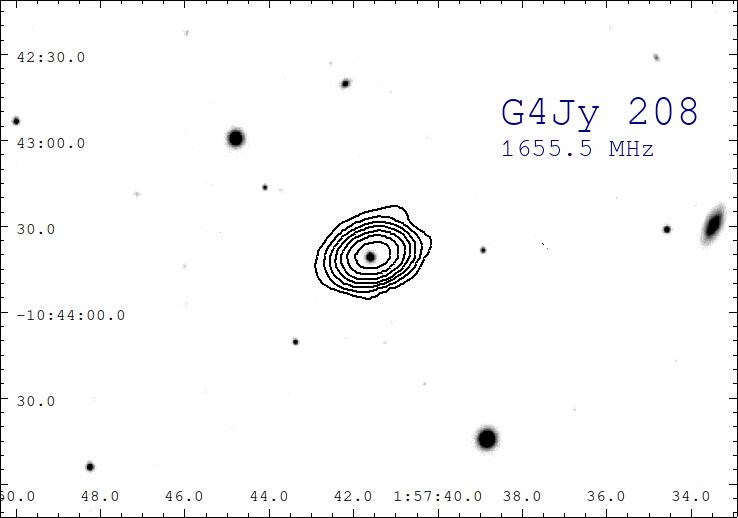}
    \caption{}
    \label{G}
\end{figure*}
\clearpage
\begin{figure*}
    \centering
    \includegraphics[scale=0.225]{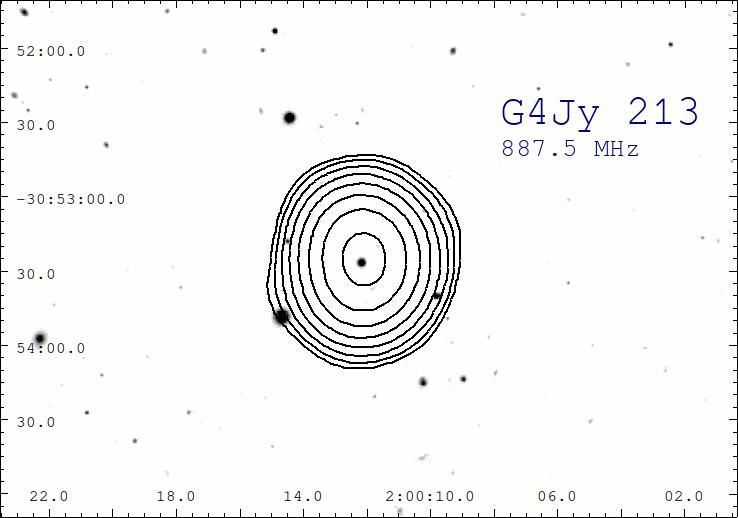}
    \includegraphics[scale=0.225]{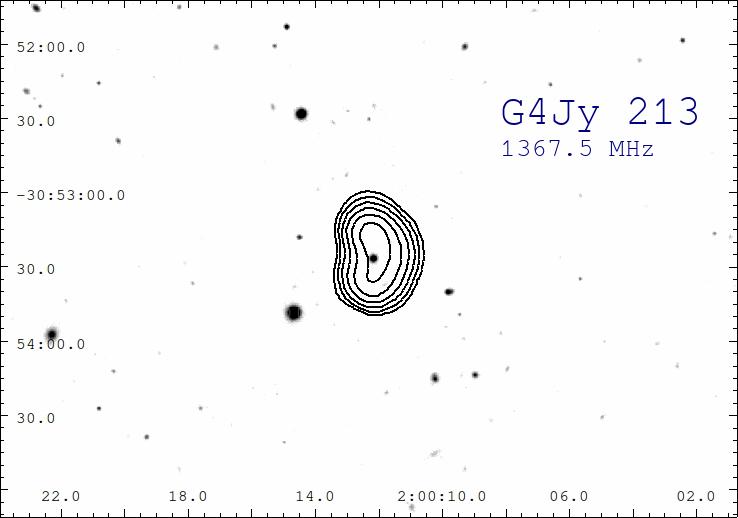}
    \includegraphics[scale=0.225]{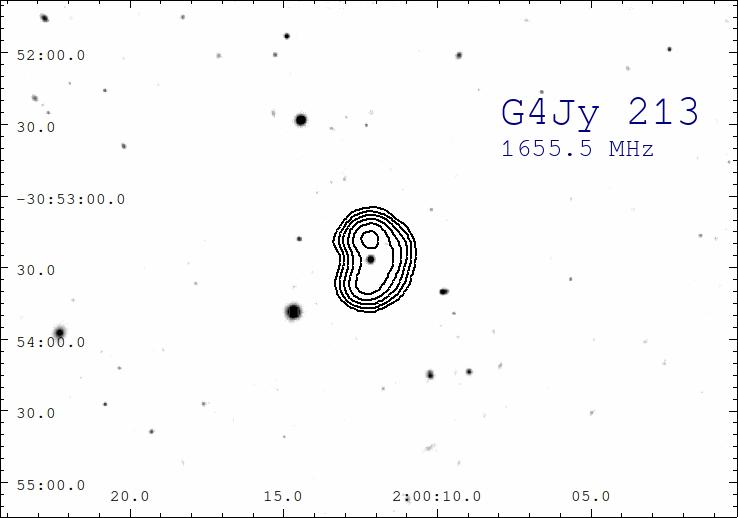}
    \includegraphics[scale=0.225]{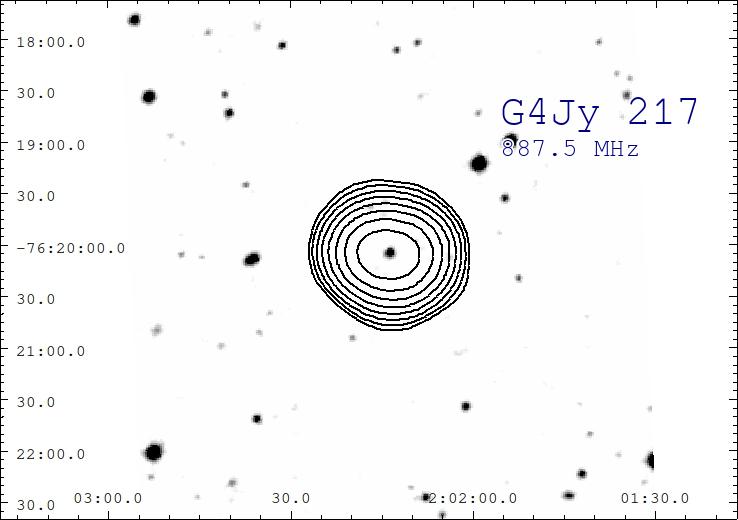}
    \includegraphics[scale=0.225]{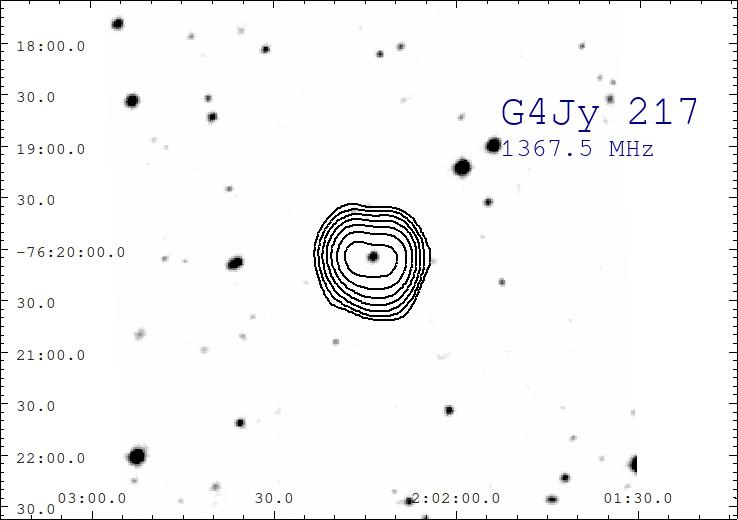}
    \includegraphics[scale=0.225]{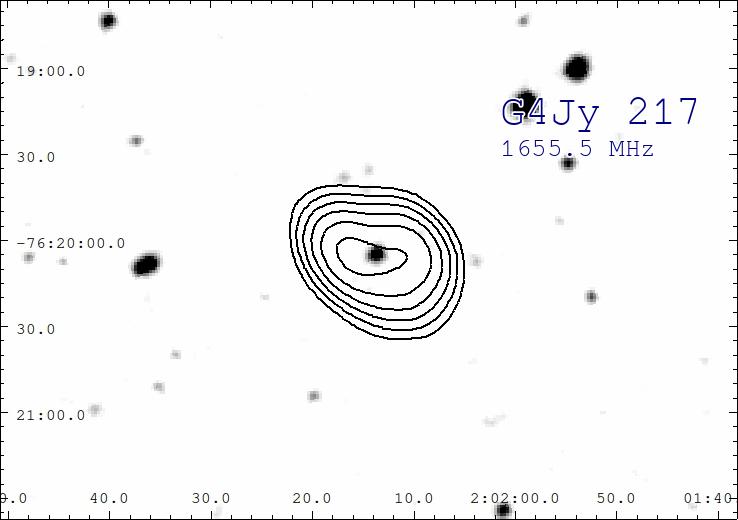}
    \includegraphics[scale=0.225]{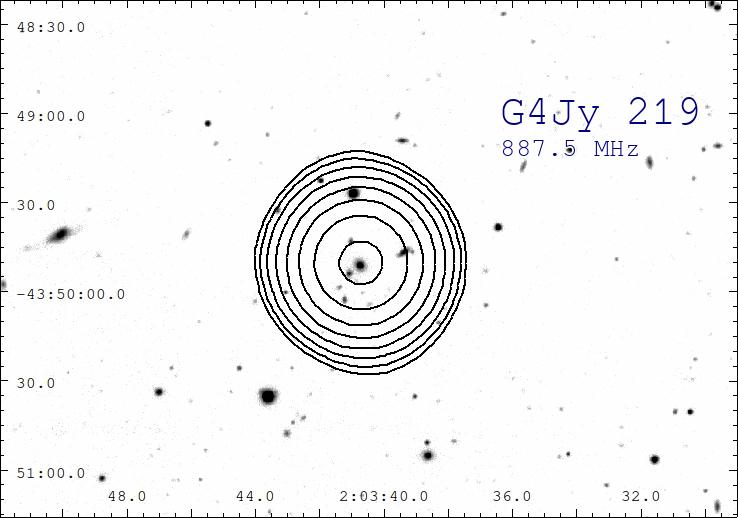}
    \includegraphics[scale=0.225]{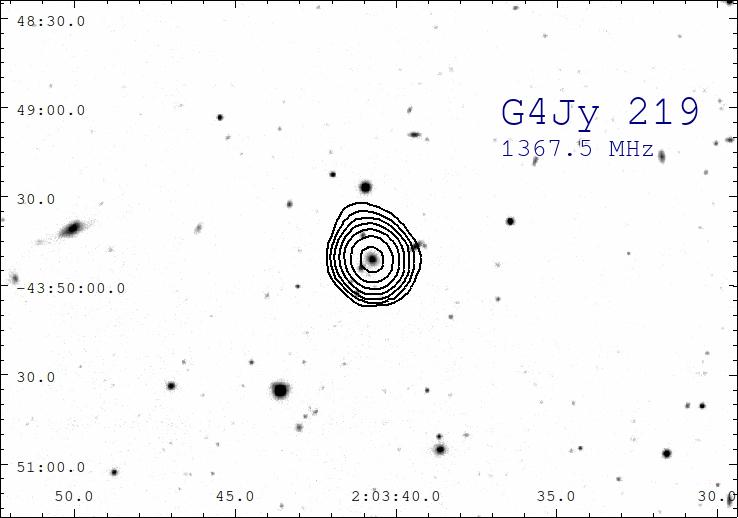}
    \includegraphics[scale=0.225]{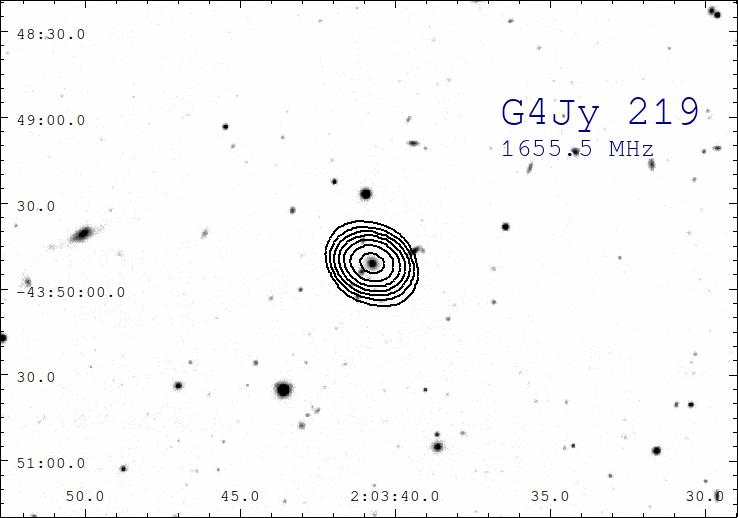}
    \includegraphics[scale=0.225]{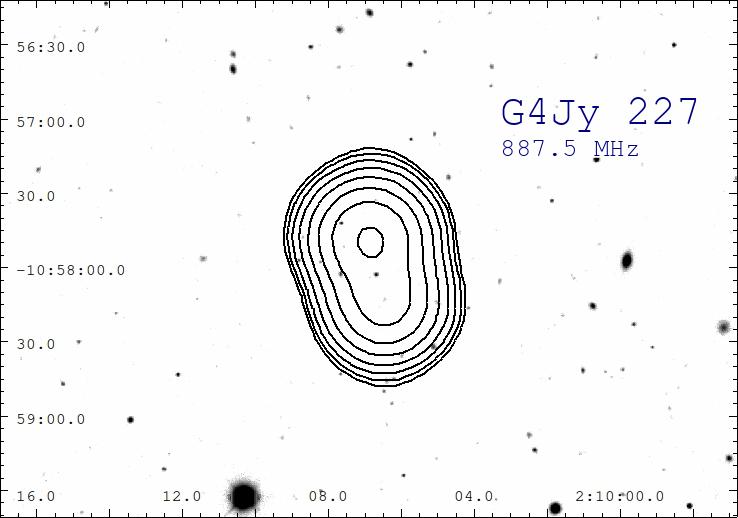}
    \includegraphics[scale=0.225]{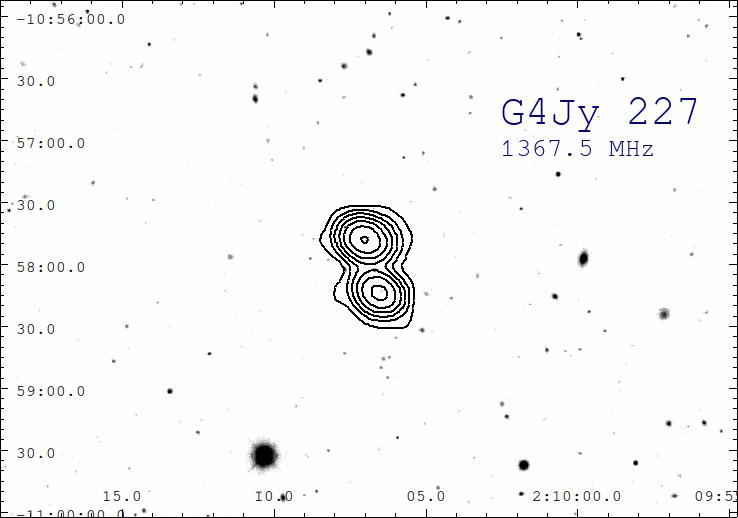}
    \includegraphics[scale=0.225]{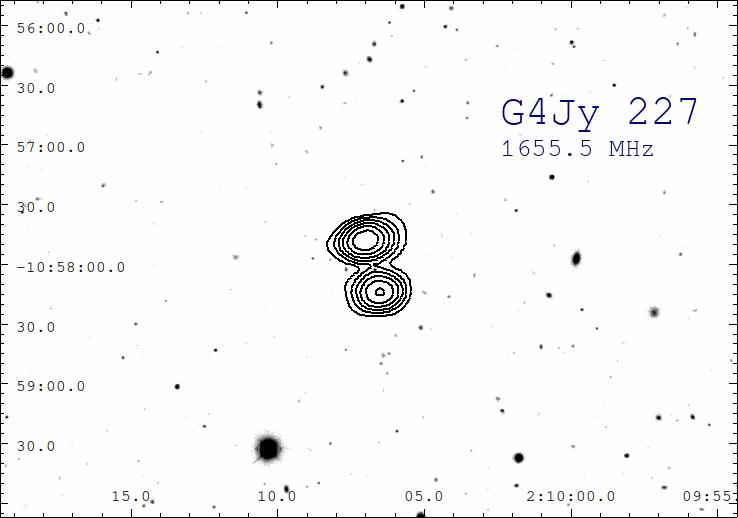}
    \includegraphics[scale=0.225]{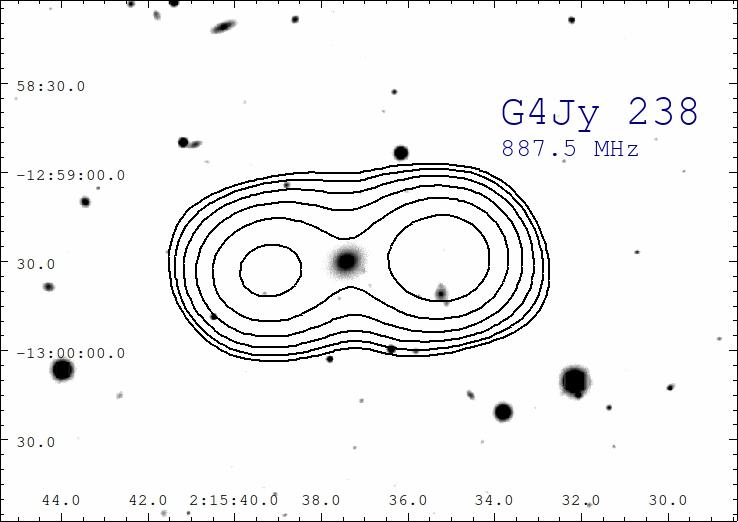}
    \includegraphics[scale=0.225]{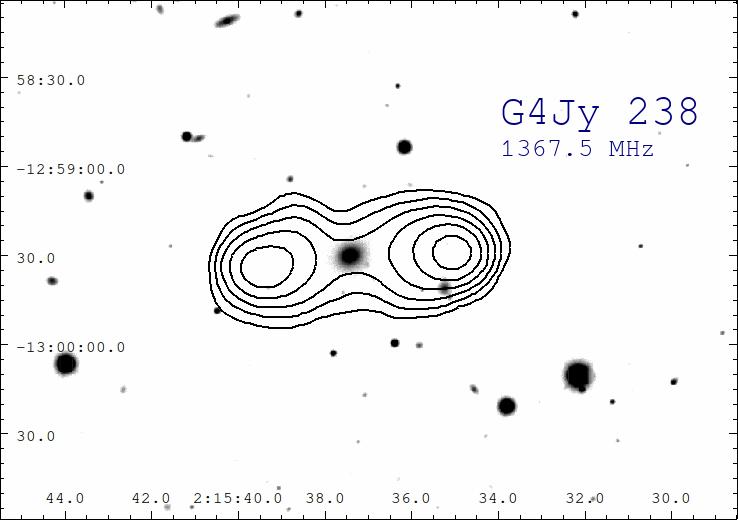}
    \includegraphics[scale=0.225]{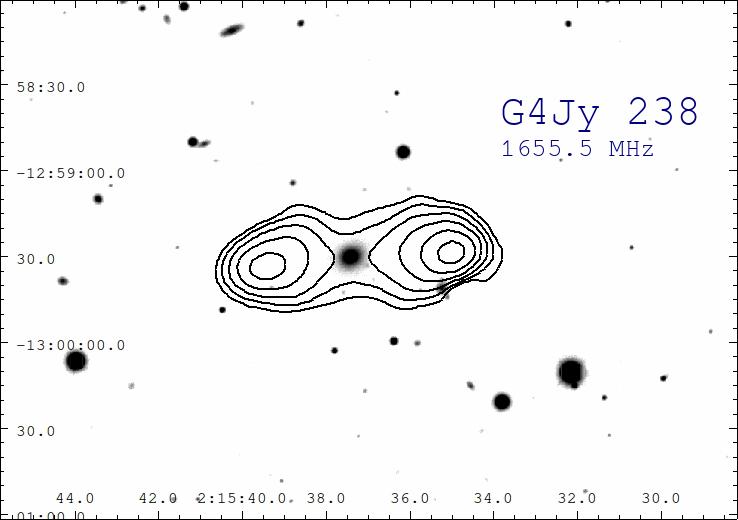}
    \caption{}
    \label{H}
\end{figure*}
\clearpage
\begin{figure*}
    \centering
    \includegraphics[scale=0.225]{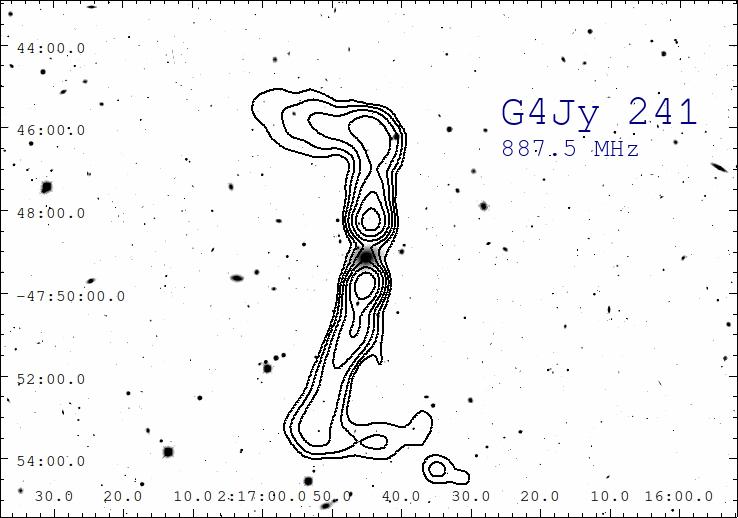}
    \includegraphics[scale=0.225]{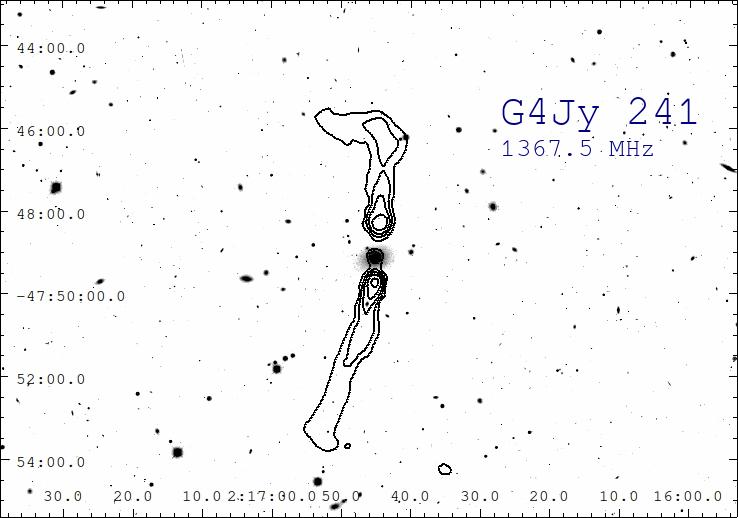}
    \includegraphics[scale=0.225]{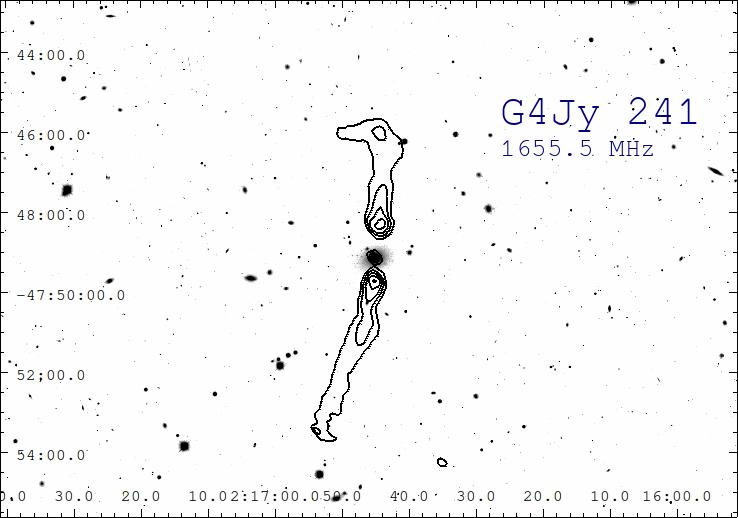}
    \includegraphics[scale=0.225]{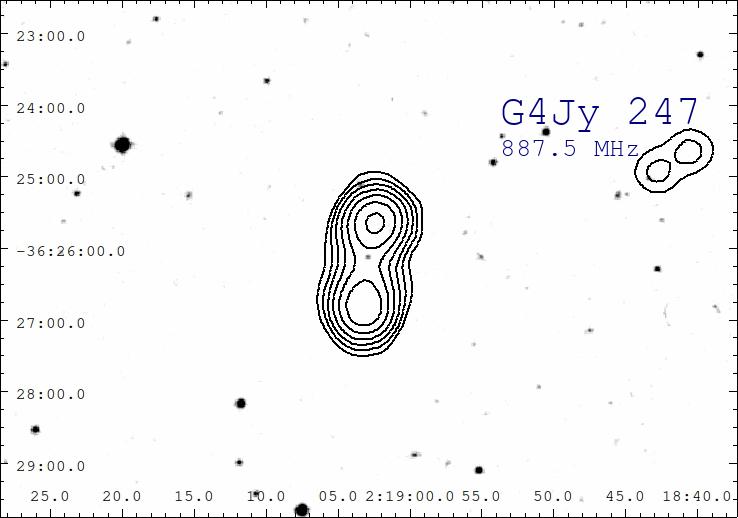}
    \includegraphics[scale=0.225]{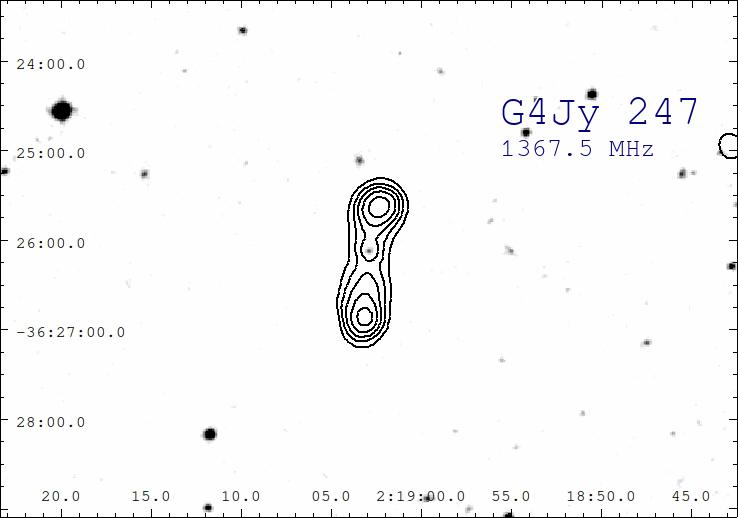}
    \includegraphics[scale=0.225]{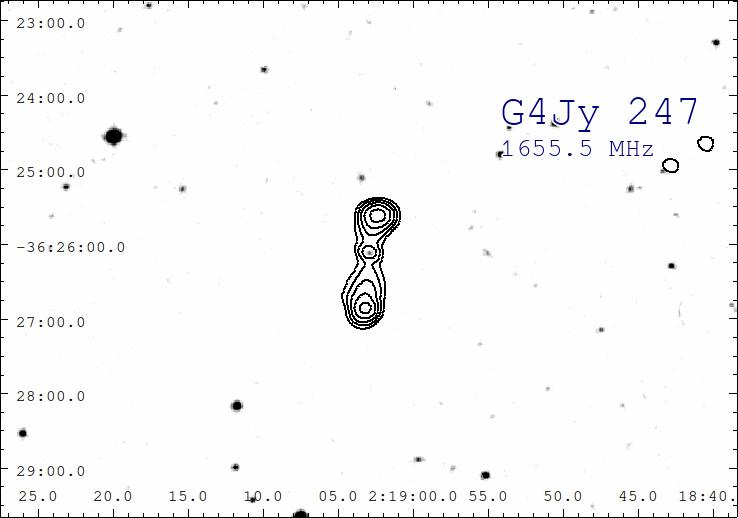}
    \includegraphics[scale=0.225]{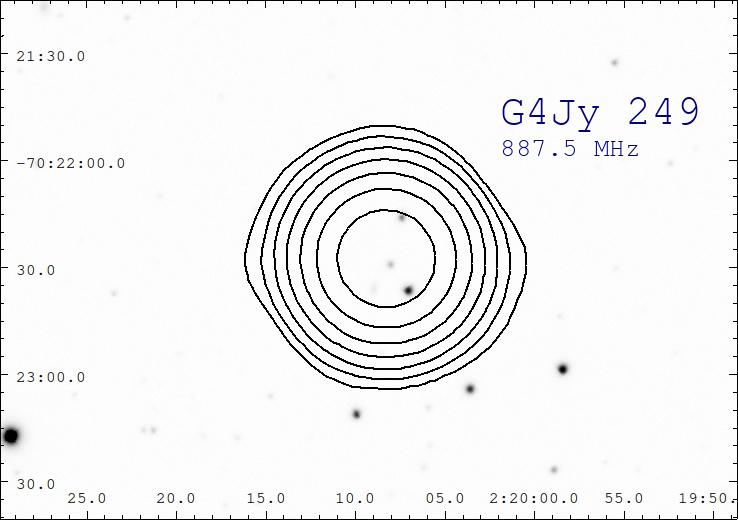}
    \includegraphics[scale=0.225]{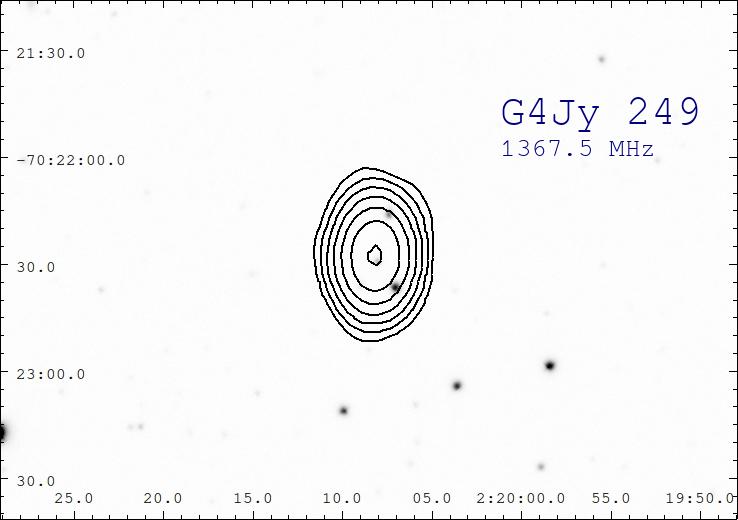}
    \includegraphics[scale=0.225]{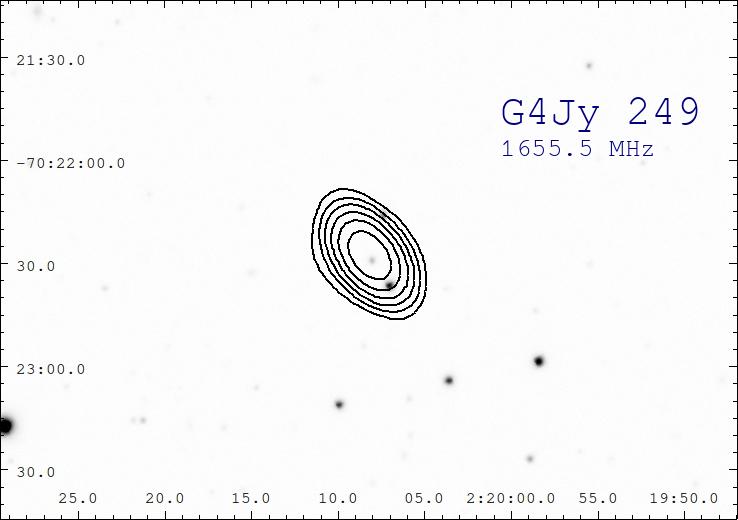}
    \includegraphics[scale=0.225]{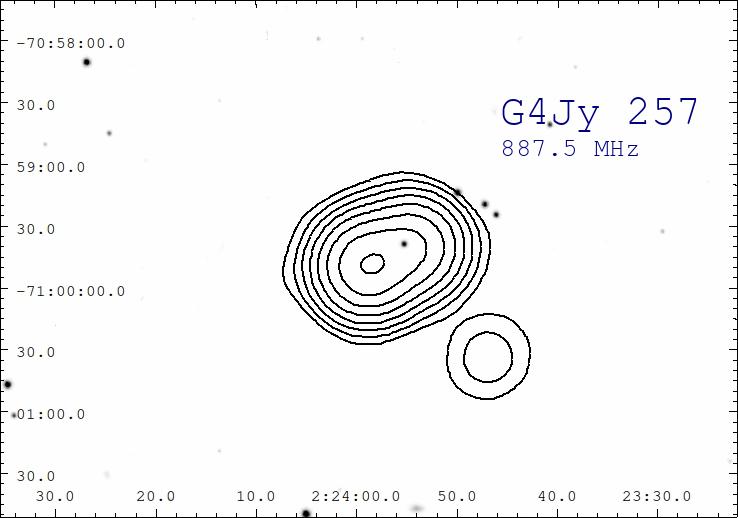}
    \includegraphics[scale=0.225]{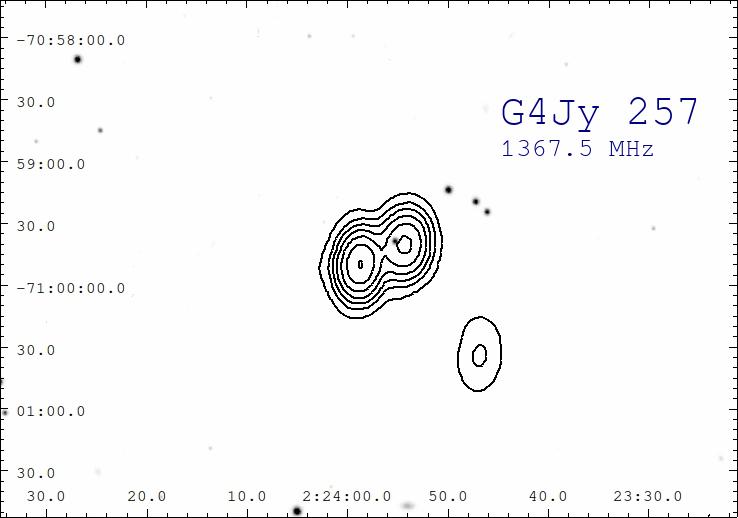}
    \includegraphics[scale=0.225]{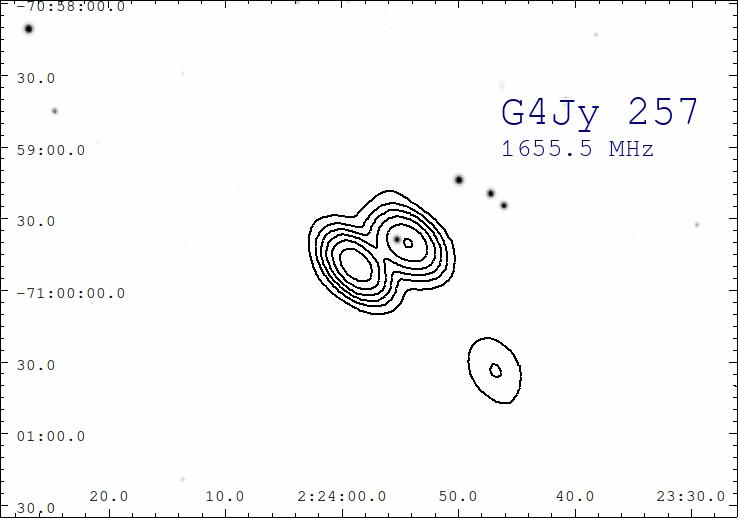}
    \includegraphics[scale=0.225]{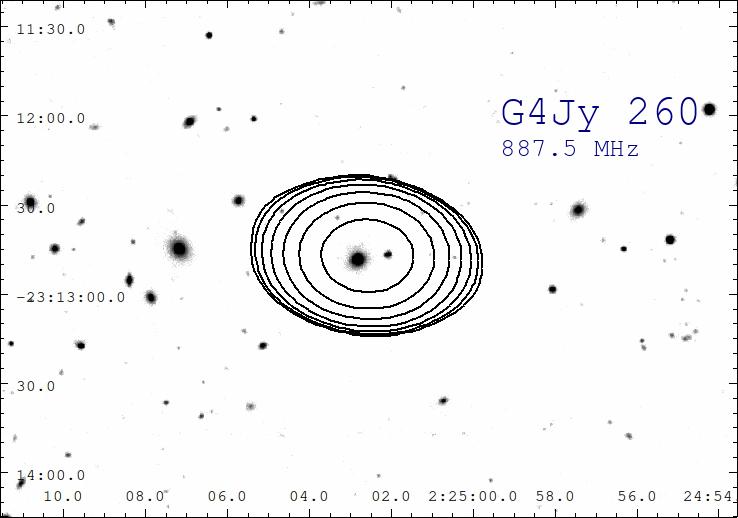}
    \includegraphics[scale=0.225]{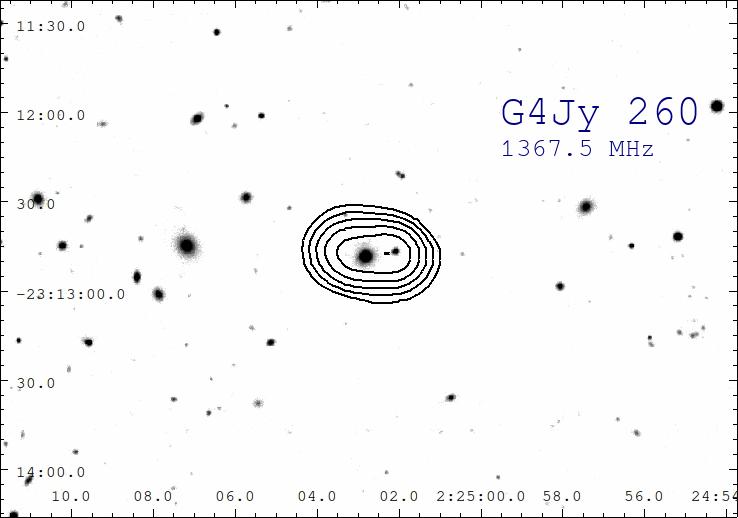}
    \includegraphics[scale=0.225]{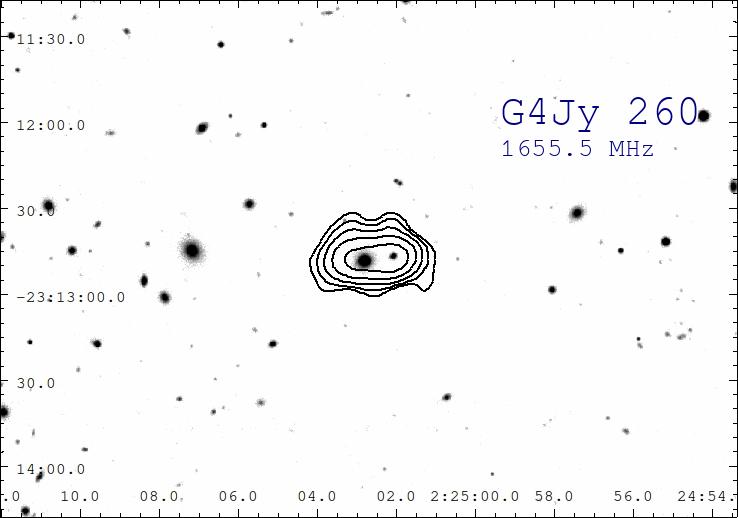}
    \caption{}
    \label{I}
\end{figure*}
\clearpage
\begin{figure*}
    \centering
    \includegraphics[scale=0.225]{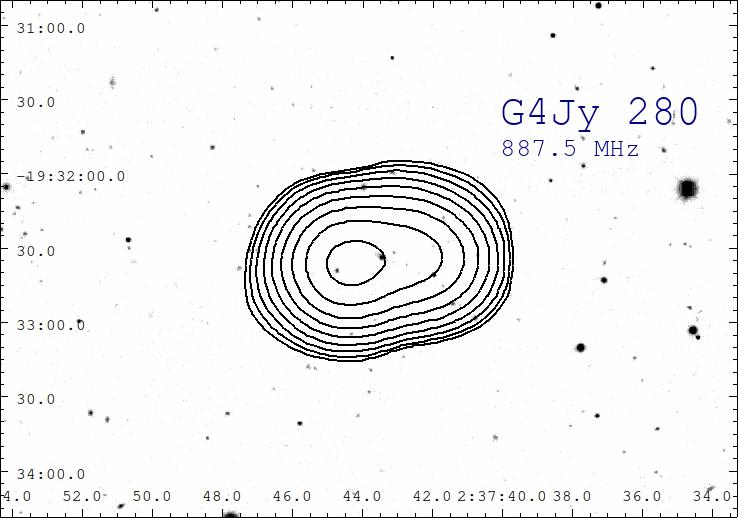}
    \includegraphics[scale=0.225]{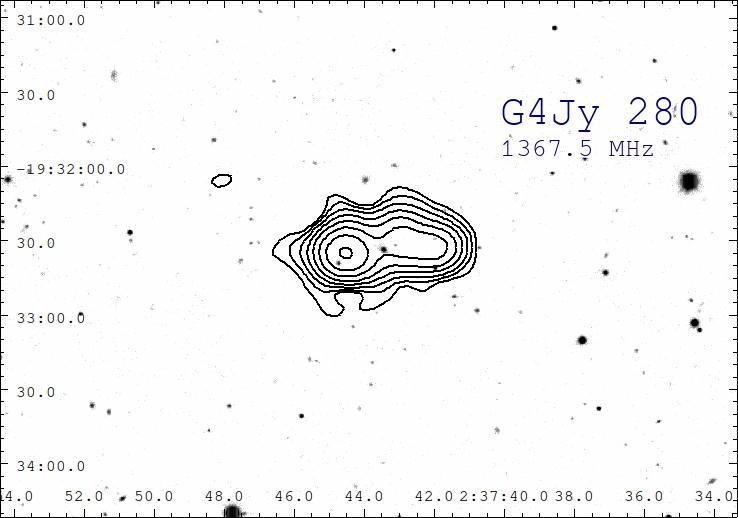}
    \includegraphics[scale=0.225]{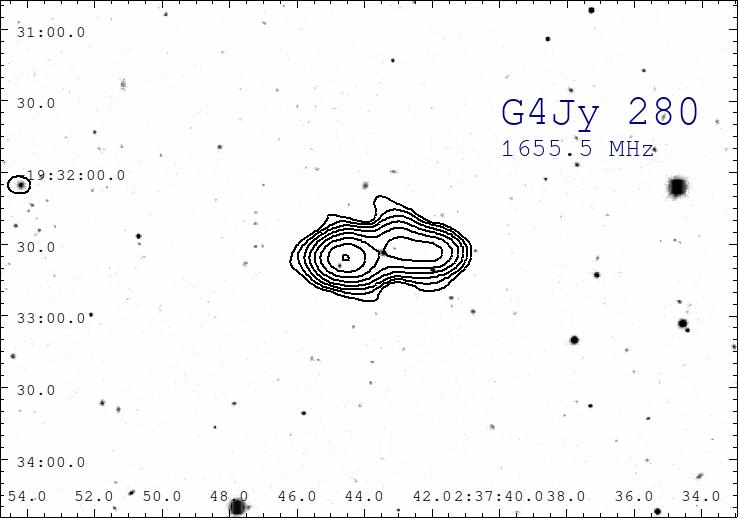}
    \includegraphics[scale=0.225]{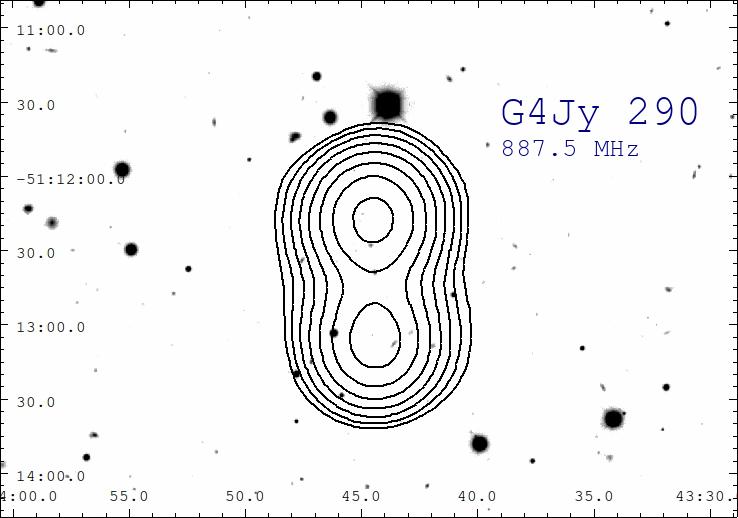}
    \includegraphics[scale=0.225]{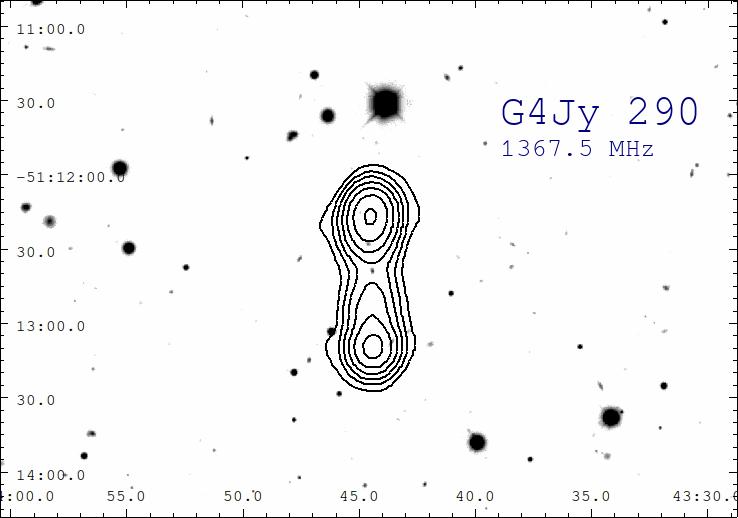}
    \includegraphics[scale=0.225]{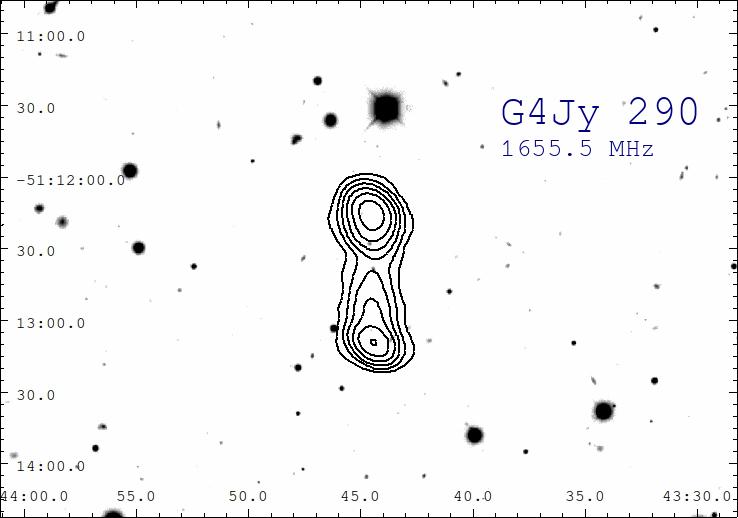}
    \includegraphics[scale=0.225]{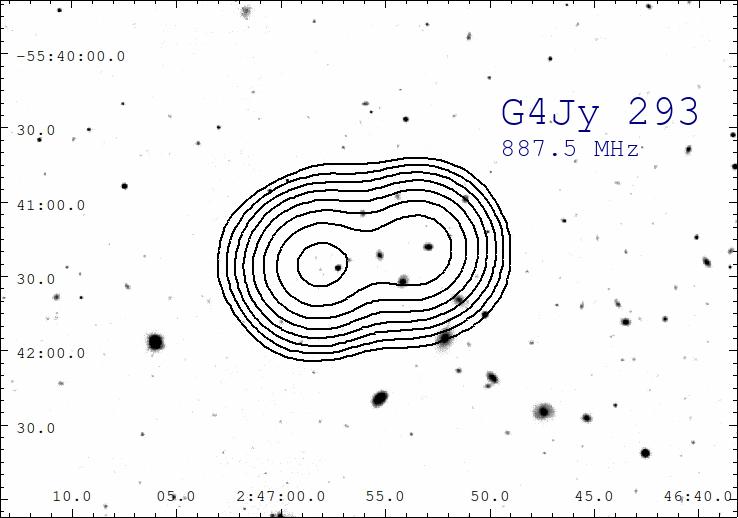}
    \includegraphics[scale=0.225]{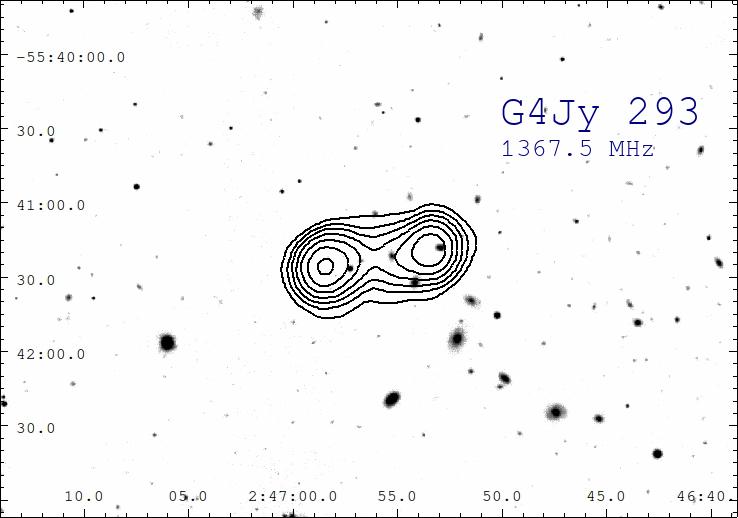}
    \includegraphics[scale=0.225]{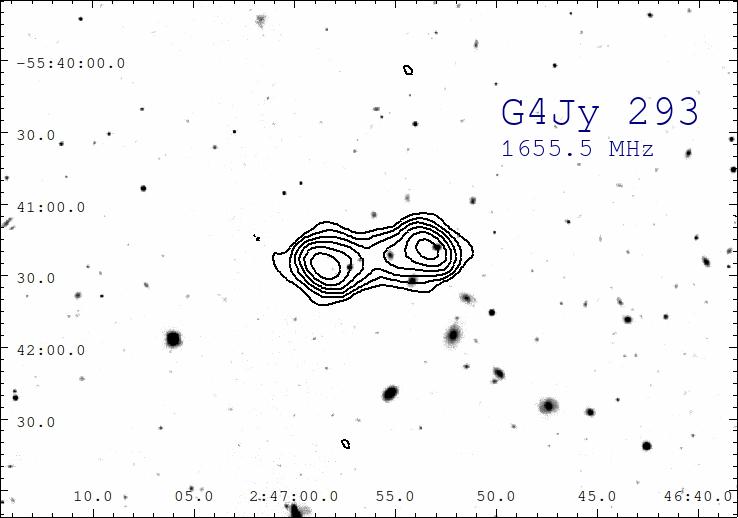}
    \includegraphics[scale=0.225]{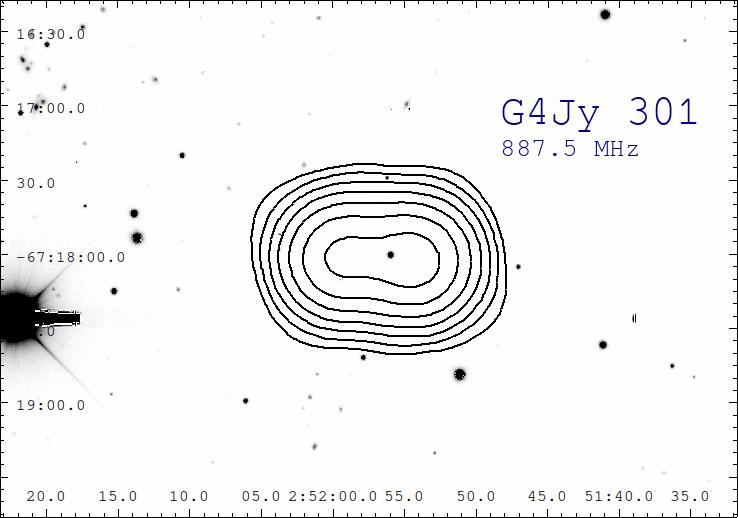}
    \includegraphics[scale=0.225]{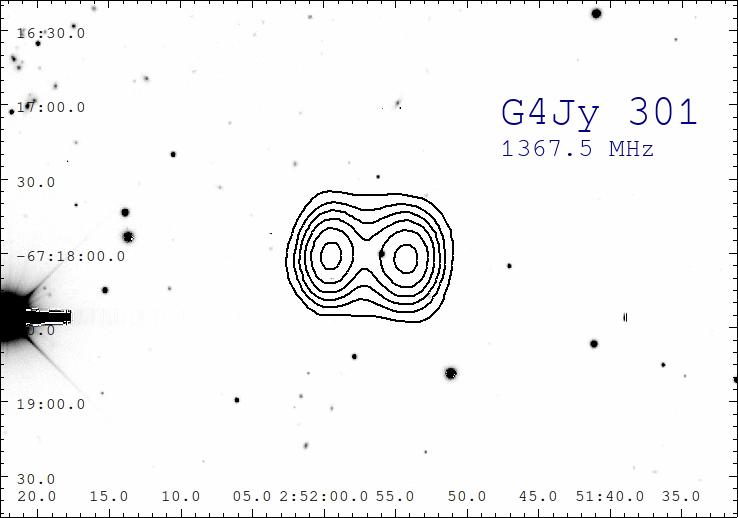}
    \includegraphics[scale=0.225]{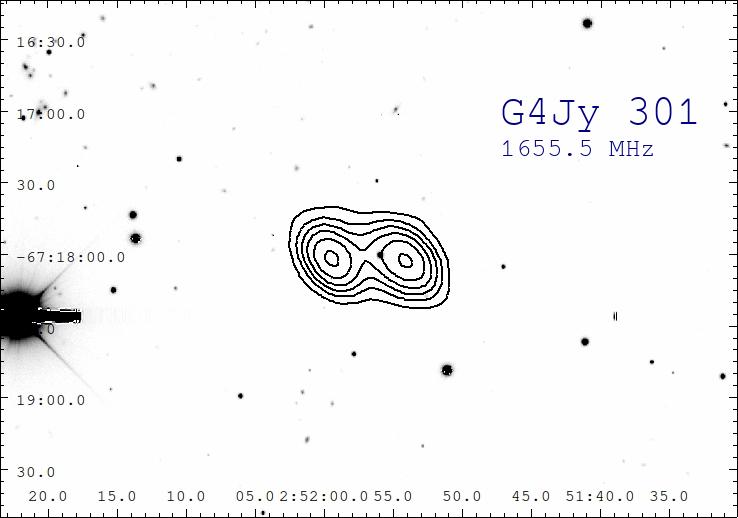}
    \includegraphics[scale=0.225]{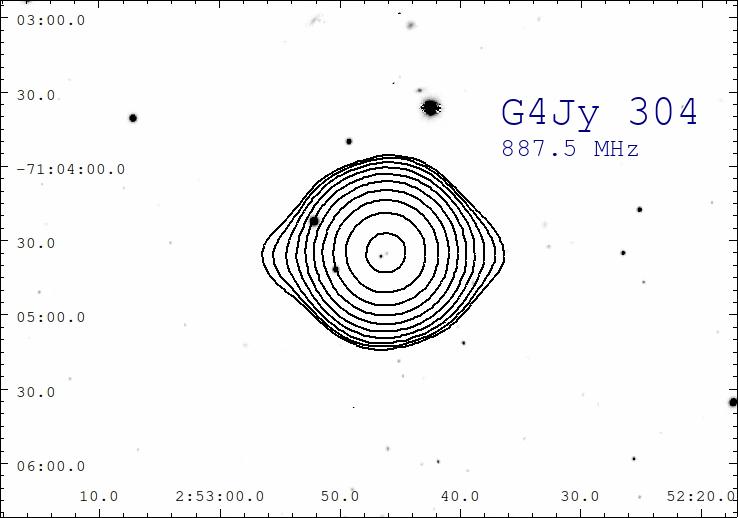}
    \includegraphics[scale=0.225]{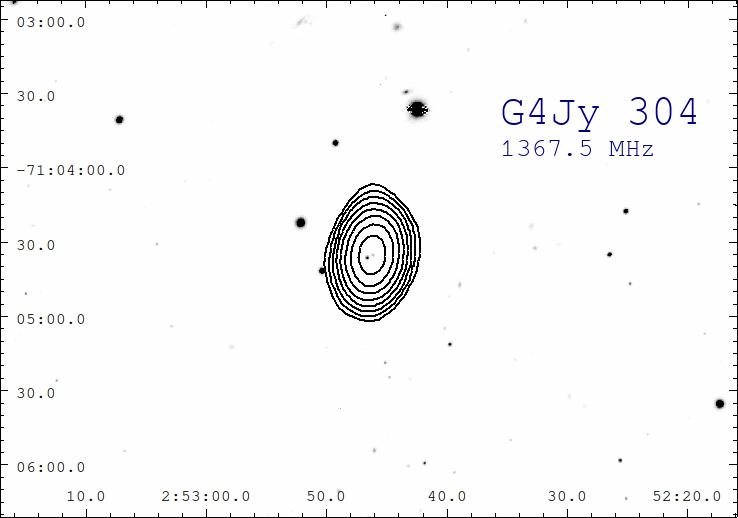}
    \includegraphics[scale=0.225]{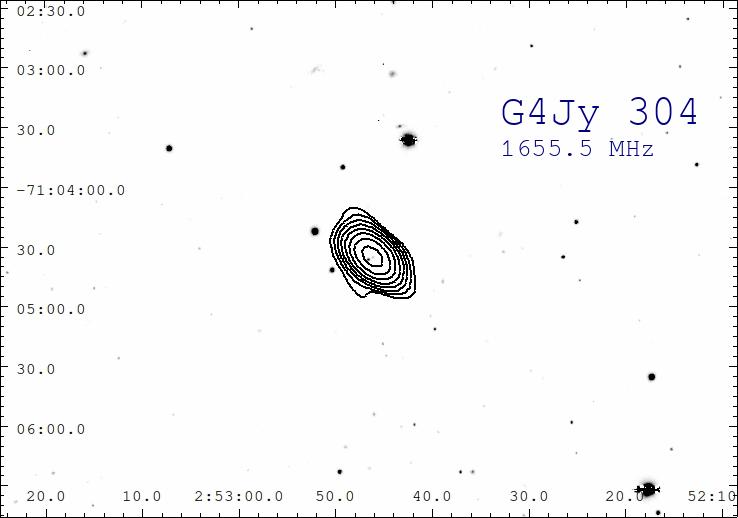}
    \caption{}
    \label{J}
\end{figure*}
\clearpage
\begin{figure*}
    \centering
    \includegraphics[scale=0.225]{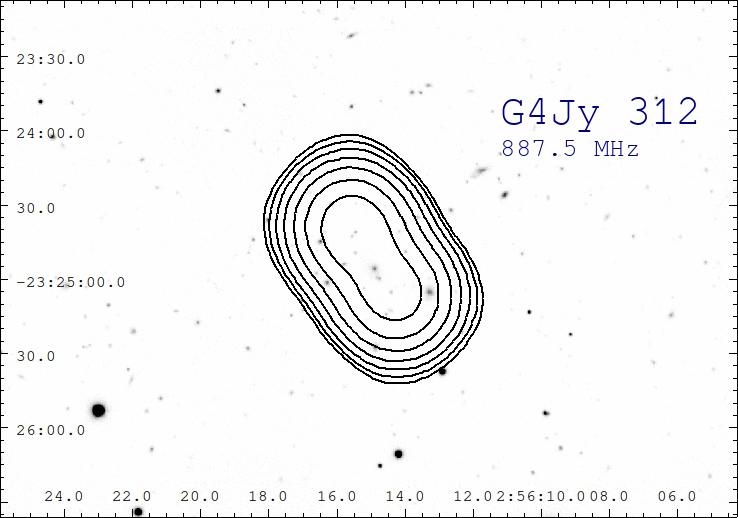}
    \includegraphics[scale=0.225]{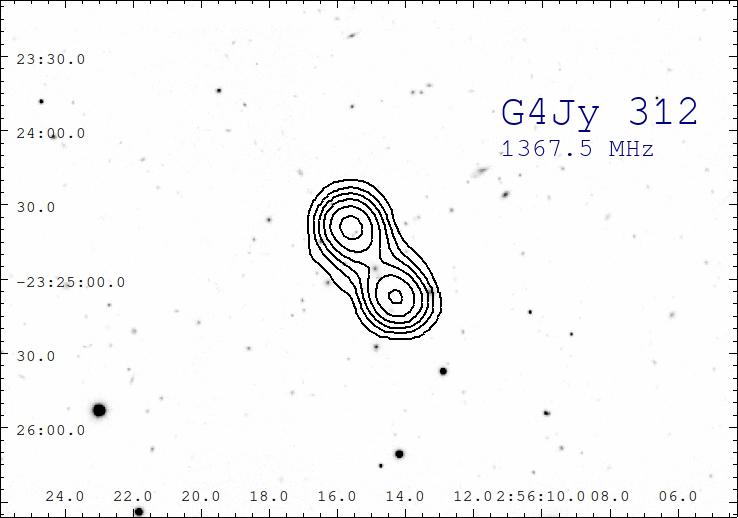}
    \includegraphics[scale=0.225]{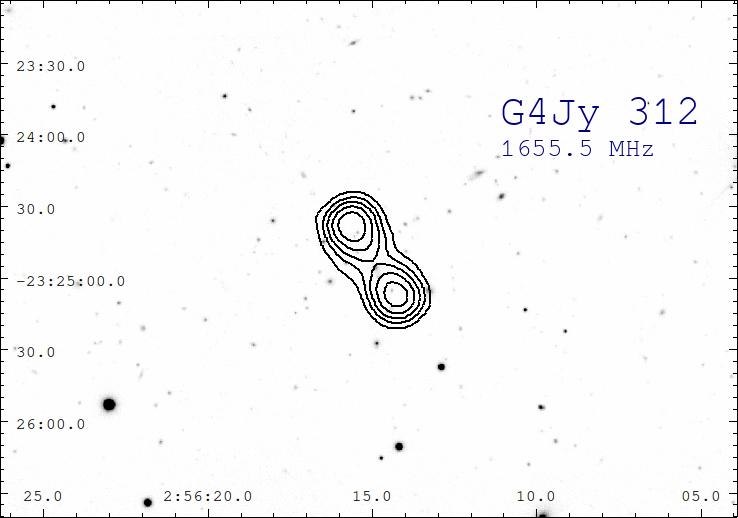}
    \includegraphics[scale=0.225]{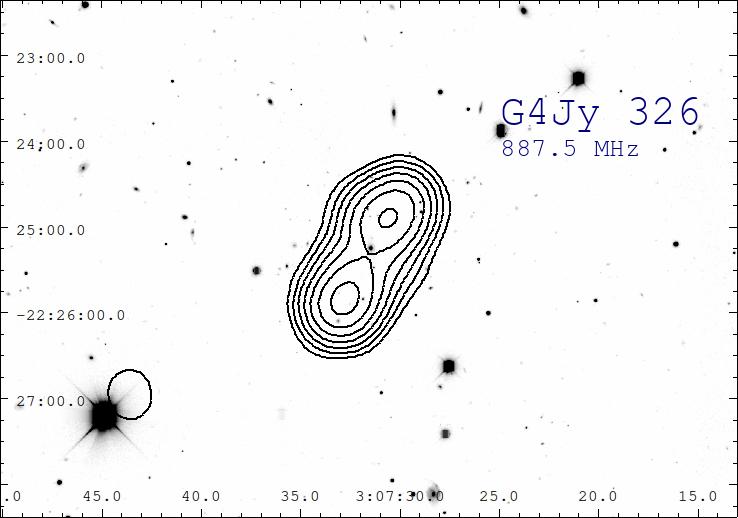}
    \includegraphics[scale=0.225]{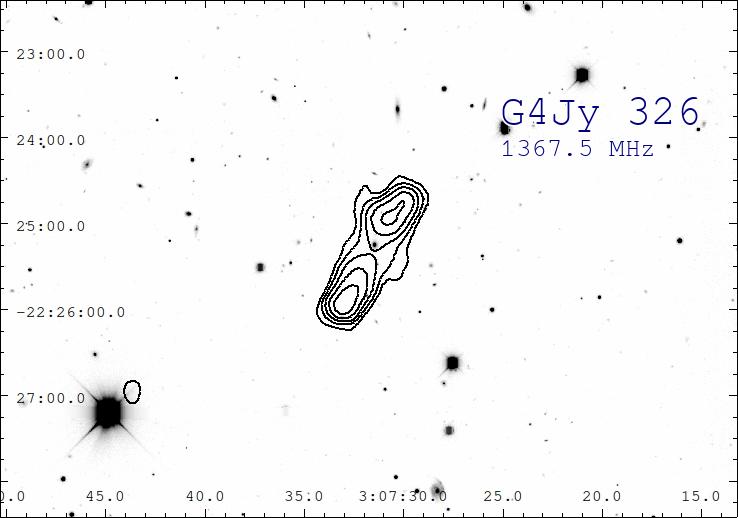}
    \includegraphics[scale=0.225]{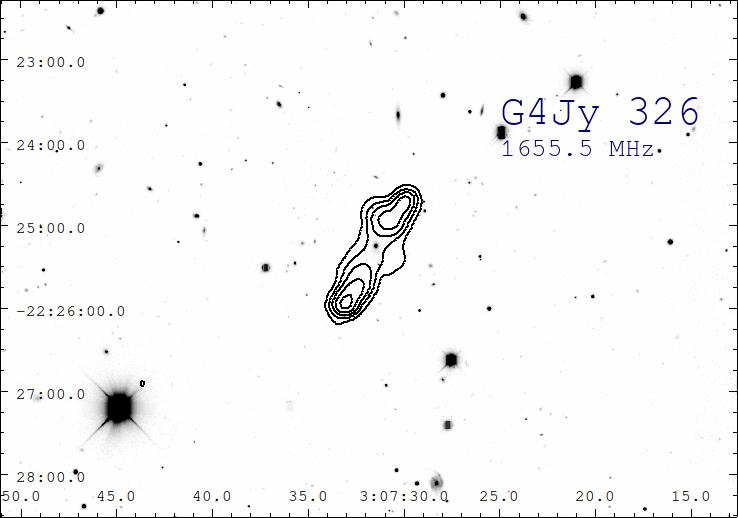}
    \includegraphics[scale=0.225]{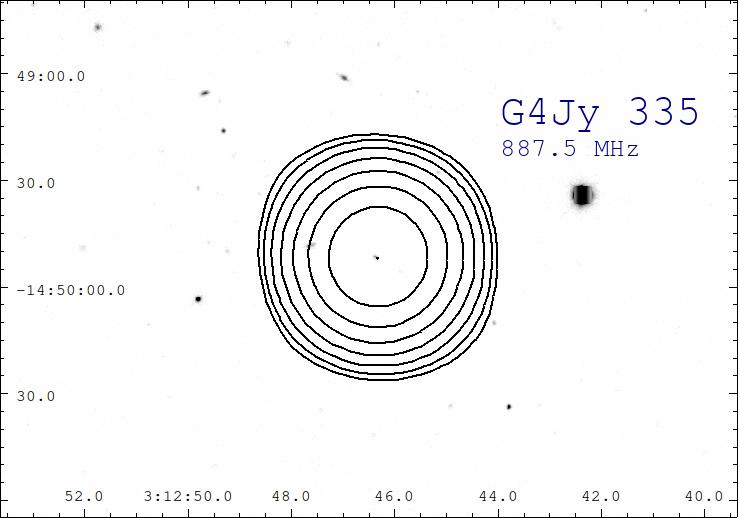}
    \includegraphics[scale=0.225]{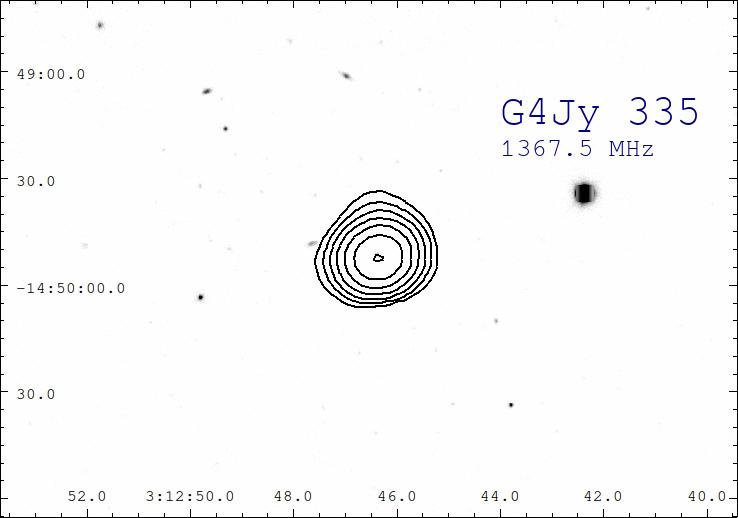}
    \includegraphics[scale=0.225]{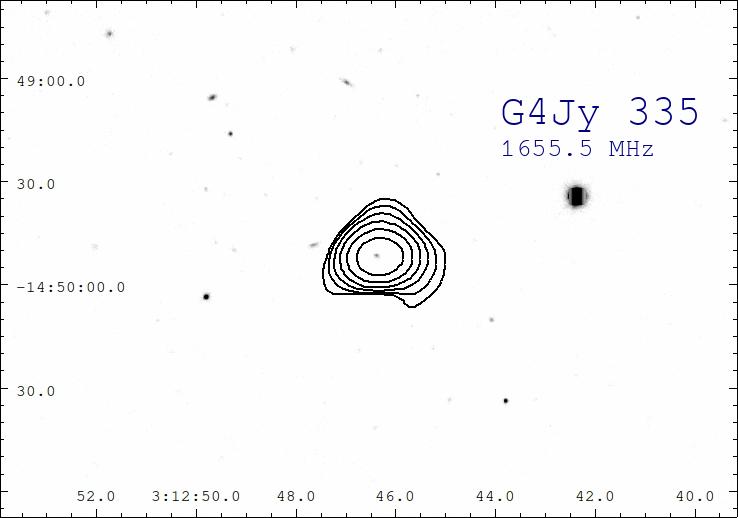}
    \includegraphics[scale=0.225]{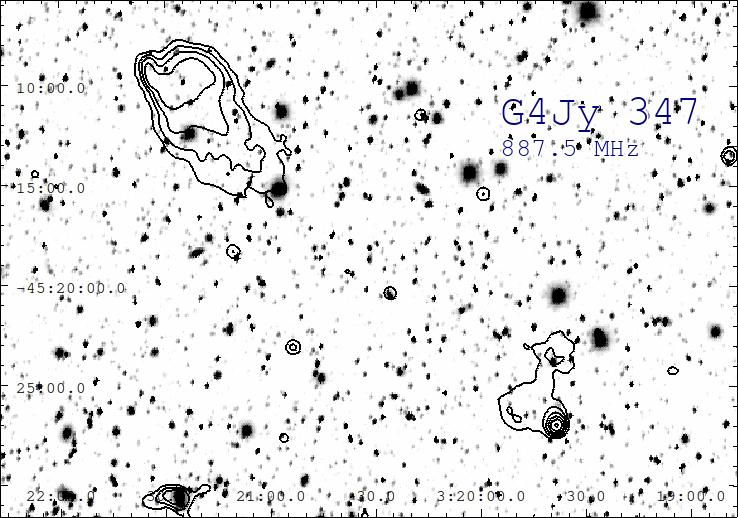}
    \includegraphics[scale=0.225]{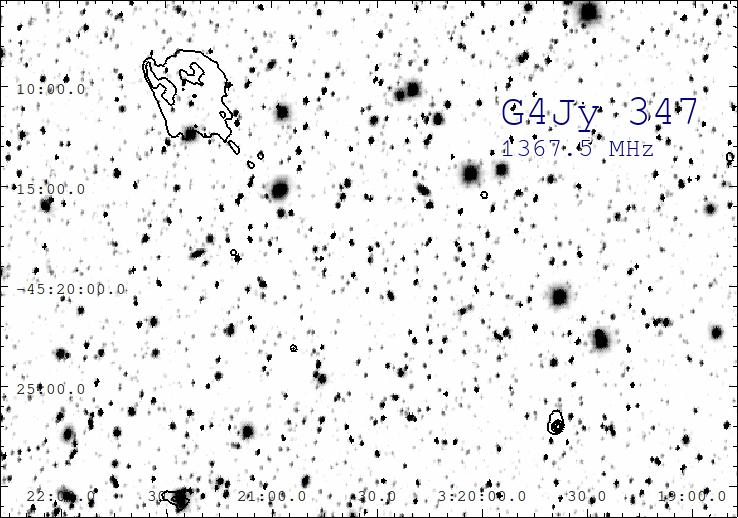}
    \includegraphics[scale=0.225]{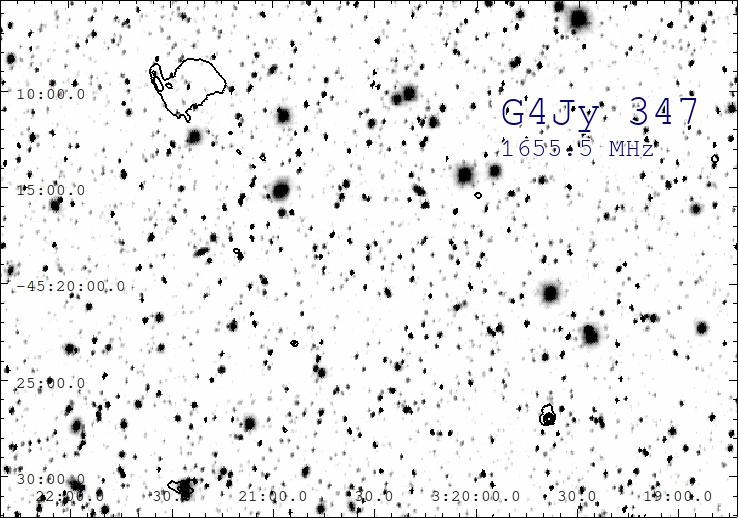}
    \includegraphics[scale=0.225]{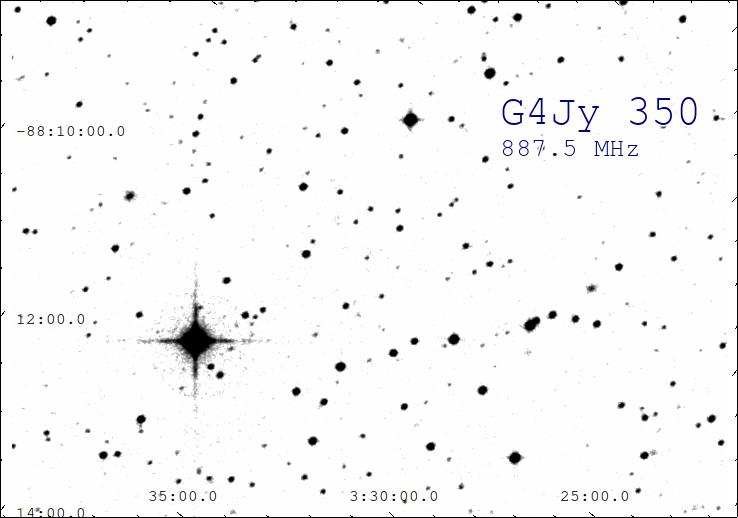}
    \includegraphics[scale=0.225]{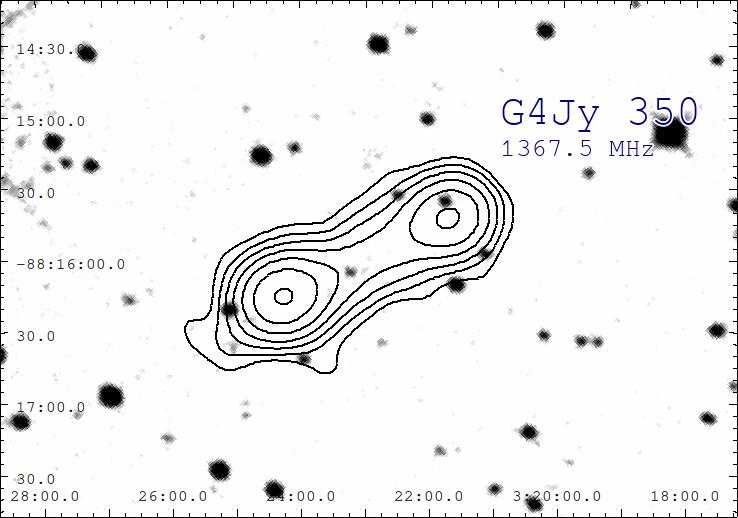}
    \includegraphics[scale=0.225]{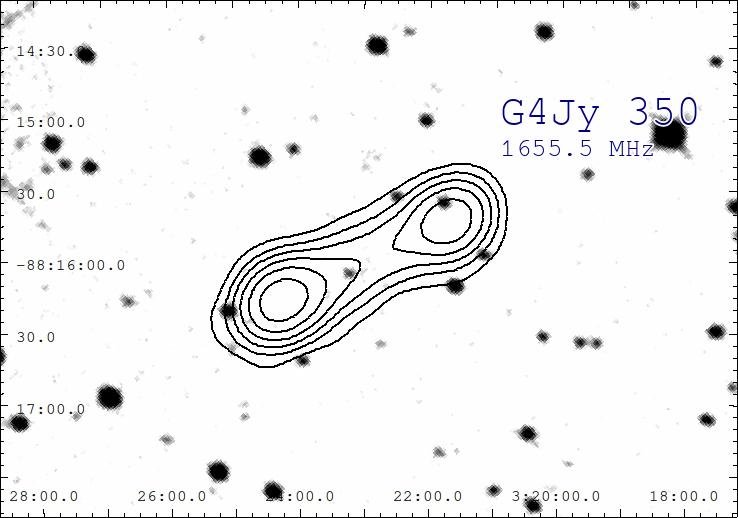}
    \caption{RACS-low data for G4Jy 350 could not be convolved to 25$\arcsec$, and therefore was not available. It may be replaced with future observations from RACS-low2 or RACS-low3 \textbf{(E.Lenc and A. Hotan, private communication)}}
    \label{K}
\end{figure*}
\clearpage
\begin{figure*}
    \centering
    \includegraphics[scale=0.225]{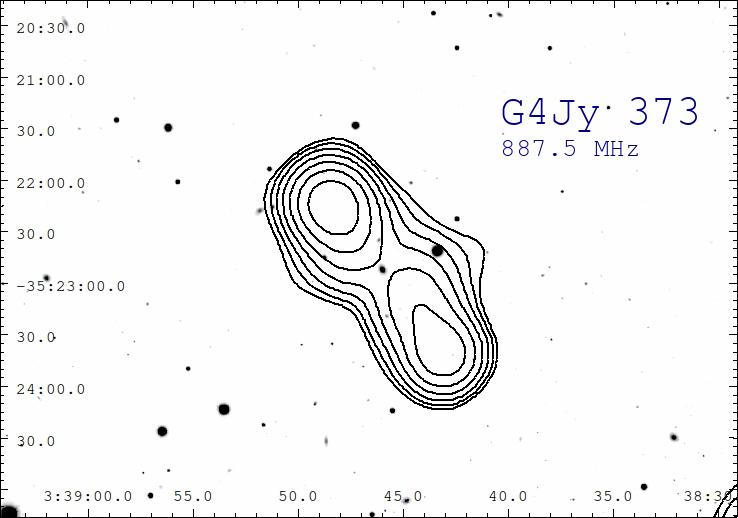}
    \includegraphics[scale=0.225]{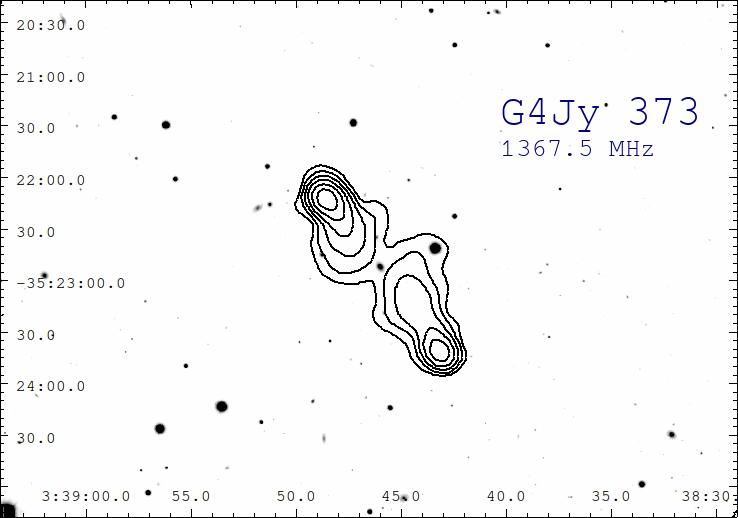}
    \includegraphics[scale=0.225]{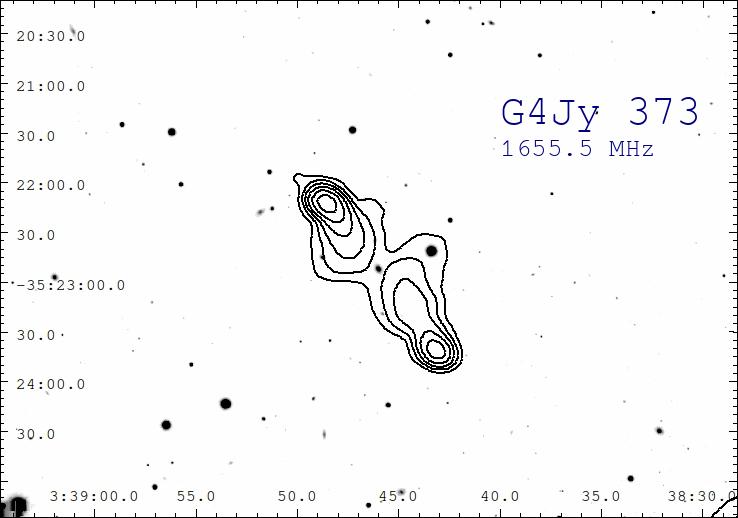}
    \includegraphics[scale=0.225]{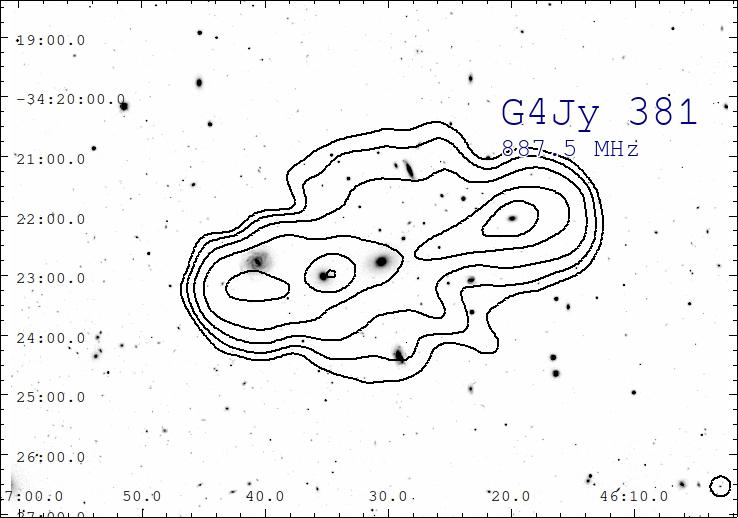}
    \includegraphics[scale=0.225]{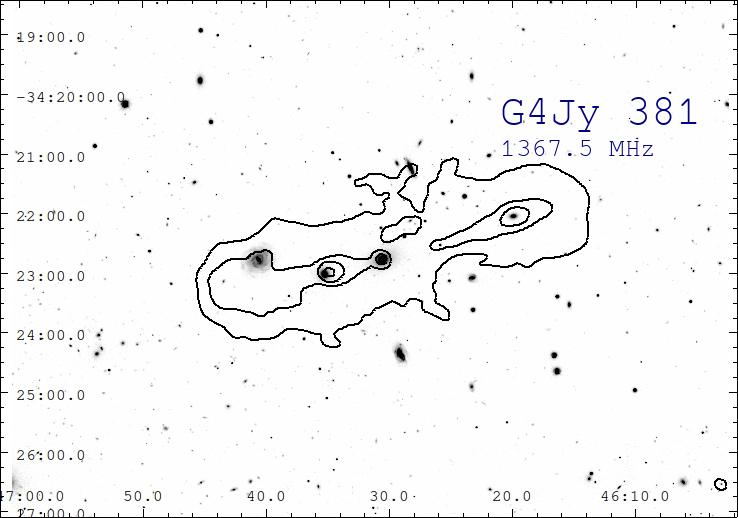}
    \includegraphics[scale=0.225]{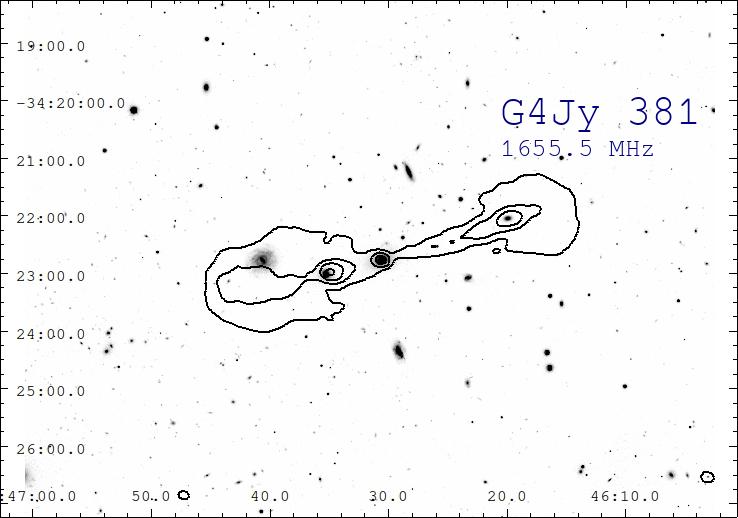}
    \includegraphics[scale=0.225]{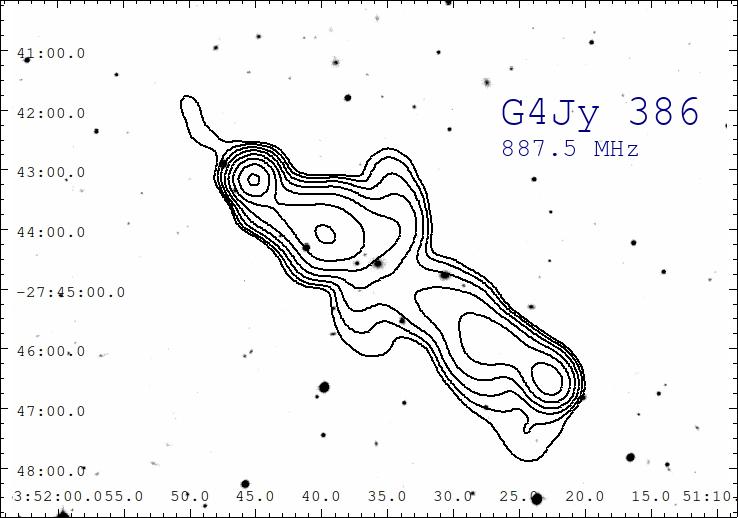}
    \includegraphics[scale=0.225]{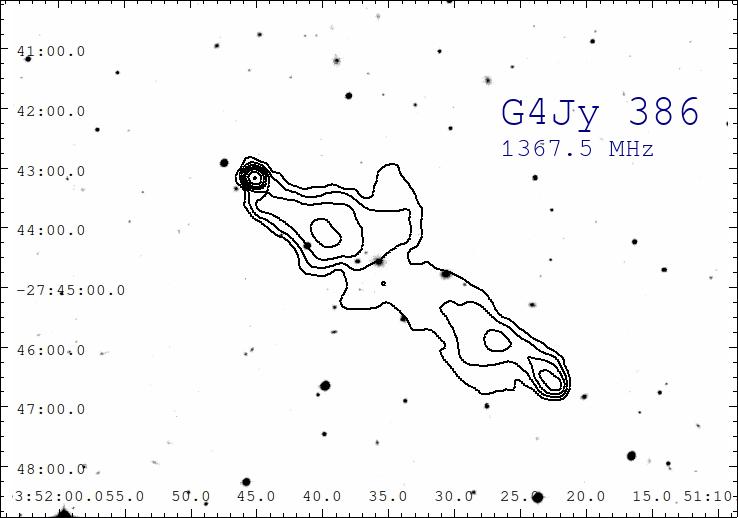}
    \includegraphics[scale=0.225]{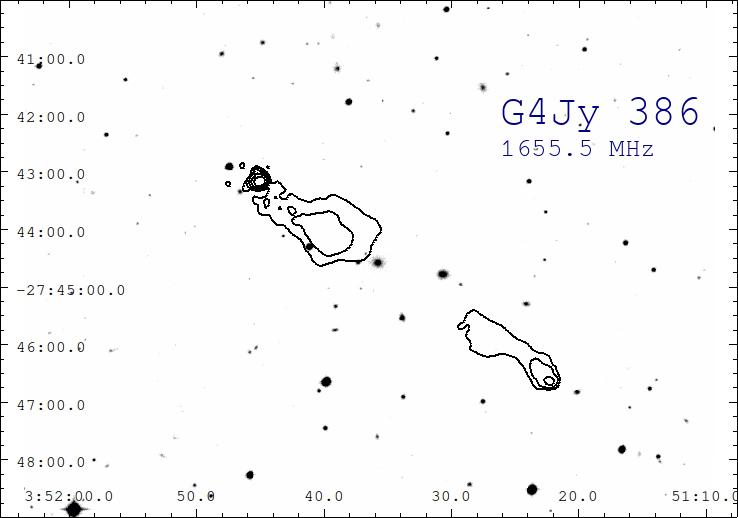}
    \includegraphics[scale=0.225]{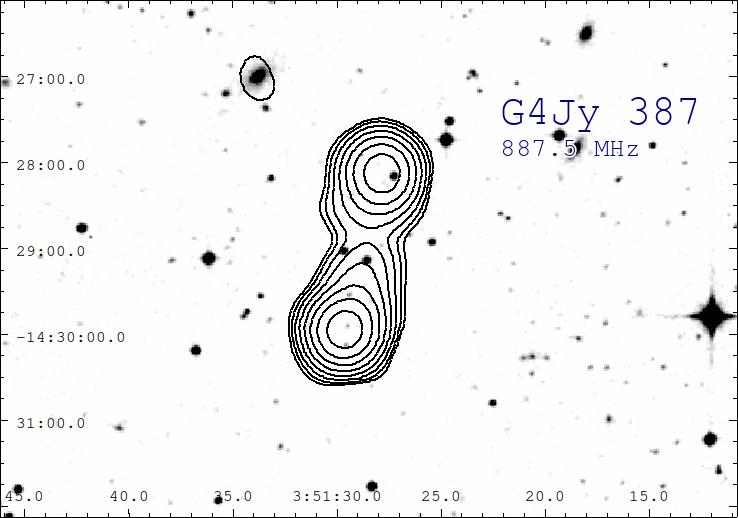}
    \includegraphics[scale=0.225]{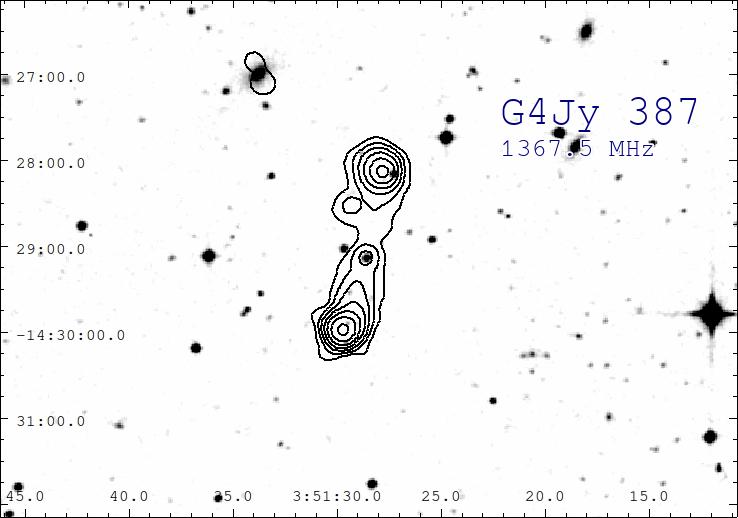}
    \includegraphics[scale=0.225]{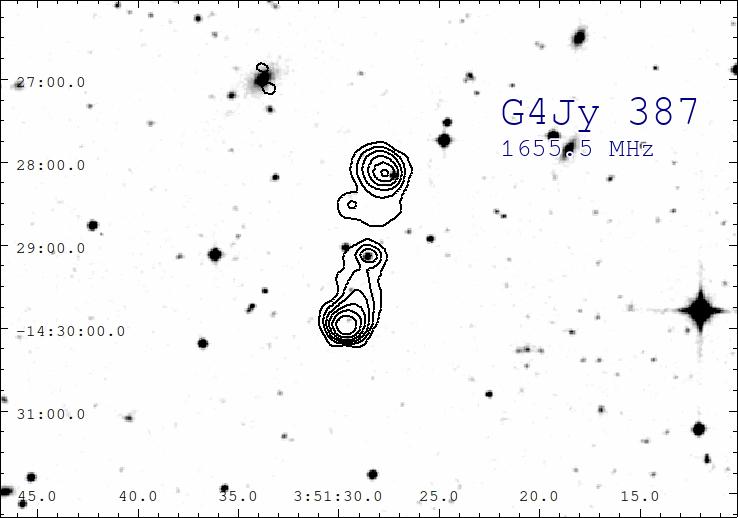}
    \includegraphics[scale=0.225]{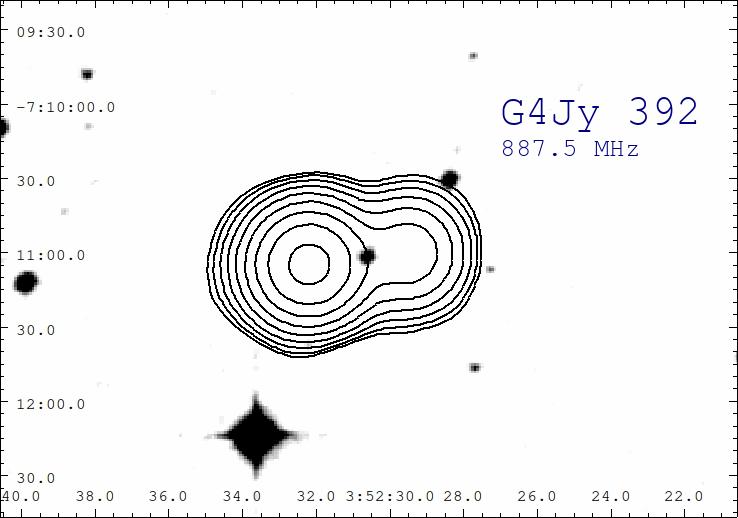}
    \includegraphics[scale=0.225]{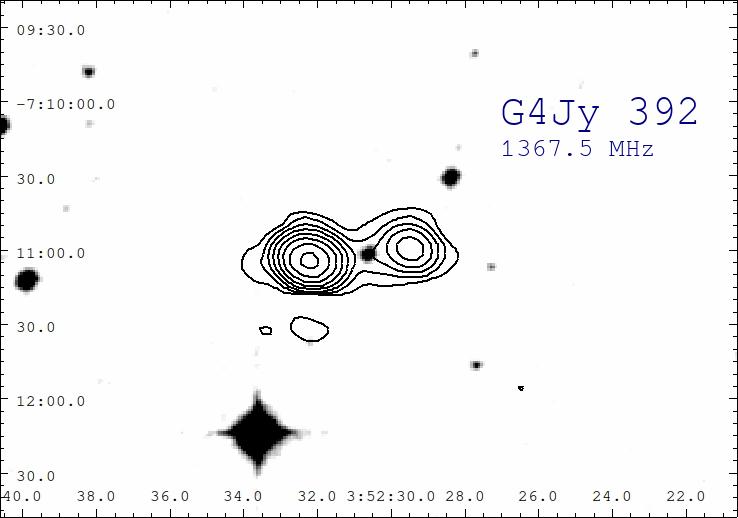}
    \includegraphics[scale=0.225]{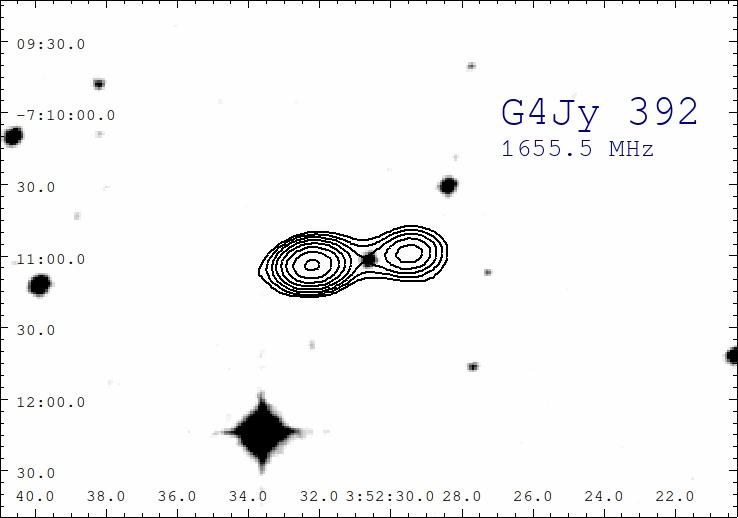}

    \caption{}
    \label{K}
\end{figure*}
\clearpage
\begin{figure*}
    \centering
    \includegraphics[scale=0.225]{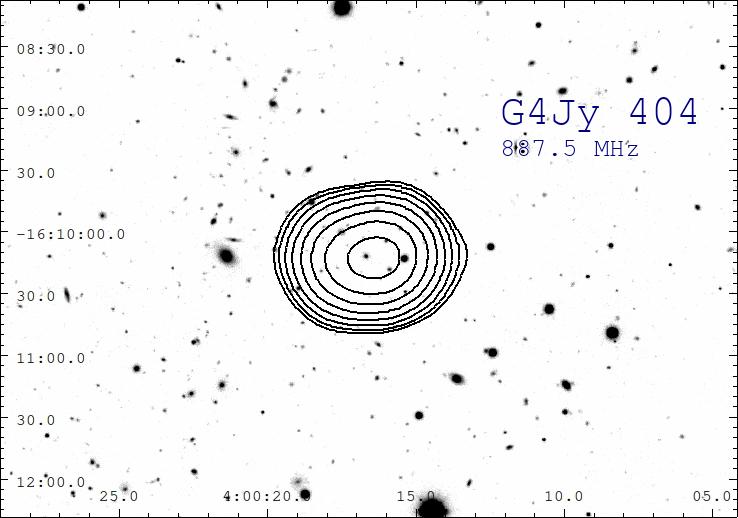}
    \includegraphics[scale=0.225]{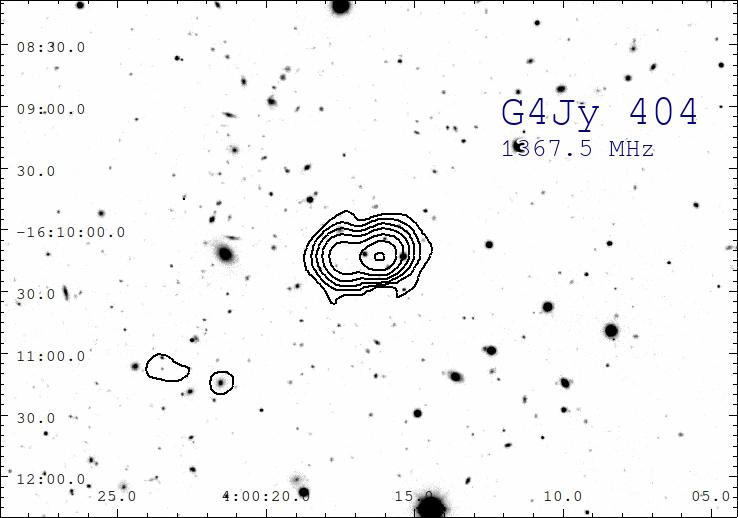}
    \includegraphics[scale=0.225]{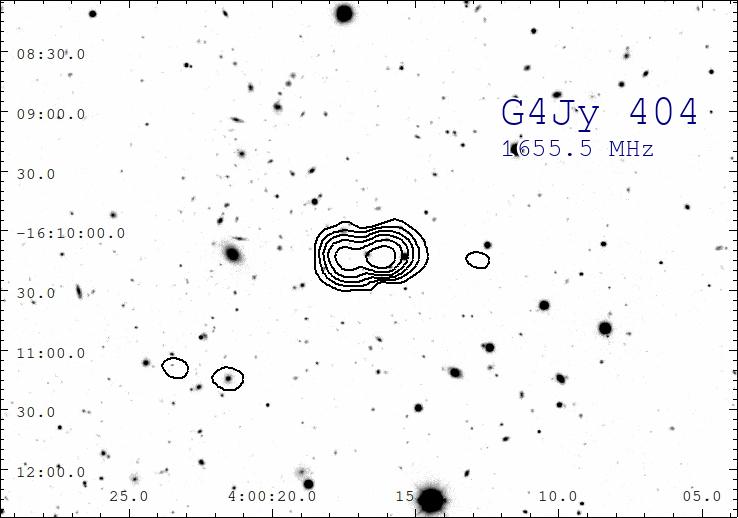}
    \includegraphics[scale=0.225]{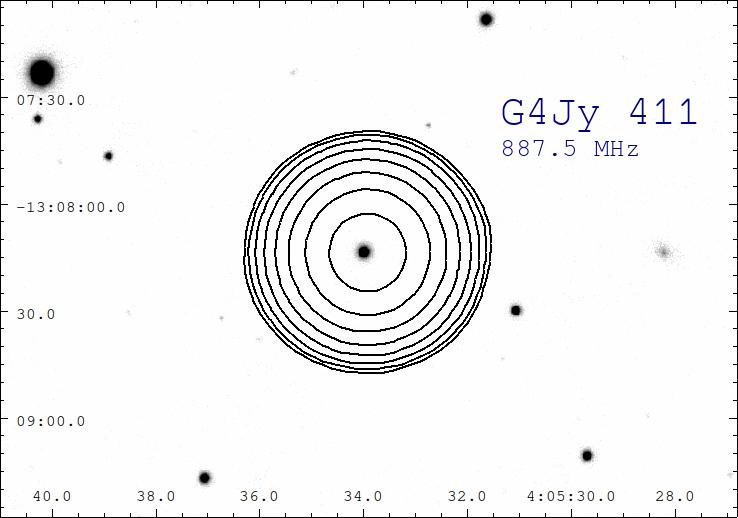}
    \includegraphics[scale=0.225]{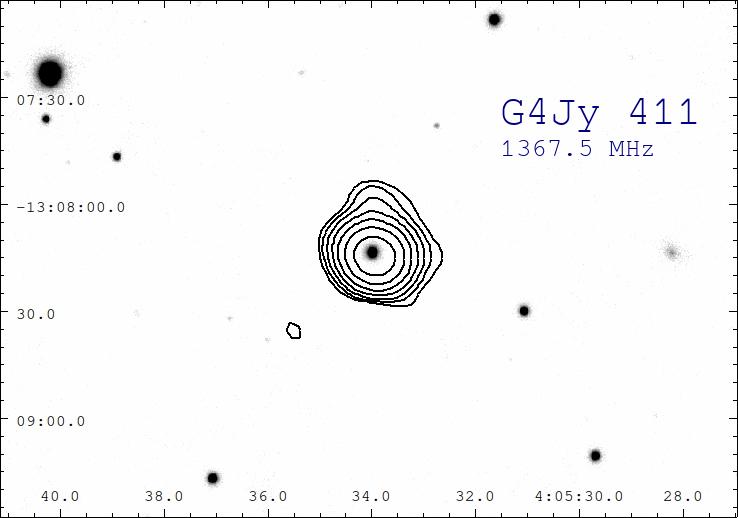}
    \includegraphics[scale=0.225]{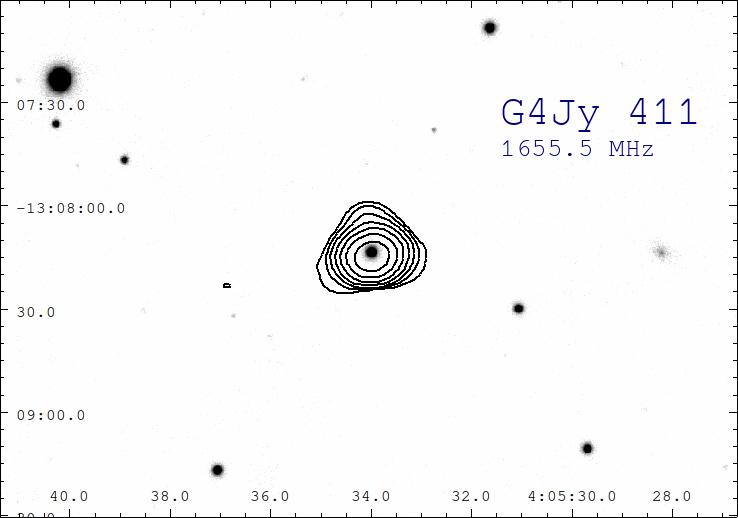}
    \includegraphics[scale=0.225]{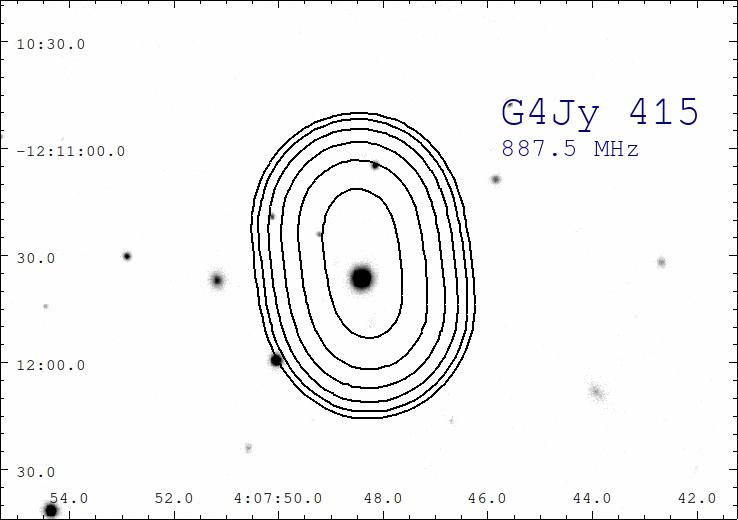}
    \includegraphics[scale=0.225]{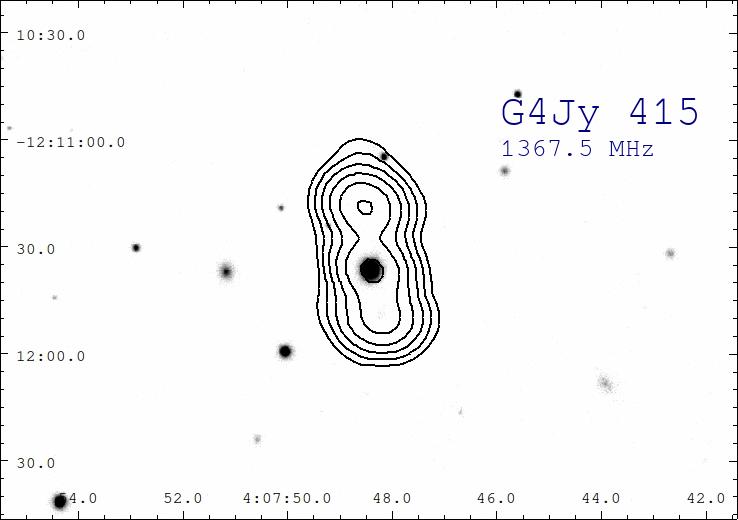}
    \includegraphics[scale=0.225]{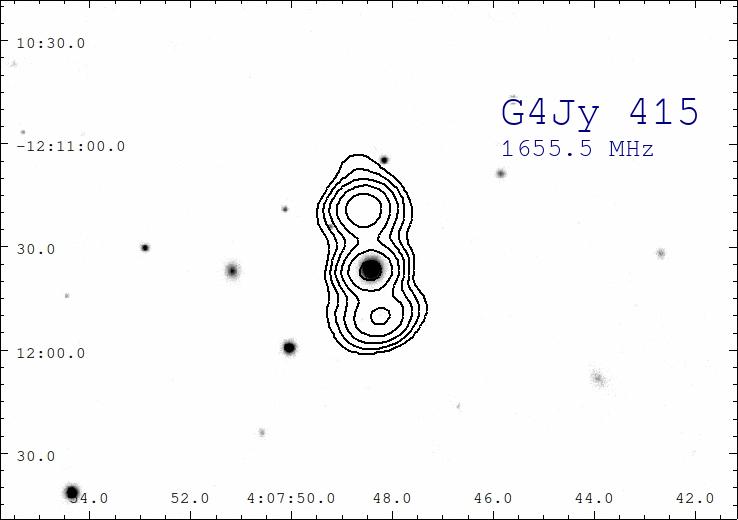}
    \includegraphics[scale=0.225]{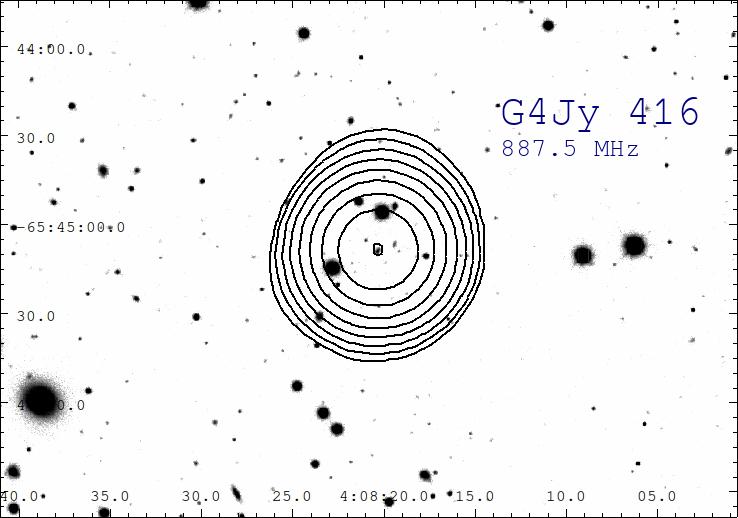}
    \includegraphics[scale=0.225]{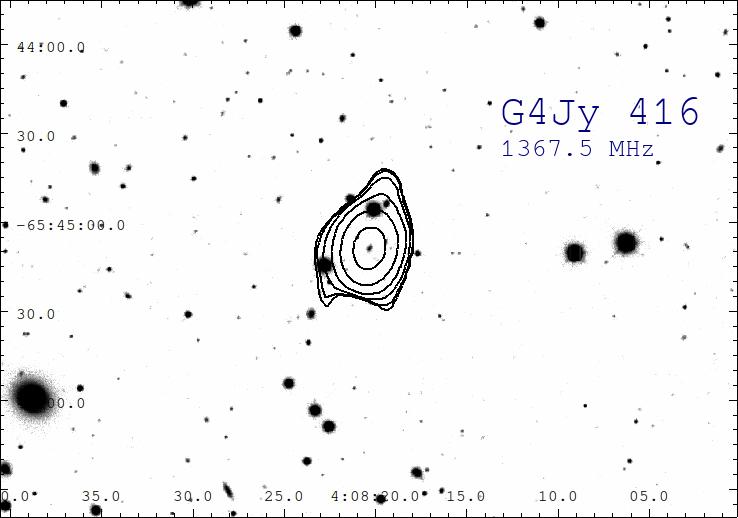}
    \includegraphics[scale=0.225]{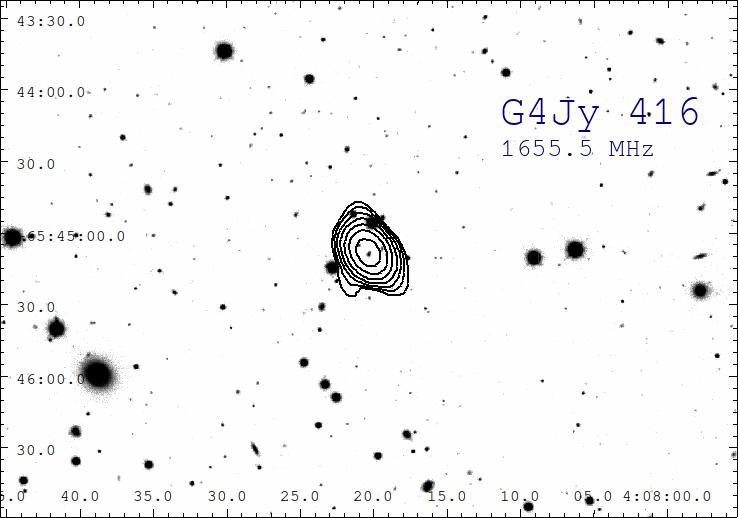}
    \includegraphics[scale=0.225]{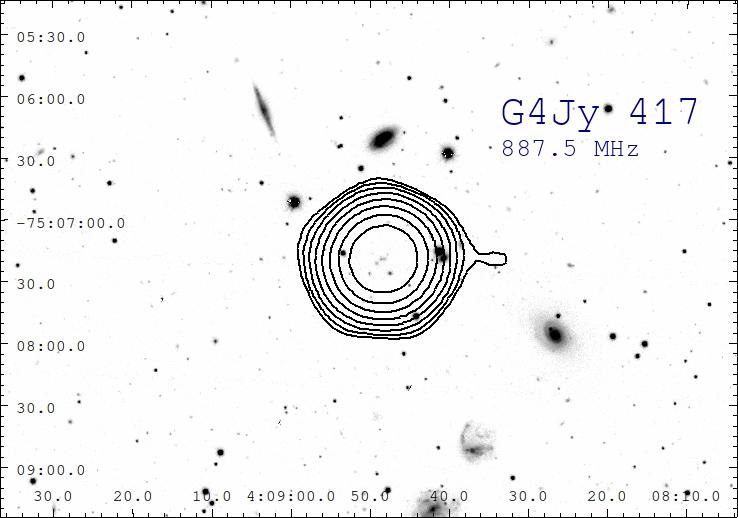}
    \includegraphics[scale=0.225]{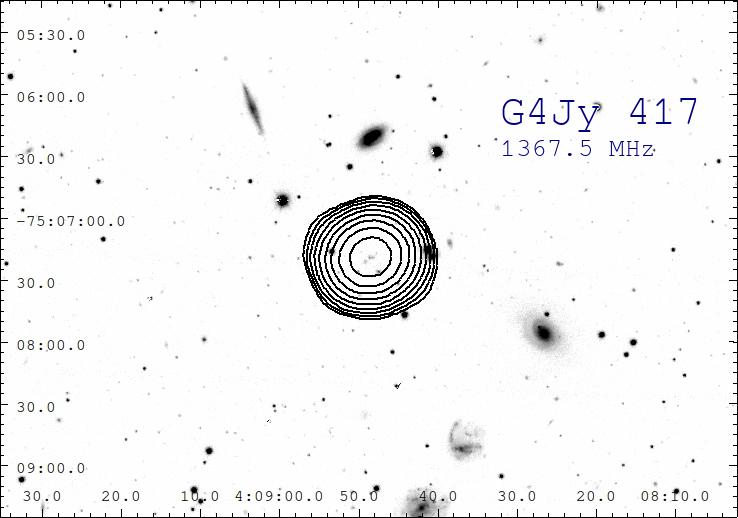}
    \includegraphics[scale=0.225]{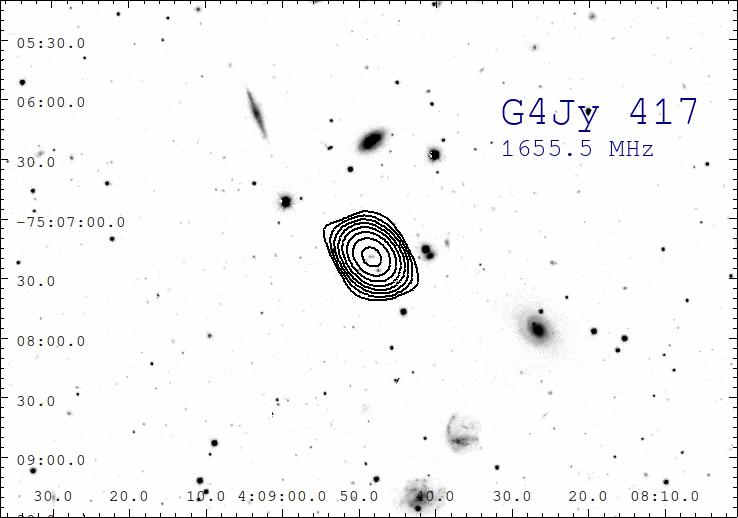}
    \caption{}
    \label{L}
\end{figure*}
\clearpage
\begin{figure*}
    \centering
    \includegraphics[scale=0.225]{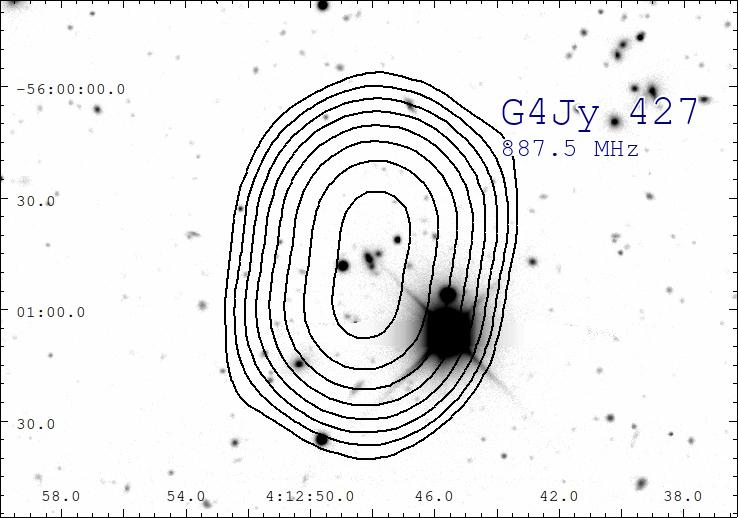}
    \includegraphics[scale=0.225]{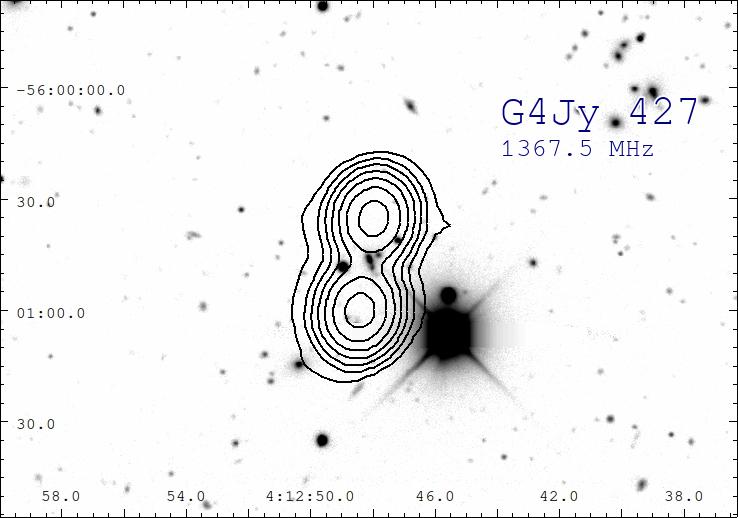}
    \includegraphics[scale=0.225]{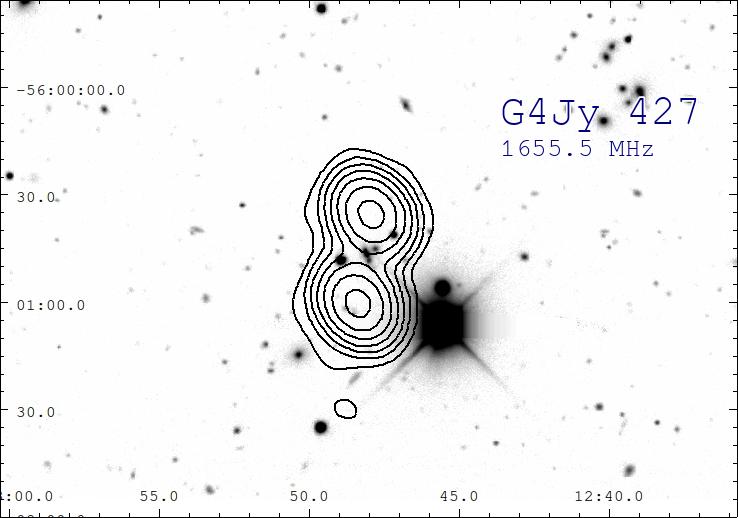}
    \includegraphics[scale=0.225]{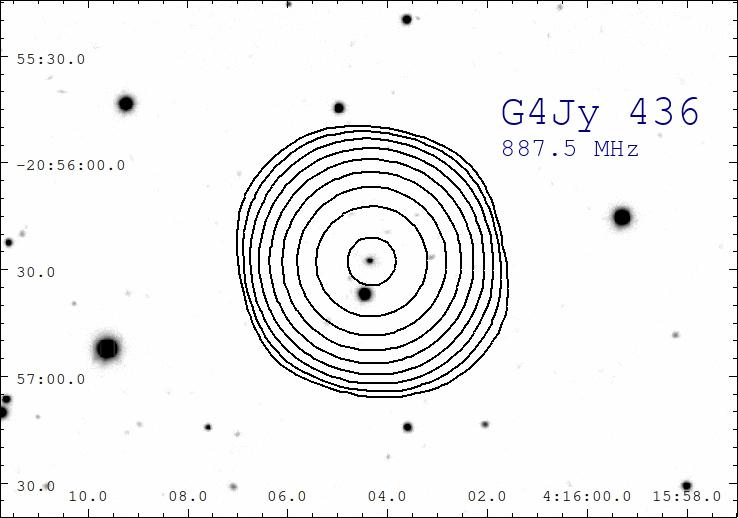}
    \includegraphics[scale=0.225]{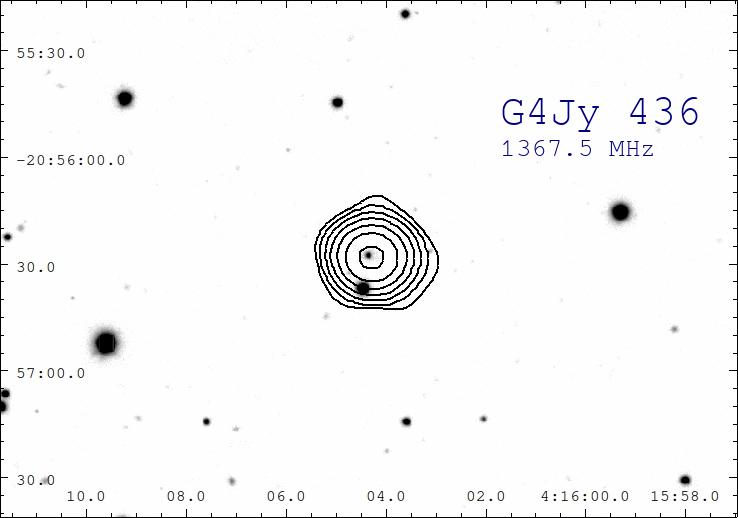}
    \includegraphics[scale=0.225]{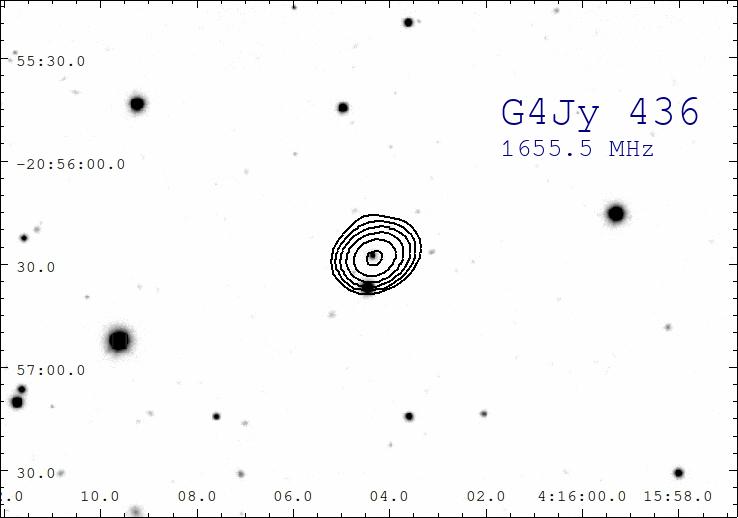}
    \includegraphics[scale=0.225]{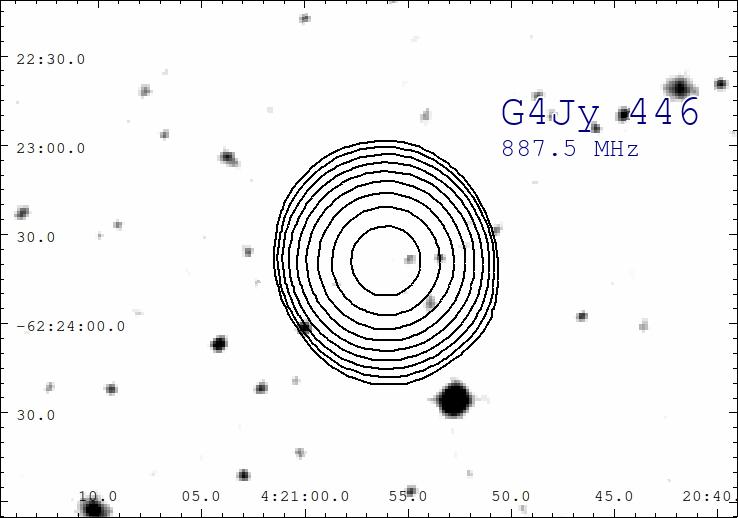}
    \includegraphics[scale=0.225]{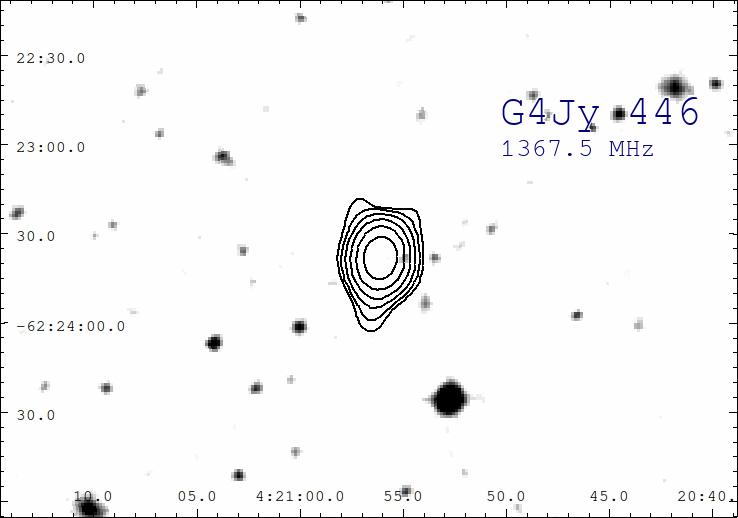}
    \includegraphics[scale=0.225]{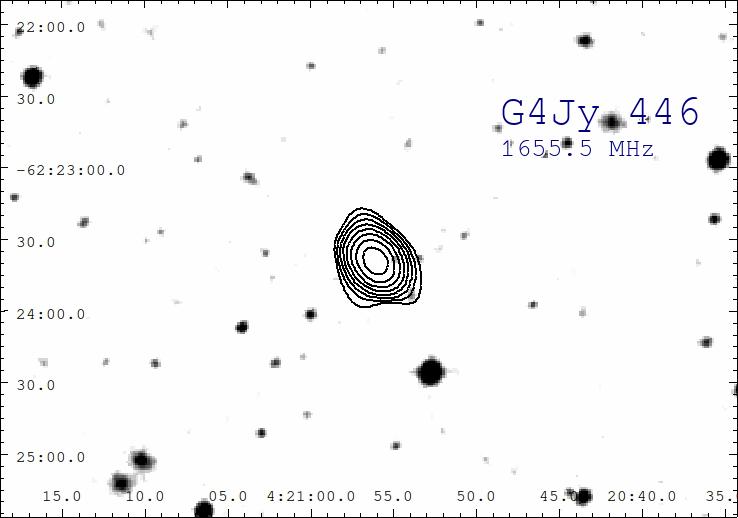}
    \includegraphics[scale=0.225]{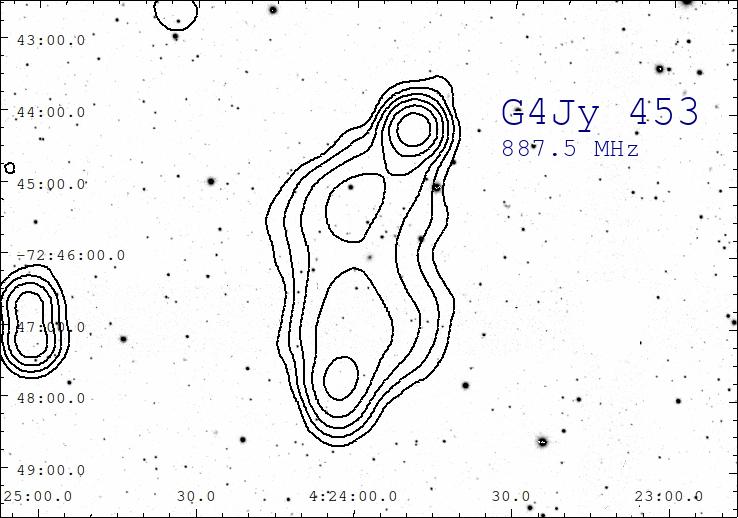}
    \includegraphics[scale=0.225]{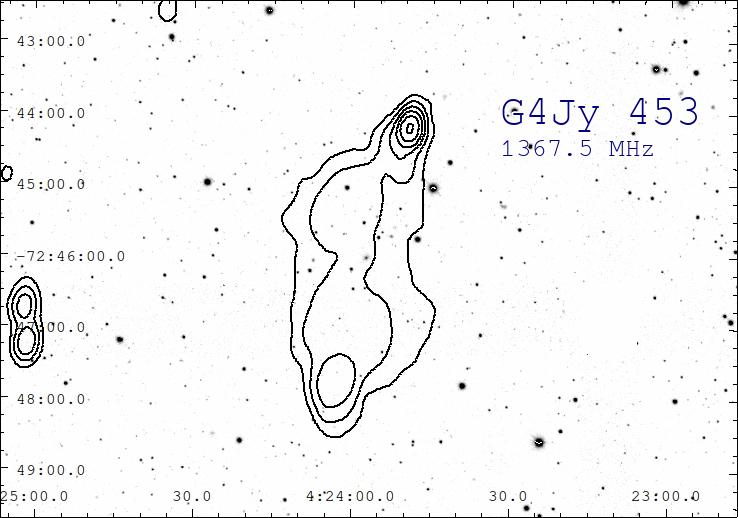}
    \includegraphics[scale=0.225]{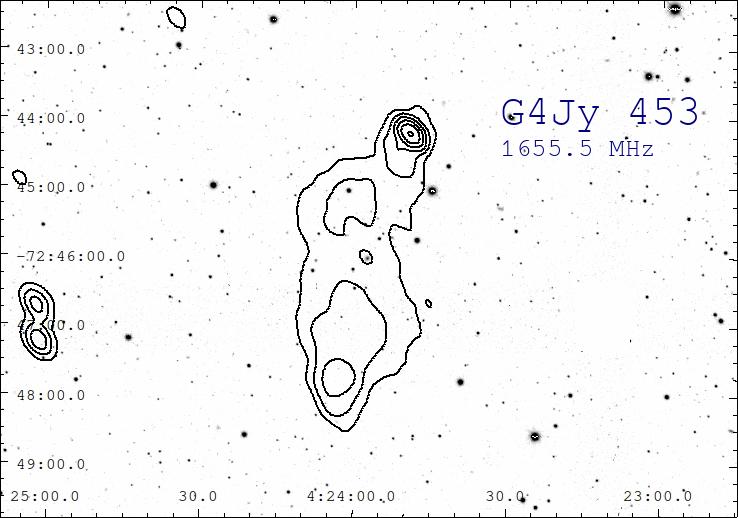}
    \includegraphics[scale=0.225]{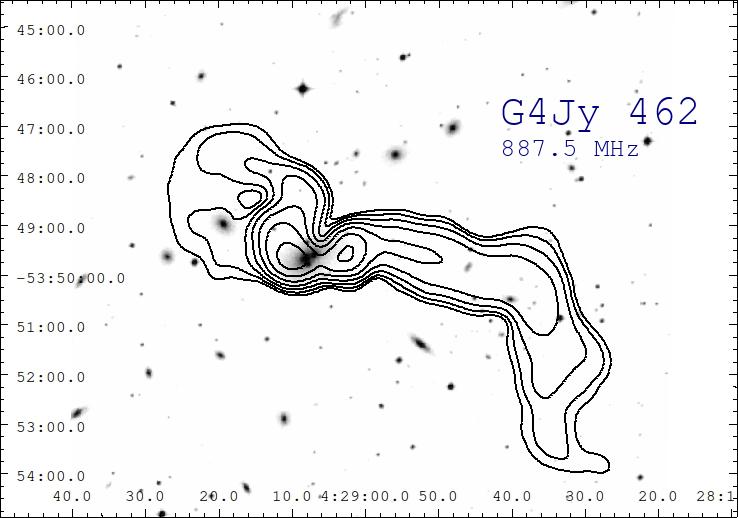}
    \includegraphics[scale=0.225]{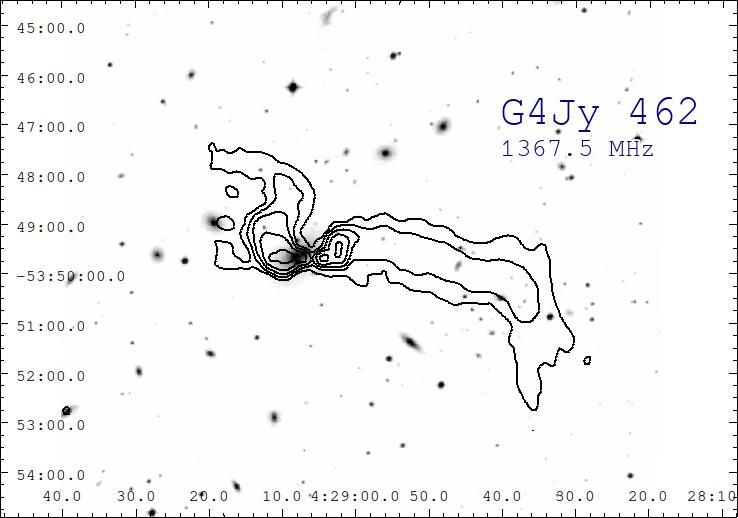}
    \includegraphics[scale=0.225]{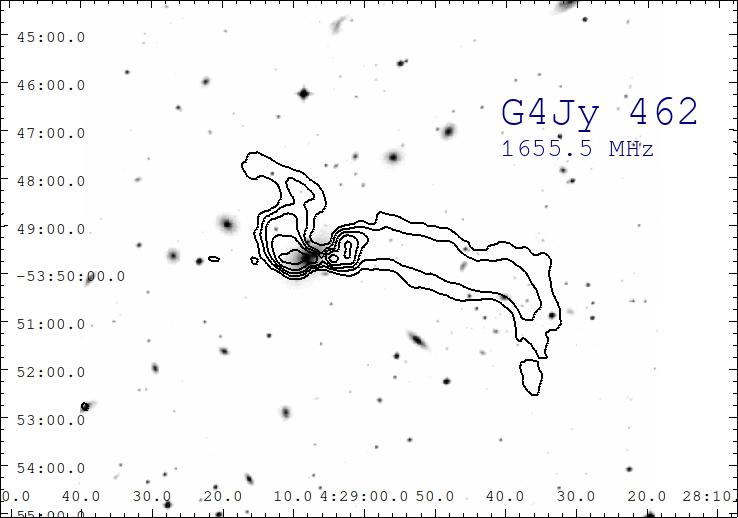}

    \caption{}
    \label{M}
\end{figure*}
\clearpage
\begin{figure*}
    \centering
    \includegraphics[scale=0.225]{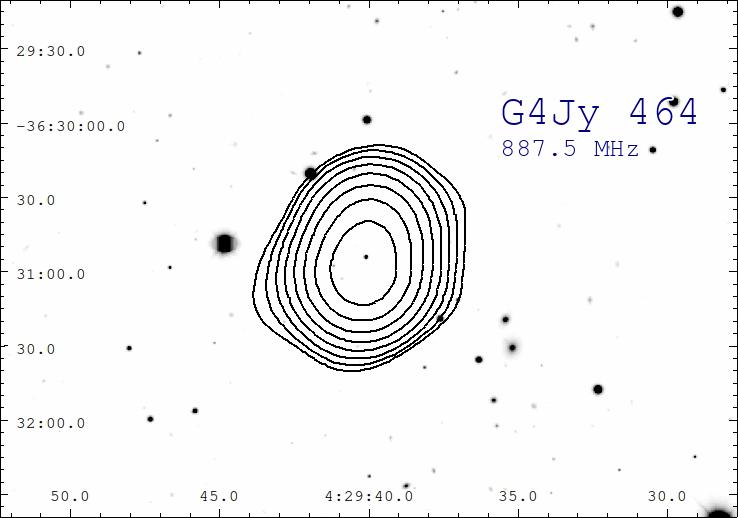}
    \includegraphics[scale=0.225]{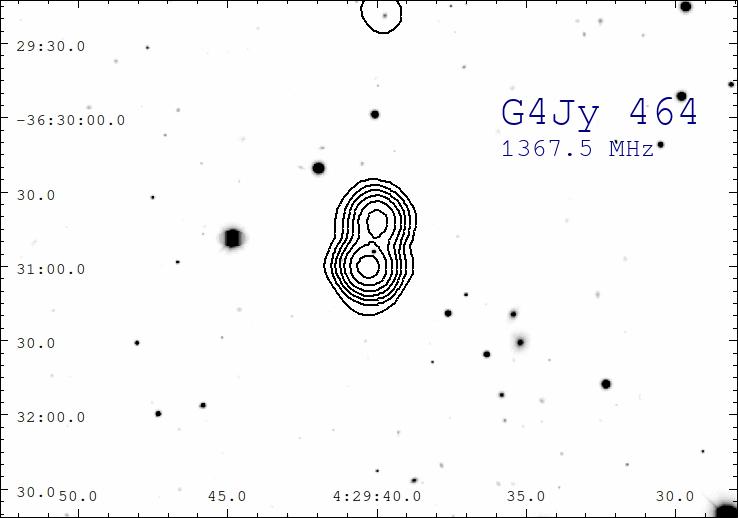}
    \includegraphics[scale=0.225]{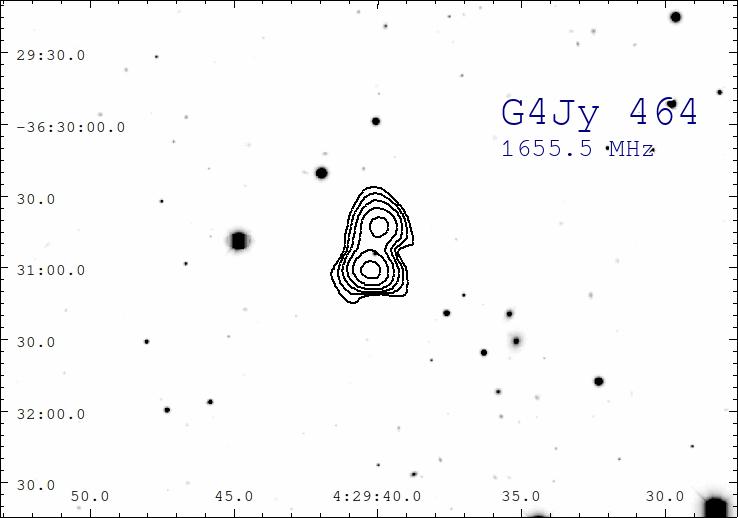}
    \includegraphics[scale=0.225]{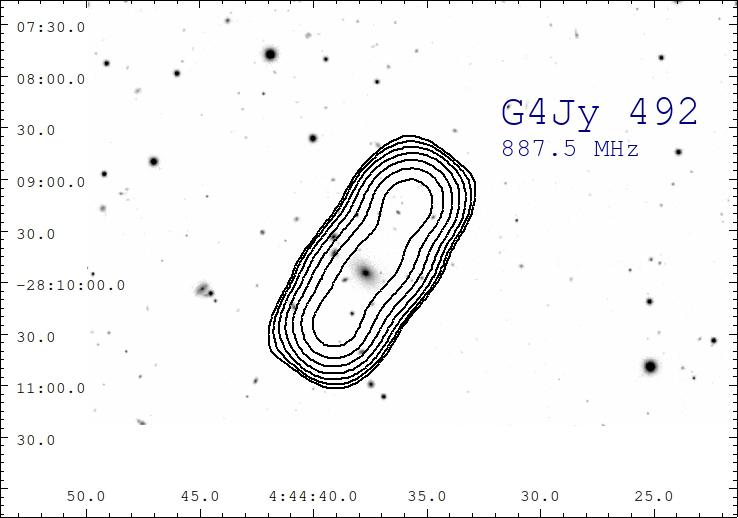}
    \includegraphics[scale=0.225]{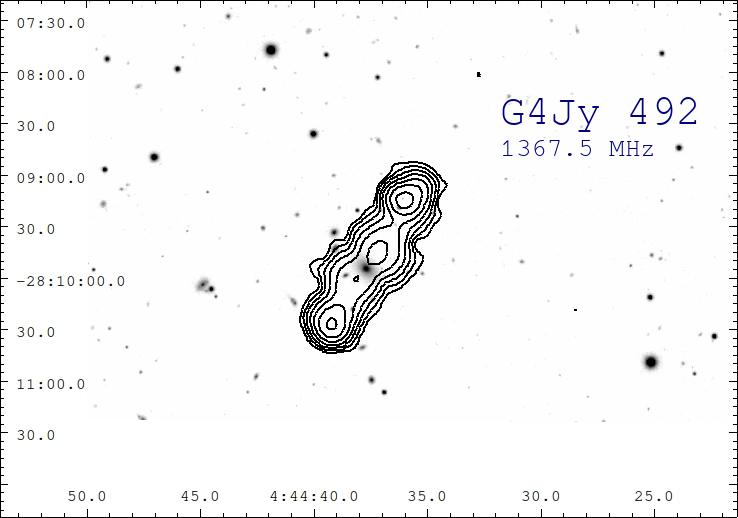}
    \includegraphics[scale=0.225]{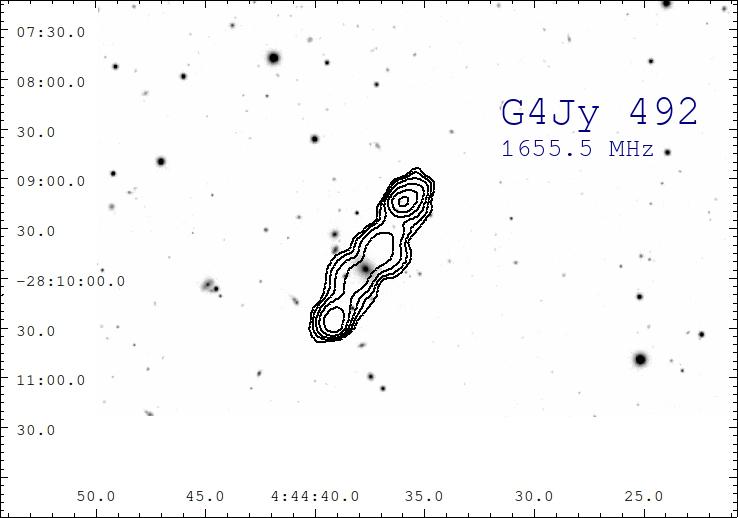}
    \includegraphics[scale=0.225]{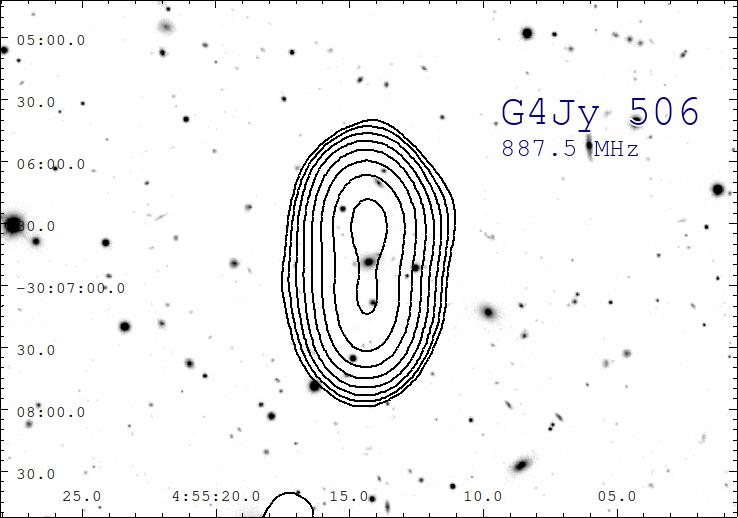}
    \includegraphics[scale=0.225]{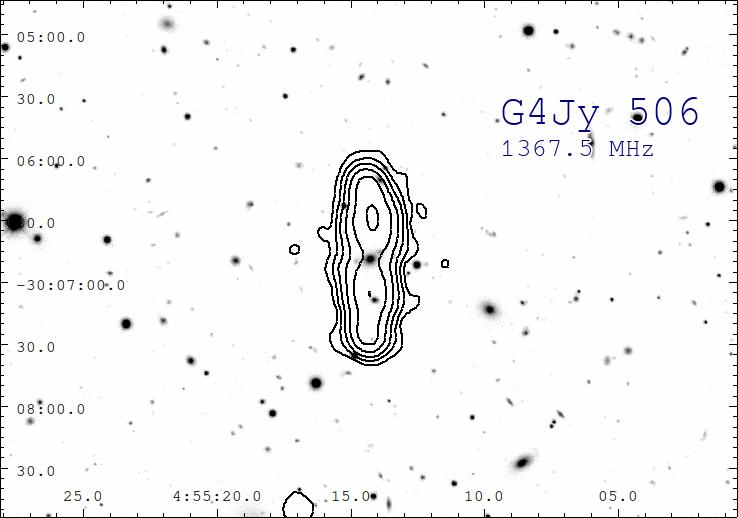}
    \includegraphics[scale=0.225]{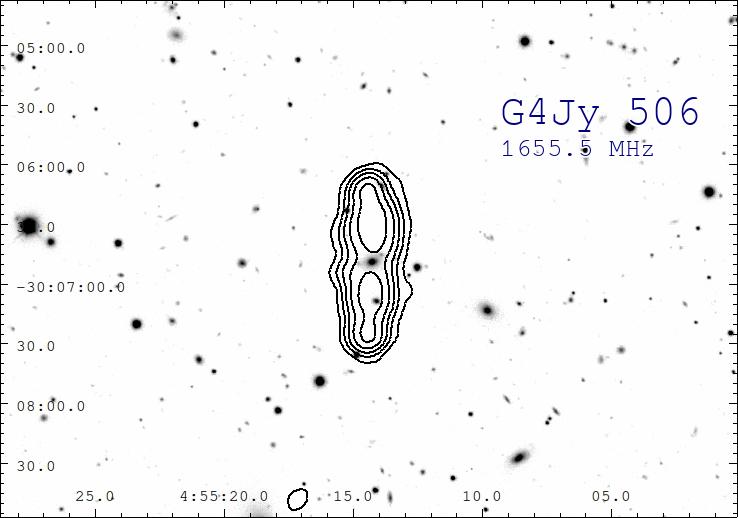}
    \includegraphics[scale=0.225]{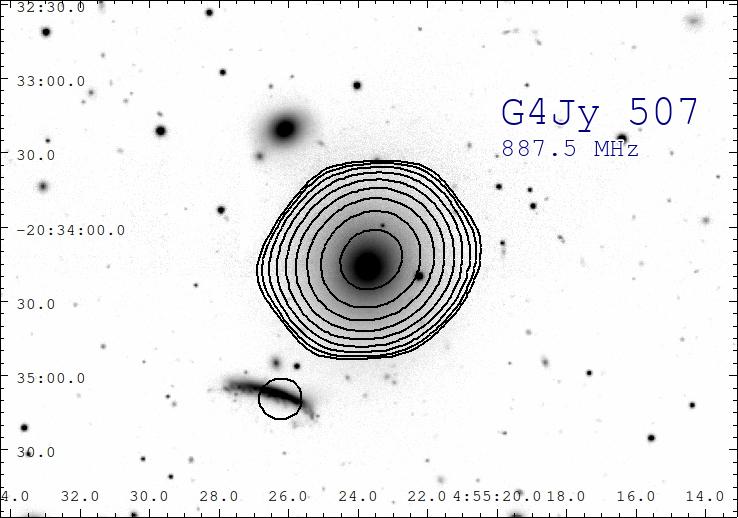}
    \includegraphics[scale=0.225]{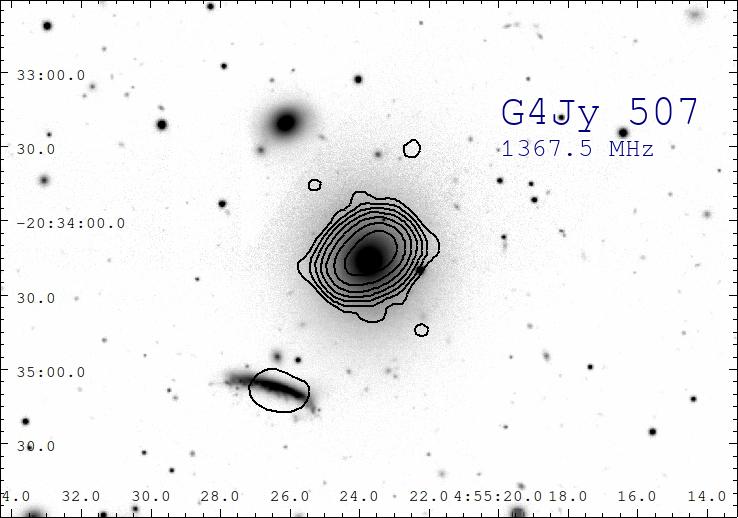}
    \includegraphics[scale=0.225]{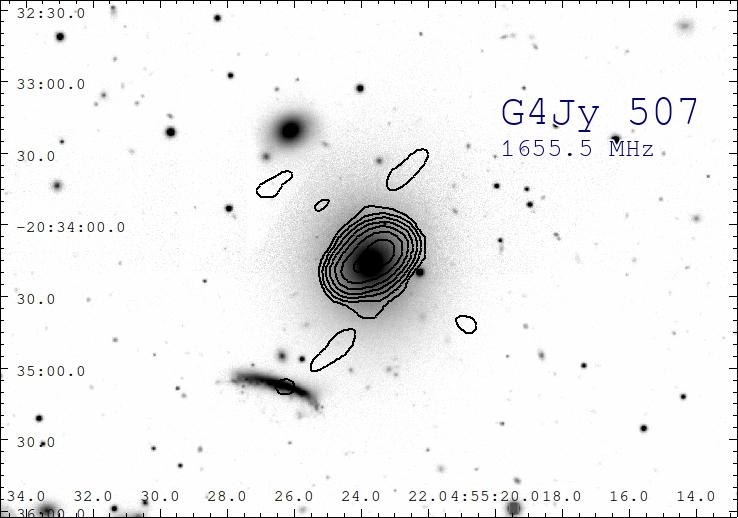}
    \includegraphics[scale=0.225]{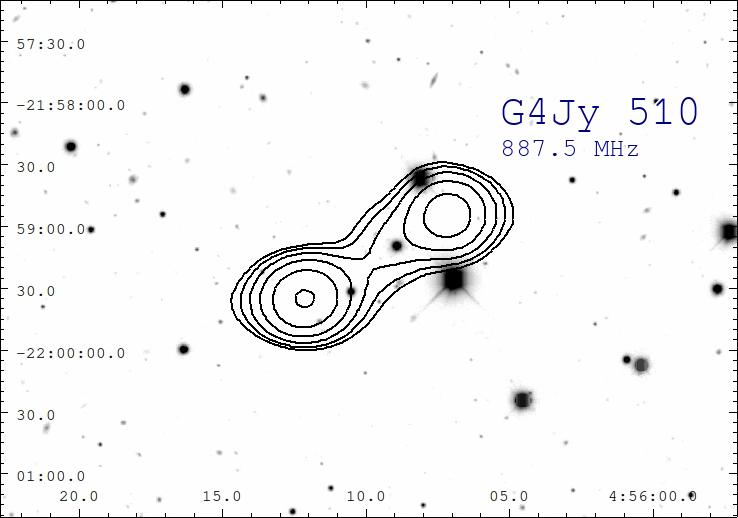}
    \includegraphics[scale=0.225]{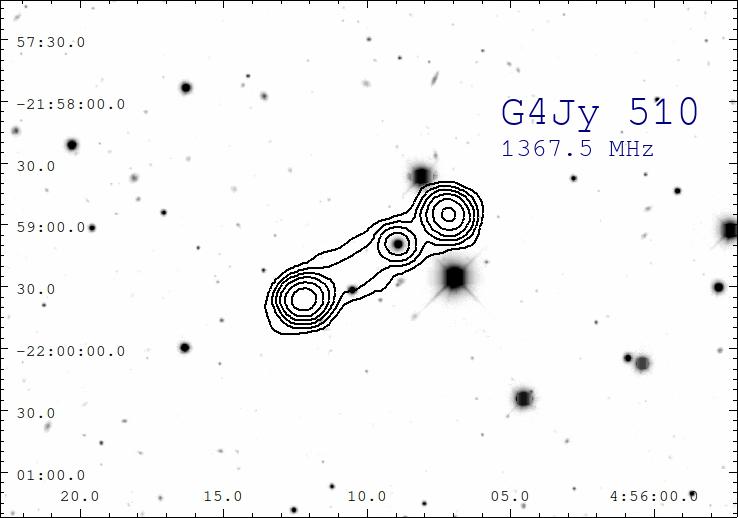}
    \includegraphics[scale=0.225]{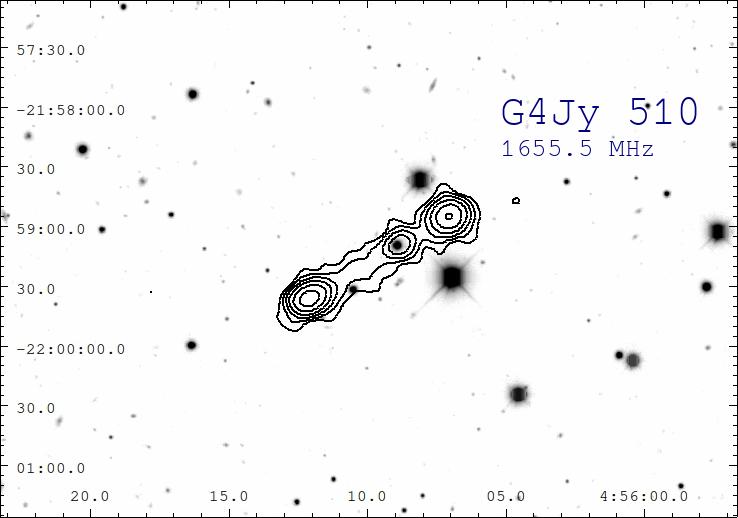}

    \caption{}
    \label{N}
\end{figure*}
\clearpage
\begin{figure*}
    \centering
    \includegraphics[scale=0.225]{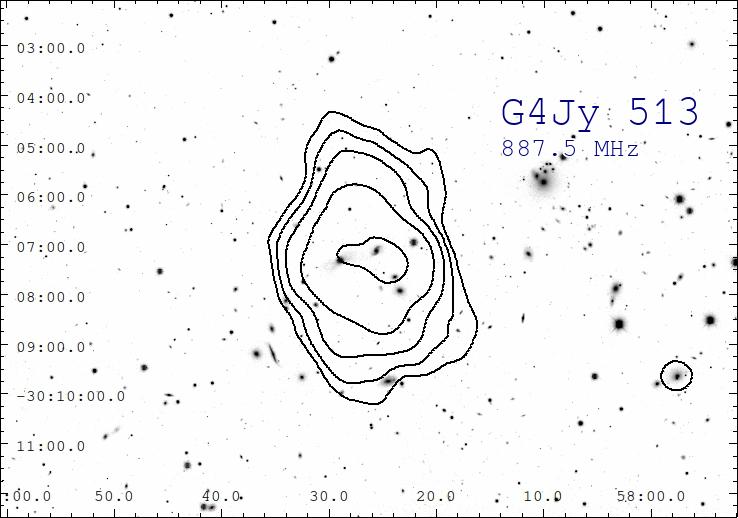}
    \includegraphics[scale=0.225]{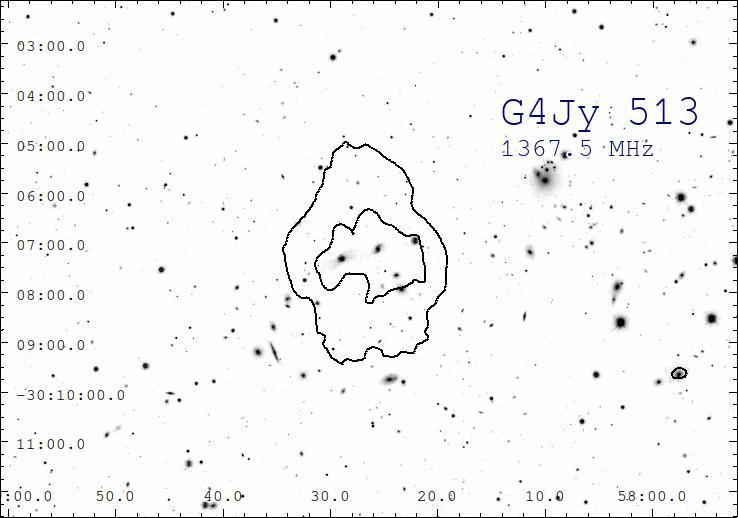}
    \includegraphics[scale=0.225]{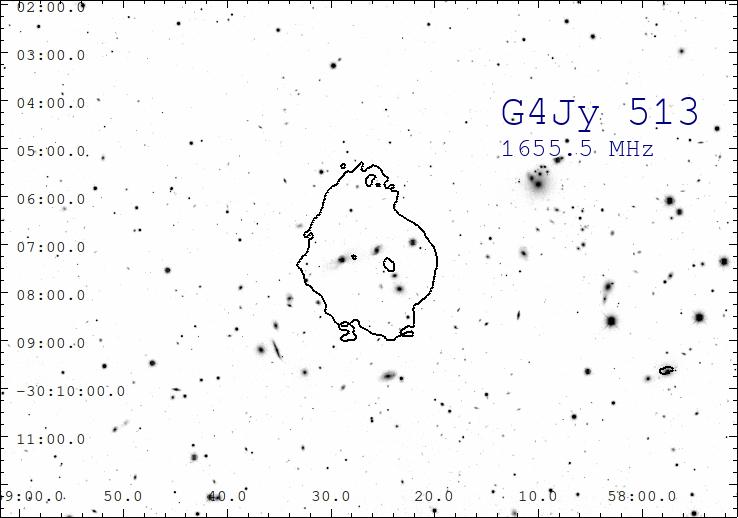}
    \includegraphics[scale=0.225]{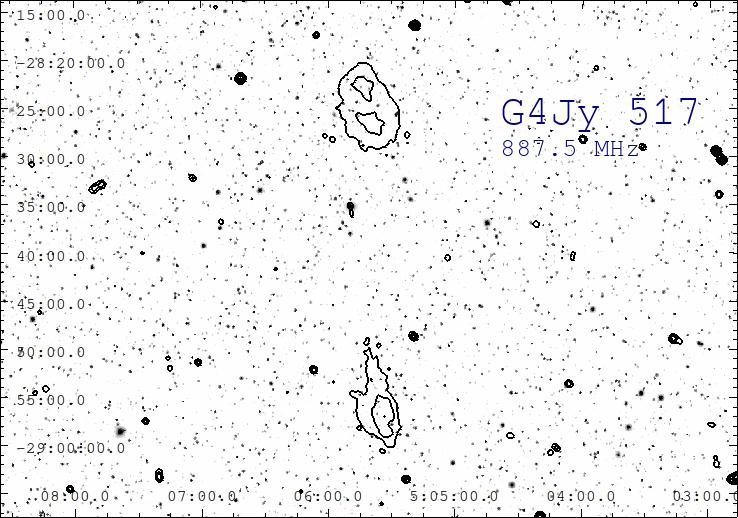}
    \includegraphics[scale=0.225]{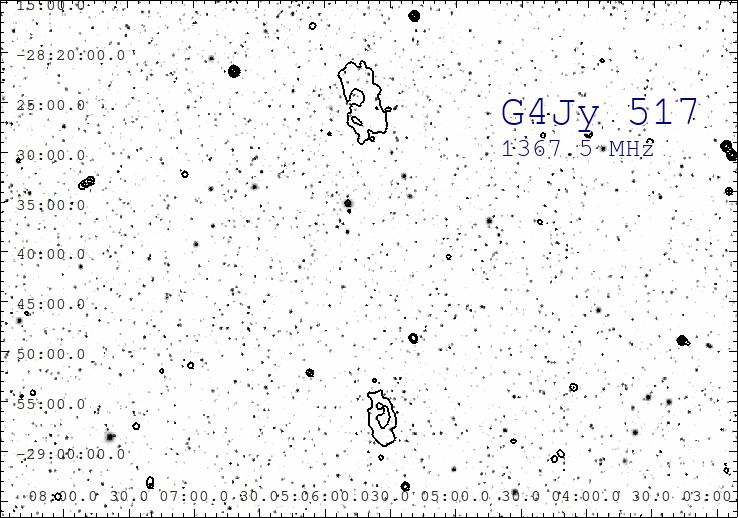}
    \includegraphics[scale=0.225]{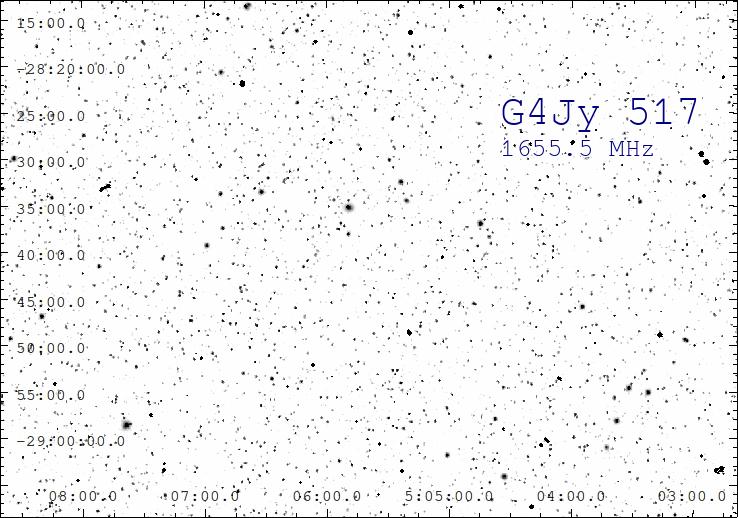}
    \includegraphics[scale=0.225]{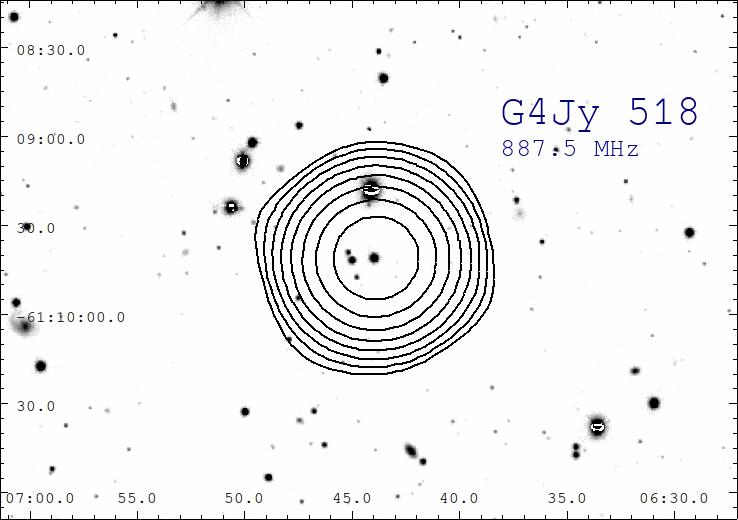}
    \includegraphics[scale=0.225]{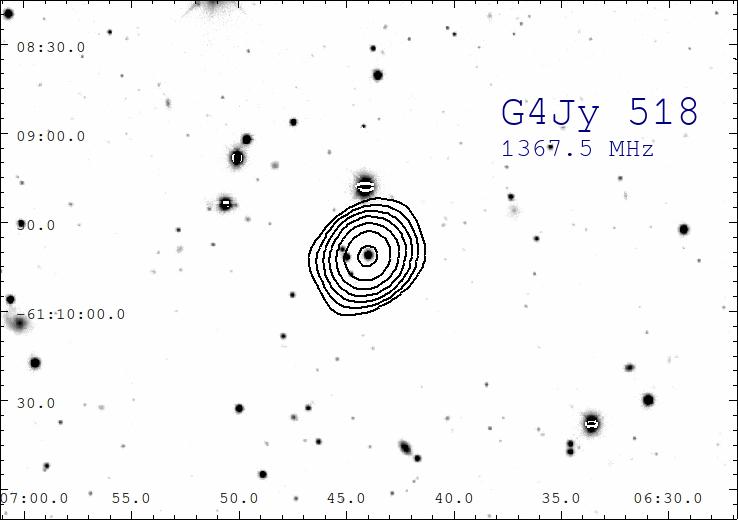}
    \includegraphics[scale=0.225]{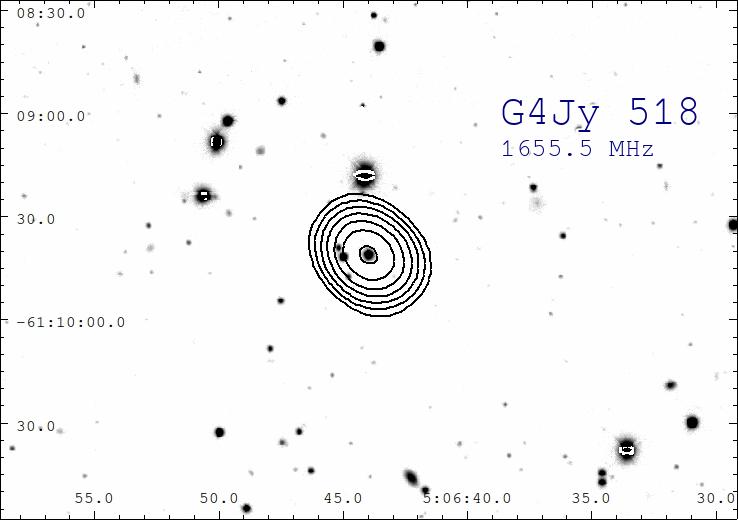}
    \includegraphics[scale=0.225]{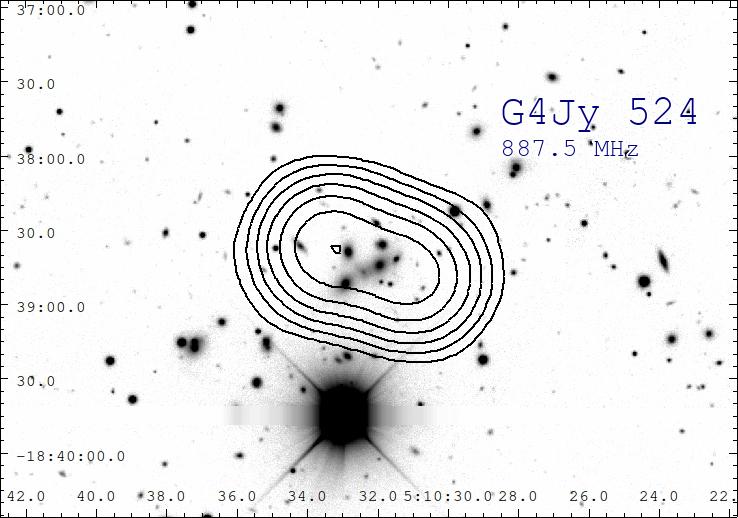}
    \includegraphics[scale=0.225]{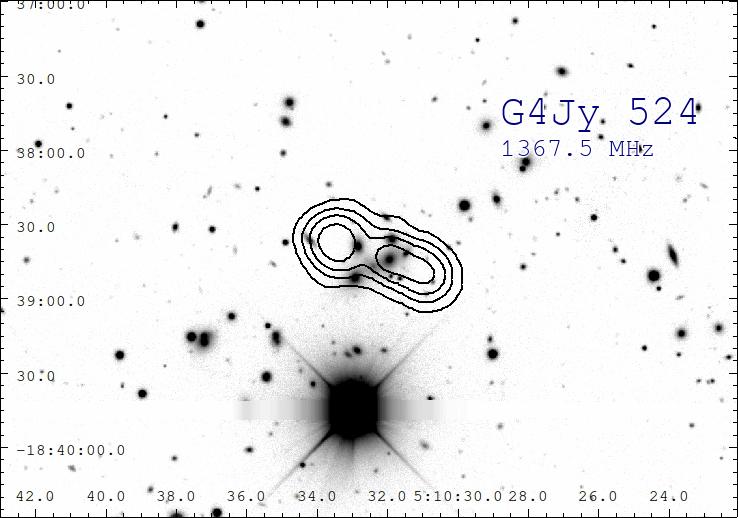}
    \includegraphics[scale=0.225]{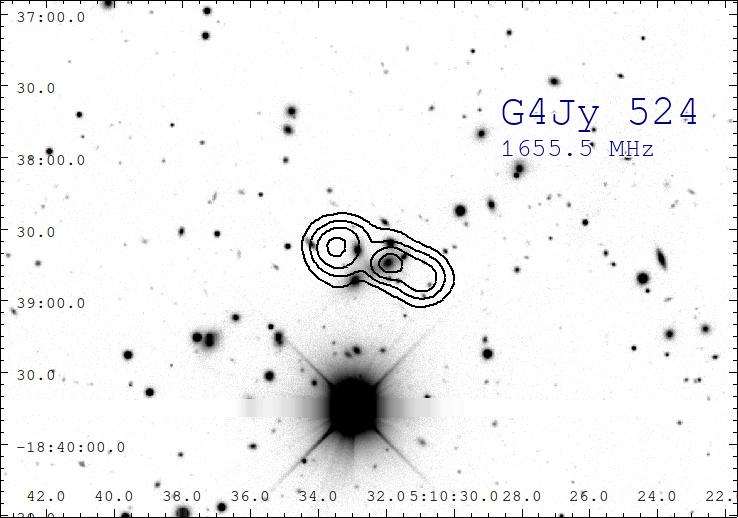}
    \includegraphics[scale=0.225]{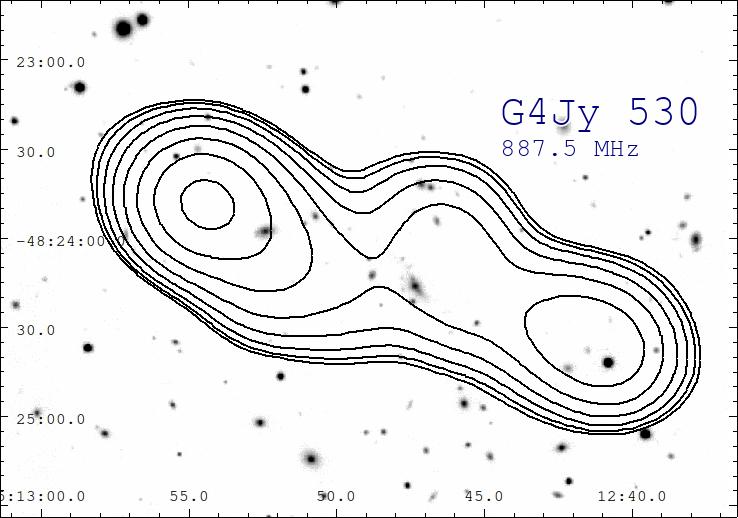}
    \includegraphics[scale=0.225]{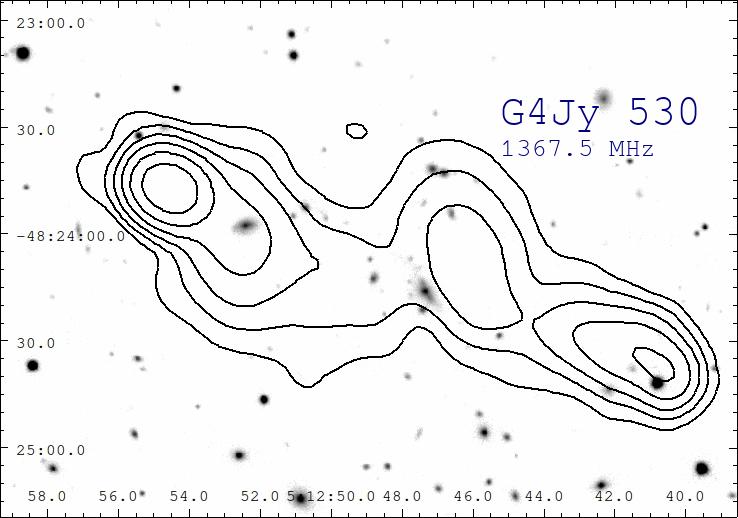}
    \includegraphics[scale=0.225]{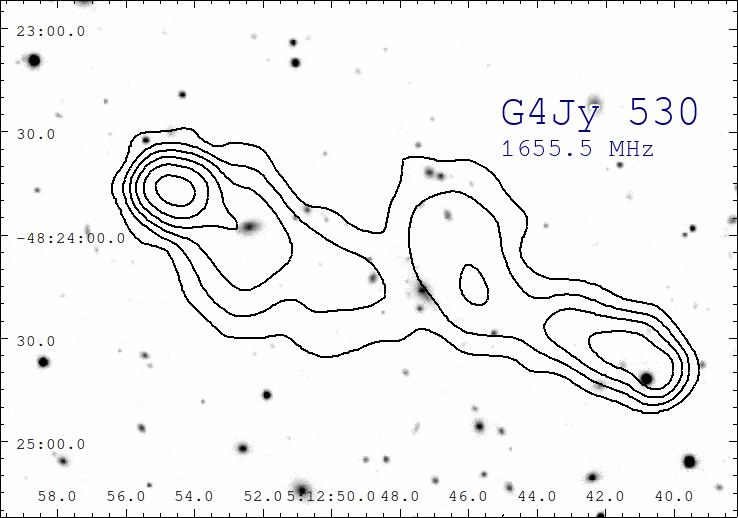}

    \caption{}
    \label{O}
\end{figure*}
\clearpage
\begin{figure*}
    \centering
    \includegraphics[scale=0.225]{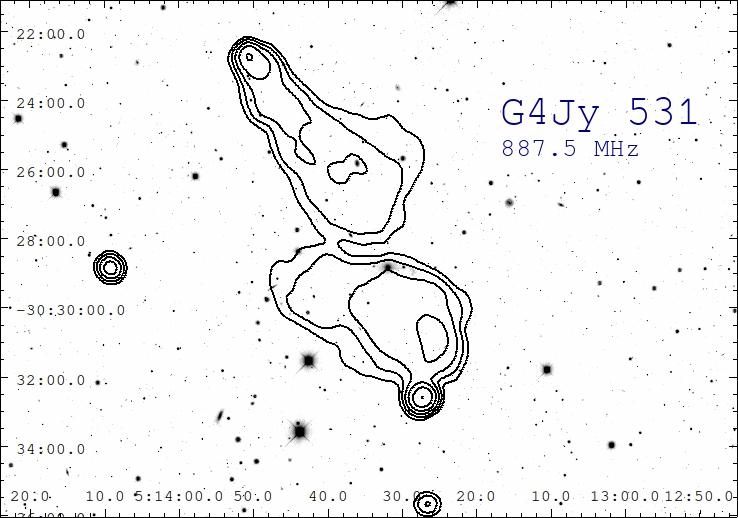}
    \includegraphics[scale=0.225]{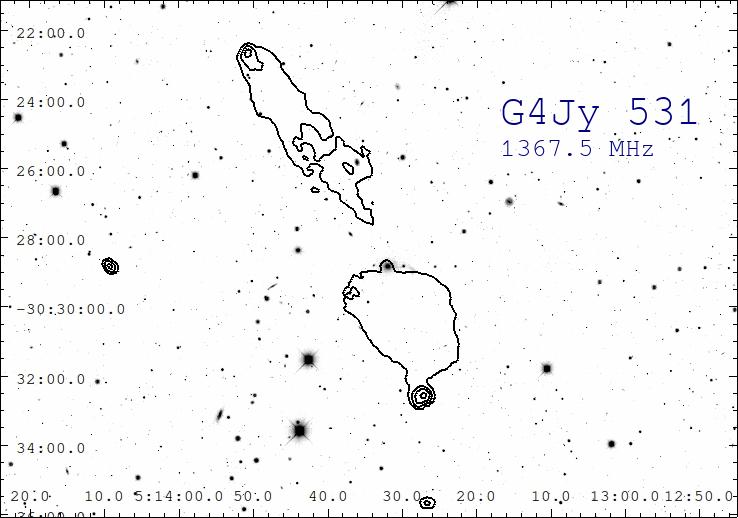}
    \includegraphics[scale=0.225]{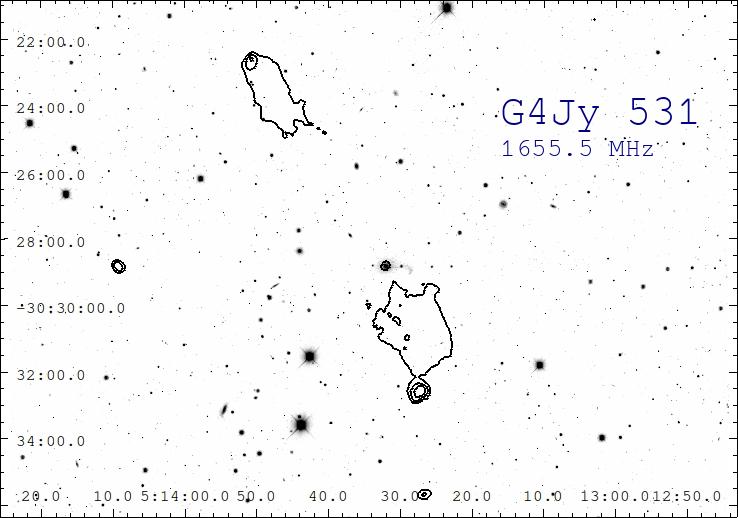}
    \includegraphics[scale=0.225]{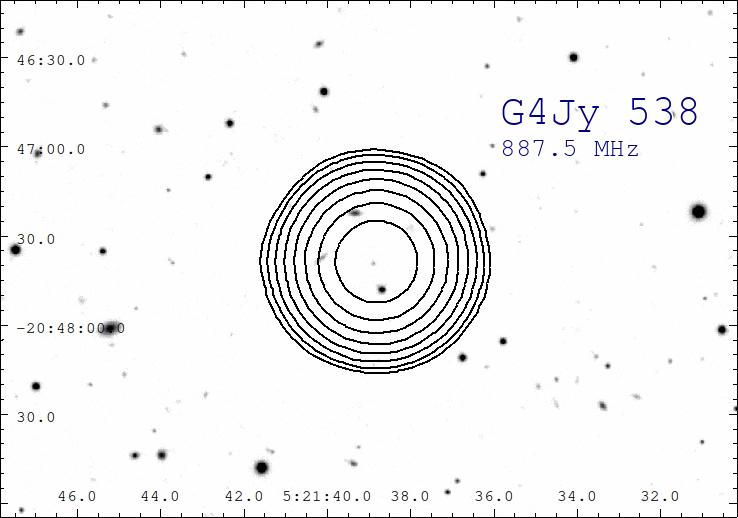}
    \includegraphics[scale=0.225]{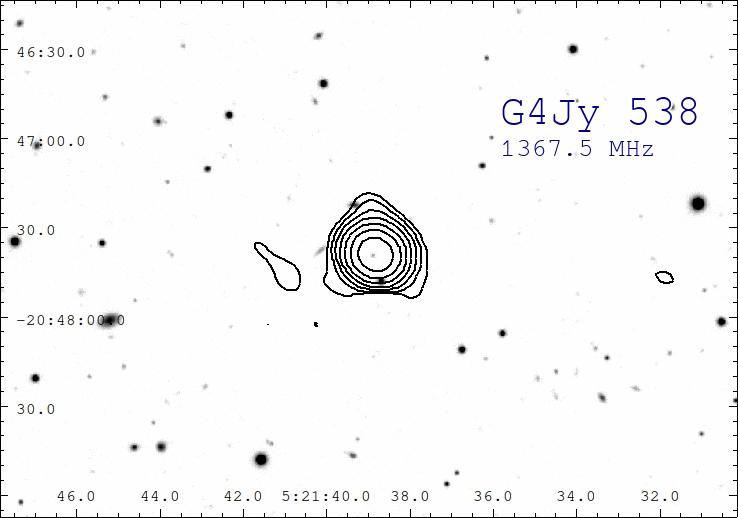}
    \includegraphics[scale=0.225]{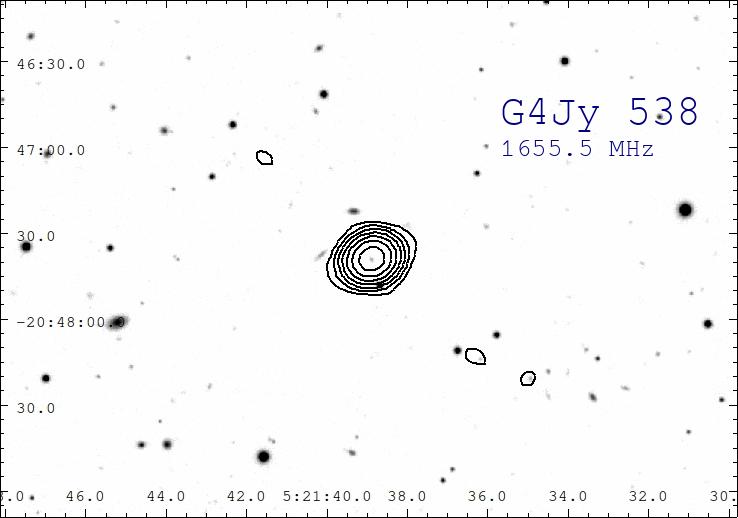}
    \includegraphics[scale=0.225]{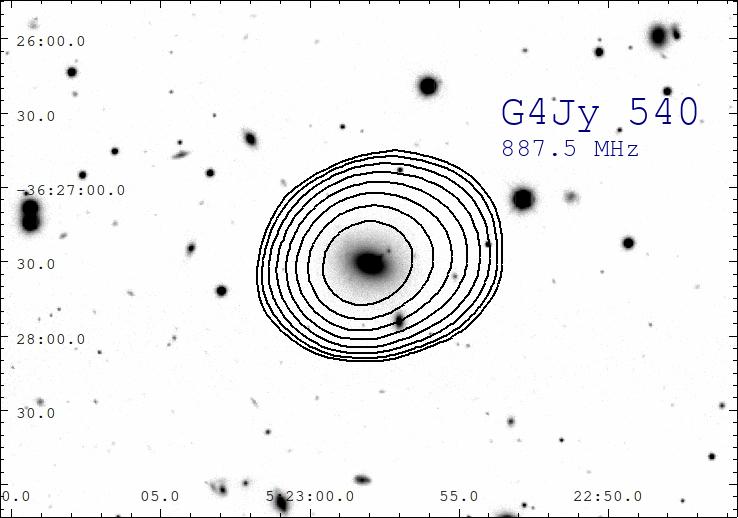}
    \includegraphics[scale=0.225]{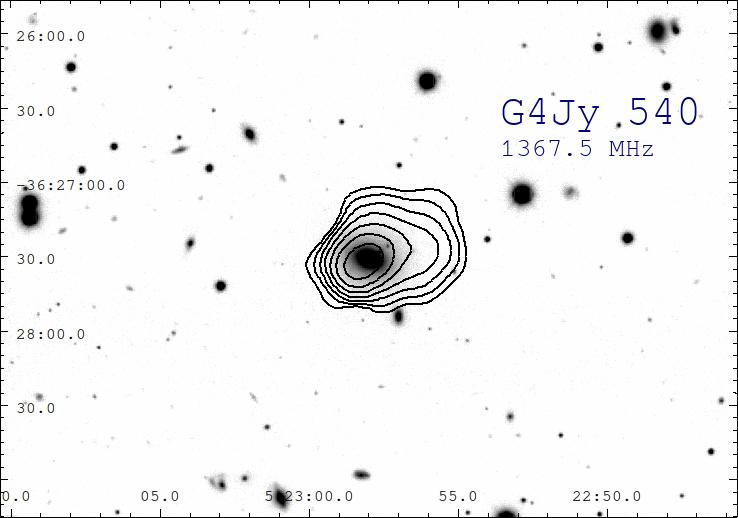}
    \includegraphics[scale=0.225]{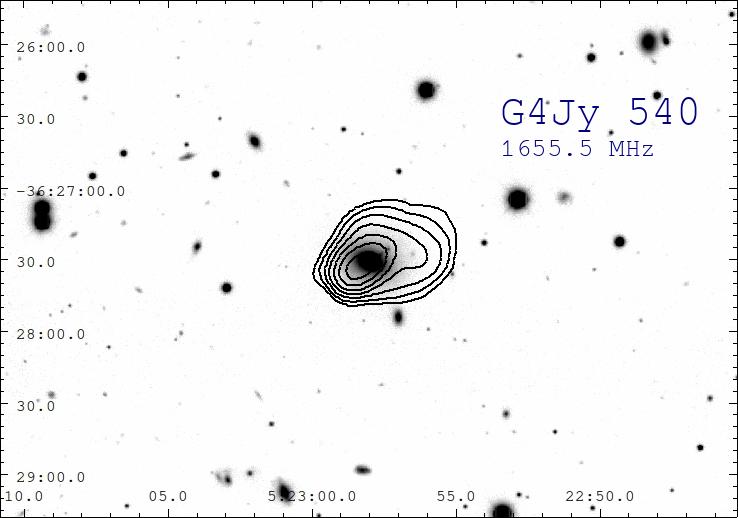}
    \includegraphics[scale=0.225]{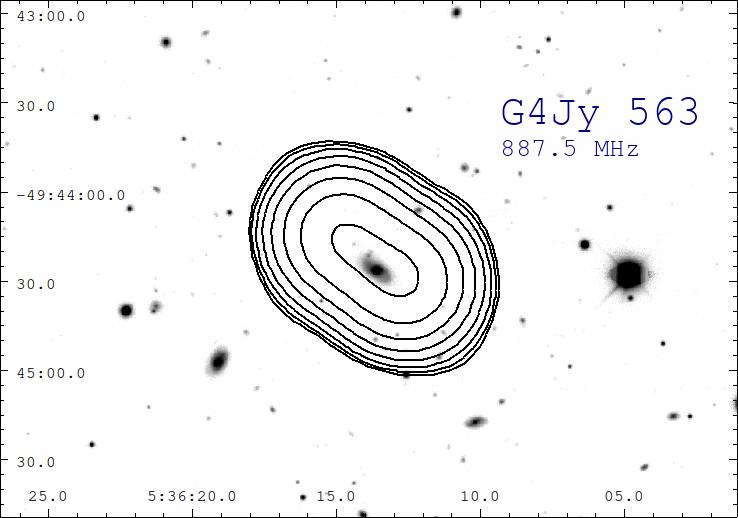}
    \includegraphics[scale=0.225]{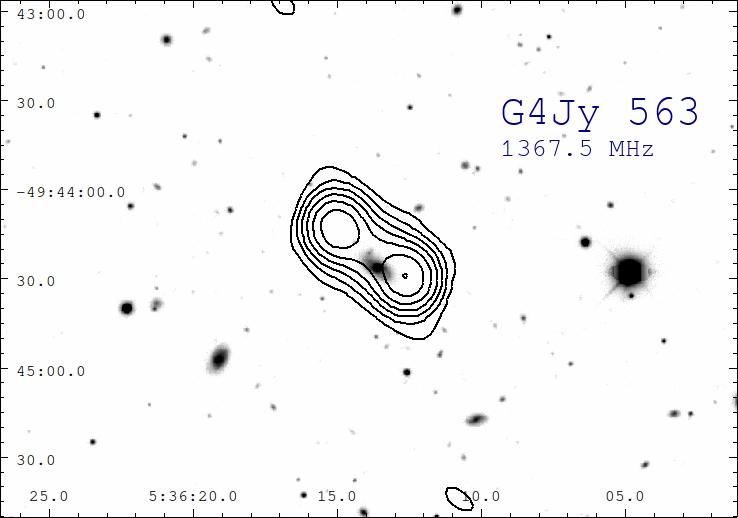}
    \includegraphics[scale=0.225]{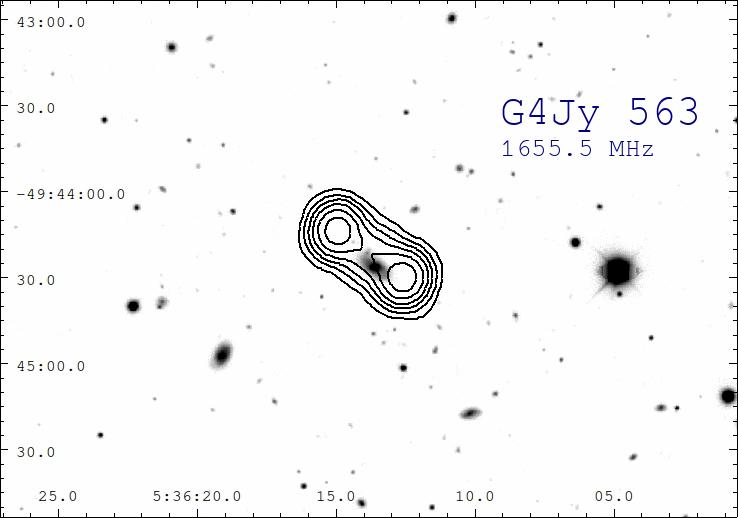}
    \includegraphics[scale=0.225]{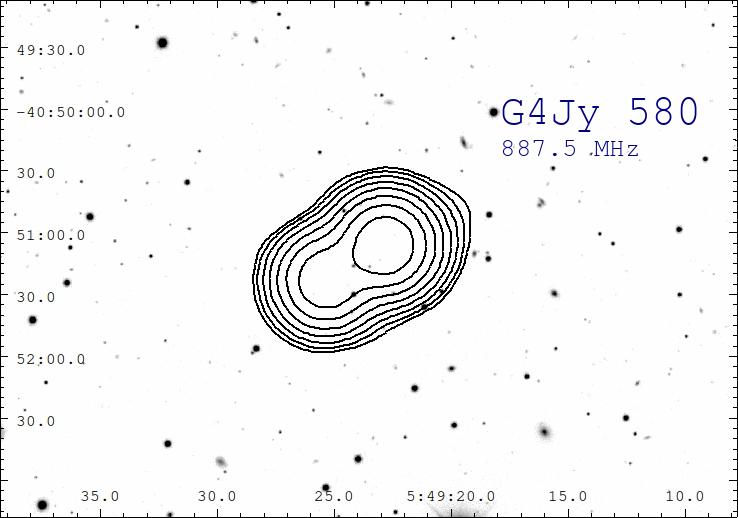}
    \includegraphics[scale=0.225]{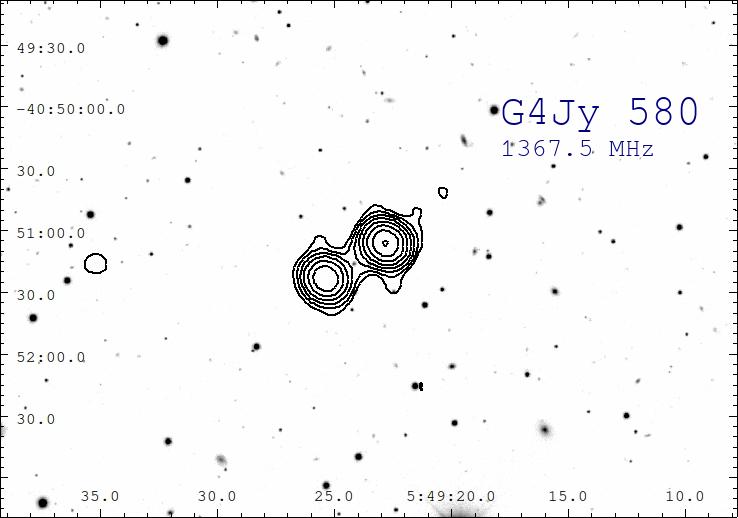}
    \includegraphics[scale=0.225]{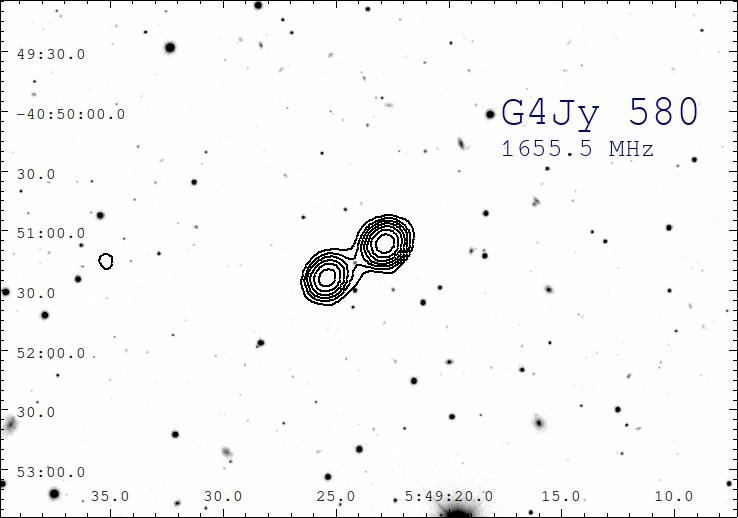}

    \caption{}
    \label{P}
\end{figure*}
\clearpage

\begin{figure*}
    \centering
    \includegraphics[scale=0.225]{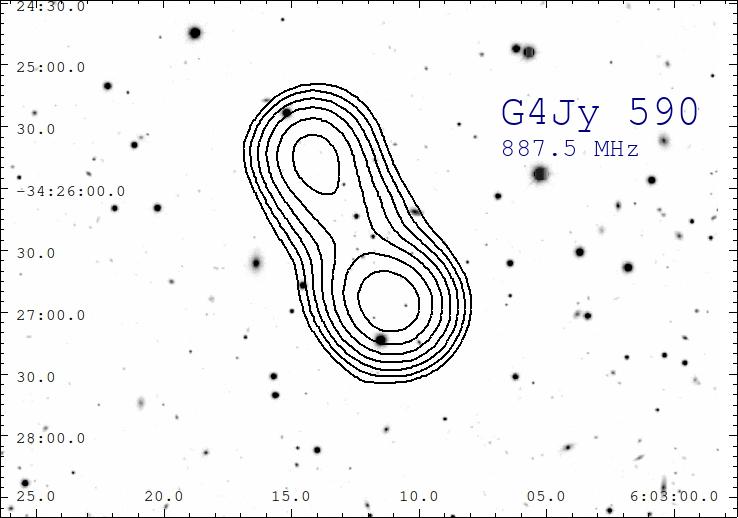}
    \includegraphics[scale=0.225]{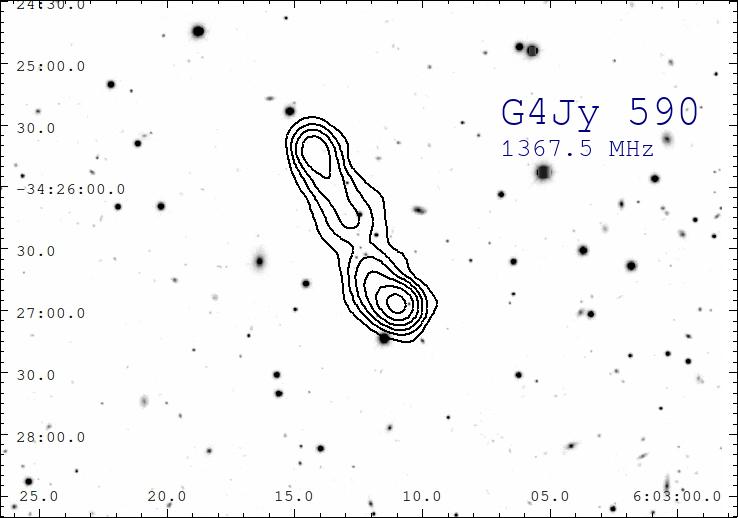}
    \includegraphics[scale=0.225]{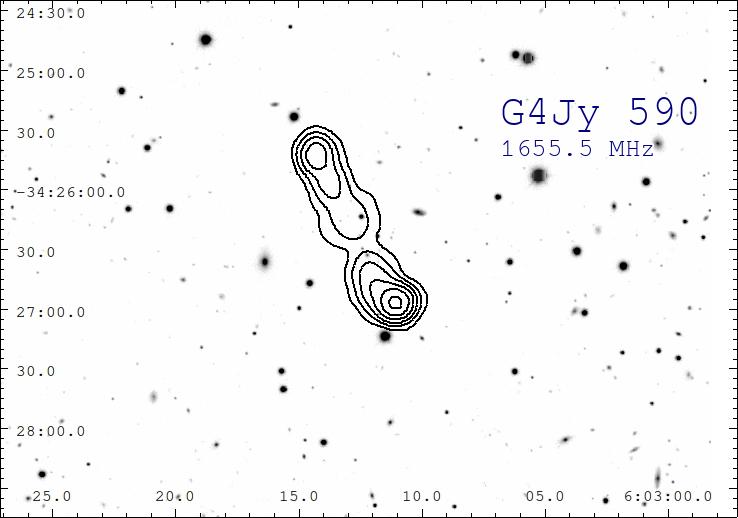}
    \includegraphics[scale=0.225]{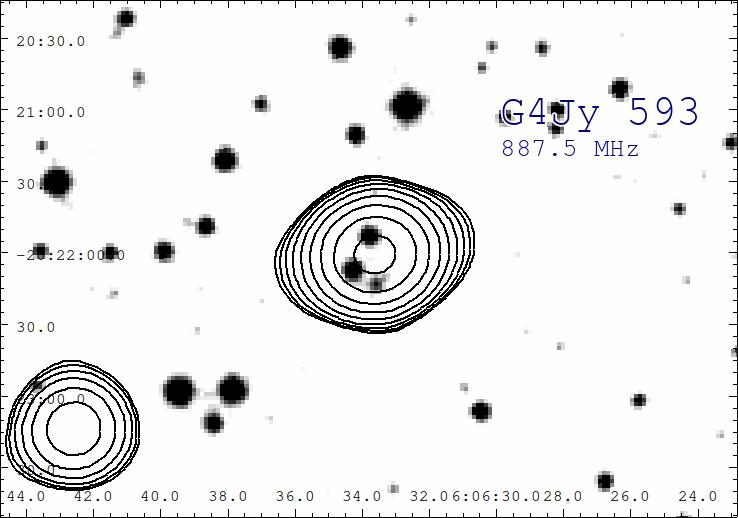}
    \includegraphics[scale=0.225]{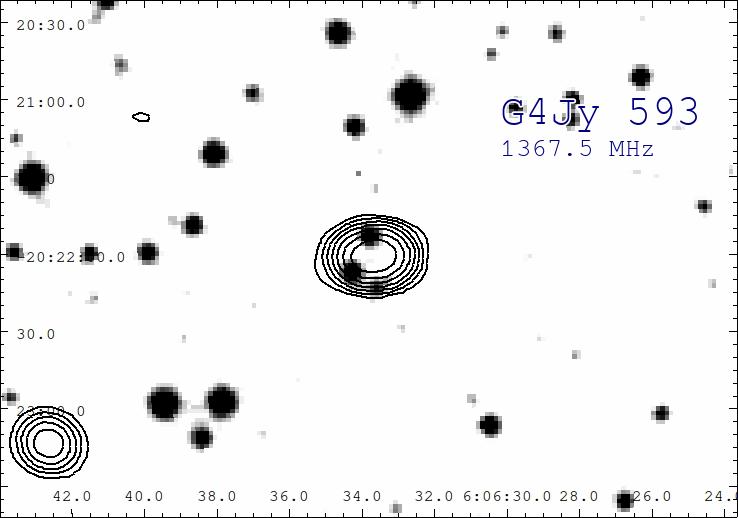}
    \includegraphics[scale=0.225]{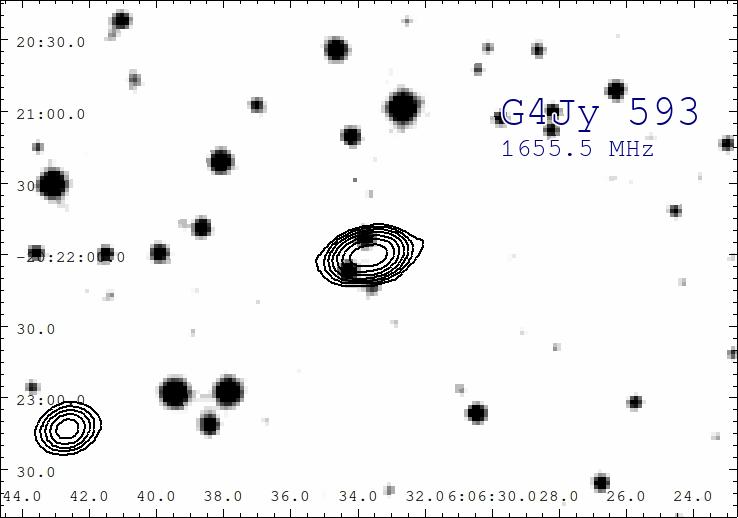}
    \includegraphics[scale=0.225]{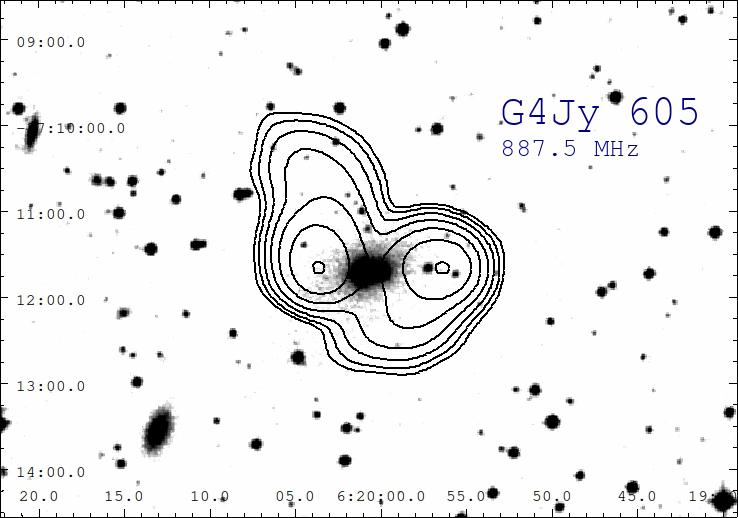}
    \includegraphics[scale=0.225]{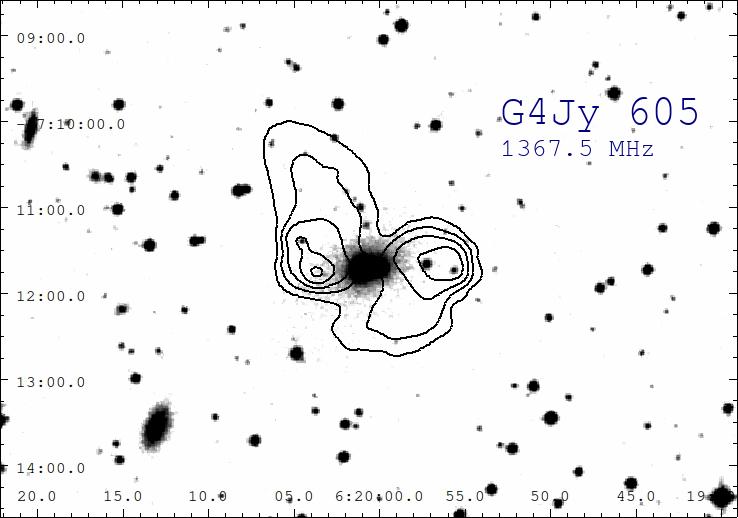}
    \includegraphics[scale=0.225]{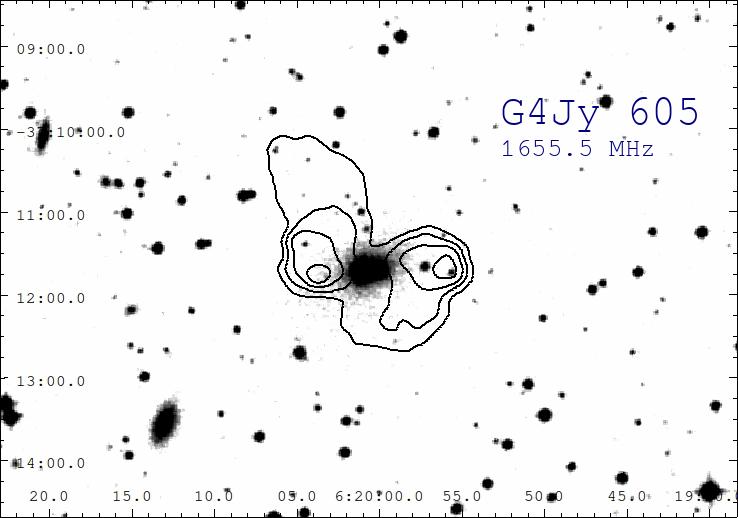}
    \includegraphics[scale=0.225]{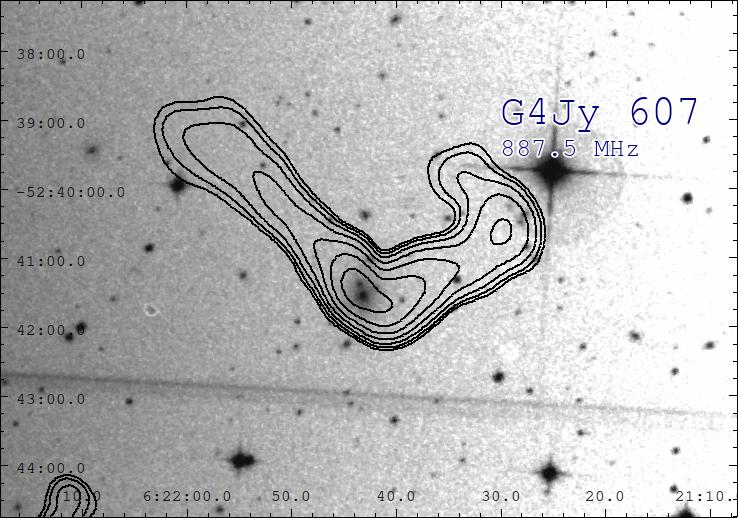}
    \includegraphics[scale=0.225]{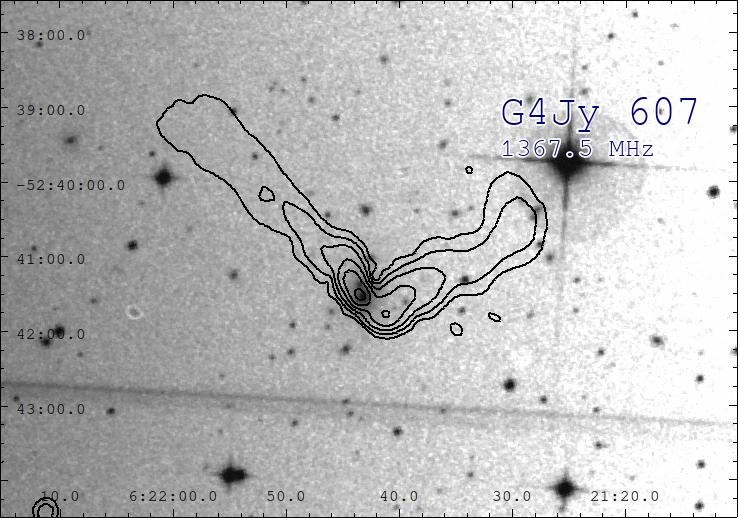}
    \includegraphics[scale=0.225]{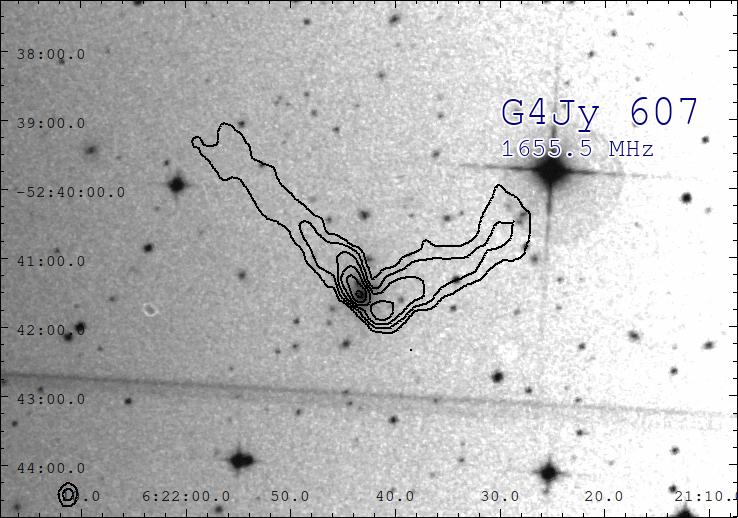}
    \includegraphics[scale=0.225]{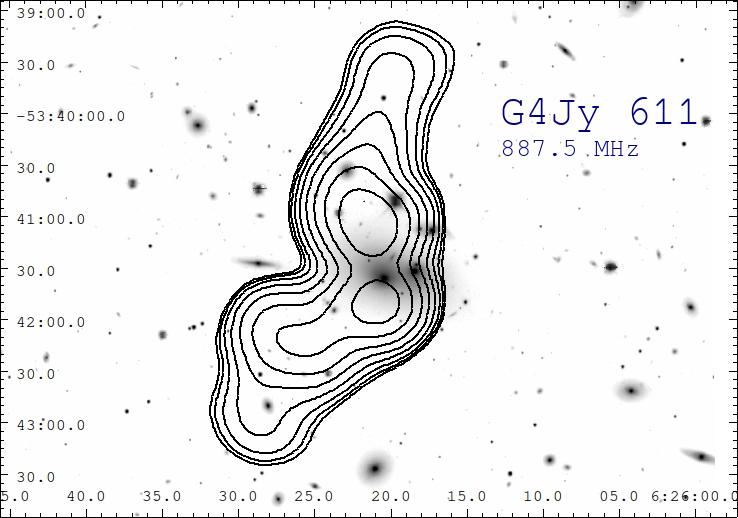}
    \includegraphics[scale=0.225]{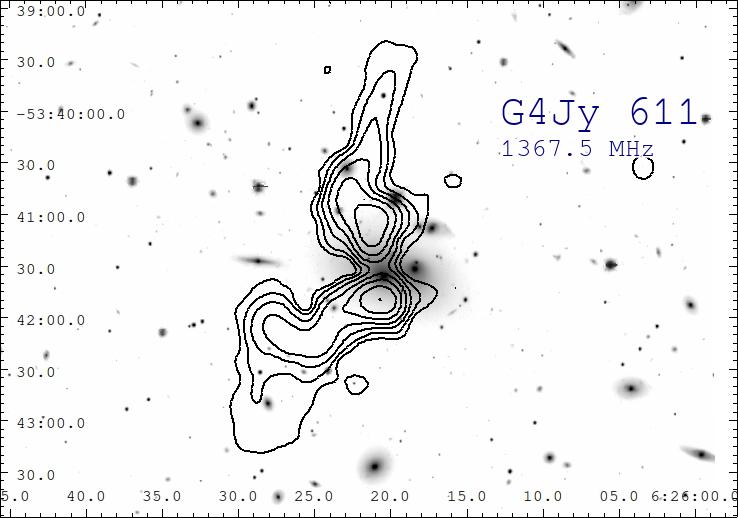}
    \includegraphics[scale=0.225]{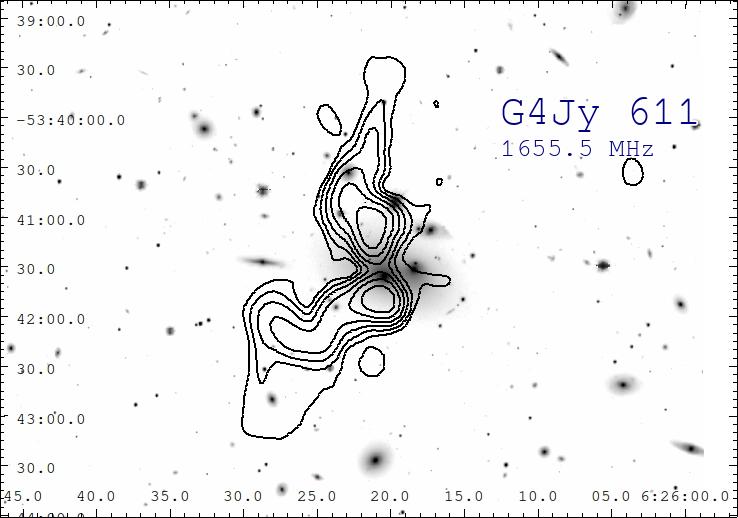}

    \caption{}
    \label{Q}
\end{figure*}
\clearpage

\begin{figure*}
    \centering
    \includegraphics[scale=0.225]{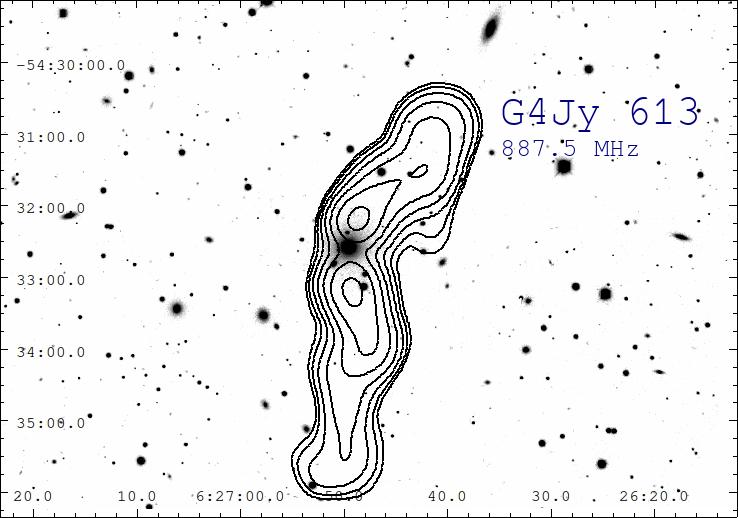}
    \includegraphics[scale=0.225]{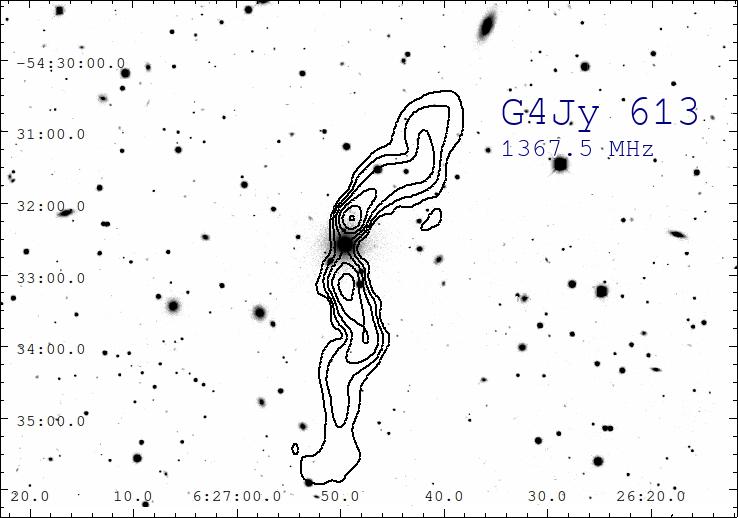}
    \includegraphics[scale=0.225]{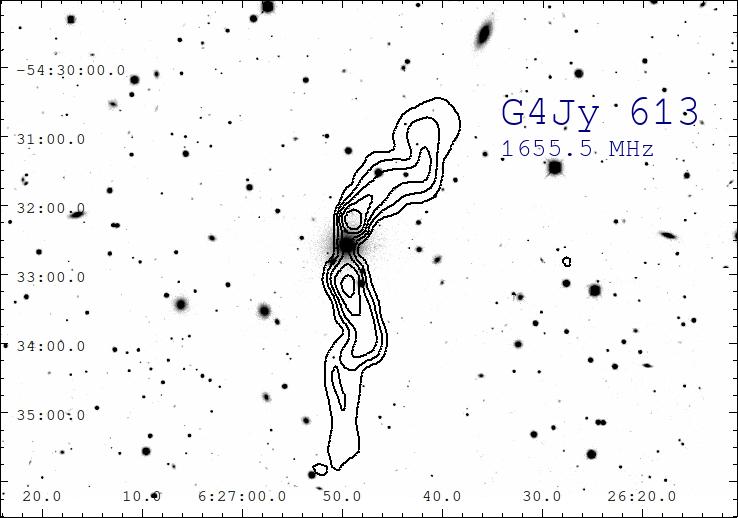}
    \includegraphics[scale=0.225]{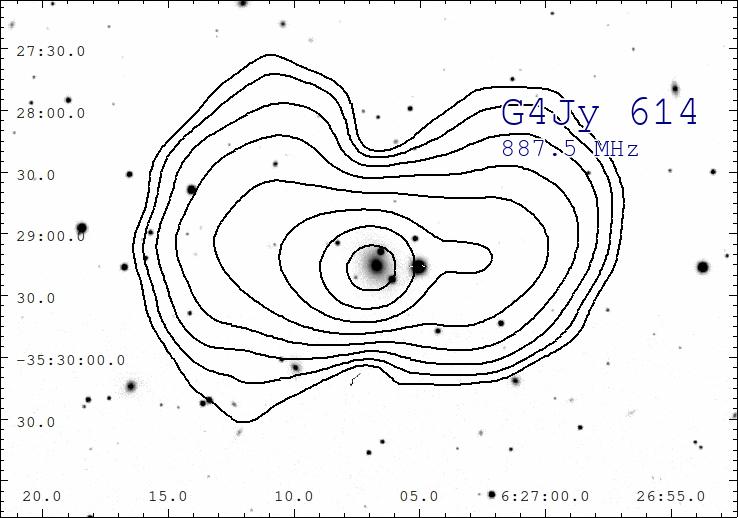}
    \includegraphics[scale=0.225]{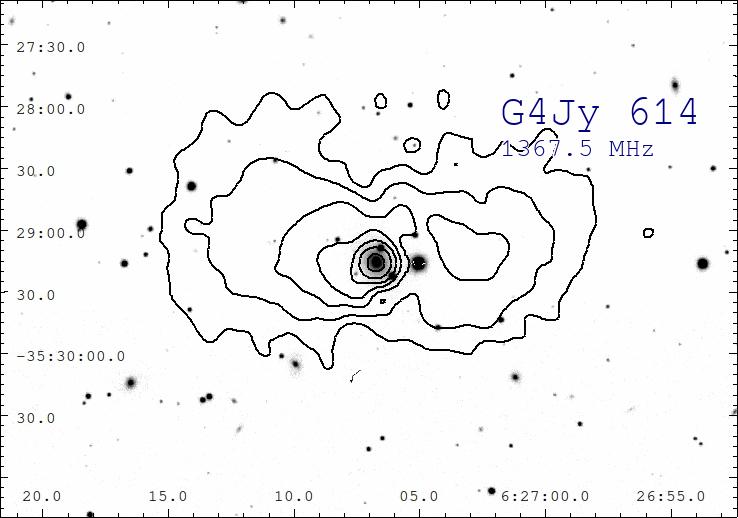}
    \includegraphics[scale=0.225]{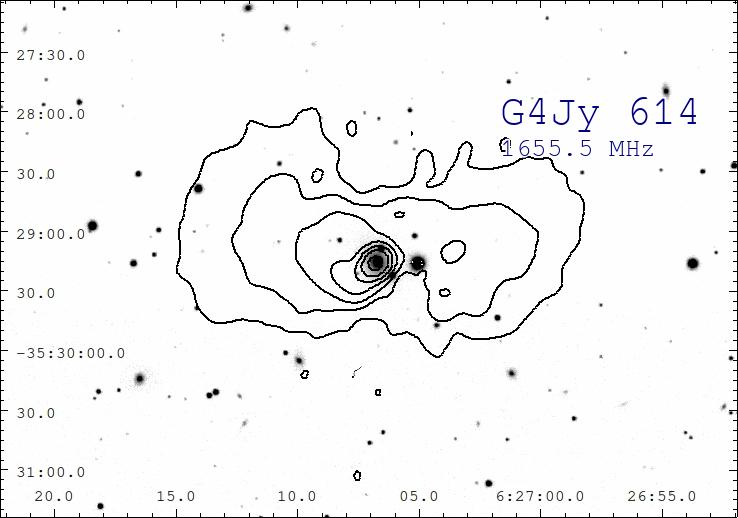}
    \includegraphics[scale=0.225]{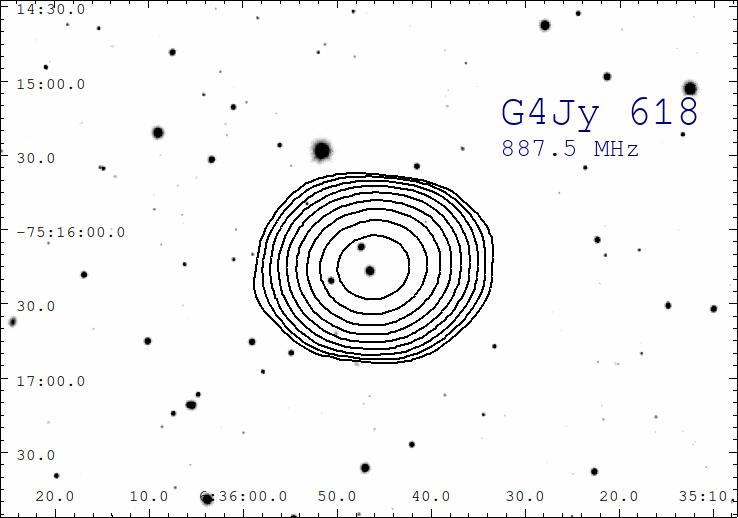}
    \includegraphics[scale=0.225]{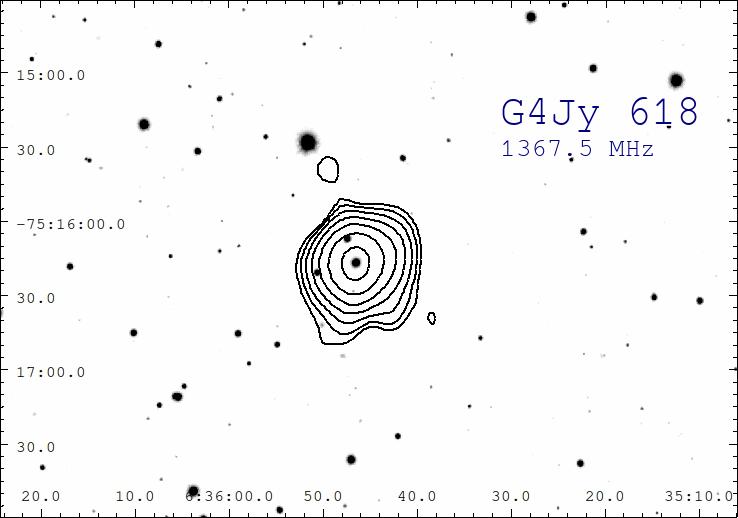}
    \includegraphics[scale=0.225]{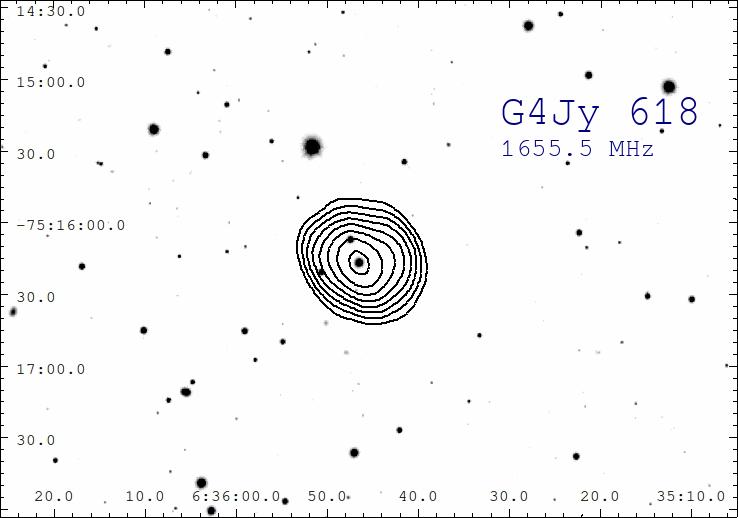}
    \includegraphics[scale=0.225]{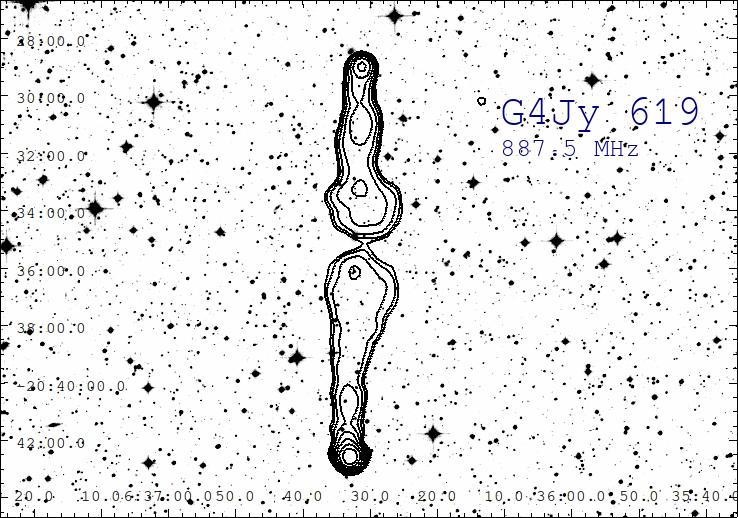}
    \includegraphics[scale=0.225]{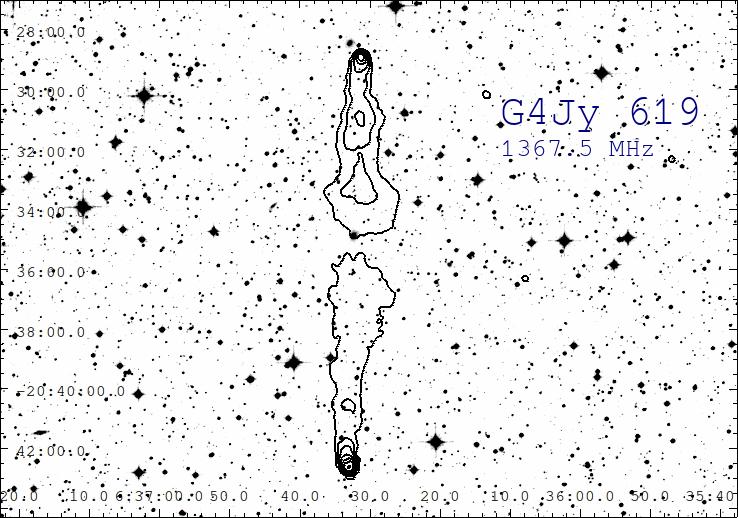}
    \includegraphics[scale=0.225]{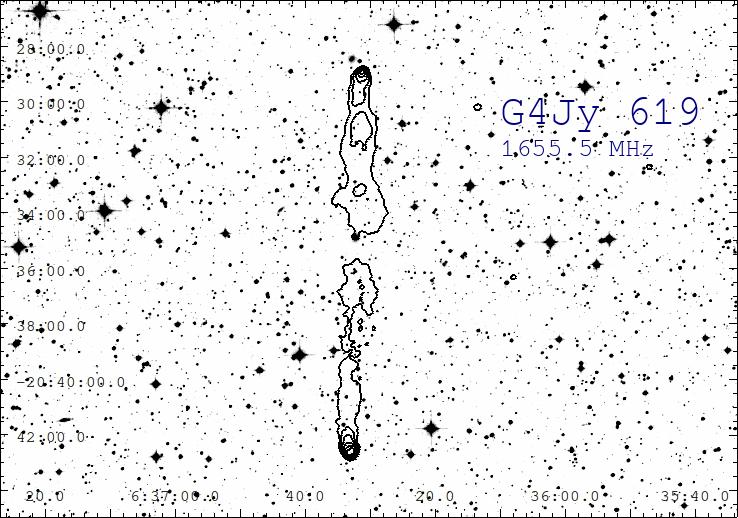}
    \includegraphics[scale=0.225]{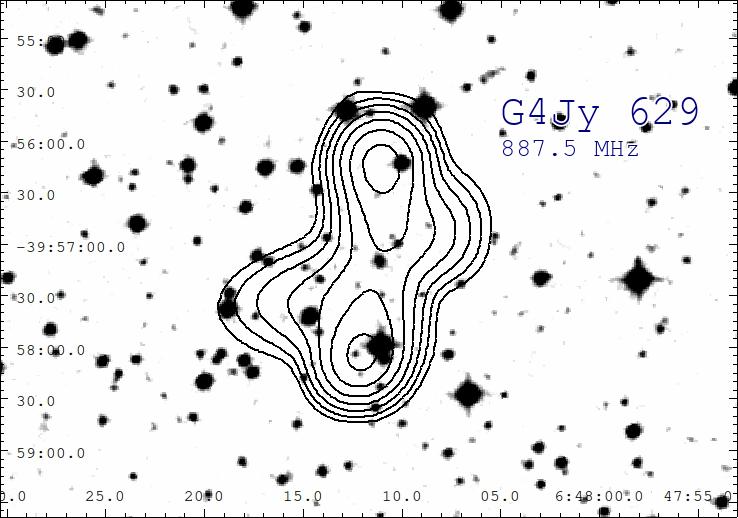}
    \includegraphics[scale=0.225]{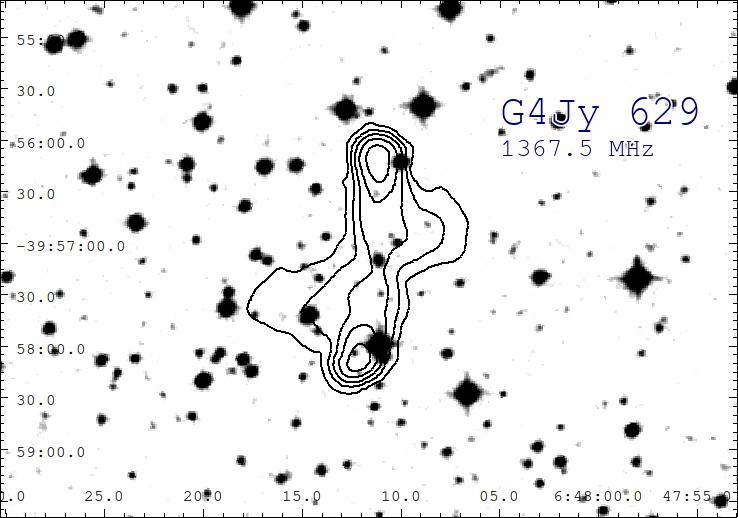}
    \includegraphics[scale=0.225]{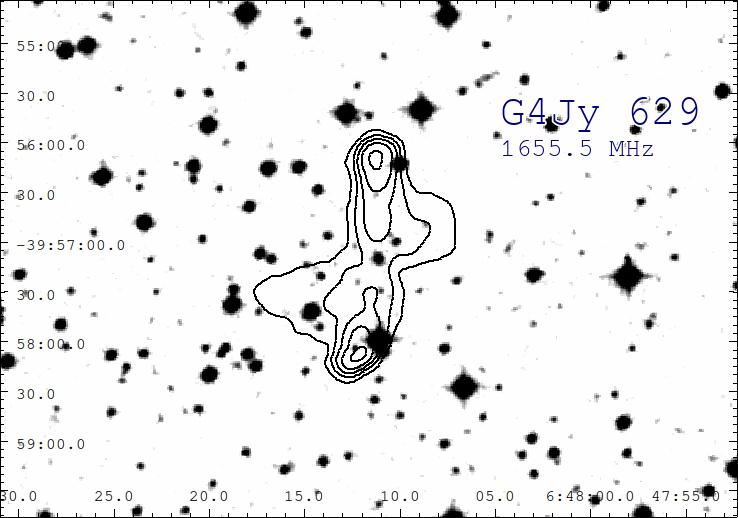}

    \caption{}
    \label{R}
\end{figure*}
\clearpage

\begin{figure*}
    \centering
    \includegraphics[scale=0.225]{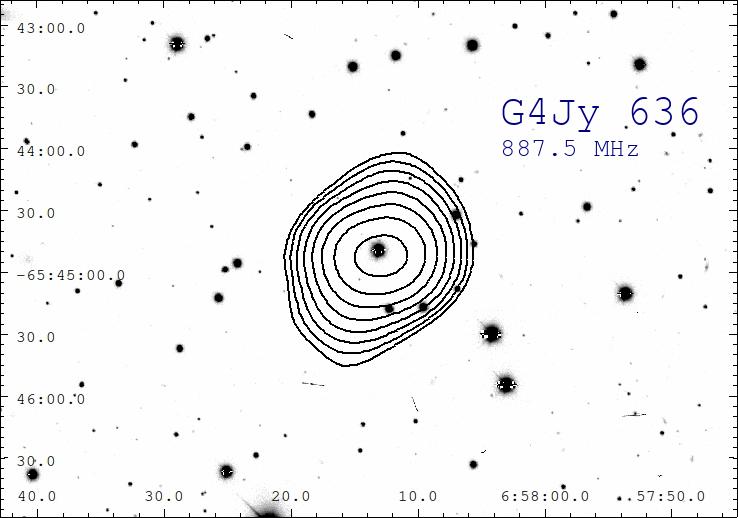}
    \includegraphics[scale=0.225]{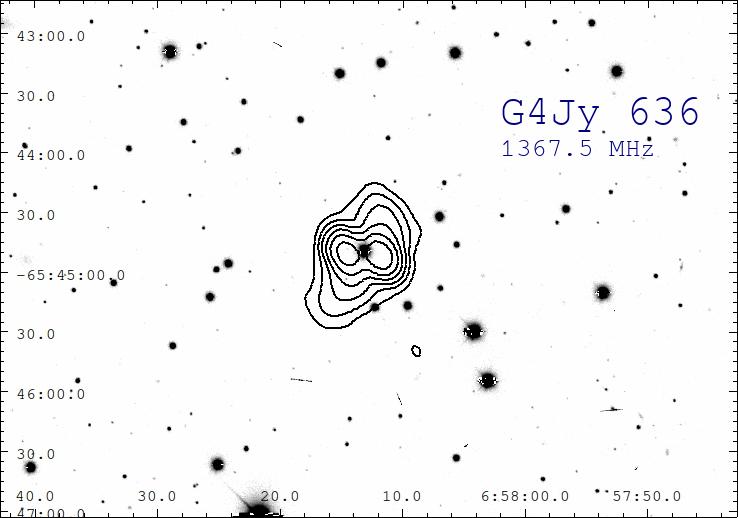}
    \includegraphics[scale=0.225]{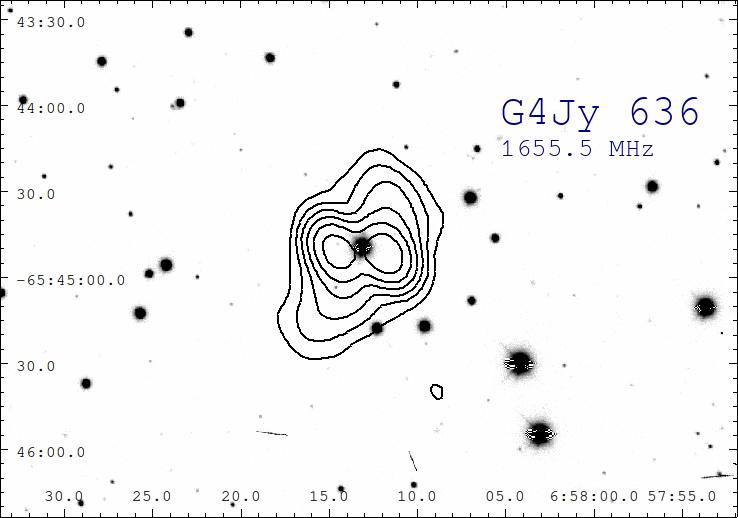}
    \includegraphics[scale=0.225]{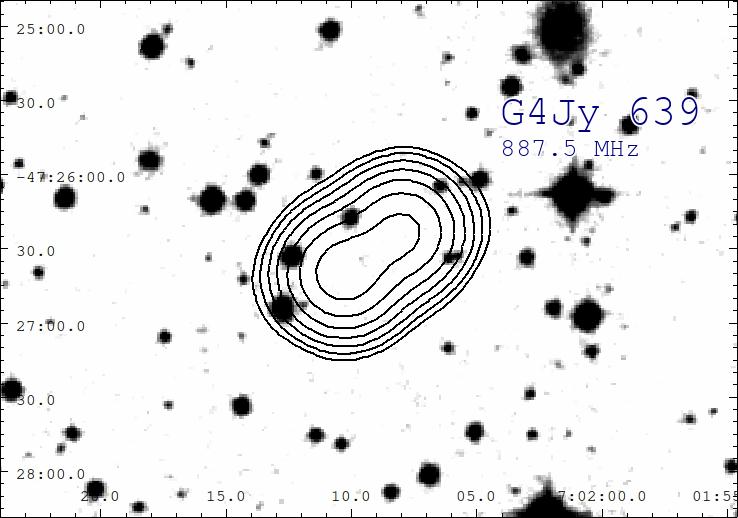}
    \includegraphics[scale=0.225]{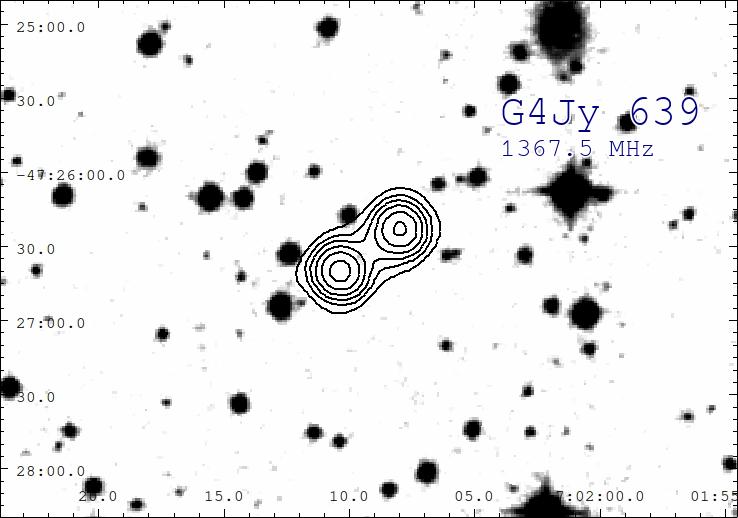}
    \includegraphics[scale=0.225]{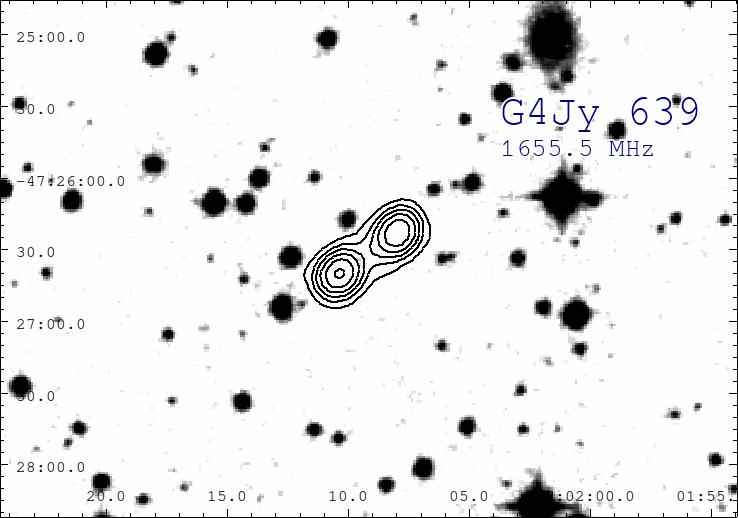}
    \includegraphics[scale=0.225]{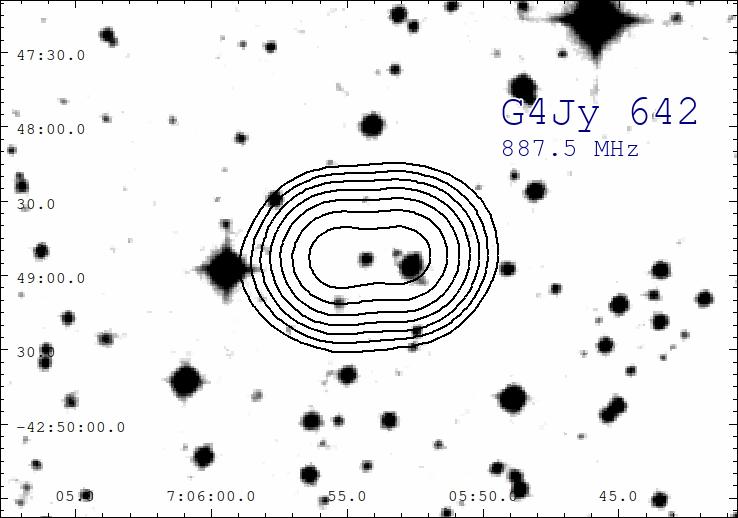}
    \includegraphics[scale=0.225]{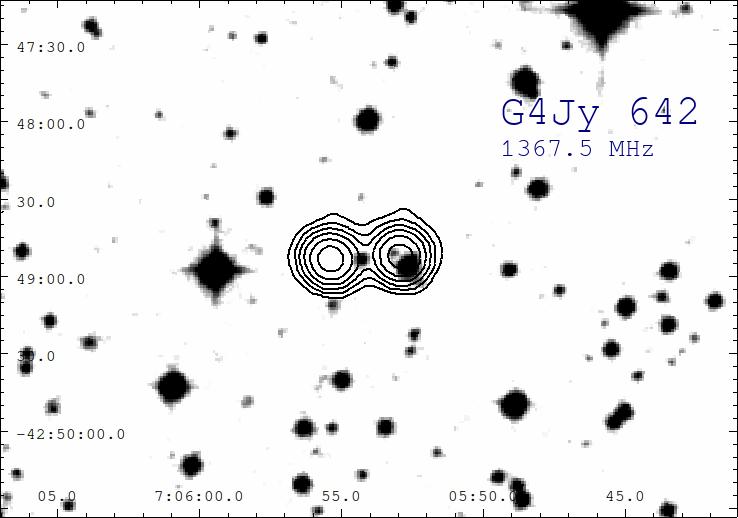}
    \includegraphics[scale=0.225]{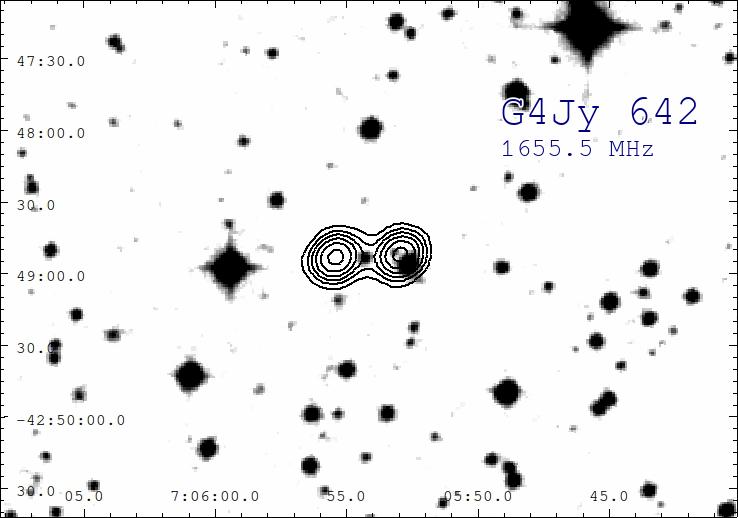}
    \includegraphics[scale=0.225]{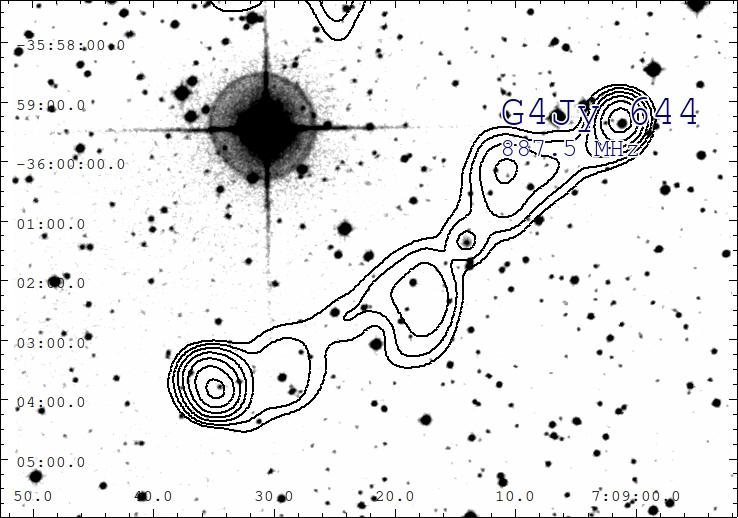}
    \includegraphics[scale=0.225]{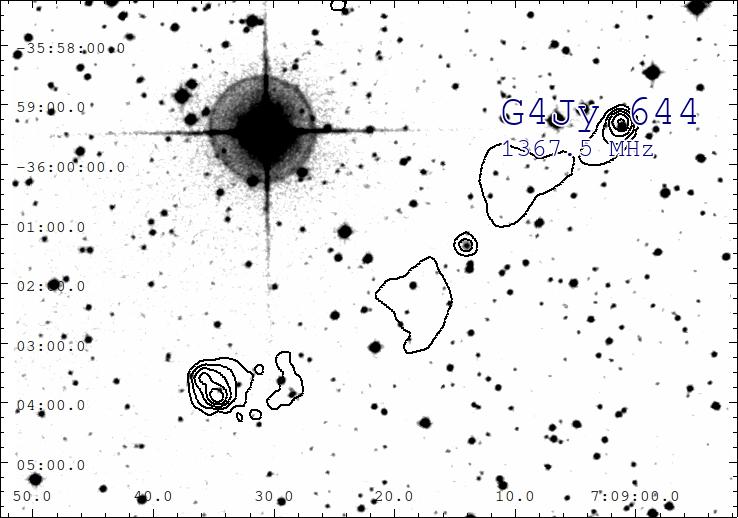}
    \includegraphics[scale=0.225]{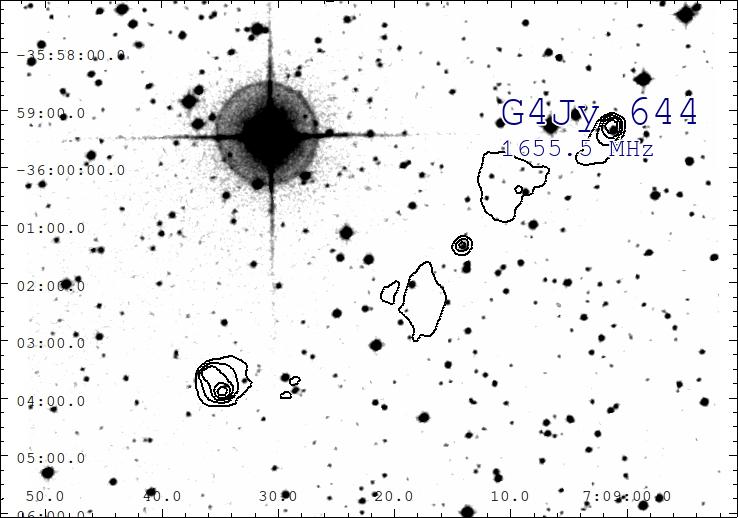}
    \includegraphics[scale=0.225]{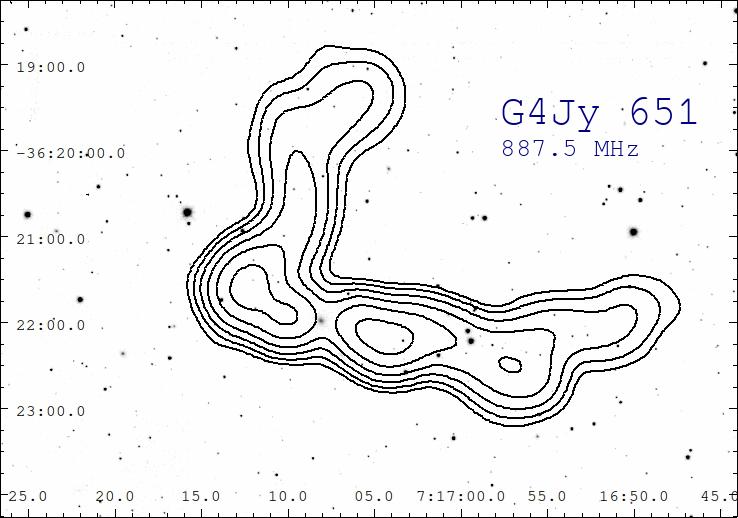}
    \includegraphics[scale=0.225]{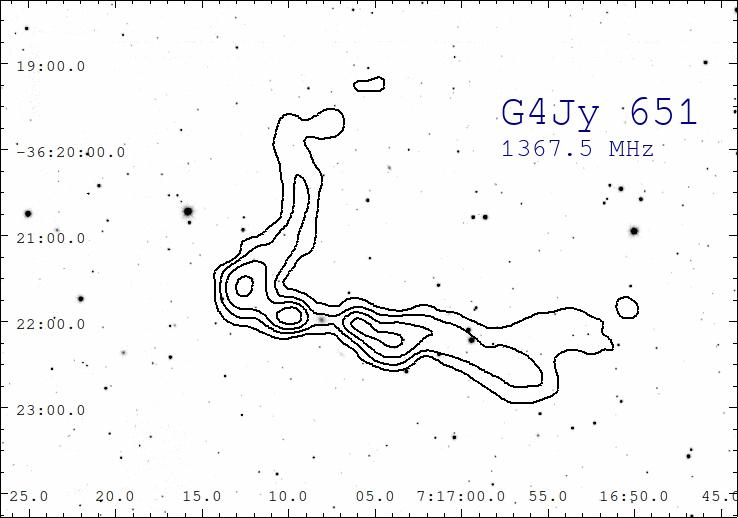}
    \includegraphics[scale=0.225]{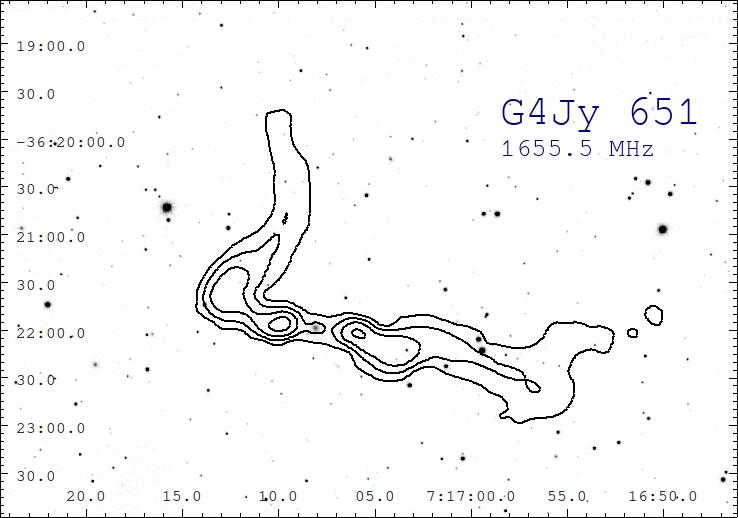}

    \caption{}
    \label{S}
\end{figure*}
\clearpage

\begin{figure*}
    \centering
    \includegraphics[scale=0.225]{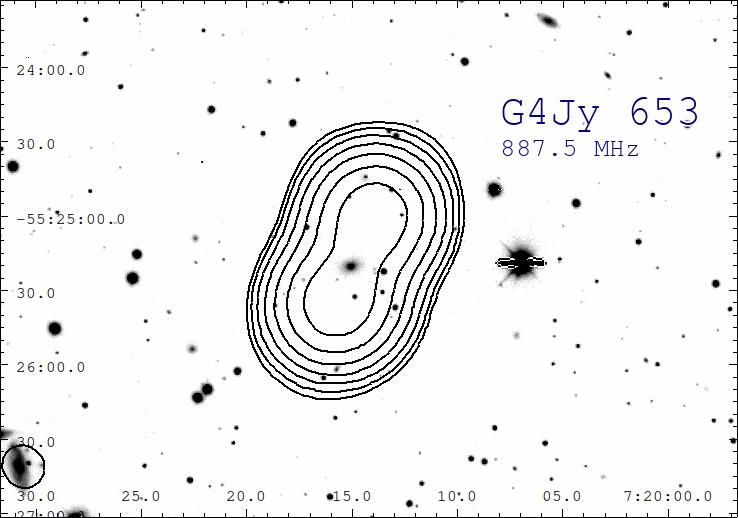}
    \includegraphics[scale=0.225]{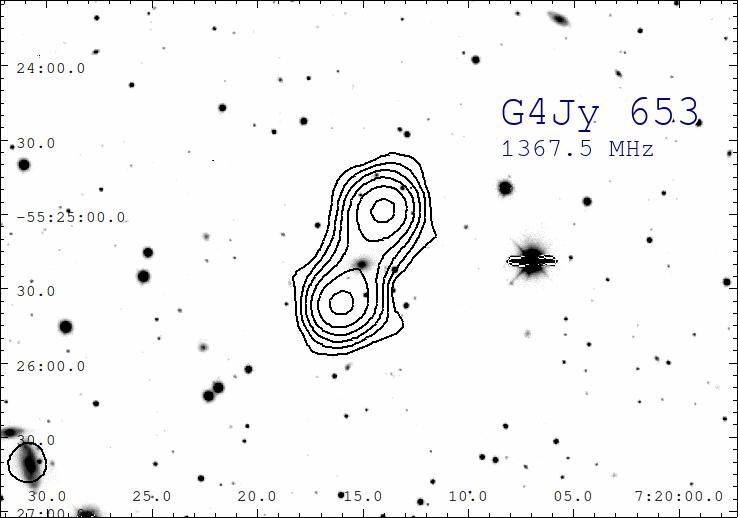}
    \includegraphics[scale=0.225]{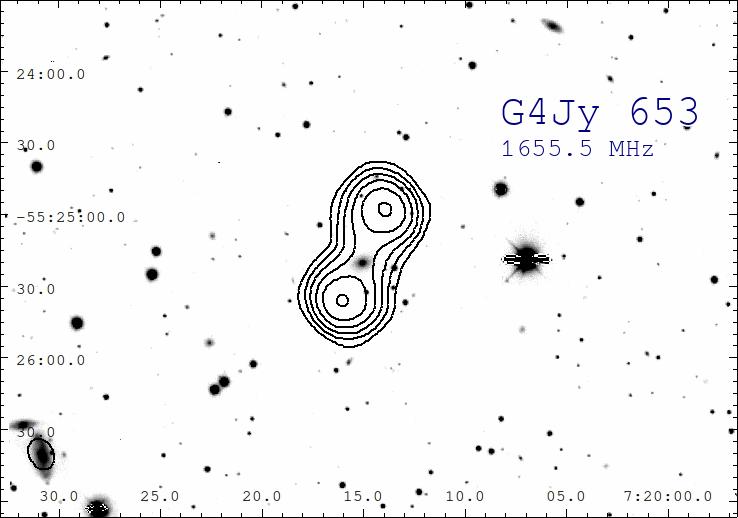}
    \includegraphics[scale=0.225]{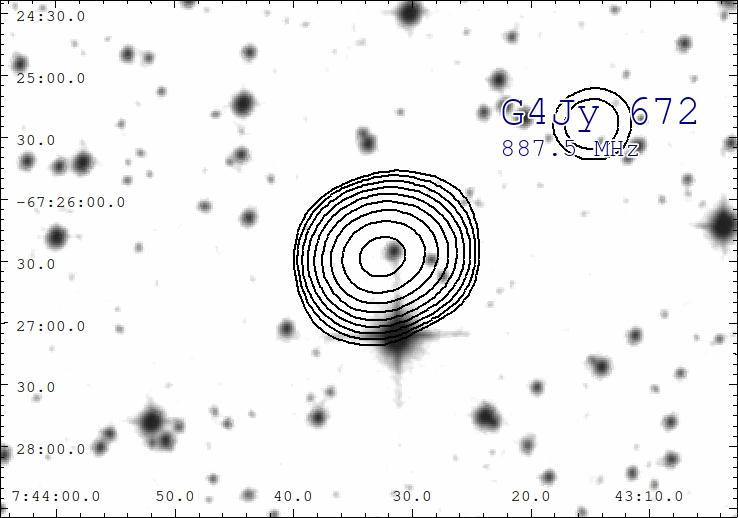}
    \includegraphics[scale=0.225]{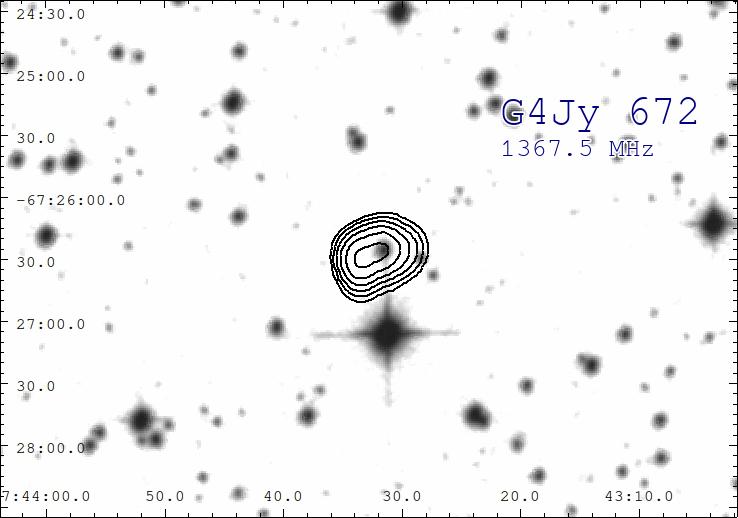}
    \includegraphics[scale=0.225]{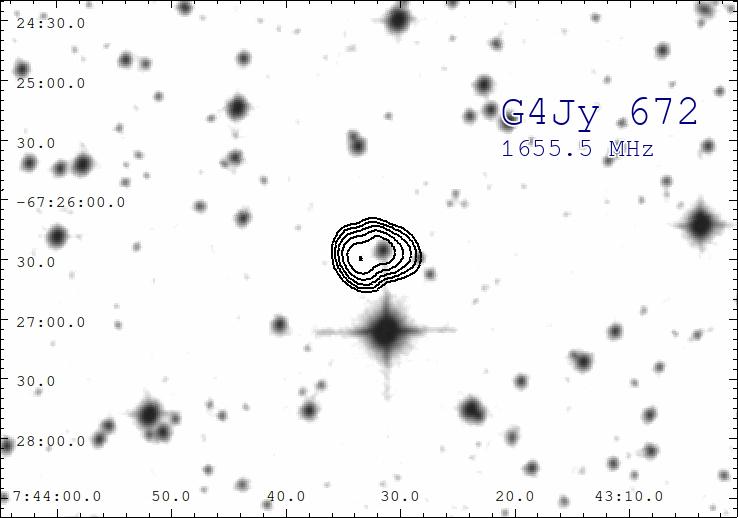}
    \includegraphics[scale=0.225]{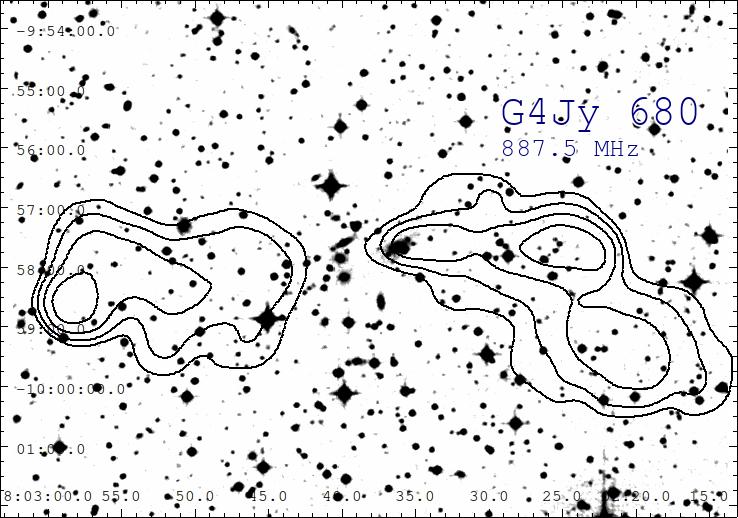}
    \includegraphics[scale=0.225]{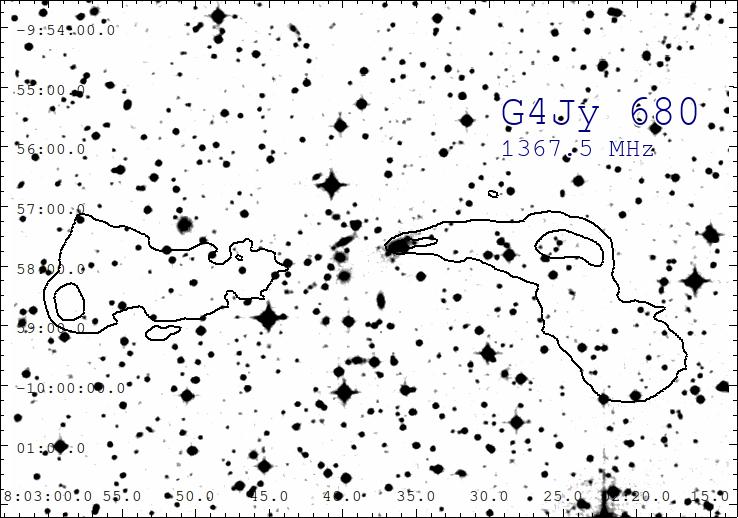}
    \includegraphics[scale=0.225]{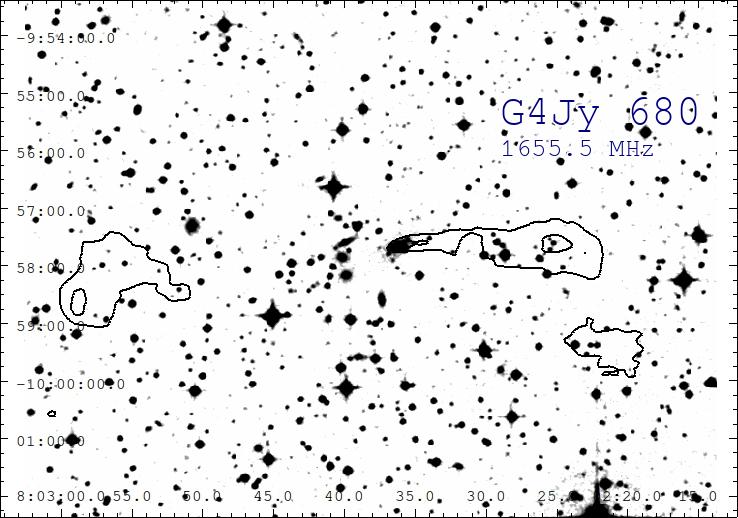}
    \includegraphics[scale=0.225]{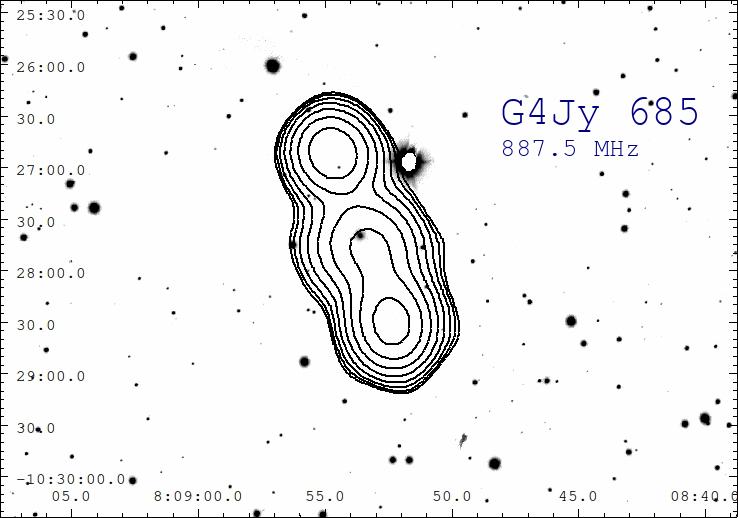}
    \includegraphics[scale=0.225]{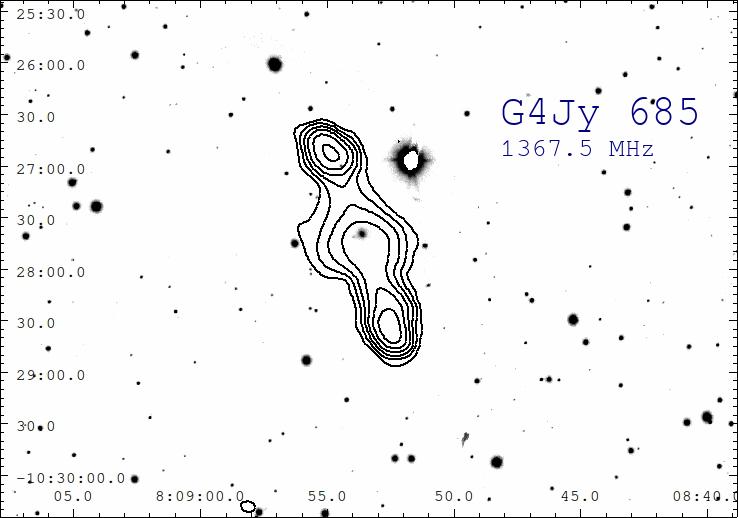}
    \includegraphics[scale=0.225]{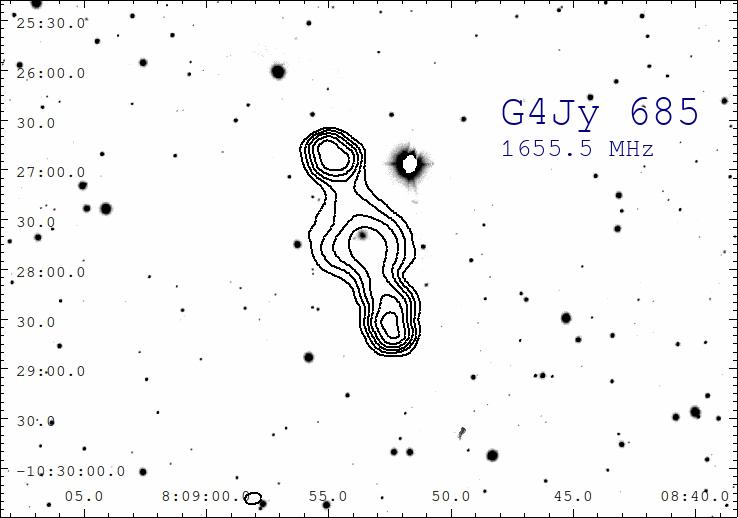}
    \includegraphics[scale=0.225]{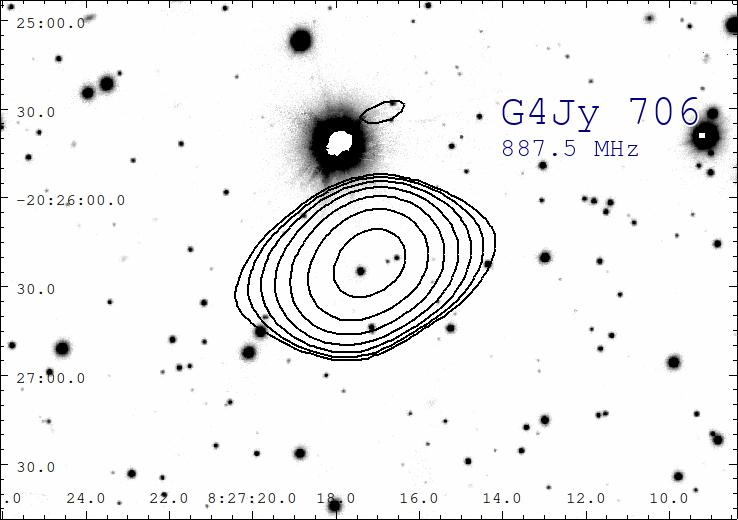}
    \includegraphics[scale=0.225]{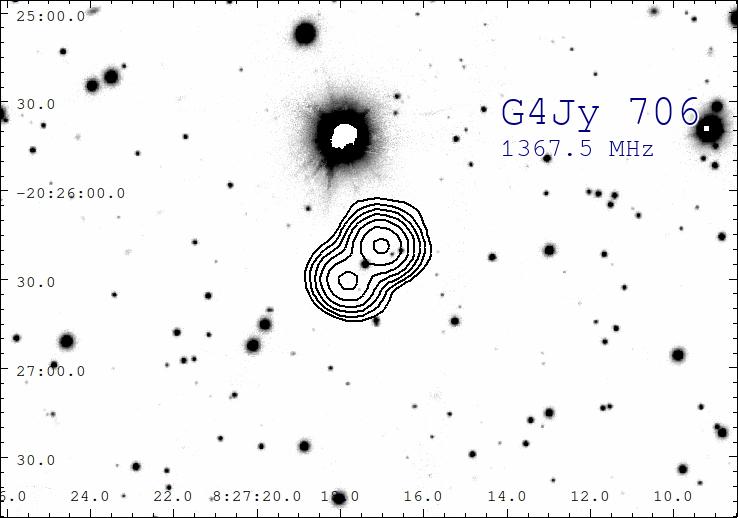}
    \includegraphics[scale=0.225]{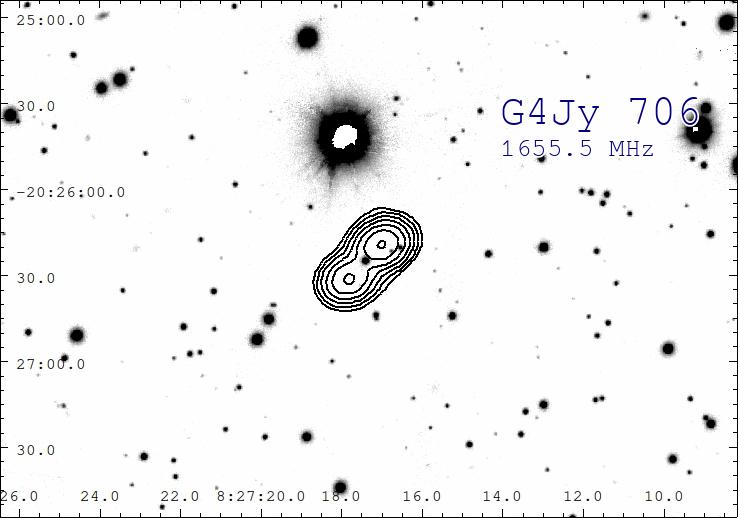}

    \caption{RACS-low data for G4Jy 718 could not be convolved to 25$\arcsec$, and therefore was not available. It may be replaced with future observations from RACS-low2 or RACS-low3 \textbf{(E.Lenc and A. Hotan, private communication)}.}
    \label{T}
\end{figure*}
\clearpage

\begin{figure*}
    \centering
    \includegraphics[scale=0.225]{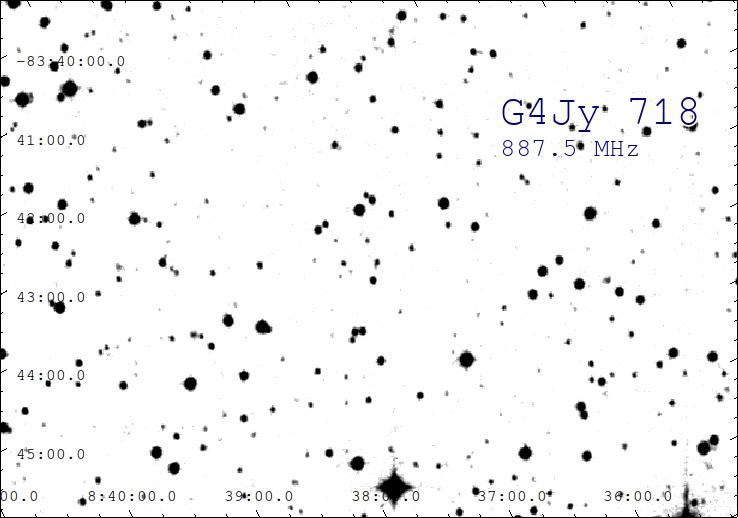}
    \includegraphics[scale=0.225]{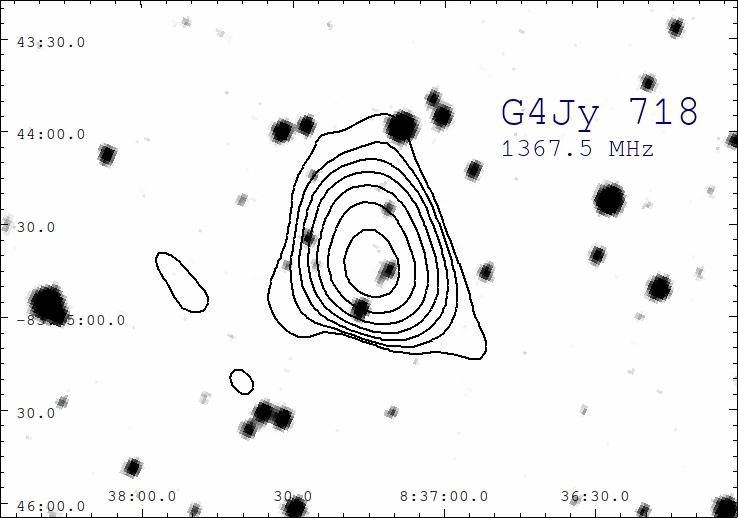}
    \includegraphics[scale=0.225]{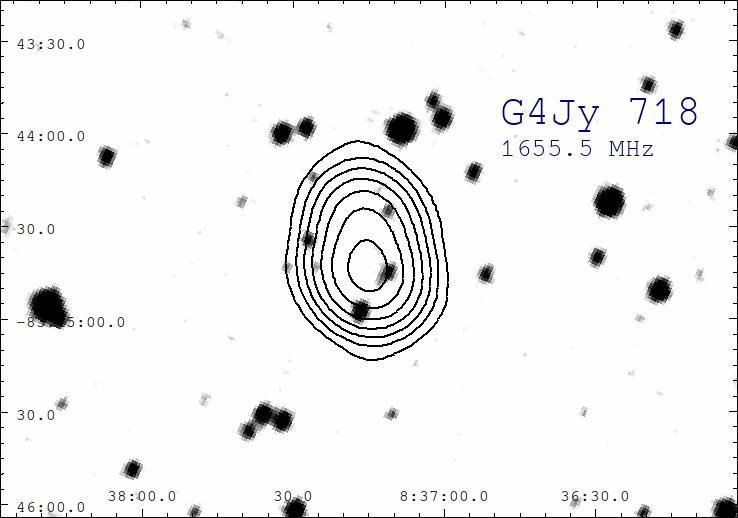}
    \includegraphics[scale=0.225]{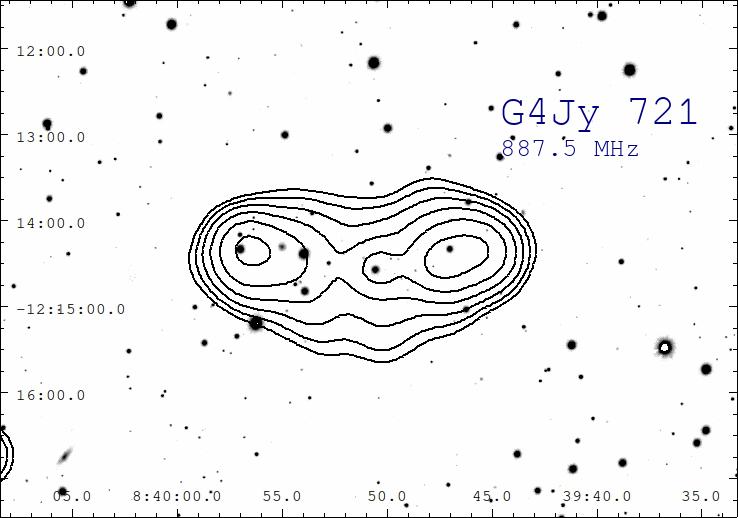}
    \includegraphics[scale=0.225]{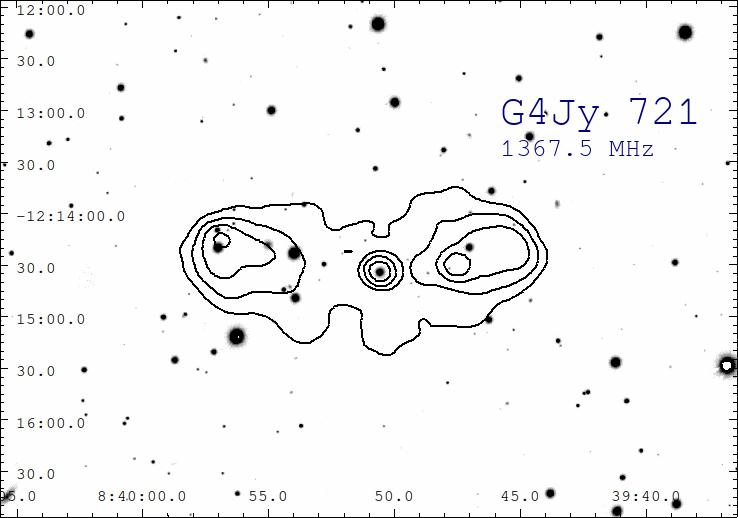}
    \includegraphics[scale=0.225]{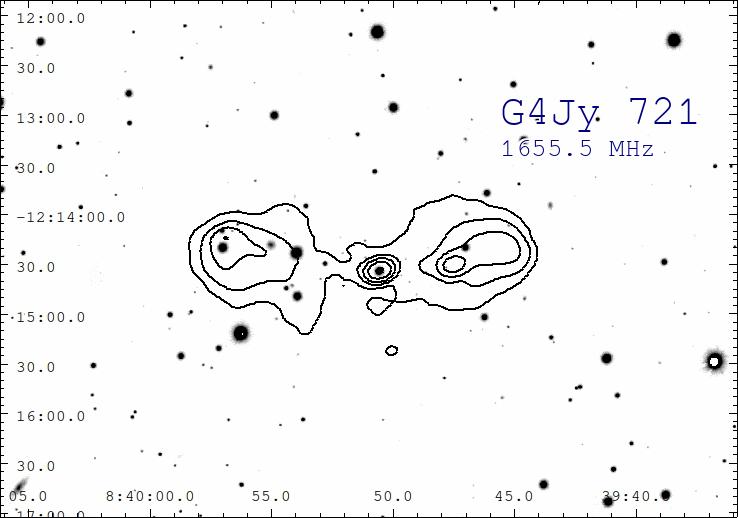}
    \includegraphics[scale=0.225]{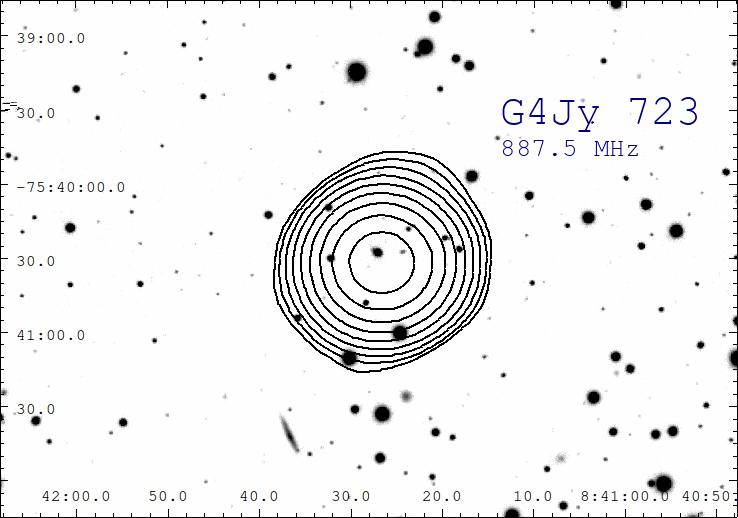}
    \includegraphics[scale=0.225]{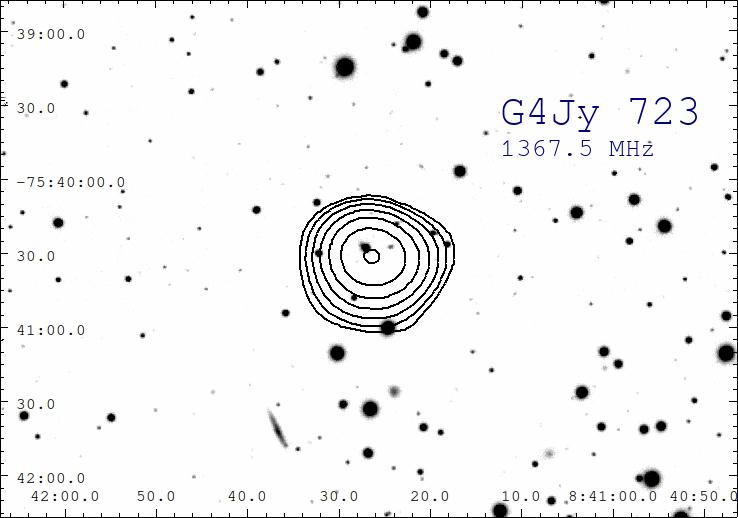}
    \includegraphics[scale=0.225]{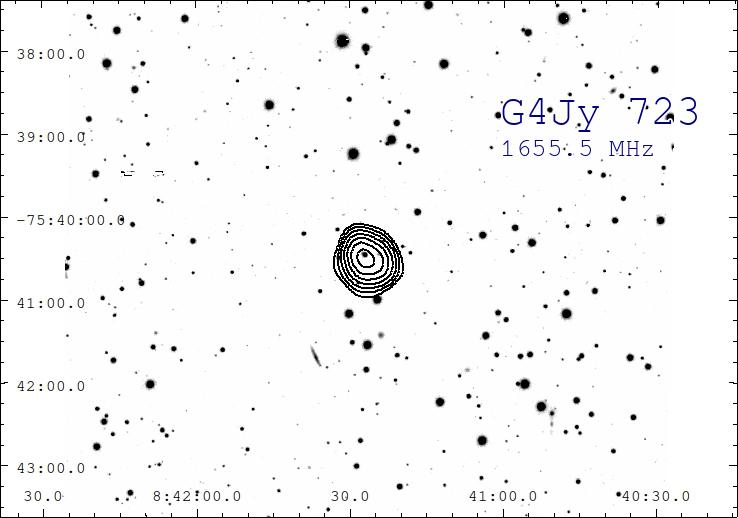}
    \includegraphics[scale=0.225]{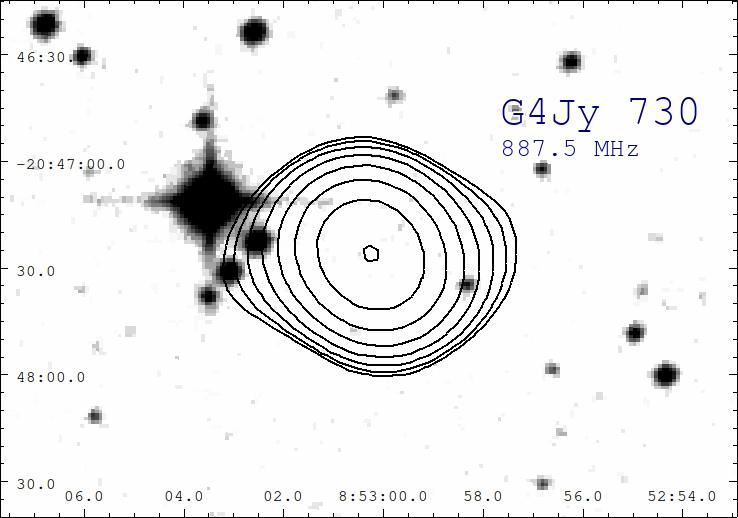}
    \includegraphics[scale=0.225]{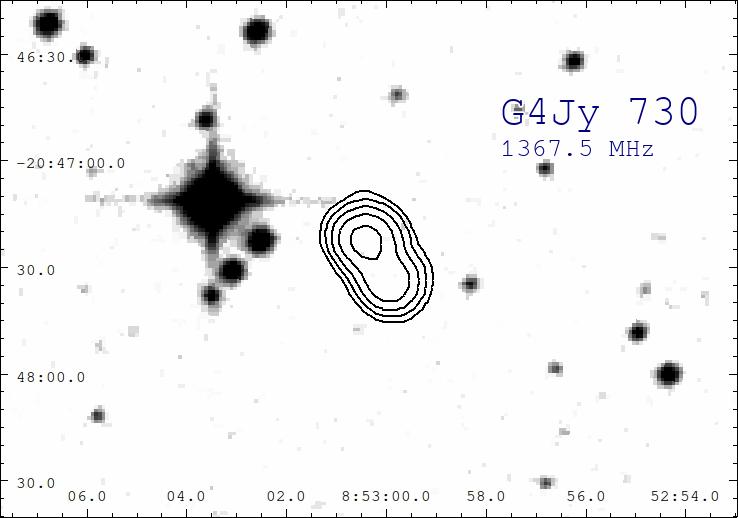}
    \includegraphics[scale=0.225]{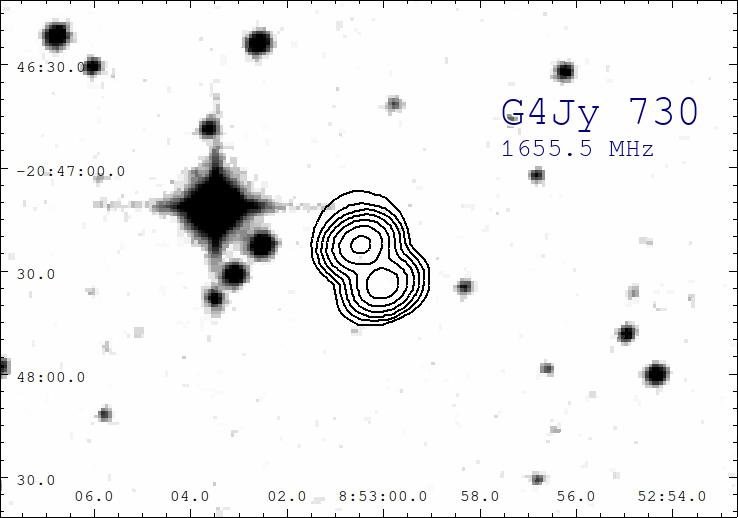}
    \includegraphics[scale=0.225]{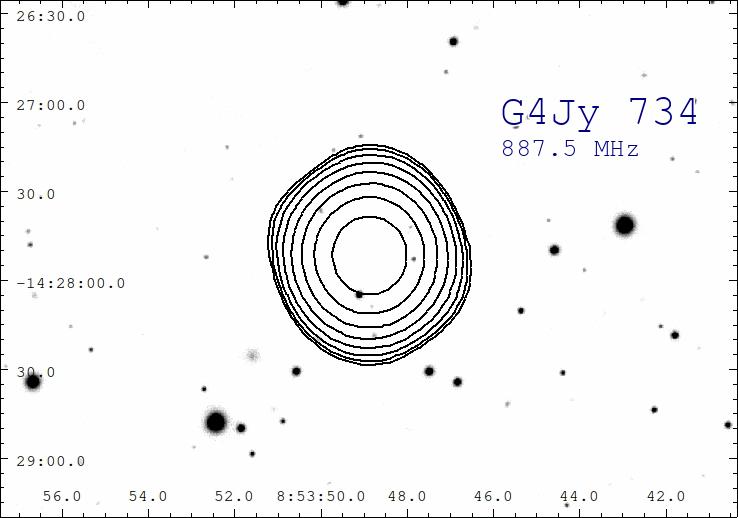}
    \includegraphics[scale=0.225]{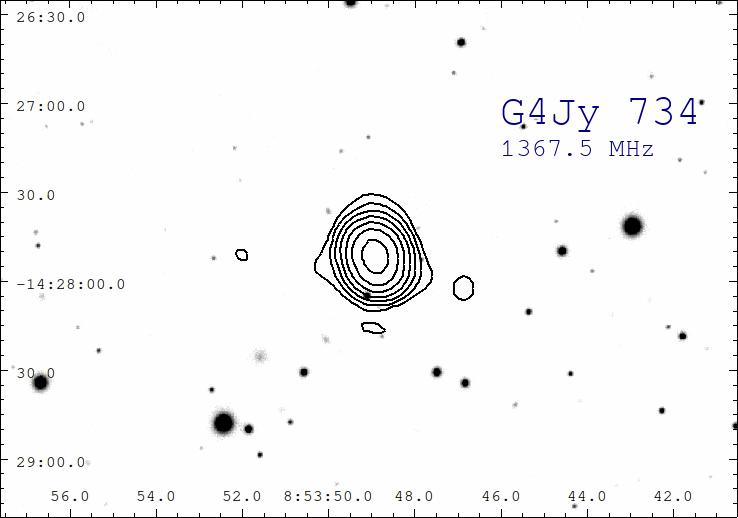}
    \includegraphics[scale=0.225]{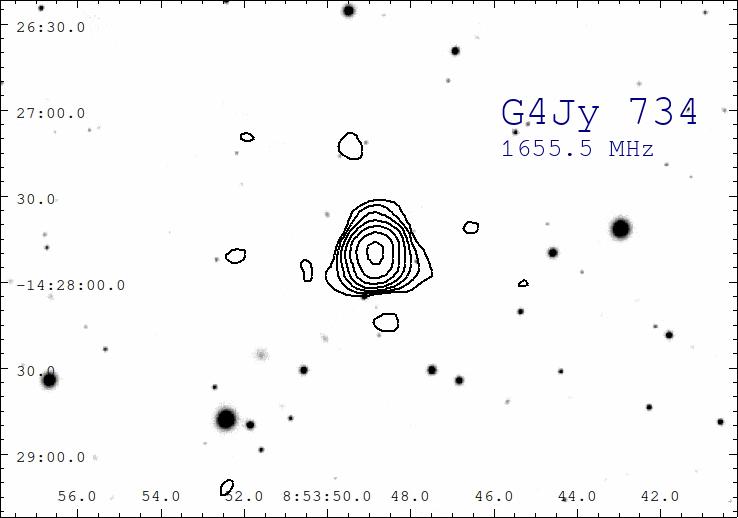}

    \caption{}
    \label{U}
\end{figure*}
\clearpage

\begin{figure*}
    \centering
    \includegraphics[scale=0.225]{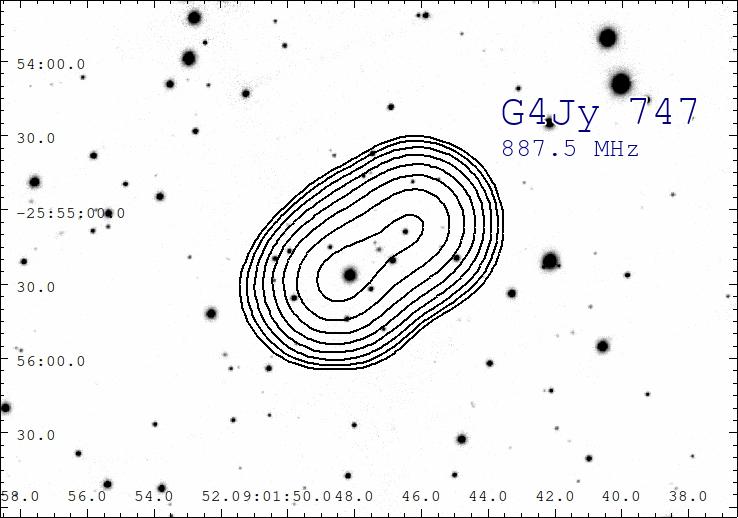}
    \includegraphics[scale=0.225]{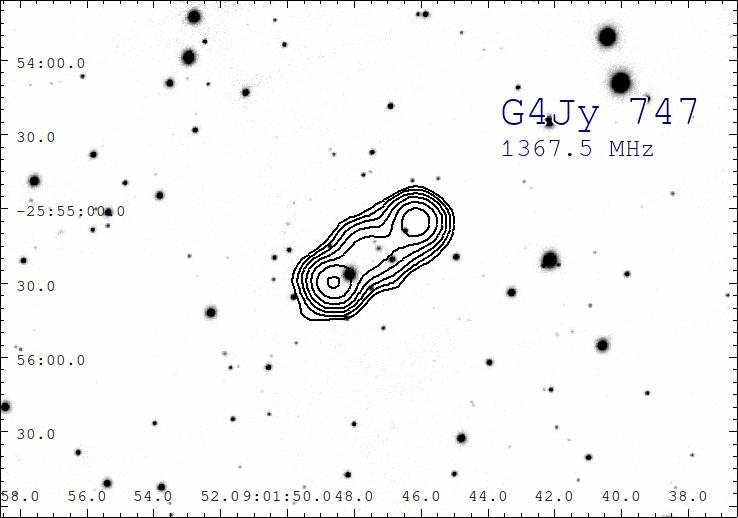}
    \includegraphics[scale=0.225]{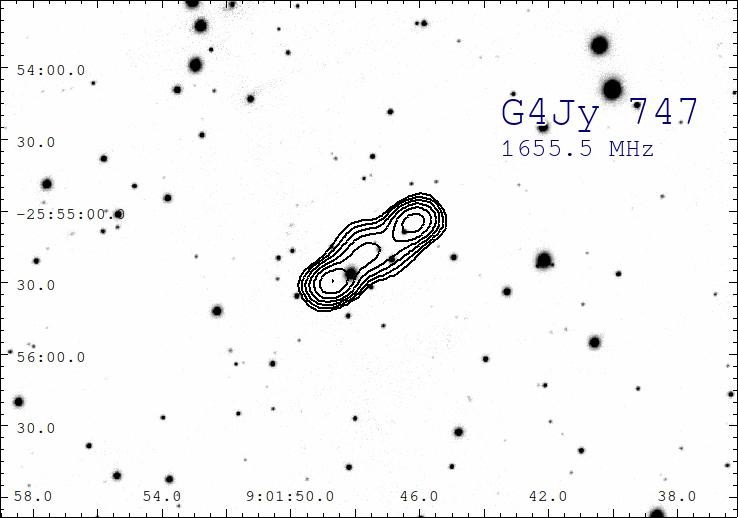}
    \includegraphics[scale=0.225]{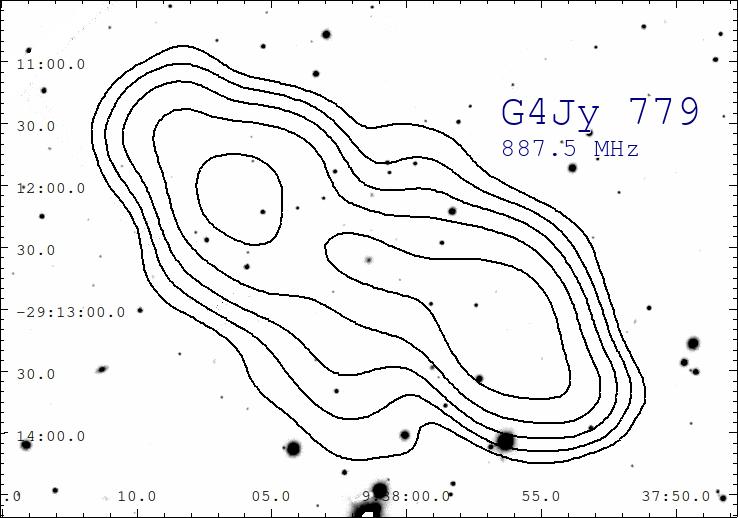}
    \includegraphics[scale=0.225]{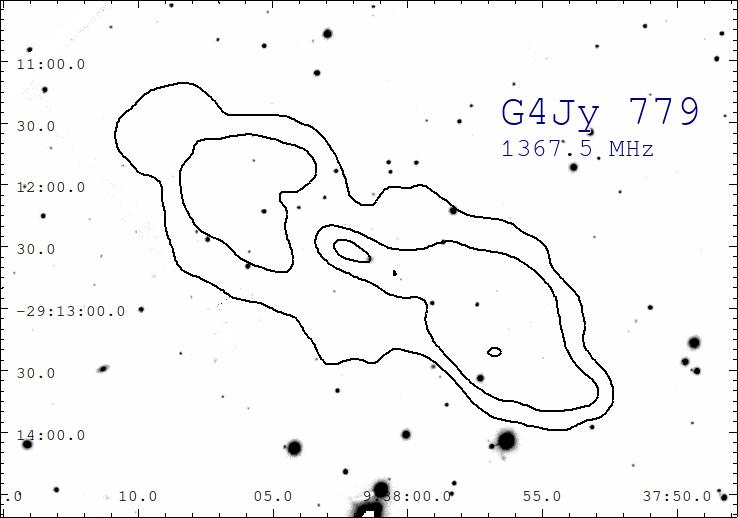}
    \includegraphics[scale=0.225]{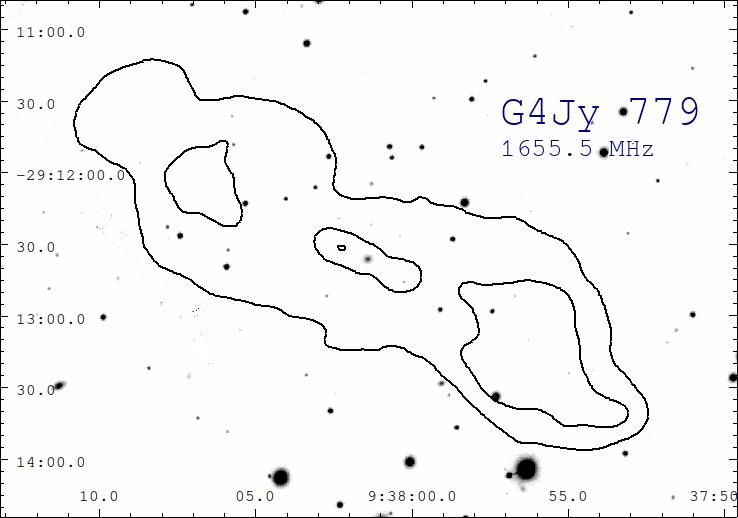}
    \includegraphics[scale=0.225]{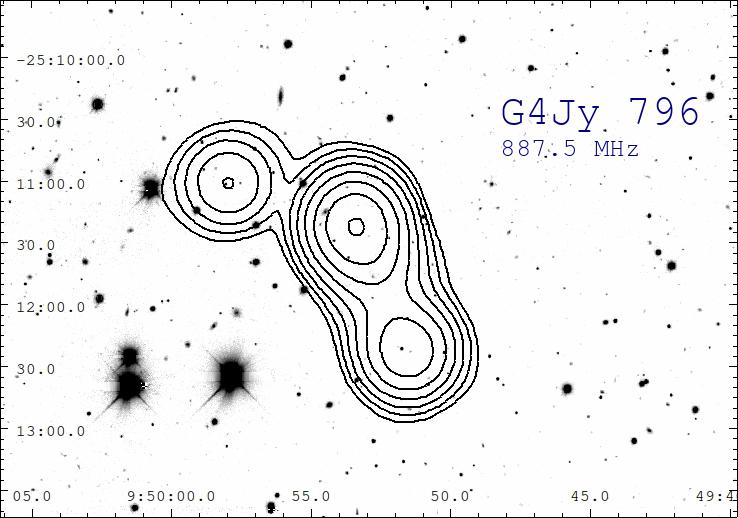}
    \includegraphics[scale=0.225]{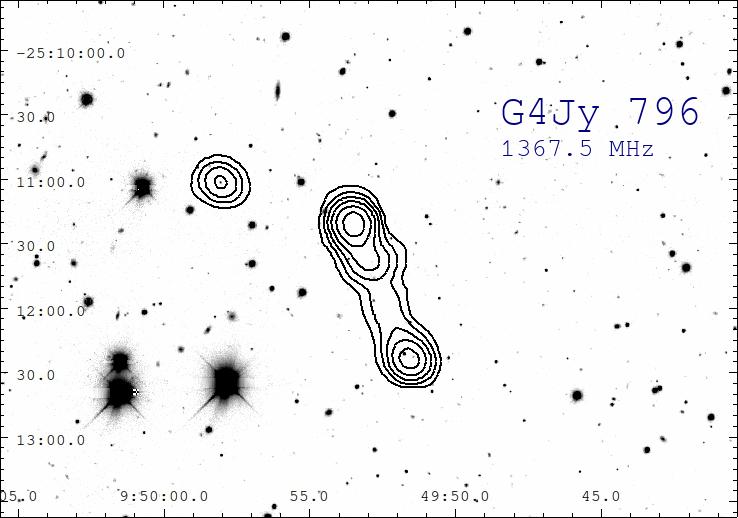}
    \includegraphics[scale=0.225]{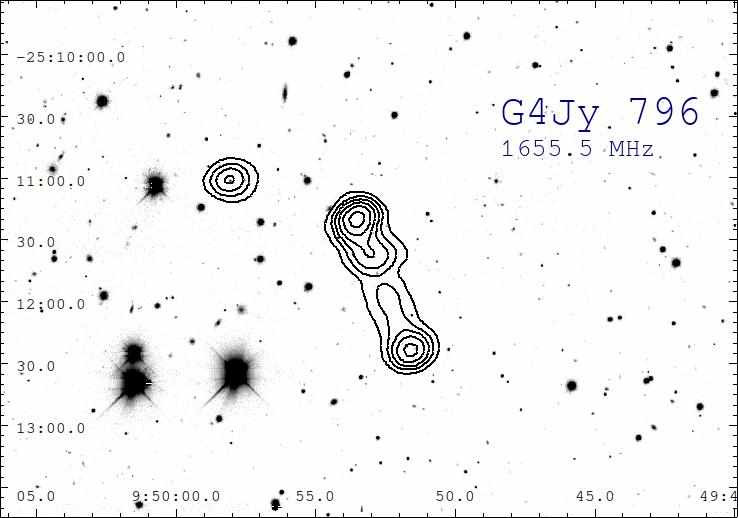}
    \includegraphics[scale=0.225]{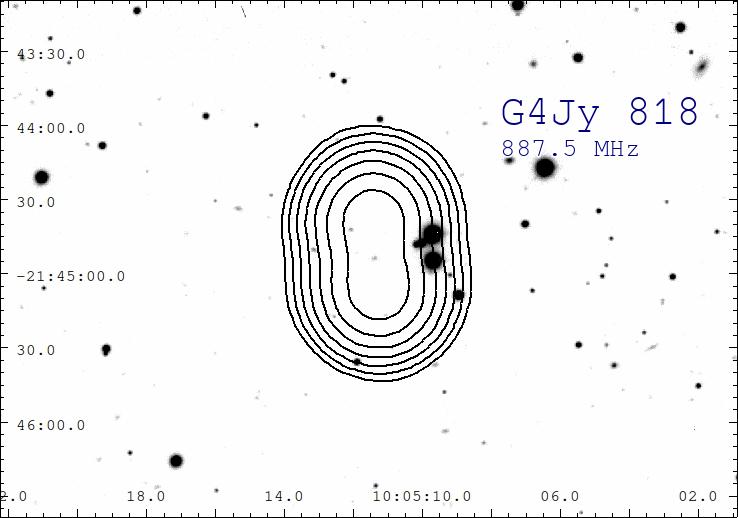}
    \includegraphics[scale=0.225]{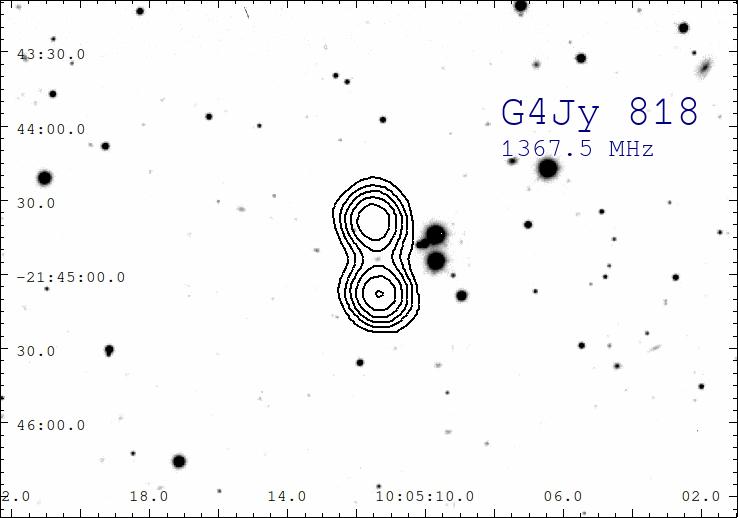}
    \includegraphics[scale=0.225]{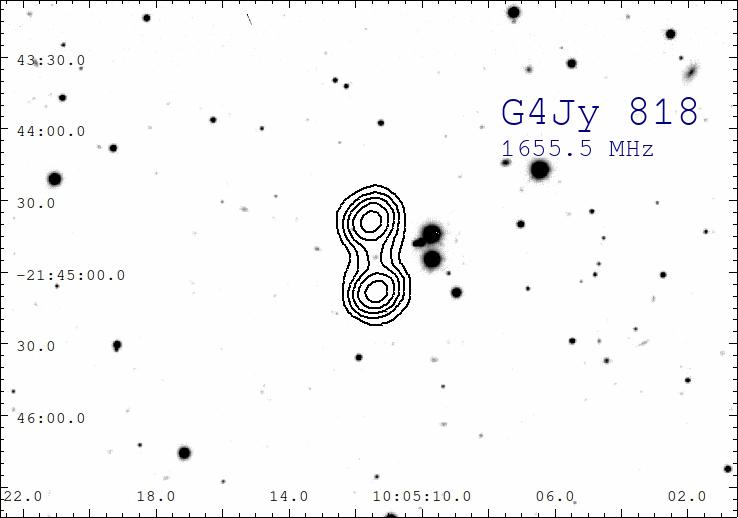}
    \includegraphics[scale=0.225]{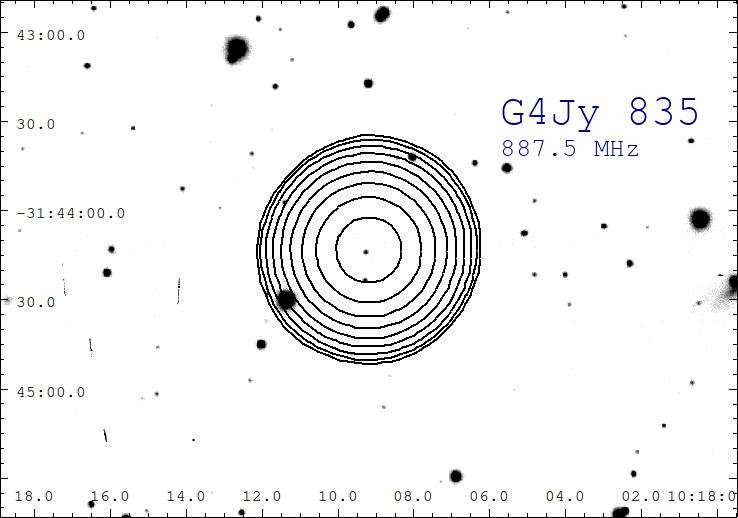}
    \includegraphics[scale=0.225]{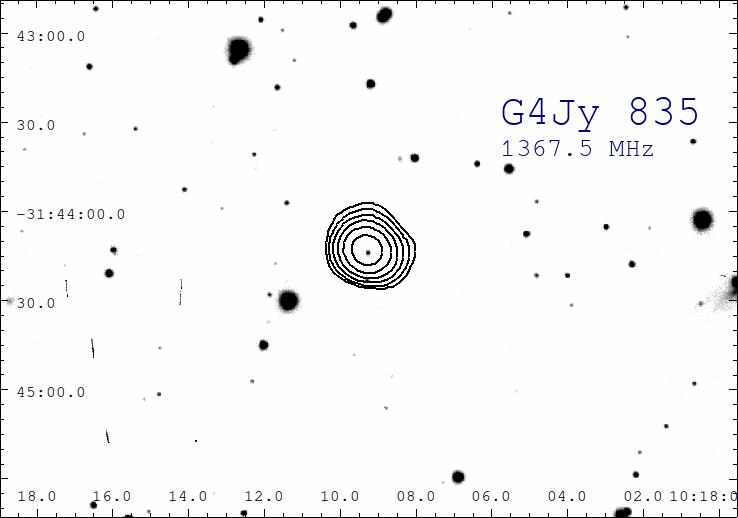}
    \includegraphics[scale=0.225]{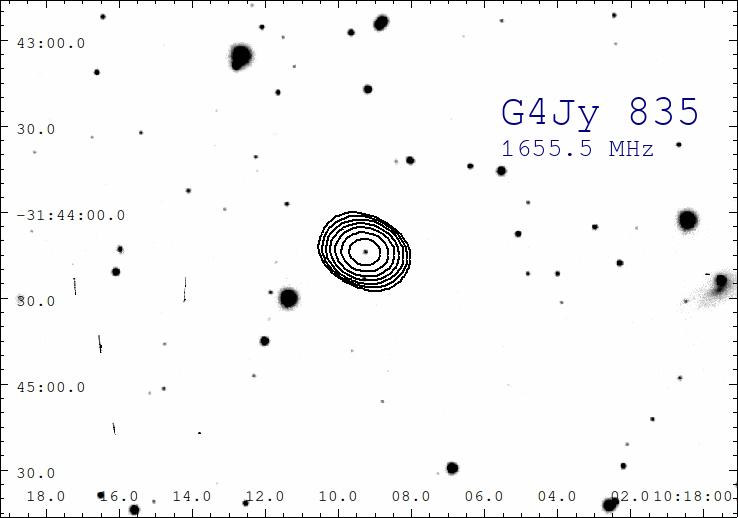}

    \caption{}
    \label{V}
\end{figure*}
\clearpage

\begin{figure*}
    \centering
    \includegraphics[scale=0.225]{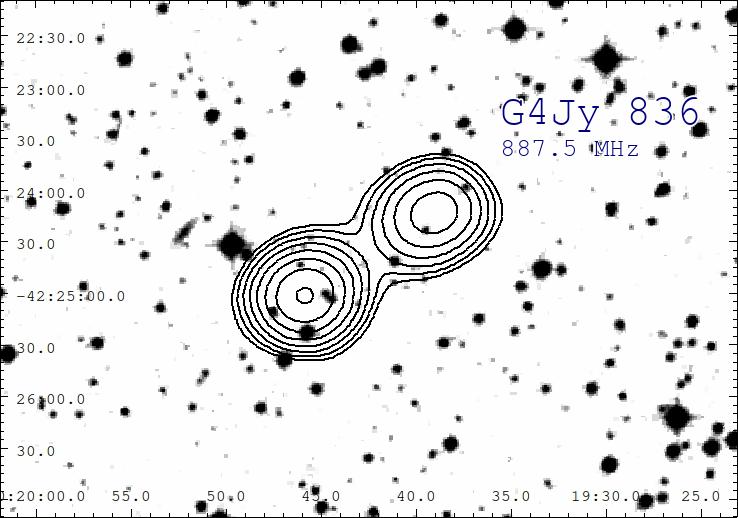}
    \includegraphics[scale=0.225]{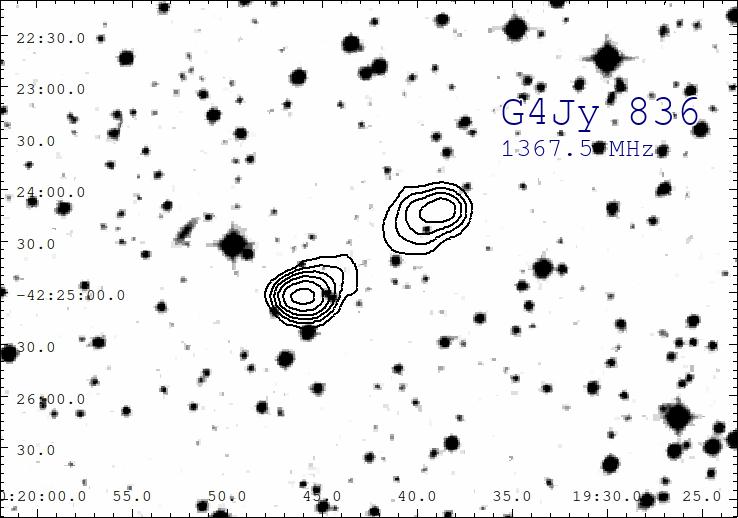}
    \includegraphics[scale=0.225]{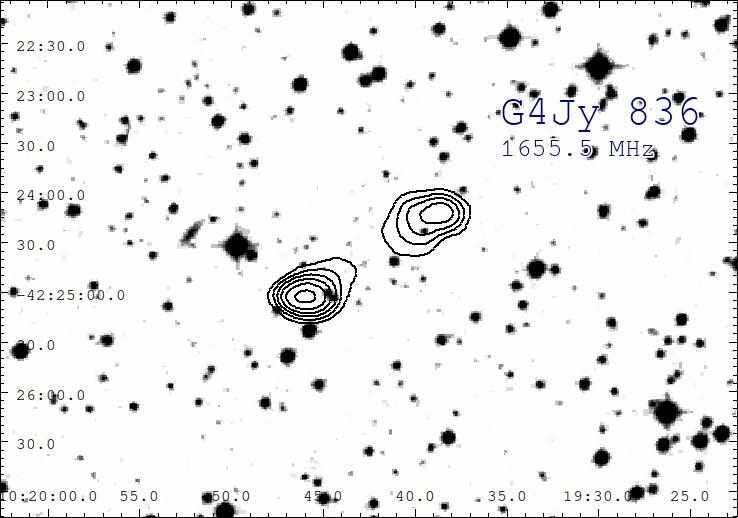}
    \includegraphics[scale=0.225]{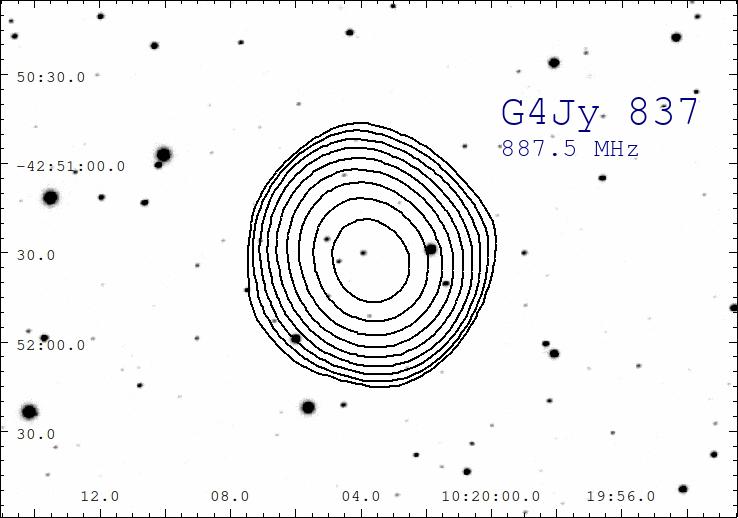}
    \includegraphics[scale=0.225]{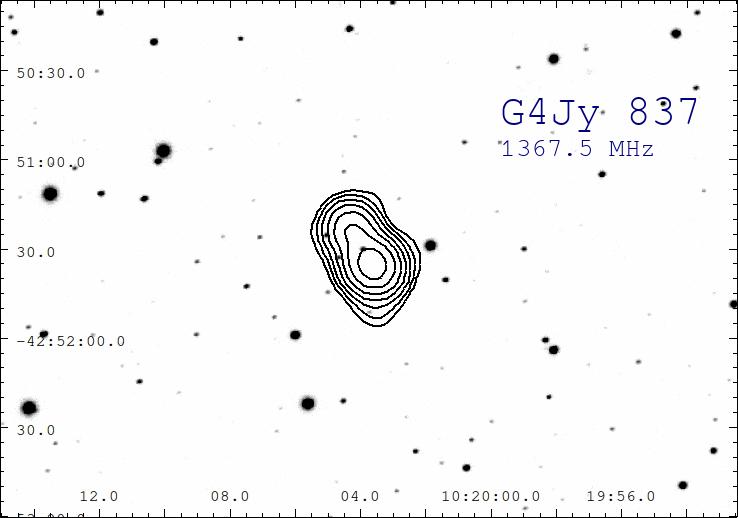}
    \includegraphics[scale=0.225]{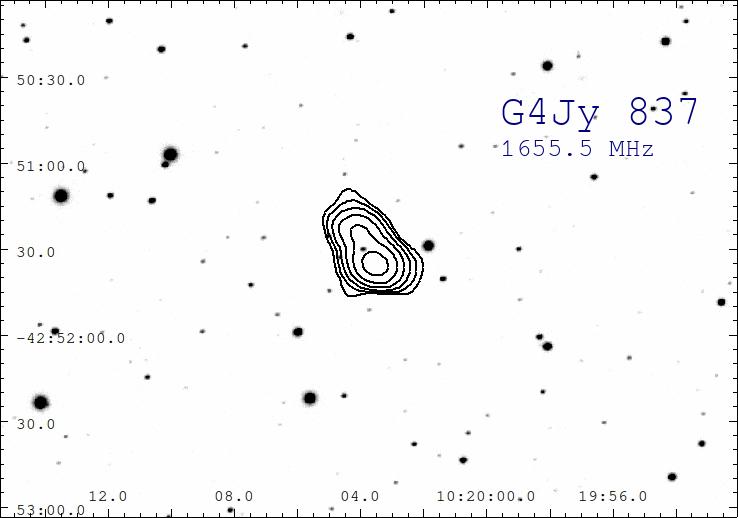}
    \includegraphics[scale=0.225]{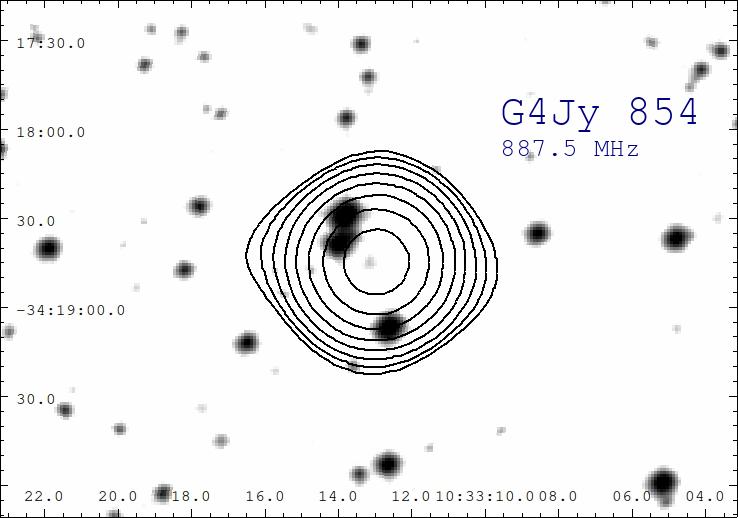}
    \includegraphics[scale=0.225]{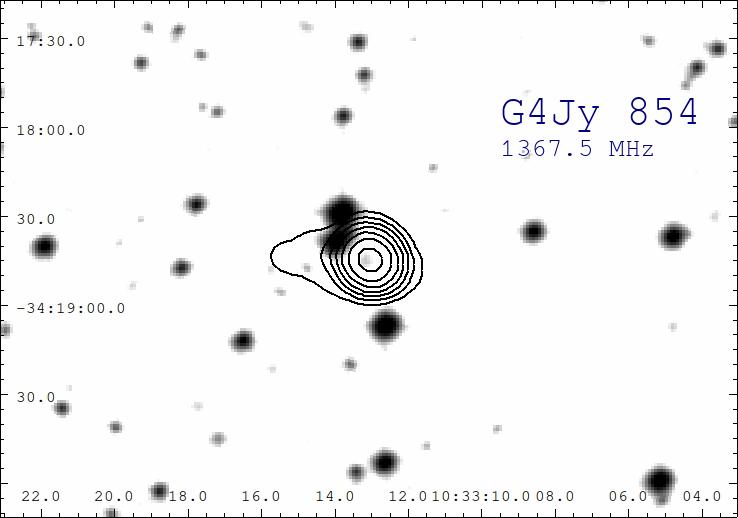}
    \includegraphics[scale=0.225]{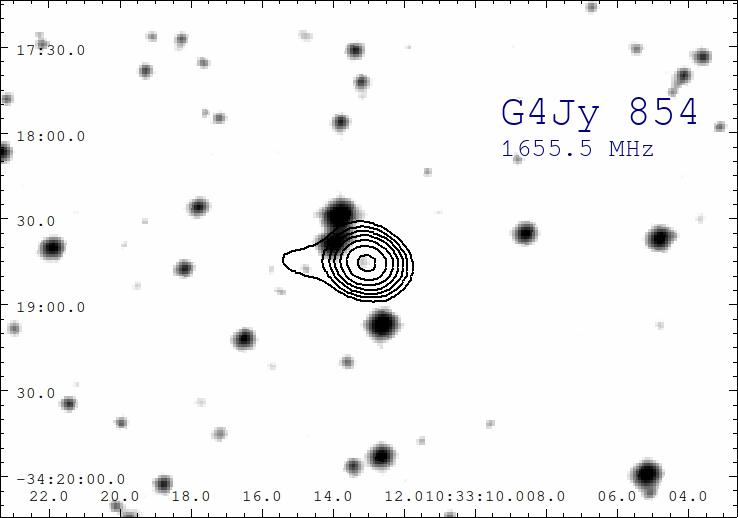}
    \includegraphics[scale=0.225]{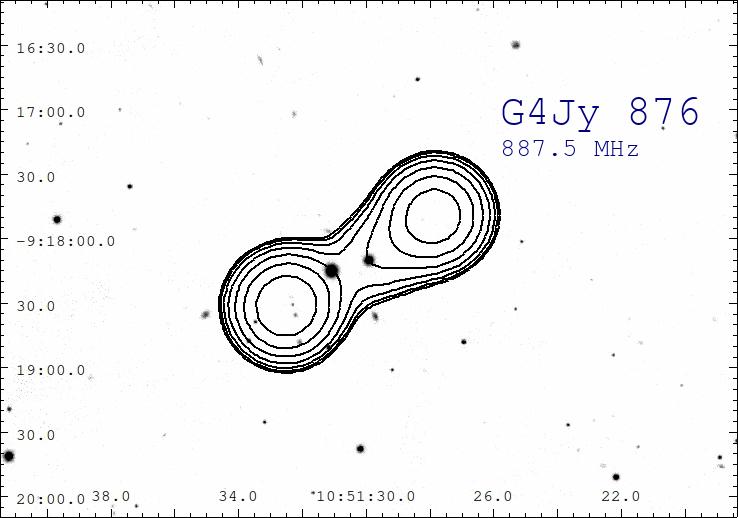}
    \includegraphics[scale=0.225]{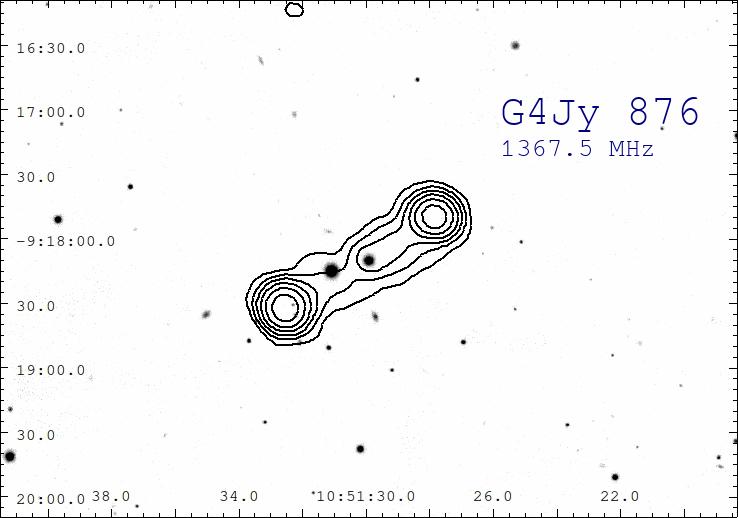}
    \includegraphics[scale=0.225]{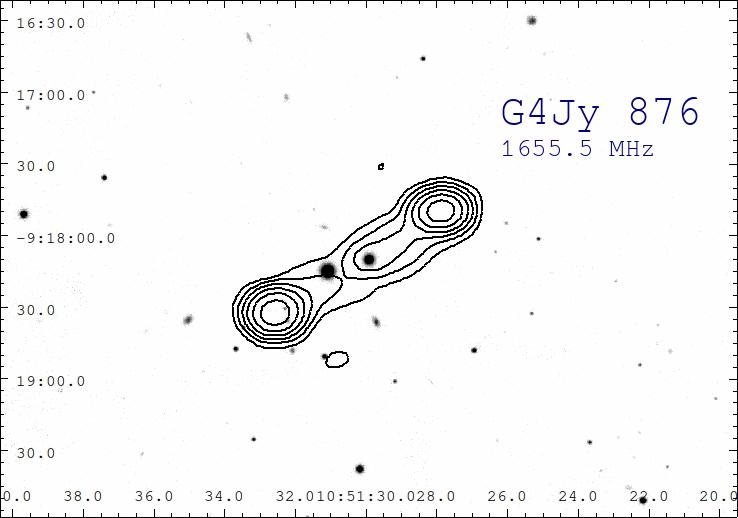}
    \includegraphics[scale=0.225]{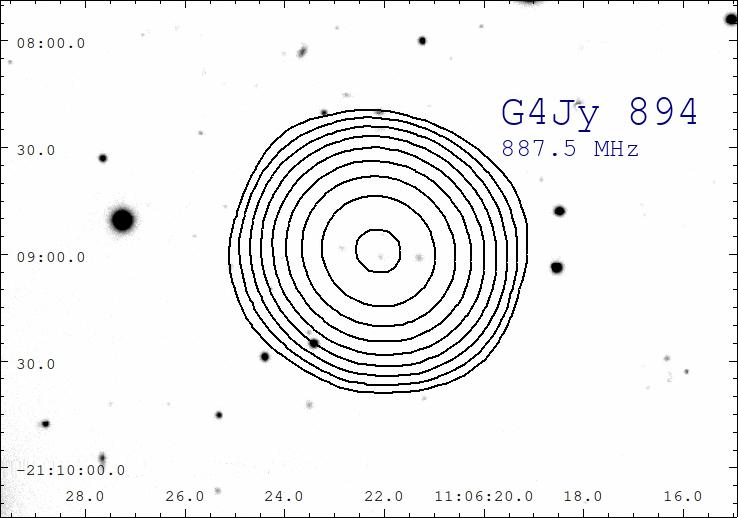}
    \includegraphics[scale=0.225]{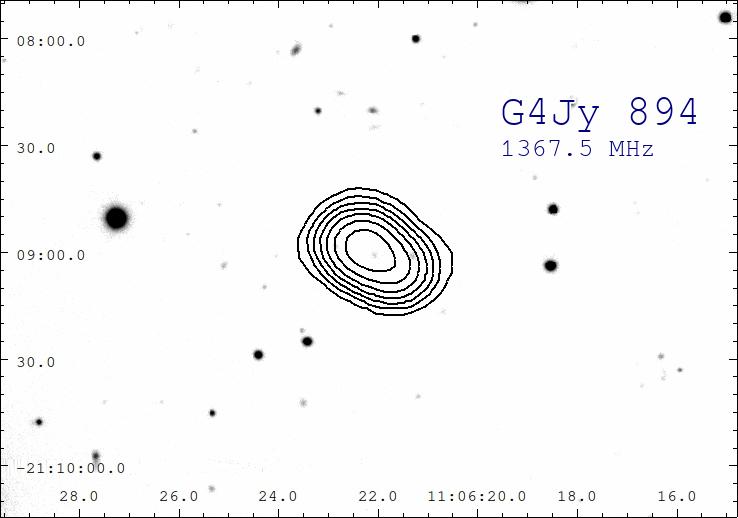}
    \includegraphics[scale=0.225]{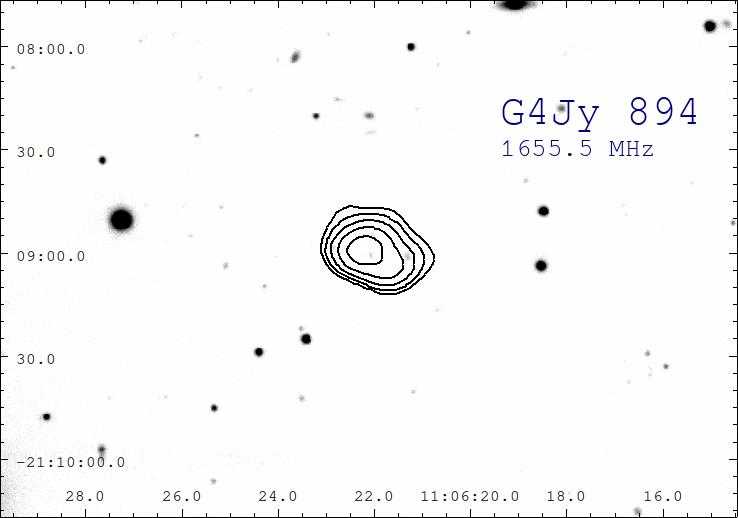}

    \caption{}
    \label{W}
\end{figure*}
\clearpage

\begin{figure*}
    \centering
    \includegraphics[scale=0.225]{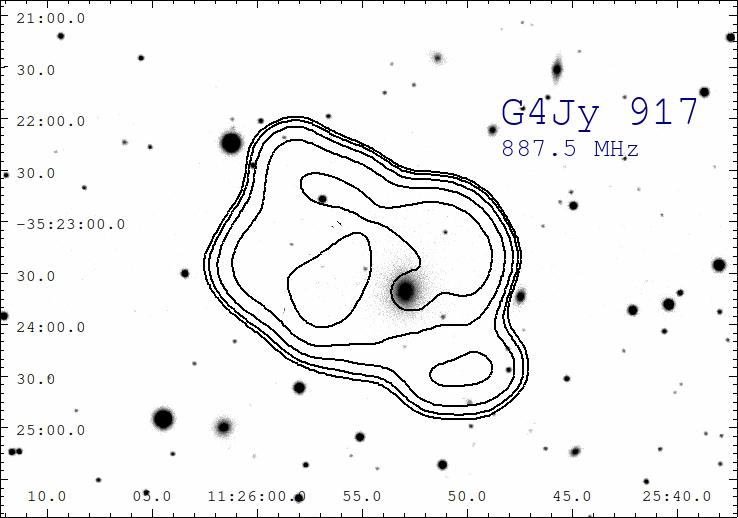}
    \includegraphics[scale=0.225]{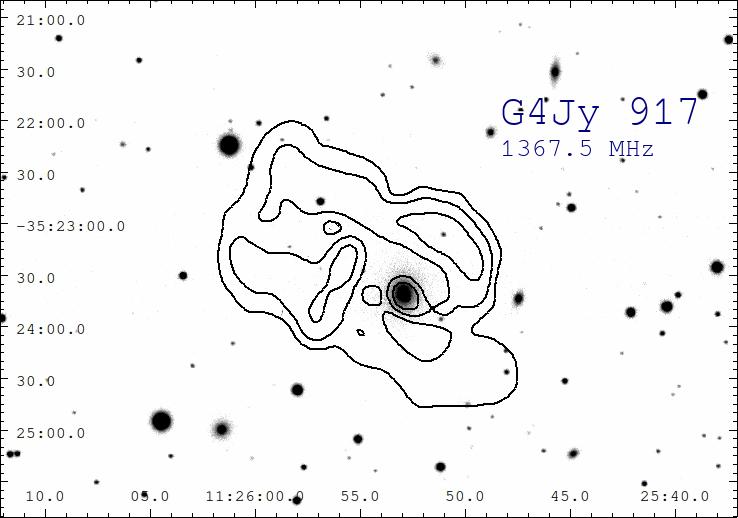}
    \includegraphics[scale=0.225]{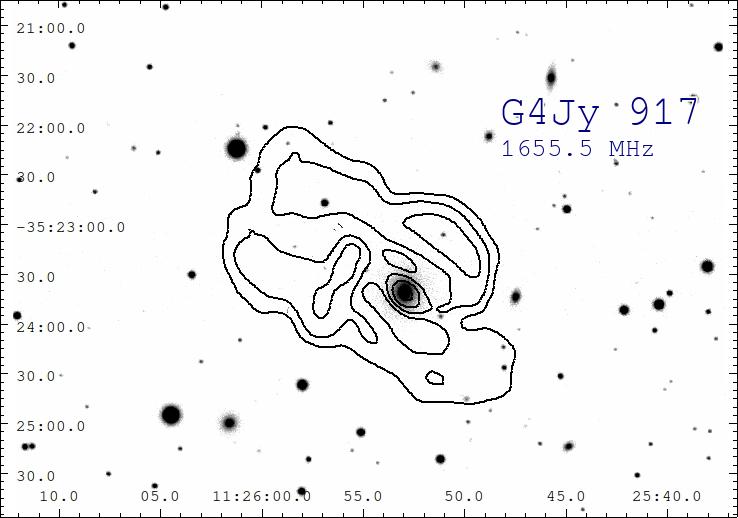}
    \includegraphics[scale=0.225]{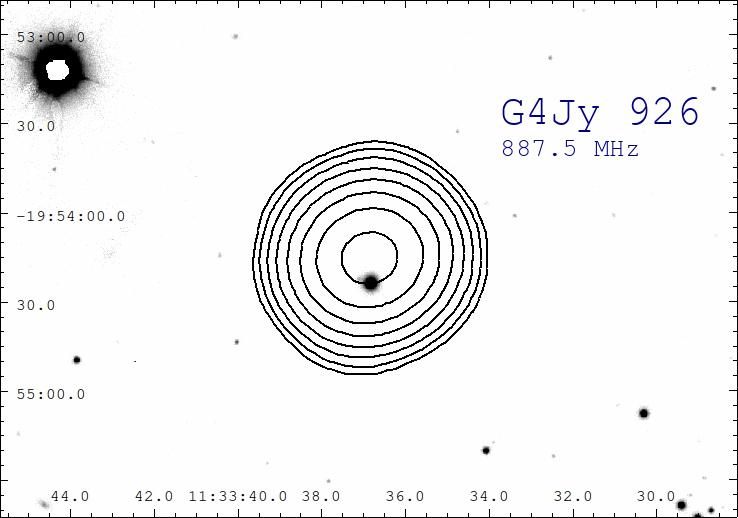}
    \includegraphics[scale=0.225]{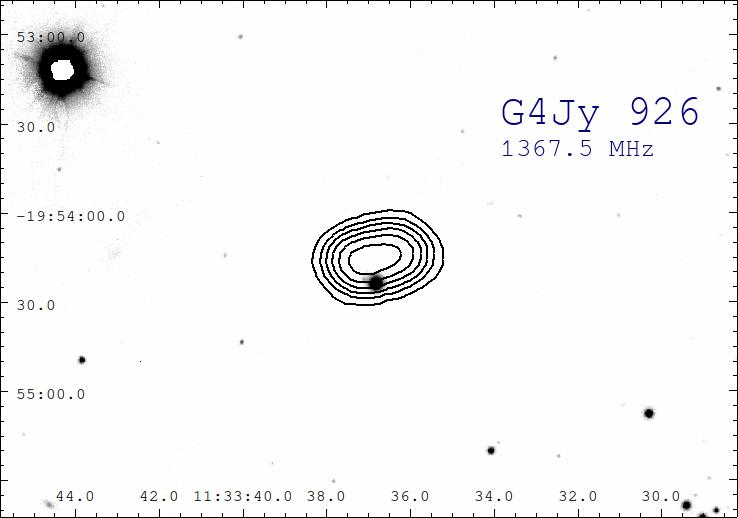}
    \includegraphics[scale=0.225]{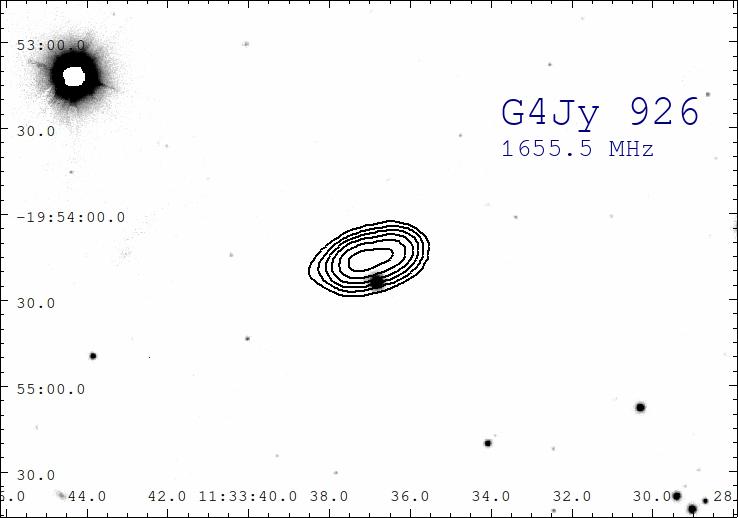}
    \includegraphics[scale=0.225]{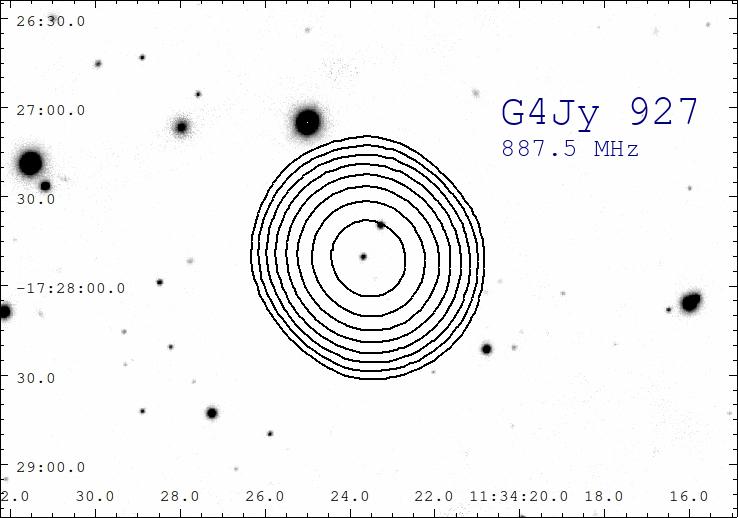}
    \includegraphics[scale=0.225]{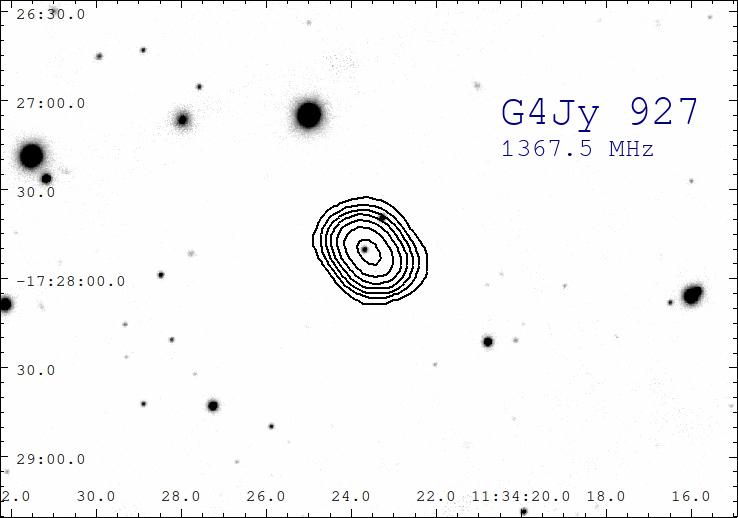}
    \includegraphics[scale=0.225]{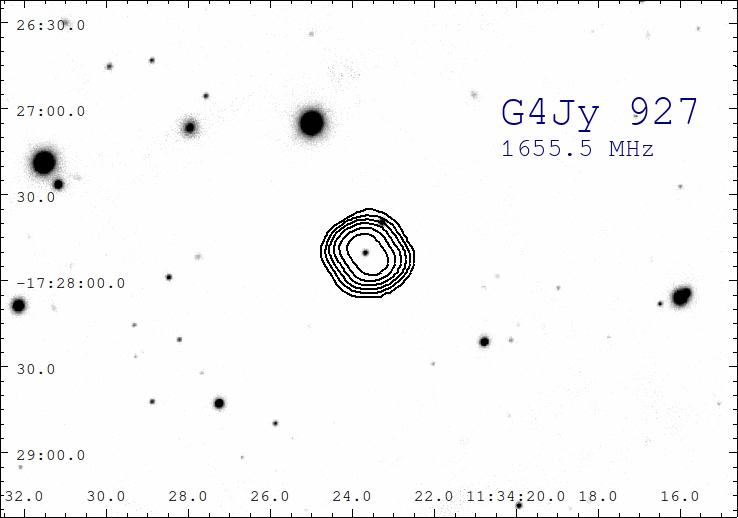}
    \includegraphics[scale=0.225]{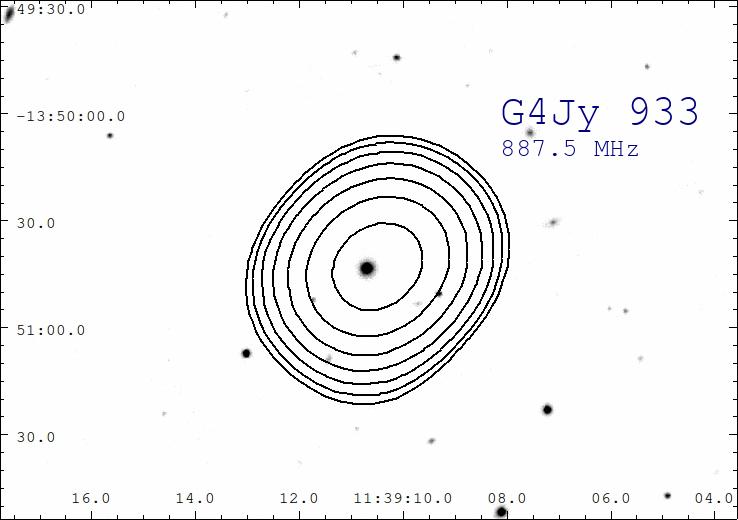}
    \includegraphics[scale=0.225]{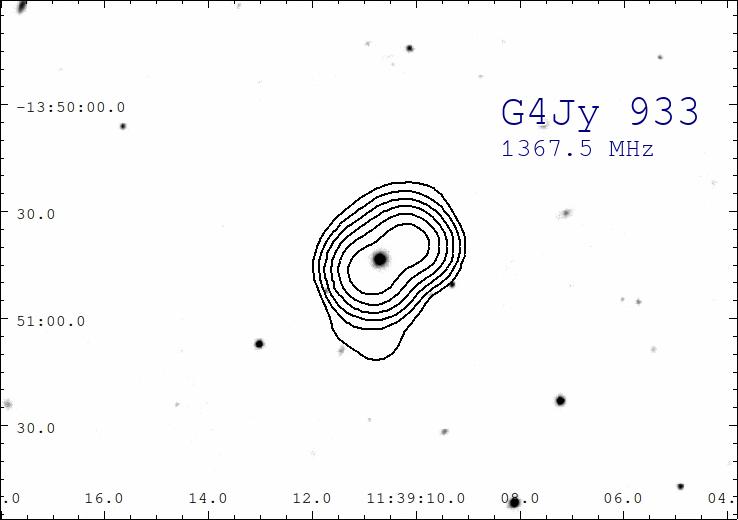}
    \includegraphics[scale=0.225]{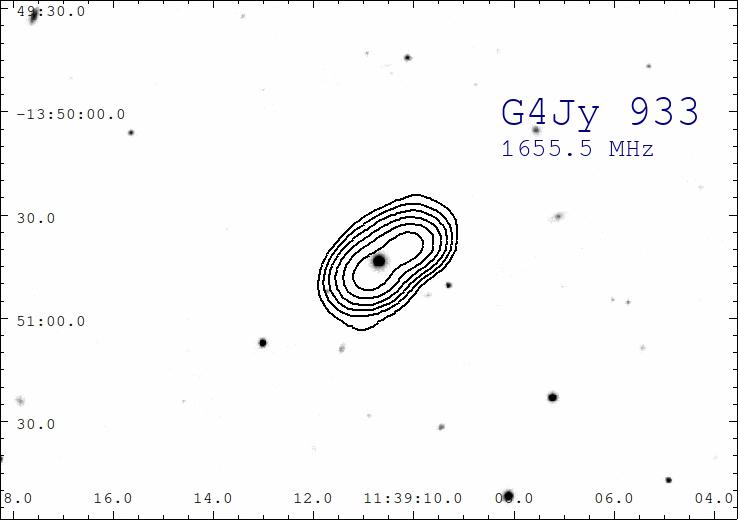}
    \includegraphics[scale=0.225]{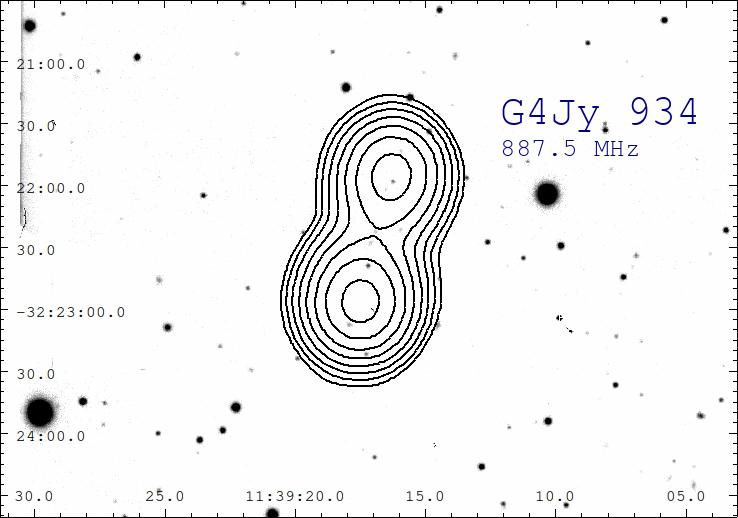}
    \includegraphics[scale=0.225]{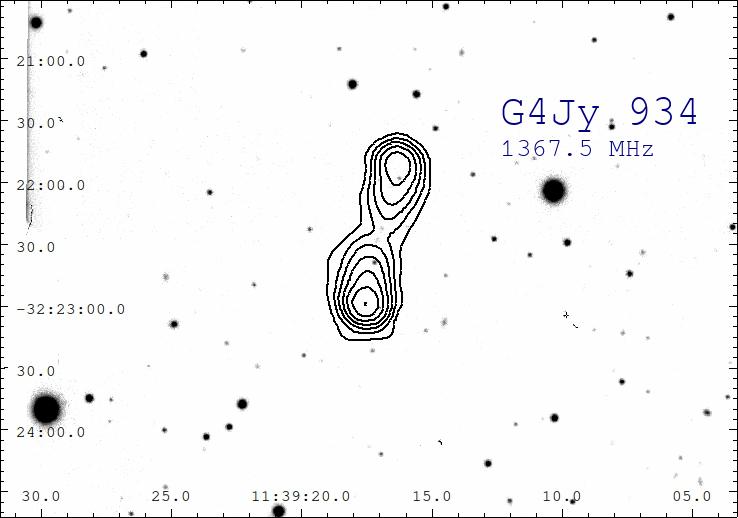}
    \includegraphics[scale=0.225]{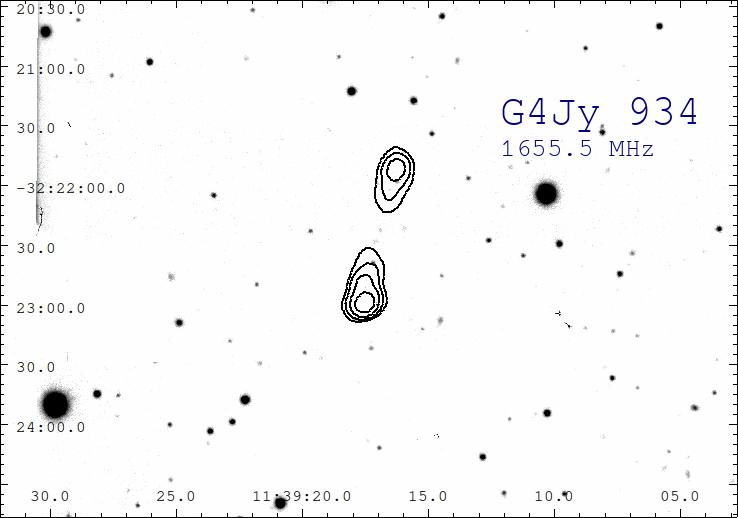}

    \caption{}
    \label{X}
\end{figure*}
\clearpage

\begin{figure*}
    \centering
    \includegraphics[scale=0.225]{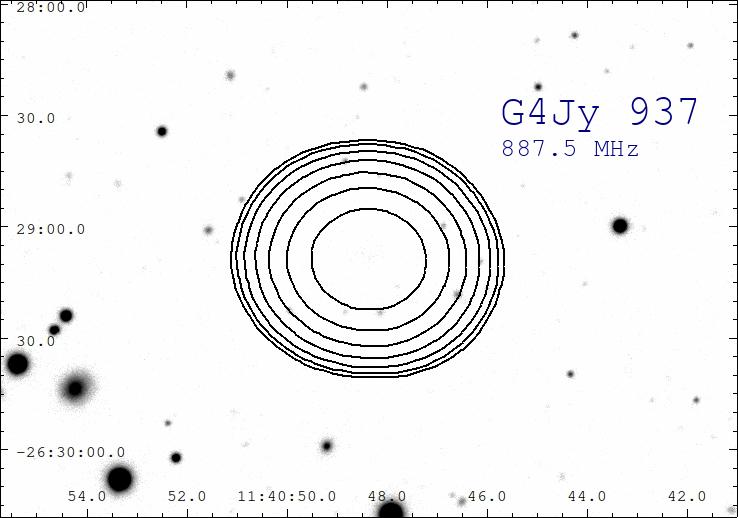}
    \includegraphics[scale=0.225]{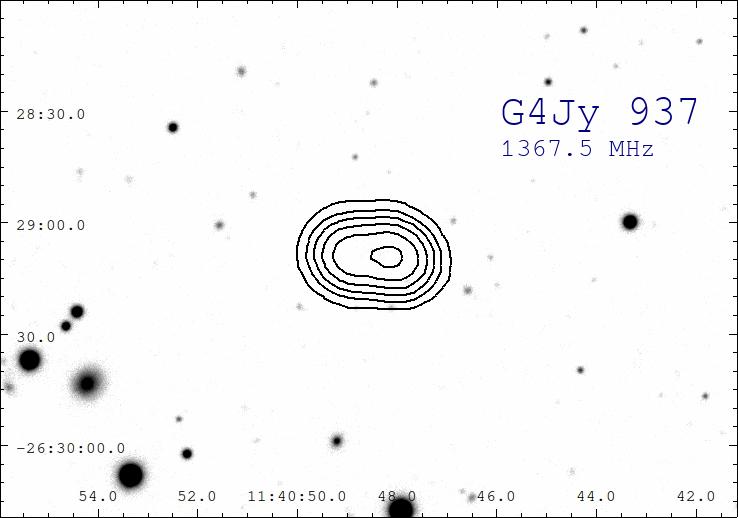}
    \includegraphics[scale=0.225]{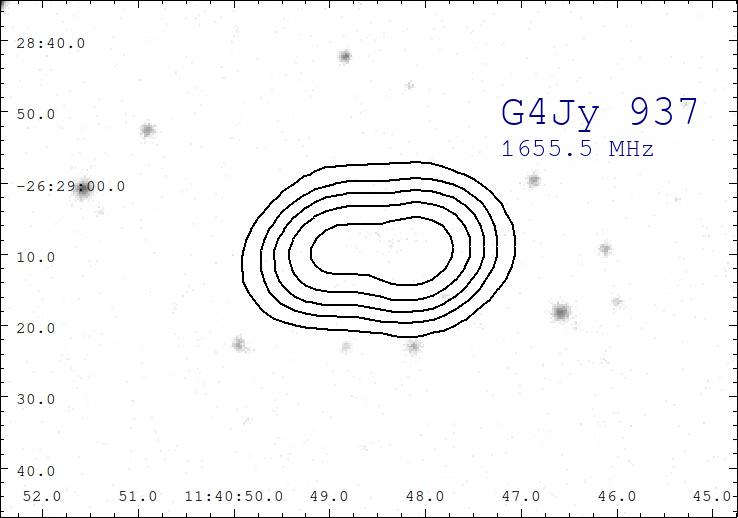}
    \includegraphics[scale=0.225]{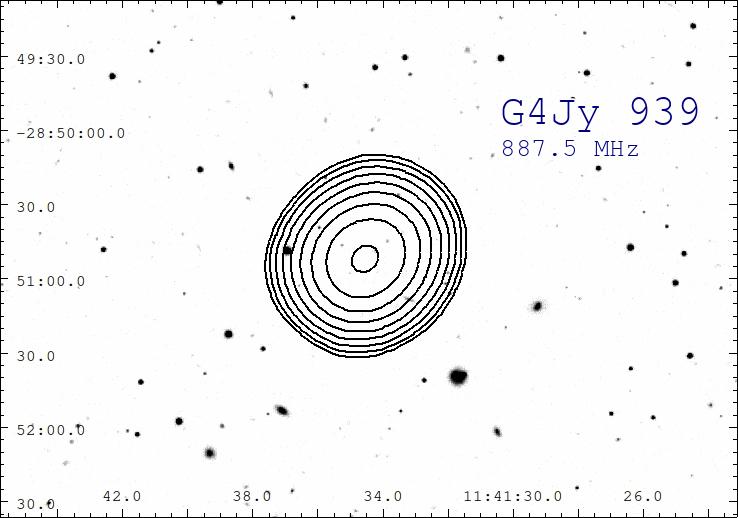}
    \includegraphics[scale=0.225]{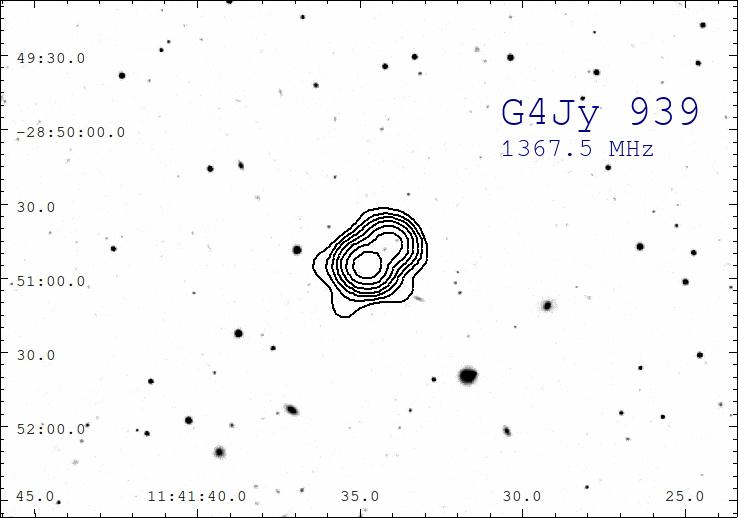}
    \includegraphics[scale=0.225]{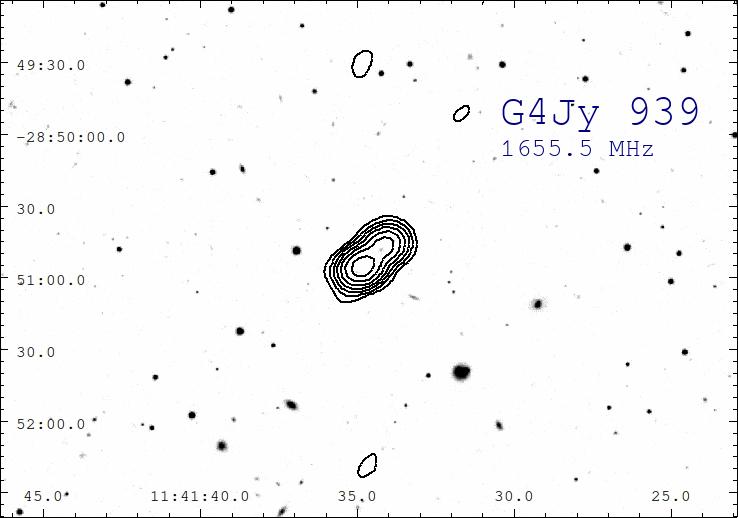}
    \includegraphics[scale=0.225]{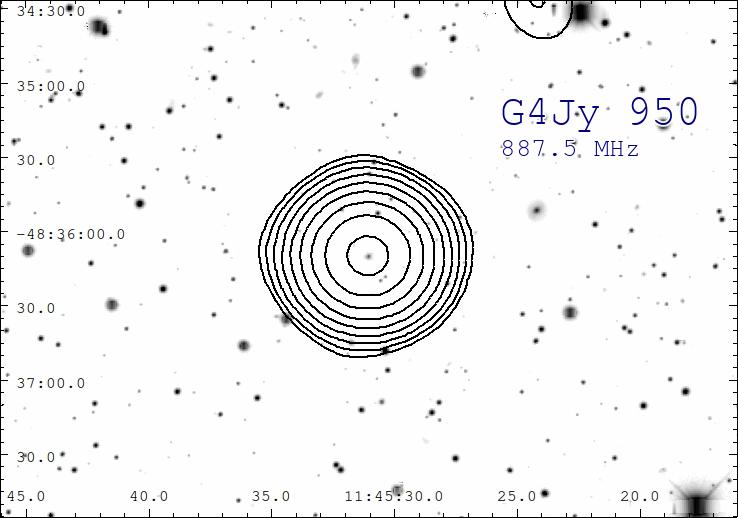}
    \includegraphics[scale=0.225]{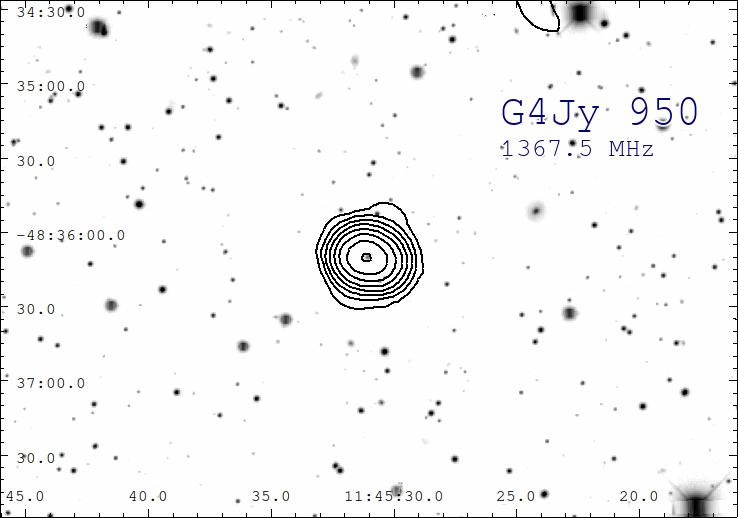}
    \includegraphics[scale=0.225]{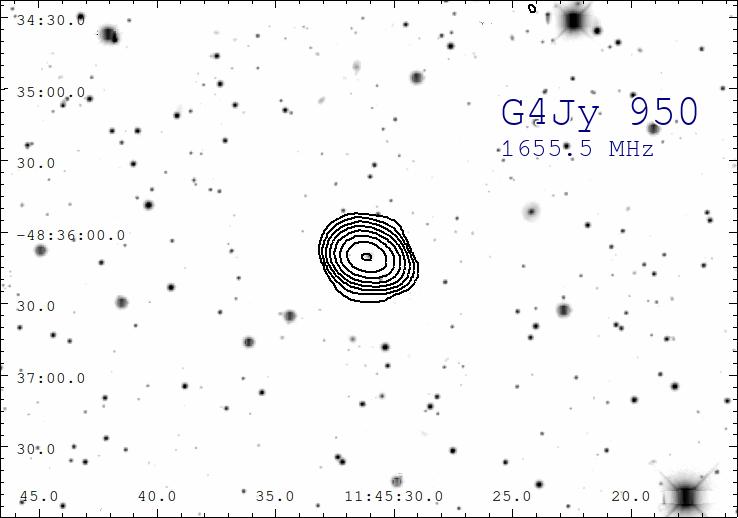}
    \includegraphics[scale=0.225]{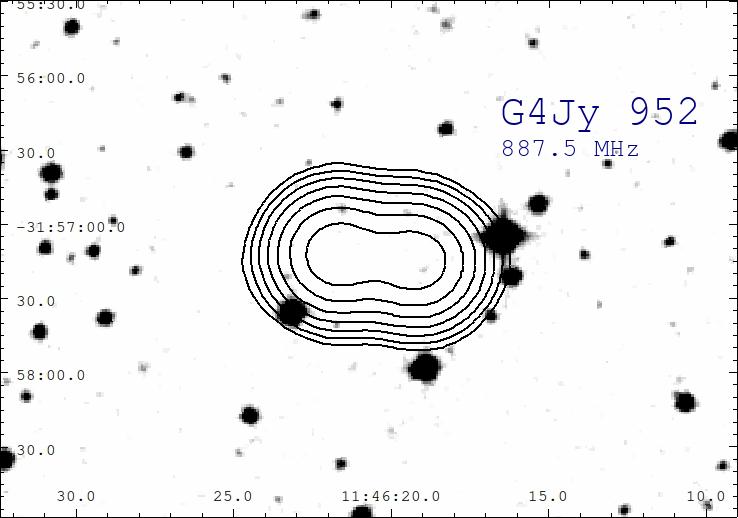}
    \includegraphics[scale=0.225]{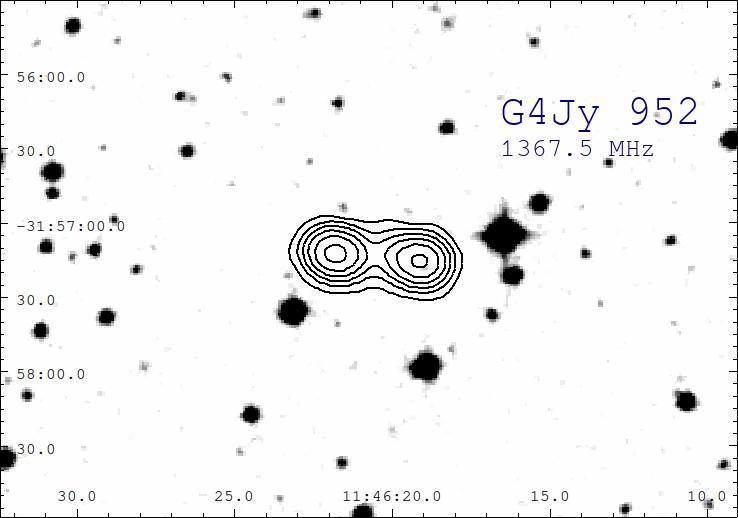}
    \includegraphics[scale=0.225]{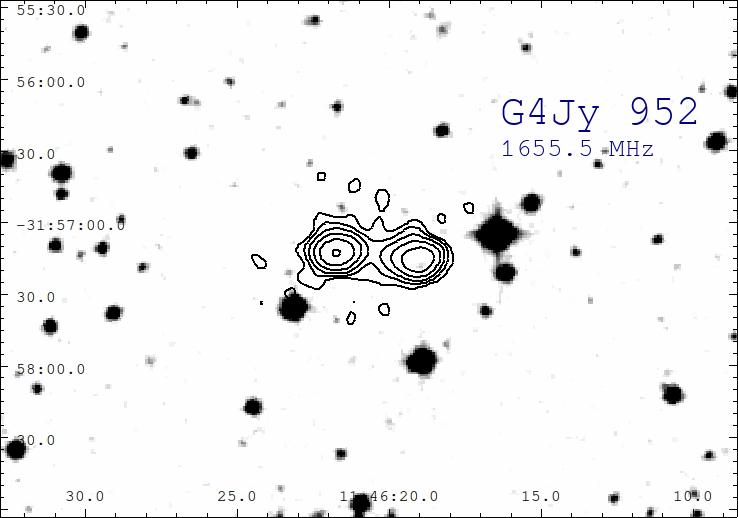}
    \includegraphics[scale=0.225]{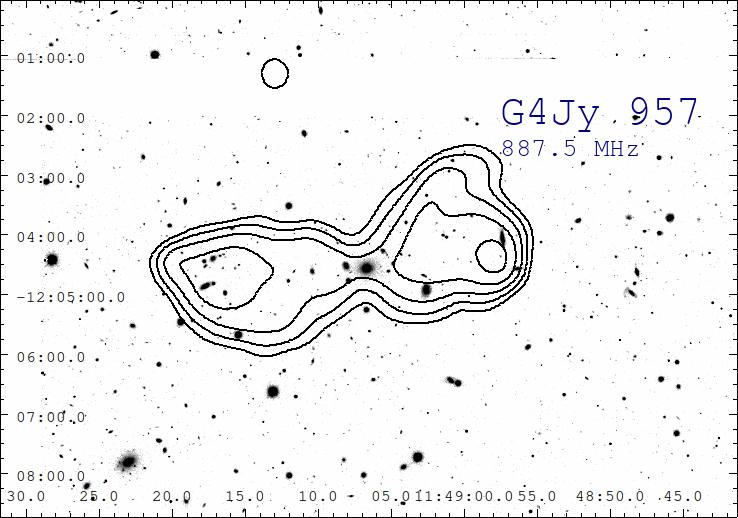}
    \includegraphics[scale=0.225]{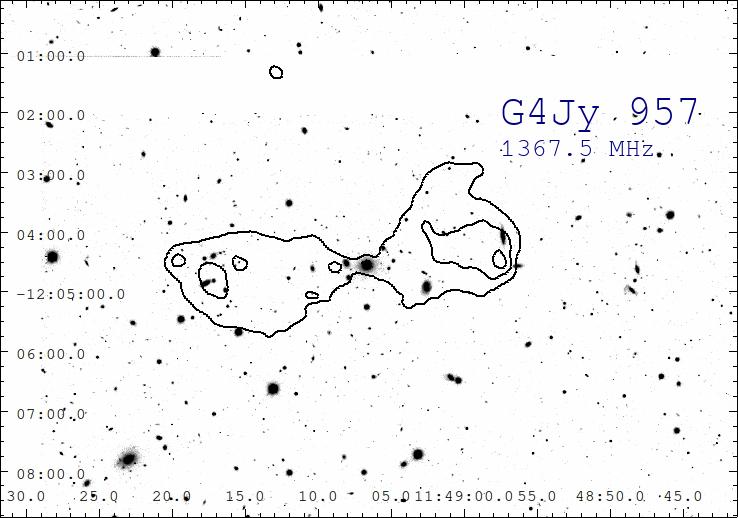}
    \includegraphics[scale=0.225]{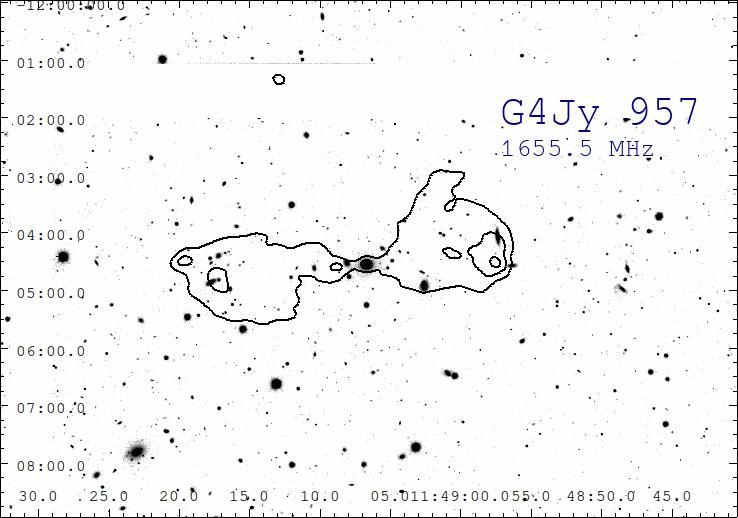}

    \caption{}
    \label{Y}
\end{figure*}
\clearpage

\begin{figure*}
    \centering
    \includegraphics[scale=0.225]{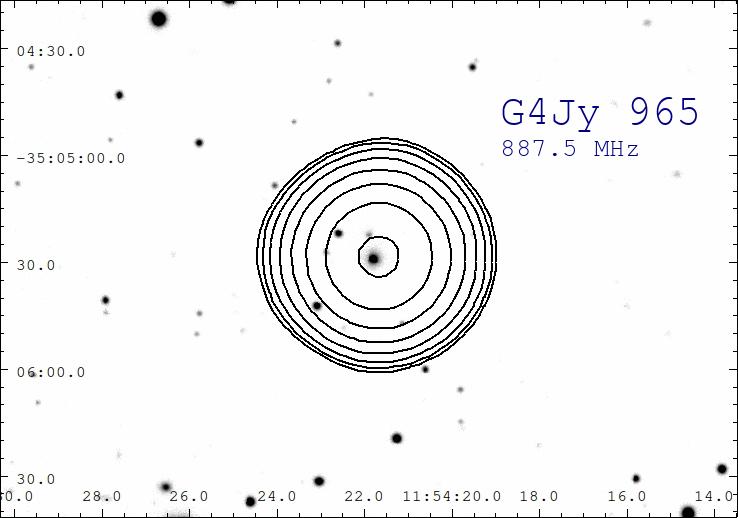}
    \includegraphics[scale=0.225]{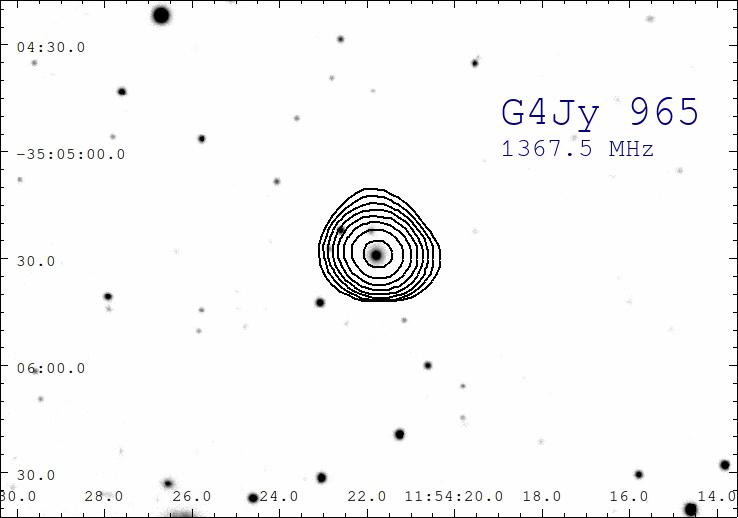}
    \includegraphics[scale=0.225]{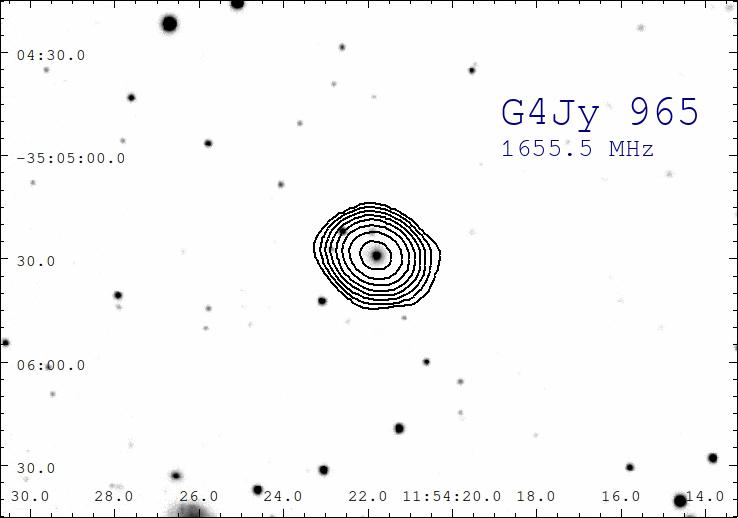}
    \includegraphics[scale=0.225]{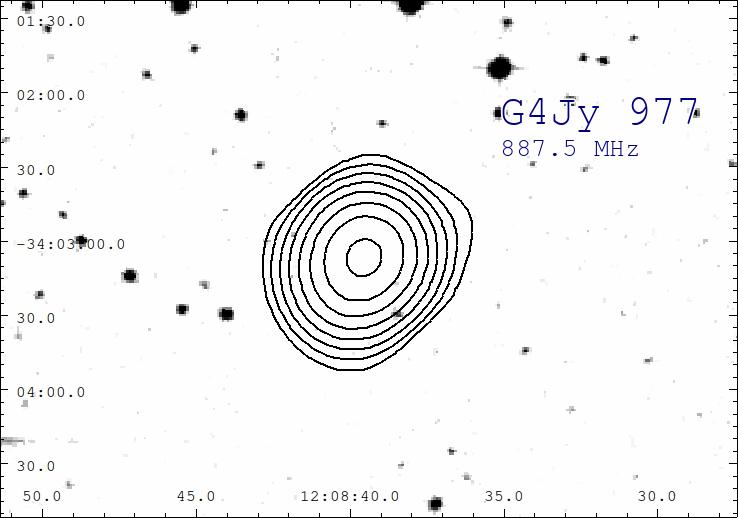}
    \includegraphics[scale=0.225]{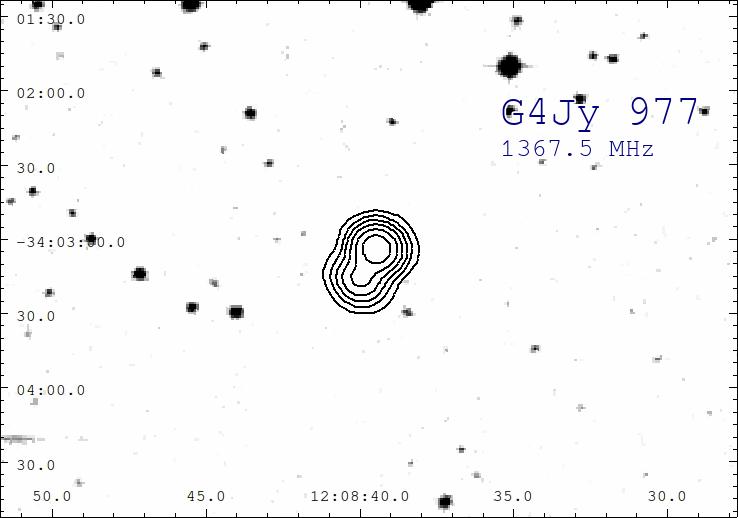}
    \includegraphics[scale=0.225]{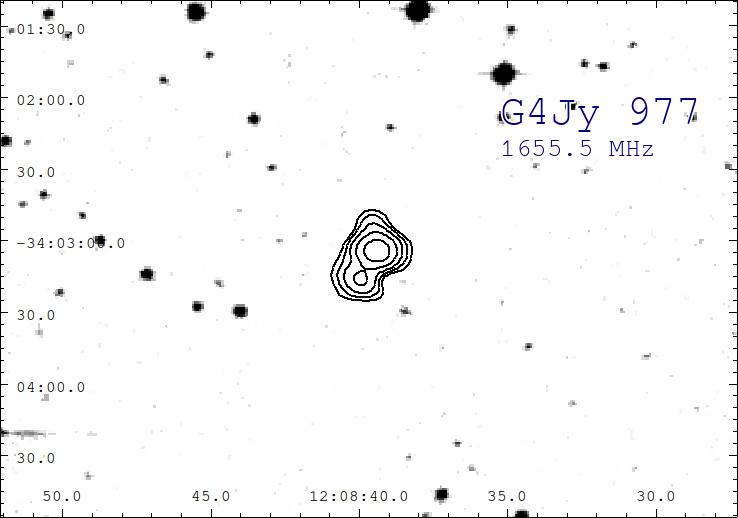}
    \includegraphics[scale=0.225]{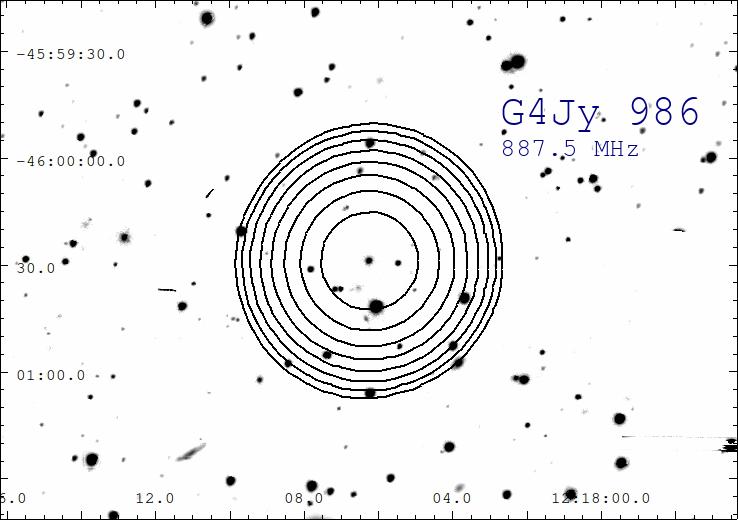}
    \includegraphics[scale=0.225]{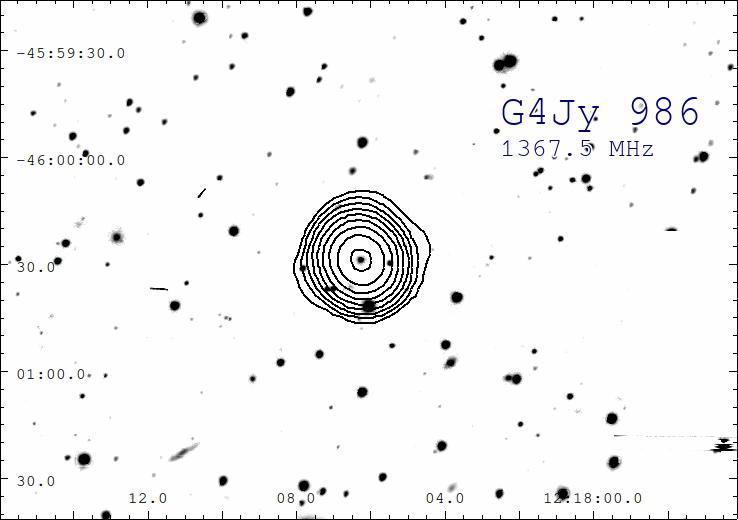}
    \includegraphics[scale=0.225]{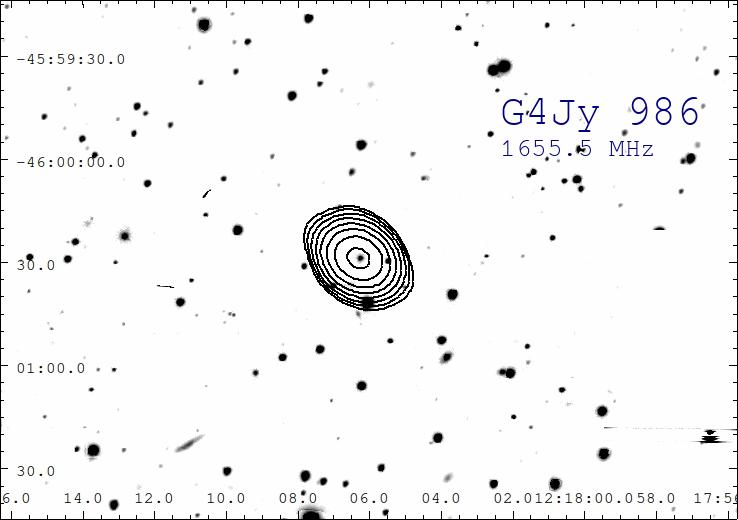}
    \includegraphics[scale=0.225]{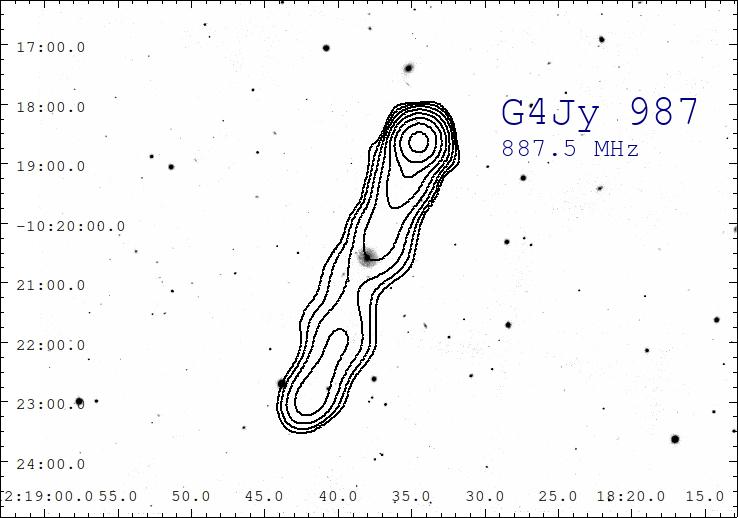}
    \includegraphics[scale=0.225]{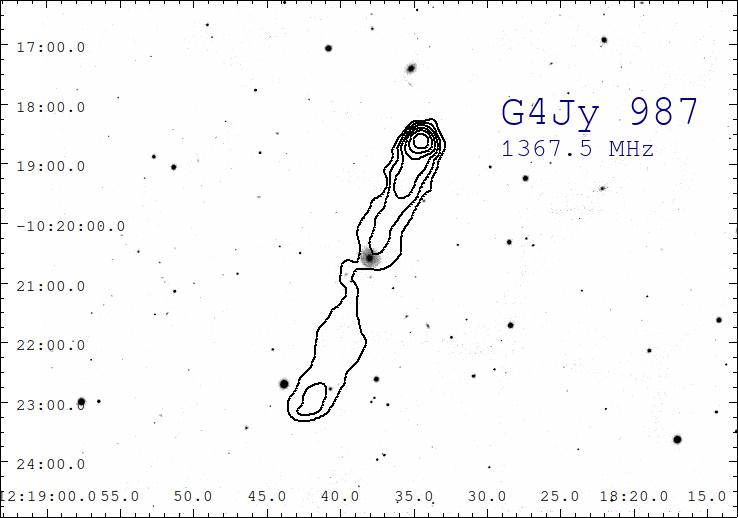}
    \includegraphics[scale=0.225]{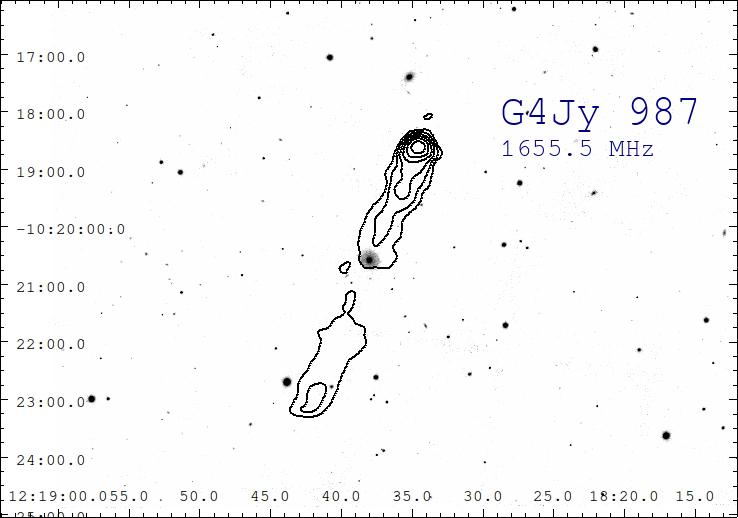}
    \includegraphics[scale=0.225]{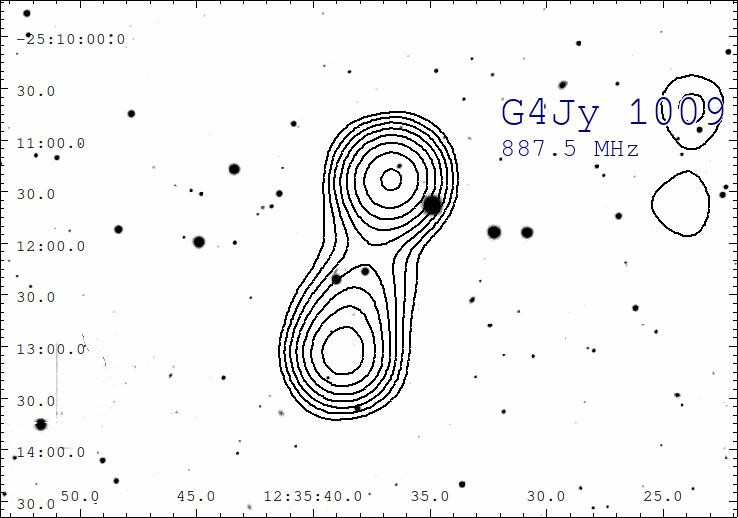}
    \includegraphics[scale=0.225]{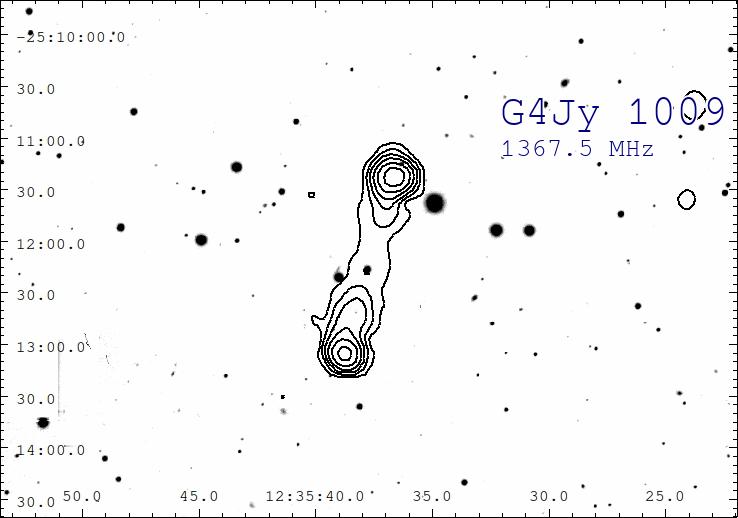}
    \includegraphics[scale=0.225]{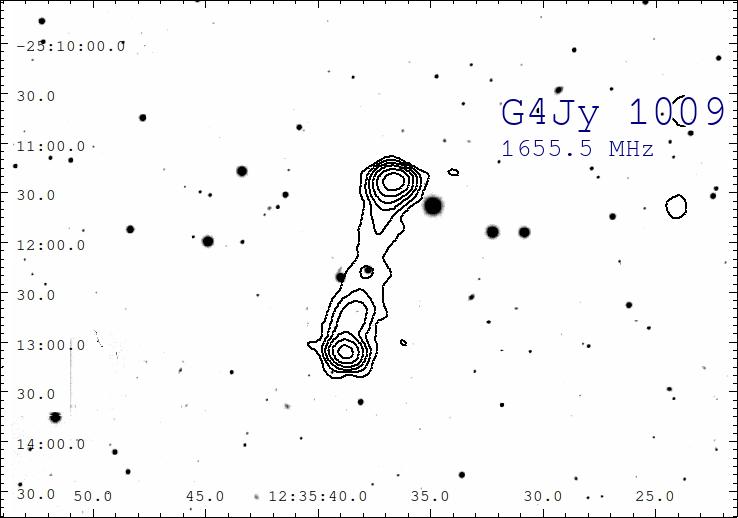}

    \caption{}
    \label{Z}
\end{figure*}
\clearpage
 
\begin{figure*}
    \centering
    \includegraphics[scale=0.225]{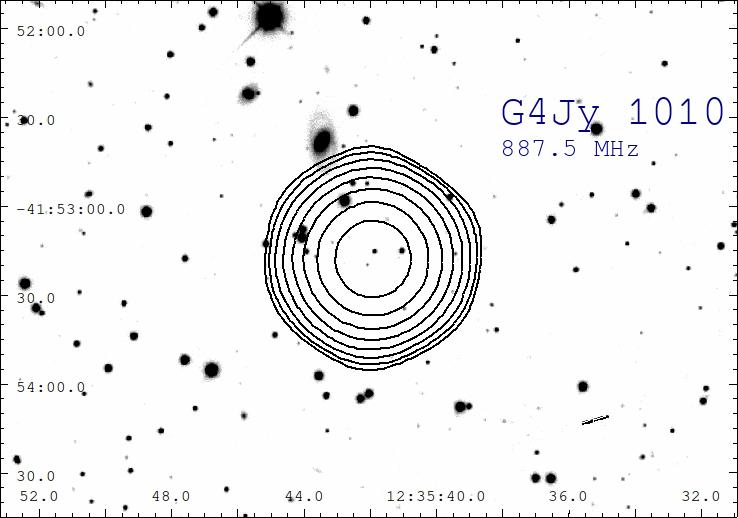}
    \includegraphics[scale=0.225]{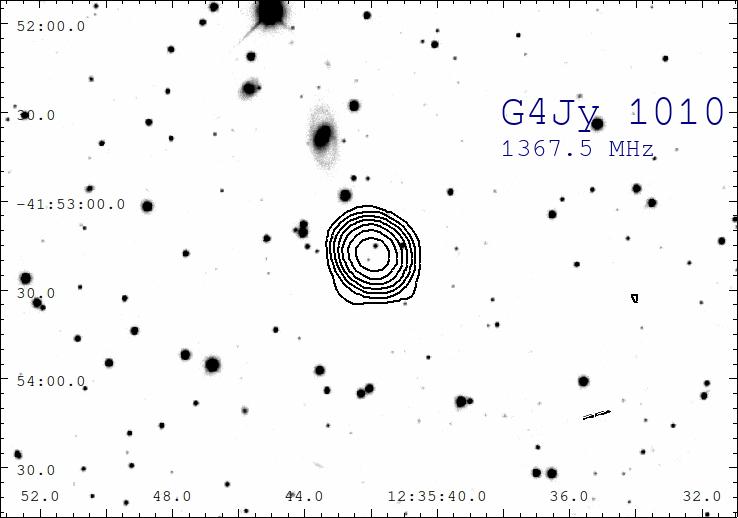}
    \includegraphics[scale=0.225]{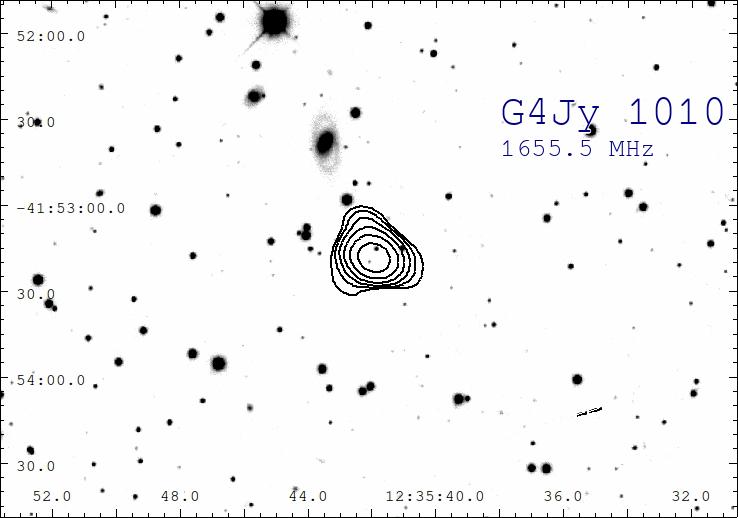}
    \includegraphics[scale=0.225]{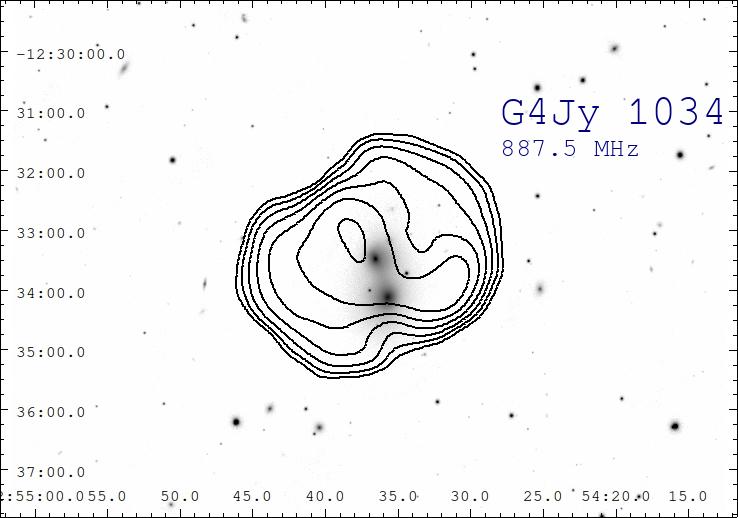}
    \includegraphics[scale=0.225]{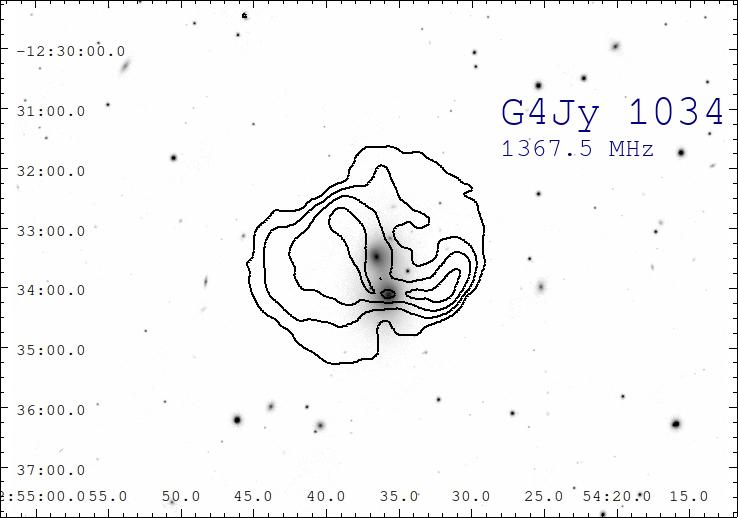}
    \includegraphics[scale=0.225]{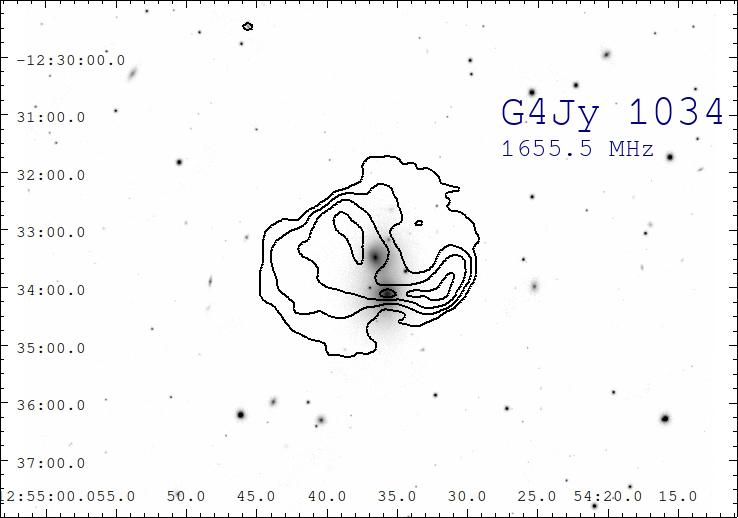}
    \includegraphics[scale=0.225]{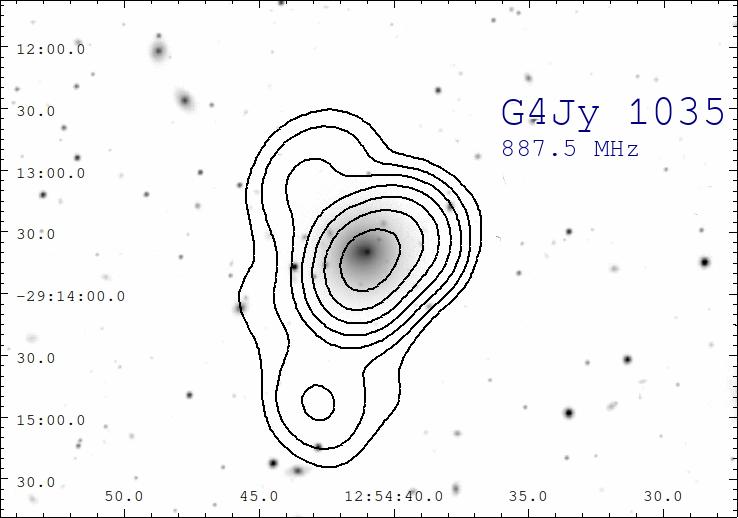}
    \includegraphics[scale=0.225]{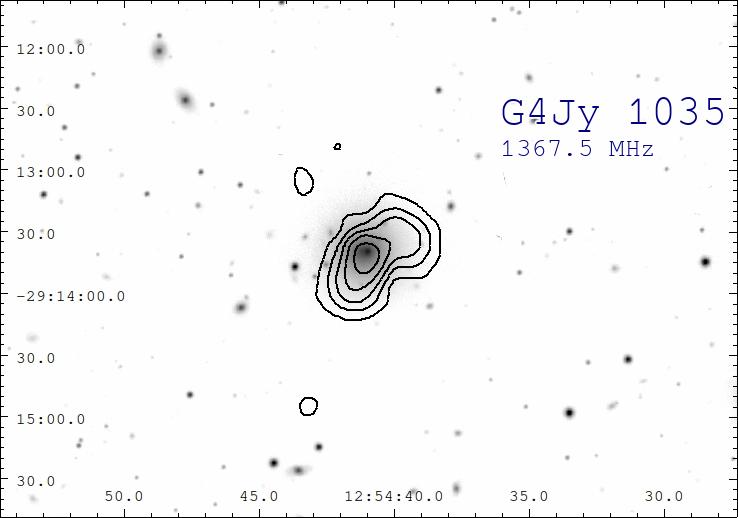}
    \includegraphics[scale=0.225]{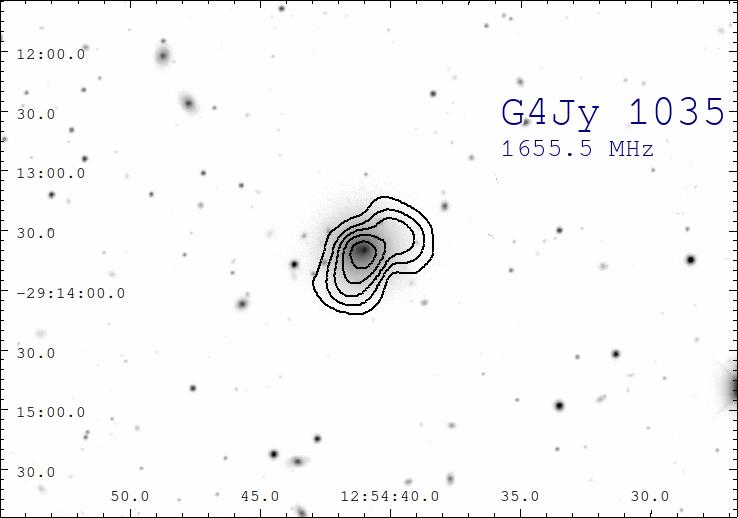}
    \includegraphics[scale=0.225]{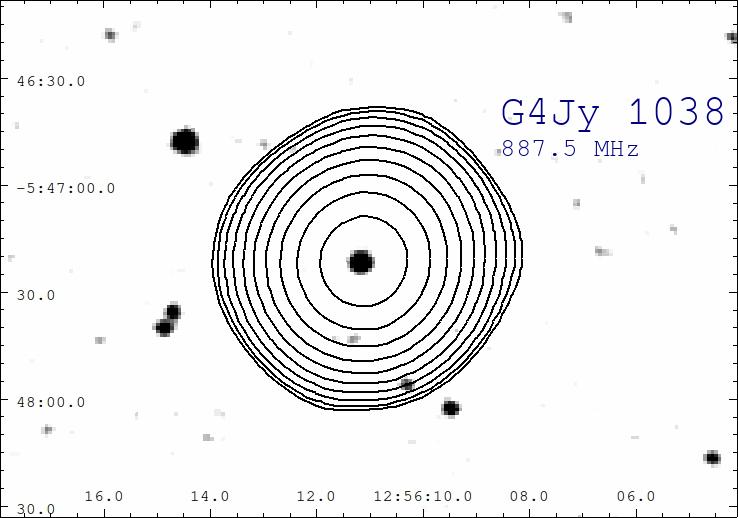}
    \includegraphics[scale=0.225]{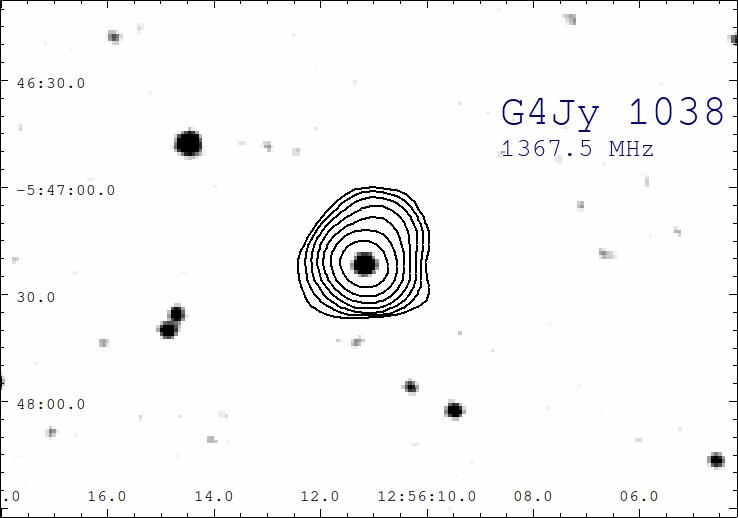}
    \includegraphics[scale=0.225]{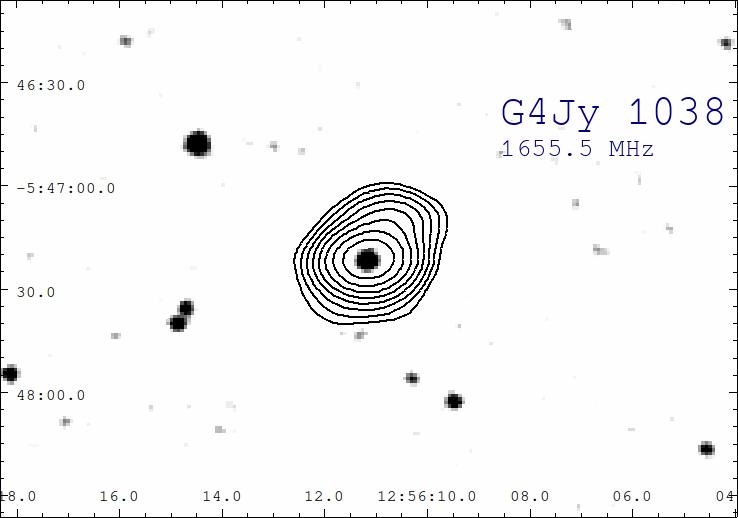}
    \includegraphics[scale=0.225]{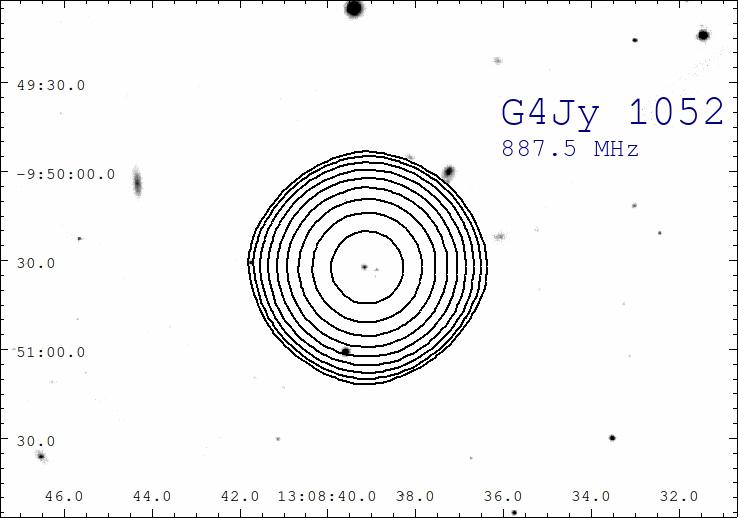}
    \includegraphics[scale=0.225]{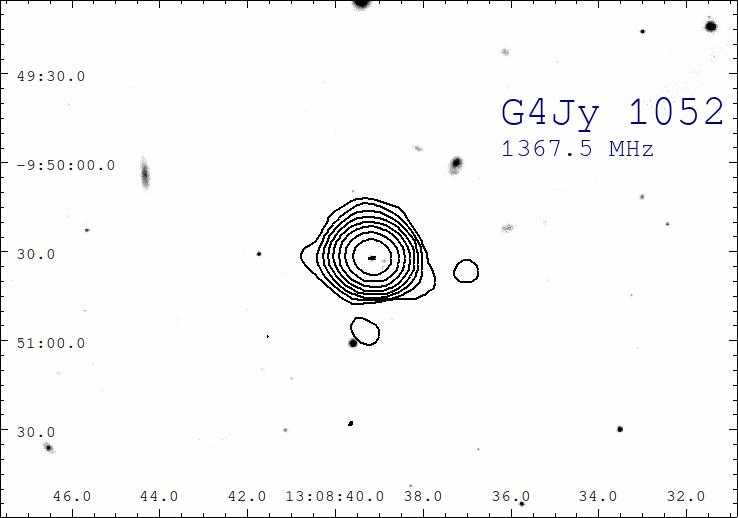}
    \includegraphics[scale=0.225]{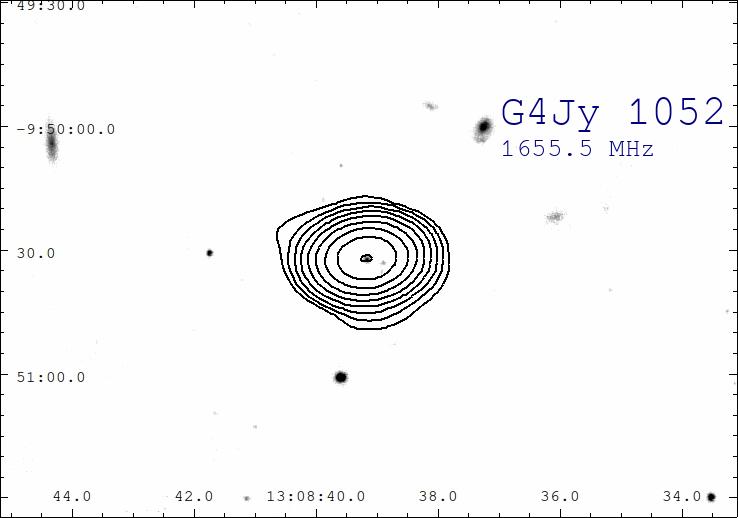}

    \caption{}
    \label{AA}
\end{figure*}
\clearpage
 
\begin{figure*}
    \centering
    \includegraphics[scale=0.225]{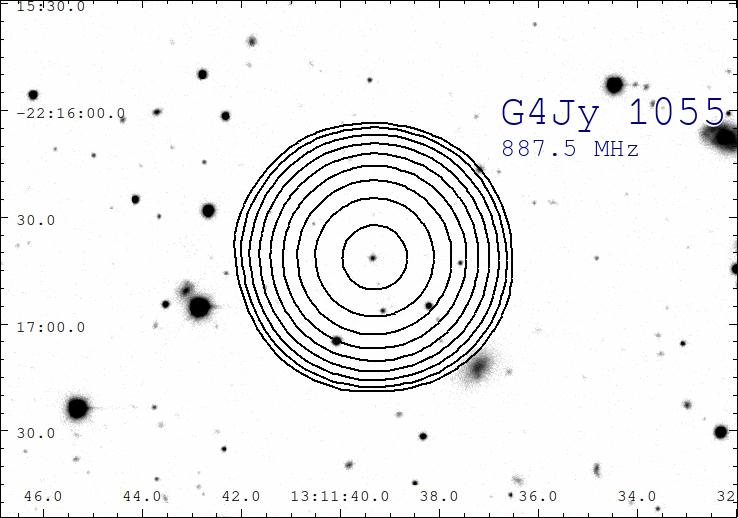}
    \includegraphics[scale=0.225]{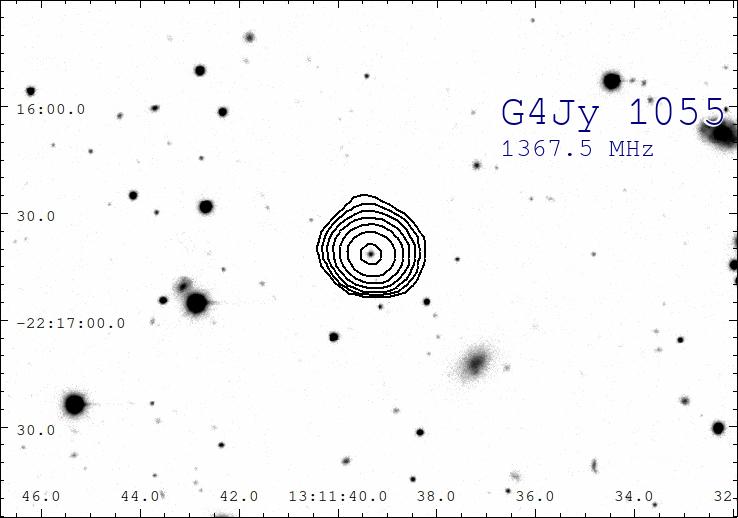}
    \includegraphics[scale=0.225]{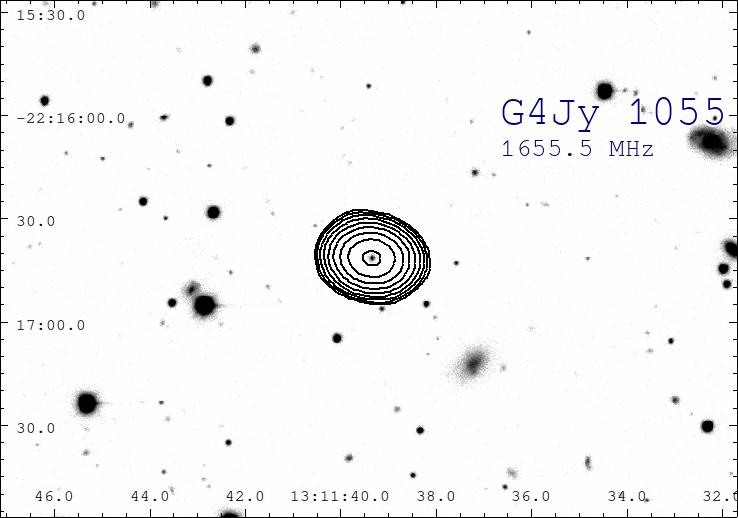}
    \includegraphics[scale=0.225]{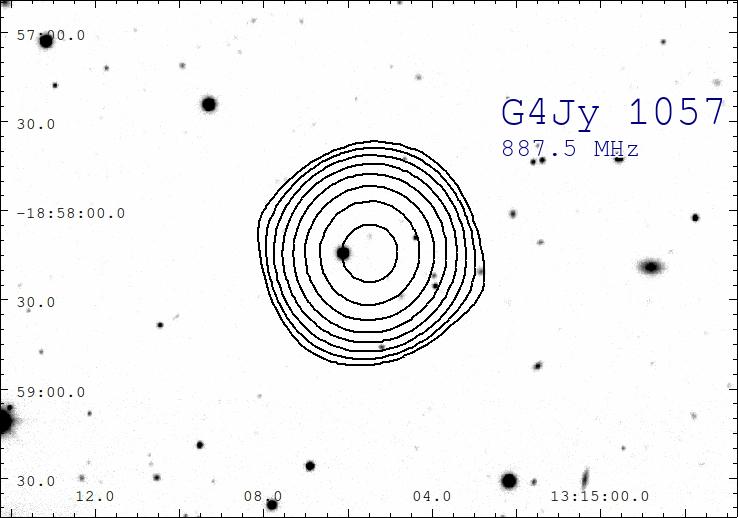}
    \includegraphics[scale=0.225]{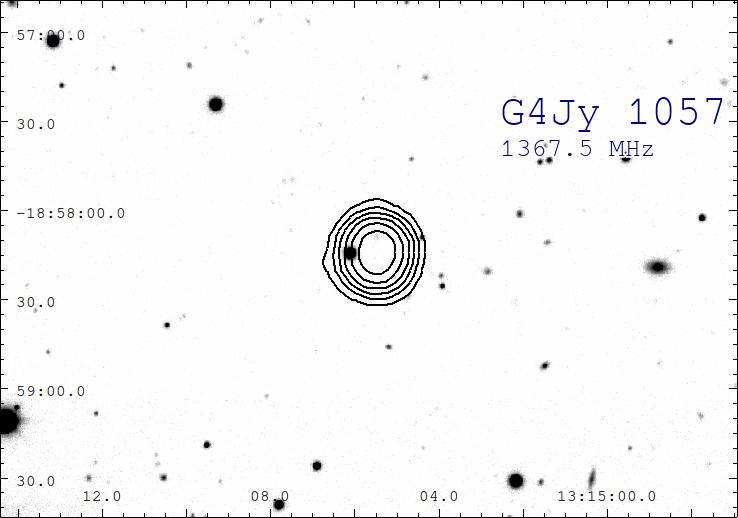}
    \includegraphics[scale=0.225]{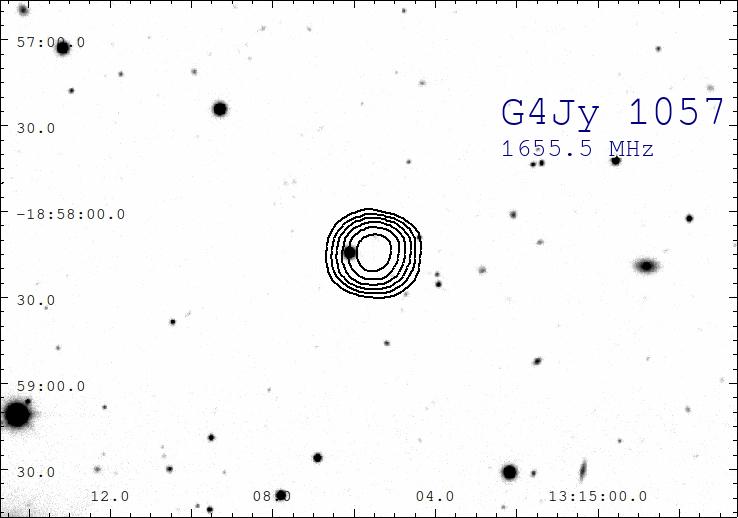}
    \includegraphics[scale=0.225]{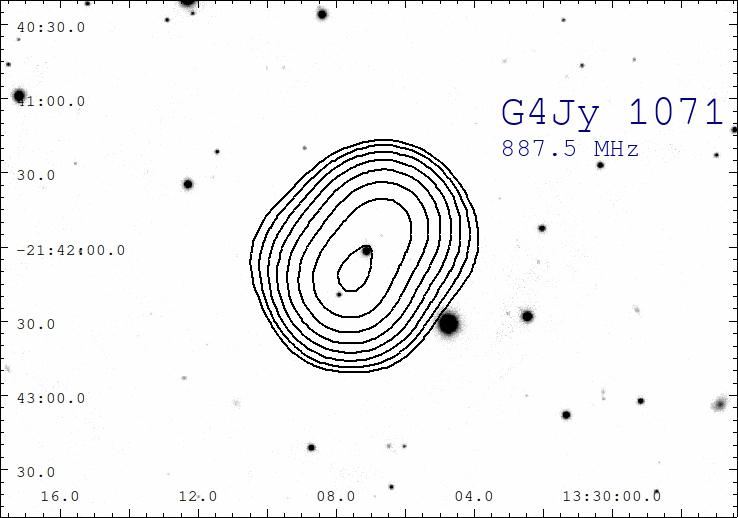}
    \includegraphics[scale=0.225]{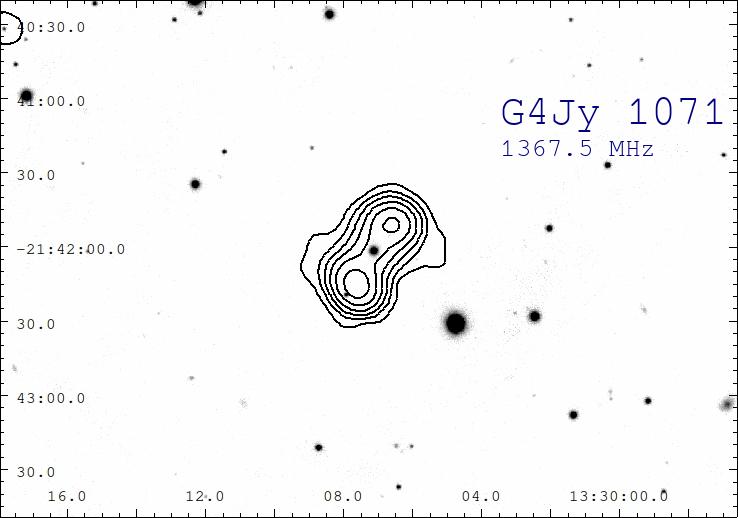}
    \includegraphics[scale=0.225]{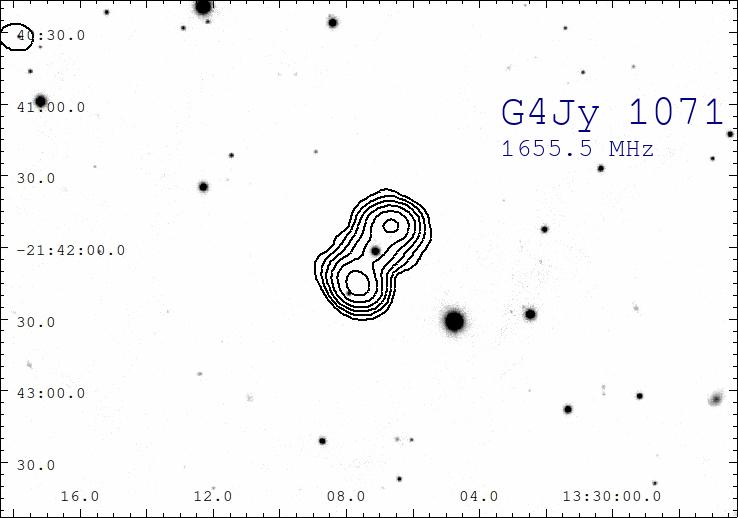}
    \includegraphics[scale=0.225]{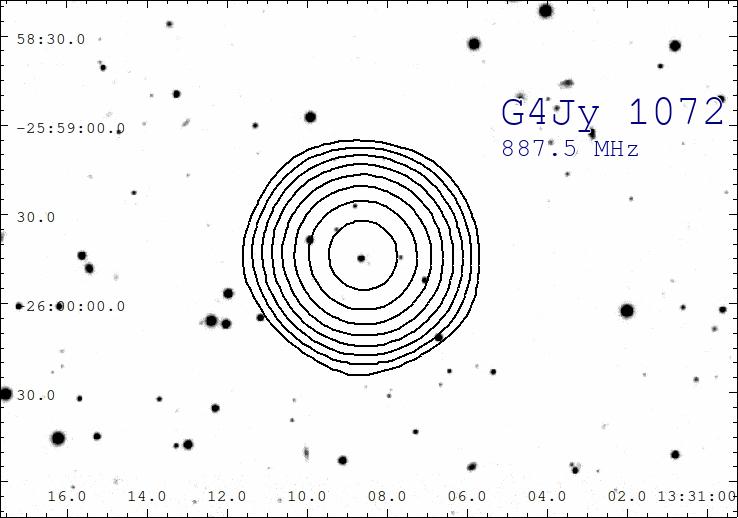}
    \includegraphics[scale=0.225]{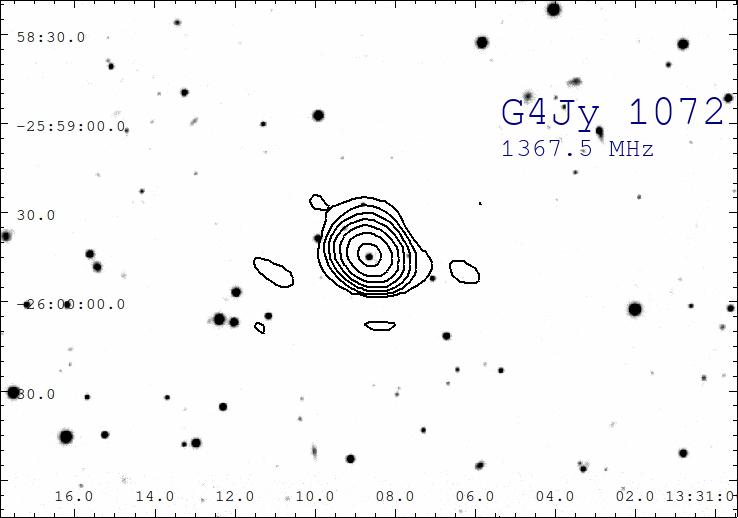}
    \includegraphics[scale=0.225]{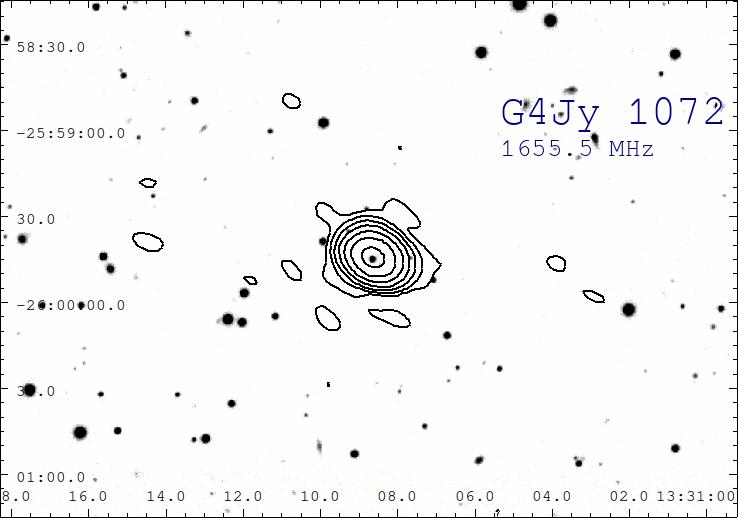}
    \includegraphics[scale=0.225]{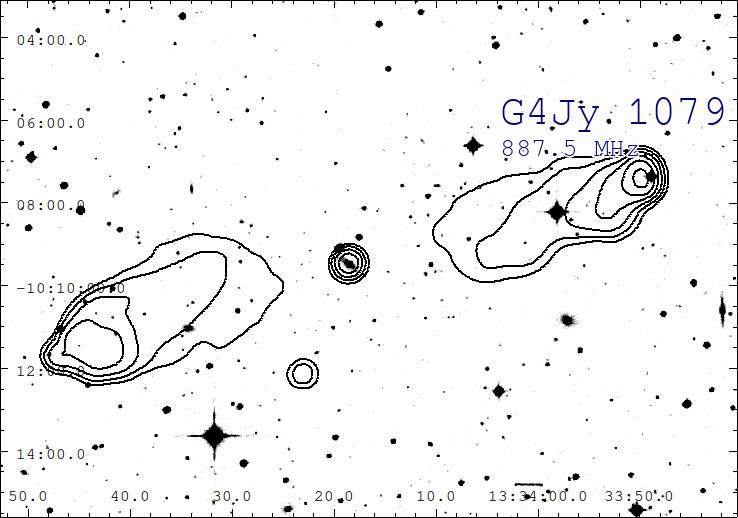}
    \includegraphics[scale=0.225]{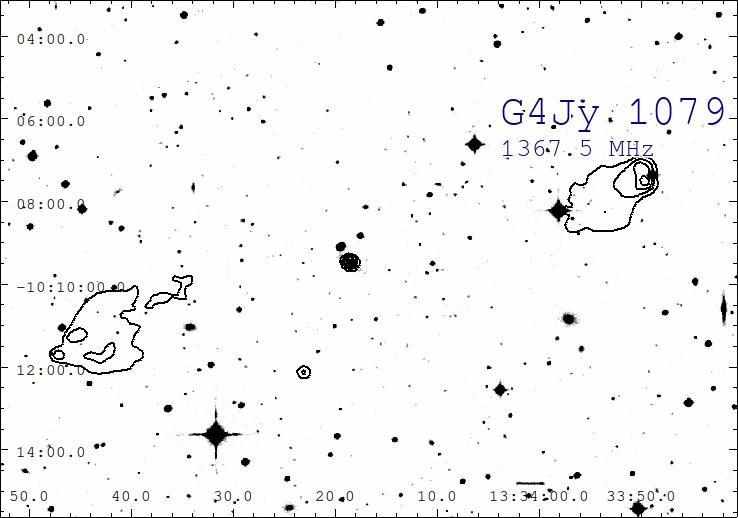}
    \includegraphics[scale=0.225]{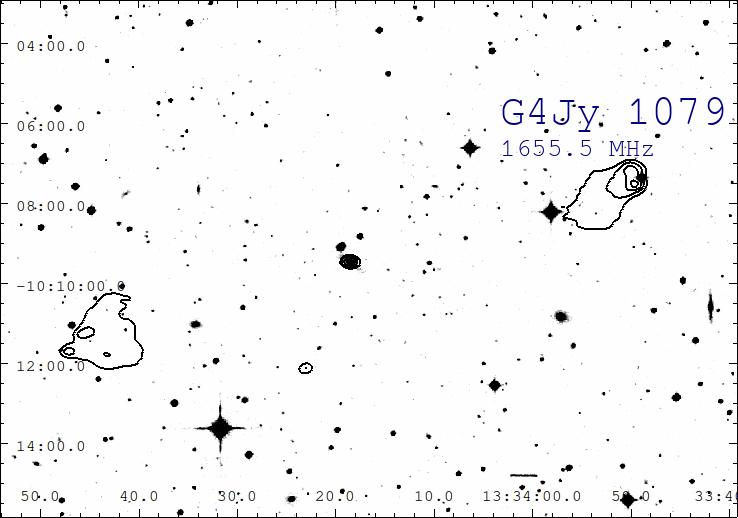}

    \caption{}
    \label{AB}
\end{figure*}
\clearpage
 
\begin{figure*}
    \centering
    \includegraphics[scale=0.225]{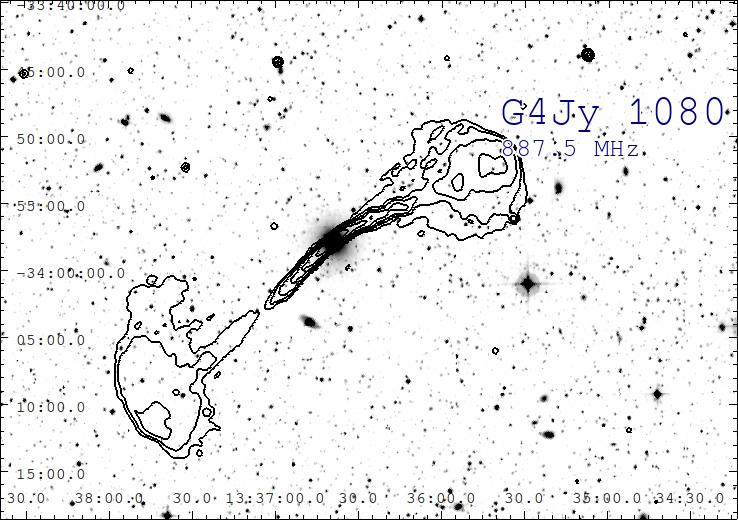}
    \includegraphics[scale=0.225]{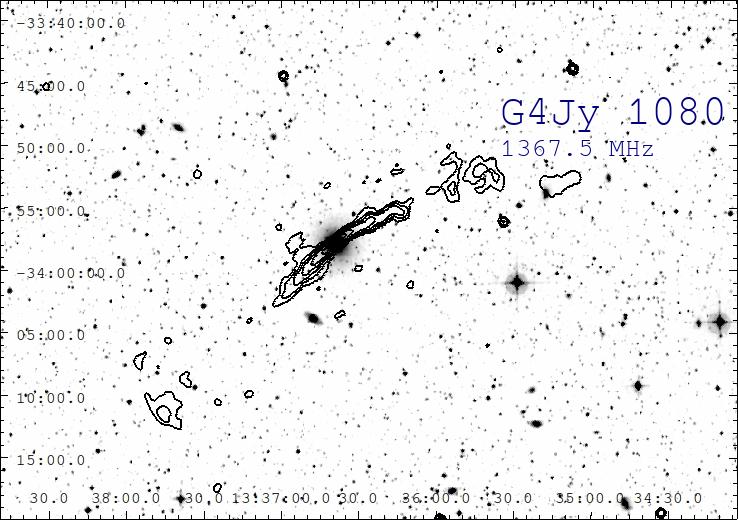}
    \includegraphics[scale=0.225]{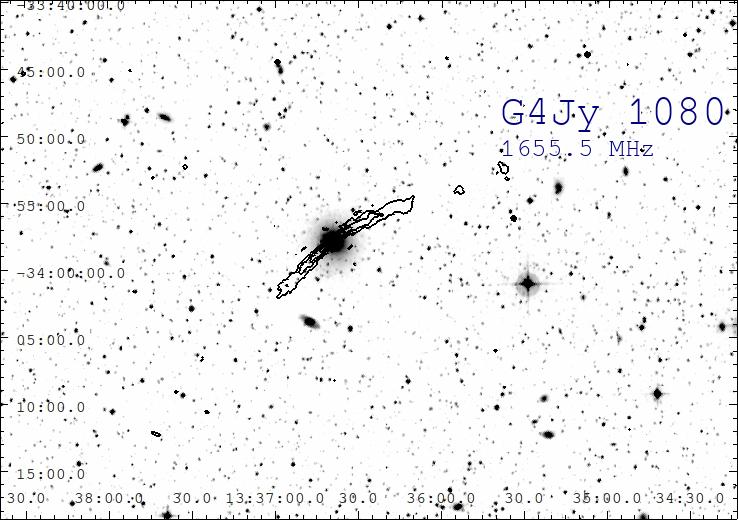}
    \includegraphics[scale=0.225]{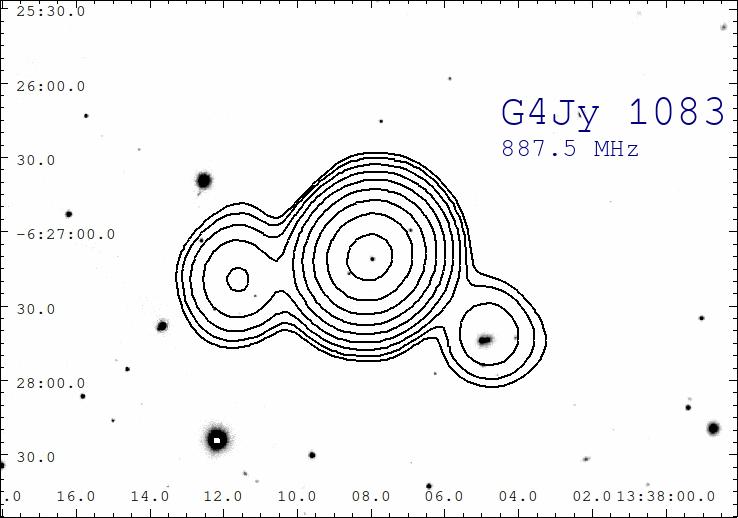}
    \includegraphics[scale=0.225]{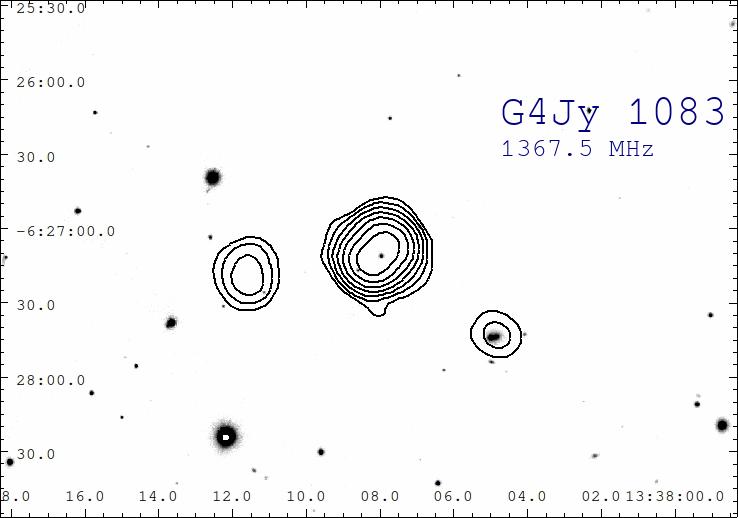}
    \includegraphics[scale=0.225]{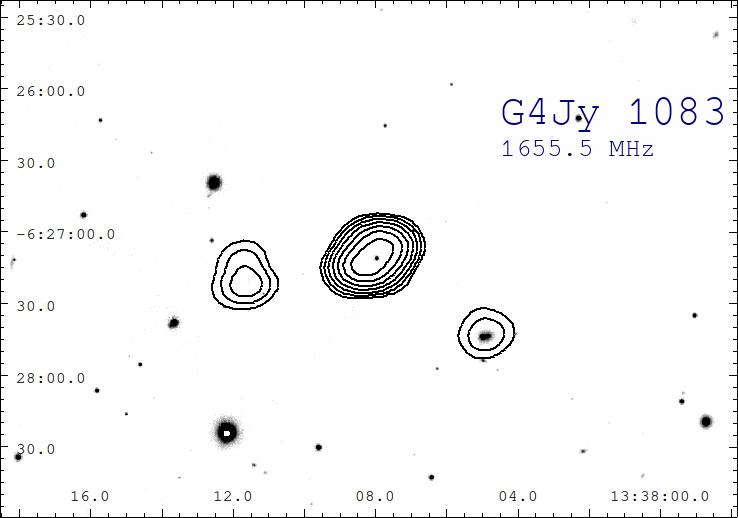}
    \includegraphics[scale=0.225]{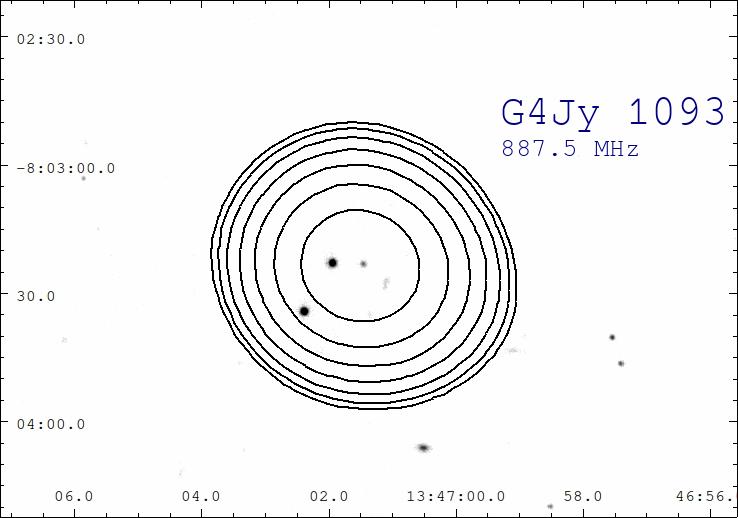}
    \includegraphics[scale=0.225]{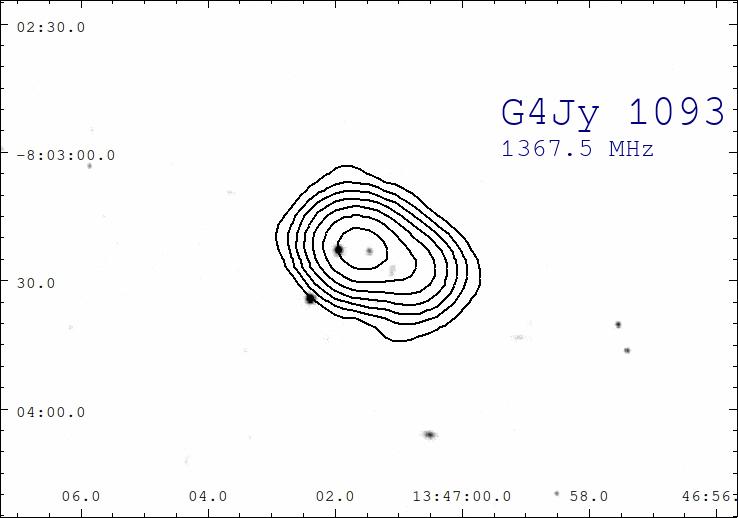}
    \includegraphics[scale=0.225]{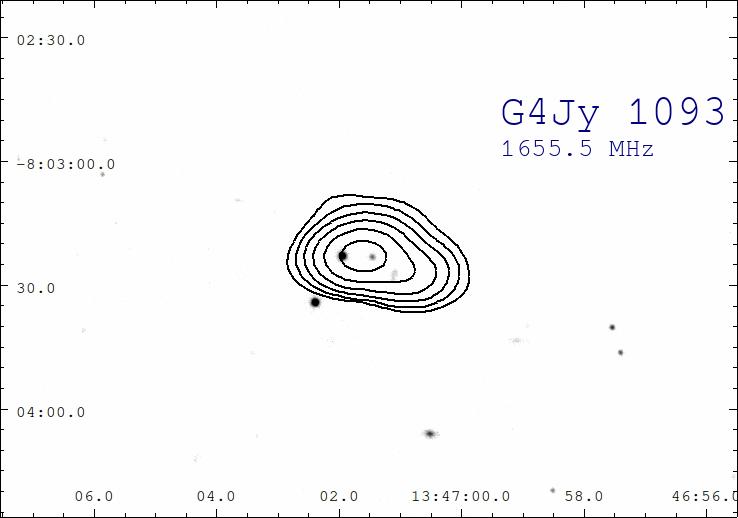}
    \includegraphics[scale=0.225]{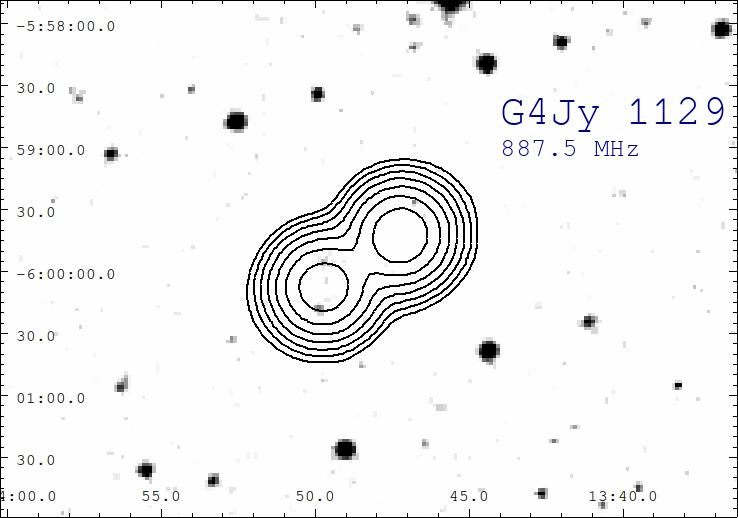}
    \includegraphics[scale=0.225]{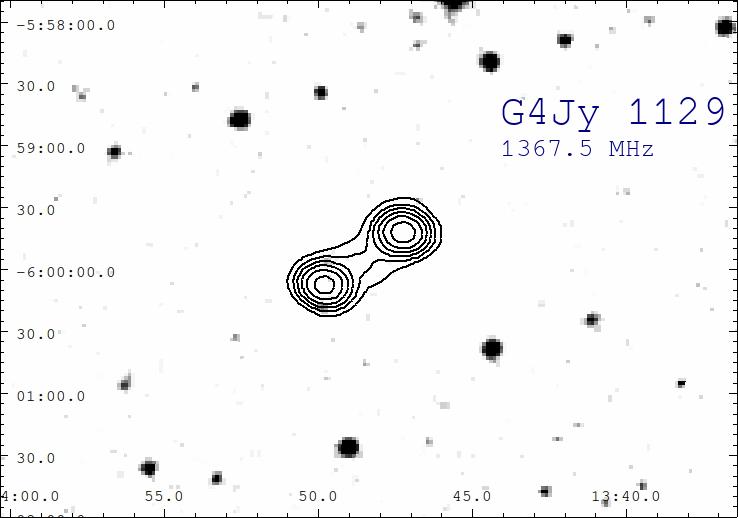}
    \includegraphics[scale=0.225]{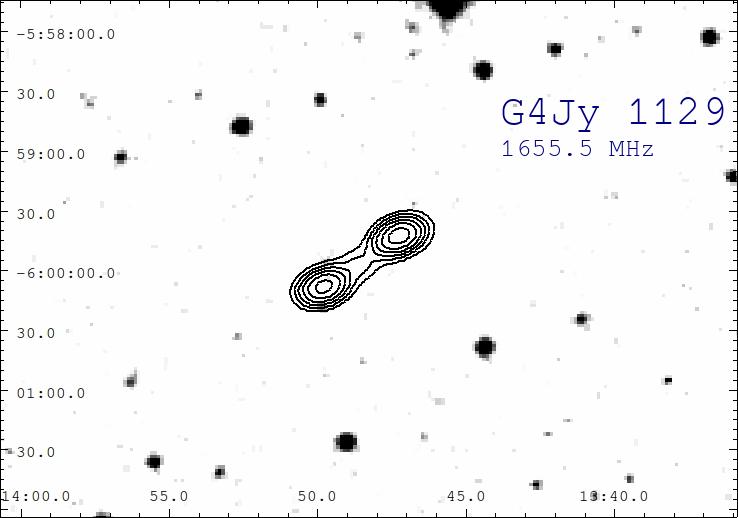}
    \includegraphics[scale=0.225]{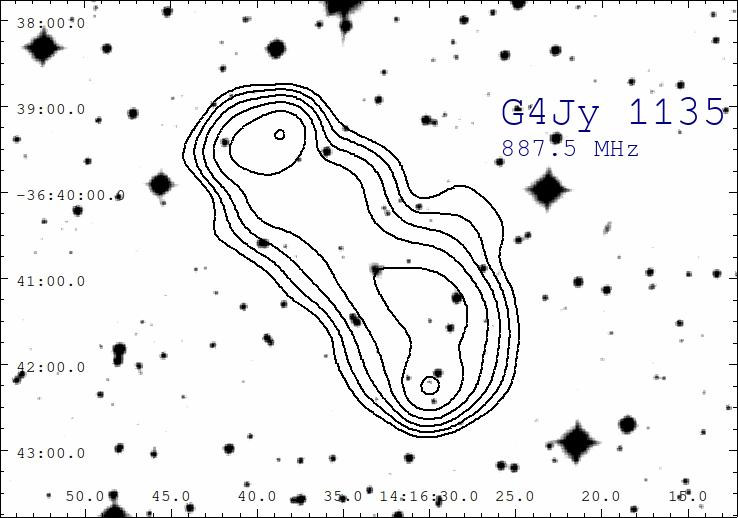}
    \includegraphics[scale=0.225]{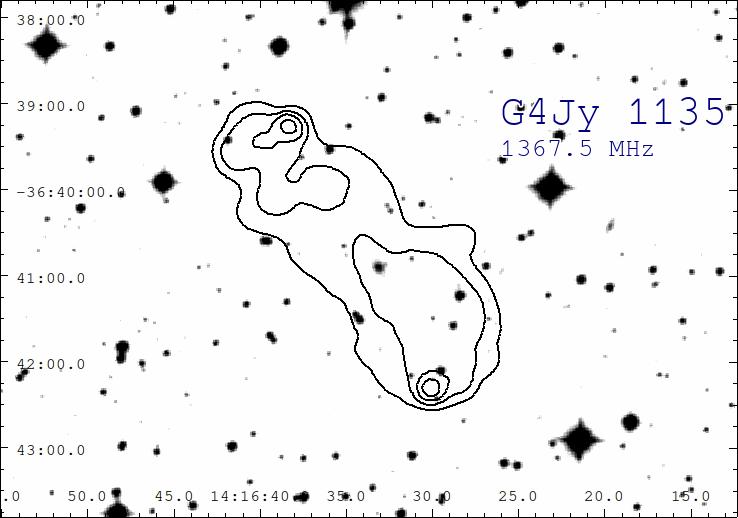}
    \includegraphics[scale=0.225]{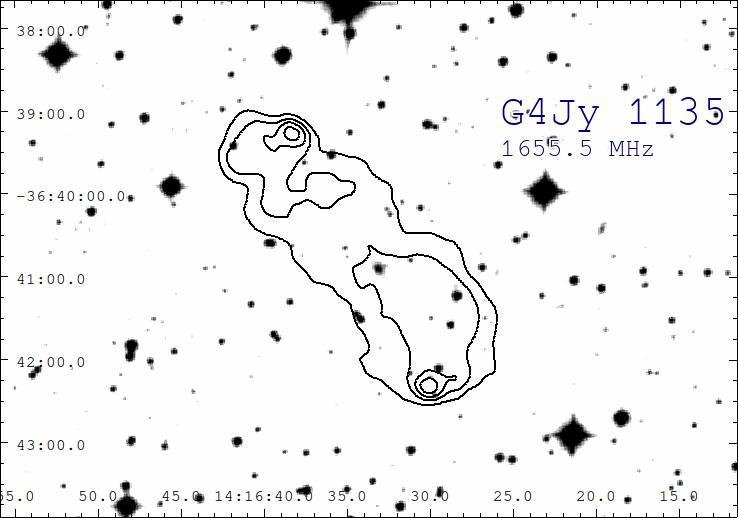}

    \caption{}
    \label{AC}
\end{figure*}
\clearpage
 
\begin{figure*}
    \centering
    \includegraphics[scale=0.225]{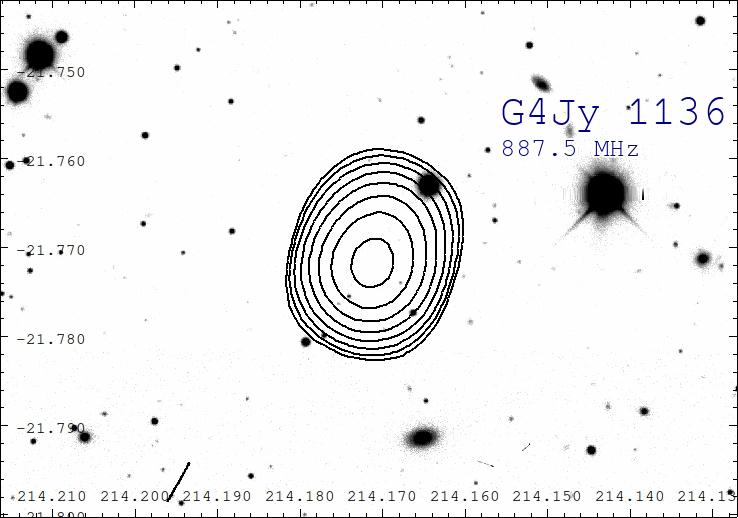}
    \includegraphics[scale=0.225]{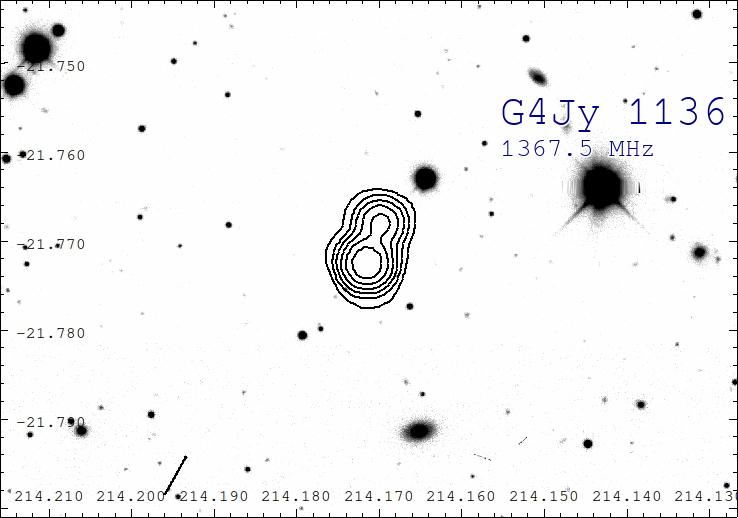}
    \includegraphics[scale=0.225]{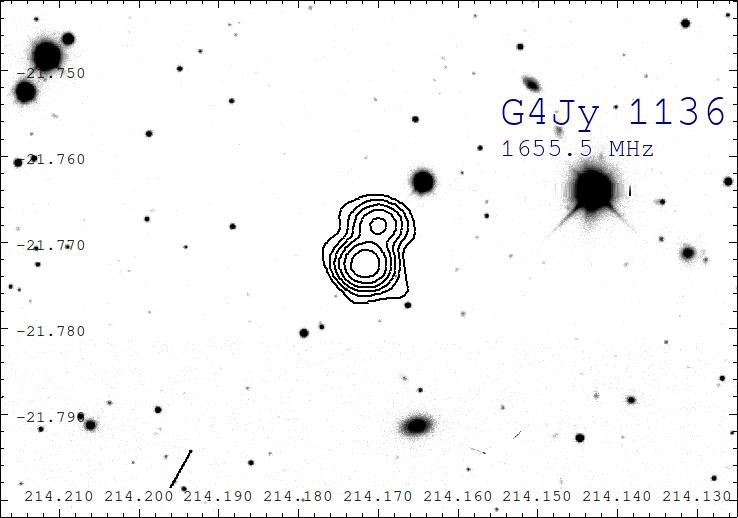}
    \includegraphics[scale=0.225]{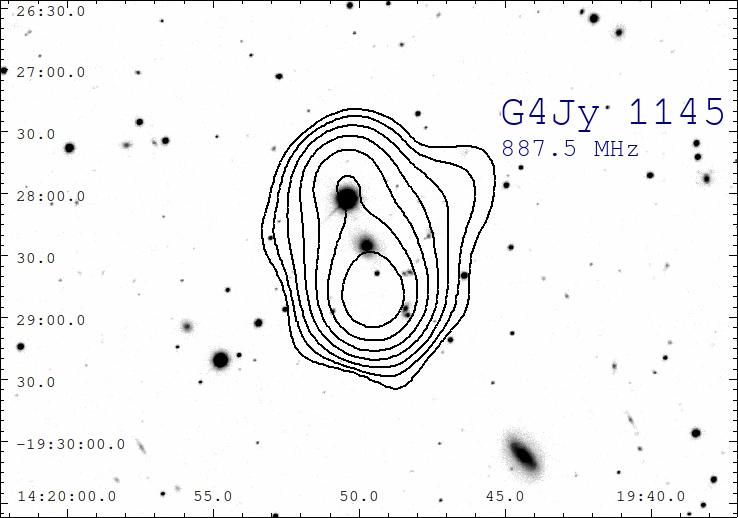}
    \includegraphics[scale=0.225]{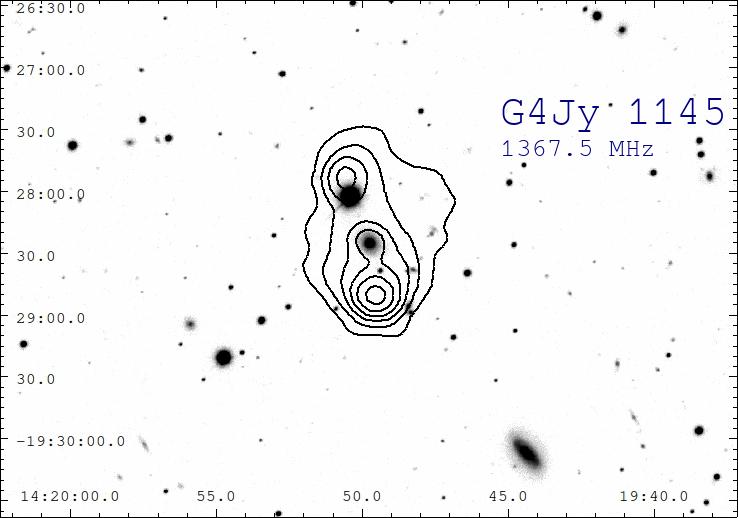}
    \includegraphics[scale=0.225]{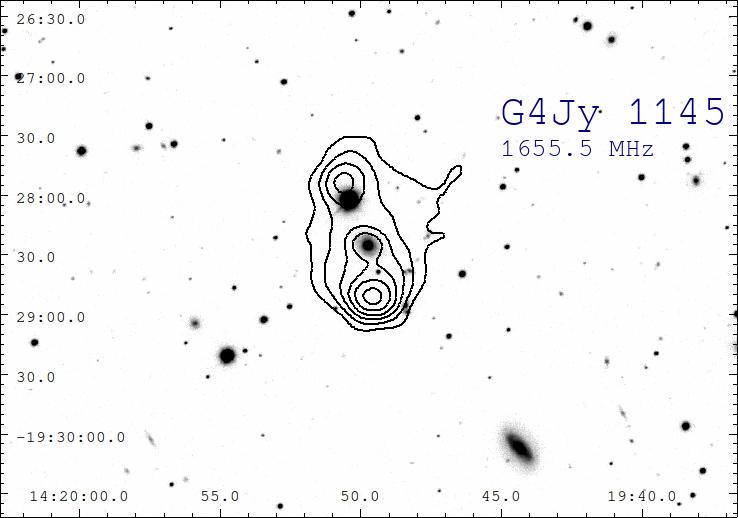}
    \includegraphics[scale=0.225]{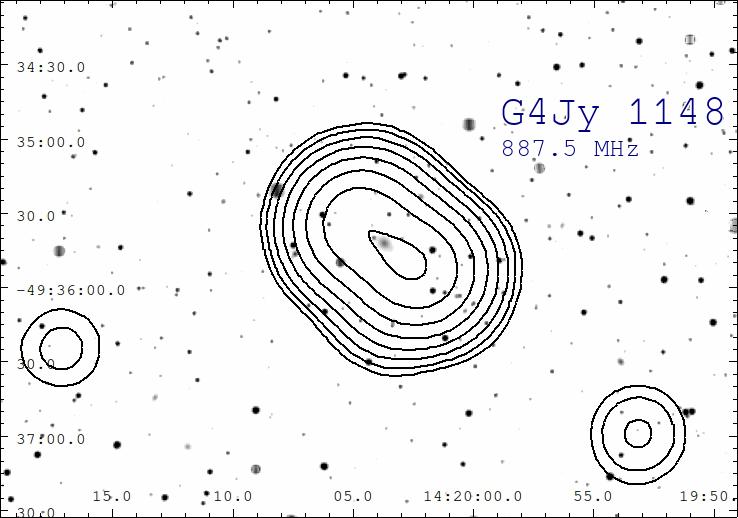}
    \includegraphics[scale=0.225]{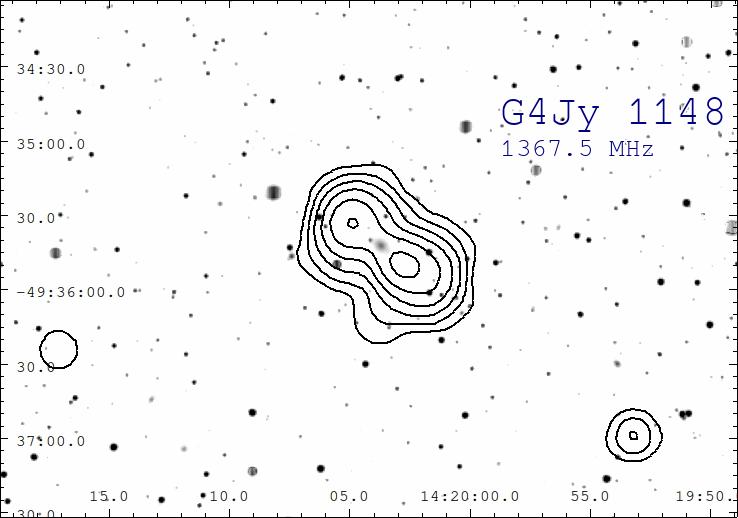}
    \includegraphics[scale=0.225]{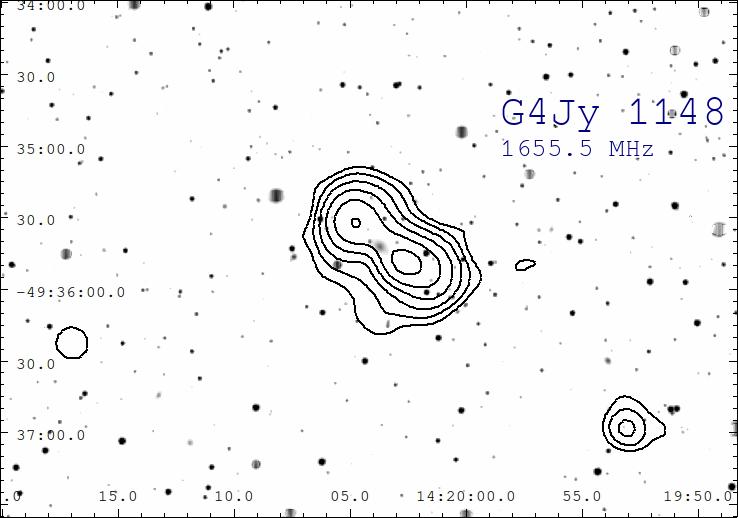}
    \includegraphics[scale=0.225]{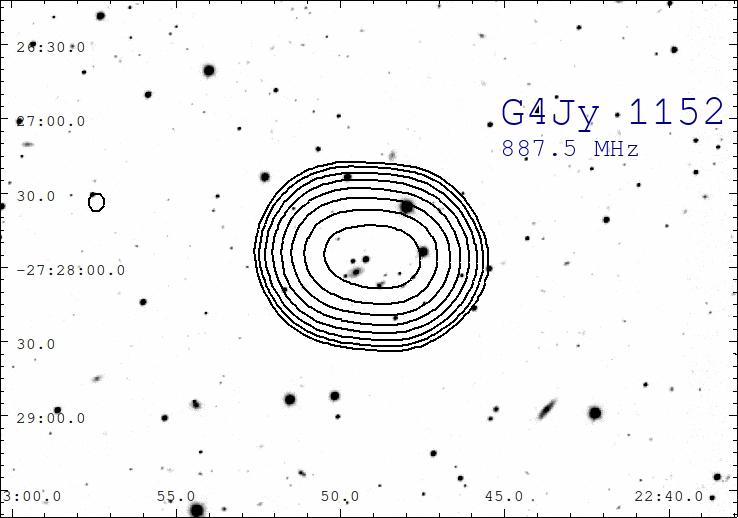}
    \includegraphics[scale=0.225]{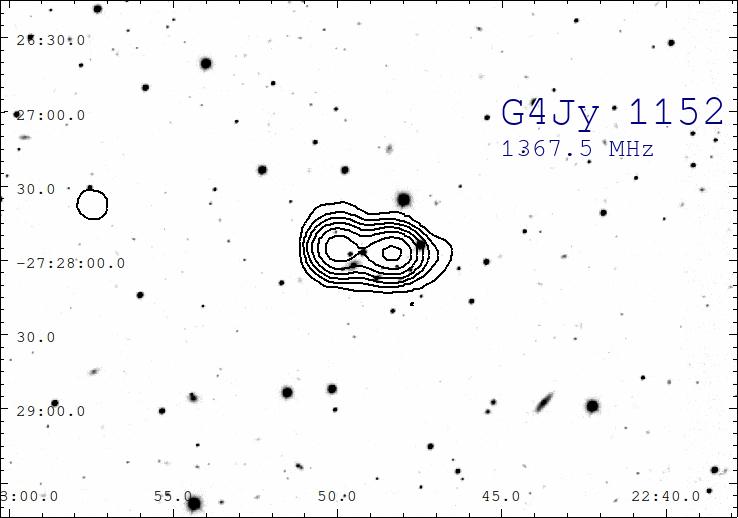}
    \includegraphics[scale=0.225]{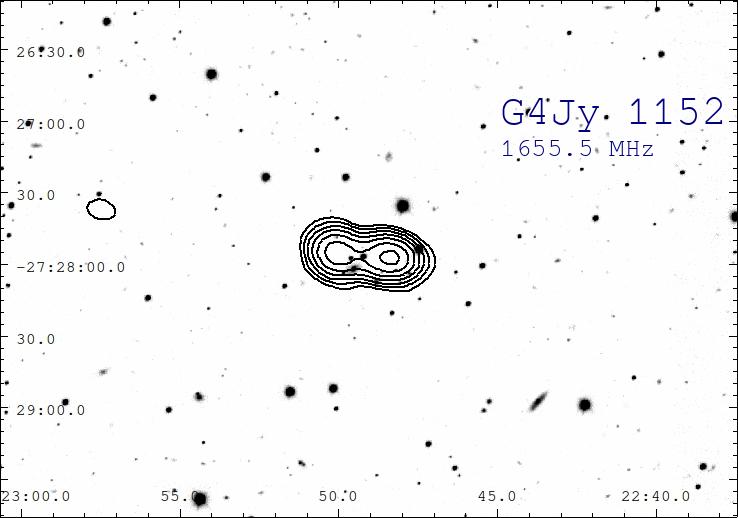}
    \includegraphics[scale=0.225]{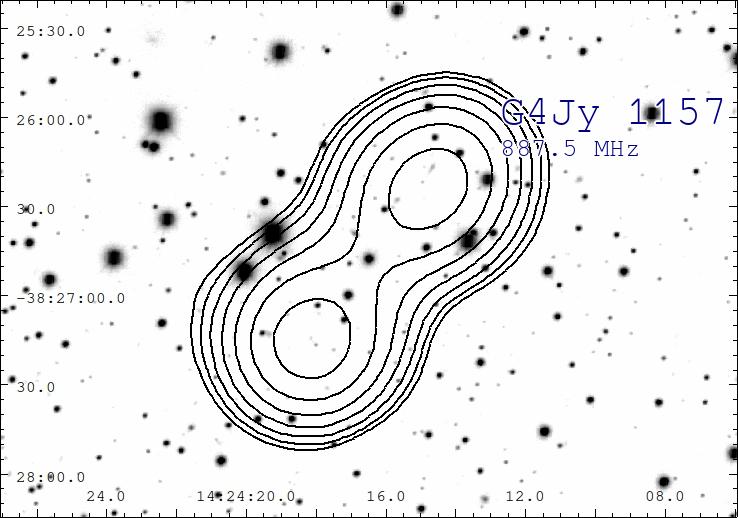}
    \includegraphics[scale=0.225]{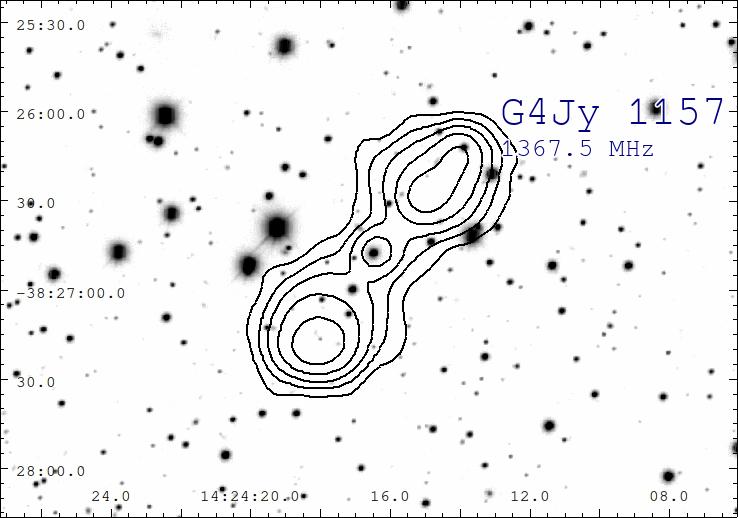}
    \includegraphics[scale=0.225]{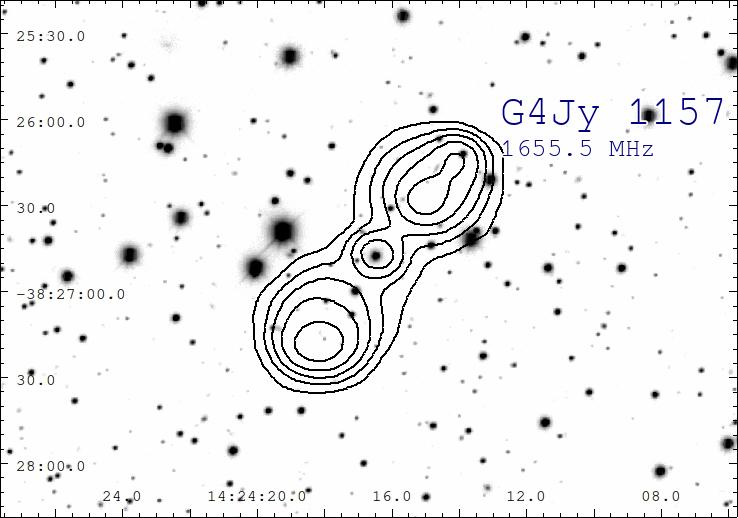}

    \caption{}
    \label{AD}
\end{figure*}
\clearpage
 
\begin{figure*}
    \centering
    \includegraphics[scale=0.225]{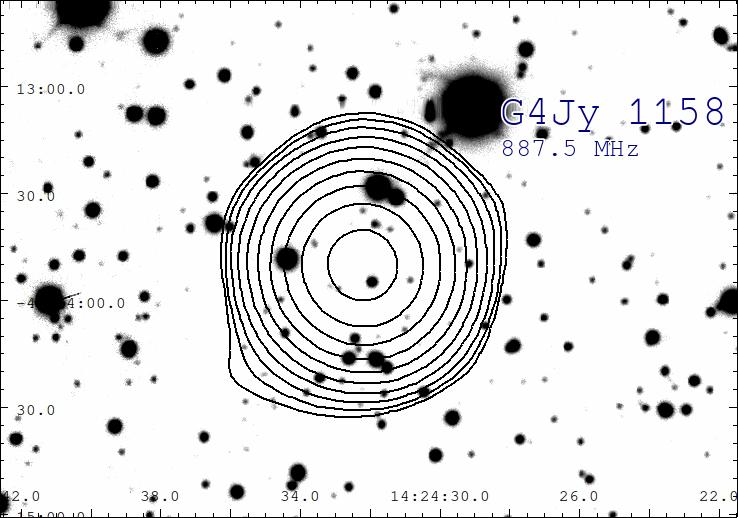}
    \includegraphics[scale=0.225]{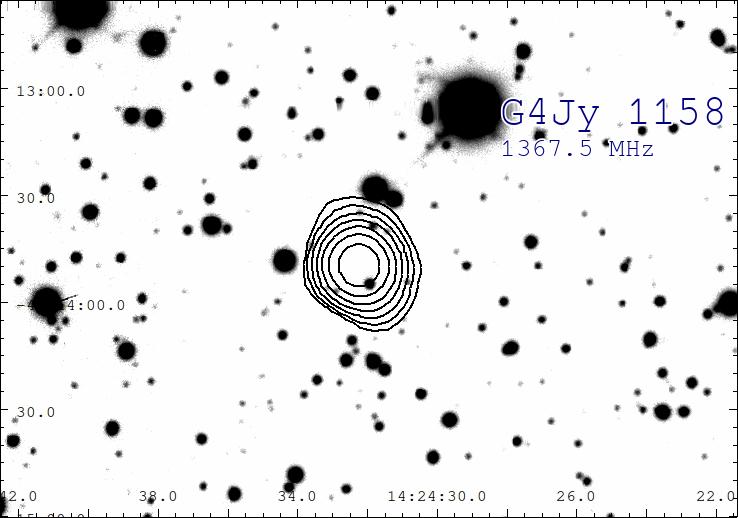}
    \includegraphics[scale=0.225]{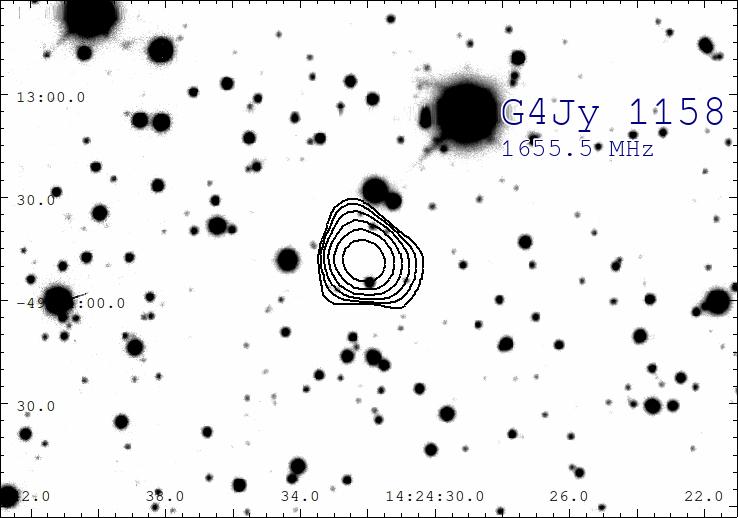}
    \includegraphics[scale=0.225]{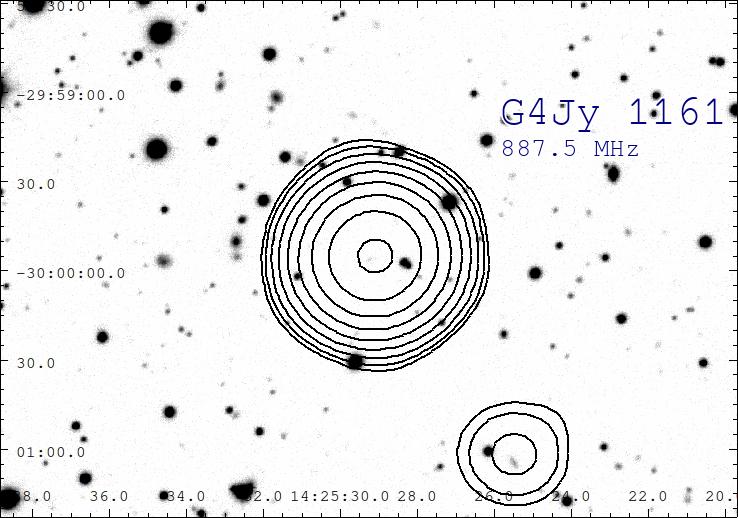}
    \includegraphics[scale=0.225]{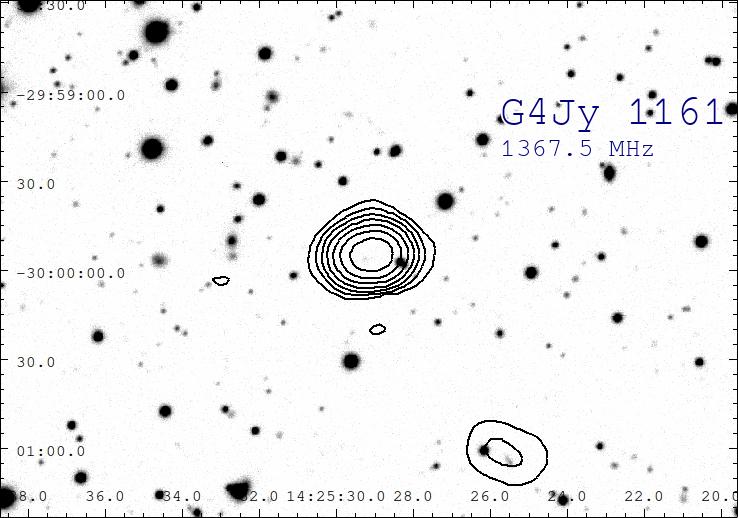}
    \includegraphics[scale=0.225]{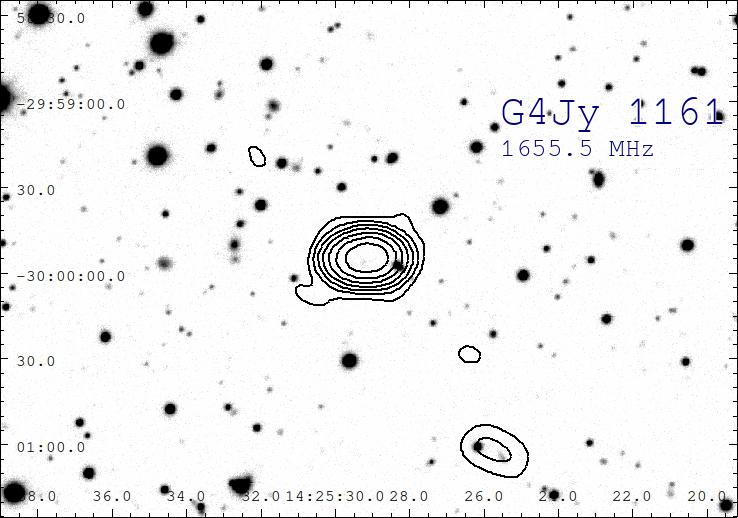}
    \includegraphics[scale=0.225]{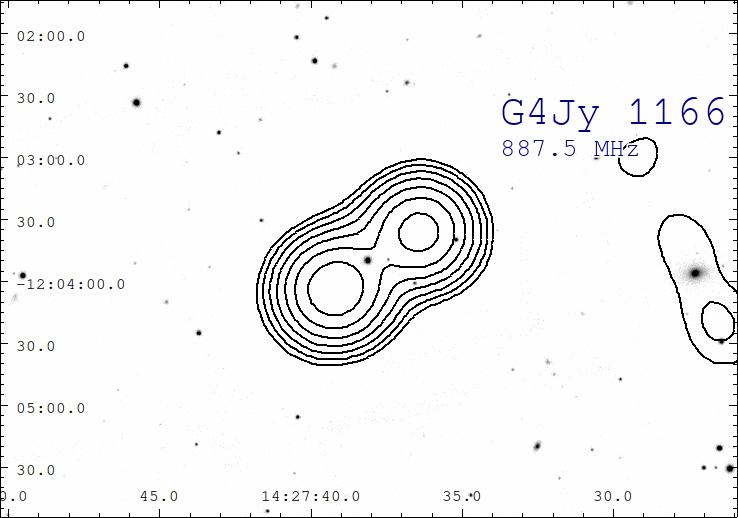}
    \includegraphics[scale=0.225]{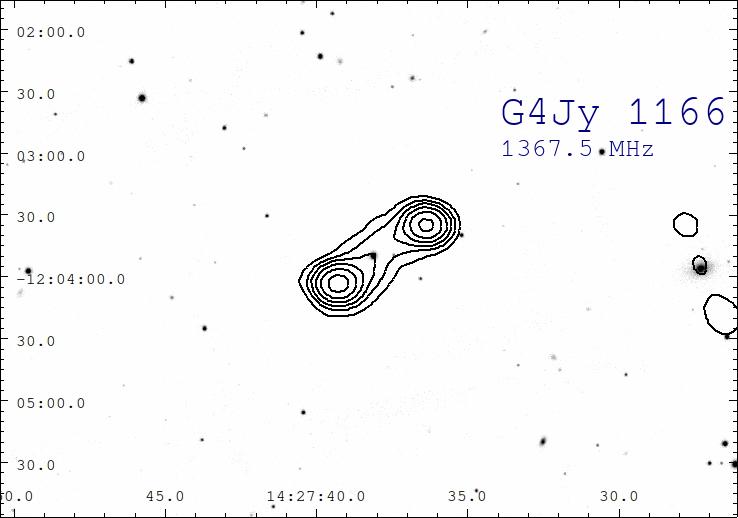}
    \includegraphics[scale=0.225]{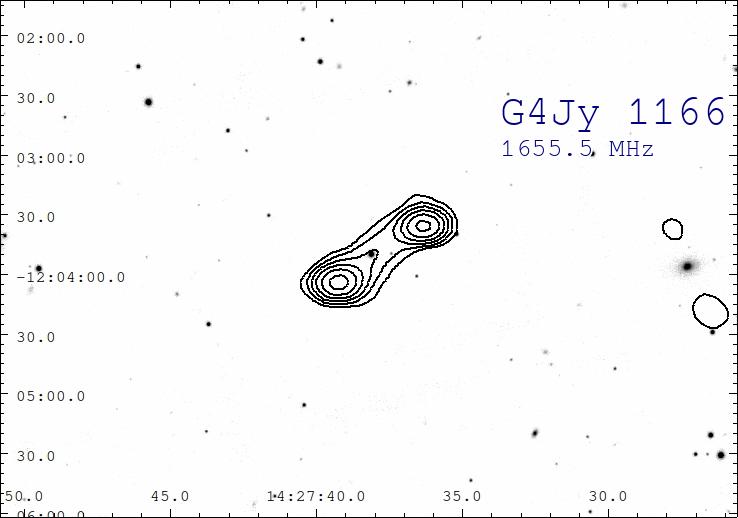}
    \includegraphics[scale=0.225]{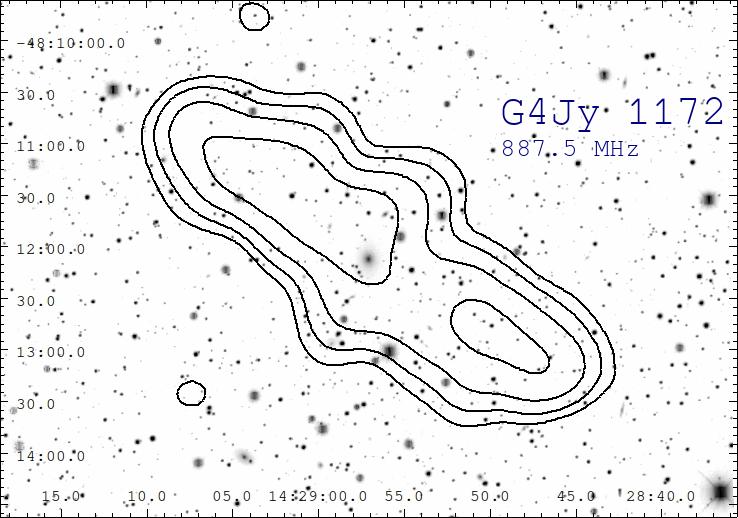}
    \includegraphics[scale=0.225]{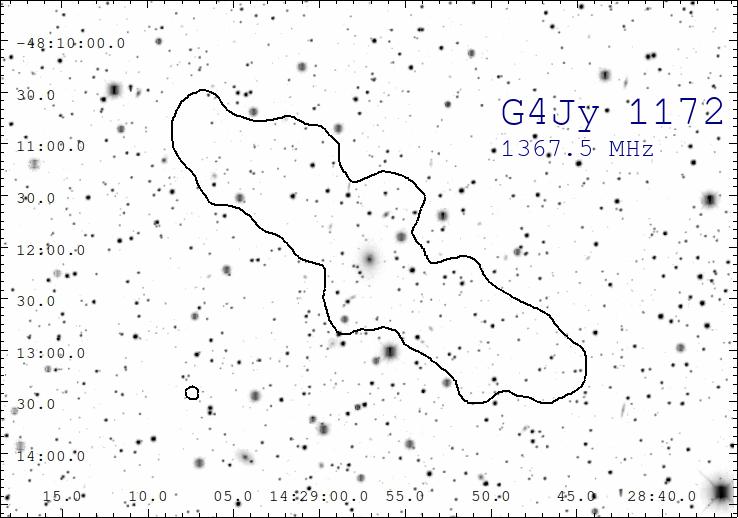}
    \includegraphics[scale=0.225]{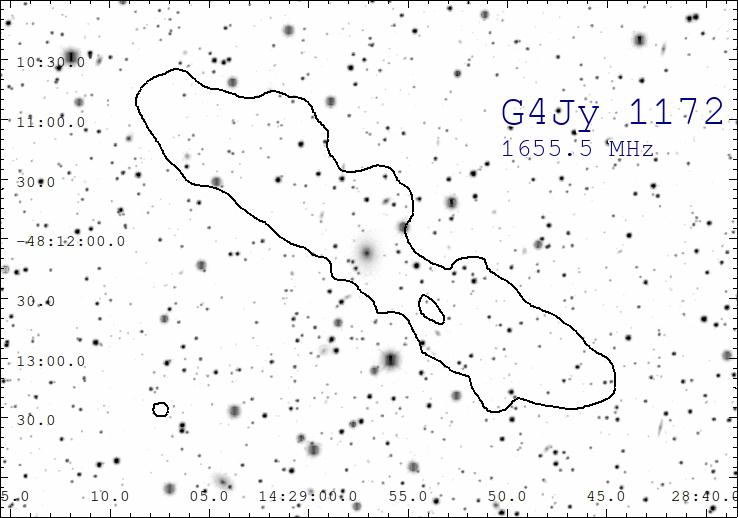}
    \includegraphics[scale=0.225]{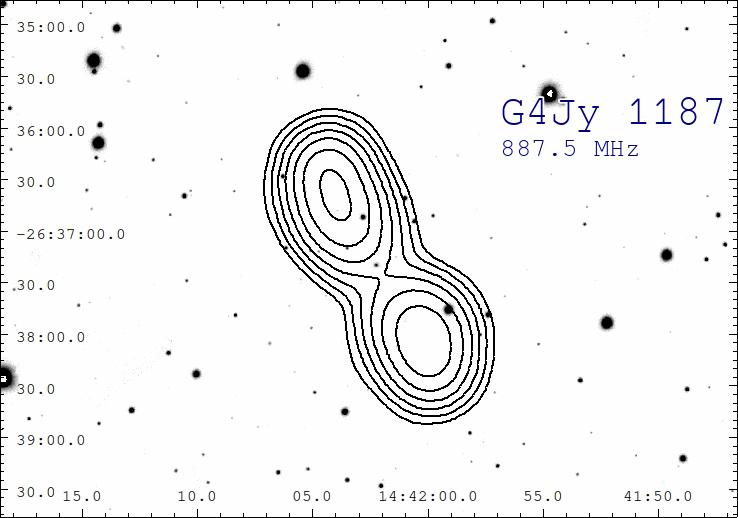}
    \includegraphics[scale=0.225]{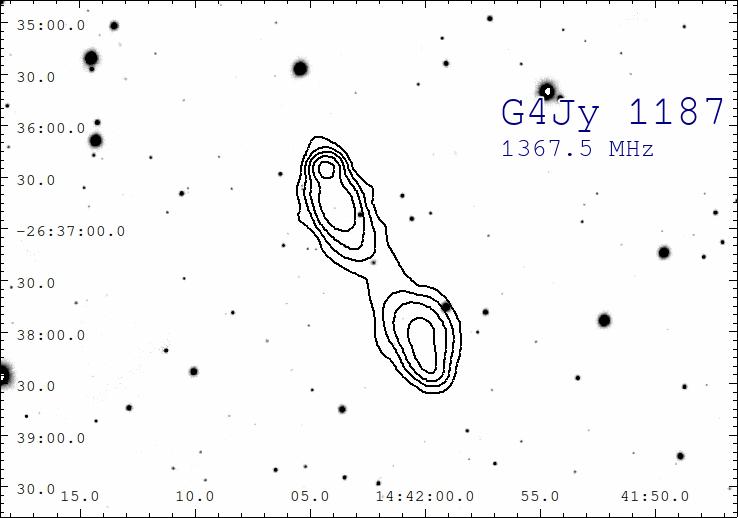}
    \includegraphics[scale=0.225]{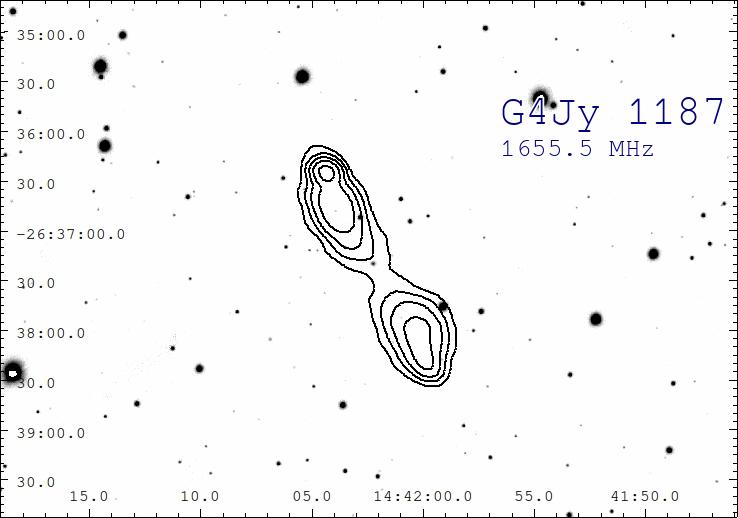}

    \caption{}
    \label{AE}
\end{figure*}
\clearpage
 
\begin{figure*}
    \centering
    \includegraphics[scale=0.225]{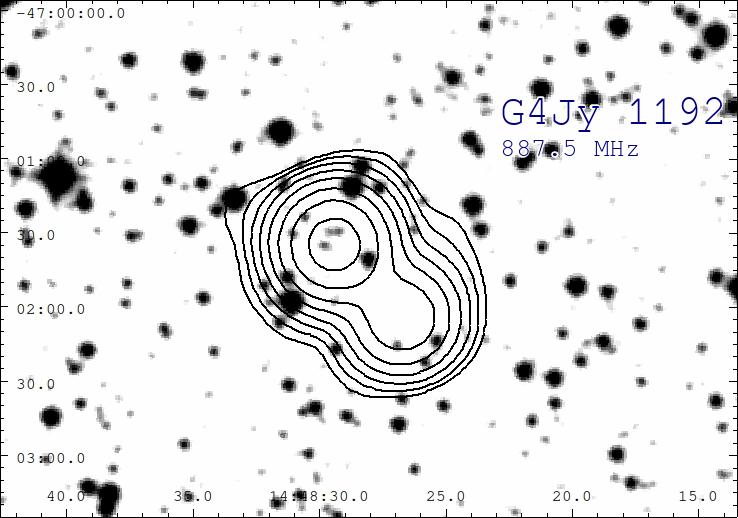}
    \includegraphics[scale=0.225]{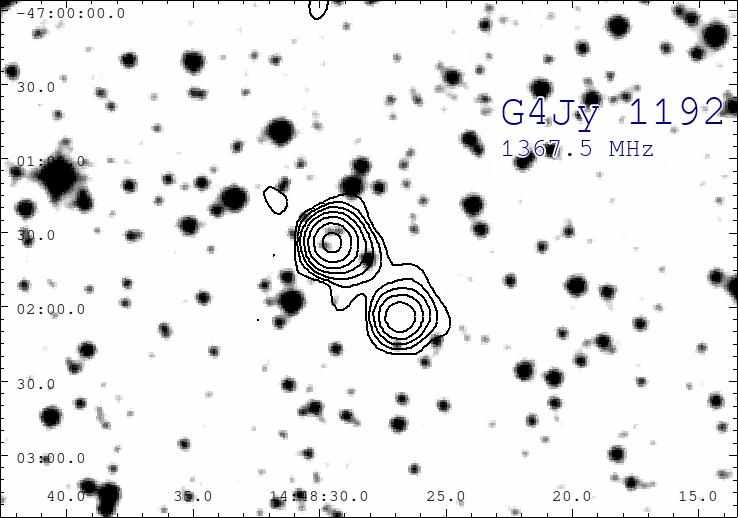}
    \includegraphics[scale=0.225]{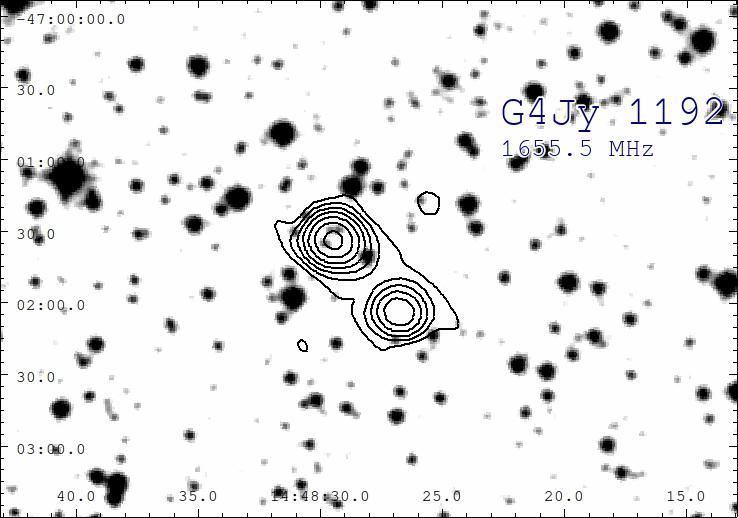}
    \includegraphics[scale=0.225]{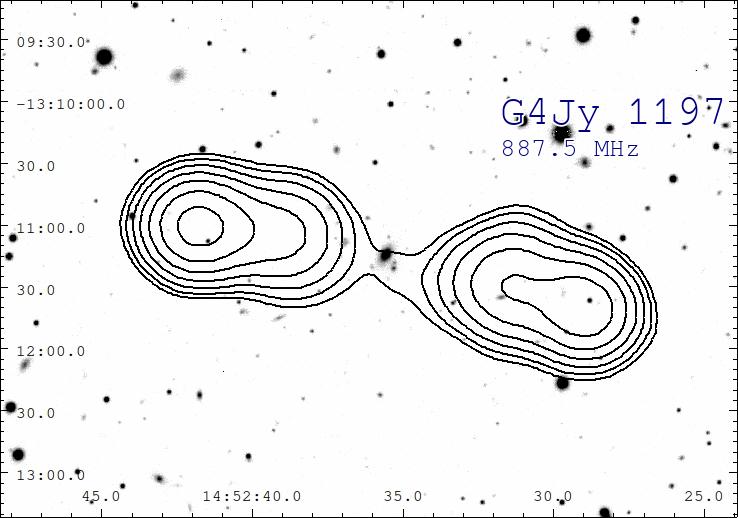}
    \includegraphics[scale=0.225]{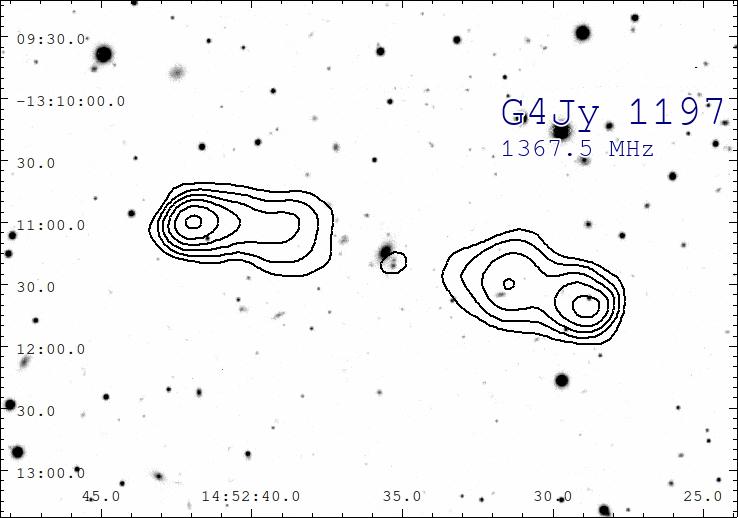}
    \includegraphics[scale=0.225]{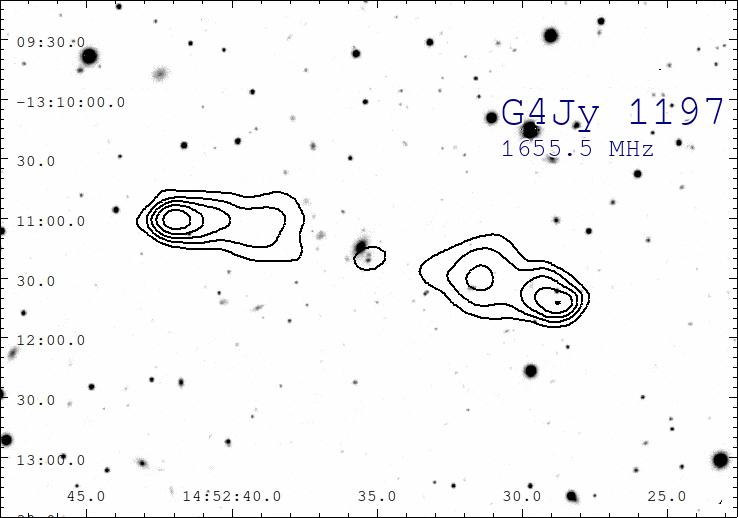}
    \includegraphics[scale=0.225]{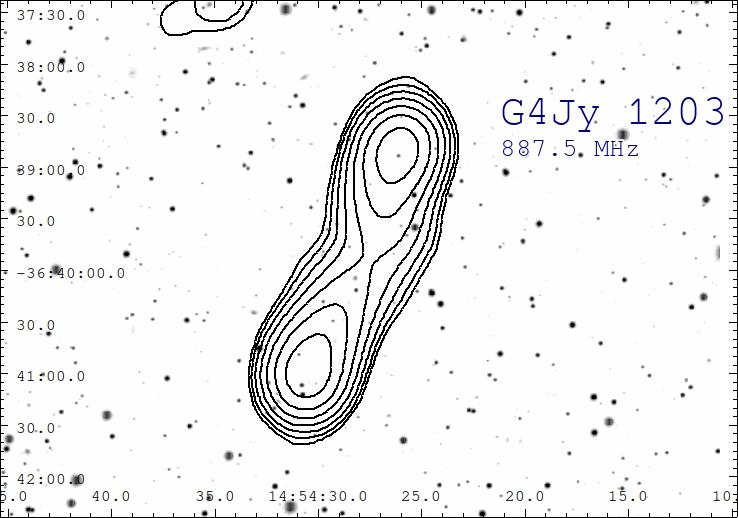}
    \includegraphics[scale=0.225]{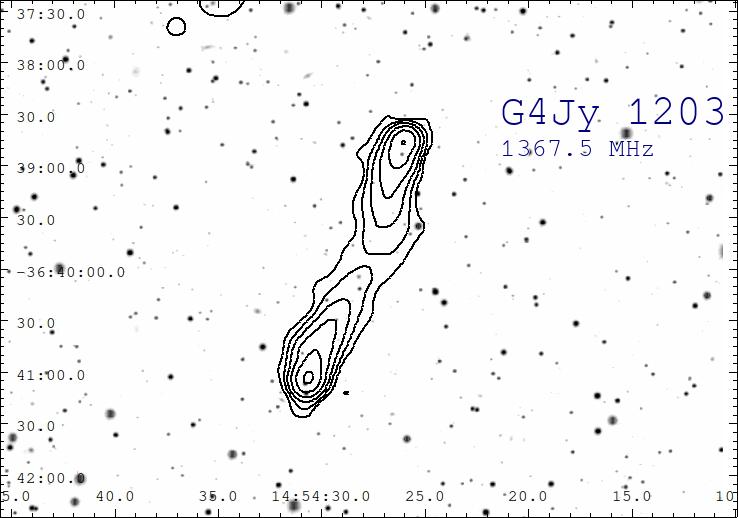}
    \includegraphics[scale=0.225]{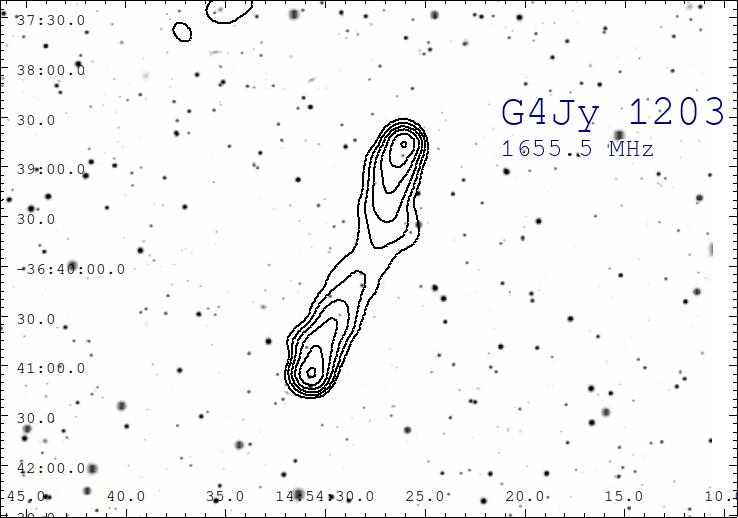}
    \includegraphics[scale=0.225]{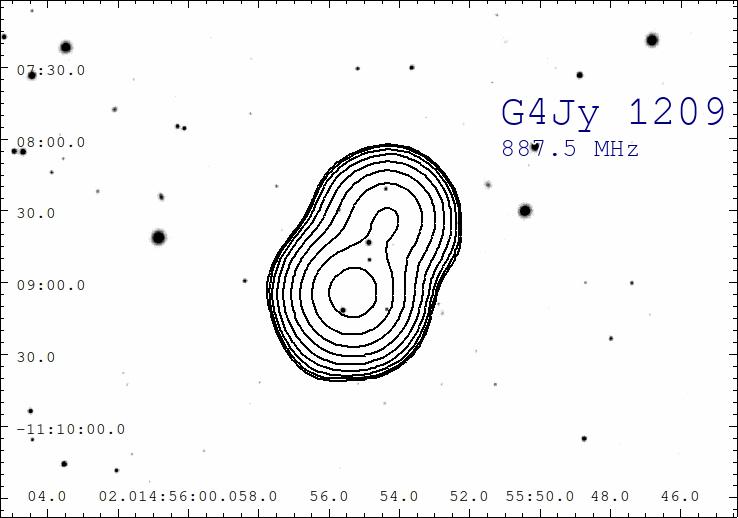}
    \includegraphics[scale=0.225]{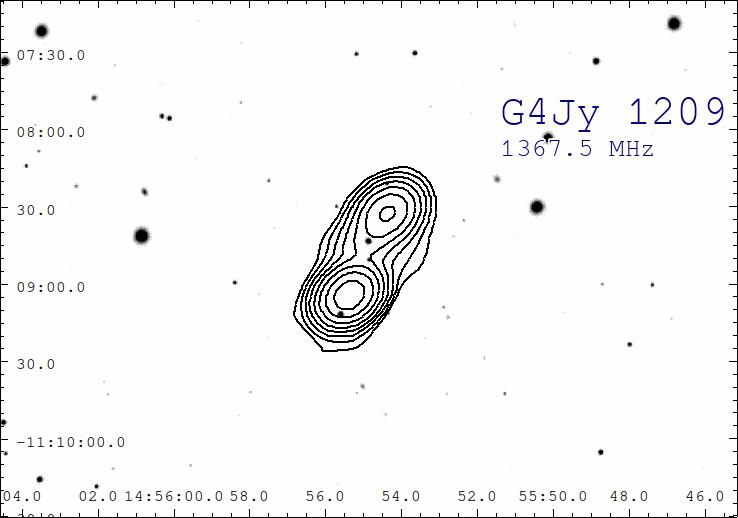}
    \includegraphics[scale=0.225]{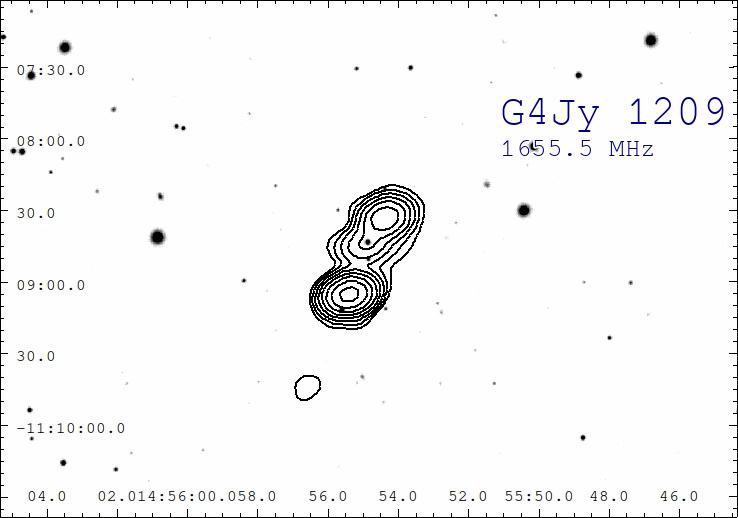}
    \includegraphics[scale=0.225]{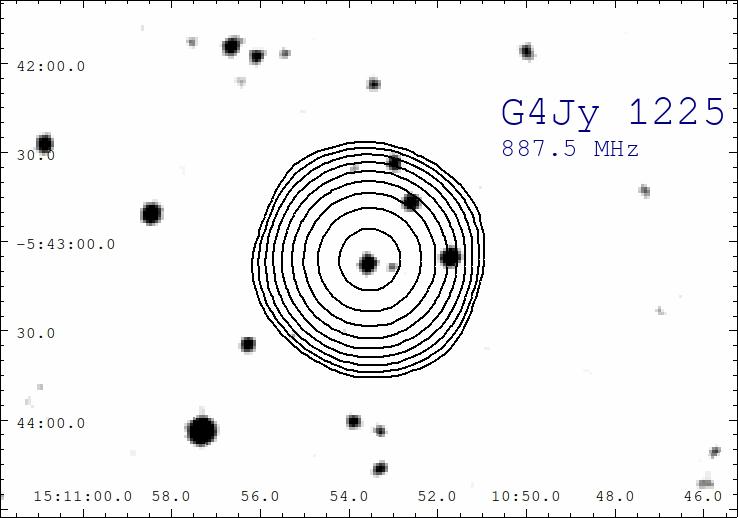}
    \includegraphics[scale=0.225]{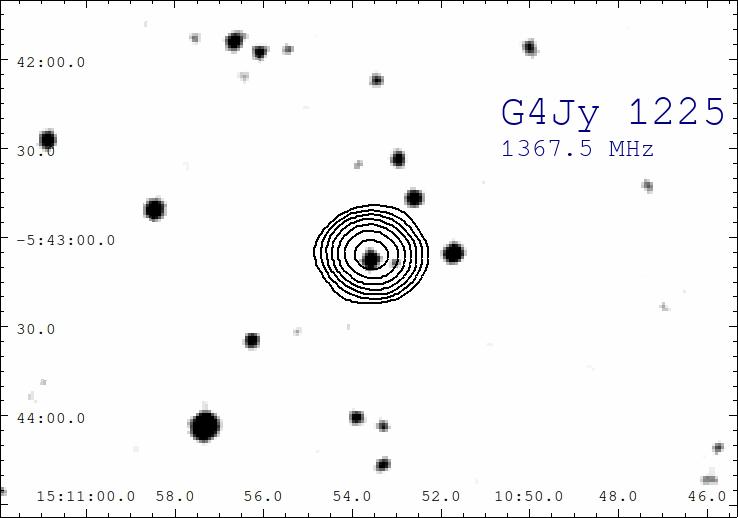}
    \includegraphics[scale=0.225]{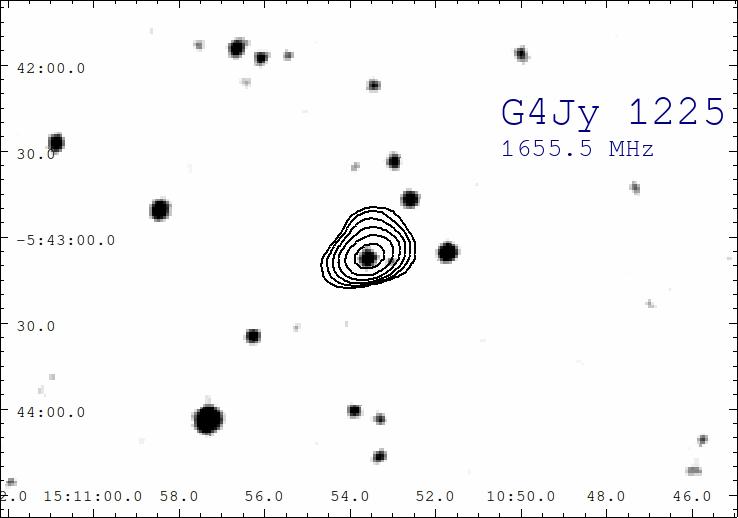}

    \caption{}
    \label{AF}
\end{figure*}
\clearpage
 
\begin{figure*}
    \centering
    \includegraphics[scale=0.225]{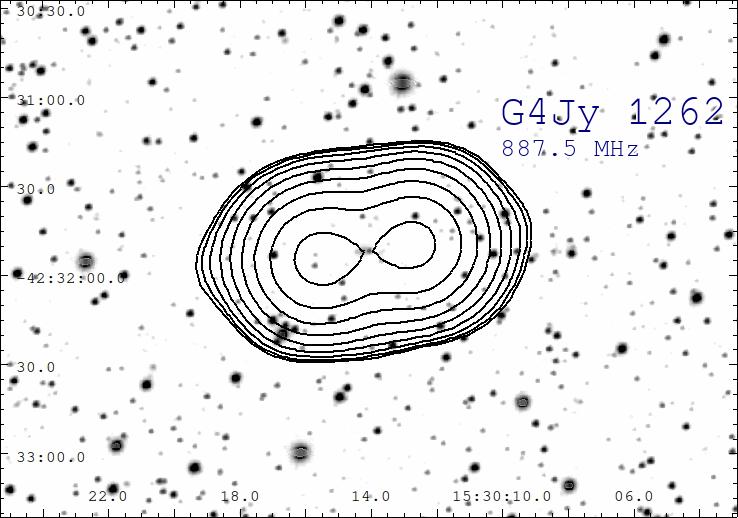}
    \includegraphics[scale=0.225]{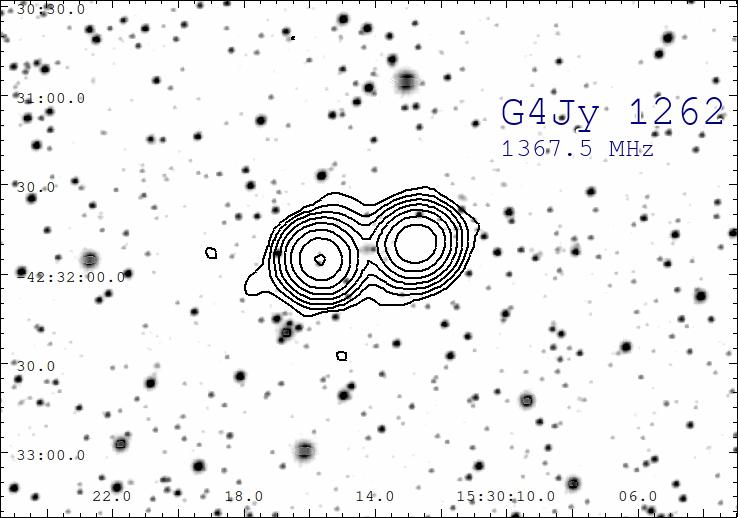}
    \includegraphics[scale=0.225]{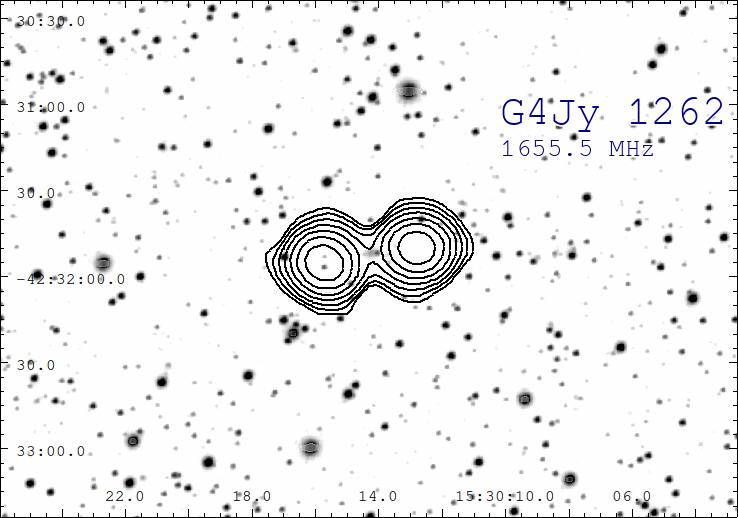}
    \includegraphics[scale=0.225]{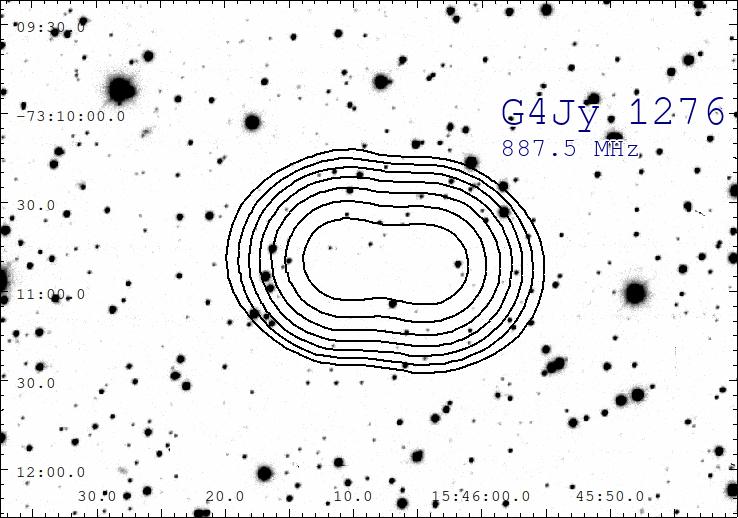}
    \includegraphics[scale=0.225]{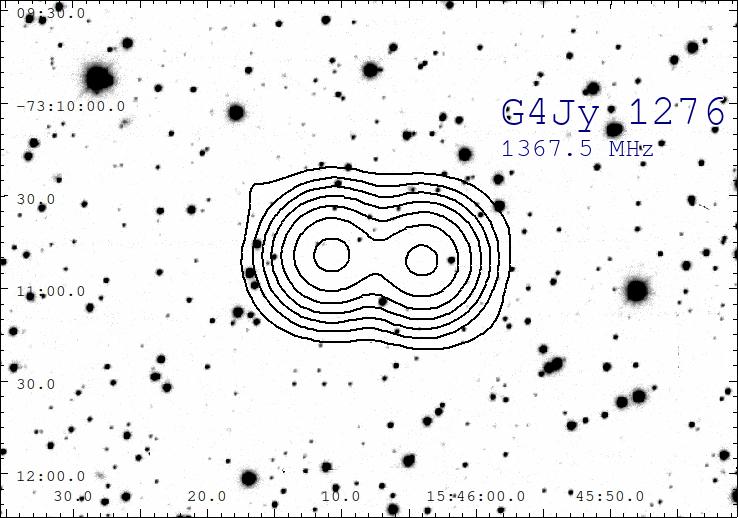}
    \includegraphics[scale=0.225]{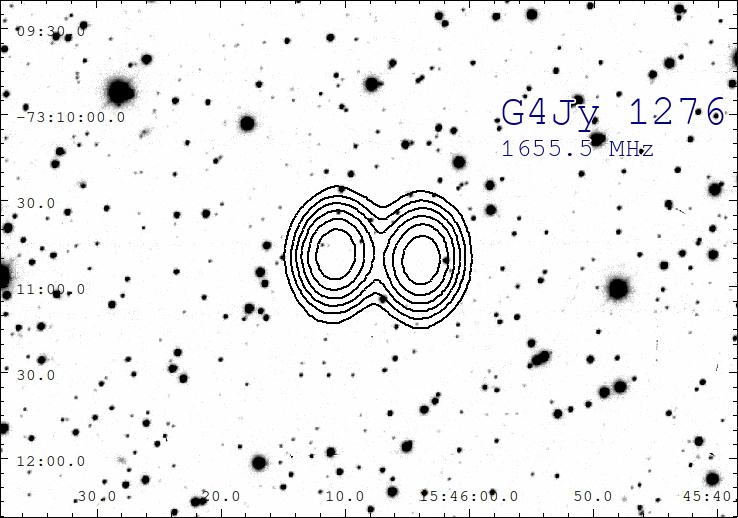}
    \includegraphics[scale=0.225]{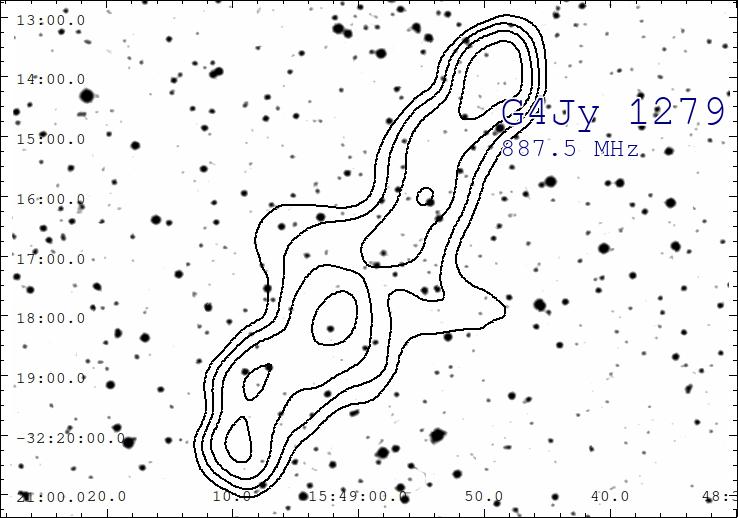}
    \includegraphics[scale=0.225]{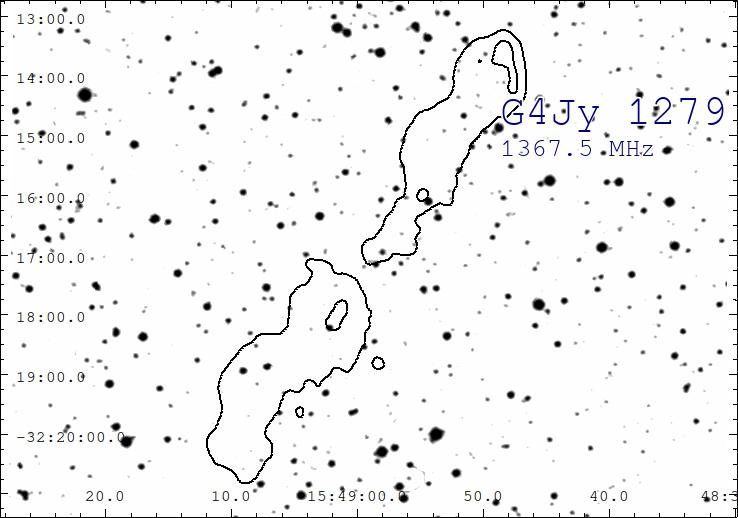}
    \includegraphics[scale=0.225]{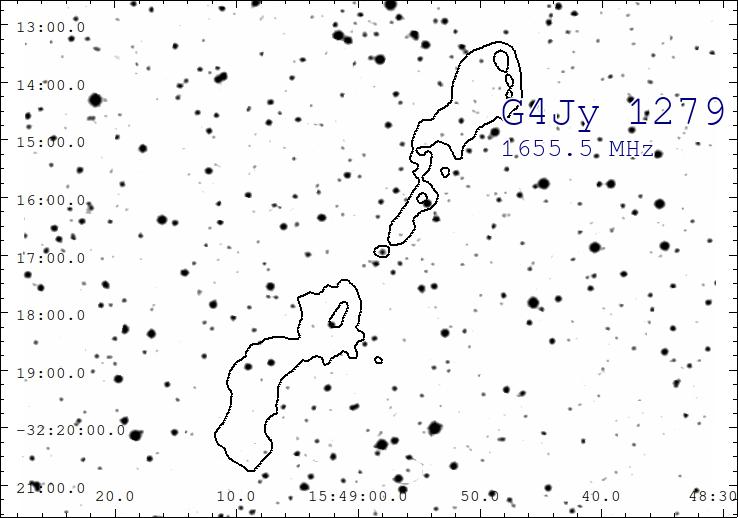}
    \includegraphics[scale=0.225]{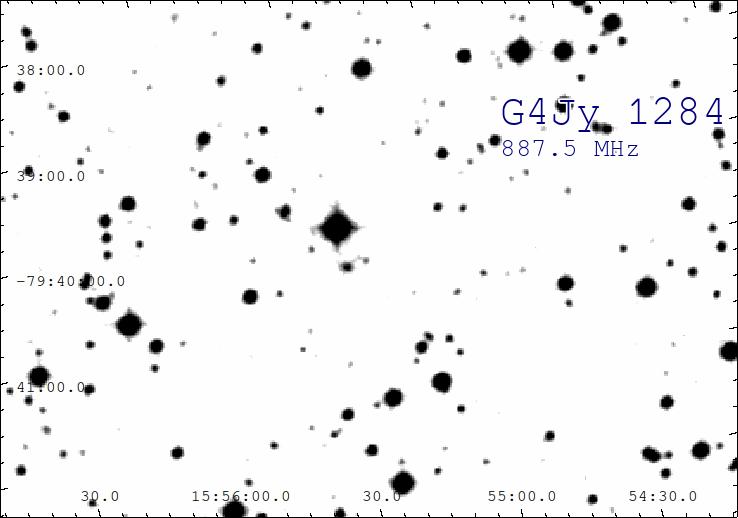}
    \includegraphics[scale=0.225]{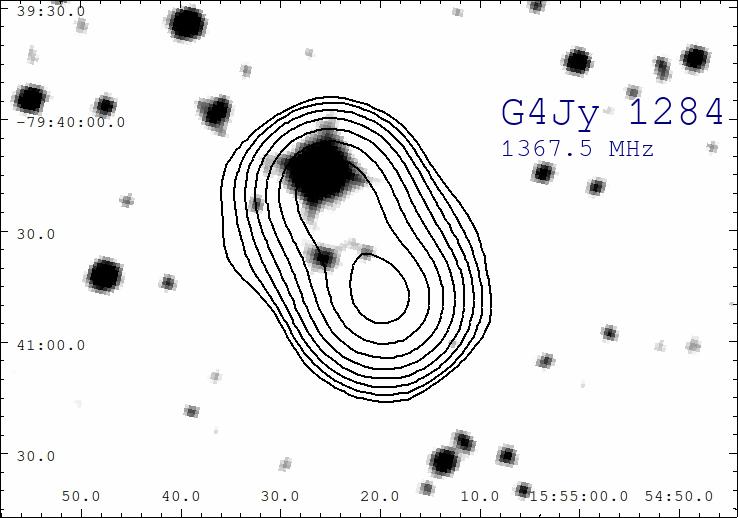}
    \includegraphics[scale=0.225]{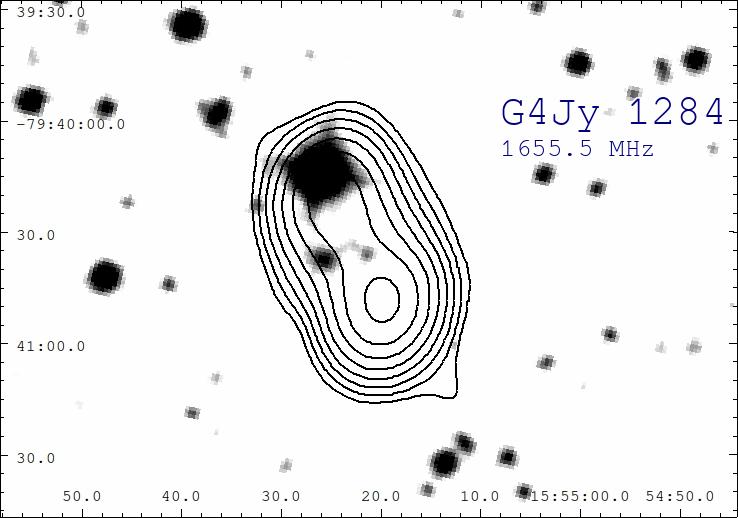}
    \includegraphics[scale=0.225]{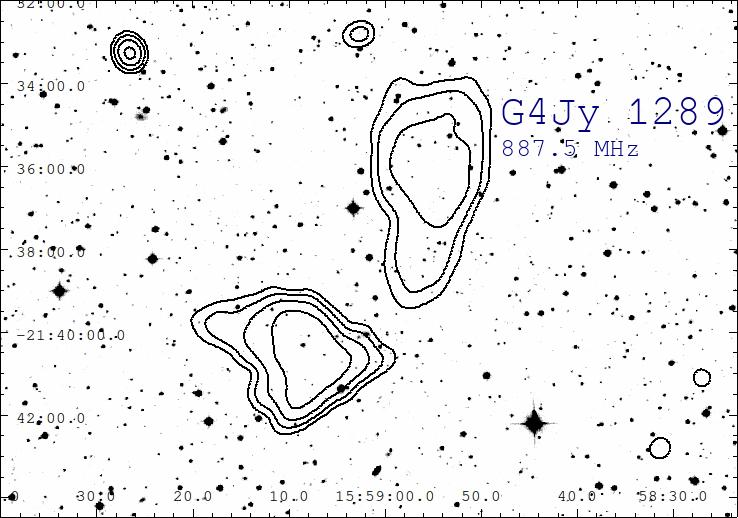}
    \includegraphics[scale=0.225]{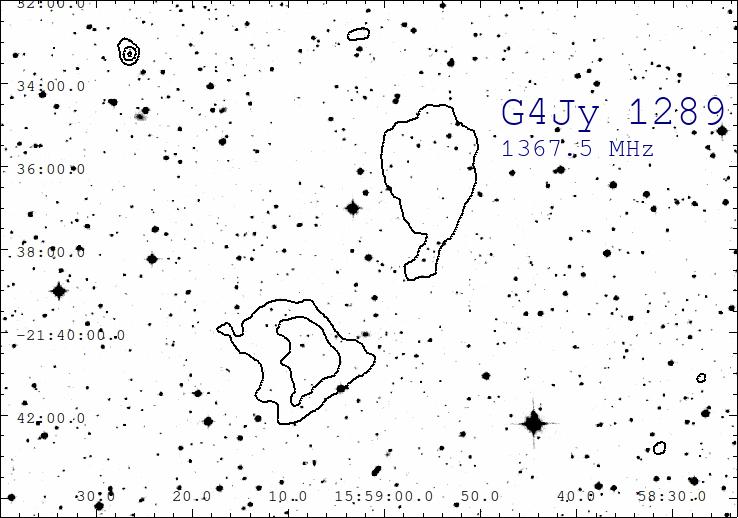}
    \includegraphics[scale=0.225]{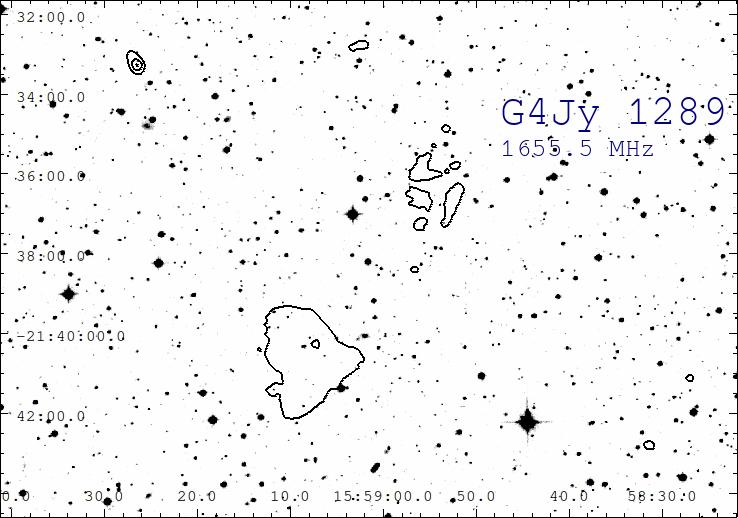}

    \caption{RACS-low data for G4Jy 1284 could not be convolved to 25$\arcsec$, and therefore was not available. It may be replaced with future observations from RACS-low2 or RACS-low3 \textbf{(E.Lenc and A. Hotan, private communication)}.}
    \label{AG}
\end{figure*}
\clearpage
 
\begin{figure*}
    \centering
    \includegraphics[scale=0.225]{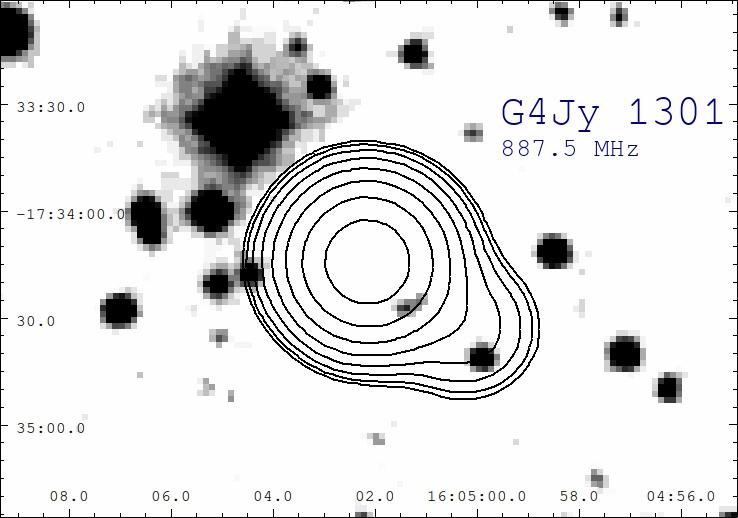}
    \includegraphics[scale=0.225]{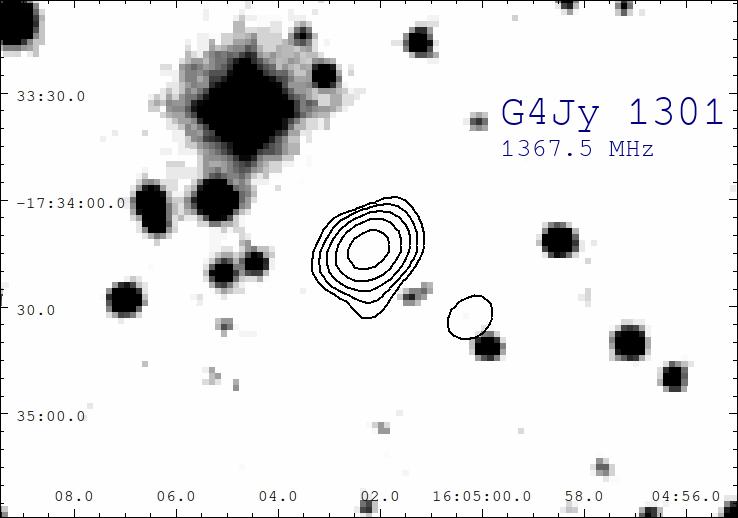}
    \includegraphics[scale=0.225]{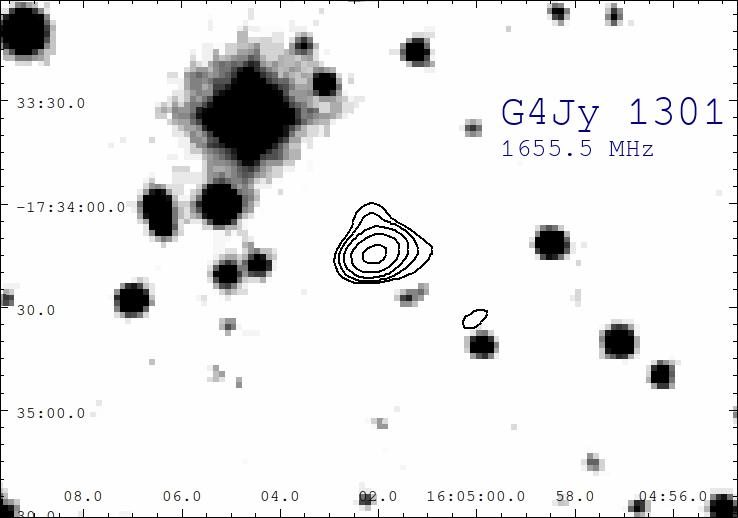}
    \includegraphics[scale=0.225]{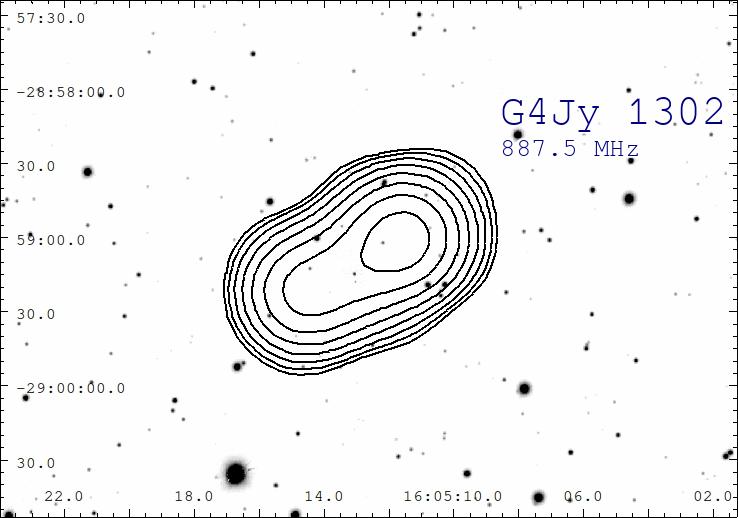}
    \includegraphics[scale=0.225]{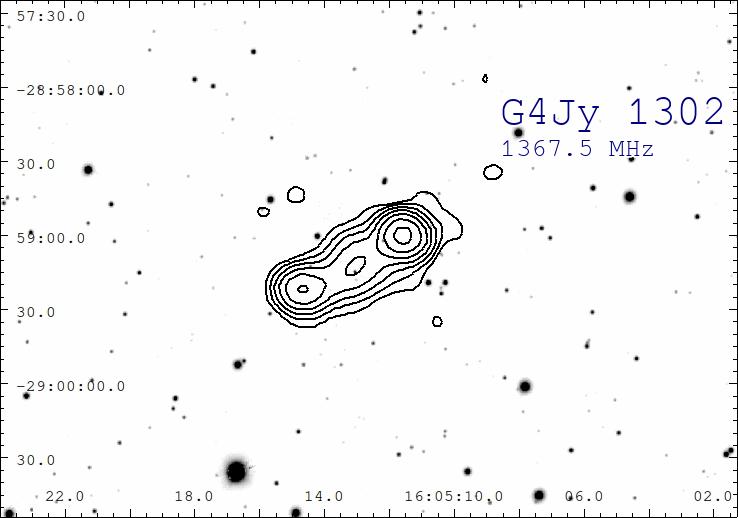}
    \includegraphics[scale=0.225]{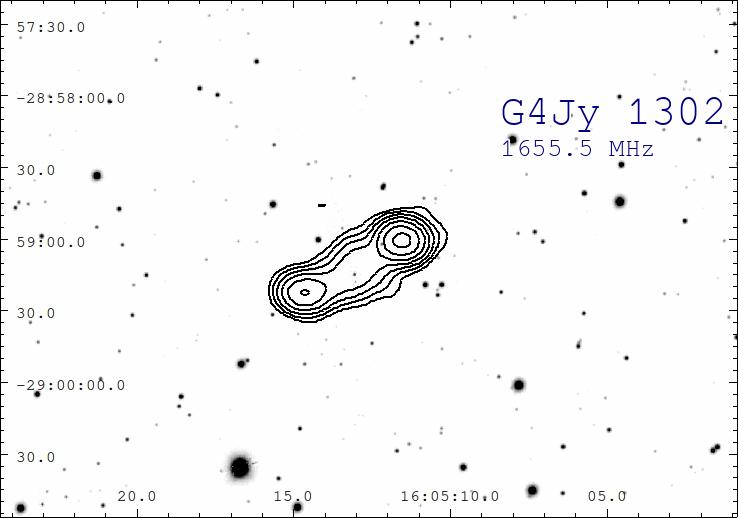}
    \includegraphics[scale=0.225]{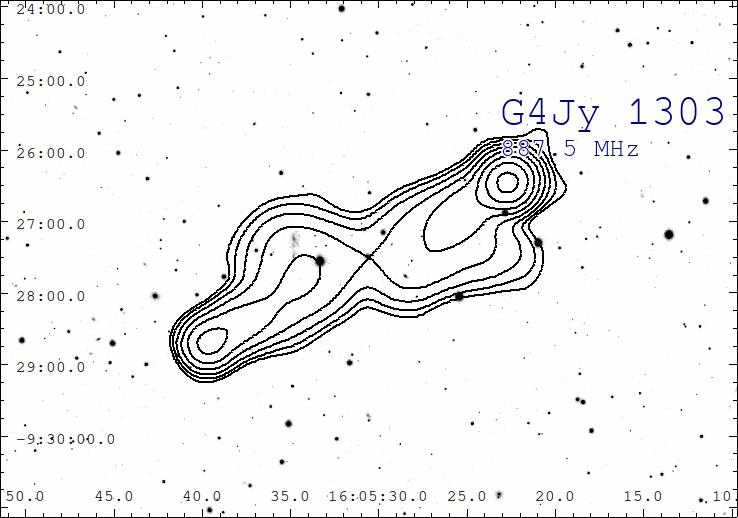}
    \includegraphics[scale=0.225]{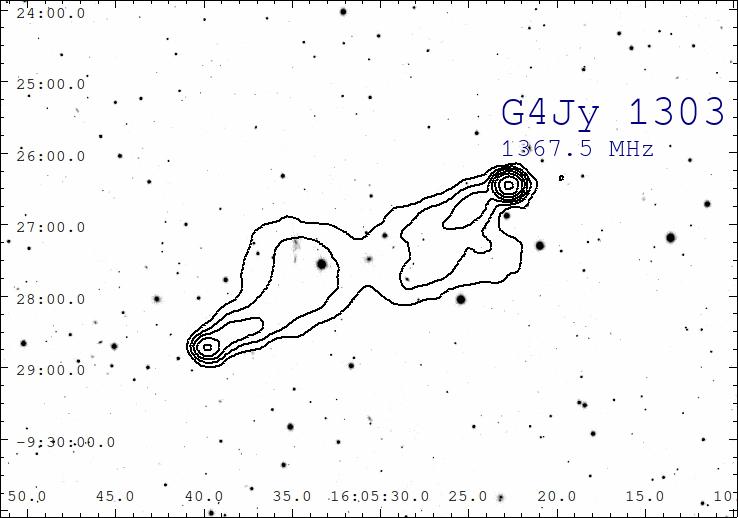}
    \includegraphics[scale=0.225]{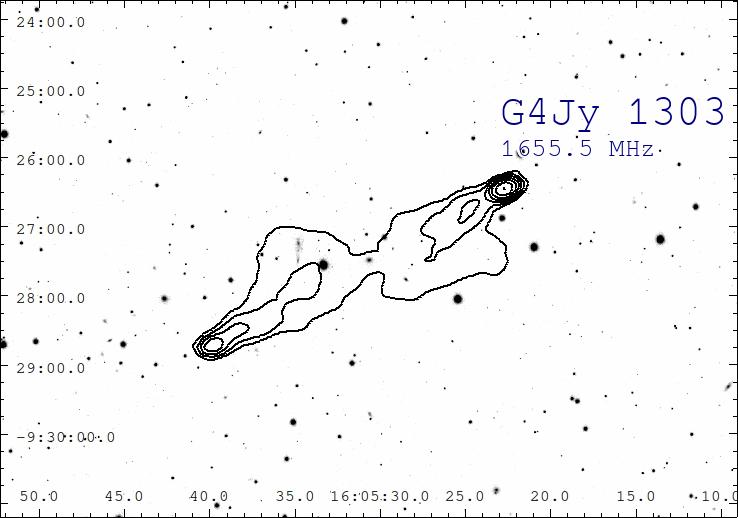}
    \includegraphics[scale=0.225]{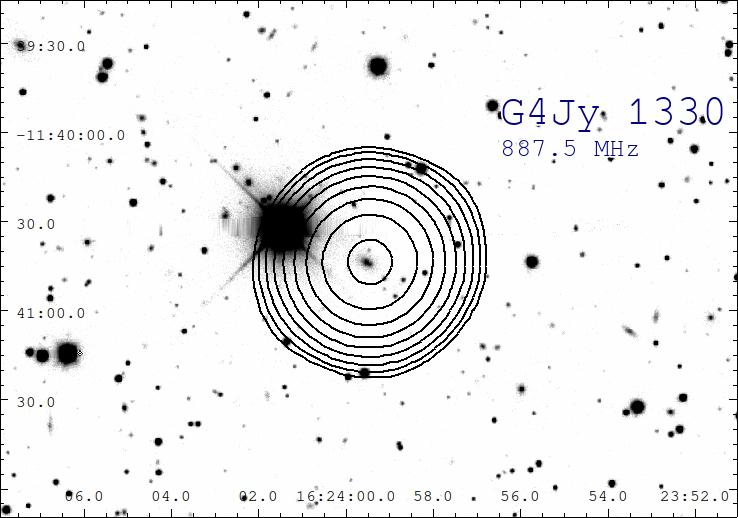}
    \includegraphics[scale=0.225]{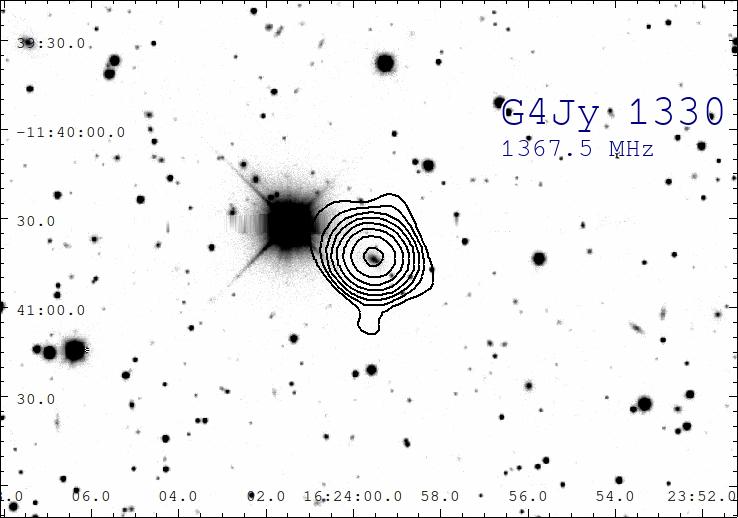}
    \includegraphics[scale=0.225]{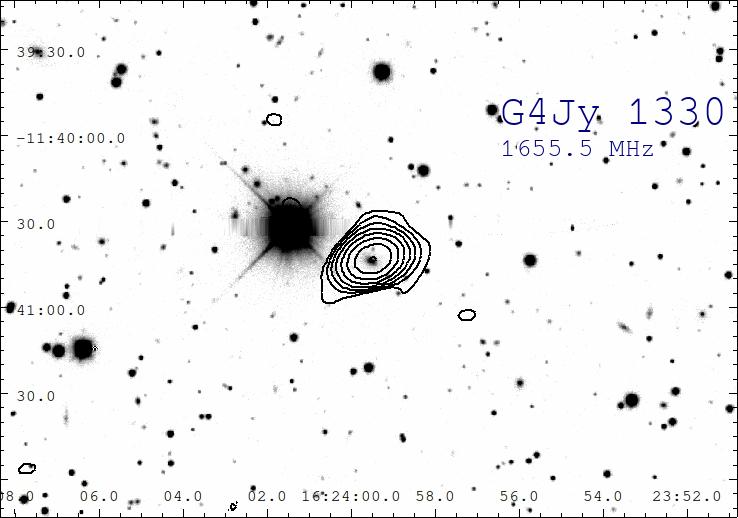}
    \includegraphics[scale=0.225]{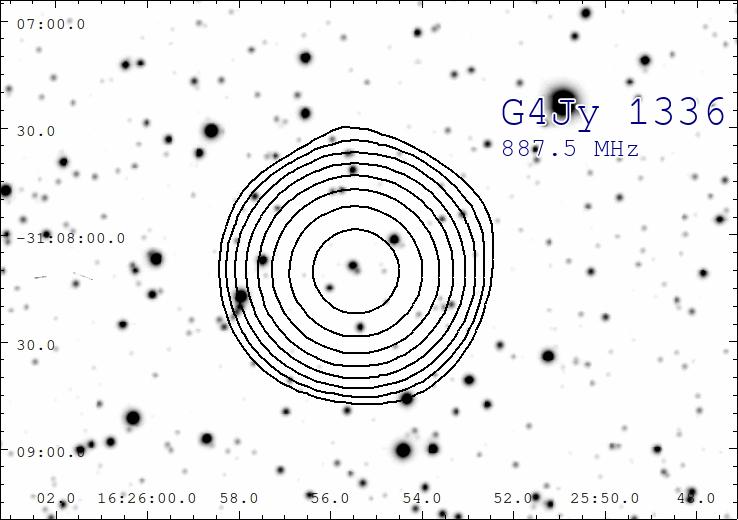}
    \includegraphics[scale=0.225]{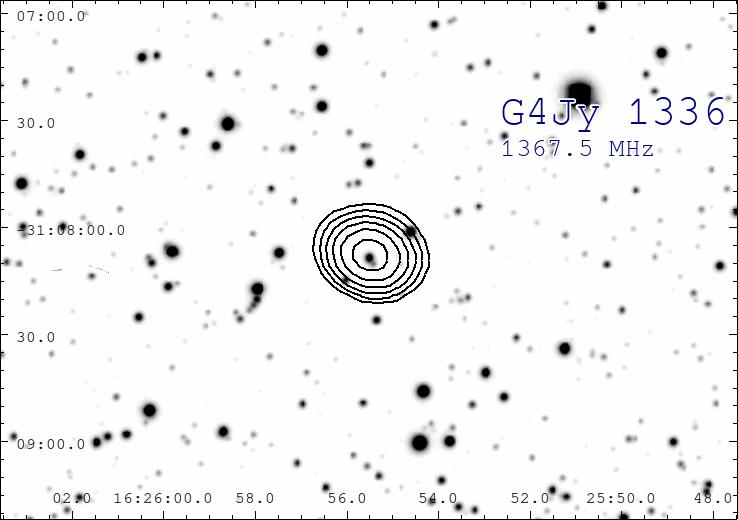}
    \includegraphics[scale=0.225]{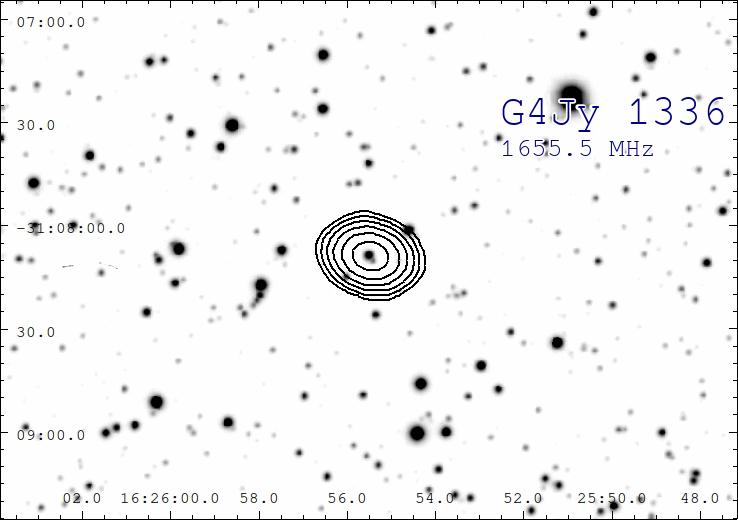}

    \caption{}
    \label{AH}
\end{figure*}
\clearpage
 
\begin{figure*}
    \centering
    \includegraphics[scale=0.225]{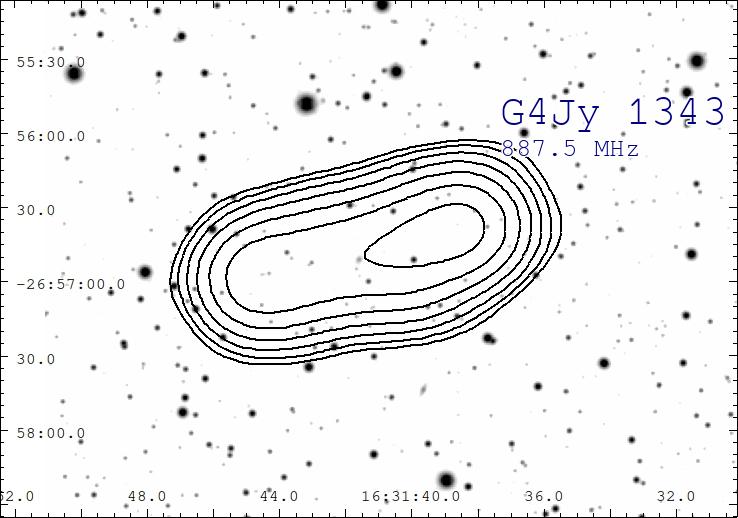}
    \includegraphics[scale=0.225]{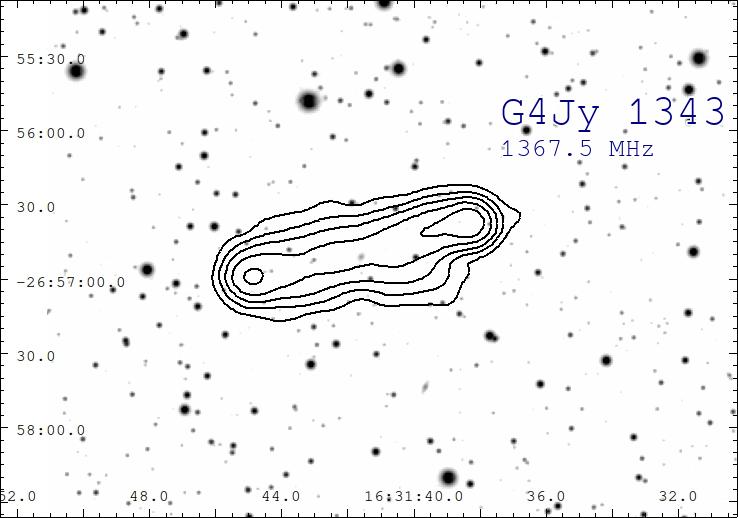}
    \includegraphics[scale=0.225]{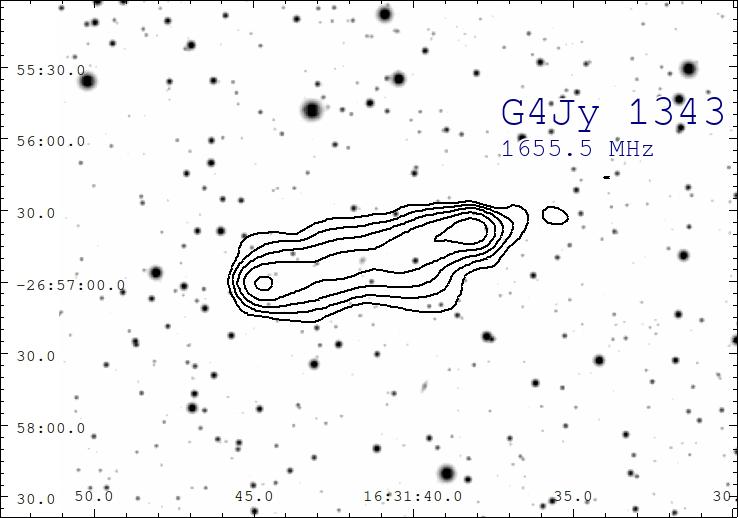}
    \includegraphics[scale=0.225]{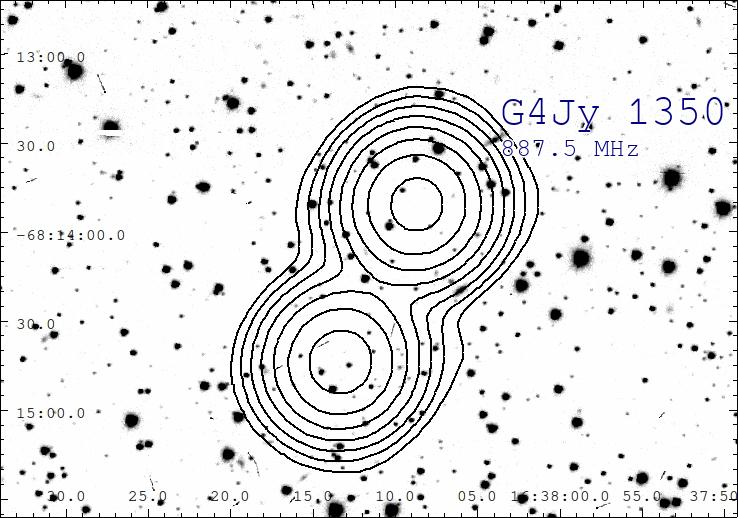}
    \includegraphics[scale=0.225]{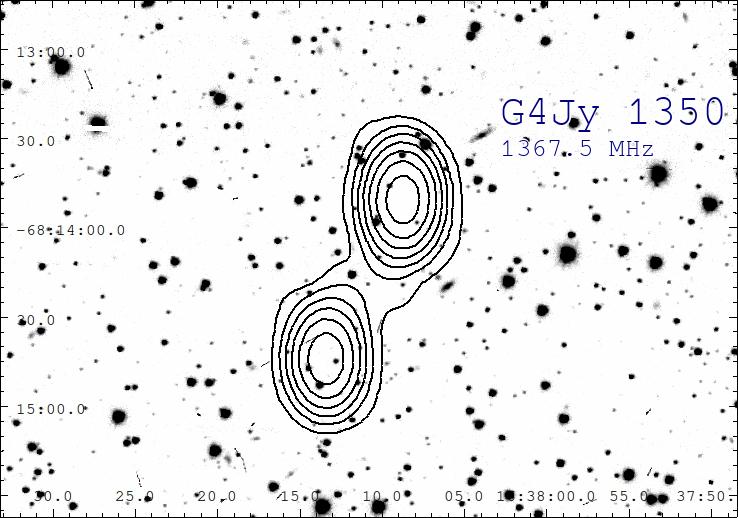}
    \includegraphics[scale=0.225]{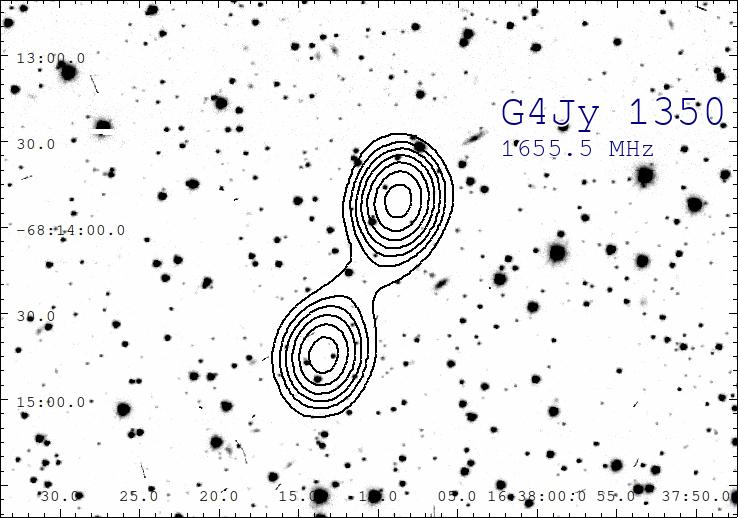}
    \includegraphics[scale=0.225]{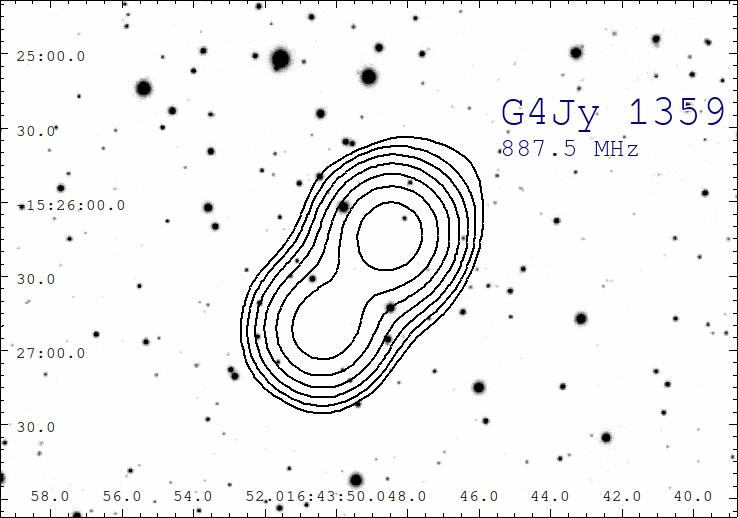}
    \includegraphics[scale=0.225]{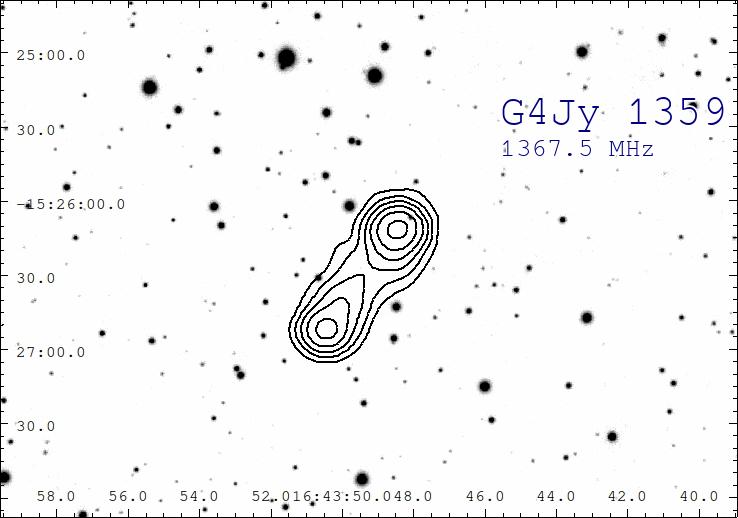}
    \includegraphics[scale=0.225]{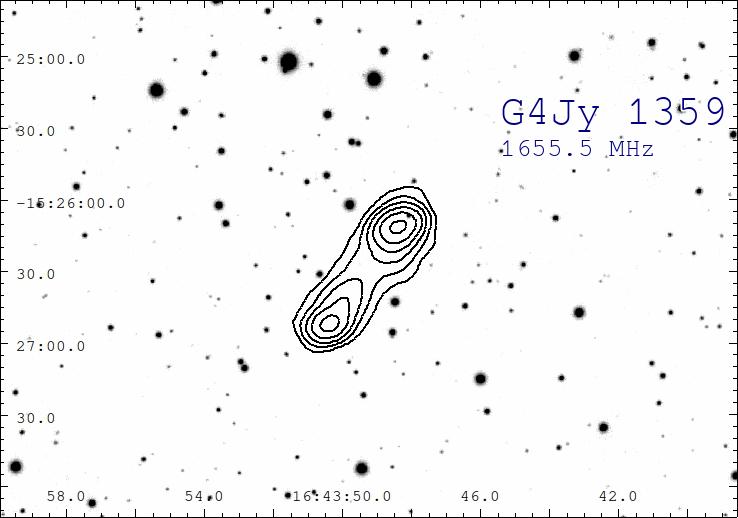}
    \includegraphics[scale=0.225]{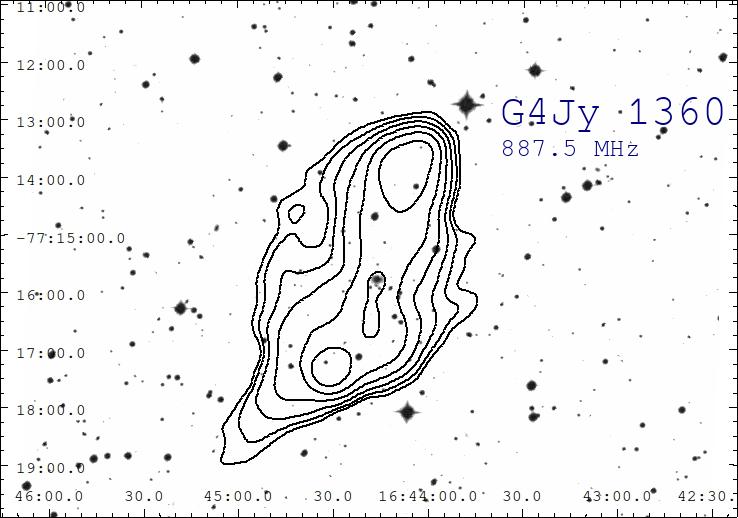}
    \includegraphics[scale=0.225]{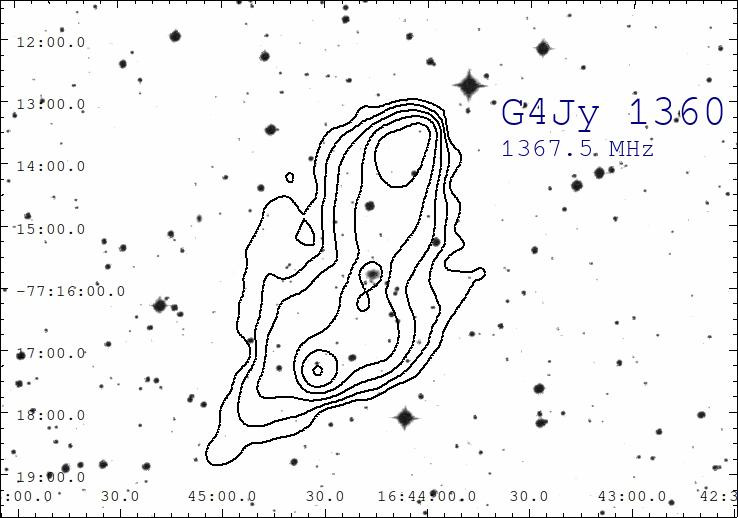}
    \includegraphics[scale=0.225]{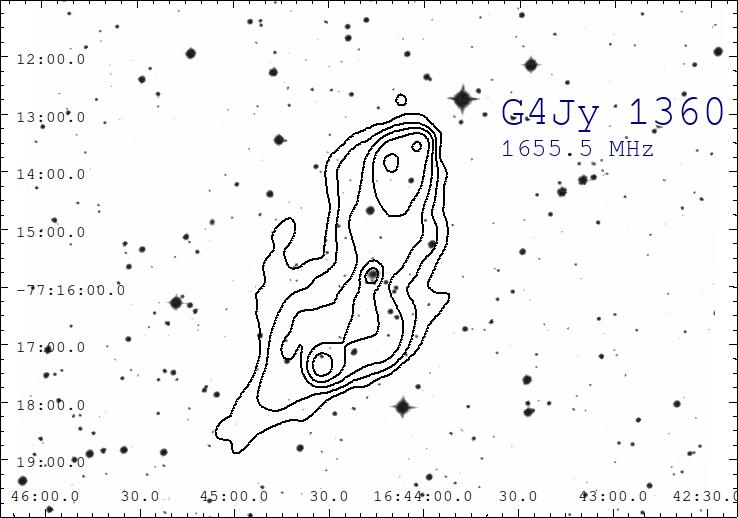}
    \includegraphics[scale=0.225]{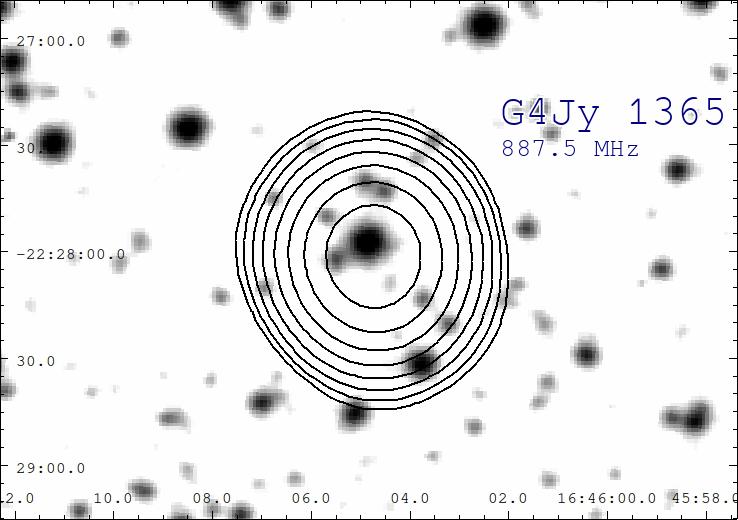}
    \includegraphics[scale=0.225]{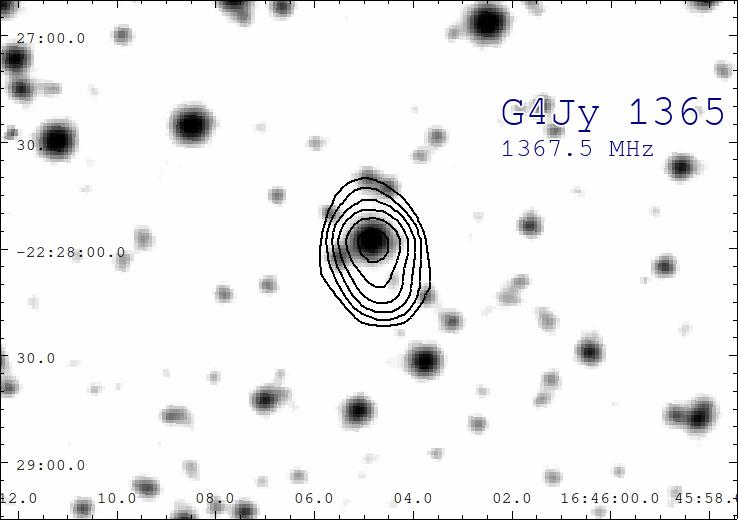}
    \includegraphics[scale=0.225]{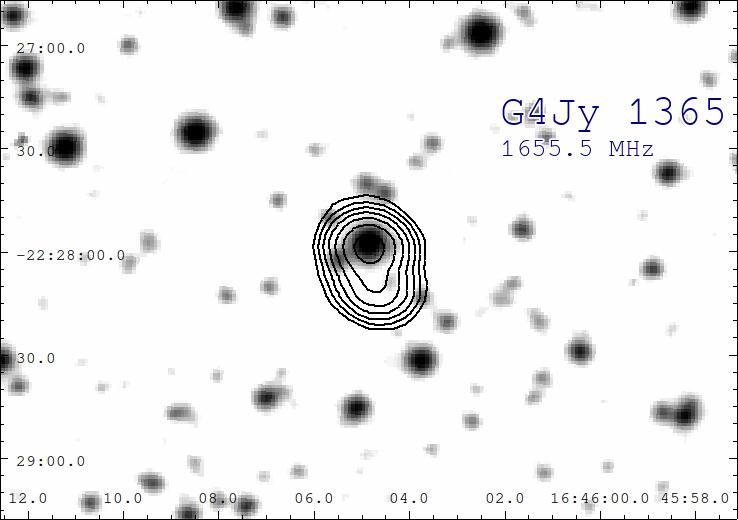}

    \caption{}
    \label{AI}
\end{figure*}
\clearpage
 
\begin{figure*}
    \centering
    \includegraphics[scale=0.225]{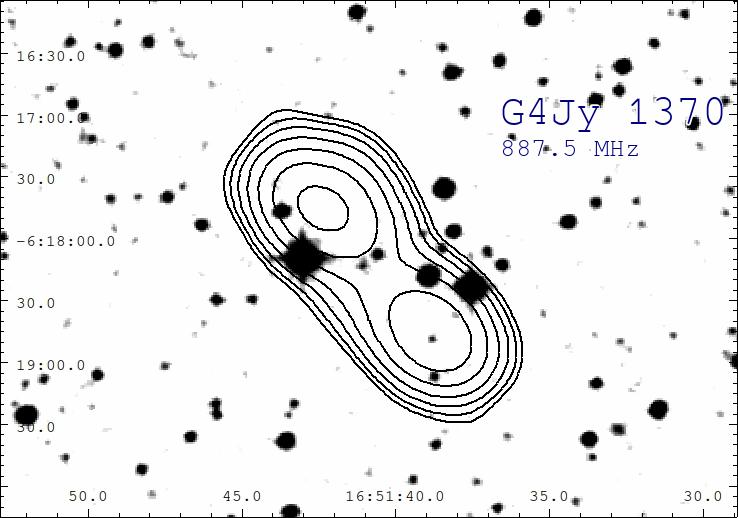}
    \includegraphics[scale=0.225]{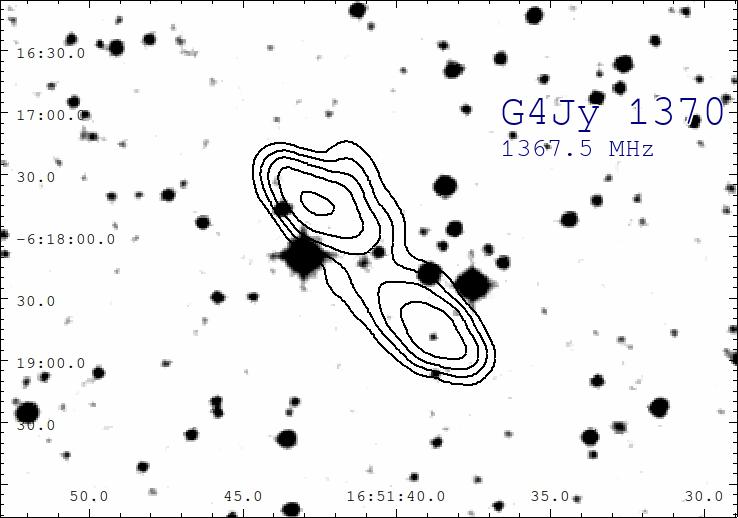}
    \includegraphics[scale=0.225]{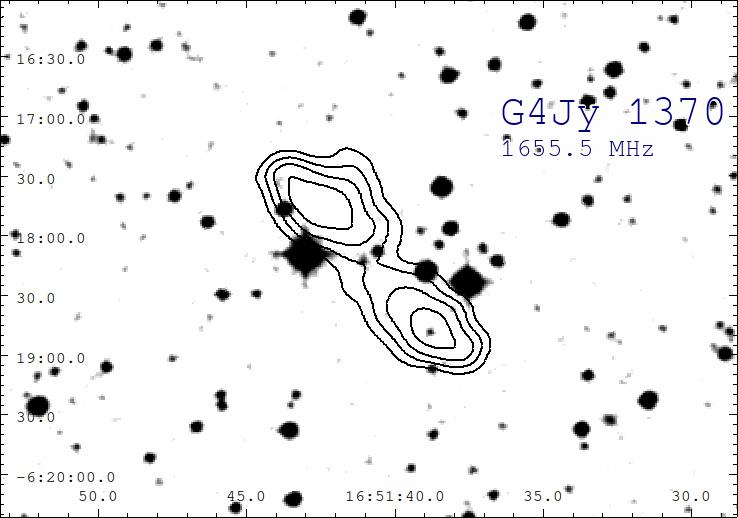}
    \includegraphics[scale=0.225]{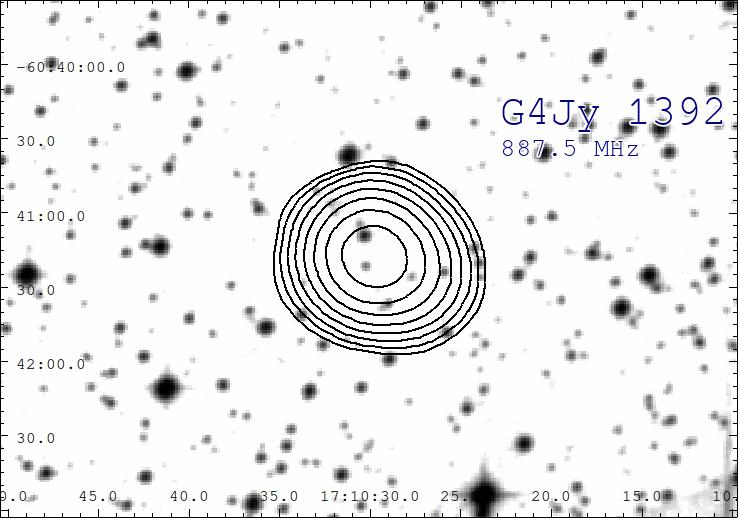}
    \includegraphics[scale=0.225]{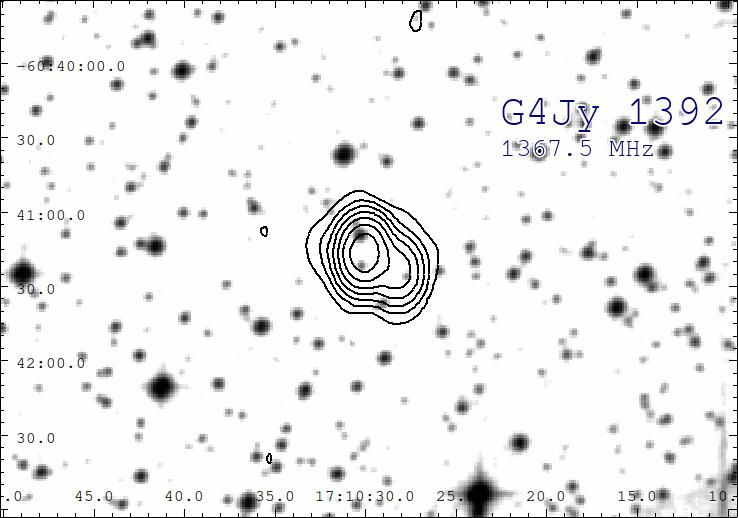}
    \includegraphics[scale=0.225]{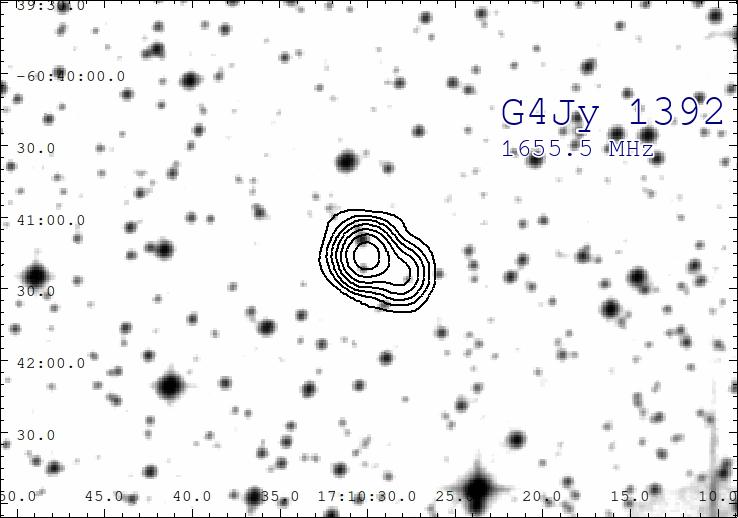}
    \includegraphics[scale=0.225]{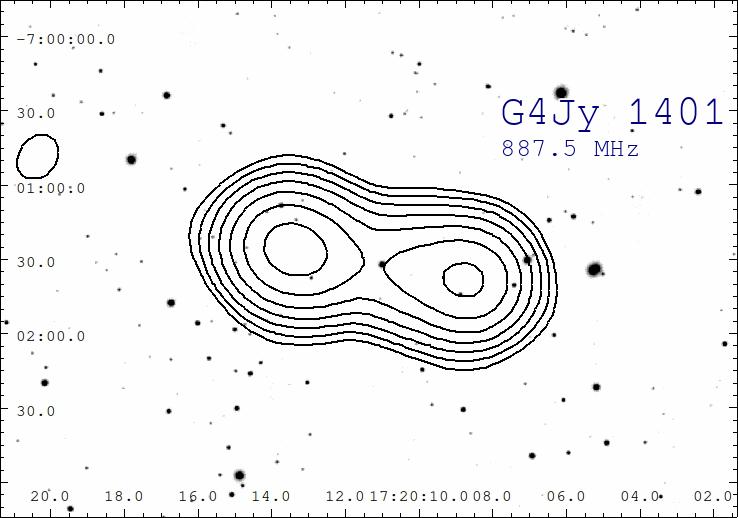}
    \includegraphics[scale=0.225]{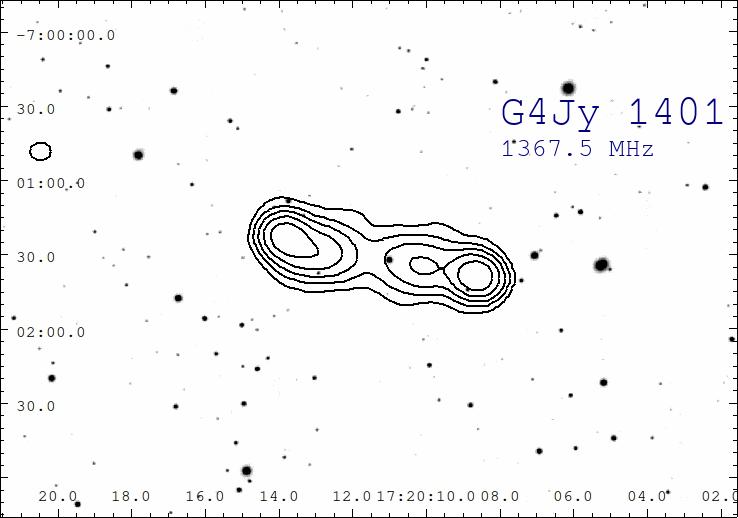}
    \includegraphics[scale=0.225]{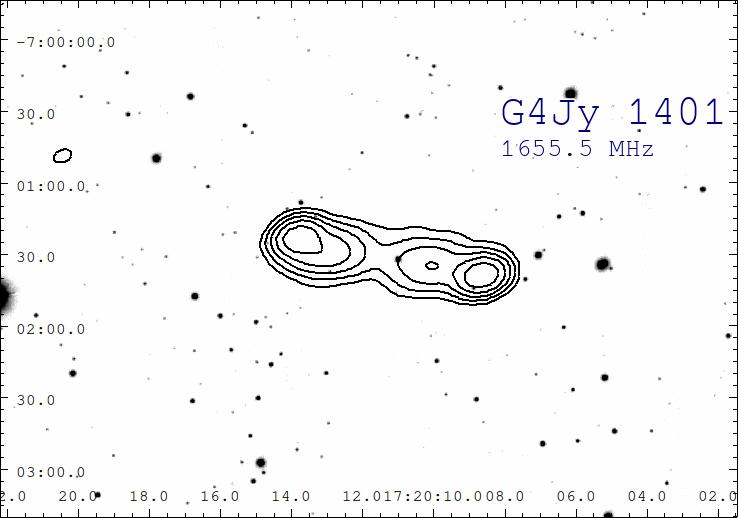}
    \includegraphics[scale=0.225]{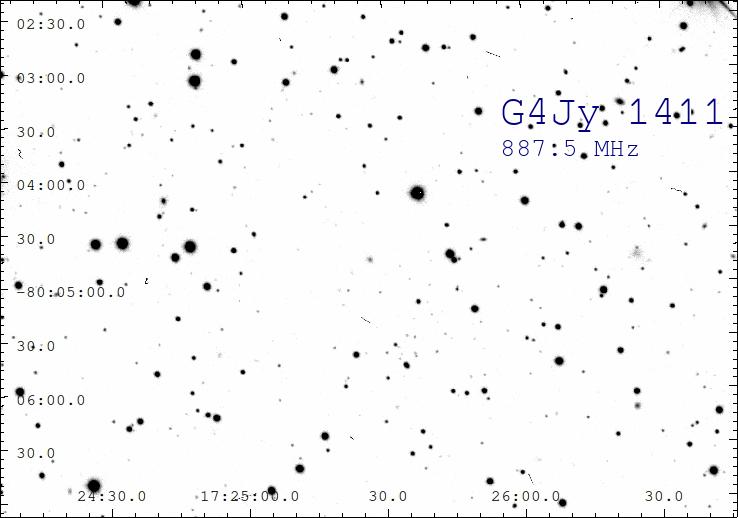}
    \includegraphics[scale=0.225]{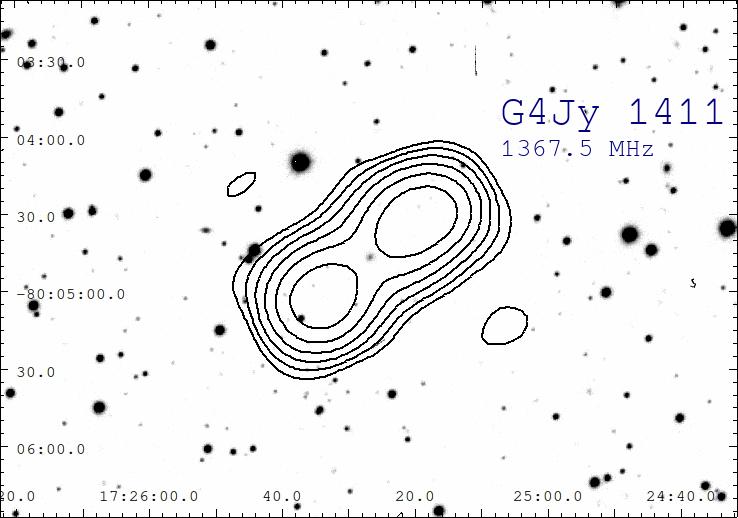}
    \includegraphics[scale=0.225]{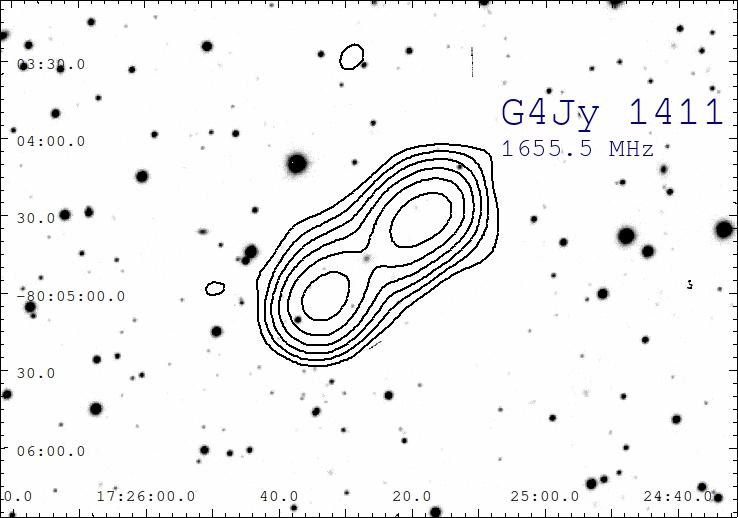}
    \includegraphics[scale=0.225]{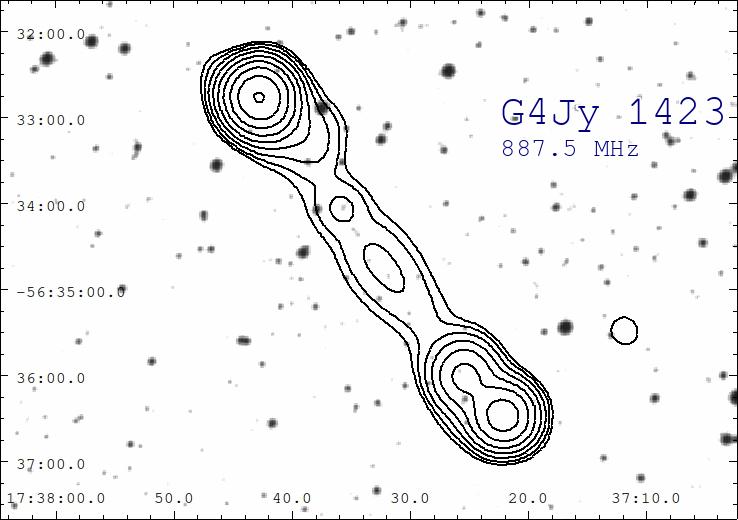}
    \includegraphics[scale=0.225]{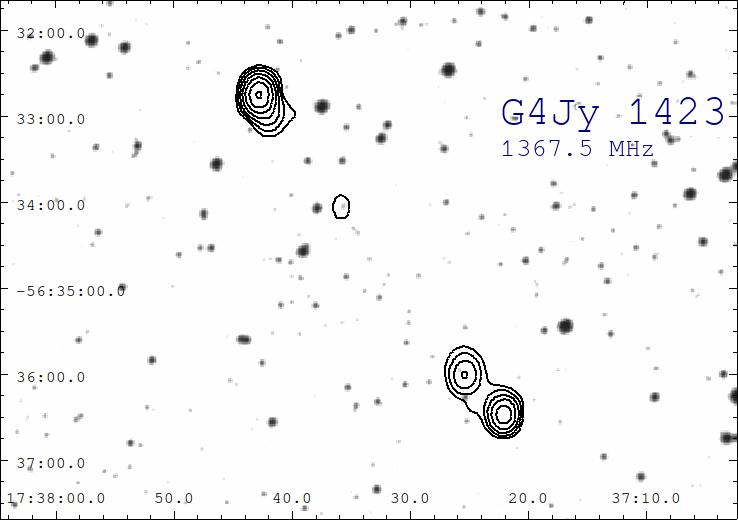}
    \includegraphics[scale=0.225]{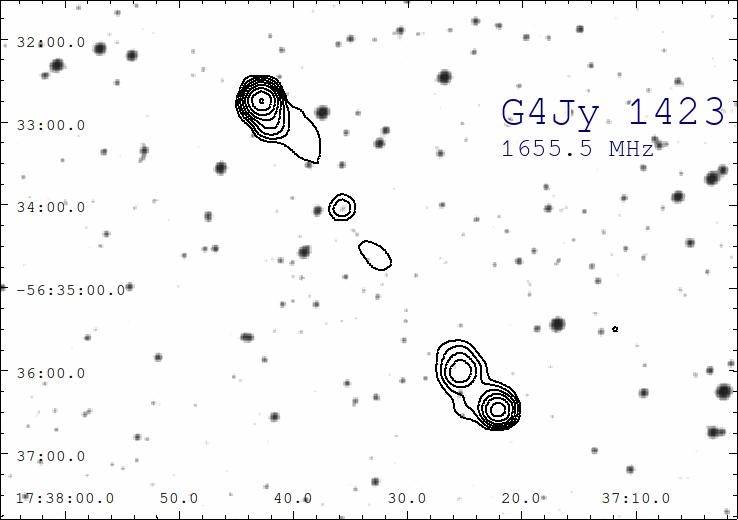}

    \caption{RACS-low data for G4Jy 1411 could not be convolved to 25$\arcsec$, and therefore was not available. It may be replaced with future observations from RACS-low2 or RACS-low3}
    \label{AJ}
\end{figure*}
\clearpage
 
\begin{figure*}
    \centering
    \includegraphics[scale=0.225]{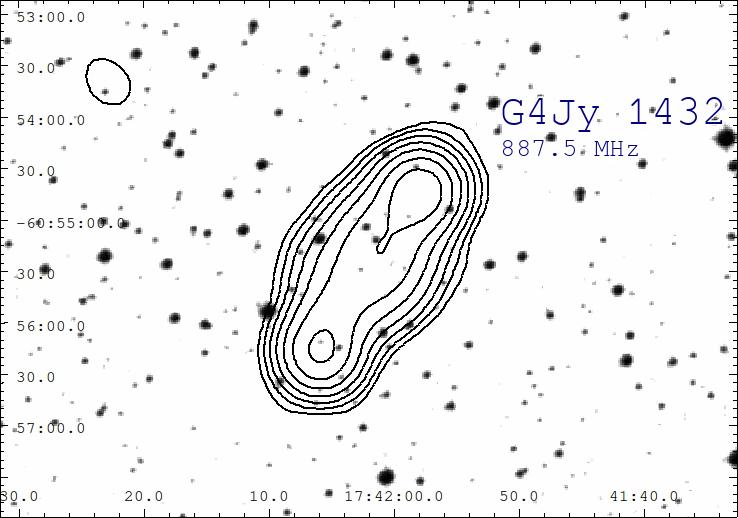}
    \includegraphics[scale=0.225]{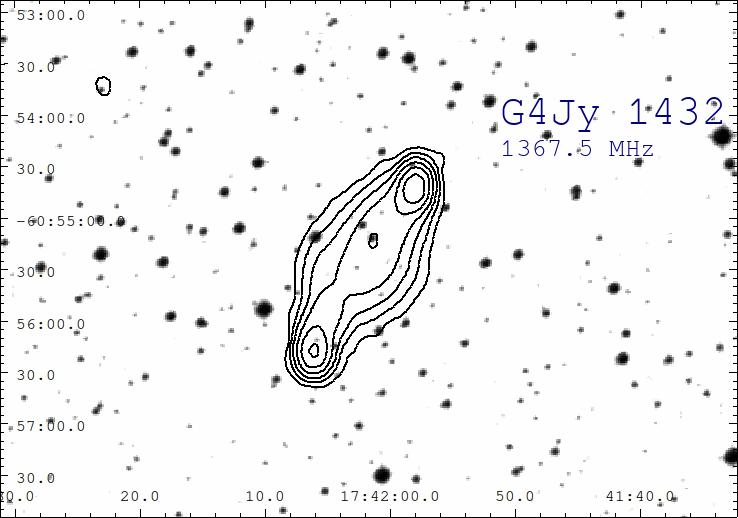}
    \includegraphics[scale=0.225]{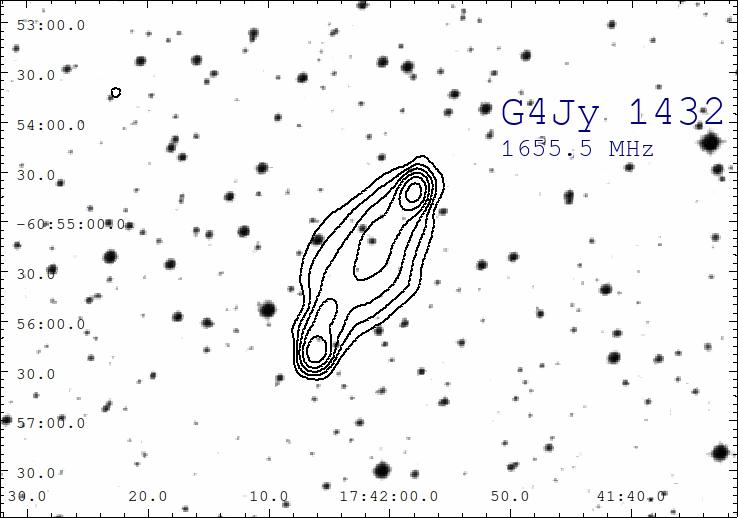}
    \includegraphics[scale=0.225]{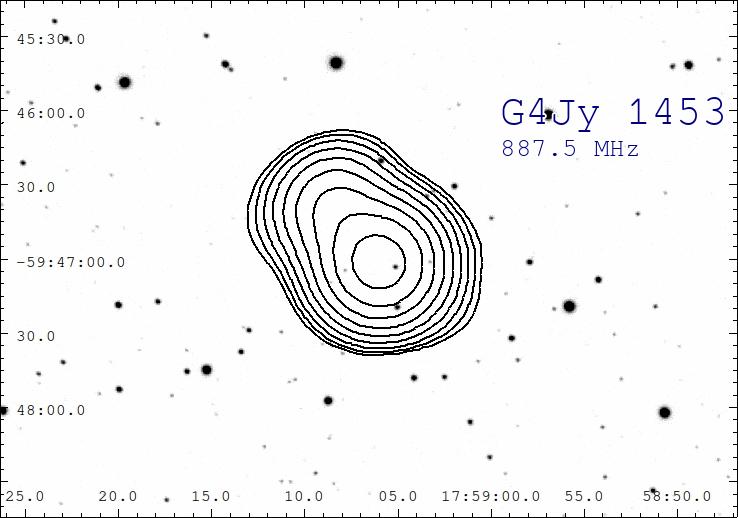}
    \includegraphics[scale=0.225]{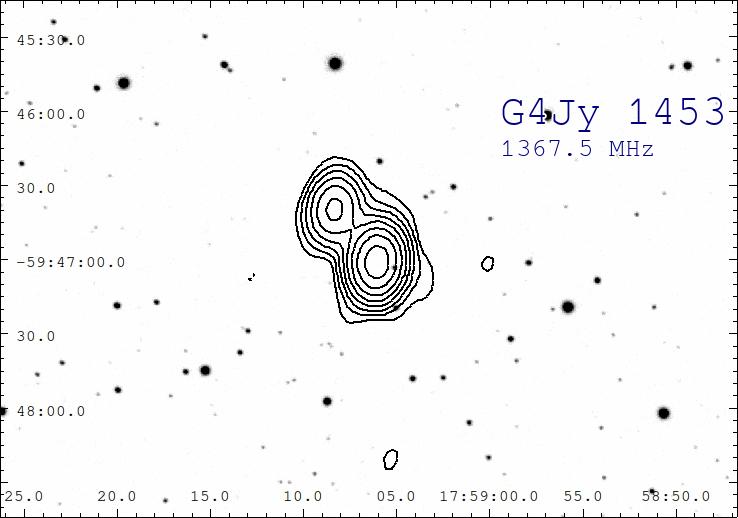}
    \includegraphics[scale=0.225]{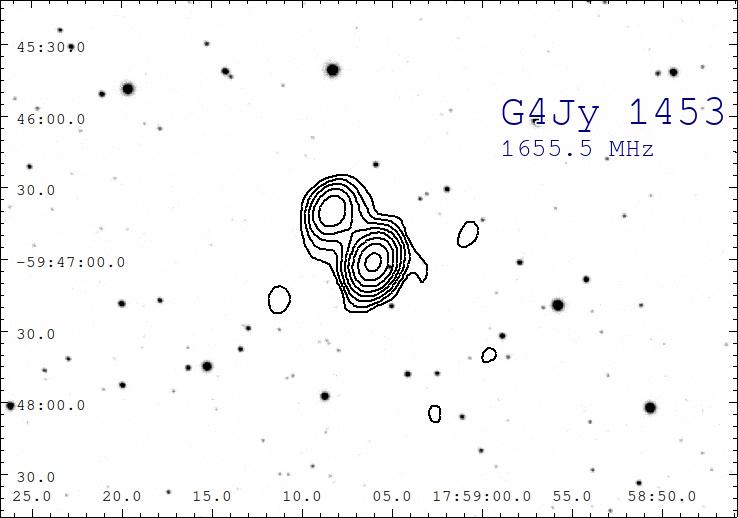}
    \includegraphics[scale=0.225]{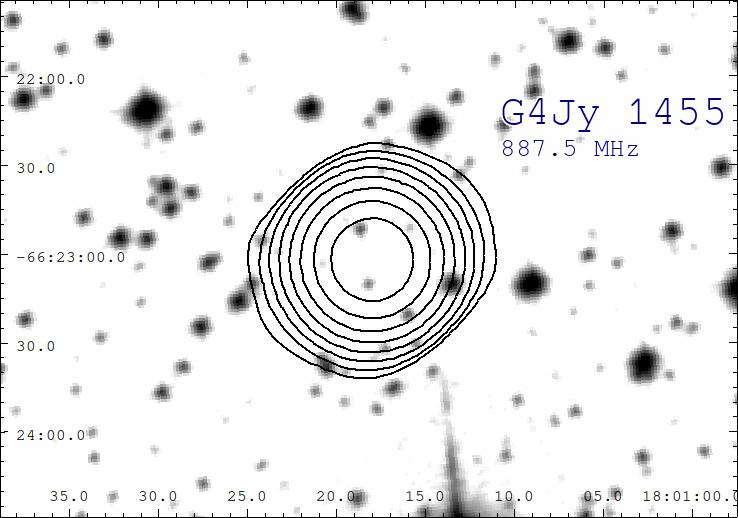}
    \includegraphics[scale=0.225]{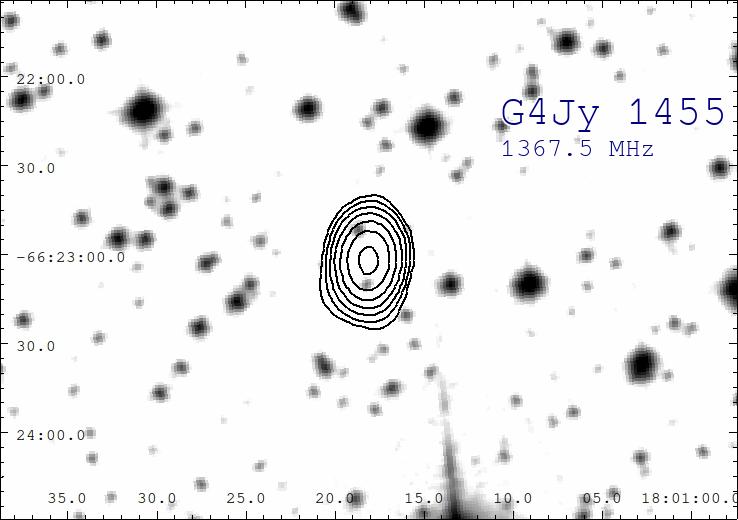}
    \includegraphics[scale=0.225]{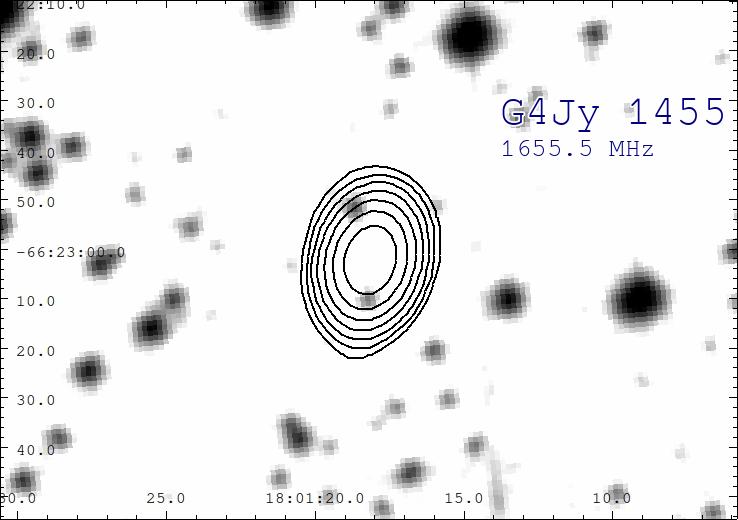}
    \includegraphics[scale=0.225]{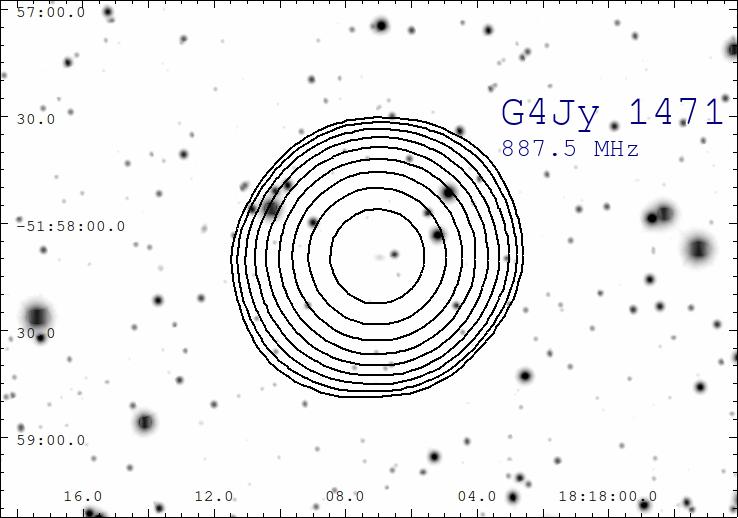}
    \includegraphics[scale=0.225]{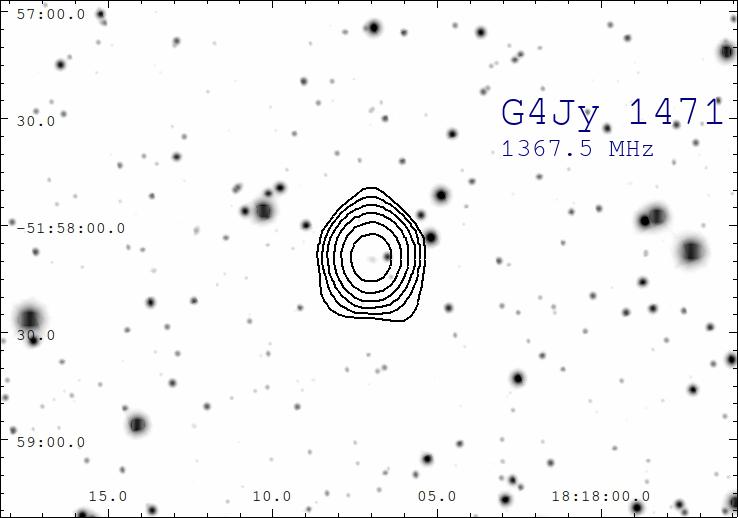}
    \includegraphics[scale=0.225]{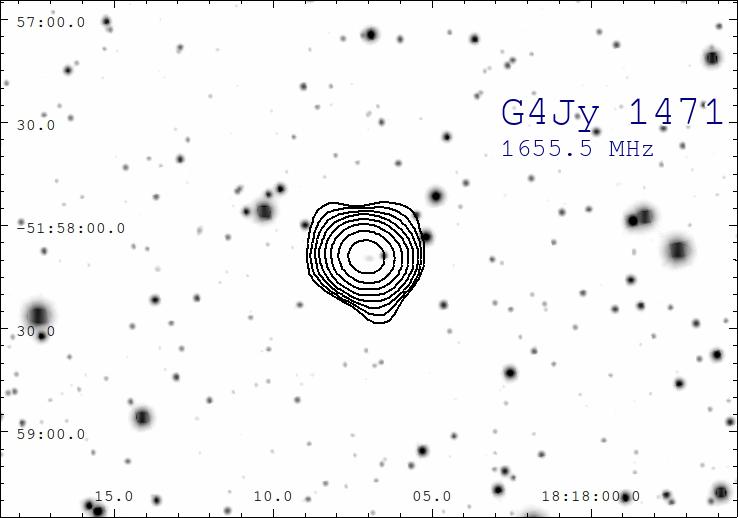}
    \includegraphics[scale=0.225]{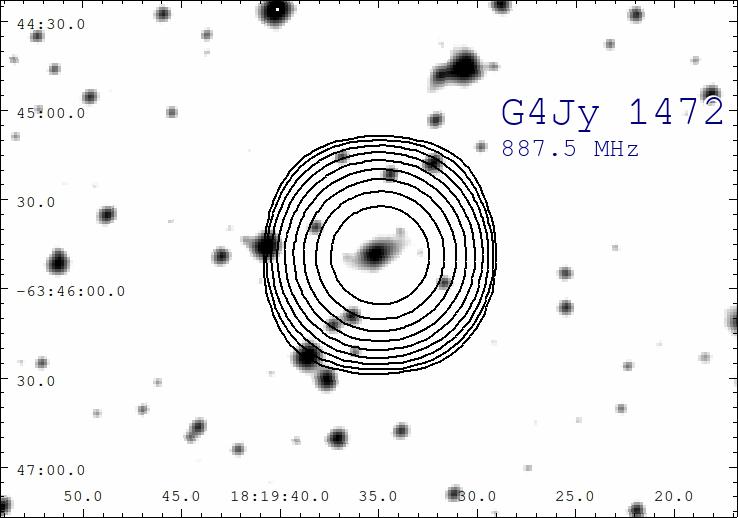}
    \includegraphics[scale=0.225]{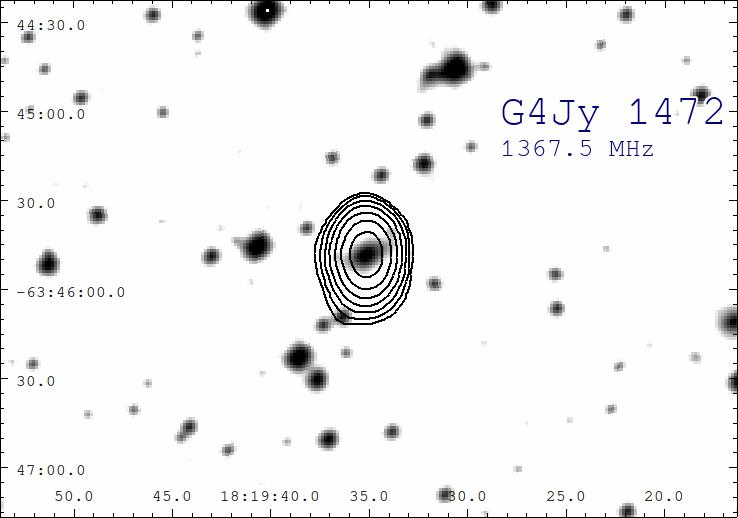}
    \includegraphics[scale=0.225]{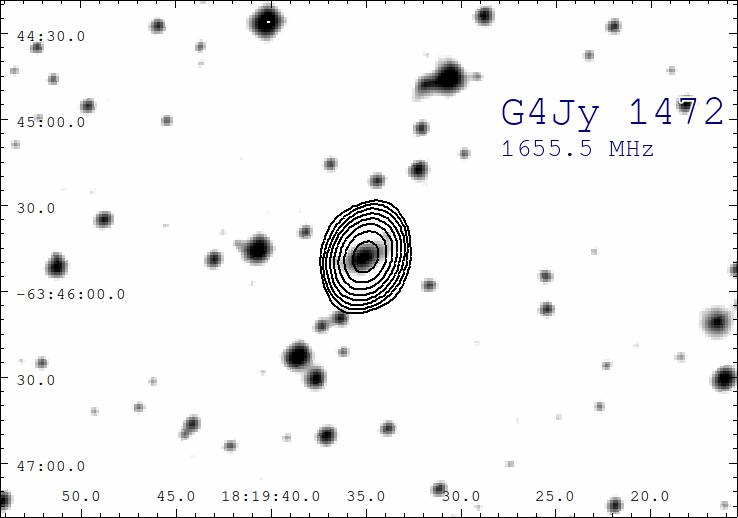}

    \caption{}
    \label{AK}
\end{figure*}
\clearpage
 
\begin{figure*}
    \centering
    \includegraphics[scale=0.225]{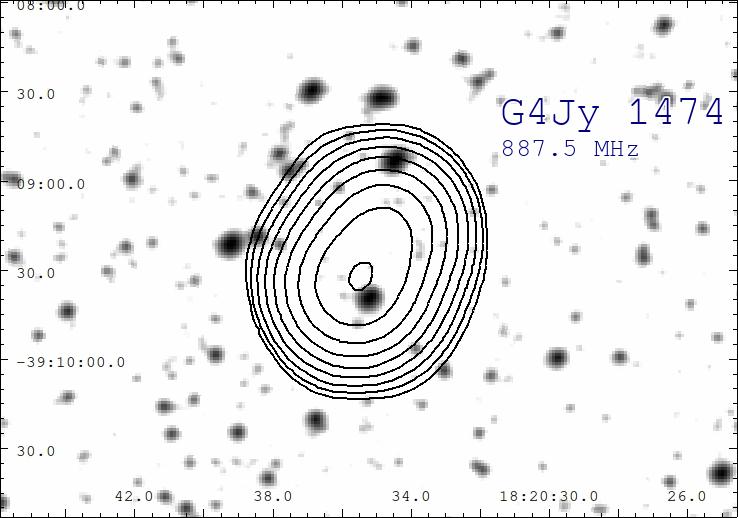}
    \includegraphics[scale=0.225]{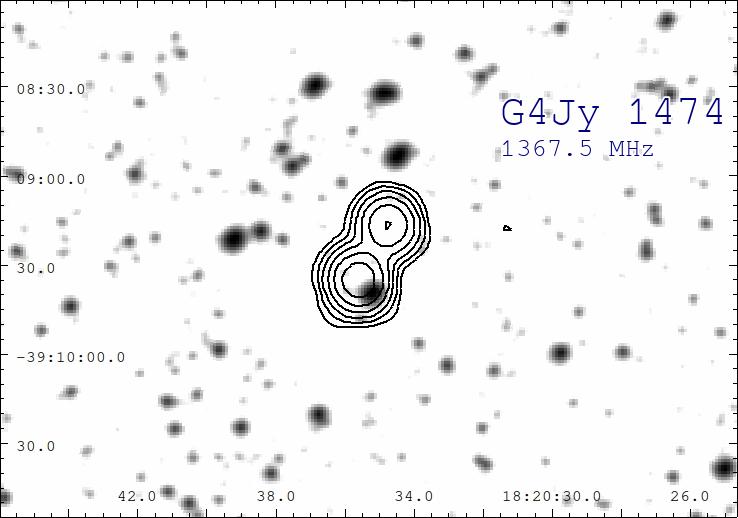}
    \includegraphics[scale=0.225]{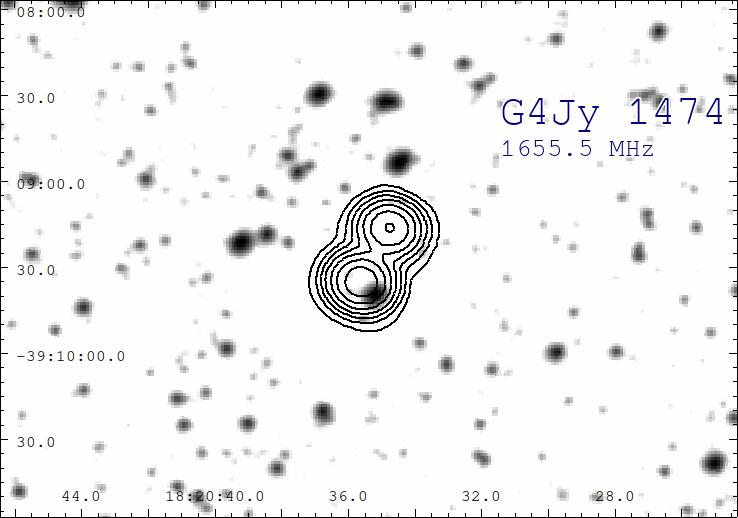}
    \includegraphics[scale=0.225]{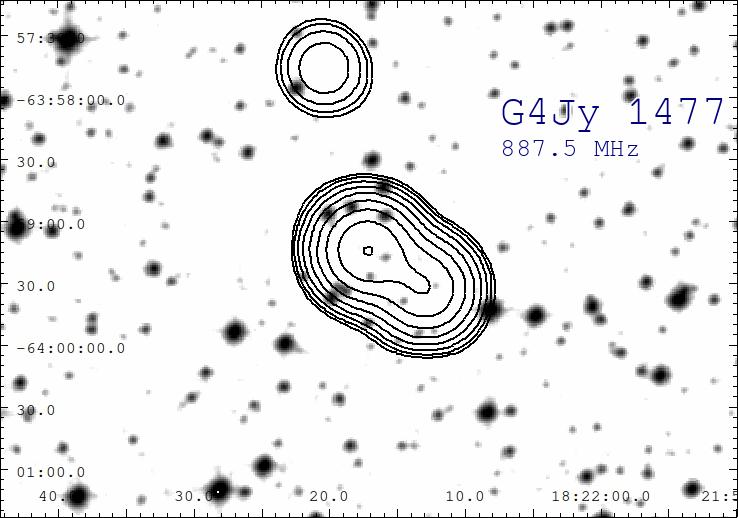}
    \includegraphics[scale=0.225]{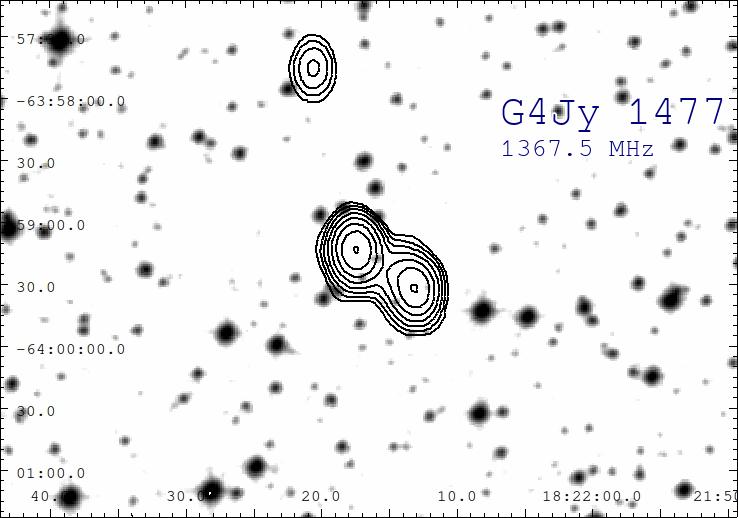}
    \includegraphics[scale=0.225]{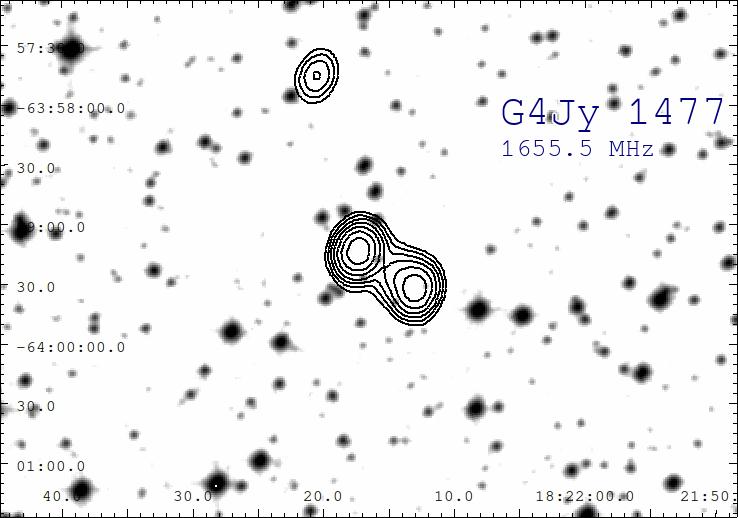}
    \includegraphics[scale=0.225]{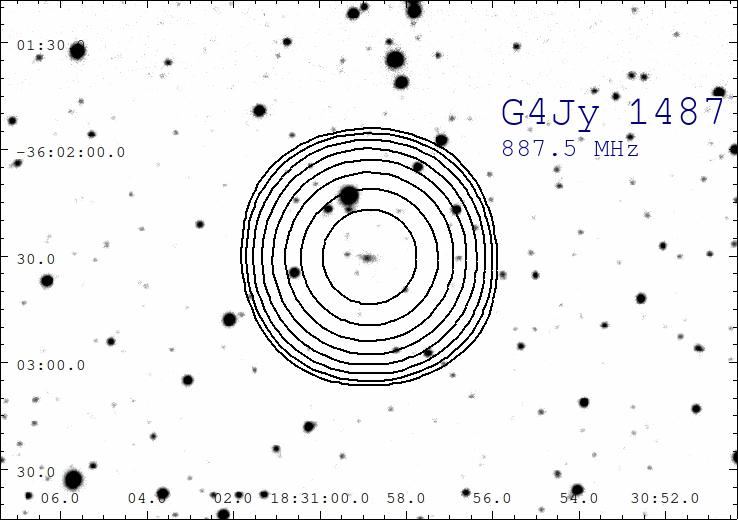}
    \includegraphics[scale=0.225]{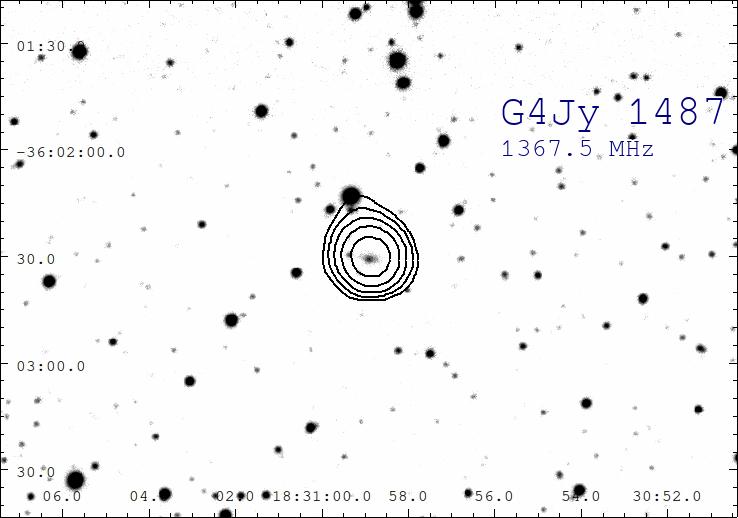}
    \includegraphics[scale=0.225]{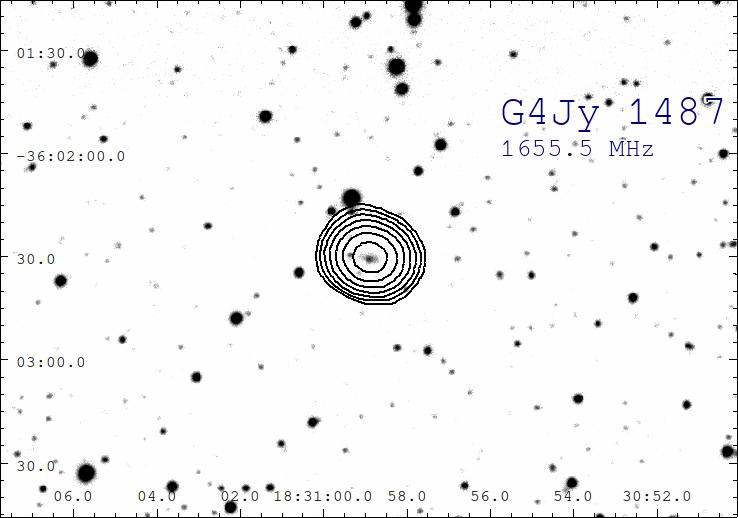}
    \includegraphics[scale=0.225]{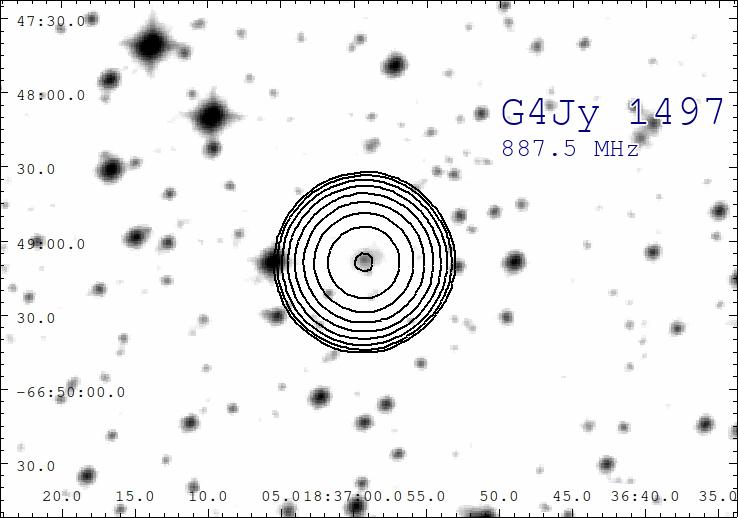}
    \includegraphics[scale=0.225]{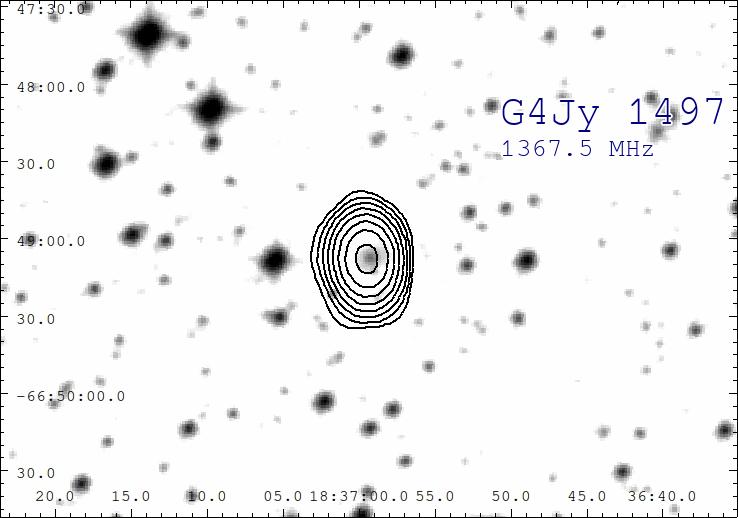}
    \includegraphics[scale=0.225]{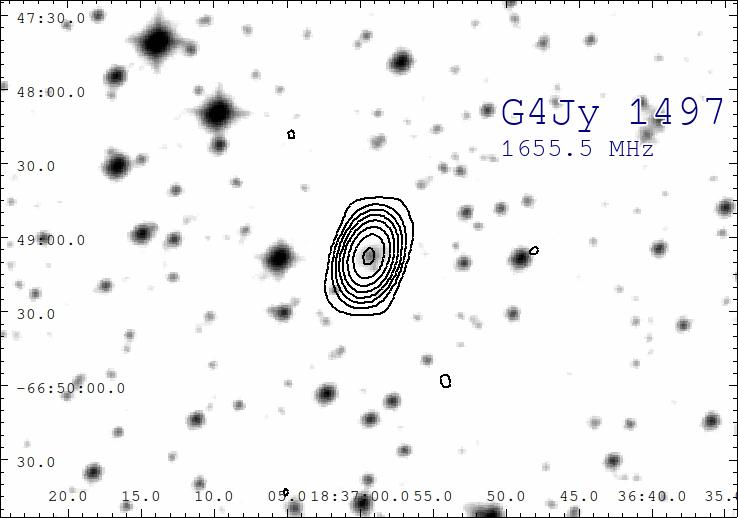}
    \includegraphics[scale=0.225]{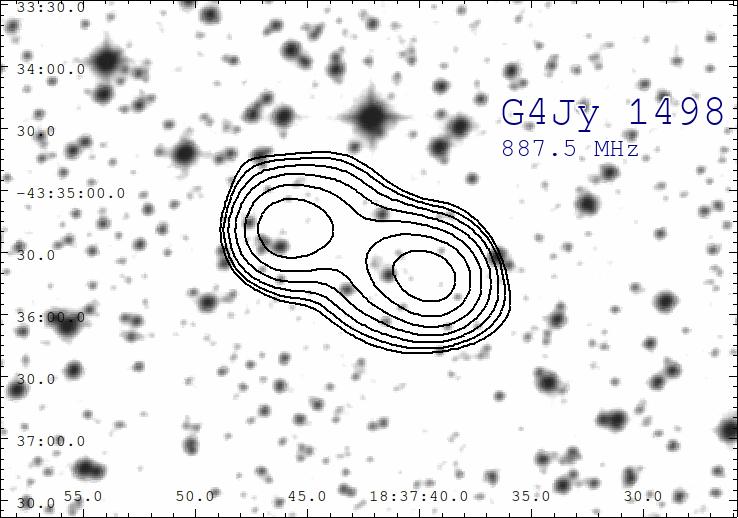}
    \includegraphics[scale=0.225]{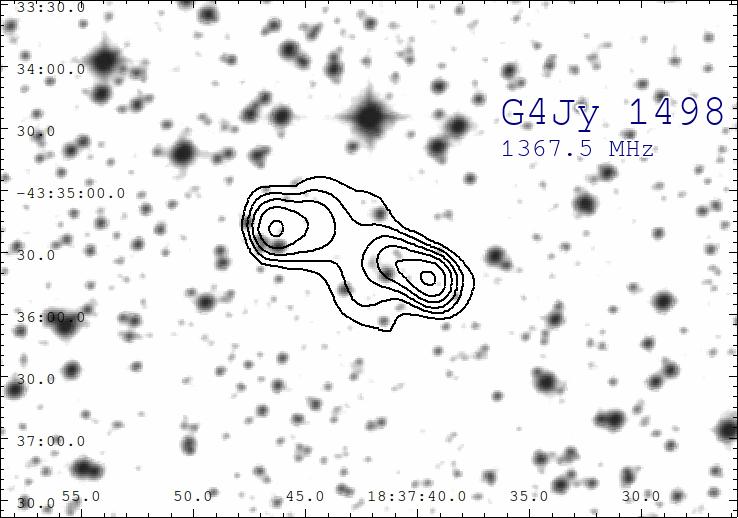}
    \includegraphics[scale=0.225]{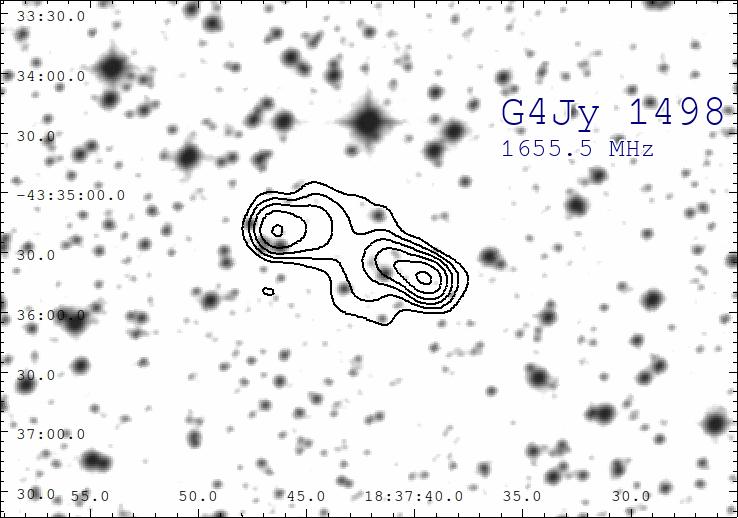}

    \caption{}
    \label{AL}
\end{figure*}
\clearpage
 
\begin{figure*}
    \centering
    \includegraphics[scale=0.225]{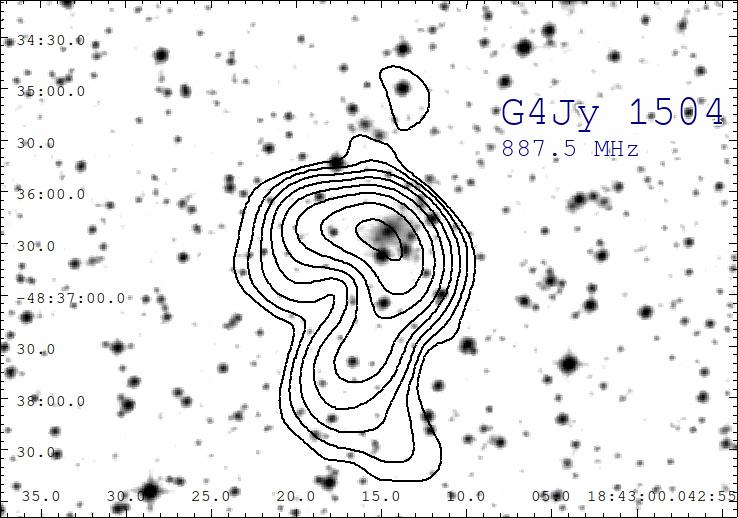}
    \includegraphics[scale=0.225]{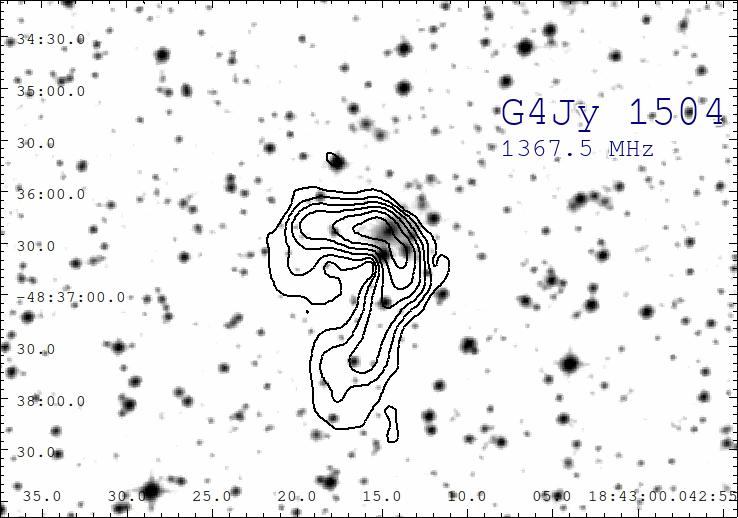}
    \includegraphics[scale=0.225]{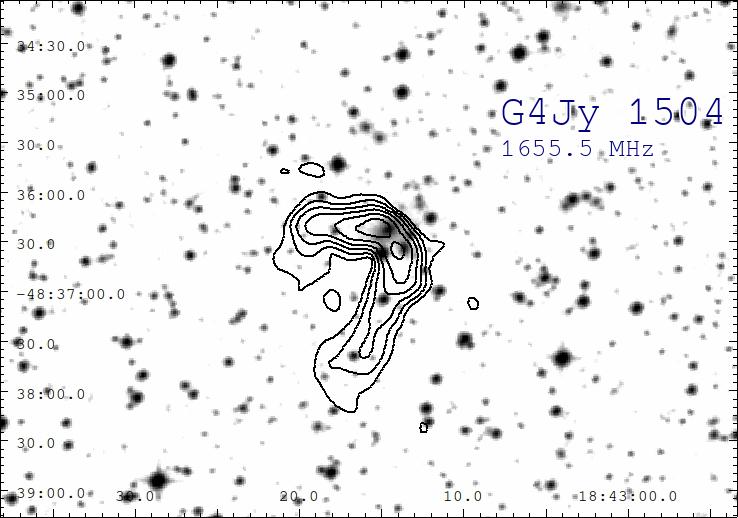}
    \includegraphics[scale=0.225]{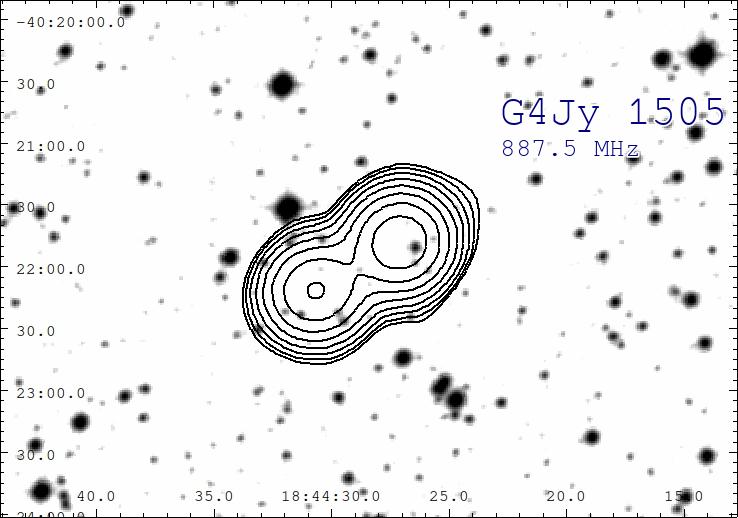}
    \includegraphics[scale=0.225]{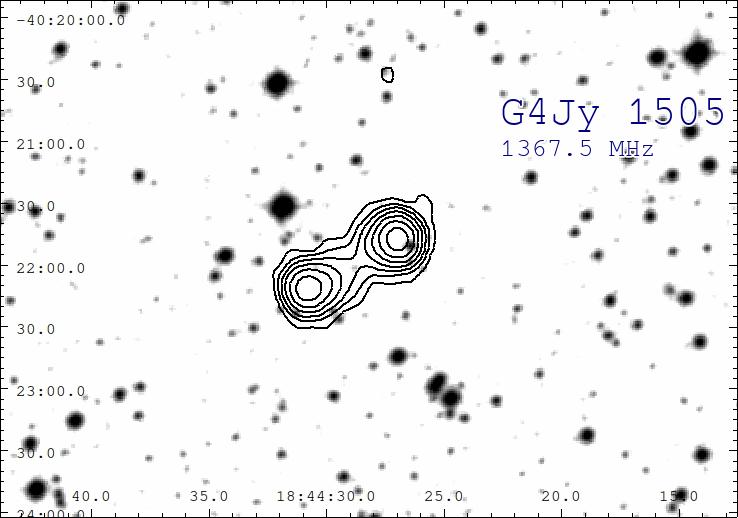}
    \includegraphics[scale=0.225]{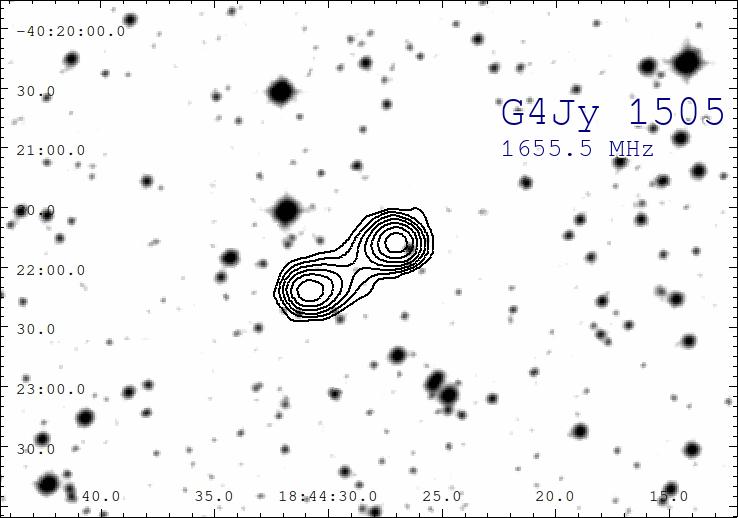}
    \includegraphics[scale=0.225]{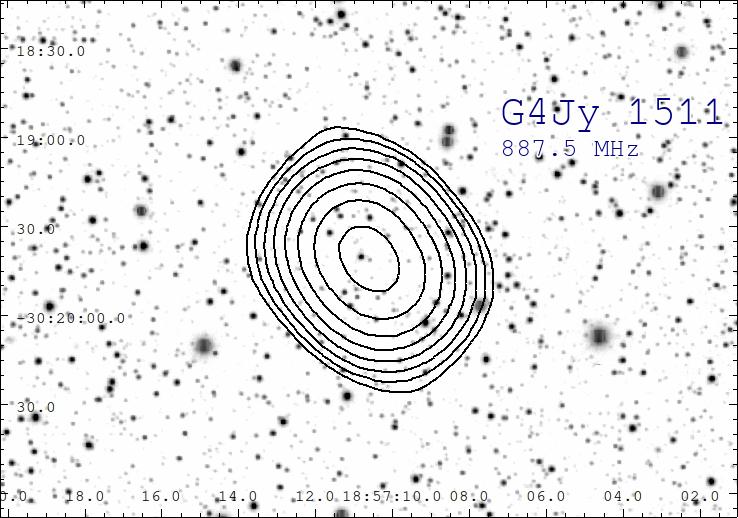}
    \includegraphics[scale=0.225]{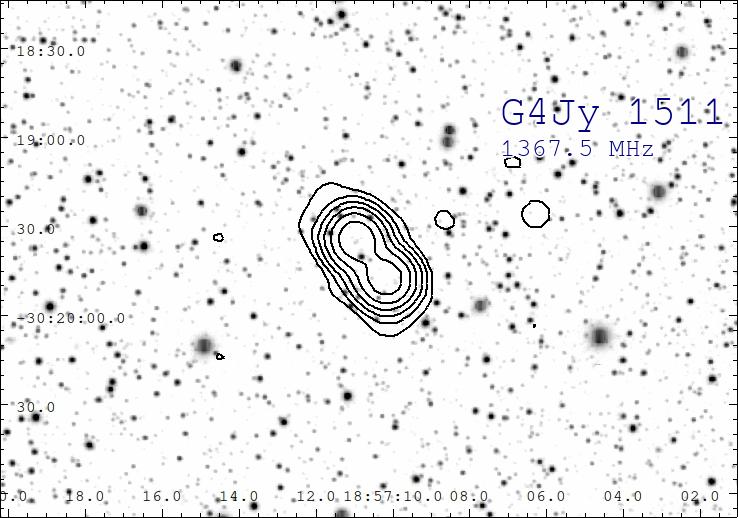}
    \includegraphics[scale=0.225]{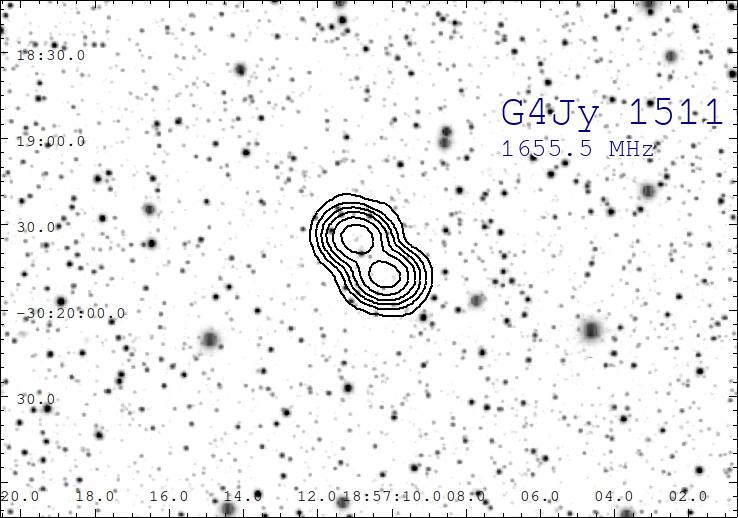}
    \includegraphics[scale=0.225]{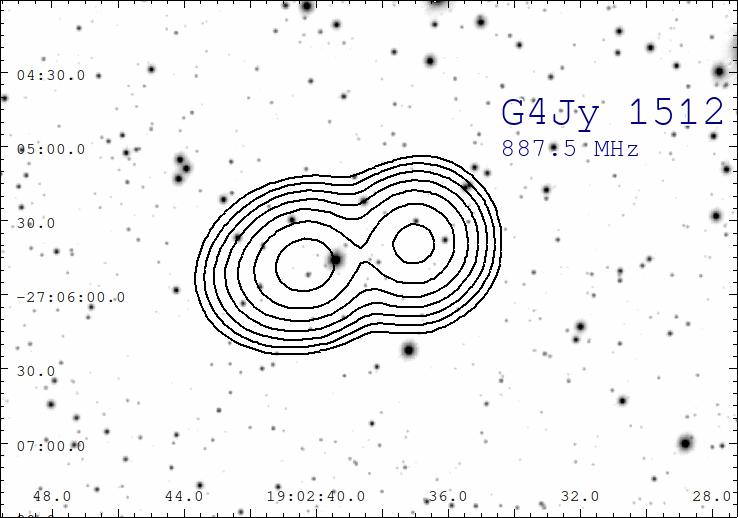}
    \includegraphics[scale=0.225]{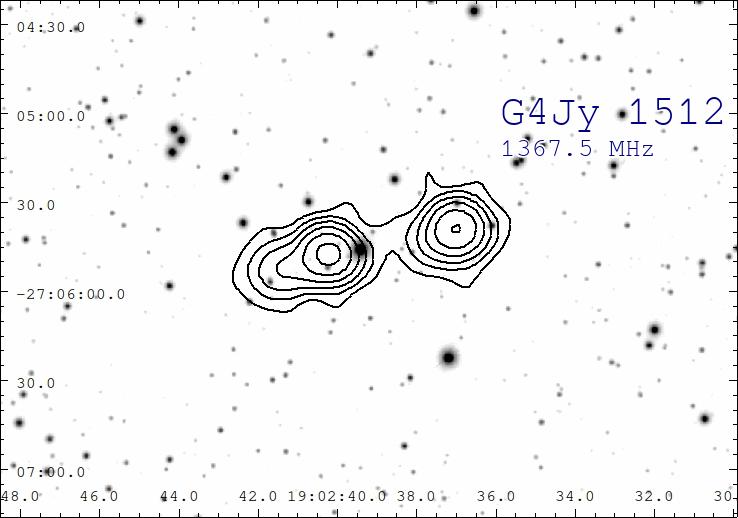}
    \includegraphics[scale=0.225]{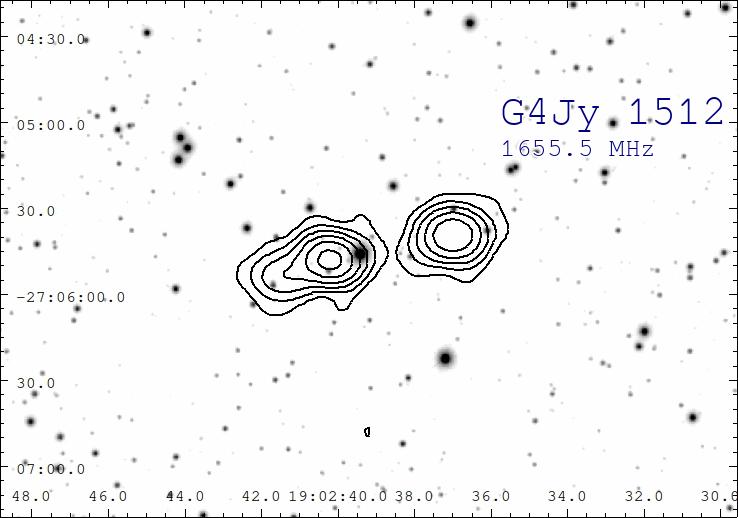}
    \includegraphics[scale=0.225]{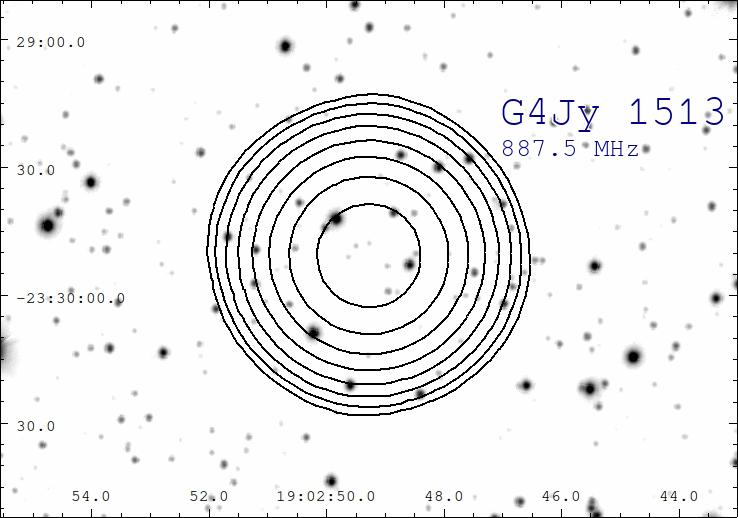}
    \includegraphics[scale=0.225]{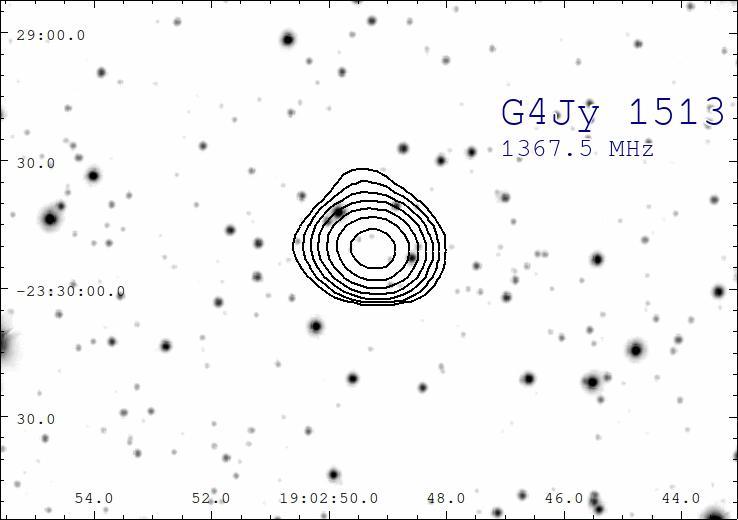}
    \includegraphics[scale=0.225]{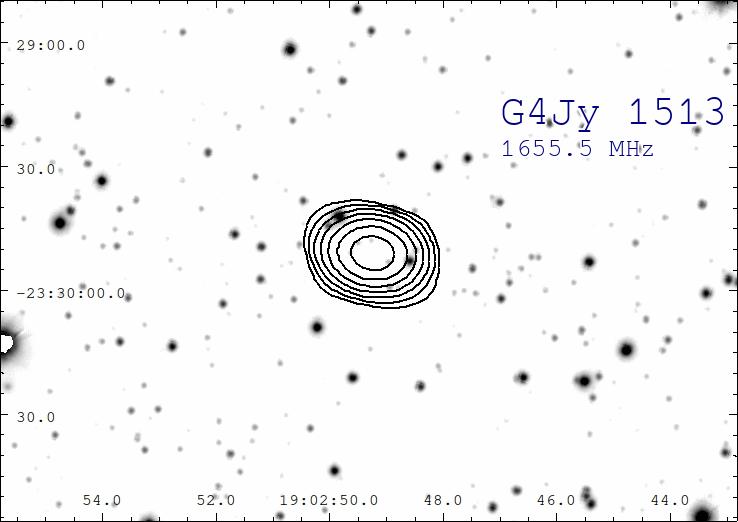}

    \caption{}
    \label{AM}
\end{figure*}
\clearpage
\begin{figure*}
    \centering
    \includegraphics[scale=0.225]{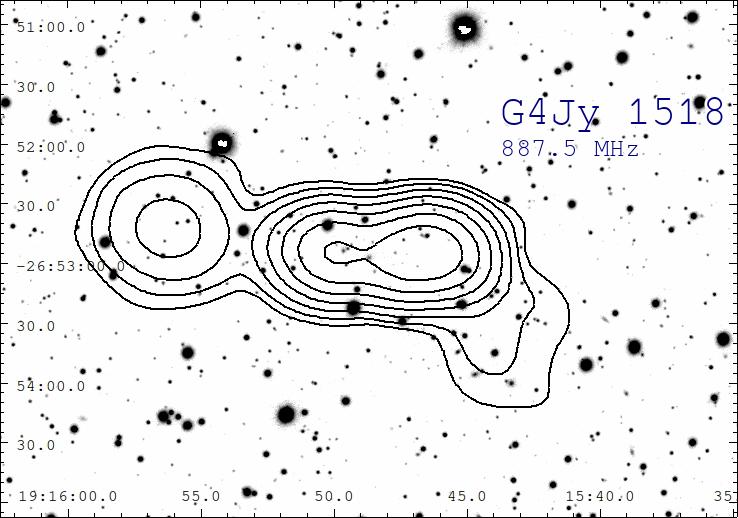}
    \includegraphics[scale=0.225]{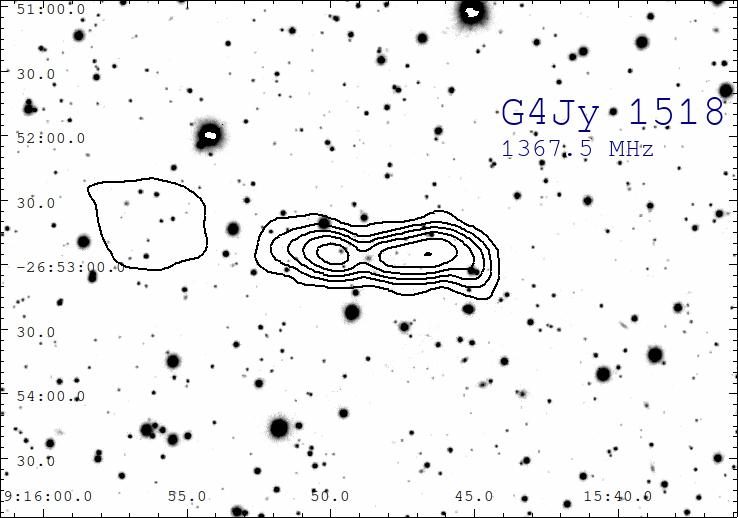}
    \includegraphics[scale=0.225]{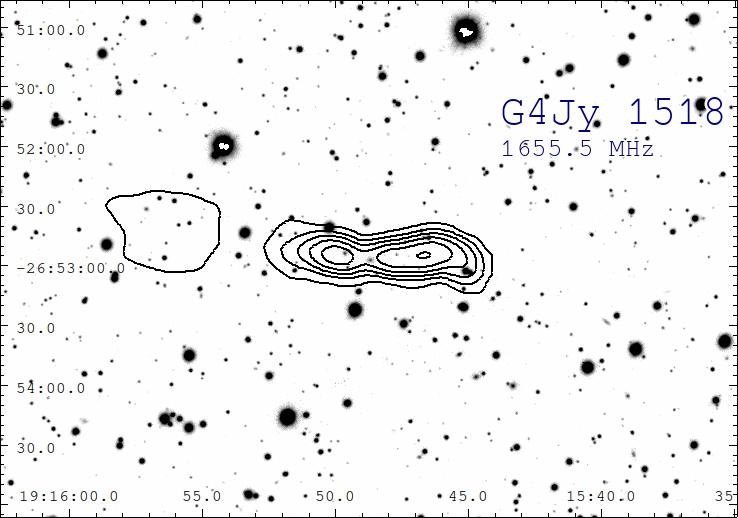}
    \includegraphics[scale=0.225]{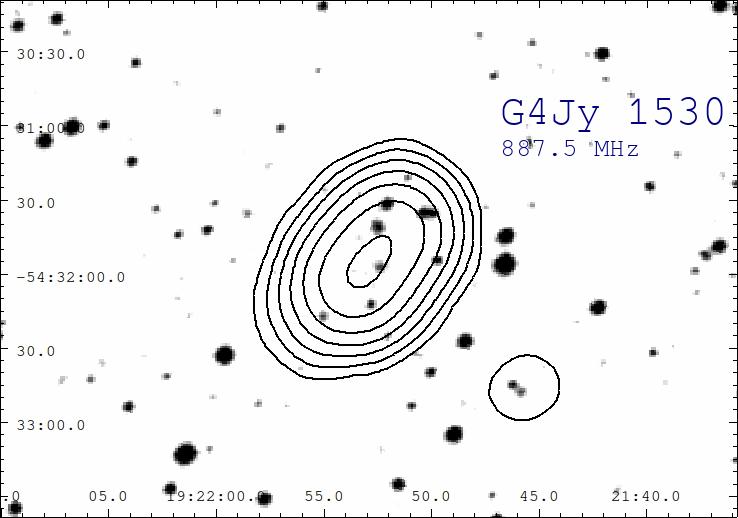}
    \includegraphics[scale=0.225]{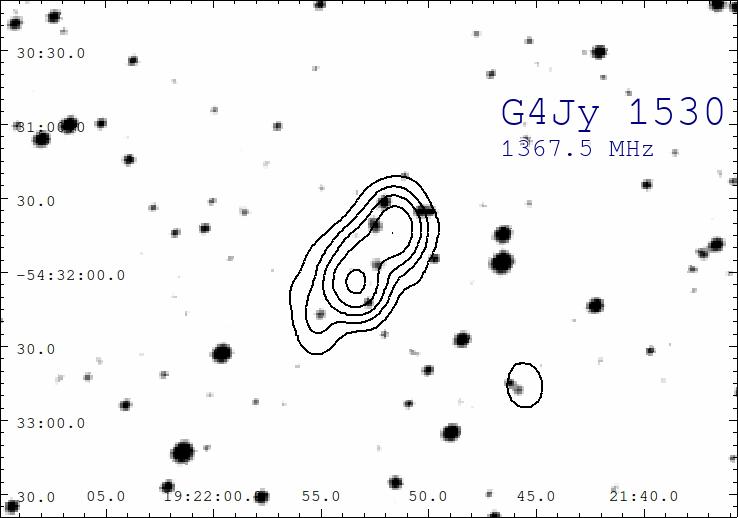}
    \includegraphics[scale=0.225]{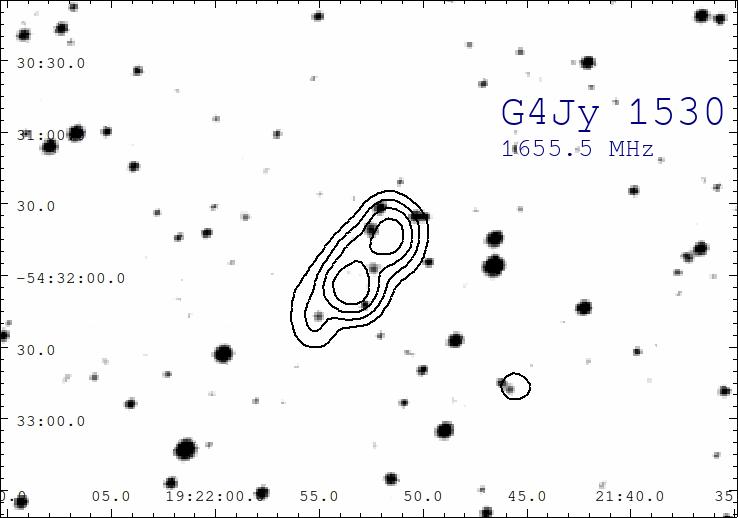}
    \includegraphics[scale=0.225]{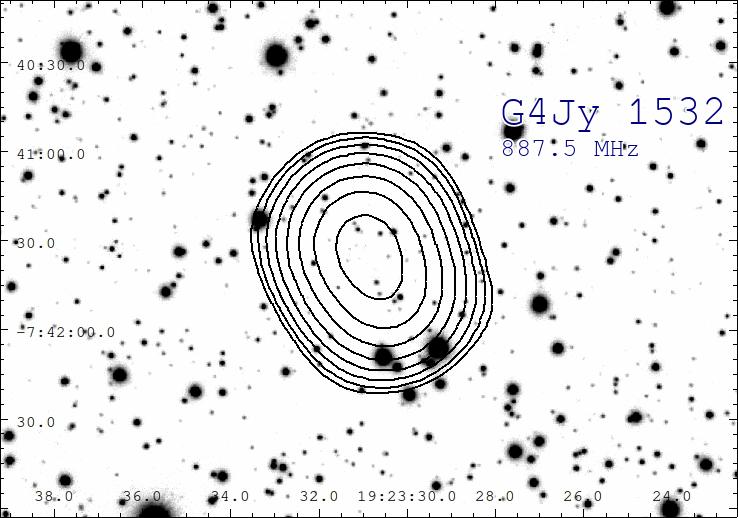}
    \includegraphics[scale=0.225]{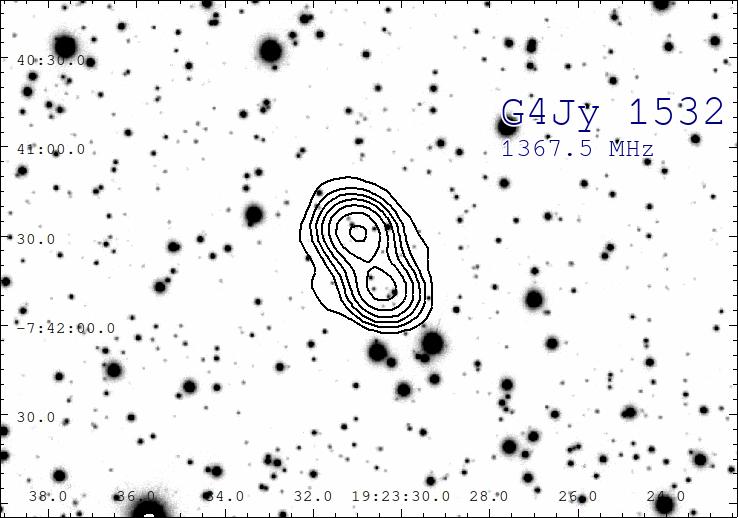}
    \includegraphics[scale=0.225]{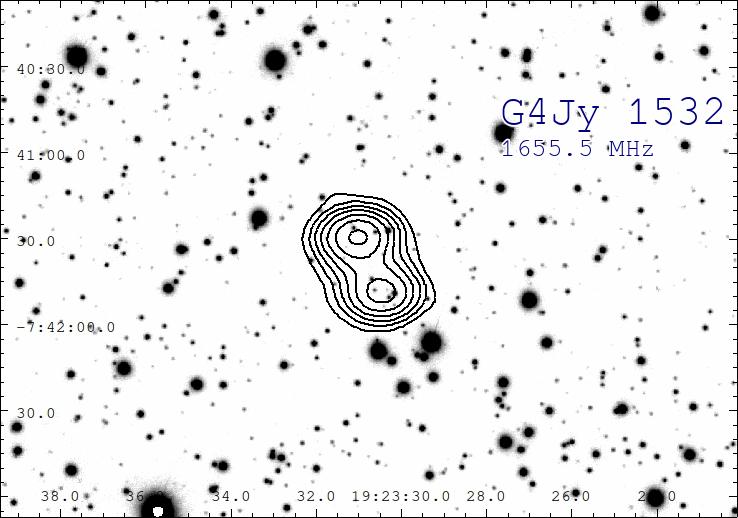}
    \includegraphics[scale=0.225]{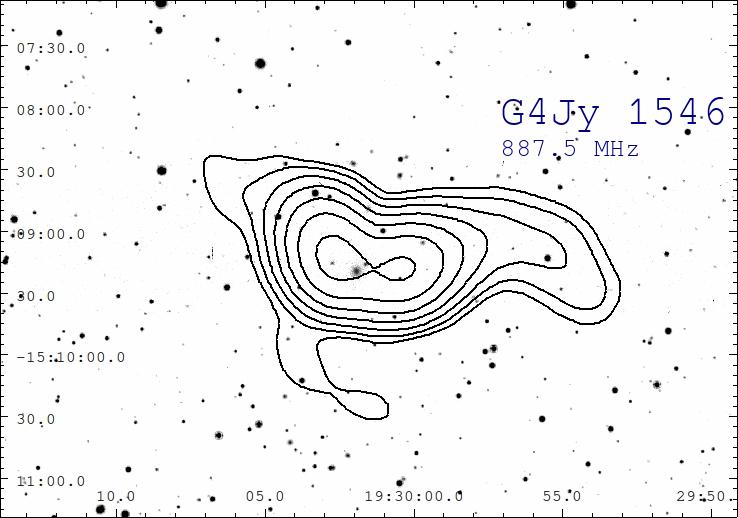}
    \includegraphics[scale=0.225]{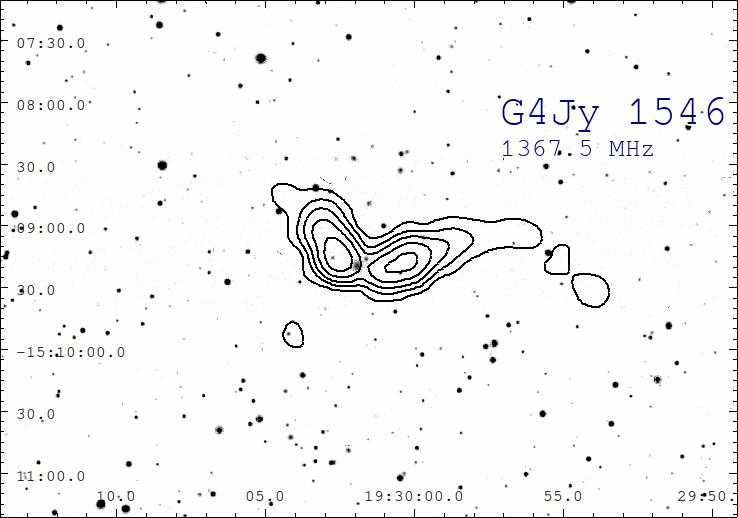}
    \includegraphics[scale=0.225]{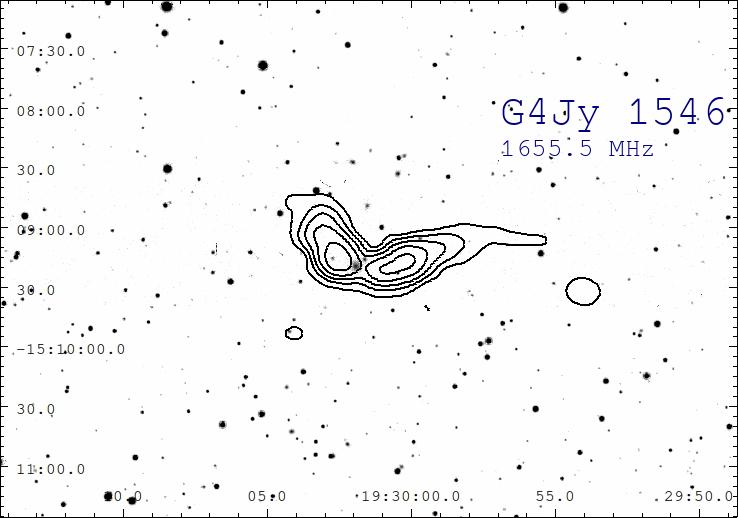}
    \includegraphics[scale=0.225]{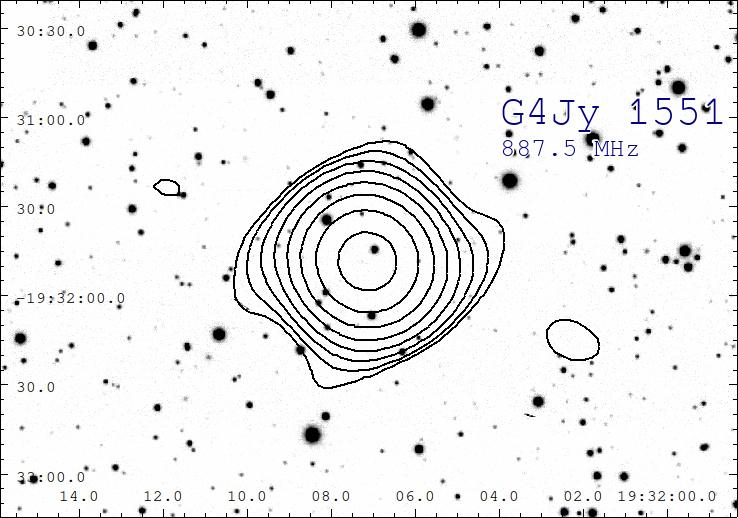}
    \includegraphics[scale=0.225]{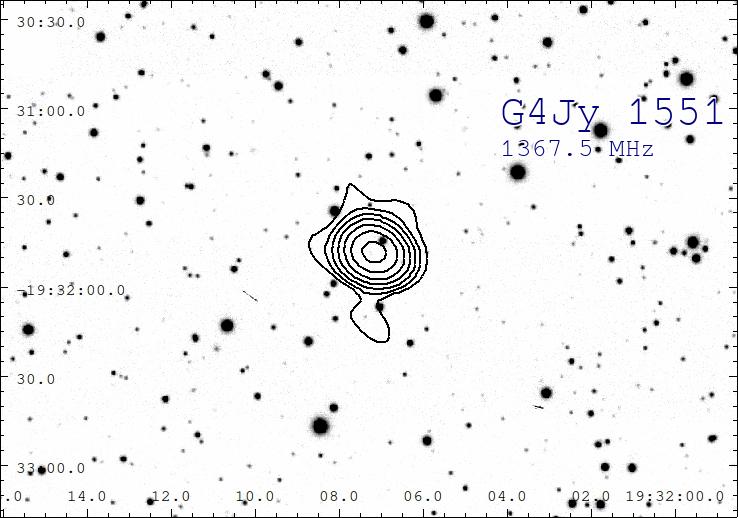}
    \includegraphics[scale=0.225]{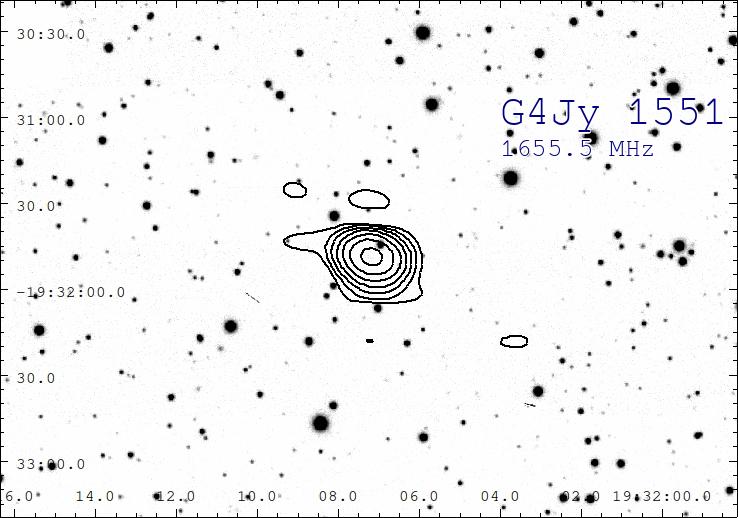}

    \caption{}
    \label{AN}
\end{figure*}
\clearpage
 \begin{figure*}
    \centering
    \includegraphics[scale=0.225]{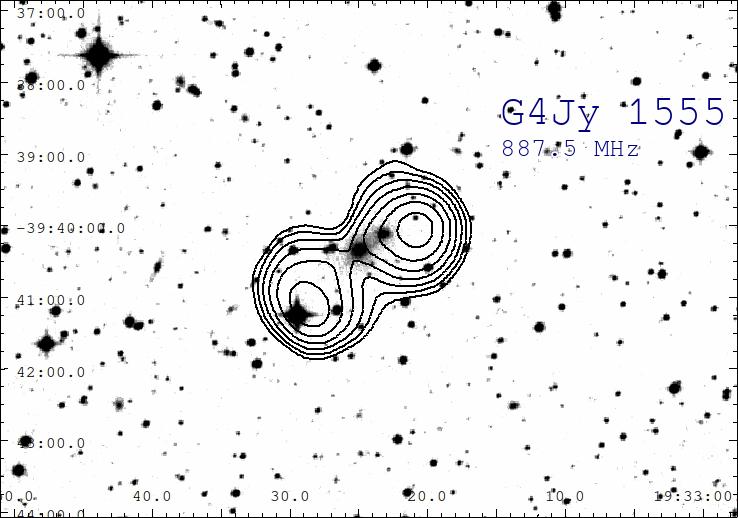}
    \includegraphics[scale=0.225]{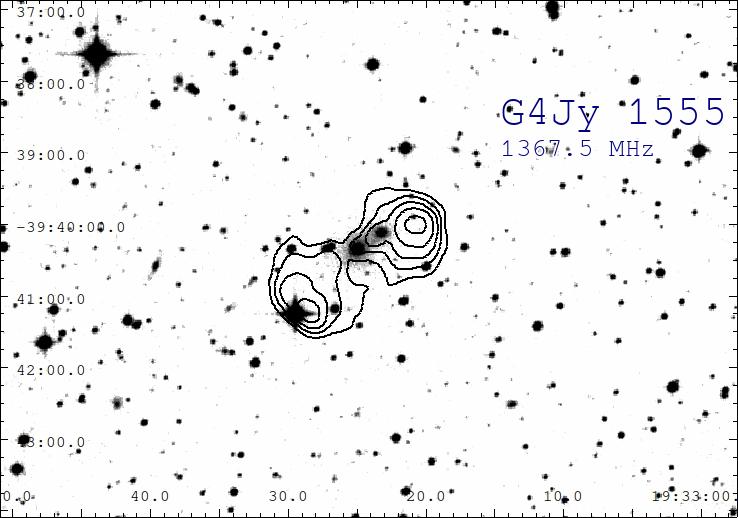}
    \includegraphics[scale=0.225]{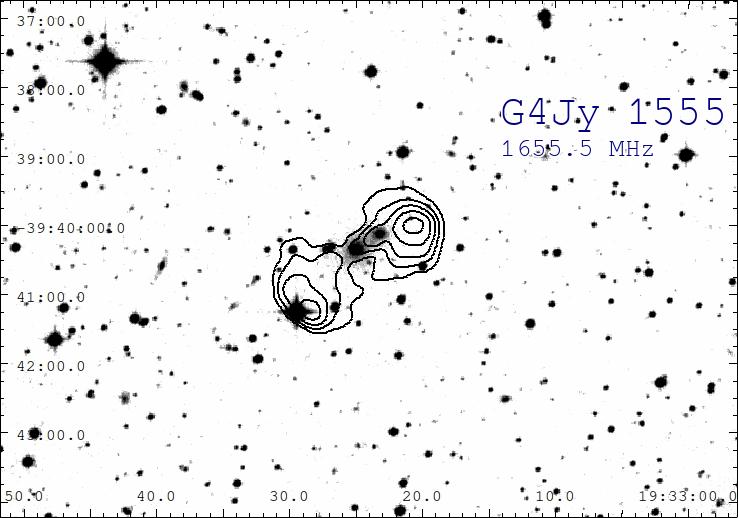}
    \includegraphics[scale=0.225]{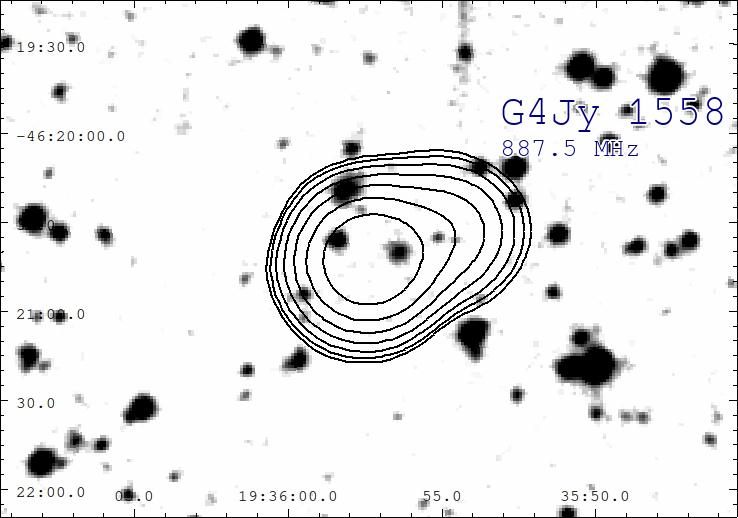}
    \includegraphics[scale=0.225]{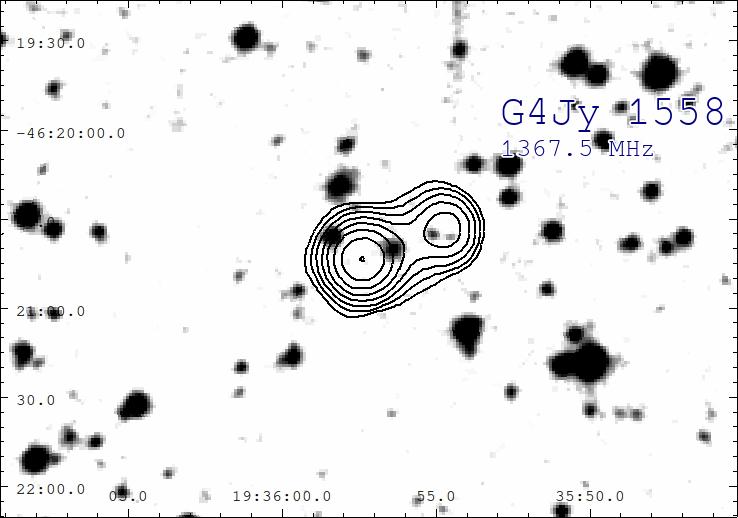}
    \includegraphics[scale=0.225]{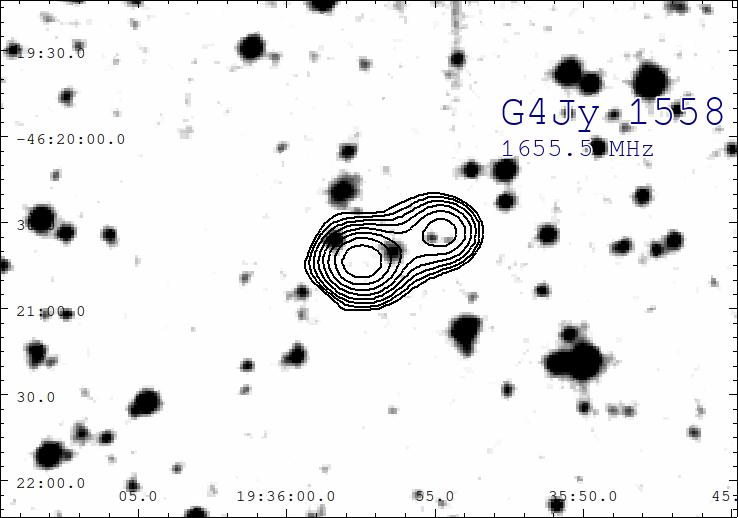}
    \includegraphics[scale=0.225]{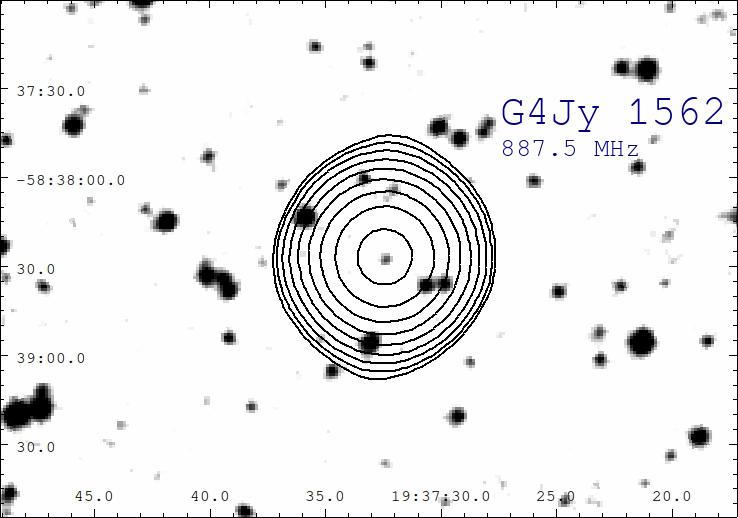}
    \includegraphics[scale=0.225]{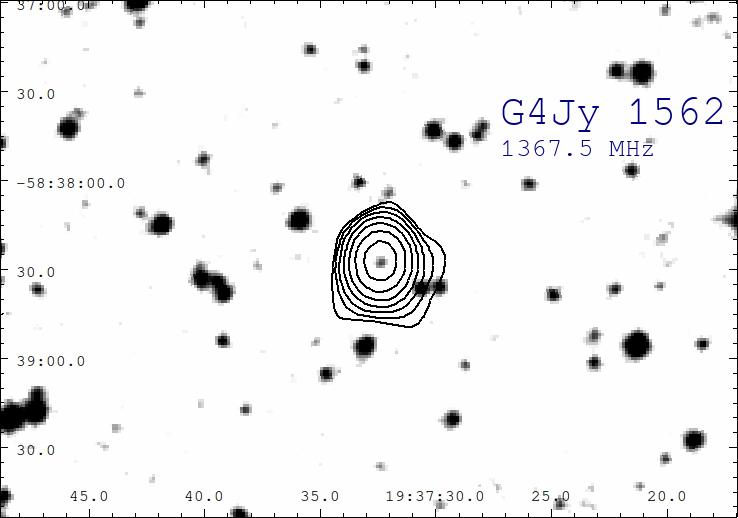}
    \includegraphics[scale=0.225]{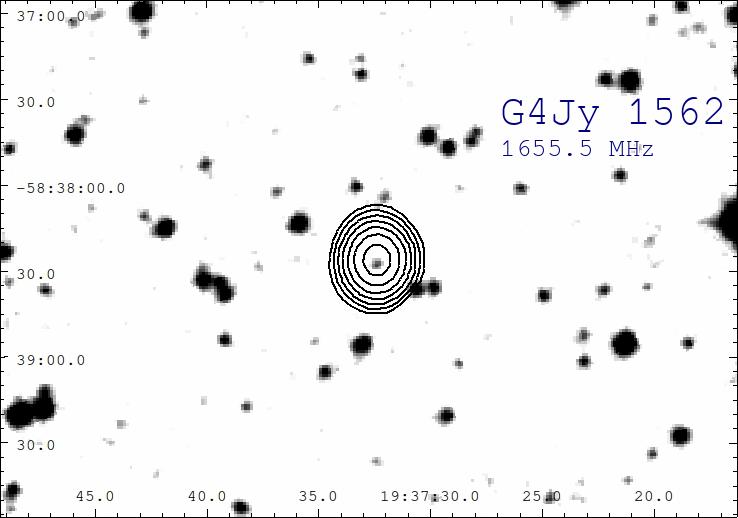}
    \includegraphics[scale=0.225]{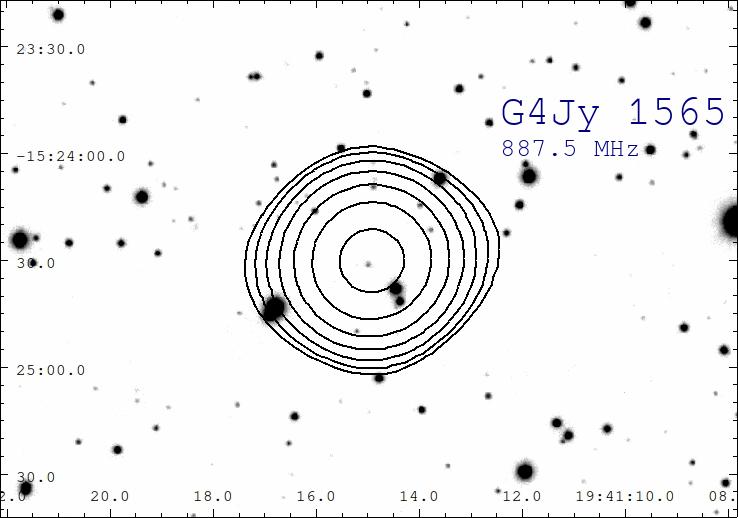}
    \includegraphics[scale=0.225]{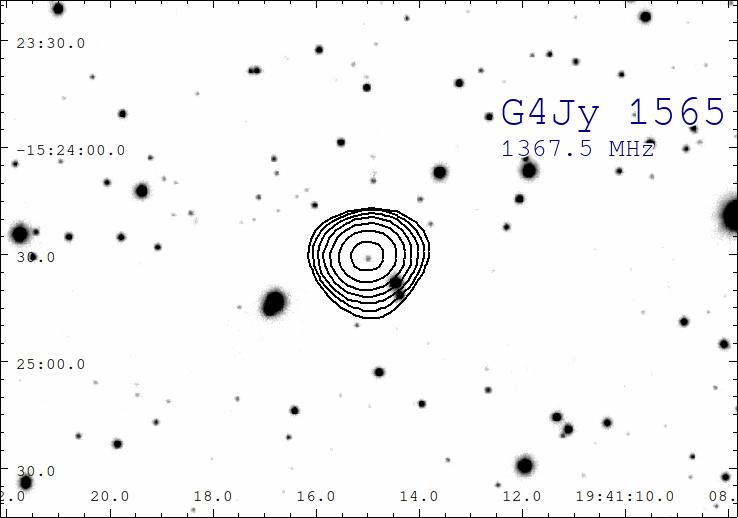}
    \includegraphics[scale=0.225]{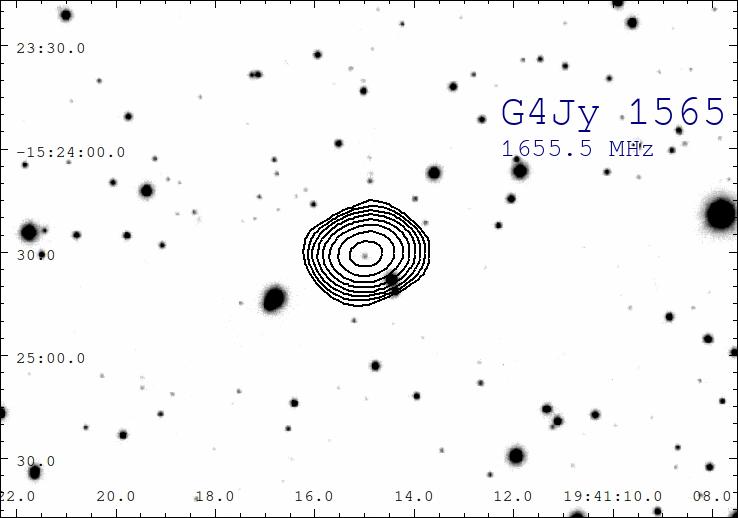}
    \includegraphics[scale=0.225]{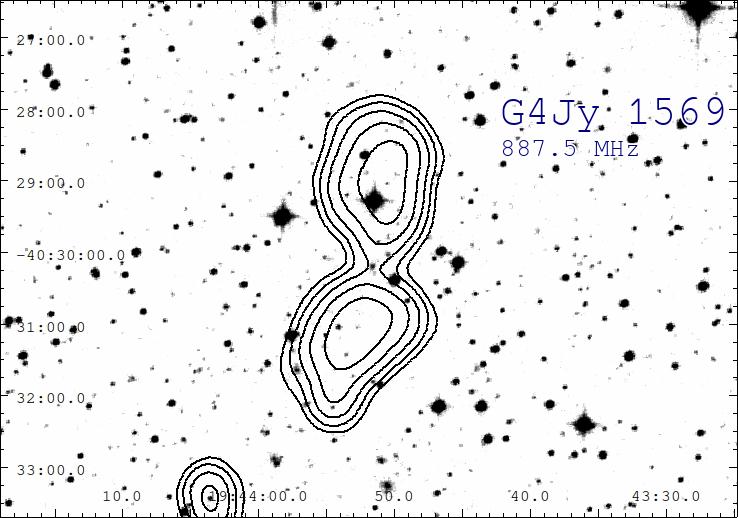}
    \includegraphics[scale=0.225]{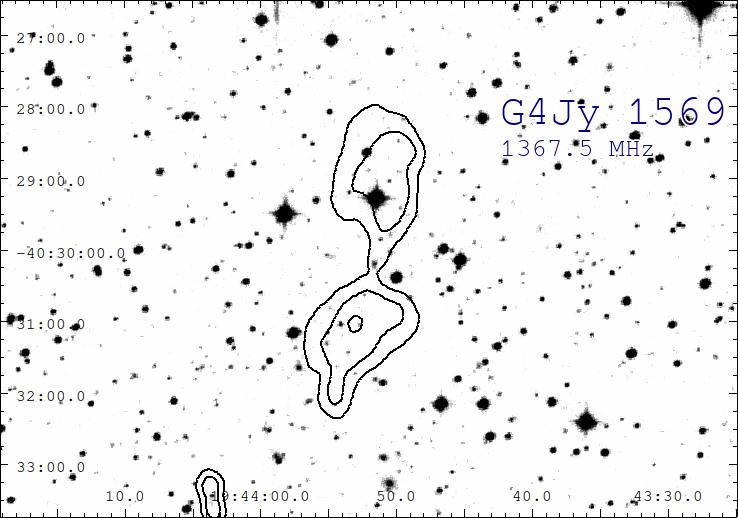}
    \includegraphics[scale=0.225]{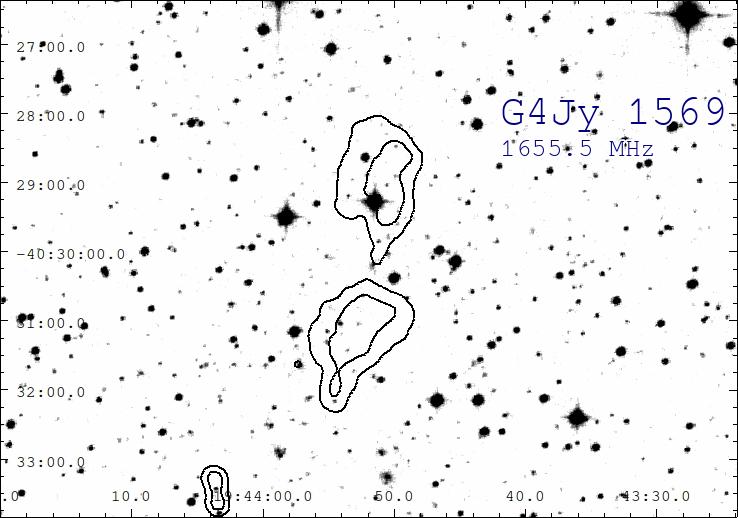}

    \caption{}
    \label{AO}
\end{figure*}
\clearpage
 \begin{figure*}
    \centering
    \includegraphics[scale=0.225]{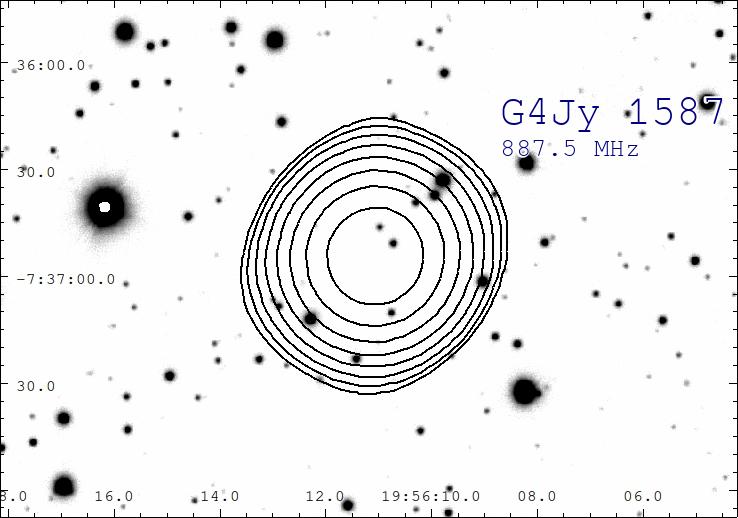}
    \includegraphics[scale=0.225]{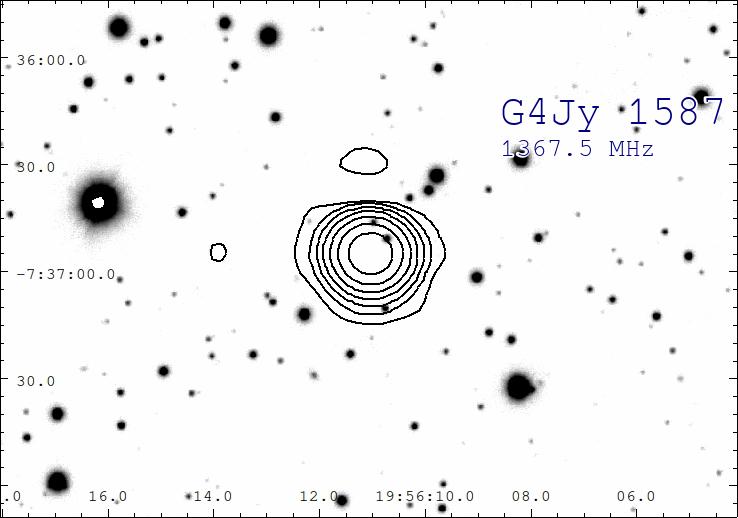}
    \includegraphics[scale=0.225]{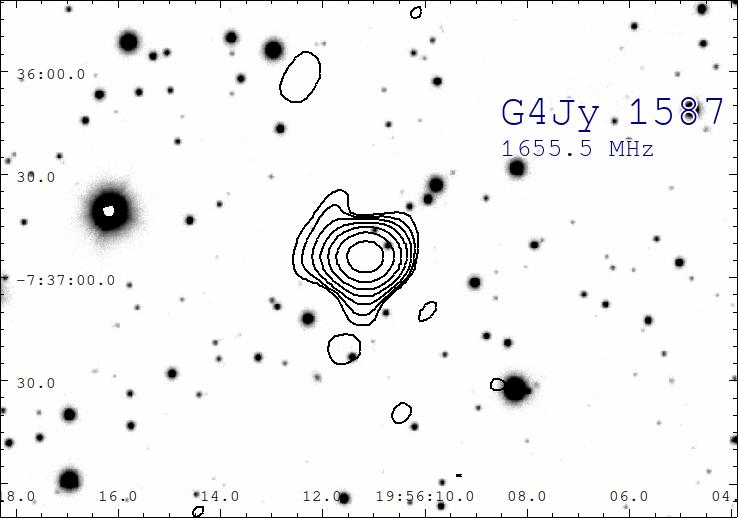}
    \includegraphics[scale=0.225]{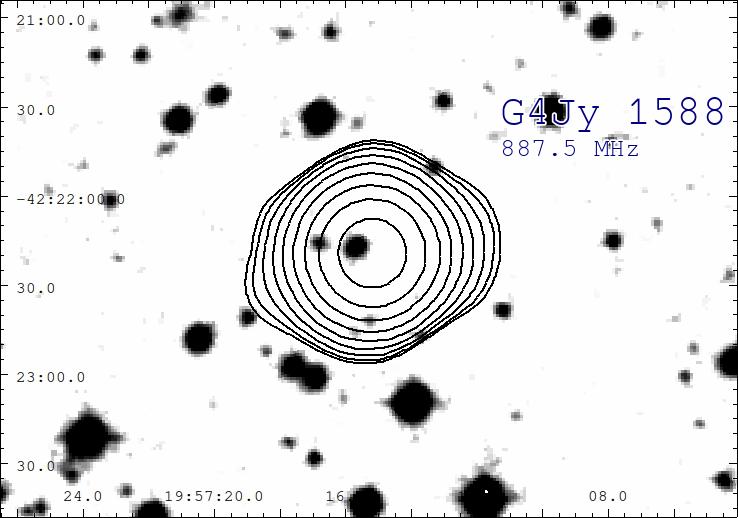}
    \includegraphics[scale=0.225]{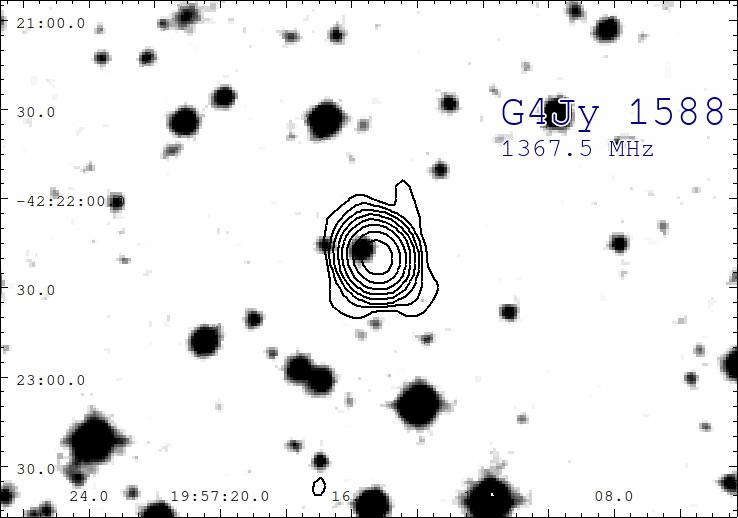}
    \includegraphics[scale=0.225]{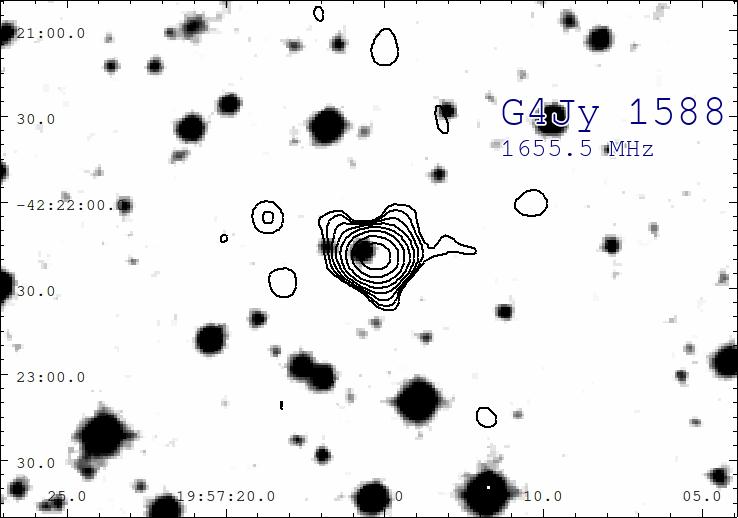}
    \includegraphics[scale=0.225]{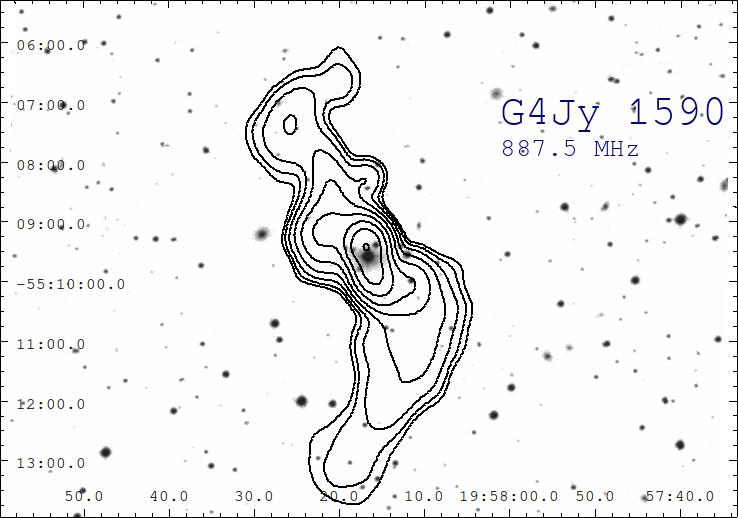}
    \includegraphics[scale=0.225]{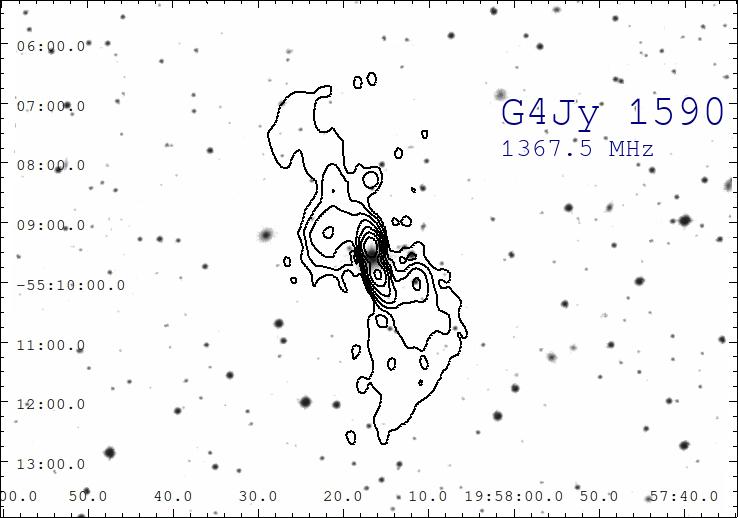}
    \includegraphics[scale=0.225]{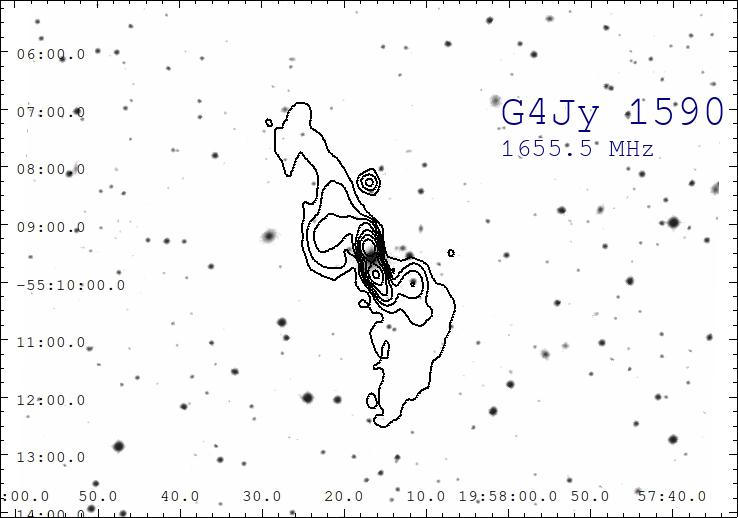}
    \includegraphics[scale=0.225]{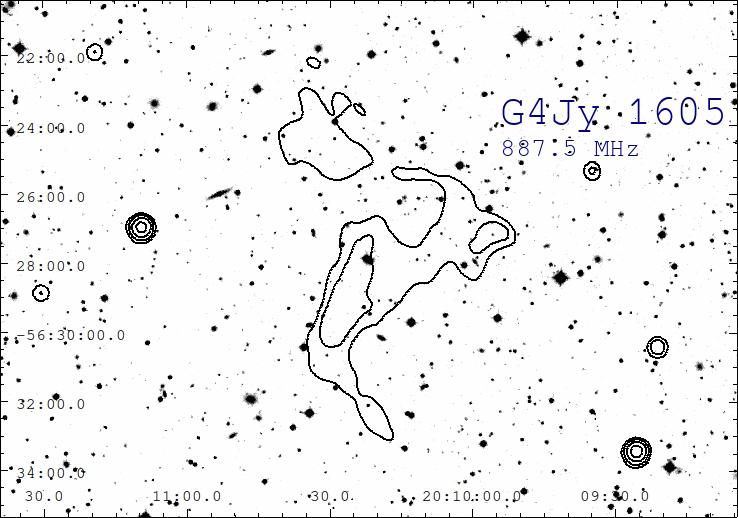}
    \includegraphics[scale=0.225]{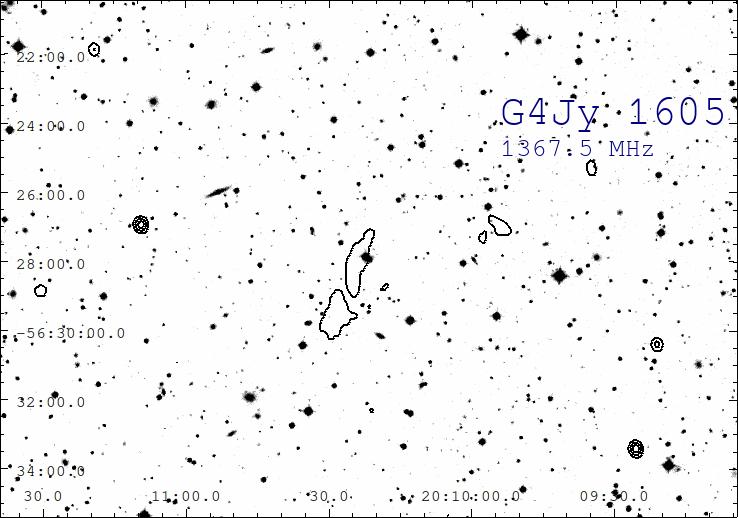}
    \includegraphics[scale=0.225]{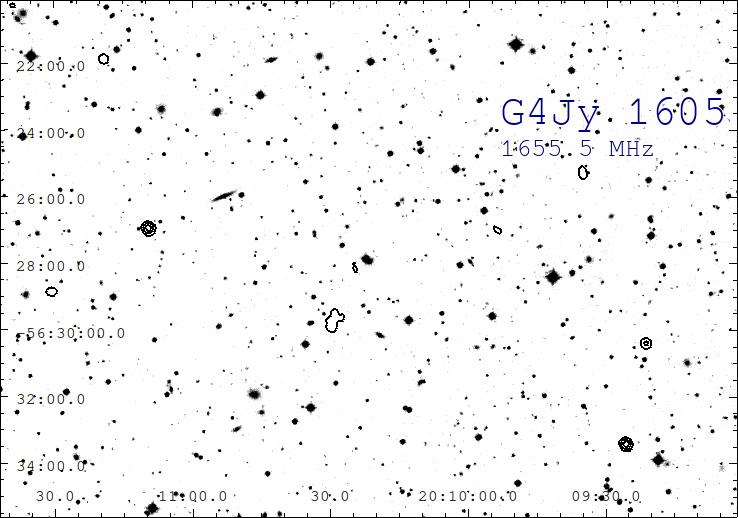}
    \includegraphics[scale=0.225]{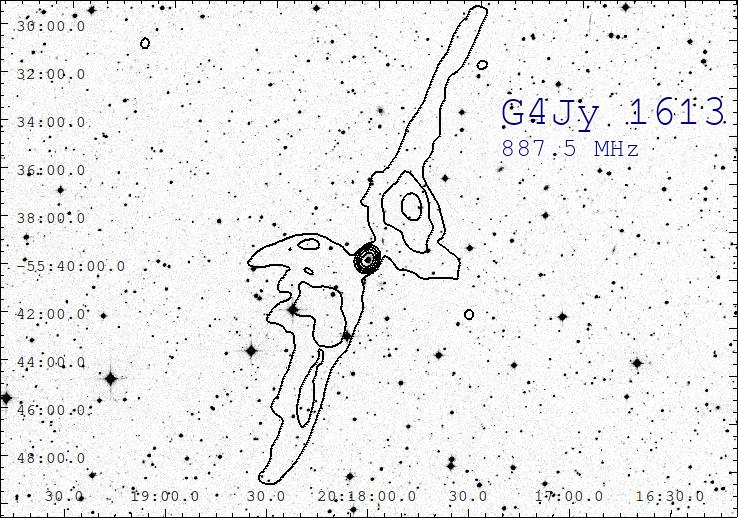}
    \includegraphics[scale=0.225]{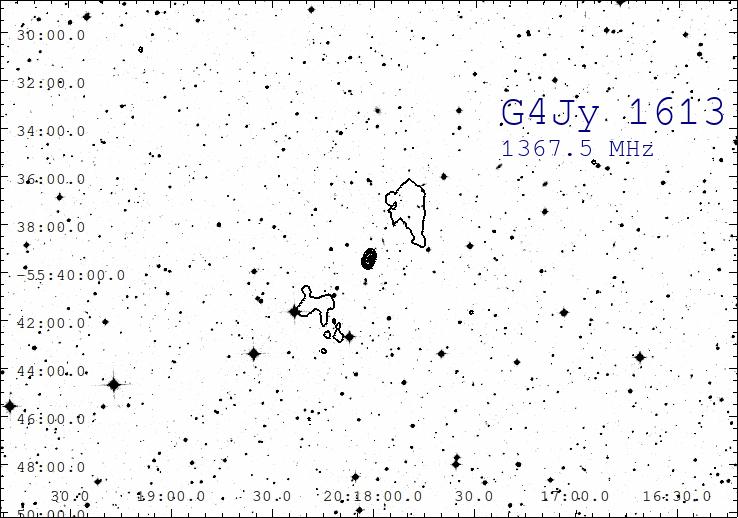}
    \includegraphics[scale=0.225]{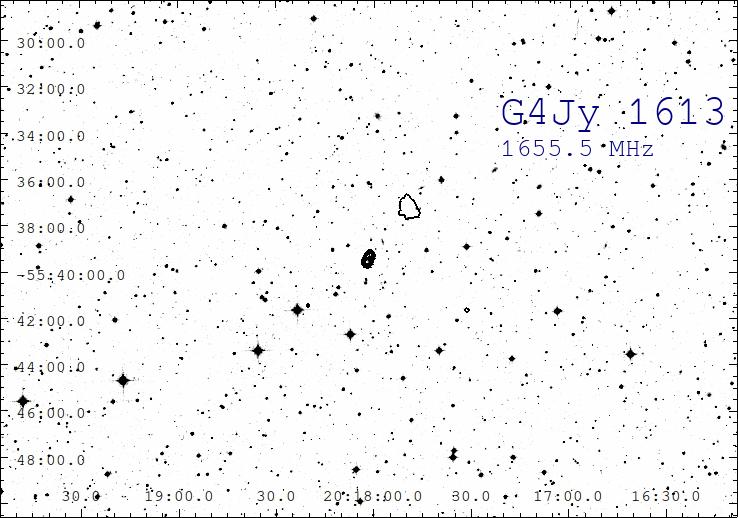}

    \caption{}
    \label{AP}
\end{figure*}
\clearpage
 \begin{figure*}
    \centering
    \includegraphics[scale=0.225]{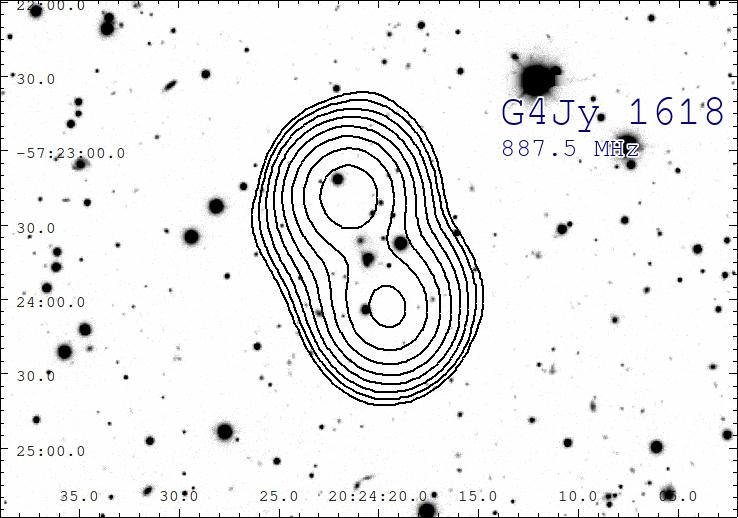}
    \includegraphics[scale=0.225]{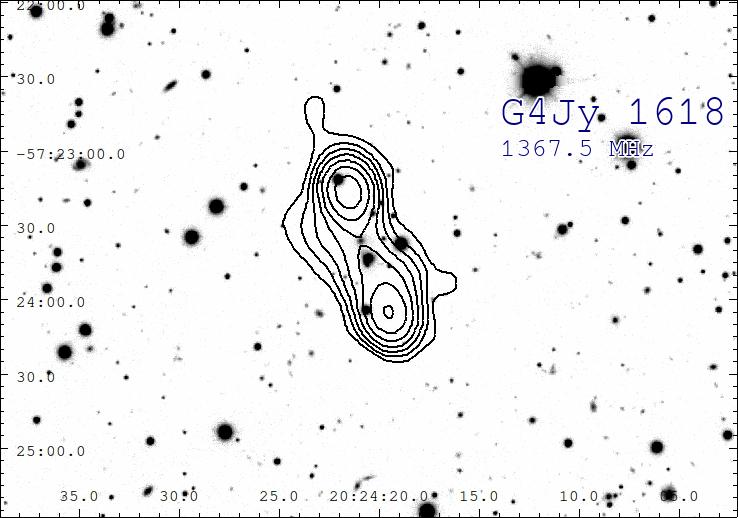}
    \includegraphics[scale=0.225]{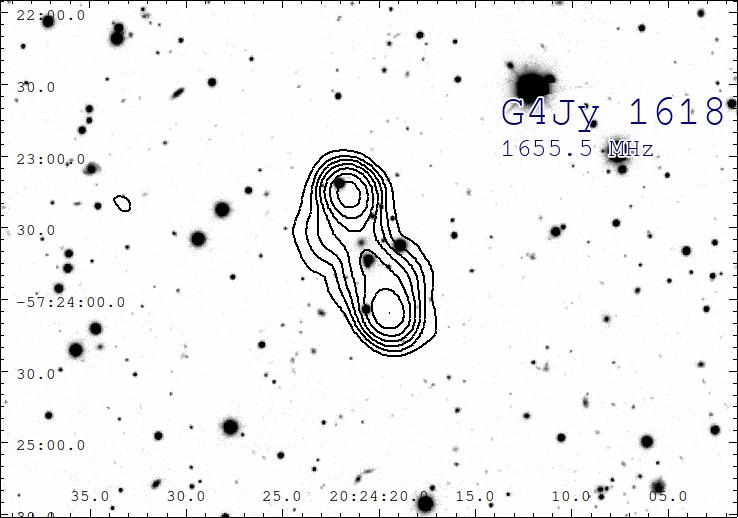}
    \includegraphics[scale=0.225]{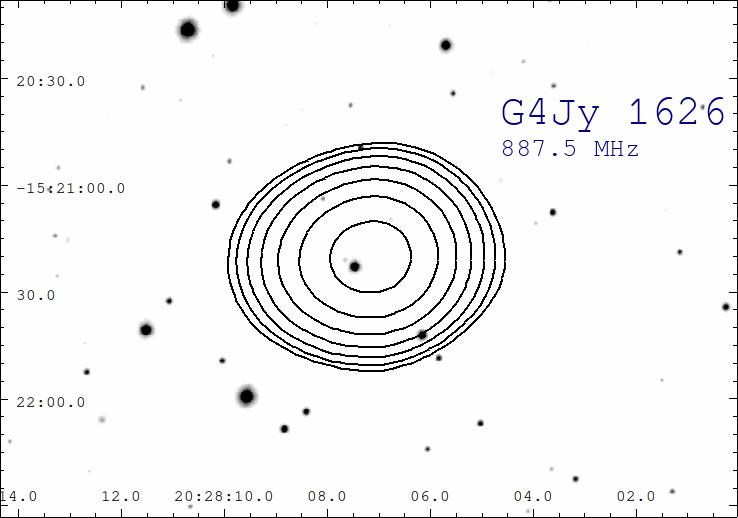}
    \includegraphics[scale=0.225]{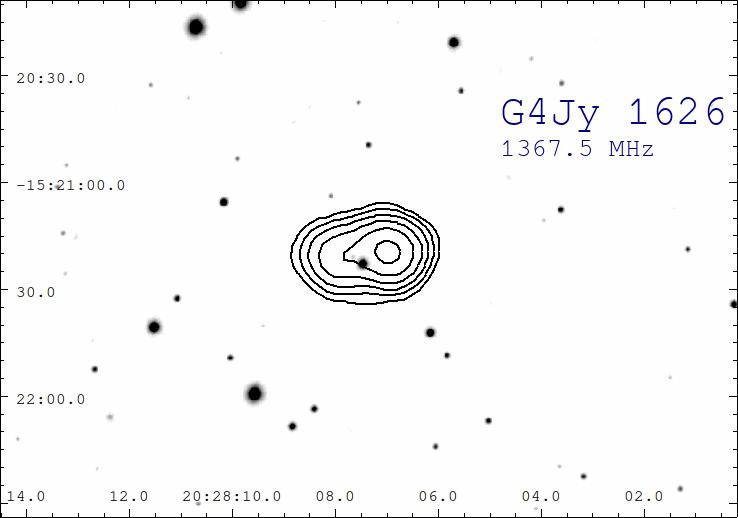}
    \includegraphics[scale=0.225]{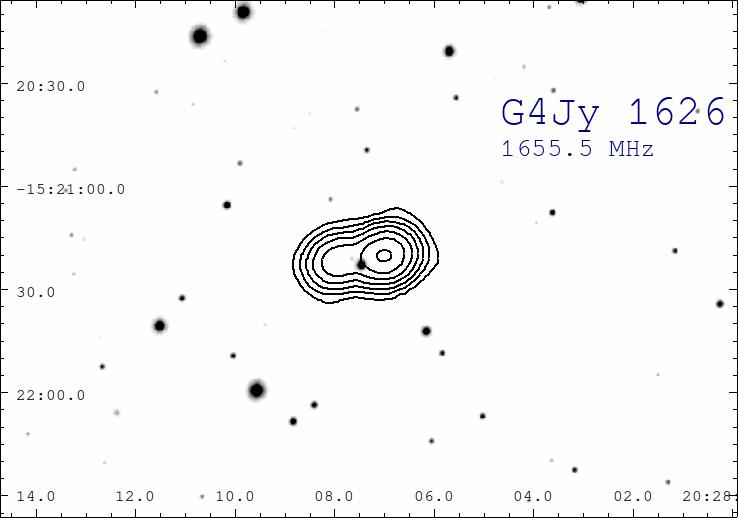}
    \includegraphics[scale=0.225]{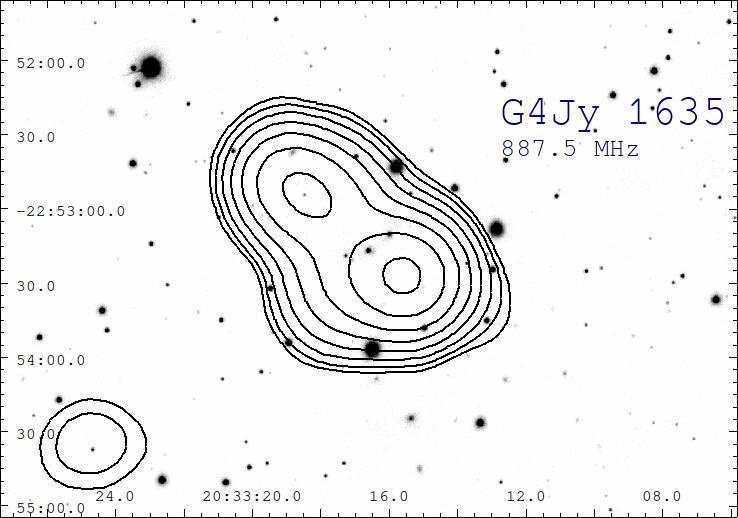}
    \includegraphics[scale=0.225]{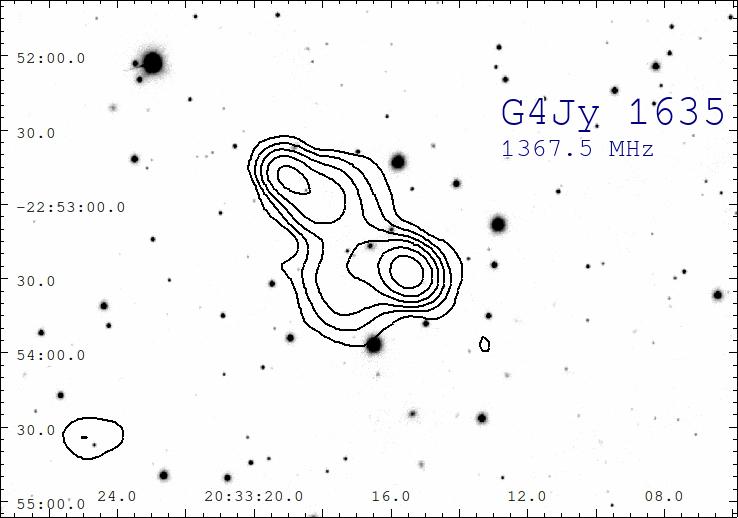}
    \includegraphics[scale=0.225]{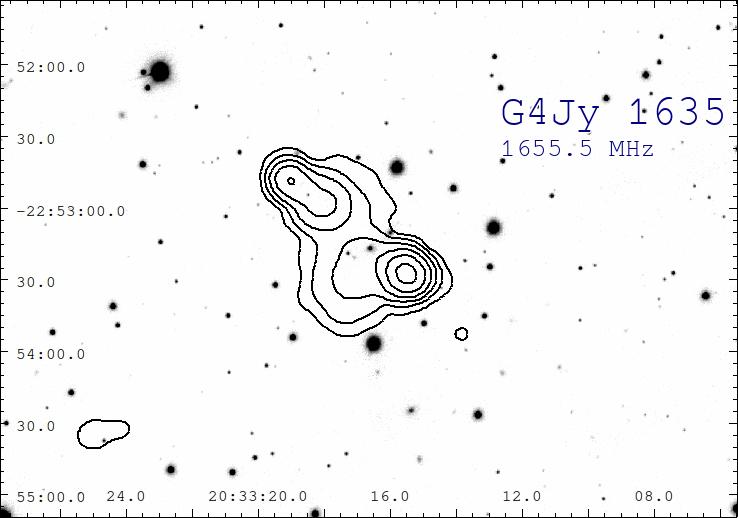}
    \includegraphics[scale=0.225]{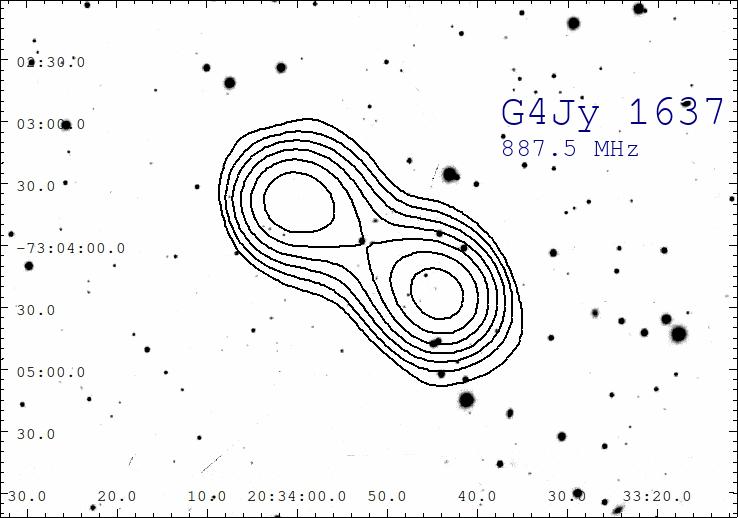}
    \includegraphics[scale=0.225]{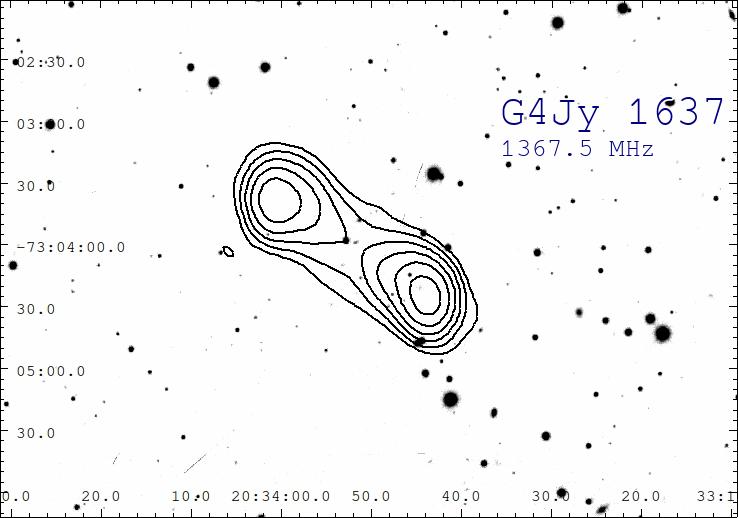}
    \includegraphics[scale=0.225]{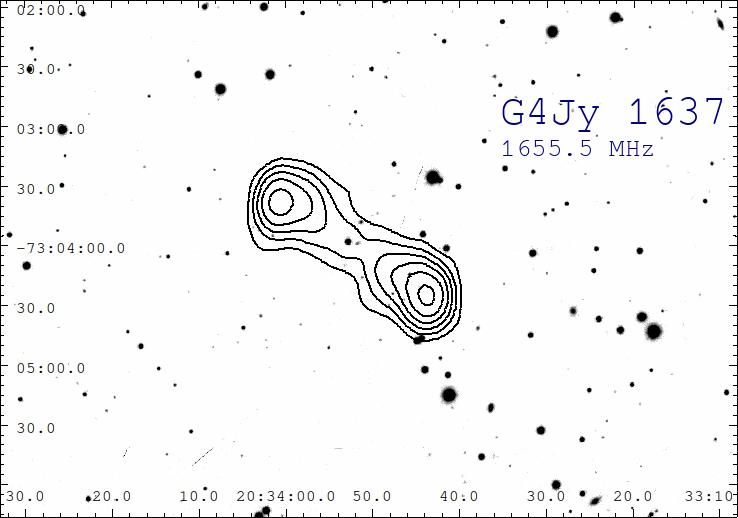}
    \includegraphics[scale=0.225]{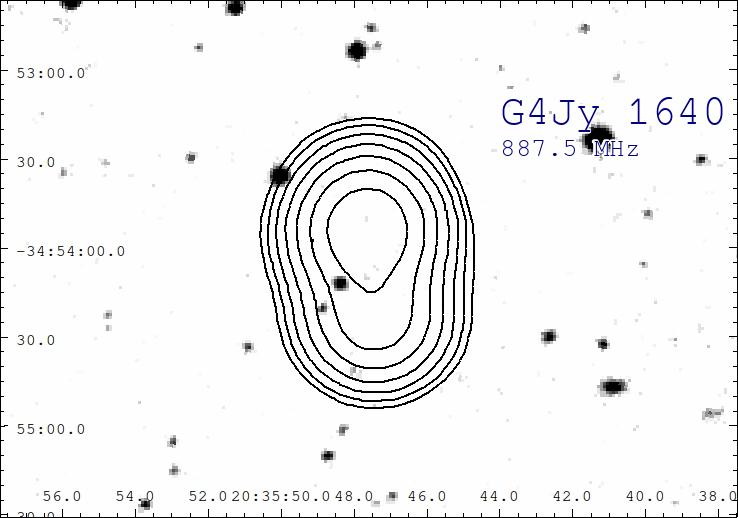}
    \includegraphics[scale=0.225]{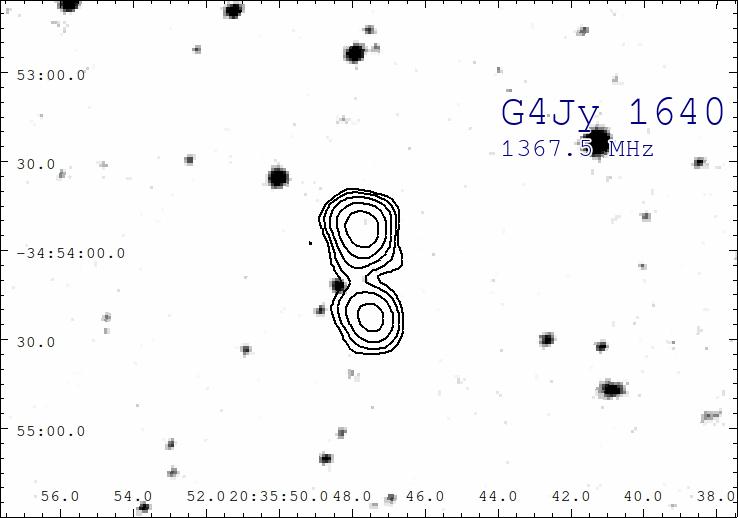}
    \includegraphics[scale=0.225]{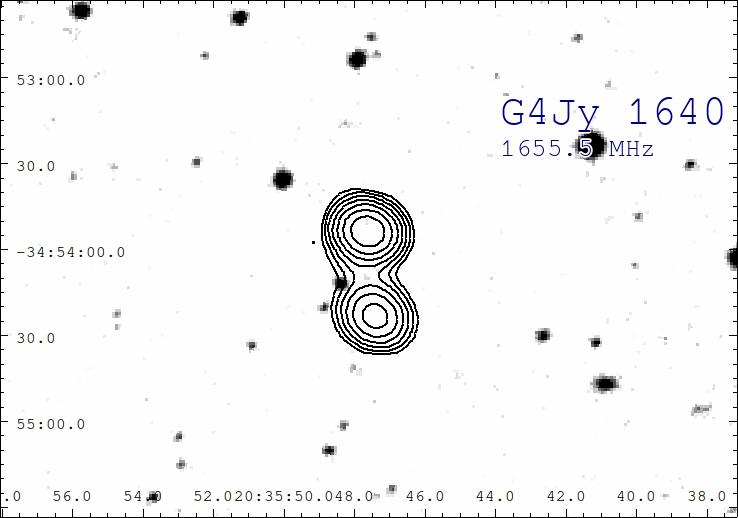}

    \caption{}
    \label{AQ}
\end{figure*}
\clearpage
 \begin{figure*}
    \centering
    \includegraphics[scale=0.225]{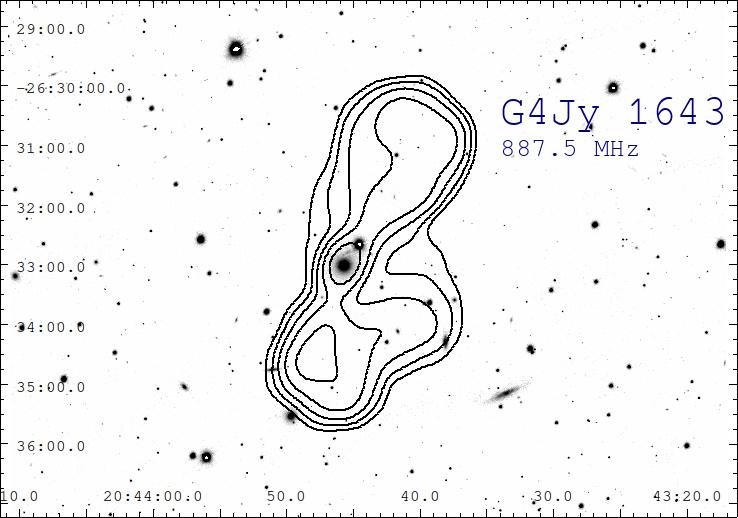}
    \includegraphics[scale=0.225]{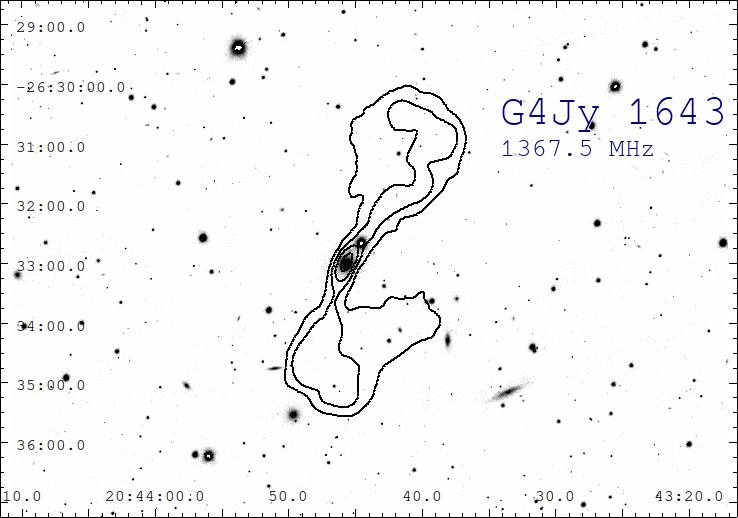}
    \includegraphics[scale=0.225]{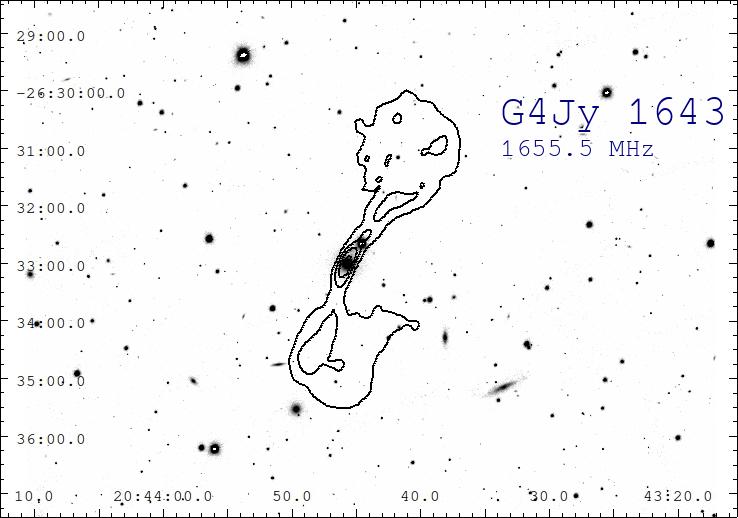}
    \includegraphics[scale=0.225]{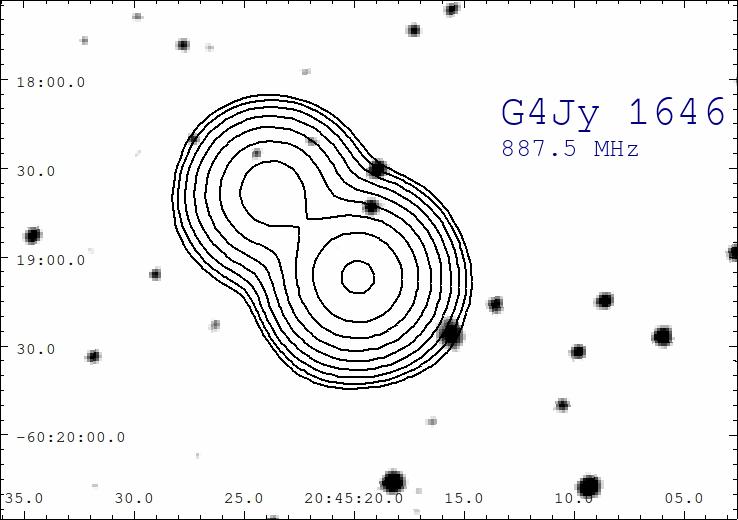}
    \includegraphics[scale=0.225]{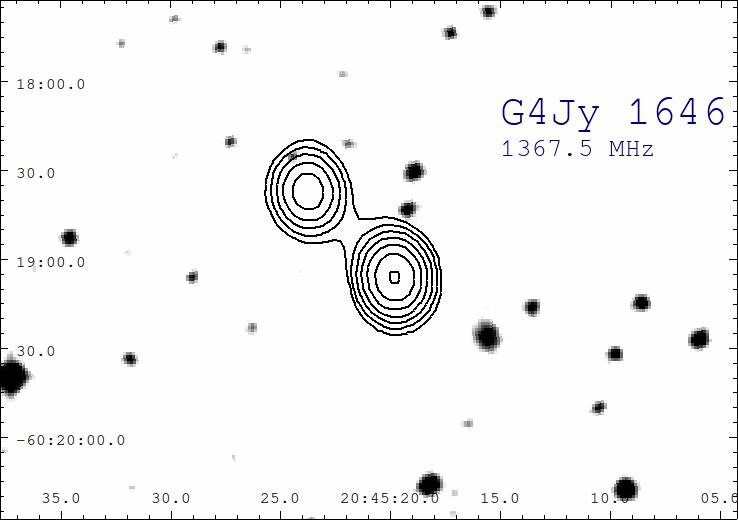}
    \includegraphics[scale=0.225]{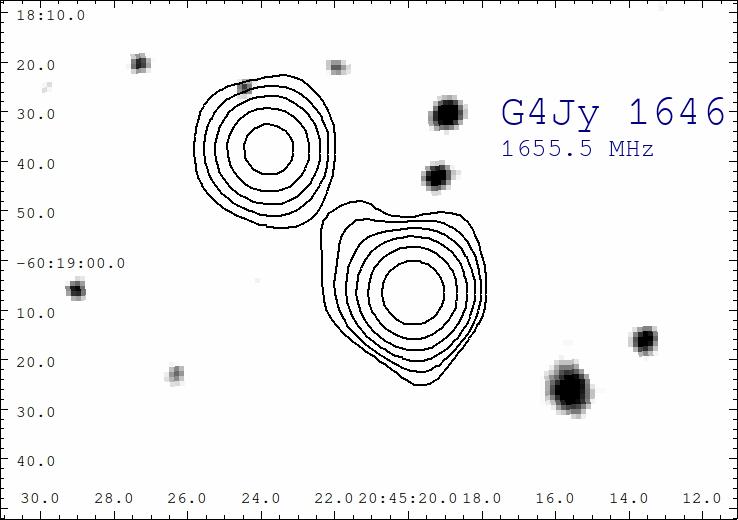}
    \includegraphics[scale=0.225]{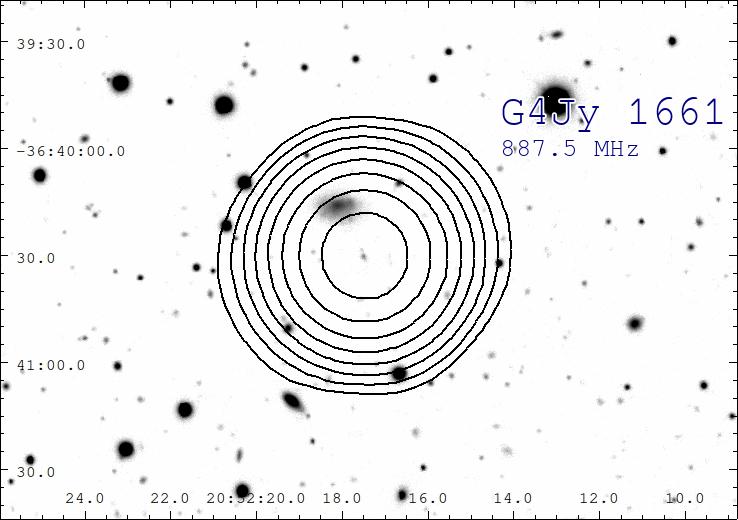}
    \includegraphics[scale=0.225]{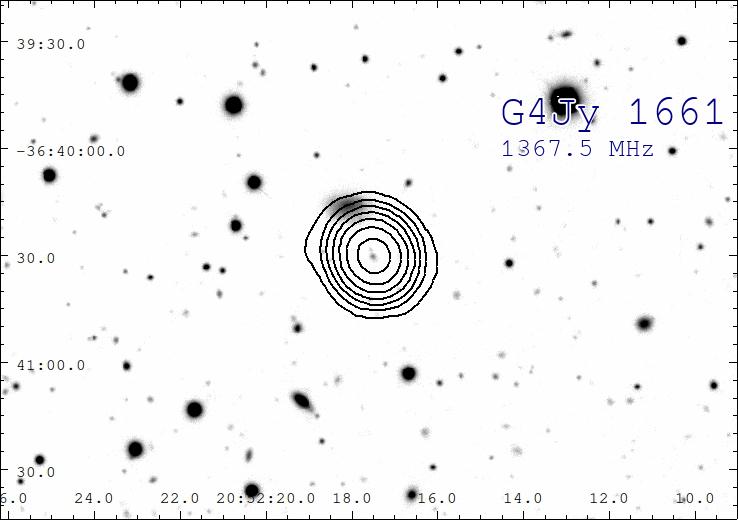}
    \includegraphics[scale=0.225]{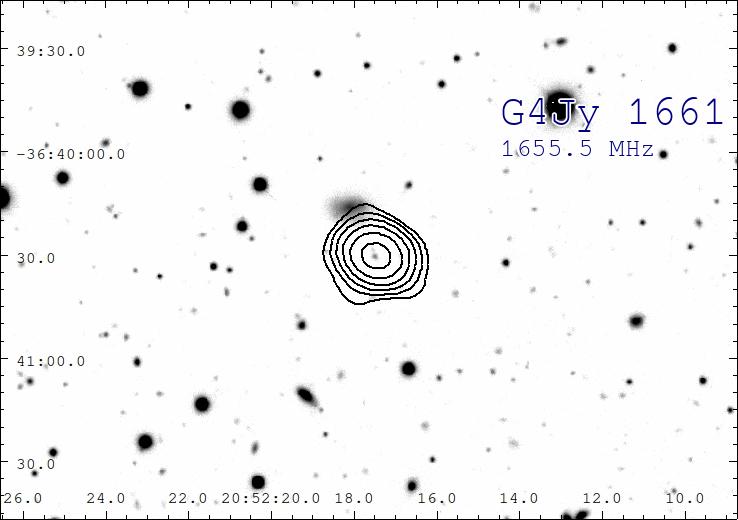}
    \includegraphics[scale=0.225]{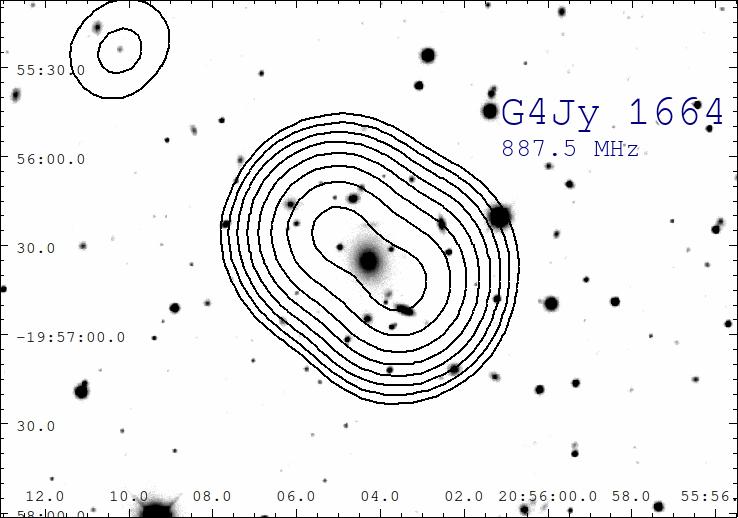}
    \includegraphics[scale=0.225]{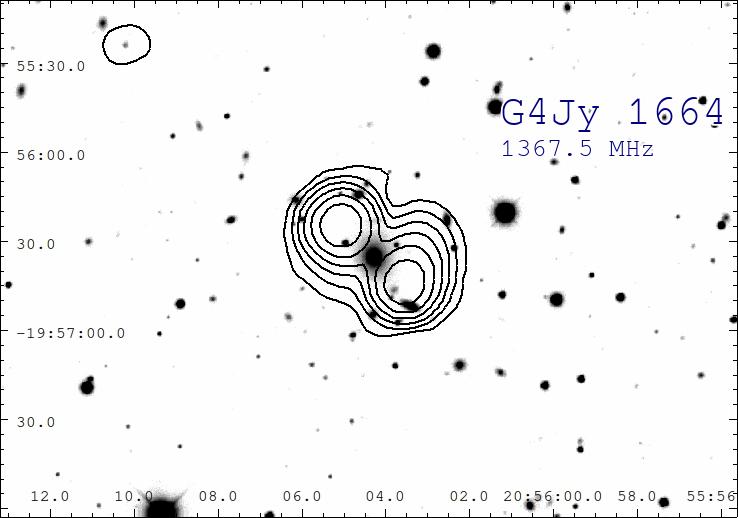}
    \includegraphics[scale=0.225]{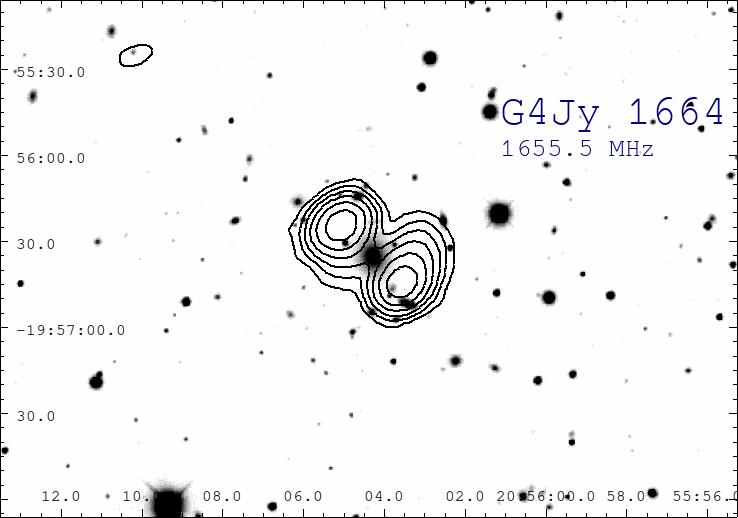}
    \includegraphics[scale=0.225]{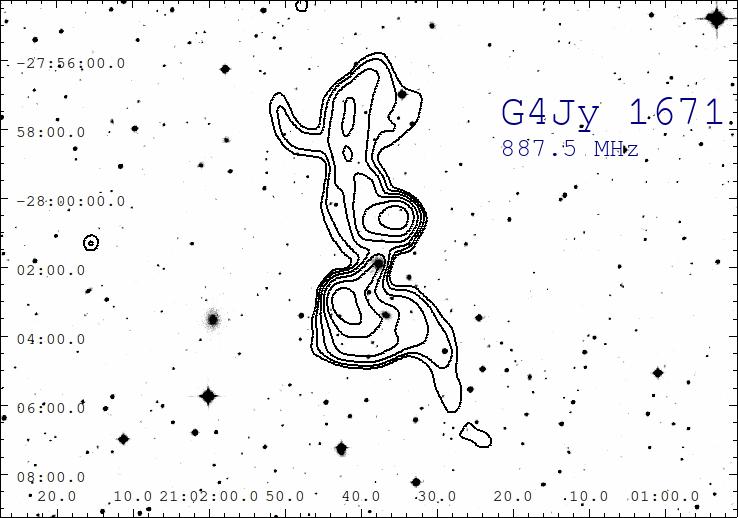}
    \includegraphics[scale=0.225]{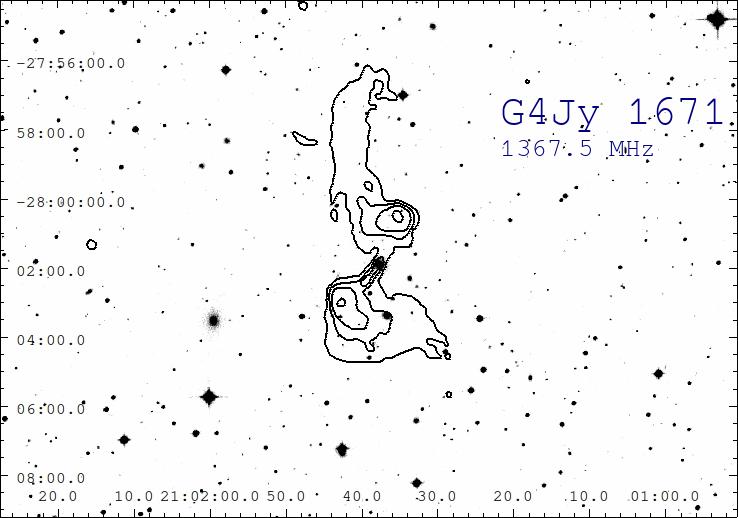}
    \includegraphics[scale=0.225]{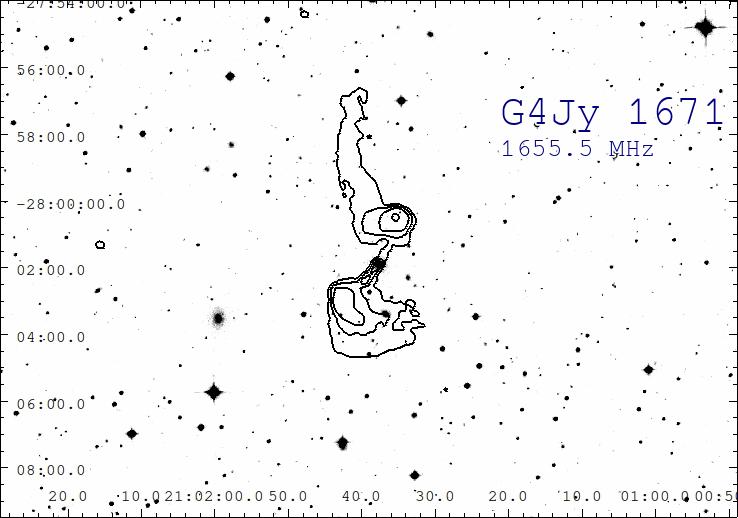}

    \caption{}
    \label{AR}
\end{figure*}
\clearpage
 \begin{figure*}
    \centering
    \includegraphics[scale=0.225]{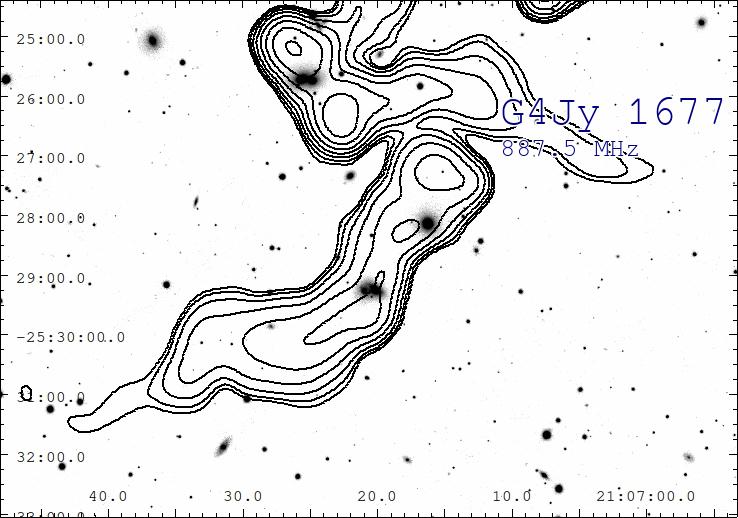}
    \includegraphics[scale=0.225]{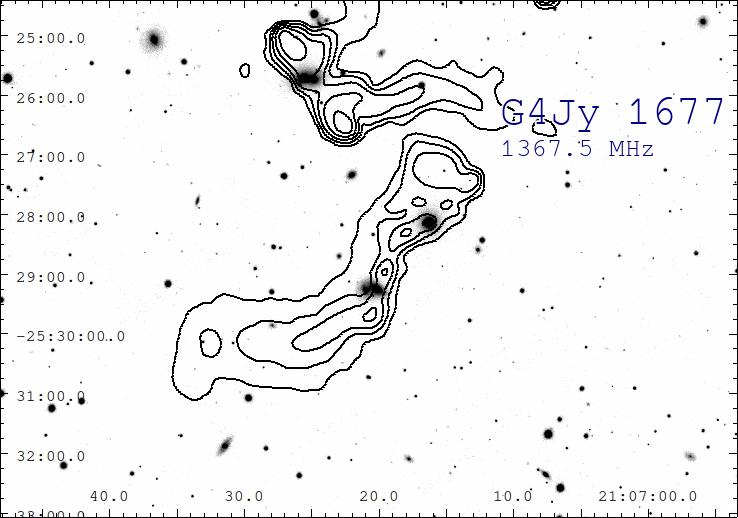}
    \includegraphics[scale=0.225]{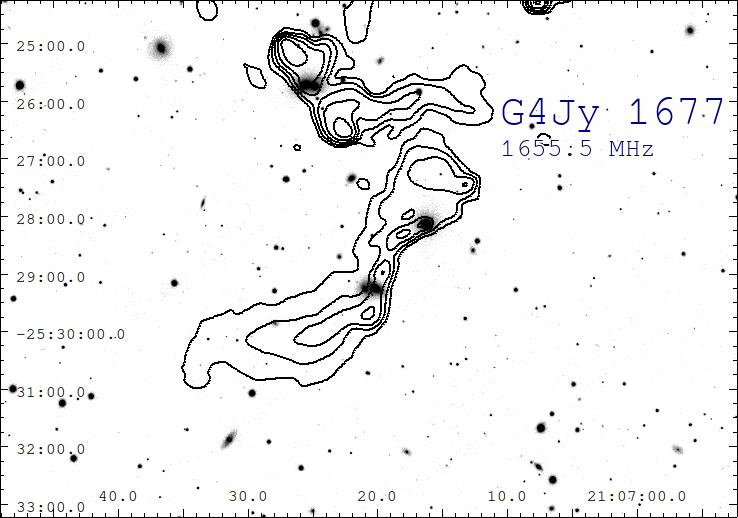}
    \includegraphics[scale=0.225]{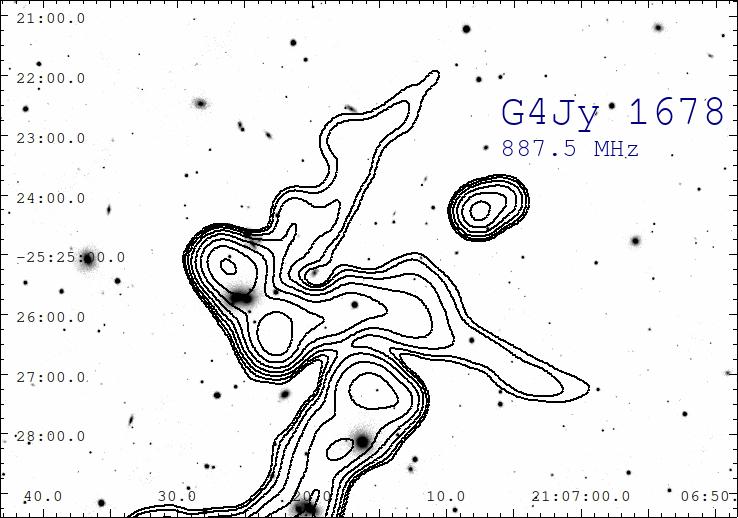}
    \includegraphics[scale=0.225]{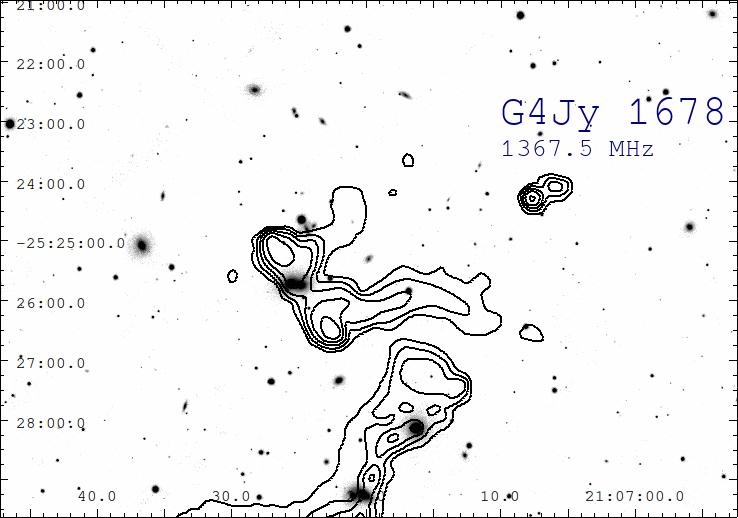}
    \includegraphics[scale=0.225]{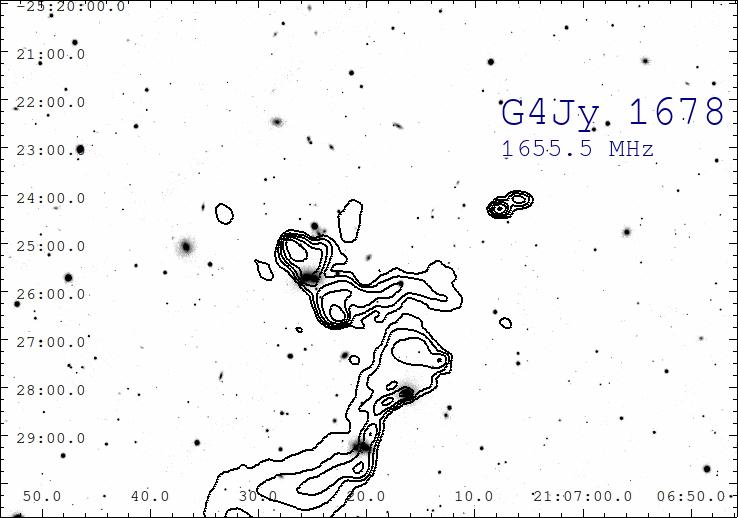}
    \includegraphics[scale=0.225]{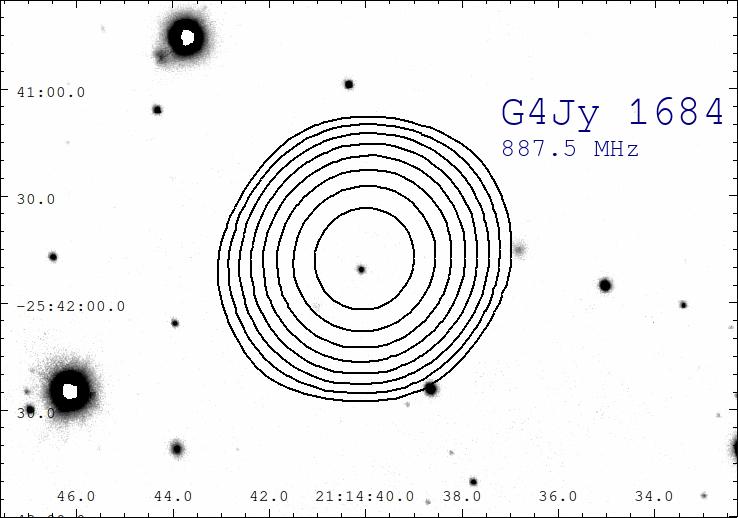}
    \includegraphics[scale=0.225]{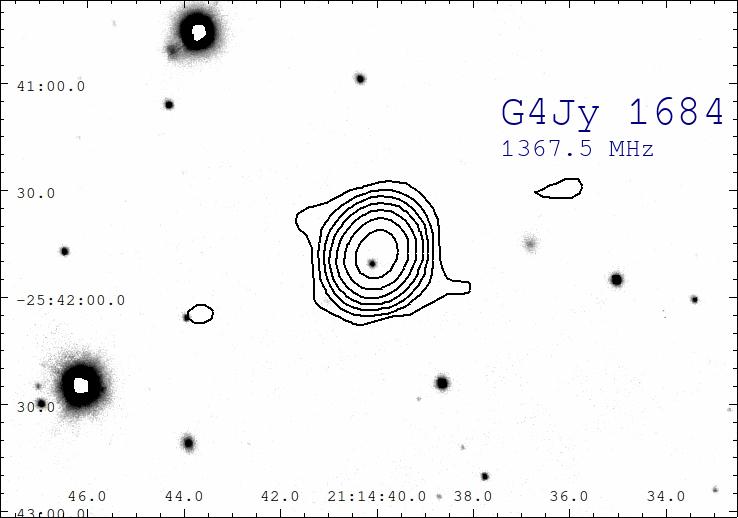}
    \includegraphics[scale=0.225]{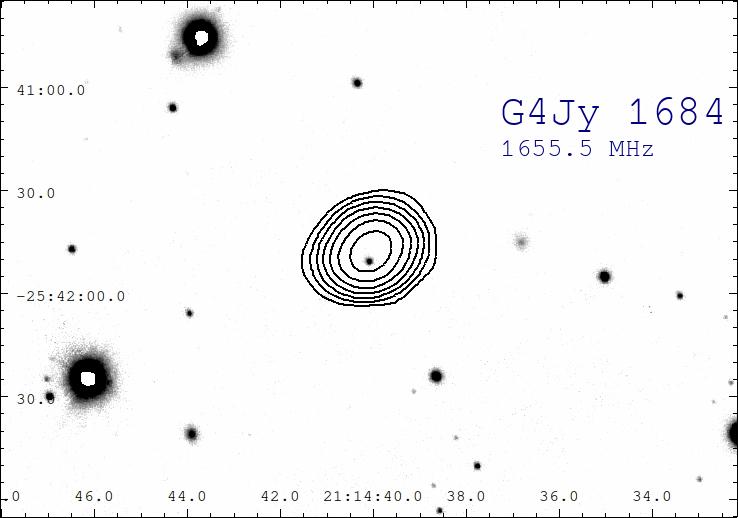}
    \includegraphics[scale=0.225]{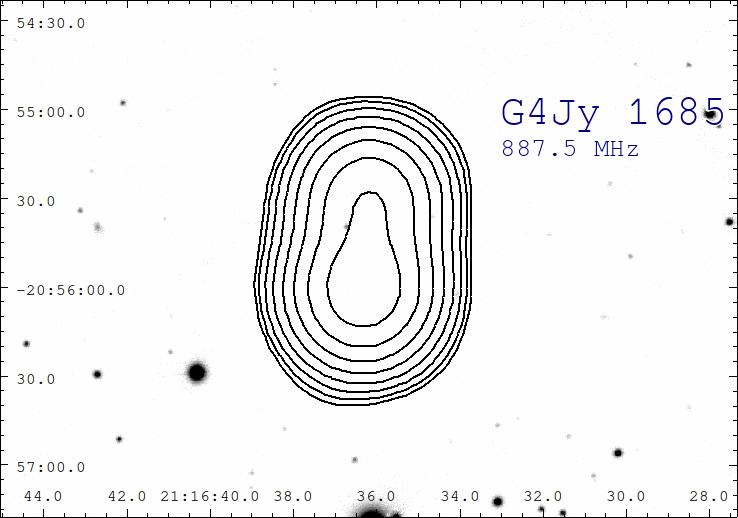}
    \includegraphics[scale=0.225]{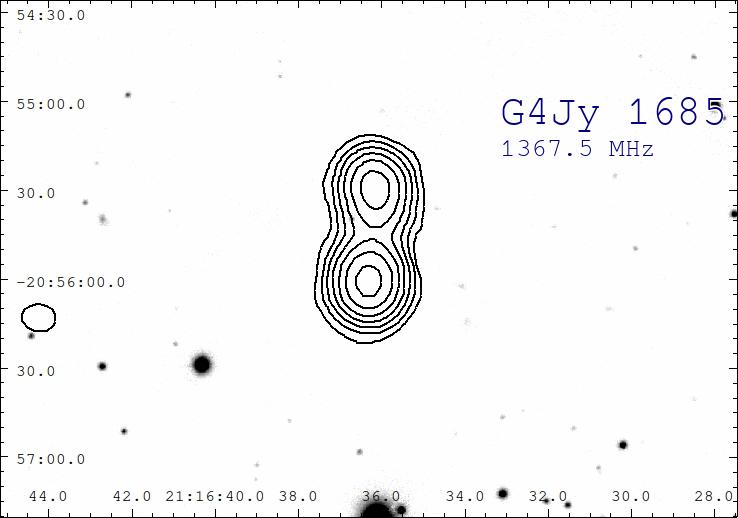}
    \includegraphics[scale=0.225]{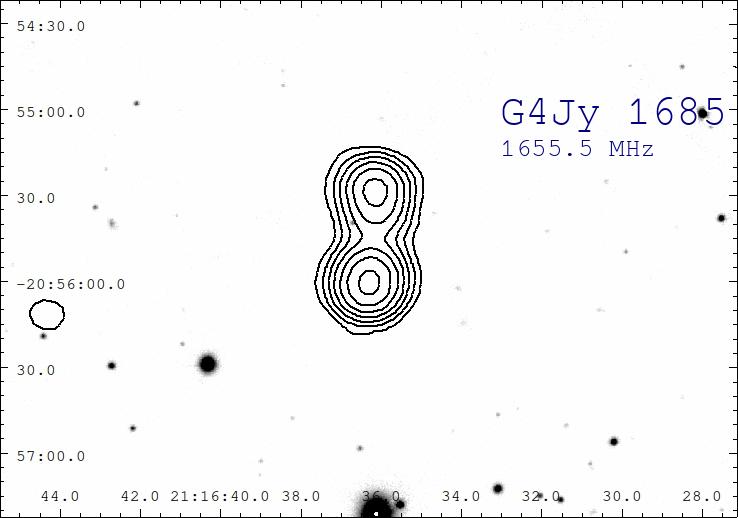}
    \includegraphics[scale=0.225]{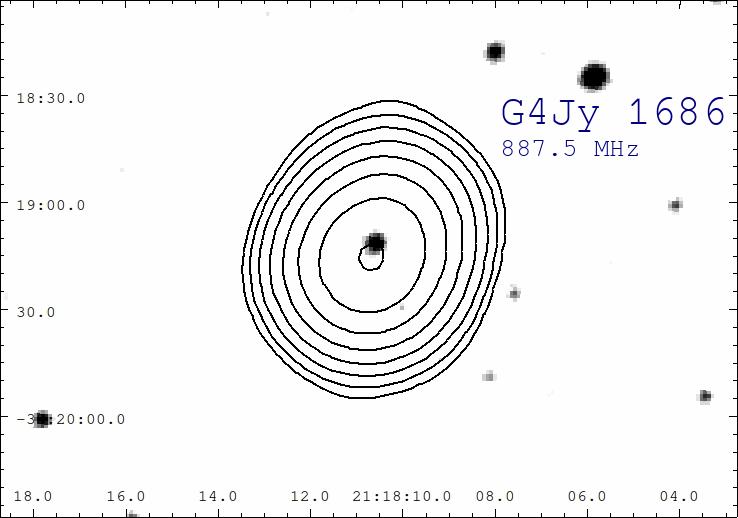}
    \includegraphics[scale=0.225]{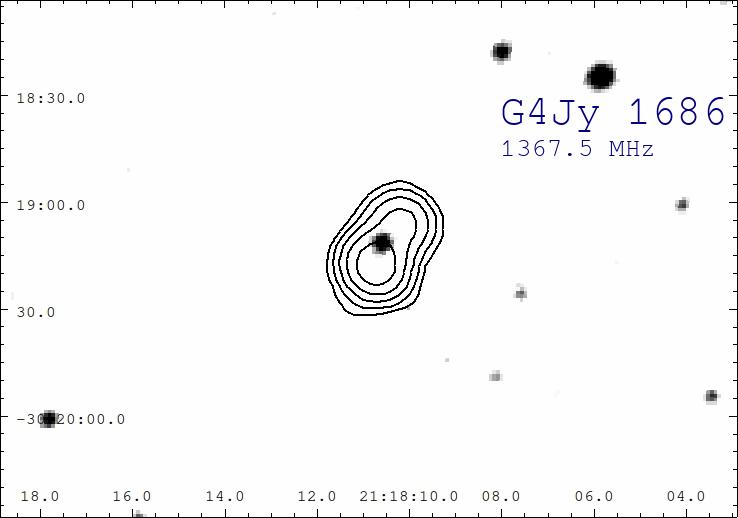}
    \includegraphics[scale=0.225]{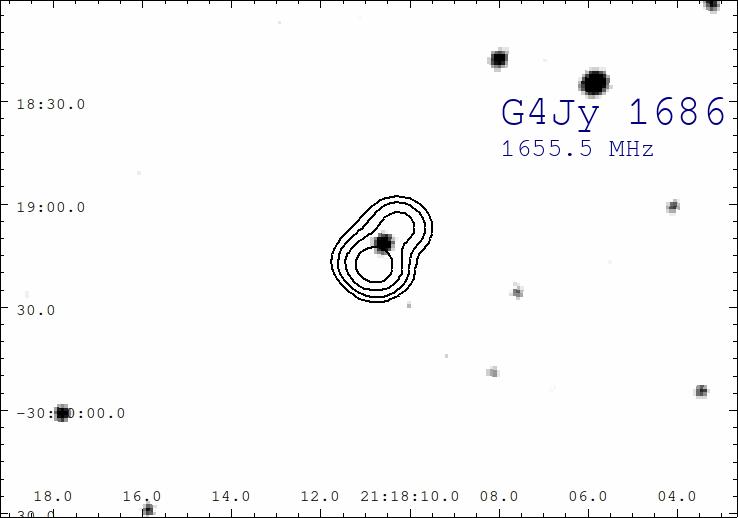}

    \caption{}
    \label{AS}
\end{figure*}
\clearpage
 \begin{figure*}
    \centering
    \includegraphics[scale=0.225]{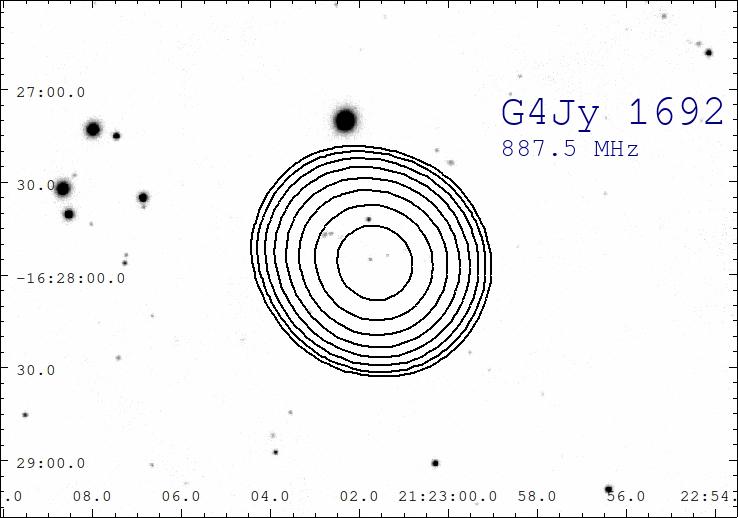}
    \includegraphics[scale=0.225]{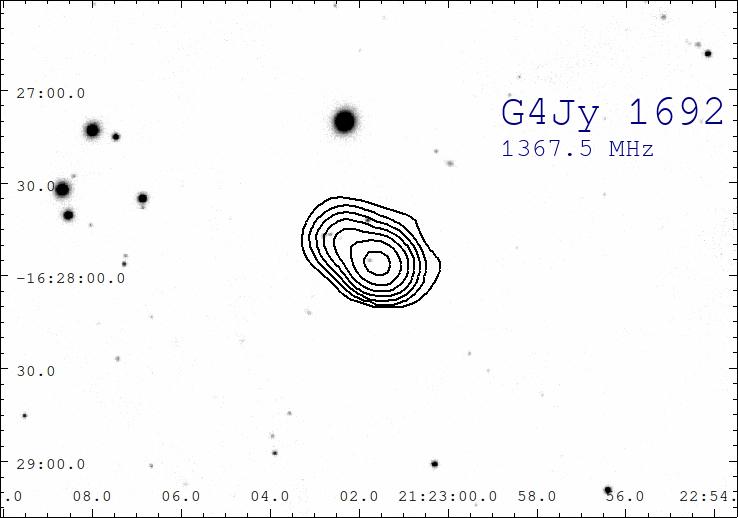}
    \includegraphics[scale=0.225]{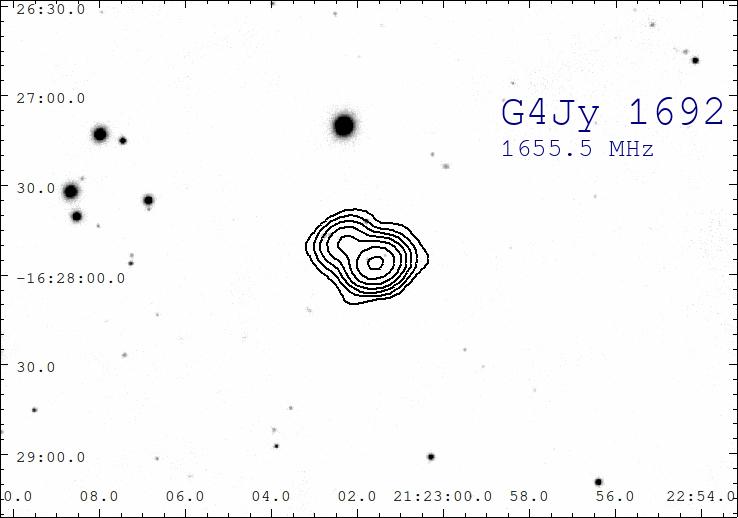}
    \includegraphics[scale=0.225]{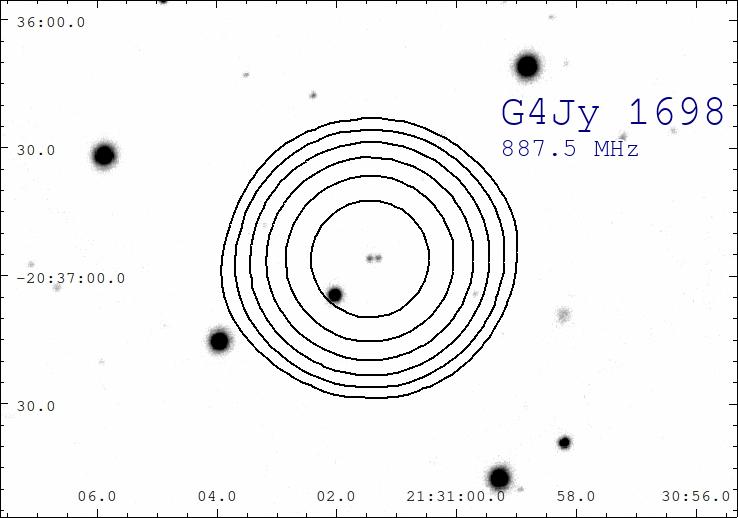}
    \includegraphics[scale=0.225]{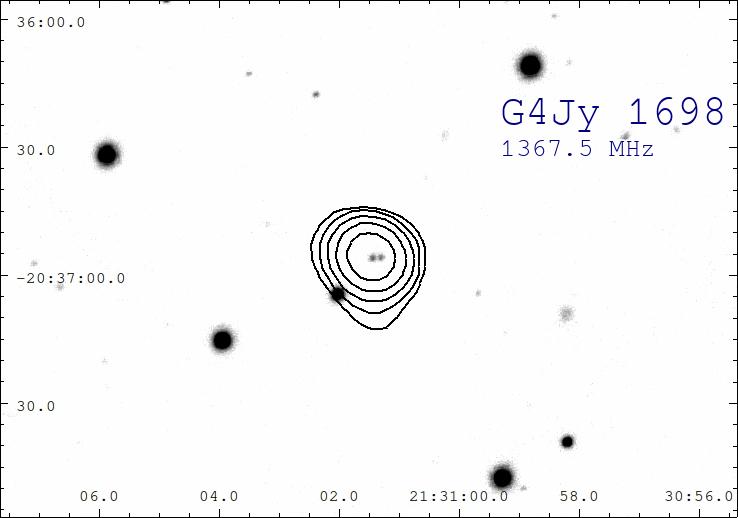}
    \includegraphics[scale=0.225]{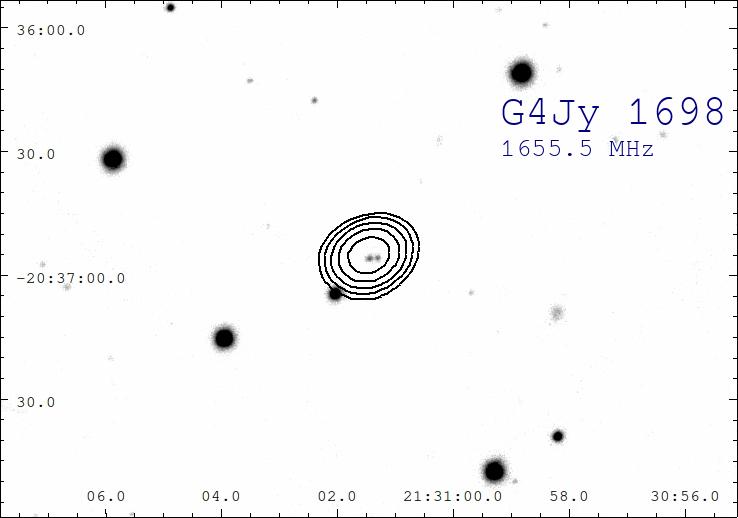}
    \includegraphics[scale=0.225]{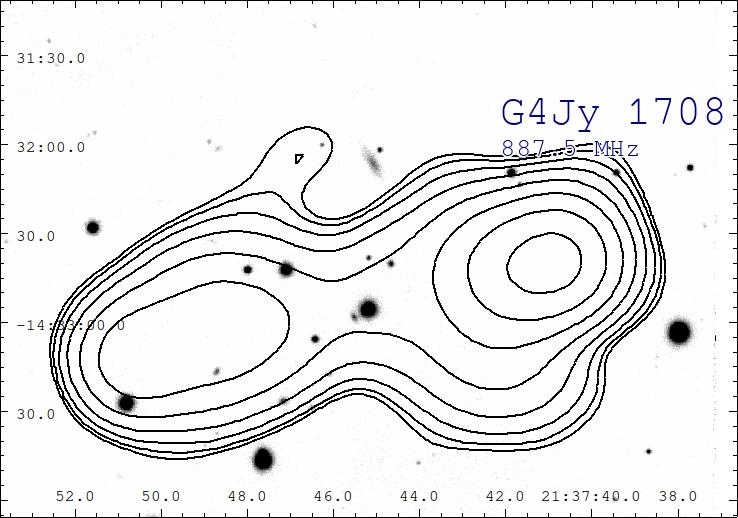}
    \includegraphics[scale=0.225]{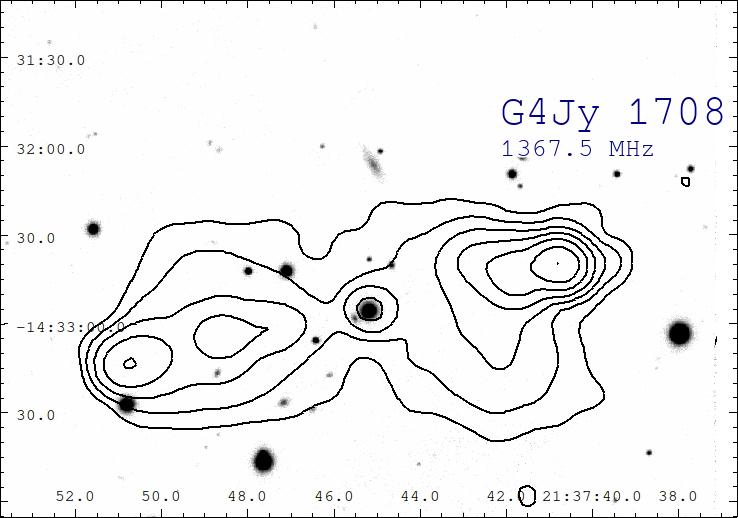}
    \includegraphics[scale=0.225]{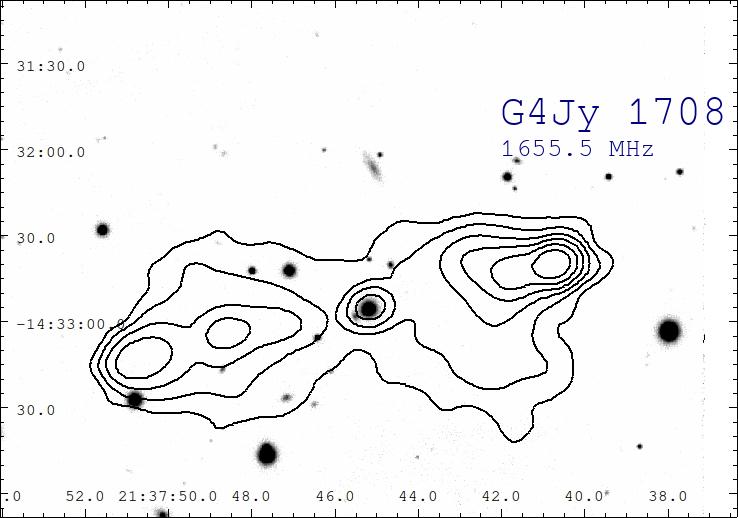}
    \includegraphics[scale=0.225]{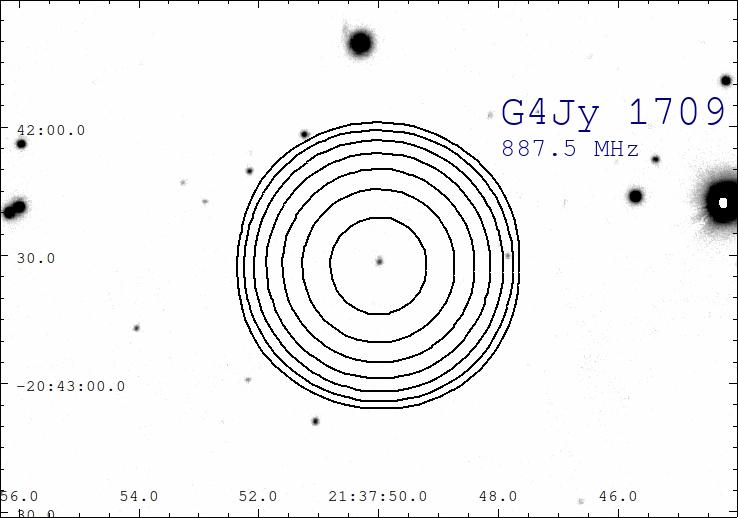}
    \includegraphics[scale=0.225]{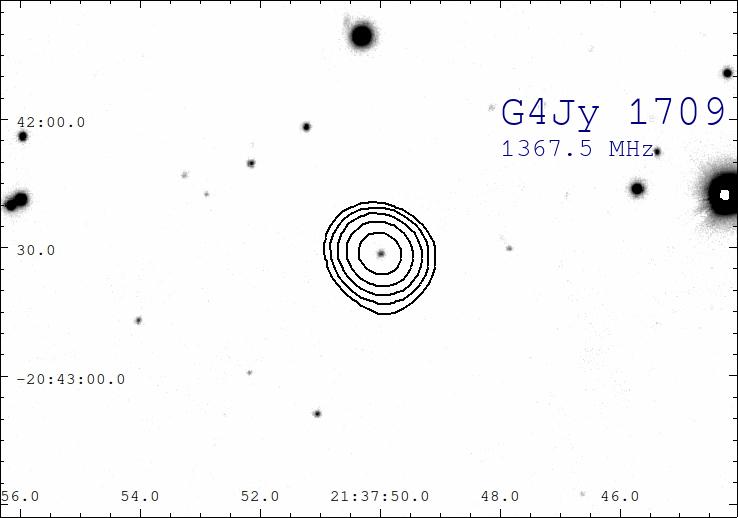}
    \includegraphics[scale=0.225]{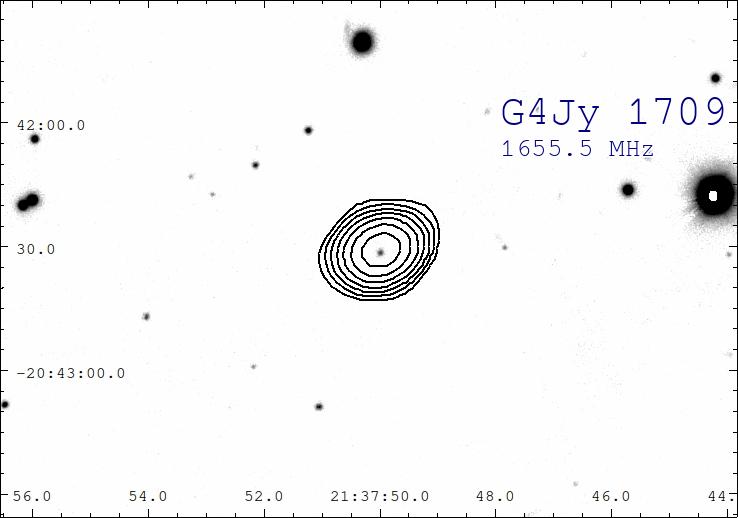}
    \includegraphics[scale=0.225]{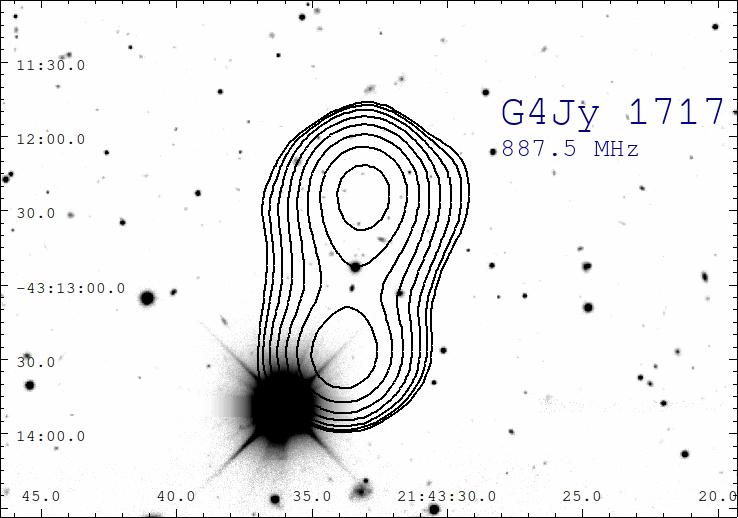}
    \includegraphics[scale=0.225]{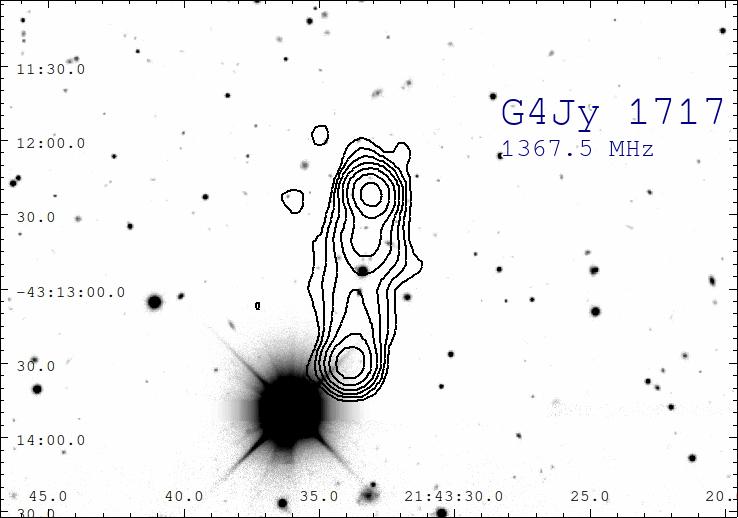}
    \includegraphics[scale=0.225]{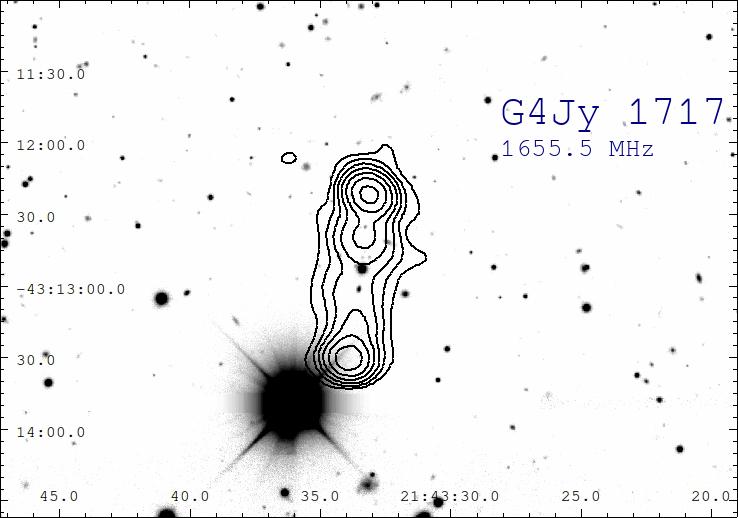}

    \caption{}
    \label{AT}
\end{figure*}
\clearpage
 \begin{figure*}
    \centering
    \includegraphics[scale=0.225]{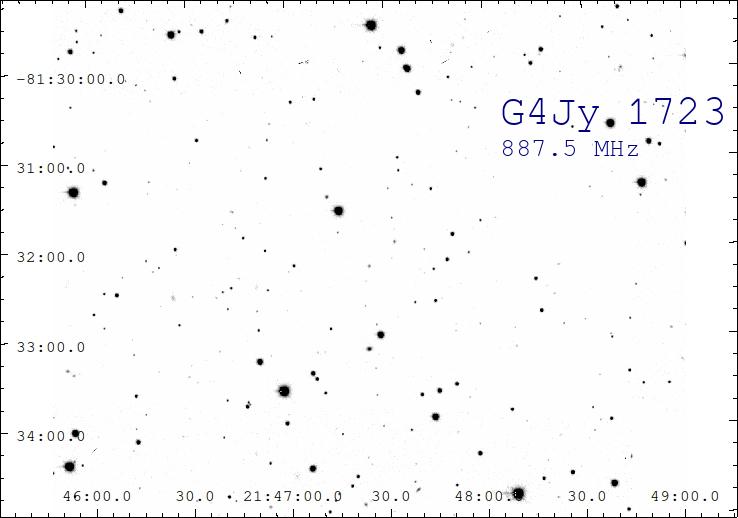}
    \includegraphics[scale=0.225]{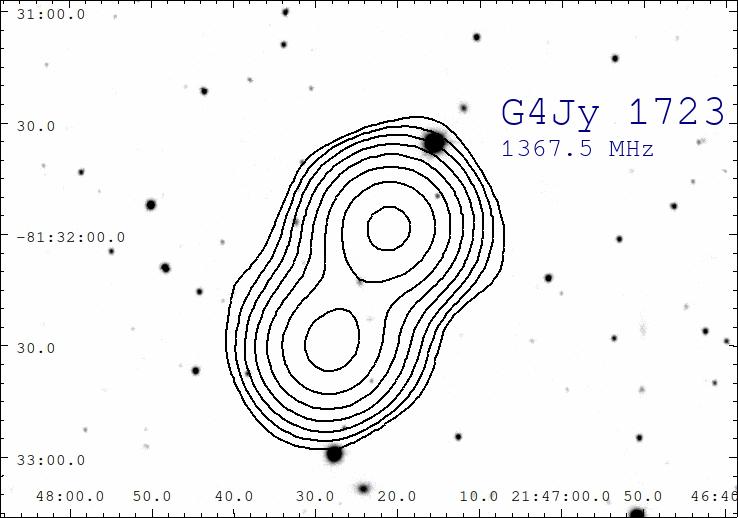}
    \includegraphics[scale=0.225]{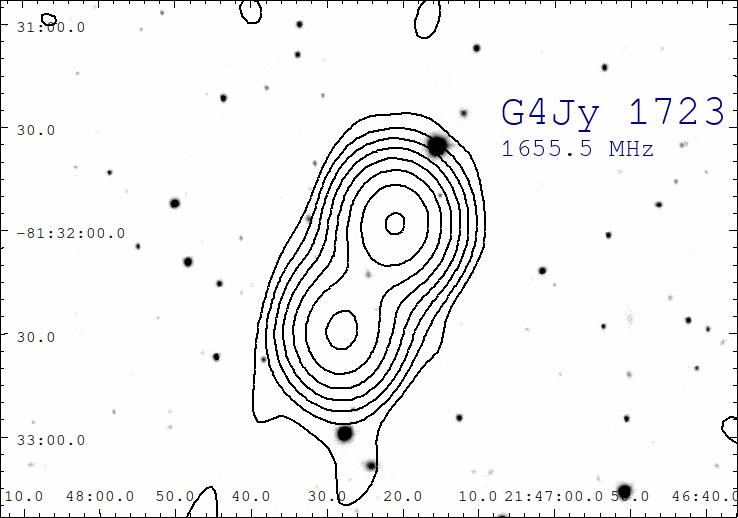}
    \includegraphics[scale=0.225]{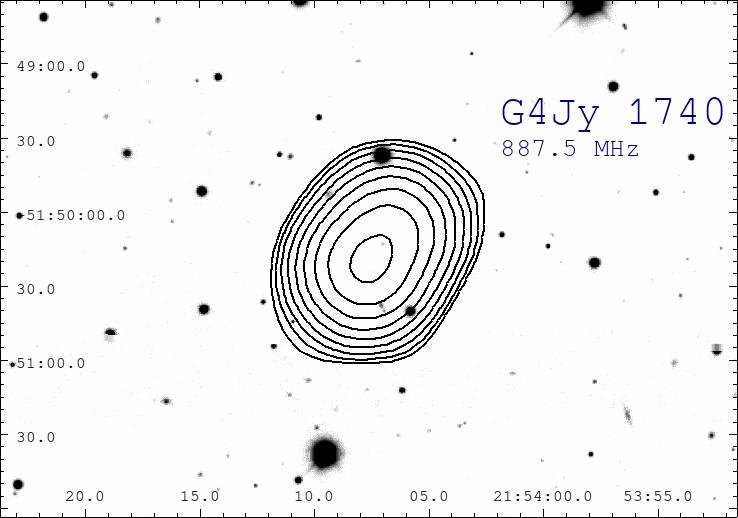}
    \includegraphics[scale=0.225]{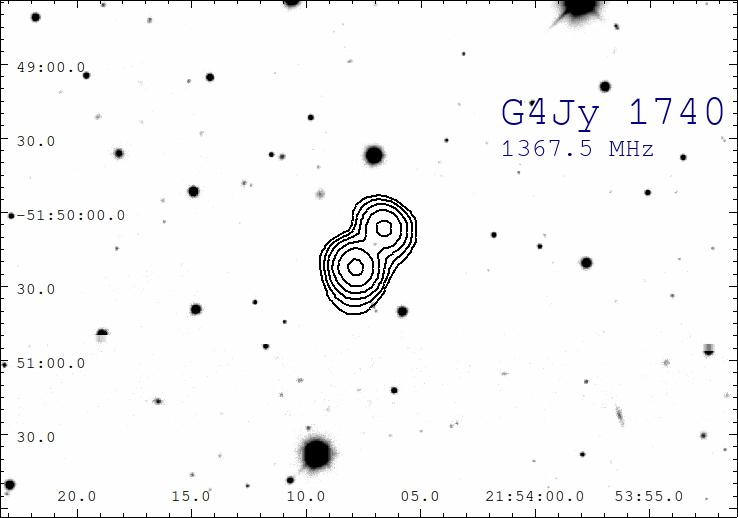}
    \includegraphics[scale=0.225]{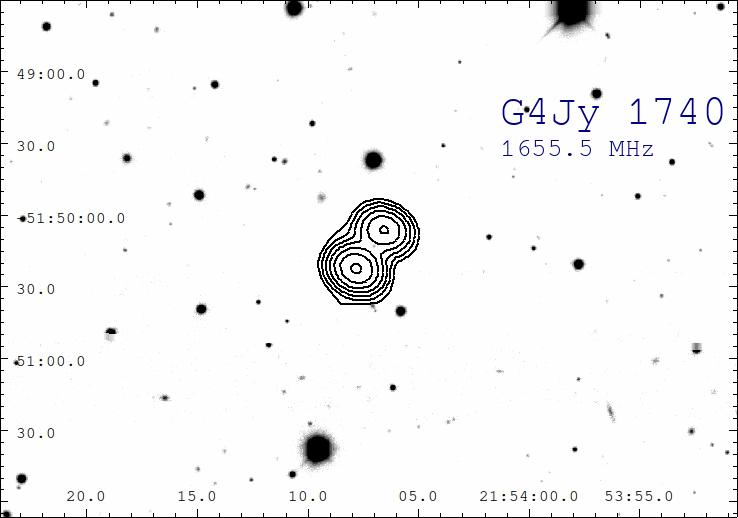}
    \includegraphics[scale=0.225]{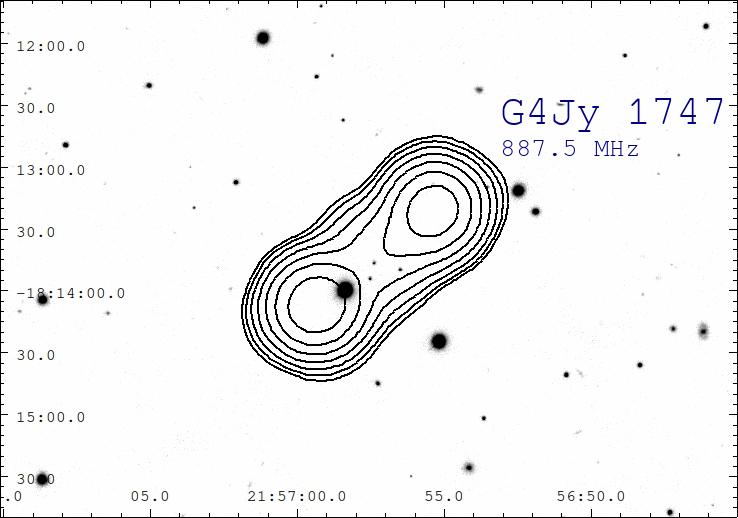}
    \includegraphics[scale=0.225]{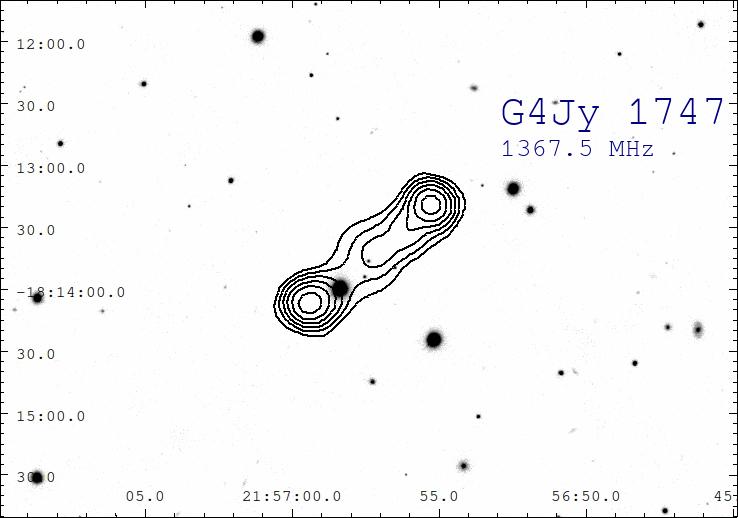}
    \includegraphics[scale=0.225]{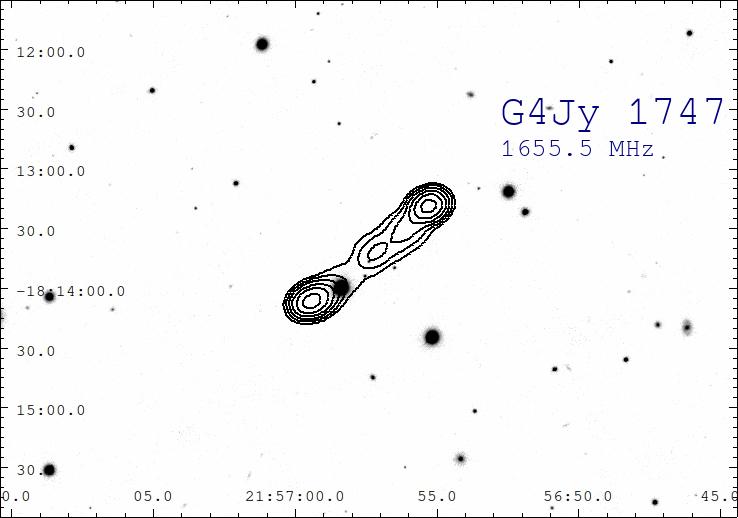}
    \includegraphics[scale=0.225]{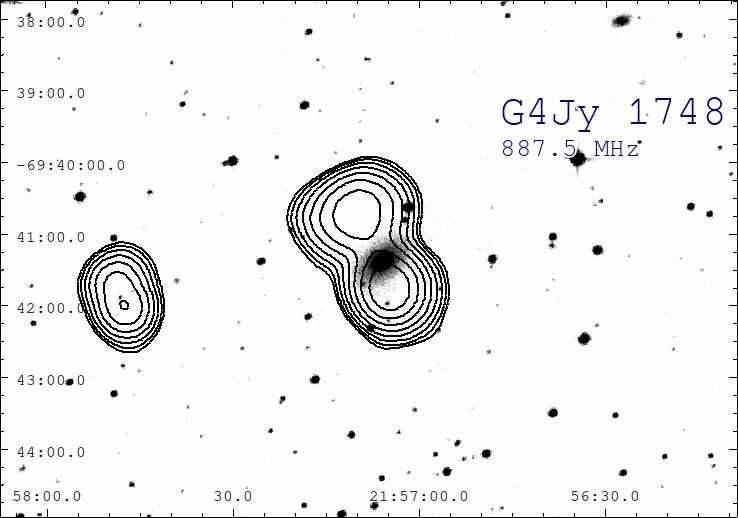}
    \includegraphics[scale=0.225]{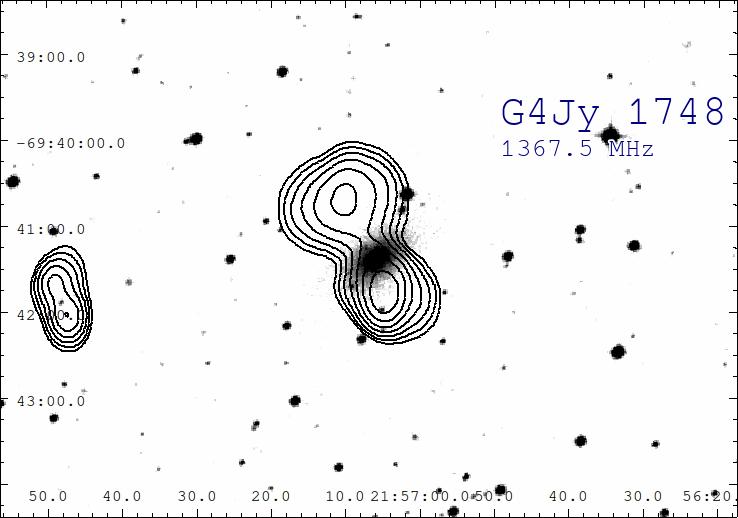}
    \includegraphics[scale=0.225]{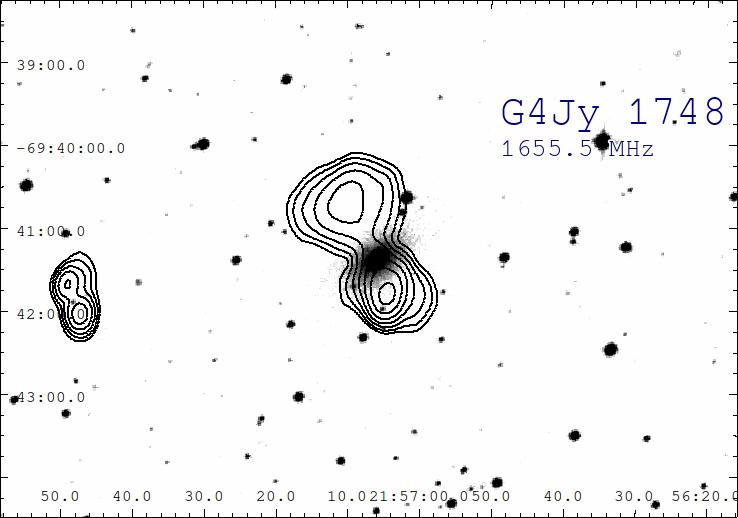}
    \includegraphics[scale=0.225]{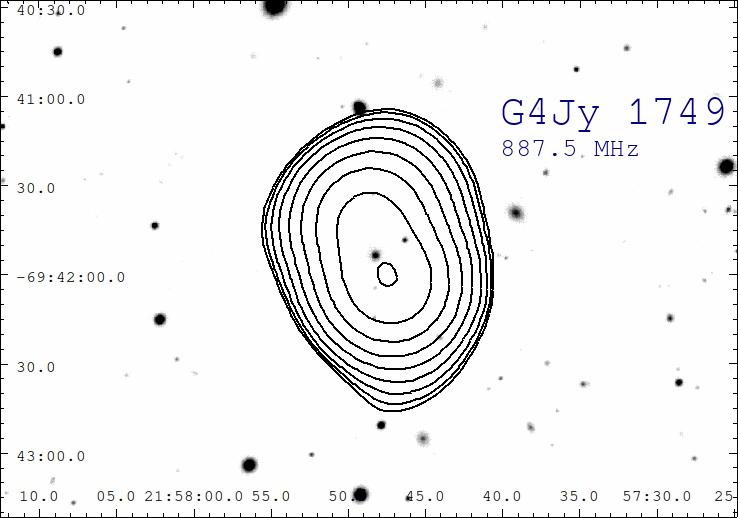}
    \includegraphics[scale=0.225]{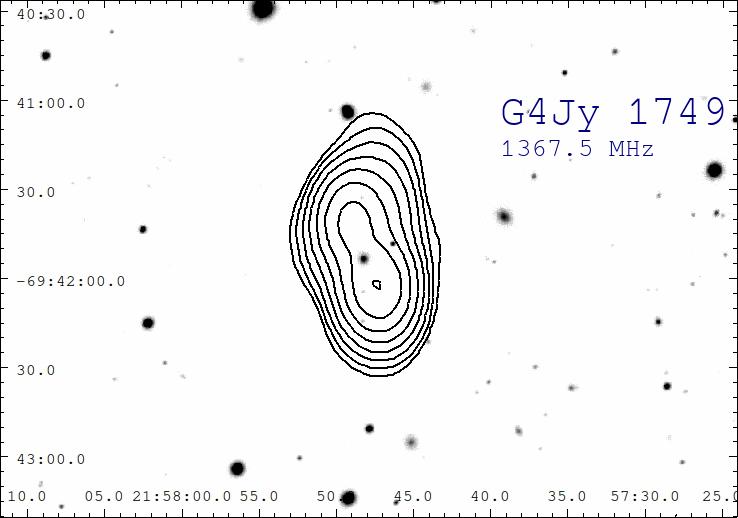}
    \includegraphics[scale=0.225]{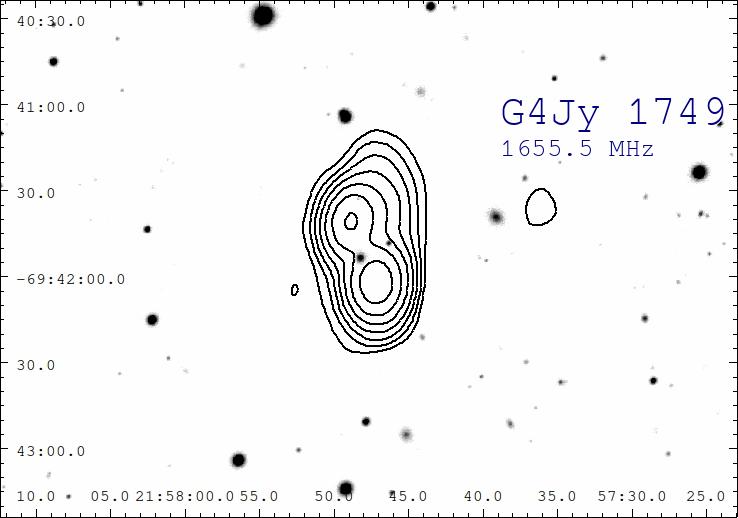}

    \caption{RACS-low data for G4Jy 1723 could not be convolved to 25$\arcsec$, and therefore was not available. It may be replaced with future observations from RACS-low2 or RACS-low3}
    \label{AU}
\end{figure*}
\clearpage
 \begin{figure*}
    \centering
    \includegraphics[scale=0.225]{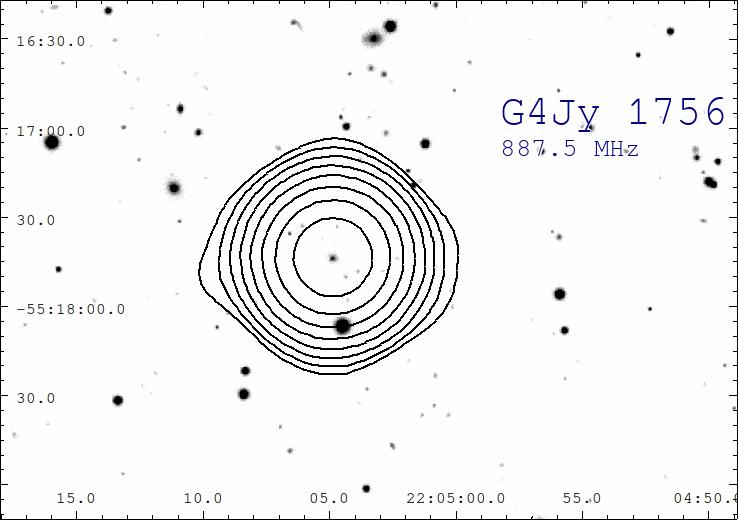}
    \includegraphics[scale=0.225]{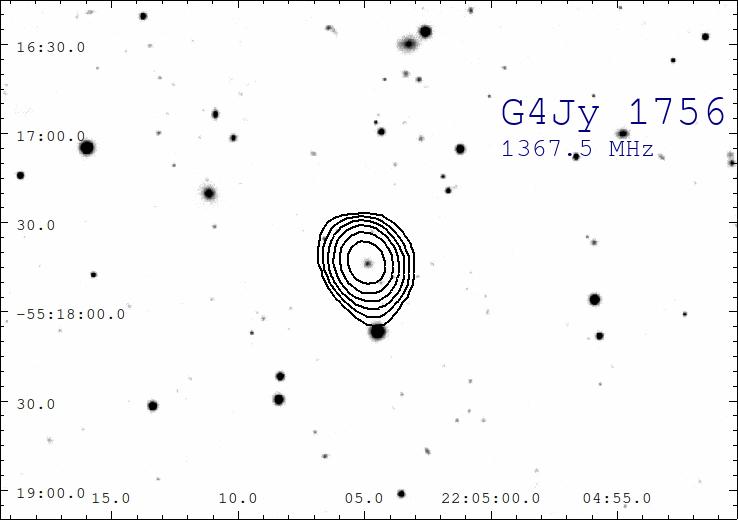}
    \includegraphics[scale=0.225]{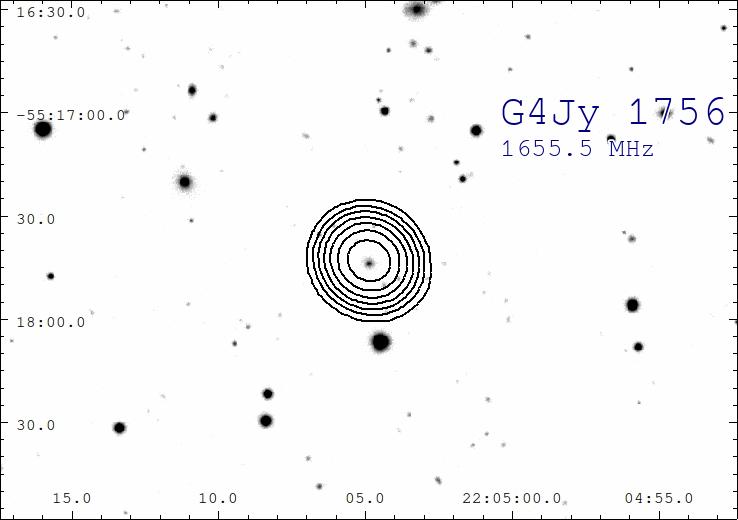}
    \includegraphics[scale=0.225]{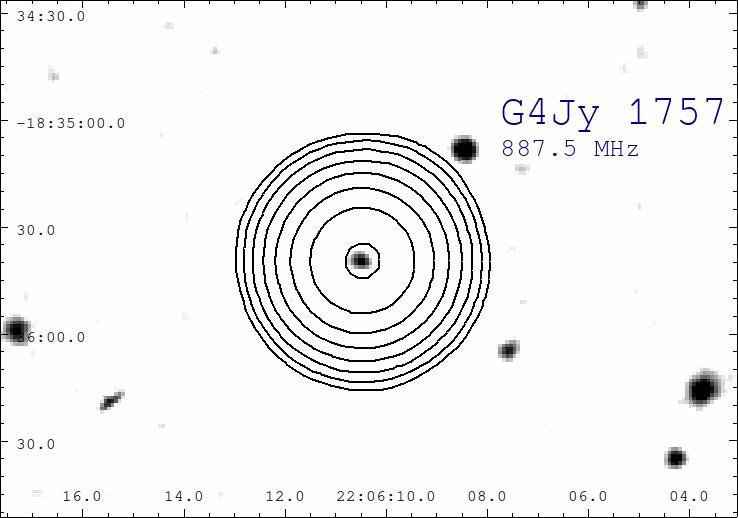}
    \includegraphics[scale=0.225]{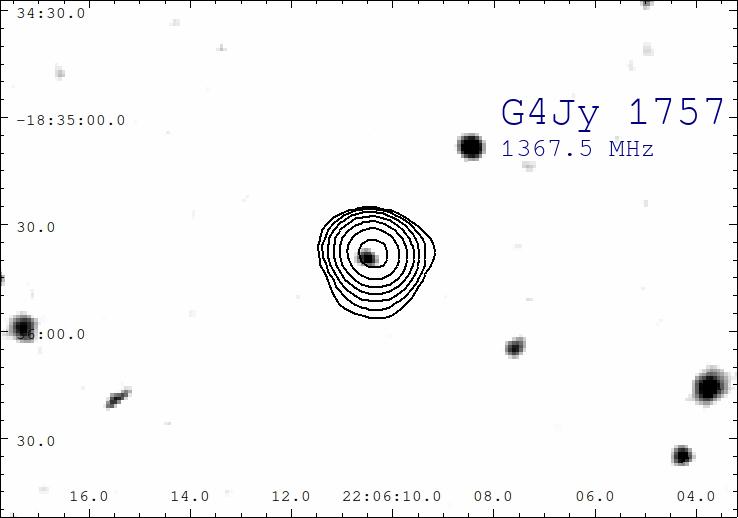}
    \includegraphics[scale=0.225]{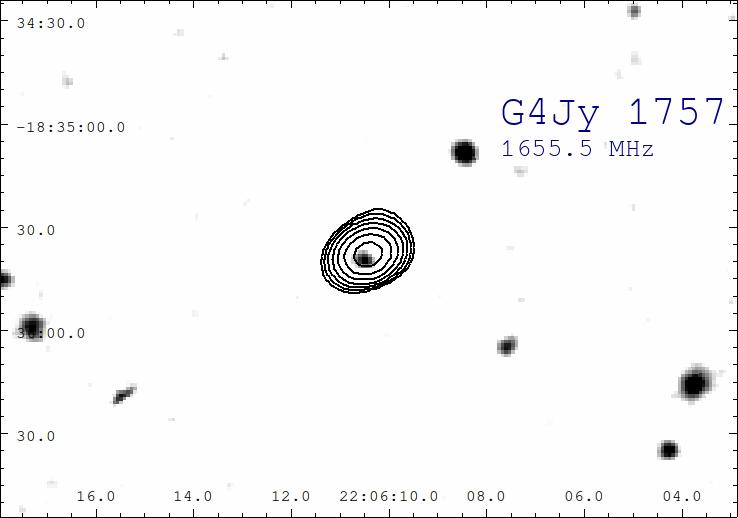}
    \includegraphics[scale=0.225]{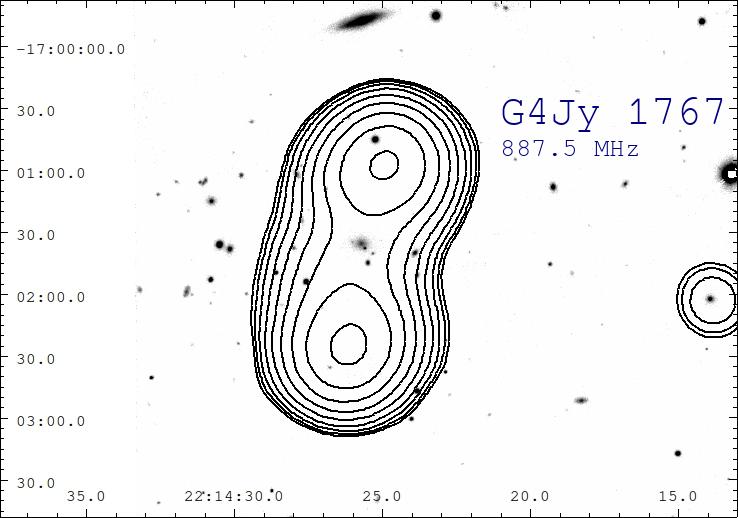}
    \includegraphics[scale=0.225]{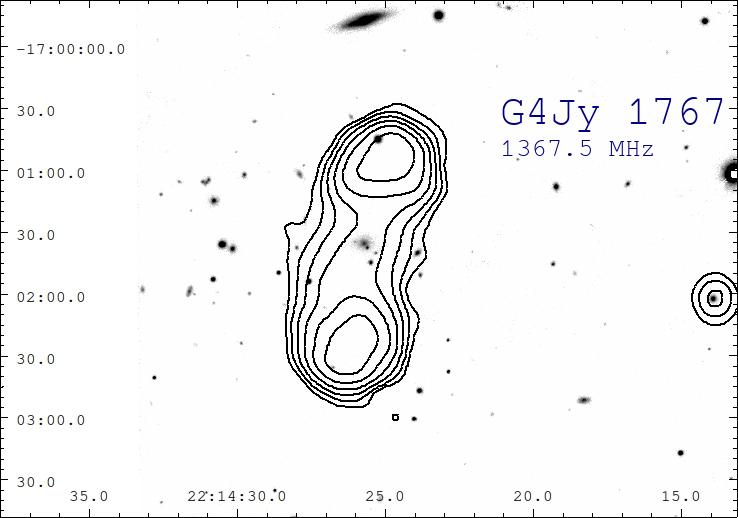}
    \includegraphics[scale=0.225]{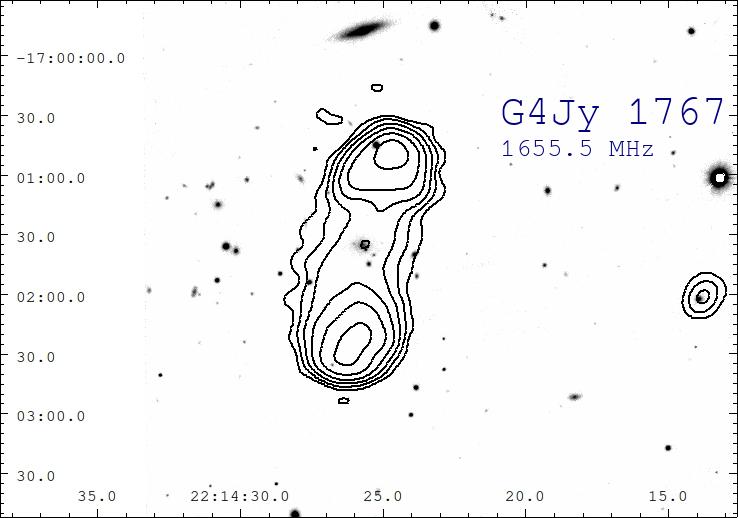}
    \includegraphics[scale=0.225]{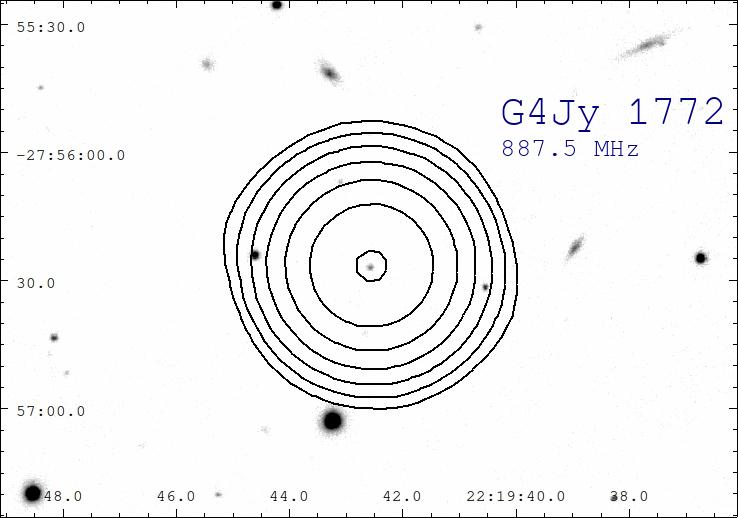}
    \includegraphics[scale=0.225]{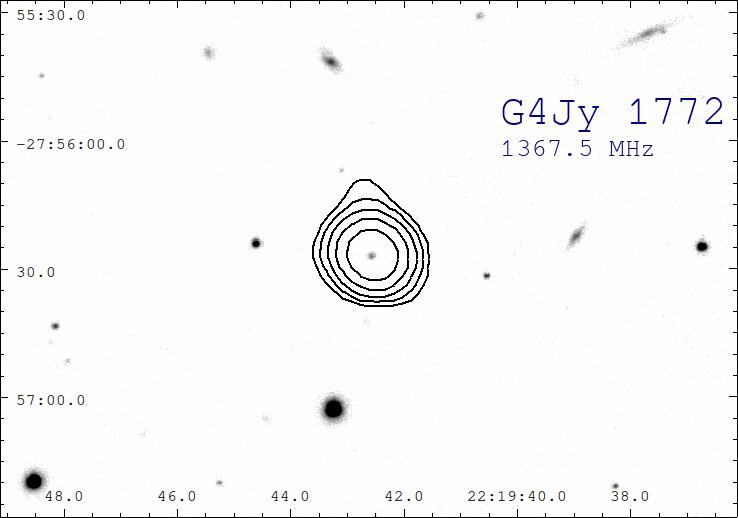}
    \includegraphics[scale=0.225]{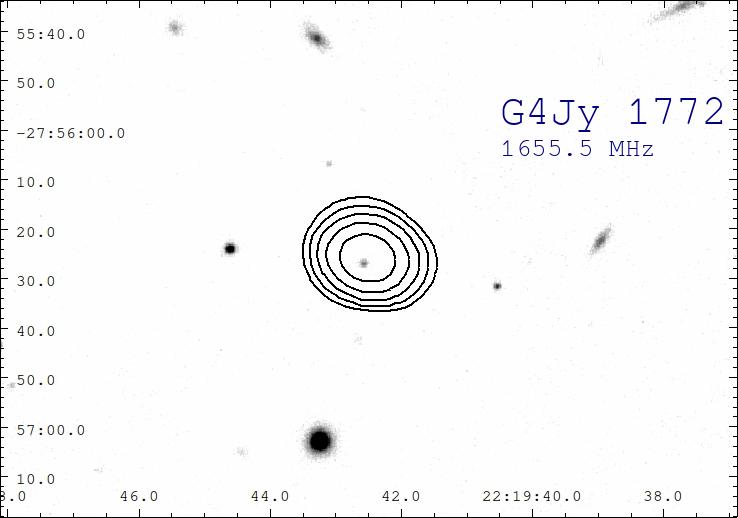}
    \includegraphics[scale=0.225]{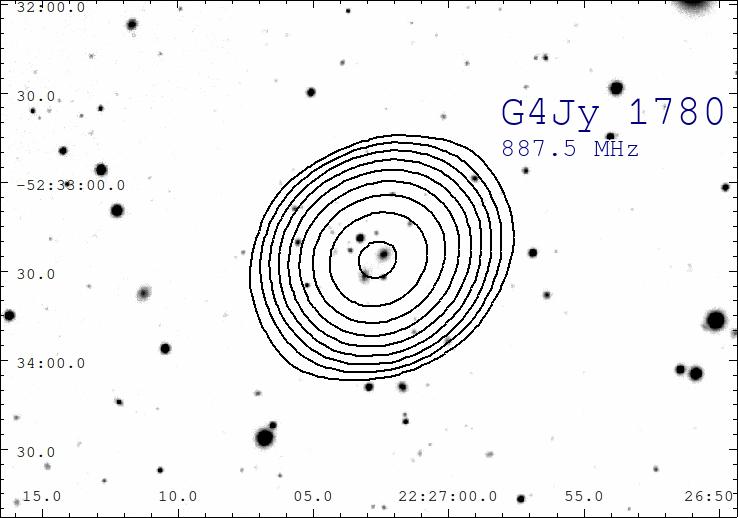}
    \includegraphics[scale=0.225]{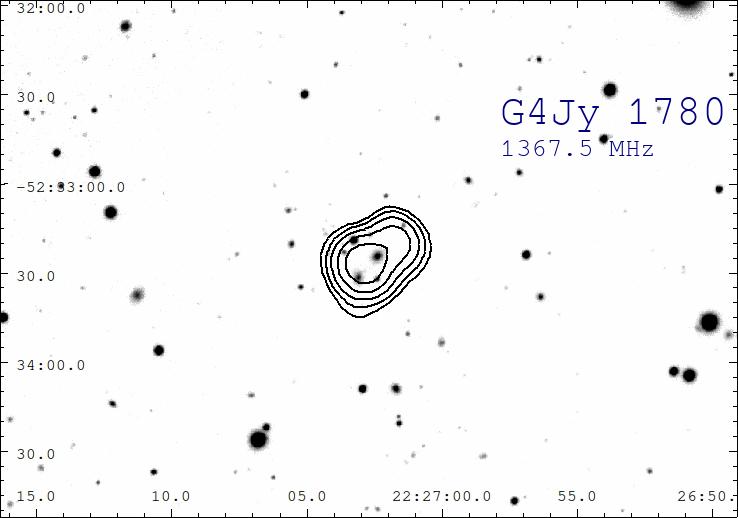}
    \includegraphics[scale=0.225]{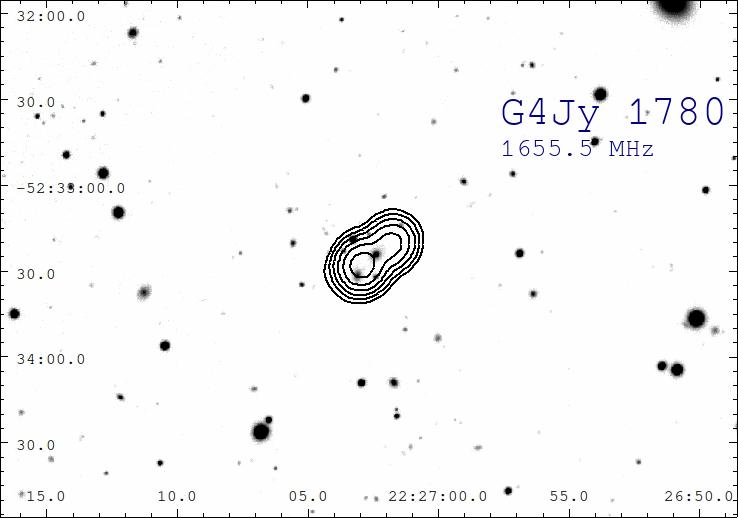}

    \caption{}
    \label{AV}
\end{figure*}
\clearpage
 \begin{figure*}
    \centering
    \includegraphics[scale=0.225]{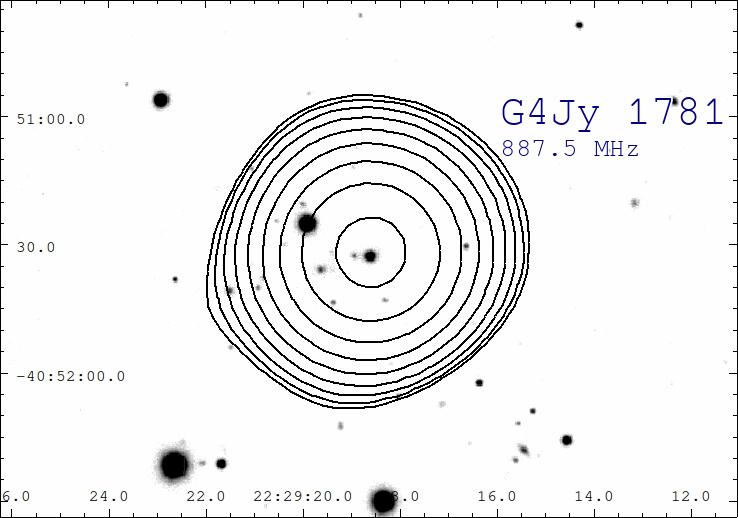}
    \includegraphics[scale=0.225]{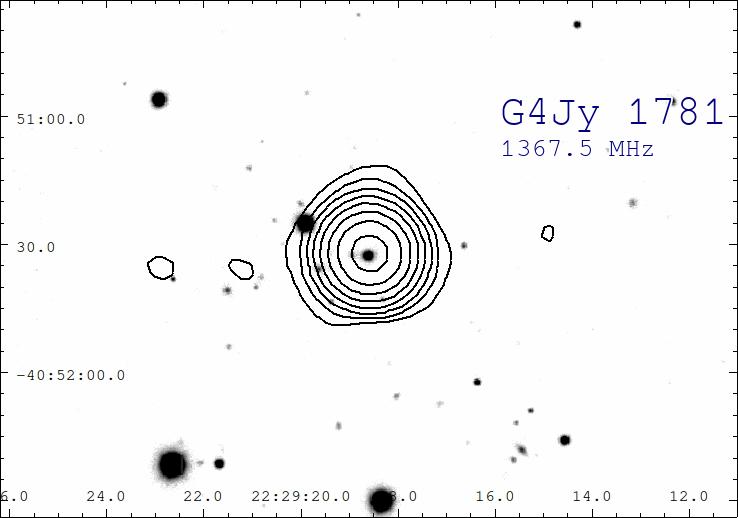}
    \includegraphics[scale=0.225]{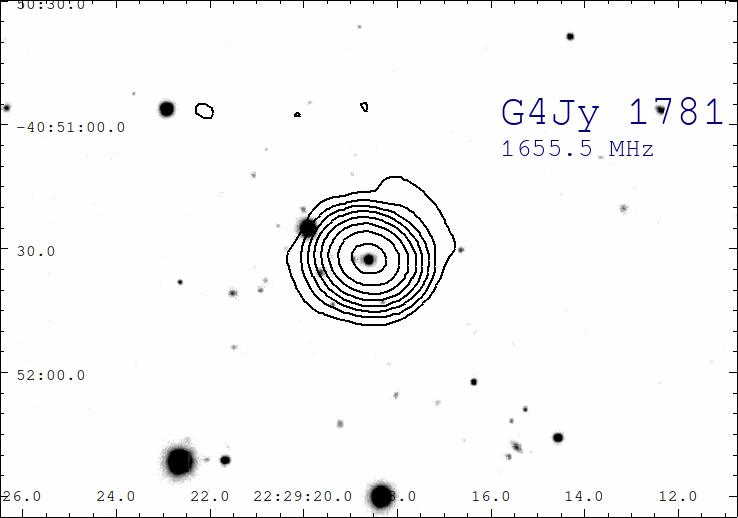}
    \includegraphics[scale=0.225]{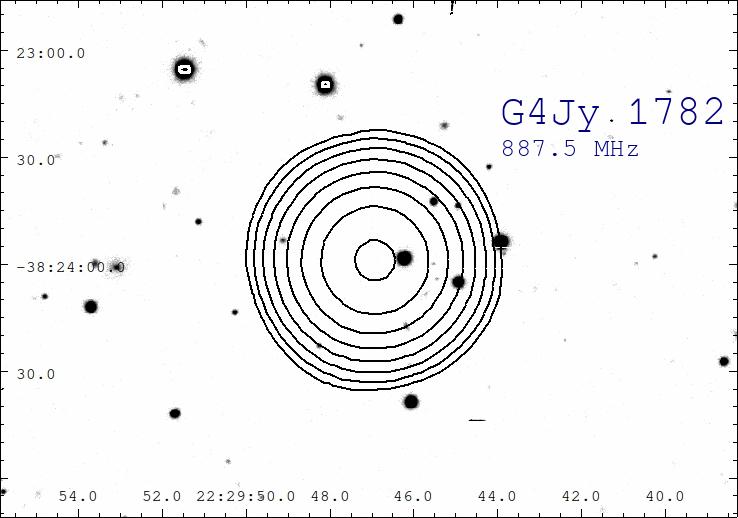}
    \includegraphics[scale=0.225]{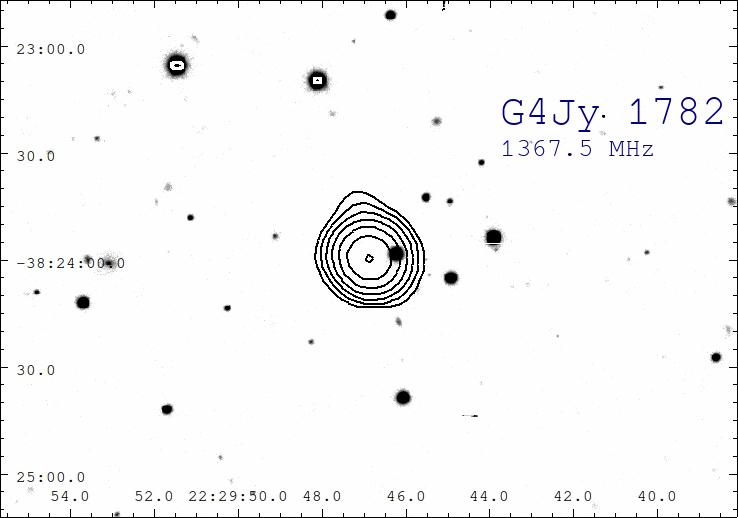}
    \includegraphics[scale=0.225]{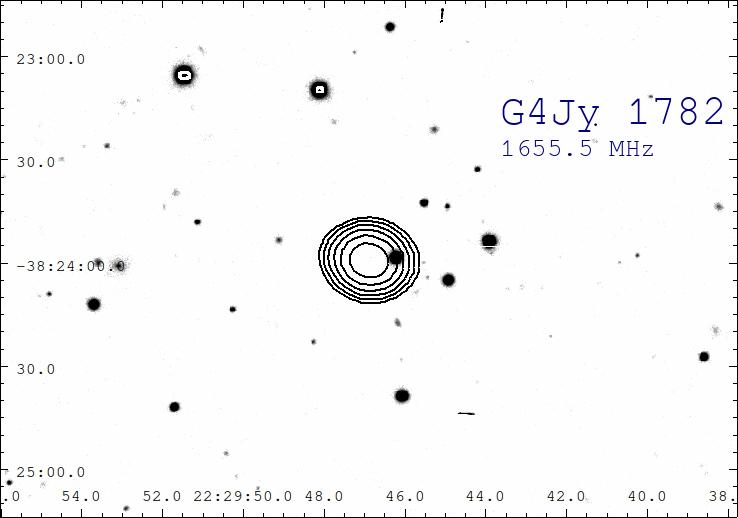}
    \includegraphics[scale=0.225]{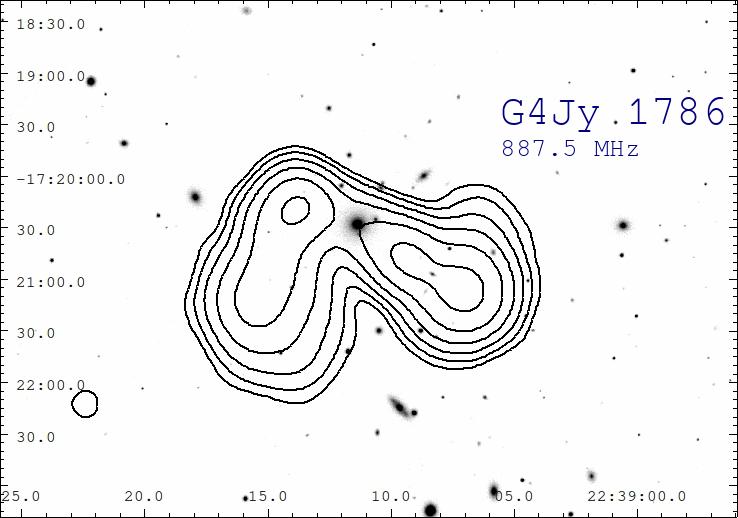}
    \includegraphics[scale=0.225]{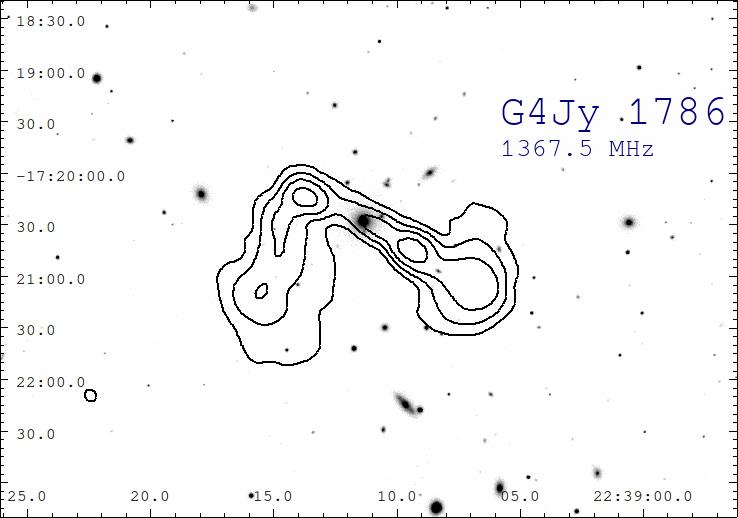}
    \includegraphics[scale=0.225]{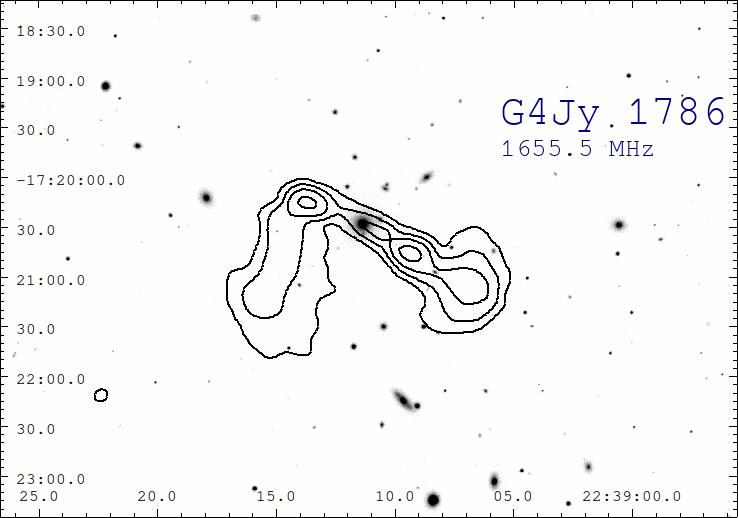}
    \includegraphics[scale=0.225]{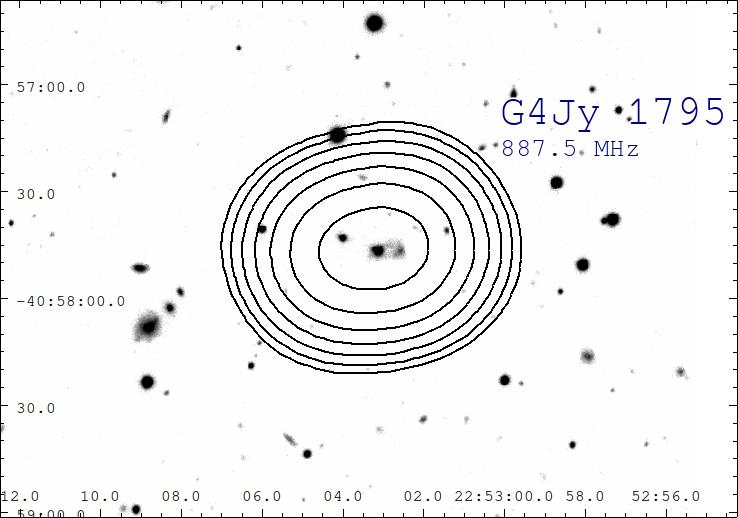}
    \includegraphics[scale=0.225]{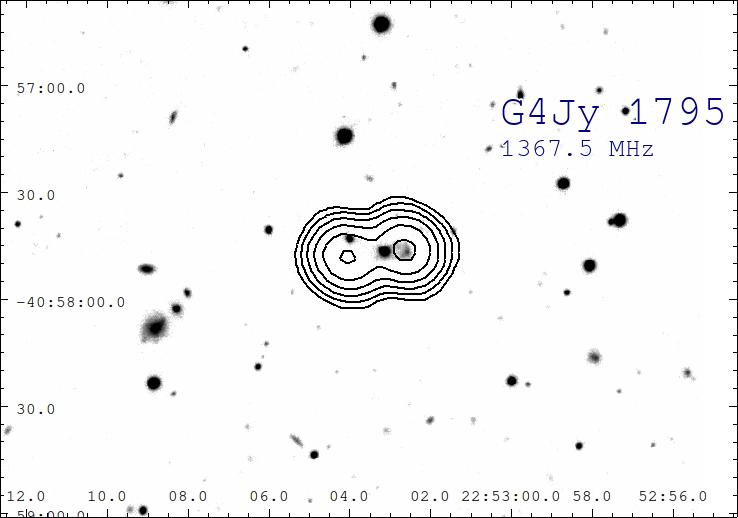}
    \includegraphics[scale=0.225]{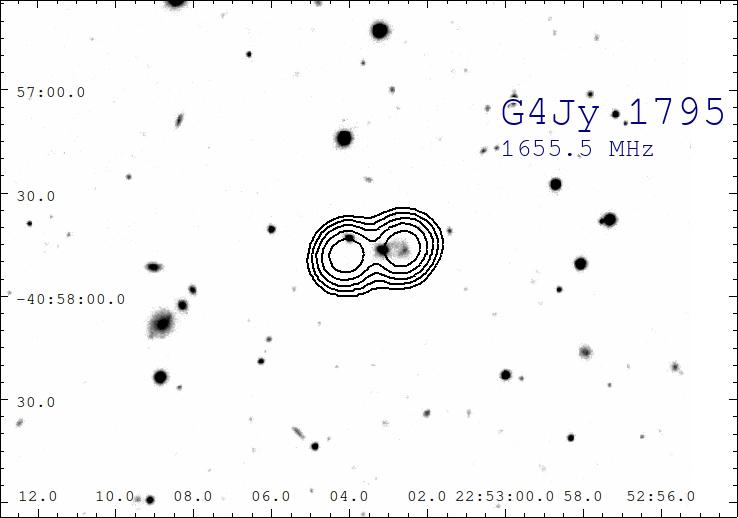}
    \includegraphics[scale=0.225]{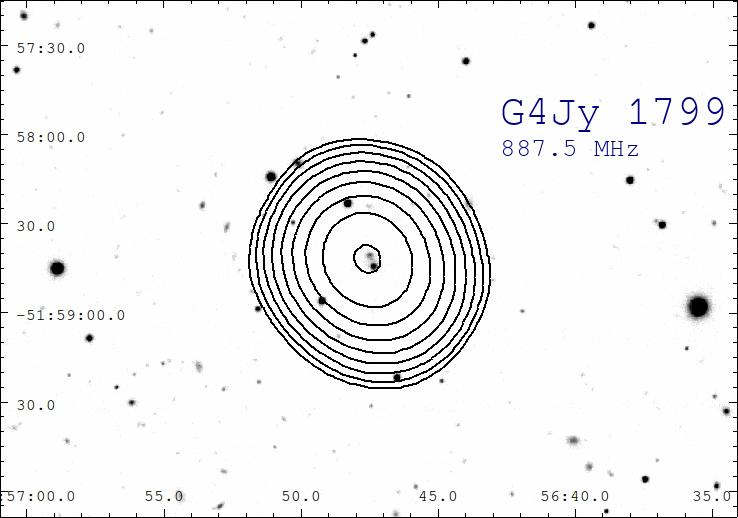}
    \includegraphics[scale=0.225]{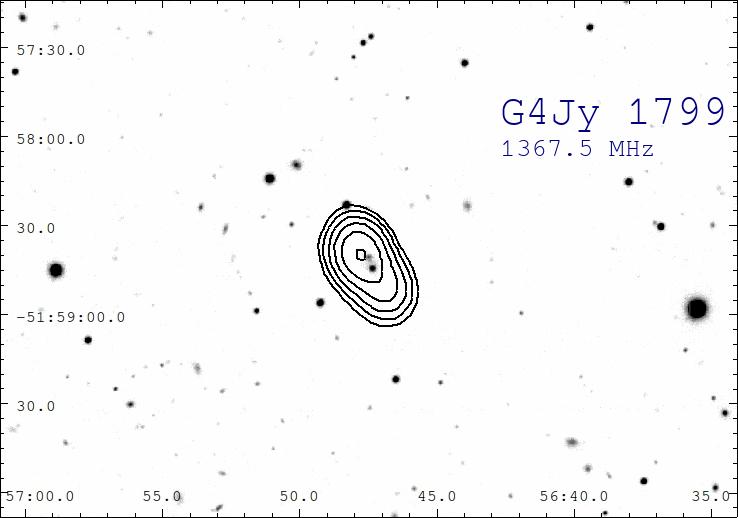}
    \includegraphics[scale=0.225]{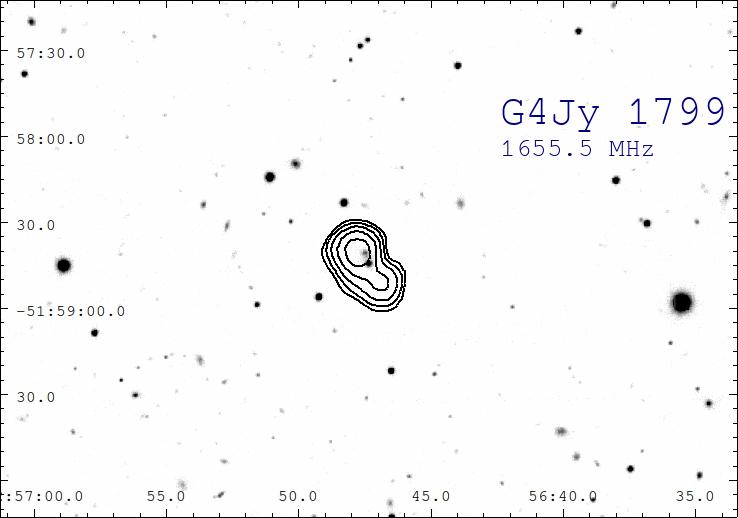}

    \caption{}
    \label{AW}
\end{figure*}
\clearpage
  \begin{figure*}
    \centering
    \includegraphics[scale=0.225]{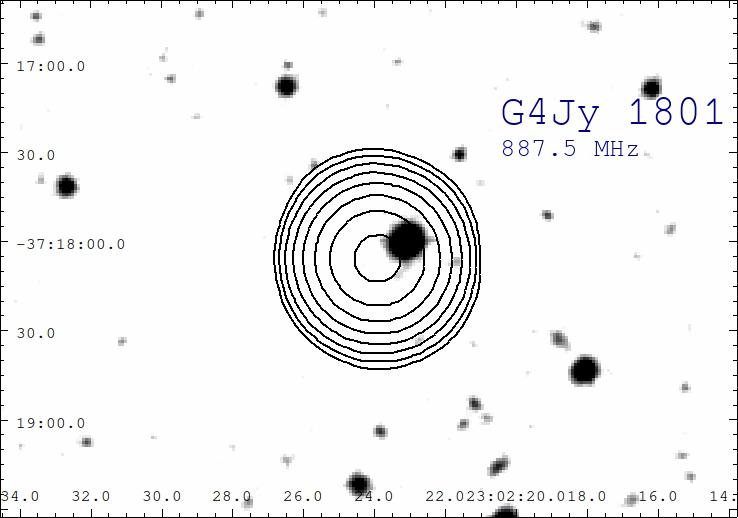}
    \includegraphics[scale=0.225]{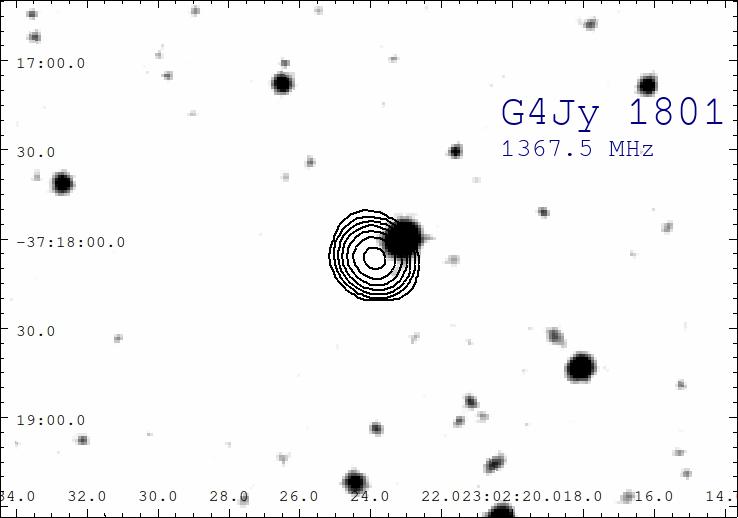}
    \includegraphics[scale=0.225]{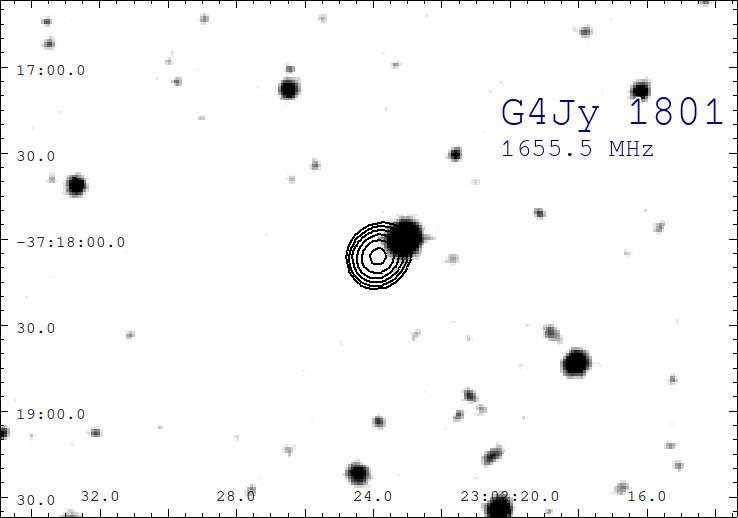}
    \includegraphics[scale=0.225]{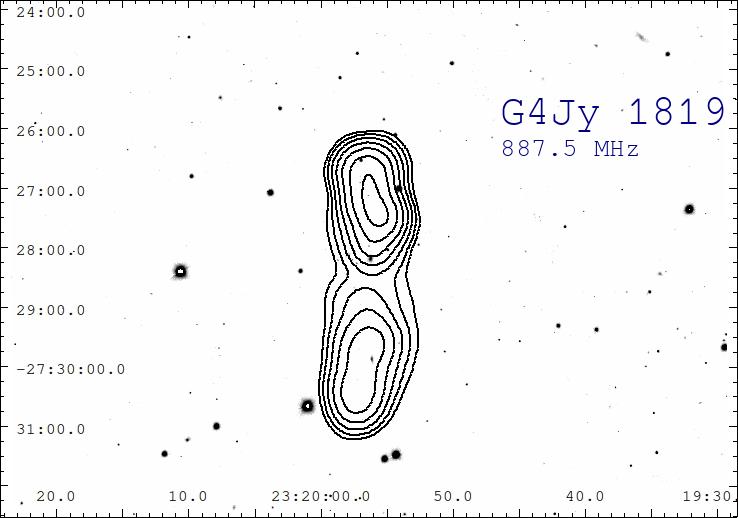}
    \includegraphics[scale=0.225]{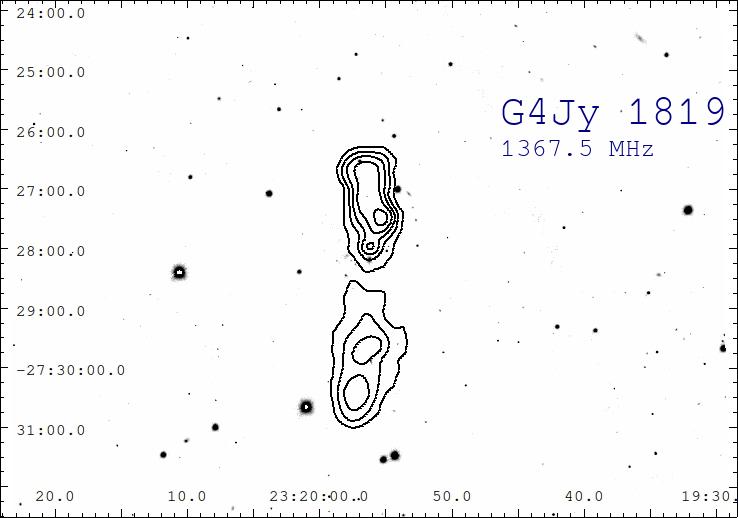}
    \includegraphics[scale=0.225]{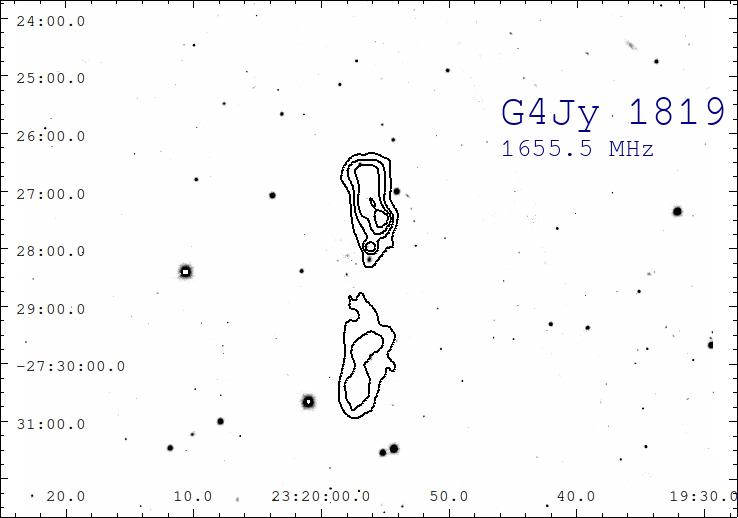}
    \includegraphics[scale=0.225]{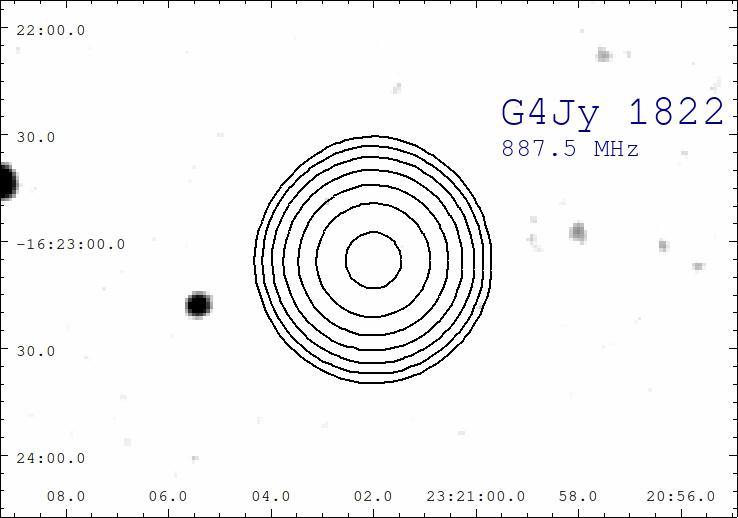}
    \includegraphics[scale=0.225]{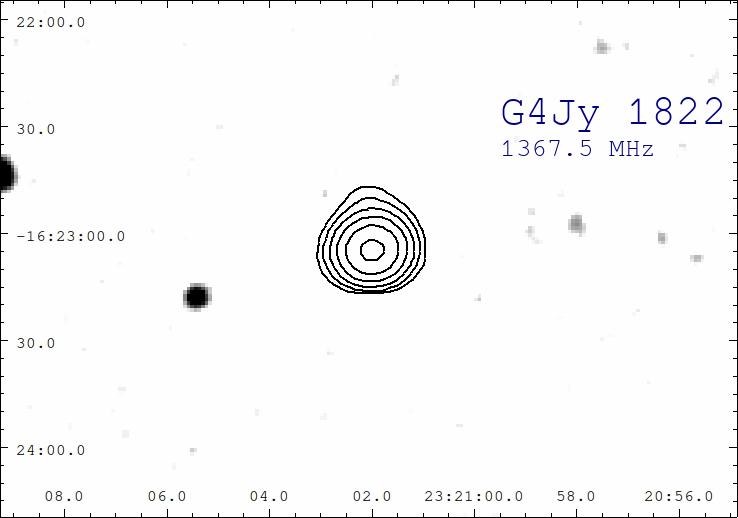}
    \includegraphics[scale=0.225]{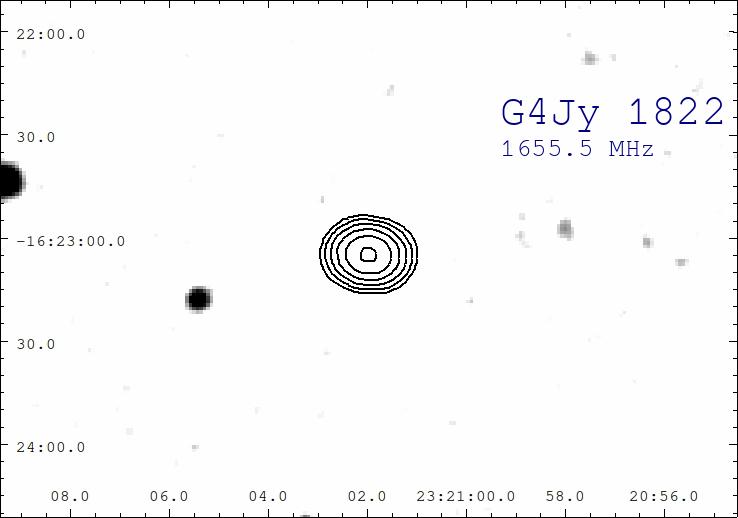}
    \includegraphics[scale=0.225]{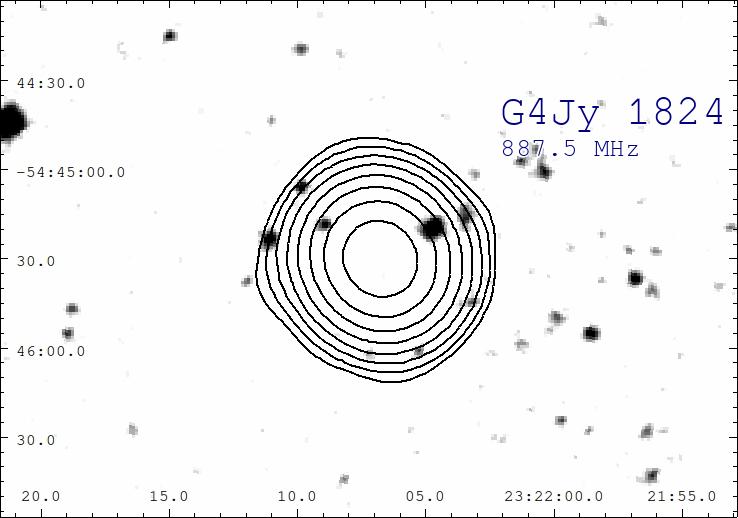}
    \includegraphics[scale=0.225]{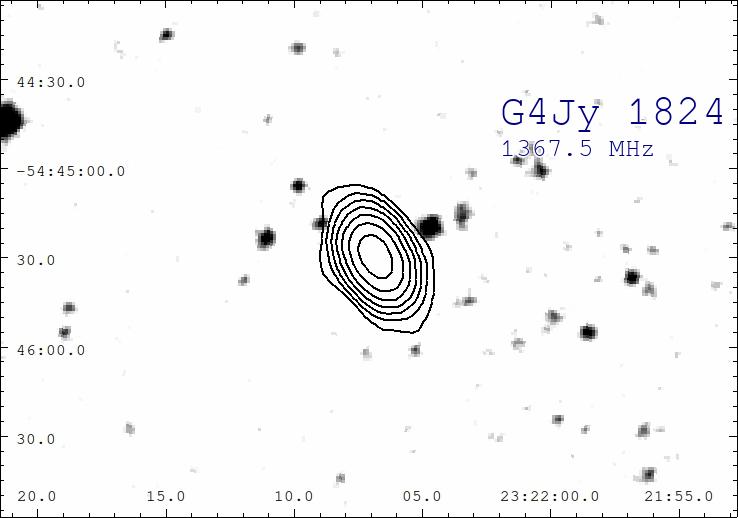}
    \includegraphics[scale=0.225]{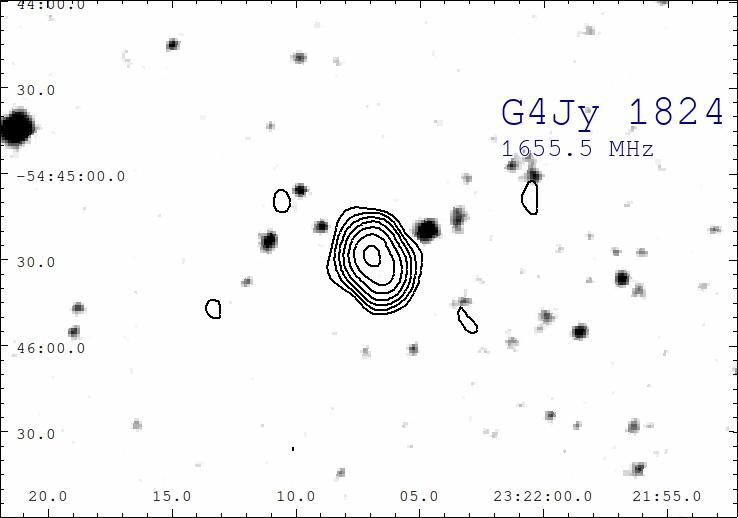}
    \includegraphics[scale=0.225]{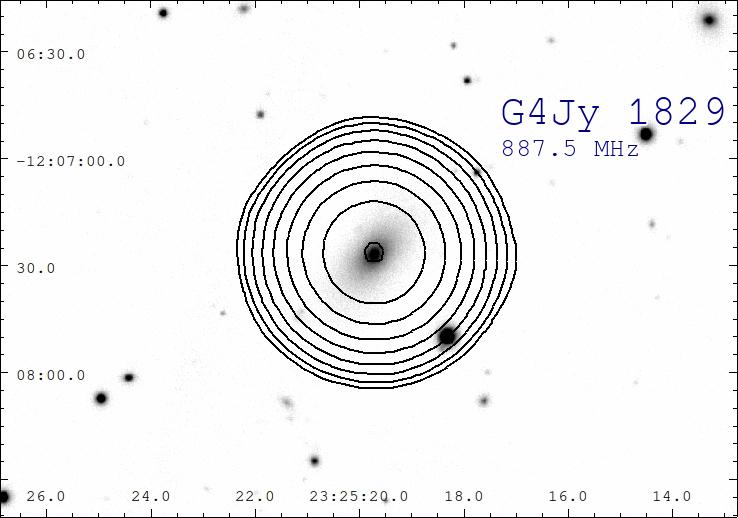}
    \includegraphics[scale=0.225]{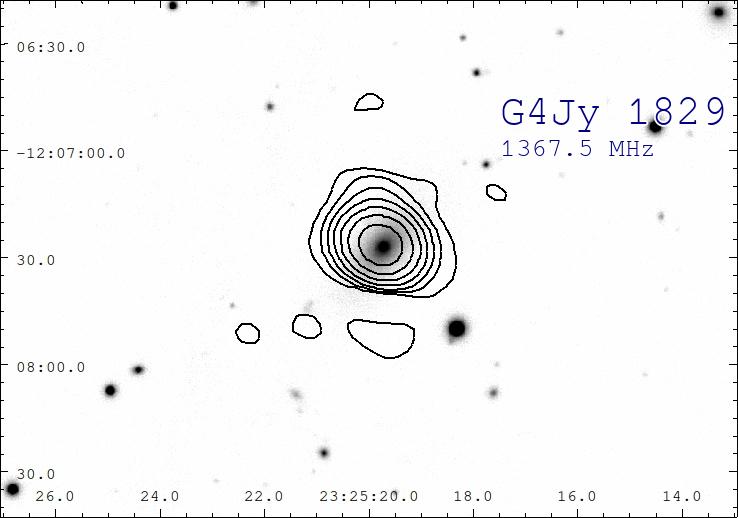}
    \includegraphics[scale=0.225]{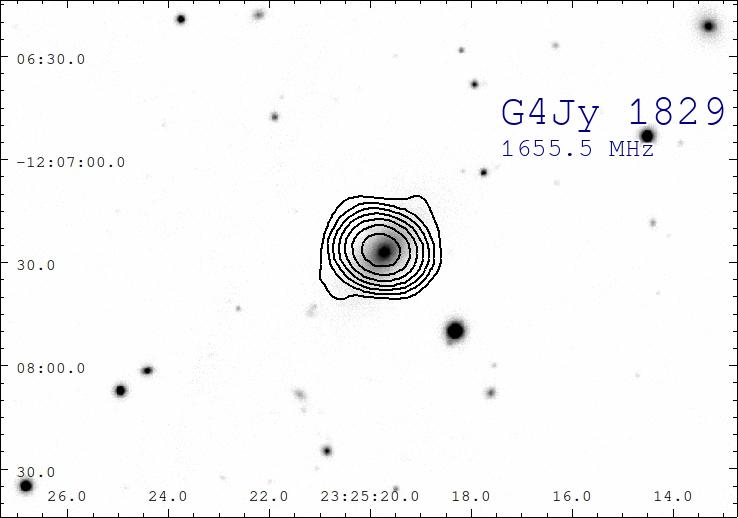}

    \caption{}
    \label{AX}
\end{figure*}
\clearpage
 
\begin{figure*}
    \centering
    \includegraphics[scale=0.225]{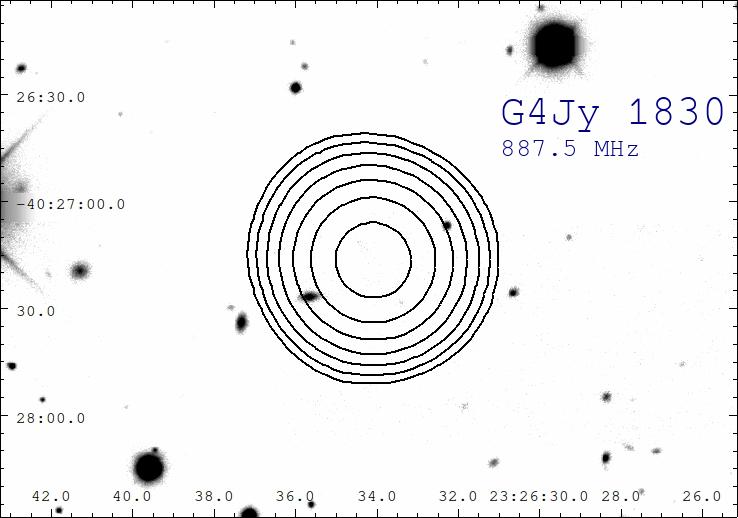}
    \includegraphics[scale=0.225]{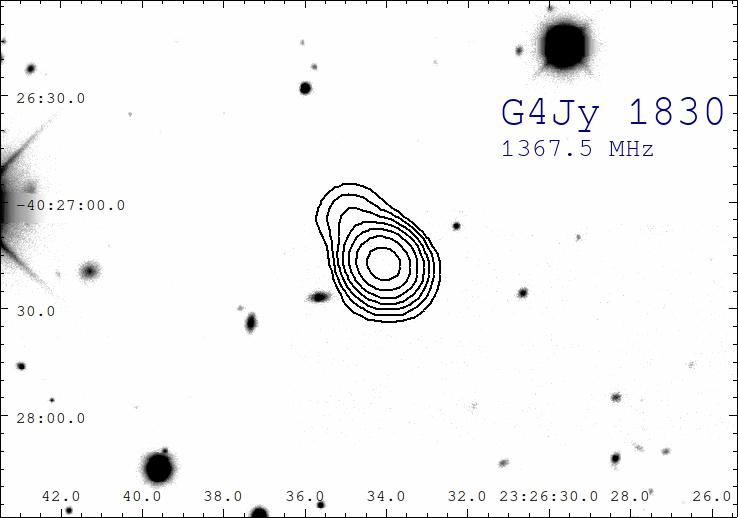}
    \includegraphics[scale=0.225]{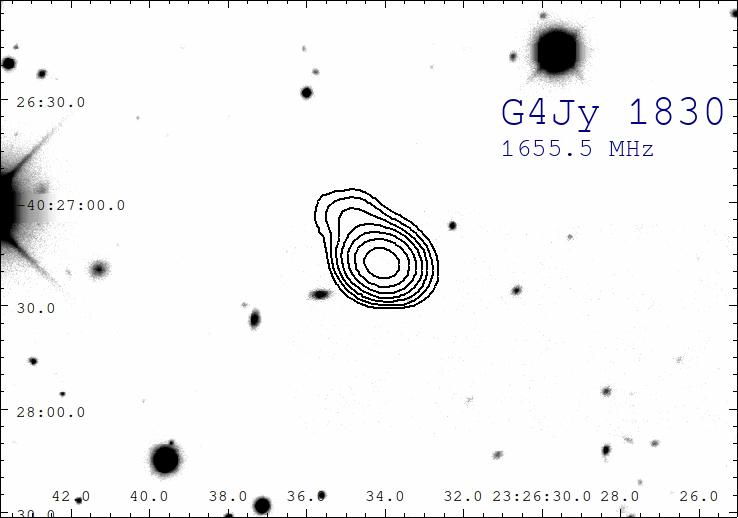}
    \includegraphics[scale=0.225]{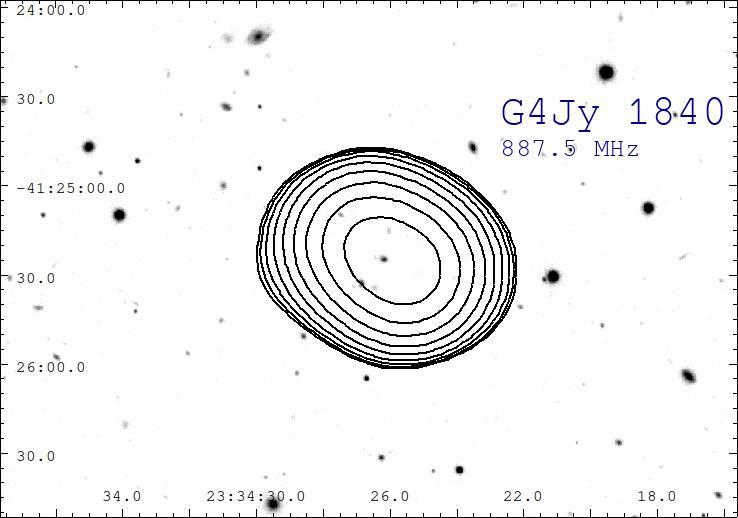}
    \includegraphics[scale=0.225]{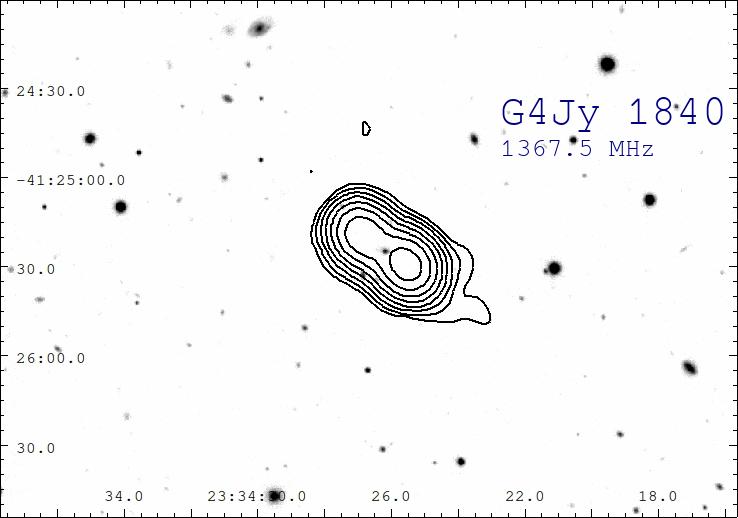}
    \includegraphics[scale=0.225]{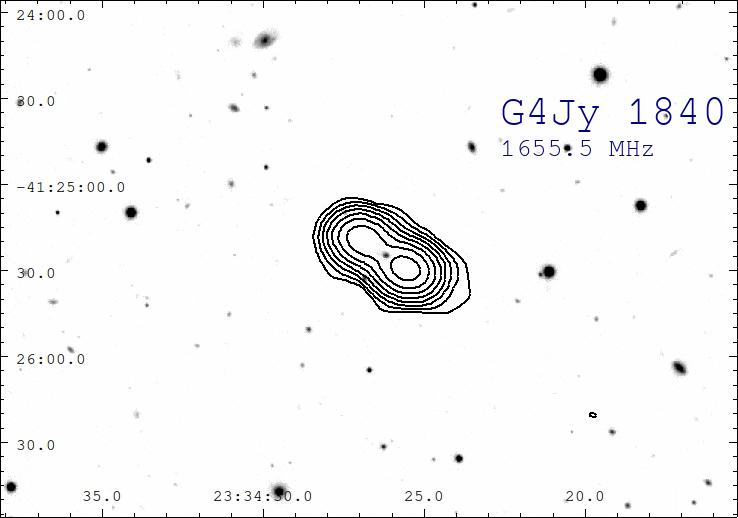}
    \includegraphics[scale=0.225]{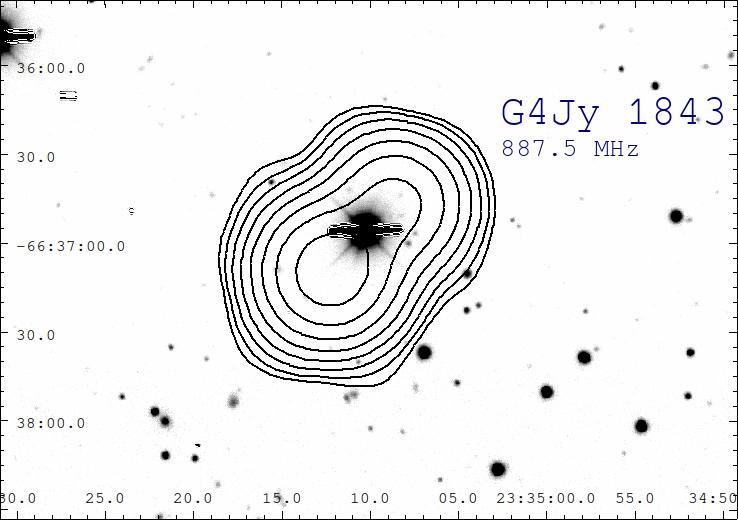}
    \includegraphics[scale=0.225]{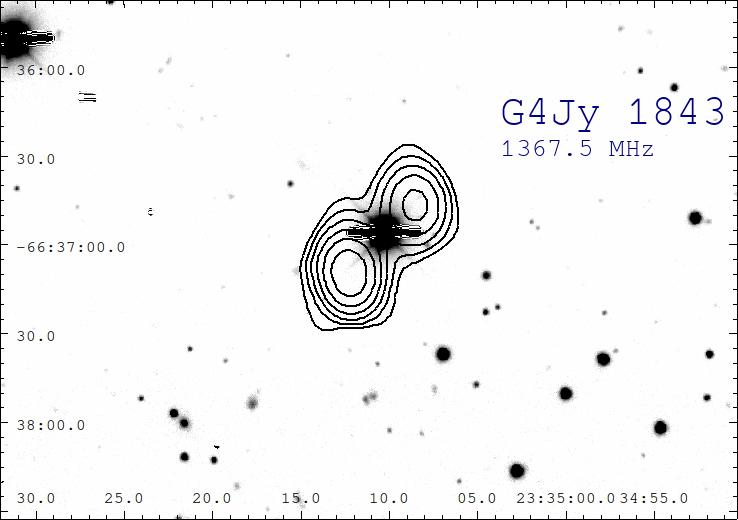}
    \includegraphics[scale=0.225]{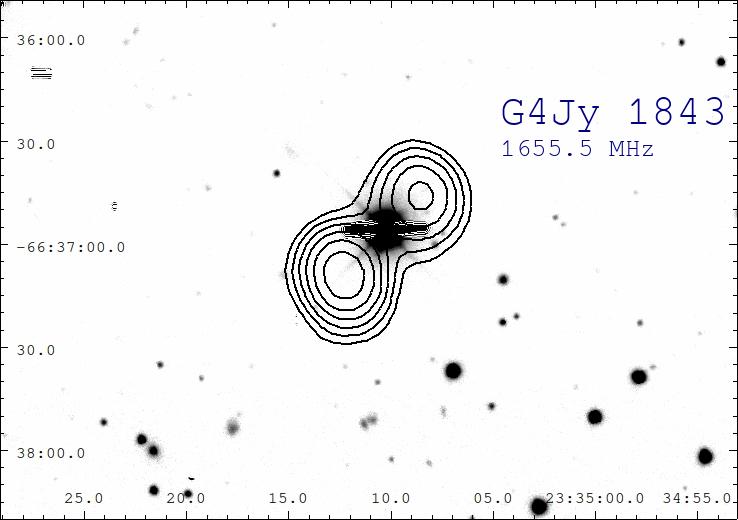}
    \includegraphics[scale=0.225]{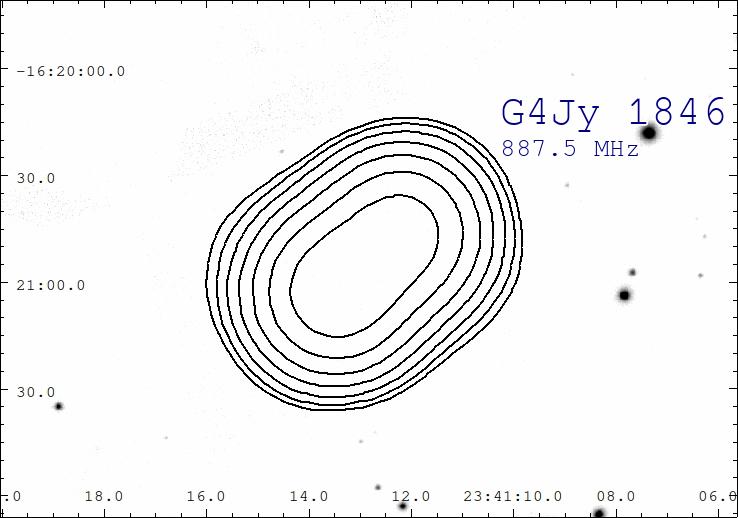}
    \includegraphics[scale=0.225]{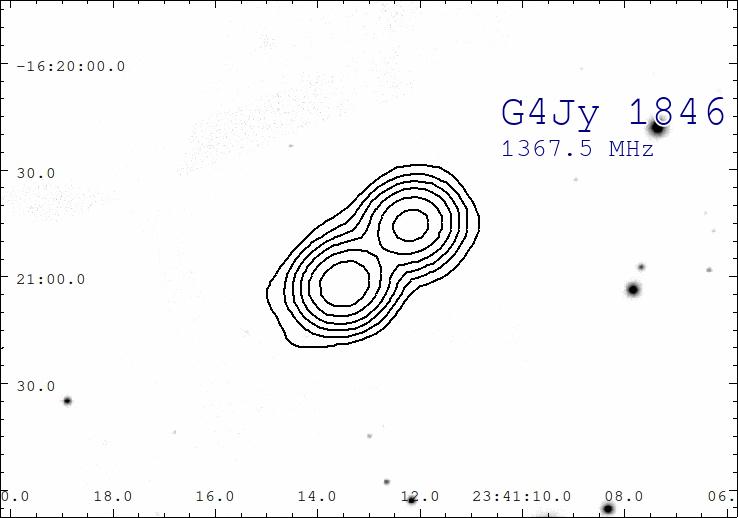}
    \includegraphics[scale=0.225]{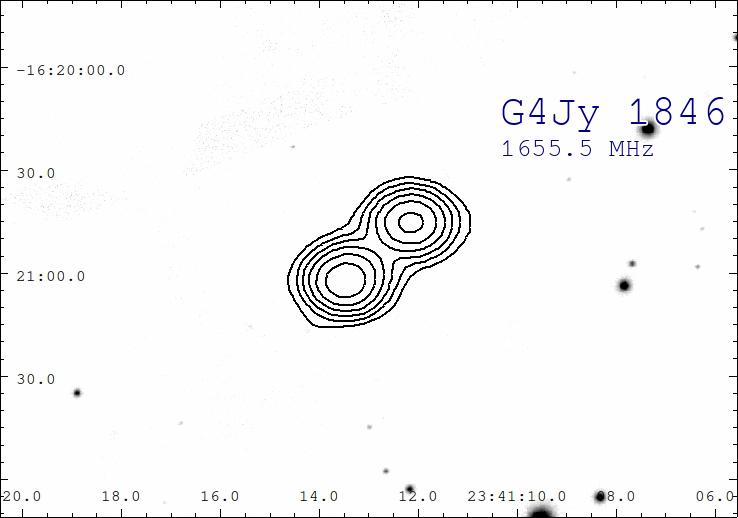}
    \includegraphics[scale=0.225]{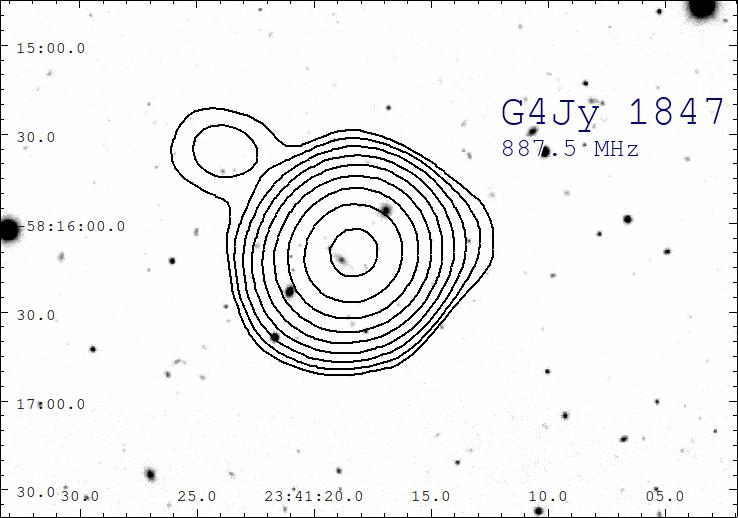}
    \includegraphics[scale=0.225]{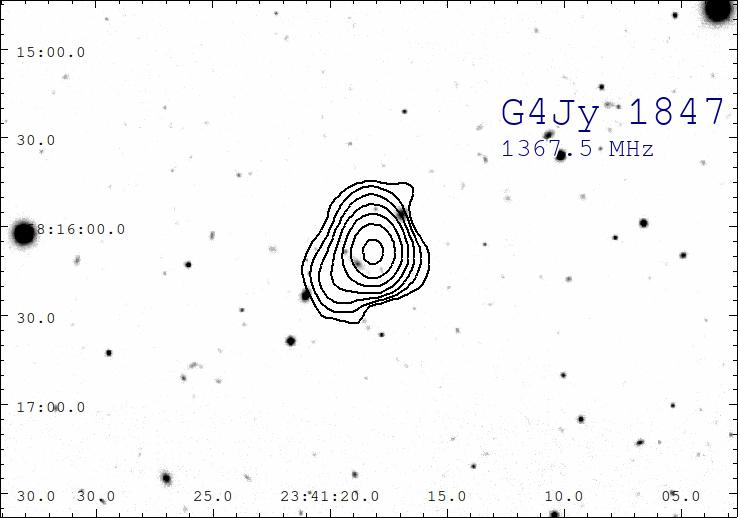}
    \includegraphics[scale=0.225]{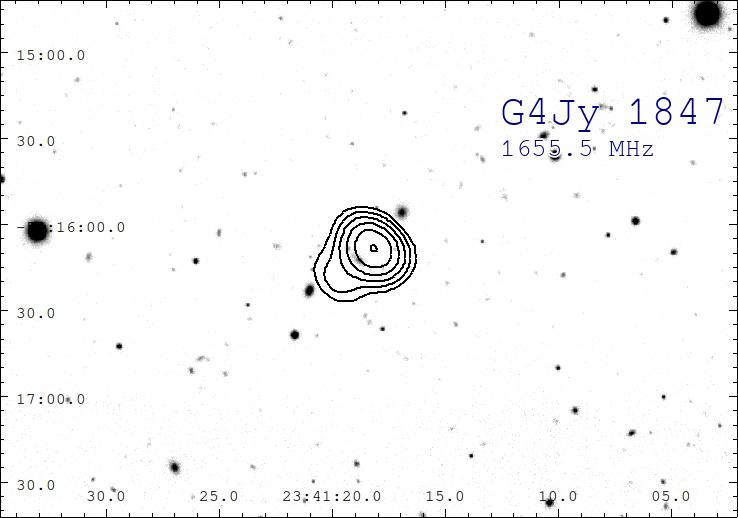}

    \caption{}
    \label{AY}
\end{figure*}
\clearpage

\begin{figure*}
    \centering
    \includegraphics[scale=0.225]{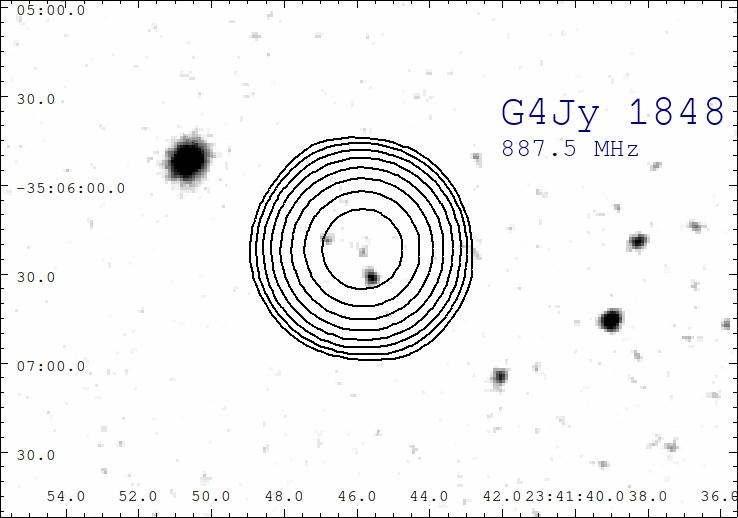}
    \includegraphics[scale=0.225]{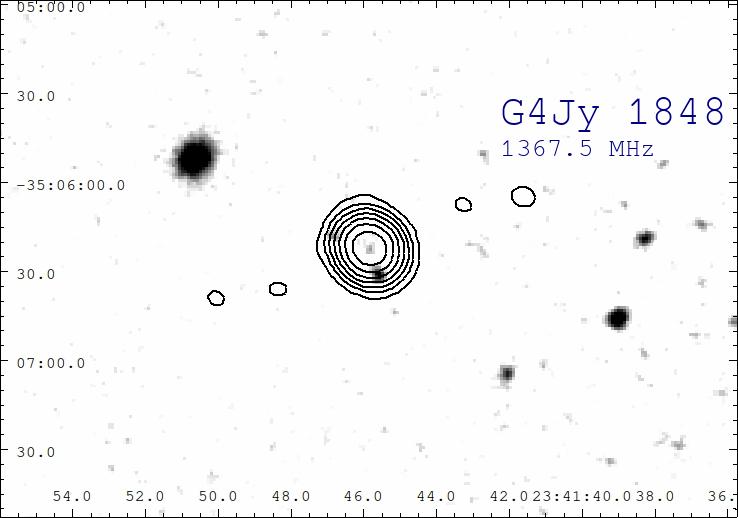}
    \includegraphics[scale=0.225]{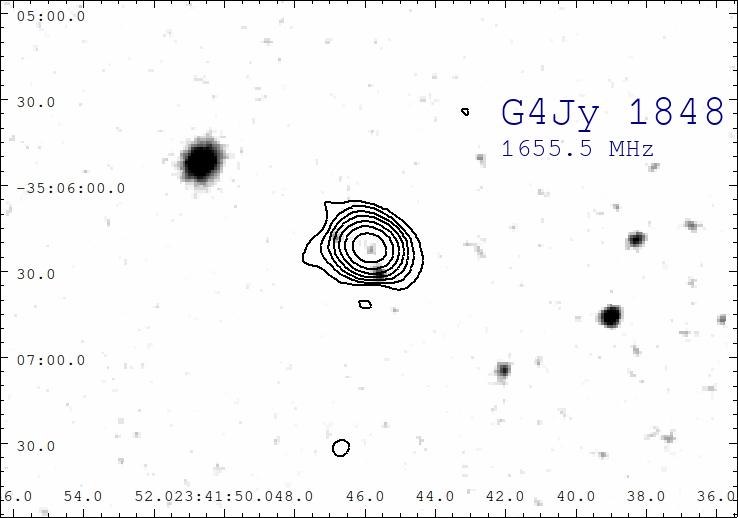}
    \includegraphics[scale=0.225]{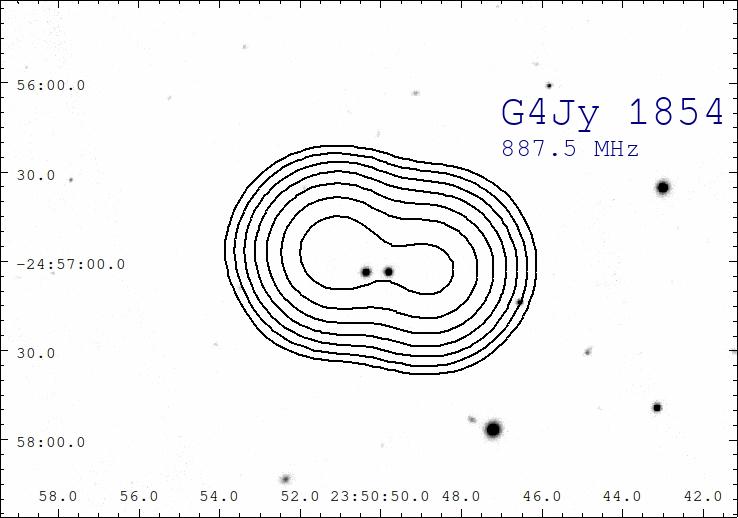}
    \includegraphics[scale=0.225]{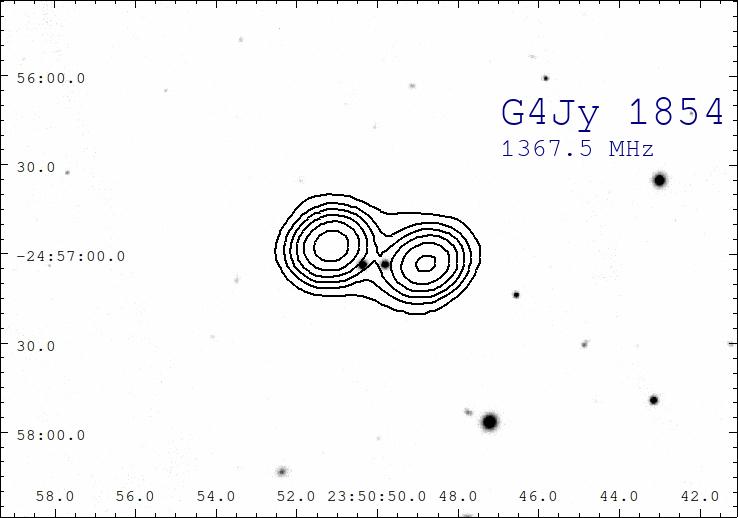}
    \includegraphics[scale=0.225]{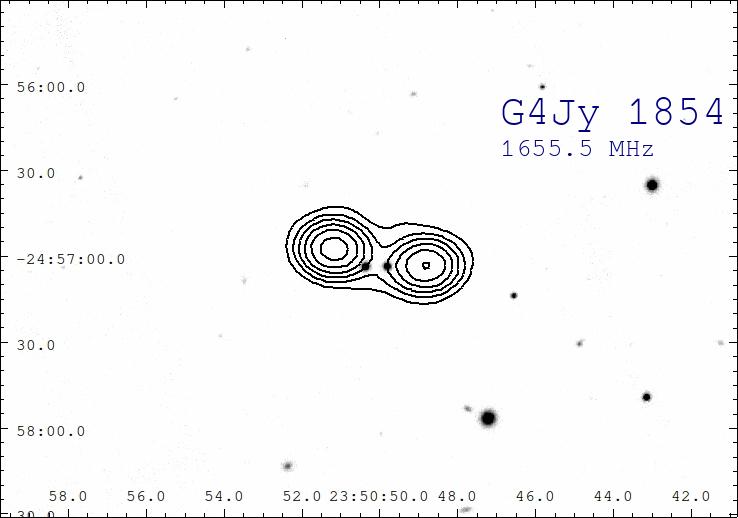}
    \includegraphics[scale=0.225]{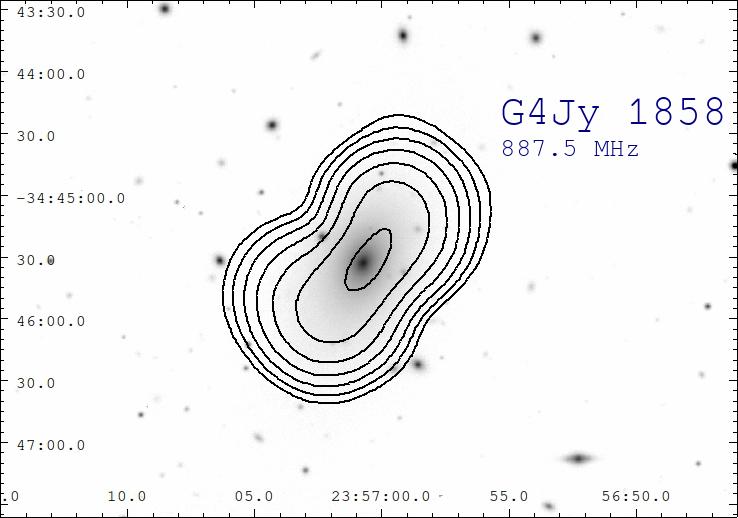}
    \includegraphics[scale=0.225]{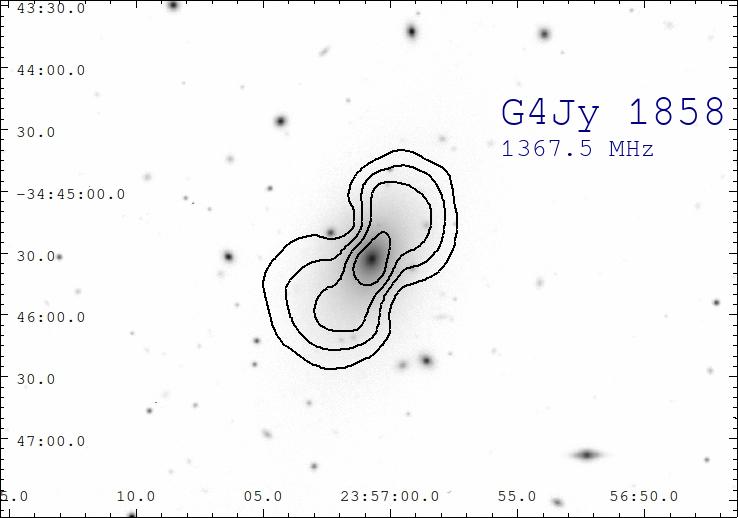}
    \includegraphics[scale=0.225]{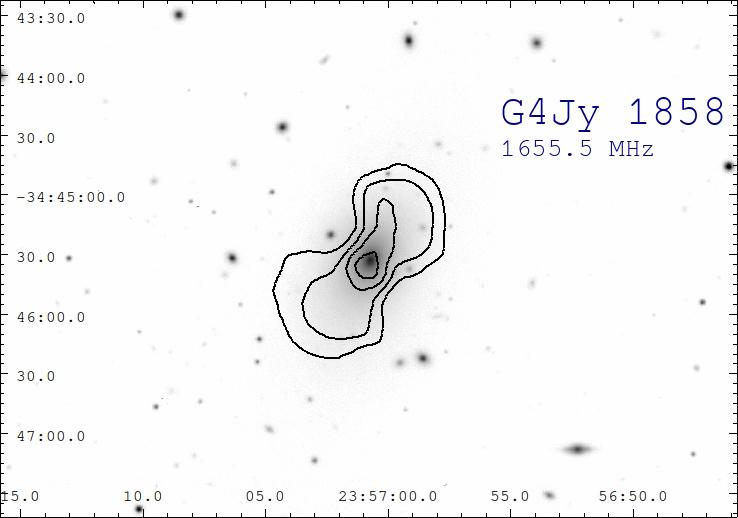}
    \includegraphics[scale=0.225]{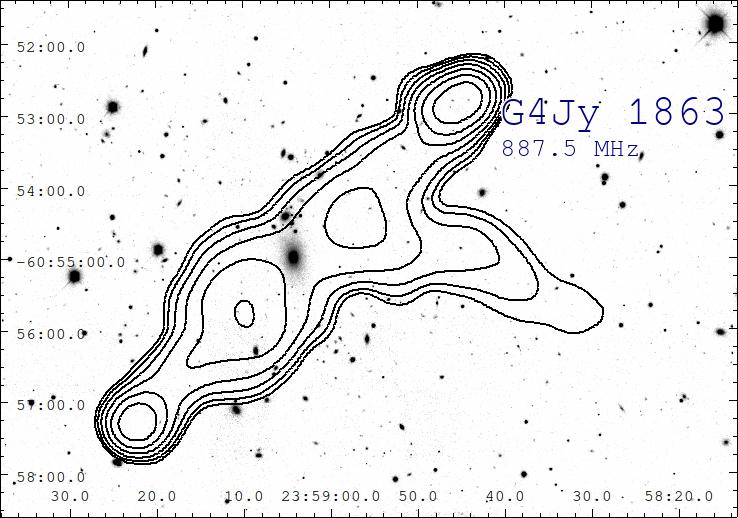}
    \includegraphics[scale=0.225]{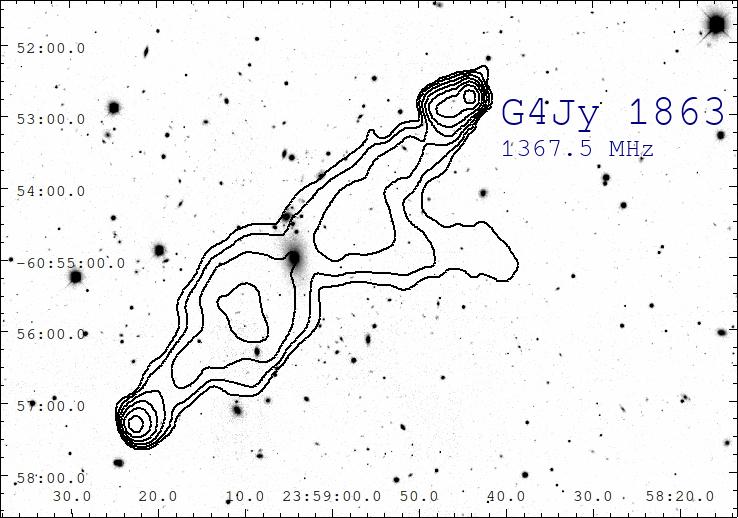}
    \includegraphics[scale=0.225]{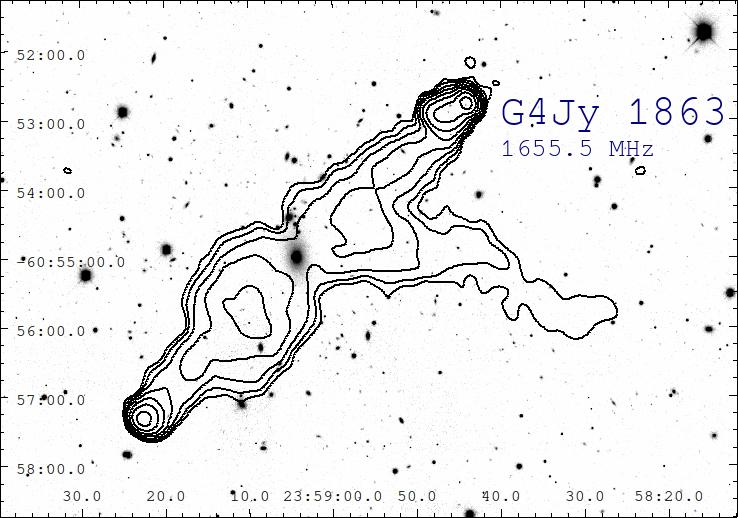}
    \caption{}
    \label{AZ}
\end{figure*}

\end{document}